\documentclass[graybox,envcountchap,sectrefs]{svmono}

\usepackage{mathptmx}
\usepackage{helvet}
\usepackage{courier}
\usepackage{type1cm}         

\usepackage{makeidx}         
\usepackage{graphicx}        
\usepackage{multicol}        
\usepackage[bottom]{footmisc}

\usepackage{amsmath}
\usepackage{amssymb,bm}
\usepackage{subfigure}
\usepackage{color} 

\usepackage{mathrsfs} 

\graphicspath{{./fig_crop/}}

\newcommand{\AmSLaTeX}{%
 $\mathcal A$\lower.4ex\hbox{$\!\mathcal M\!$}$\mathcal S$-\LaTeX}
\newcommand{\AmSLaTeXe}{%
 $\mathcal A$\lower.4ex\hbox{$\!\mathcal M\!$}$\mathcal S$-\LaTeXe}

\usepackage{macro_natsuume}
\def\nameofpaper{Springer} 

\newcommand{\margin}[1]{}

\newcommand{\head}[1]{\runinhead{#1}}
\renewcommand{\subsubsection}[1]{\runinhead[#1]}

\begin{document}

\author{Makoto Natsuume}
\title{AdS/CFT Duality User Guide}
\maketitle

\frontmatter

%
%

\preface

This book describes ``real-world" applications of the AdS/CFT duality for beginning graduate students in particle physics and for researchers in the other fields. 

The AdS/CFT duality is a powerful tool for analyzing strongly-coupled gauge theories using classical gravitational theories. The duality originated from string theory, so it has been actively investigated in particle physics. In recent years, however, the duality has been discussed beyond theoretical particle physics.  In fact, the original AdS/CFT paper by Maldacena has been cited in all physics arXivs. This is because the duality is becoming a powerful tool to analyze the ``real world." 
For example, it turns out that one prediction of AdS/CFT is indeed close to the experimental results of the real quark-gluon plasma. Since then, the  duality has been applied to various fields of physics; examples are QCD, nuclear physics, nonequilibrium physics, and condensed-matter physics.

In order to carry out such researches, one has to know many materials such as string theory, general relativity, nuclear physics, nonequilibrium physics, and condensed-matter physics. The aim of this book is to provide these background materials as well as some key applications of the AdS/CFT duality in a single volume. The emphasis throughout the book is on a pedagogical and intuitive approach focusing on the underlying physical concepts. Yet it also includes step-by-step computations for important results which are useful for beginners. Most of them are contained in the appendices.

Following conventions of many textbooks, I often do not refer to original research papers and refer only to the other textbooks and reviews that may be more useful to readers. Also, the choice of references reflects my knowledge, and I apologize in advance for possible omissions.

Initially, this project was begun for a book that was published in Japanese (Saiensu-sha Co., Ltd, 2012), and this is the ``translated" one. But I used this opportunity to improve many explanations and to add more materials to the Japanese edition. So, this book is the ``second edition" in this sense. 

I would like to thank many people who helped me with this book. This book is based on review talks at various conferences and on 
courses I taught at various graduate schools (Tohoku University, Ochanomizu University, Graduate University of Advanced Studies, and Rikkyo University). I thank the organizers and the participants of the conferences and the courses. I also like to thank Elena C\'{a}ceres, Koji Hashimoto, Tetsuo Hatsuda, Gary Horowitz, and Joe Polchinski for encouraging me to write this English edition. 
I  also thank Tetsufumi Hirano, Akihiro Ishibashi, and Takeshi Morita for useful comments and discussion.
I especially would like to thank Takashi Okamura who clarified my understanding of the subjects in this book through collaboration for many years. He also gave many suggestions for improvement. I thank my editor Hisako Niko at Springer Japan and the staff of the Lecture Notes in Physics. 
Of course, the responsibility for any remaining mistake is solely mine. I will be happy to receive comments on this book. Please send them to \href{mailto:makoto.natsuume@icloud.com}{makoto.natsuume@icloud.com}. 

An updated list of corrections will be posted on my website. The tentative address is \url{http://research.kek.jp/people/natsuume/ads-real-world.html}. Even if the address changes in future, you can probably easily search the website because my family name is rather rare. 

I hope that this book will help readers to explore new applications of the AdS/CFT duality.

\newcommand{\prefacearXiv}[1][\prefacearXivname]{\chapter*{#1}\markboth{#1}{#1}}
\def\prefacearXivname{Preface for the arXiv edition}%

\prefacearXiv

This arXiv manuscript is the updated version of Lecture Notes in Physics 903 published from Springer (\url{http://link.springer.com/book/10.1007/978-4-431-55441-7}); originally arXiv editions (v1-3) were  draft versions. 

How this manuscript differs from the published version:
\begin{itemize}
\item
Omitted: \sect{others_details} (``Appendix: Explicit form of other AdS spacetimes") It is not allowed to upload the entire draft.
\item
Changes after publication (v4)
   \begin{itemize}
   \item \sect{YM} added (``Appendix: Review of gauge theory")
   \item \chap{exercise} added (``Exercises")
   \item Minor improvements, \eg,
      \begin{itemize}
      \item list of acronyms and symbols
      \item \sect{textbooks}: prerequisites on general relativity and quantum field theory, textbooks on quantum field theory.
       \end{itemize}
   \end{itemize}
\end{itemize}



\tableofcontents

%
%

\extrachap{Acronyms \& Symbols}

\begin{description}[CABR]
\item[AdS]{anti-de~Sitter (spacetime)}
\item[AdS$_5$]{five-dimensional anti-de~Sitter (spacetime)}
\item[BF]{Breitenlohner-Freedman (bound)}
\item[CFT]{conformal field theory}
\item[dS]{de~Sitter (spacetime)}
\item[GKP]{Gubser-Klebanov-Polyakov}
\item[GL]{Ginzburg-Landau (theory)}
\item[IR]{infrared}
\item[LHC]{Large Hadron Collider}
\item[QCD]{quantum chromodynamics}
\item[QGP]{quark-gluon plasma}
\item[RHIC]{Relativistic Heavy Ion Collider}
\item[RN]{Reissner-Nordstr\"{o}m (black hole)}
\item[SAdS]{Schwarzschild-AdS (black hole)}
\item[SYM]{super-Yang-Mills (theory)}
\item[TDGL]{time-dependent Ginzburg-Landau (equation)}
\item[UV]{ultraviolet}
\end{description}

\begin{description}[CABR]
\item[$:=$]{left-hand side defined by right-hand side}
\item[$=:$]{right-hand side defined by left-hand side}
\item[$\approx$]{approximately equal (numerically)}
\item[$\simeq$]{approximately equal,\\equal up to an overall coefficient}
\item[$\sim$]{equal up to linear order,\\ equality valid asymptotically}
\item[$\xrightarrow{u\rightarrow0}{}$]{expression in the $u\to 0$ limit (at the AdS boundary)}
\end{description}

\mainmatter
\ifx\nameofpaper\undefined 
  \usepackage{macro_natsuume} 
  \def\beginsection{\section*}
  \def\endofsection{\end{document}} 
  \input draft_header.tex
\else 
  \def\beginsection{\chapter}
  \def\endofsection{ } 
\fi

\beginsection{Introduction}


\section{Overview of AdS/CFT}

In this chapter, we shall describe an overall picture of the AdS/CFT duality. Many terms appear below, but we will explain each in later chapters, so readers should not worry about them too much for the time being. 

The AdS/CFT duality is an idea that originated from \keyword{superstring theory}. Superstring theory is the prime candidate of the unified theory which unify four fundamental forces in nature, namely gravity, the electromagnetic force, the weak force, and the strong force (\chap{string}). 


Roughly speaking, the AdS/CFT duality claims the following equivalence between two theories:
\be
\fbox{
\begin{tabular}{l}
Strongly-coupled 4-dimensional gauge theory \\
$=$
Gravitational theory in 5-dimensional AdS spacetime
\end{tabular}
}
\label{eq:statement1}
\ee
%
%
AdS/CFT claims that four-dimensional physics is related to five-dimensional physics. In this sense, AdS/CFT is often called a \keyword{holographic theory}. An optical hologram encodes a three-dimensional image on a two-dimensional object. Similarly, a holographic theory encodes a five-dimensional theory by a four-dimensional theory. 

The gauge theory (on the first line) describes all forces except gravity, namely the electromagnetic force, the weak force, and the strong force. For example, the electromagnetic force is described by a $U(1)$ gauge theory, and the strong force is described by a $SU(3)$ gauge theory which is known as 
\keyword{quantum chromodynamics}, or QCD (\sect{QCD}). The theoretical foundation behind these three forces is understood as gauge theory, but it is not an easy task to compute a gauge theory at strong coupling. So, when the strong force is literally strong, we do not understand the strong force well enough. The AdS/CFT duality claims that one can analyze a strongly-coupled gauge theory using a curved spacetime, namely the AdS spacetime.

The \keyword{AdS spacetime} (on the second line) stands for the \keyword{anti-de~Sitter spacetime} (\chap{AdS}). A sphere is a space with constant positive curvature. In contrast, the AdS spacetime is a \textit{spacetime} with constant \textit{negative} curvature. De~Sitter was a Dutch astronomer who in 1917 found a solution of the Einstein equation with a constant \textit{positive} curvature (de~Sitter spacetime). The AdS spacetime instead has a constant \textit{negative} curvature; this explains the prefix ``anti." The AdS spacetime has a natural notion of a spatial boundary (\keyword{AdS boundary}). The gauge theory is supposed to ``live" on the four-dimensional boundary%
\footnote{The gauge theory is often called the ``boundary theory" whereas the gravitational theory is called the ``bulk theory."}.

Typically, a \keyword{duality} states the equivalence between two theories that look different at first glance. In the AdS/CFT duality, the gauge theory and the gravitational theory look different; even their spacetime dimensions differ. However, in a duality, one theory is strongly-coupled when the other theory is weakly-coupled%
\footnote{Normally, one would not use the word  ``weakly-coupled gravity," but this means that the spacetime curvature is small. The gravitational theory satisfies this condition when the gauge theory is strongly-coupled. Conversely, when the gravitational theory is strongly-coupled in the above sense, the gauge theory is weakly-coupled.}.
This has two consequences:
\begin{itemize}
\item The strong/weak-coupling relation suggests why two superficially different theories can be ever equivalent under the duality. When one theory (\eg, gauge theory) is strongly-coupled, it may not be appropriate to use its weakly-coupled variables, gauge fields. Rather, it may be more appropriate to use different variables. The duality claims that the appropriate variables are the weakly-coupled variables of the gravitational theory, gravitational fields. 
\item Because the duality relates two different theories, it is conceptually interesting, but it is practically important as well. Even if the gauge theory is strongly-coupled, one can use the weakly-coupled gravitational theory instead, which makes analysis much easier. 
\end{itemize}

The above relation corresponds to the case at zero temperature. At finite temperature, it is replaced by
\be
\fbox{
\begin{tabular}{l}
Strongly-coupled gauge theory at finite temperature \\
$=$
Gravitational theory in AdS black hole
\end{tabular}
}
\label{eq:statement2}
\ee
In the gravitational theory, a \keyword{black hole} appears since a black hole is also a thermal system. A black hole has a notion of temperature because of the \keyword{Hawking radiation} (\chap{BH_thermodynamics}). \textit{The aim of this book is to analyze nonequilibrium phenomena using the finite-temperature AdS/CFT. }

Using the black hole, one can get a glimpse of holography, namely why a five-dimensional gravitational theory  corresponds to a four-dimensional field theory. As a finite-temperature system, a black hole has the notion of the entropy (\chap{BH_thermodynamics}), but the black hole entropy is proportional to the ``area" of the \textit{black hole horizon} (\chap{BH}).\index{horizon} This behavior is very different from the usual statistical entropy which is proportional to the ``volume" of the system. But an ``area" in five dimensions is a ``volume" in four dimensions. This implies that if a black hole can be ever described by a four-dimensional field theory, the black hole must live in five-dimensional spacetime.

Rewriting relations \eqref{eq:statement1} and \eqref{eq:statement2} more explicitly, AdS/CFT claims that generating functionals (or partition functions) of two theories are equivalent:
\be
\boxeq{
Z_\text{gauge} = Z_\text{AdS}~,
}
\label{eq:GKPW_intro}
\ee
where $Z_\text{gauge}$ is the generating functional of a gauge theory, and $Z_\text{AdS}$ is the generating functional of a gravitational theory. In brief, this book discusses what this relation means and discusses what kinds of physical quantities one can compute from the relation.

AdS/CFT enables one to analyze a strongly-coupled gauge theory using the AdS spacetime. However, there are several important differences between the realistic $SU(3)$ gauge theory and the gauge theory studied in AdS/CFT. First, AdS/CFT typically considers a $SU(N_c)$ gauge theory. In such a theory, $N_c$ plays a role of a parameter, and the ``strong coupling" is the so-called \keyword{large-$N_c$ limit}, where one tunes $N_c$ in an appropriate way (\sect{large_N}).

Second, AdS/CFT typically considers a \keyword{supersymmetric gauge theory} which has supersymmetry. In particular, the \textit{${\cal N}=4$ super-Yang-Mills theory}\index{N=4 super-Yang-Mills theory@${\cal N}=4$ super-Yang-Mills theory} (SYM) provides the simplest example of AdS/CFT (\sect{scale_inv}). Here, ${\cal N}=4$ denotes the number of supersymmetry the theory has%
\footnote{Note that ${\cal N}$ denotes the number of supersymmetry whereas $N_c$ denotes the number of ``colors" in the $SU(N_c)$ gauge theory.}. 
The theory also has the scale invariance since the theory has no dimensionful parameter. Furthermore, the theory has a larger symmetry known as the conformal invariance which contains the Poincar\'{e} invariance and the scale invariance. Such a theory is in general known as a \keyword{conformal field theory}, or \keyword{CFT}. This is the reason why the duality is called AdS/CFT.

However, the use of AdS/CFT is not limited to CFTs. One can discuss various theories with dimensionful parameters, so they lack the scale invariance.
Thus, the name ``AdS/CFT" is not very appropriate and has only the historical meaning%
\footnote{
Since the name AdS/CFT is not really appropriate, various alternative names have been proposed. The name ``holographic theory" is one of them. The other strong candidate is the ``gauge/gravity duality," but this name becomes inconvenient since one started to apply AdS/CFT not only to QCD but to field theory in general. One may see phrases such as ``bulk/boundary duality" or ``field theory/gravity duality." The name ``gauge/string duality" is adopted in the PACS (Physics and Astronomy Classification Scheme) code by American Institute of Physics. This name is also appropriate since the dual of a gauge theory is really a string theory as we will see later. Anyway, they all mean the same. In this book, we use the most common name ``AdS/CFT duality" without making a new name since it is not clear at this moment how far AdS/CFT will be applied.  A digression: AdS/CFT keeps having troubles with name from the beginning. It was originally called the ``Maldacena's conjecture." But it is common that a duality is hard to prove and first starts as a conjecture, so the name became obsolete. }.
When one chooses a particular gauge theory on the left-hand side of \eq{GKPW_intro}, one has to choose an appropriate spacetime on the right-hand side (\sect{others}). But one typically considers spacetime which approaches the AdS spacetime at radial infinity. 

The gauge theories analyzed by AdS/CFT are not very realistic. However, this is typical of any analytic method. One often encounters strongly-coupled problems in field theory, but only a few analytic methods and exactly solvable models are available. But those examples played vital roles to develop our intuitions on field theory. Such techniques are valuable and brought important progress to field theory even if they only apply to a restricted class of theories. One would regard AdS/CFT as another example of such techniques. 

Finally, we should stress that AdS/CFT has not been proven and it remains a conjecture although there are many circumstantial evidences. We will not discuss these evidences much in this book.

\section{AdS/real-world}

\begin{figure}[tb]
\centering
\scalebox{0.75}{ \includegraphics{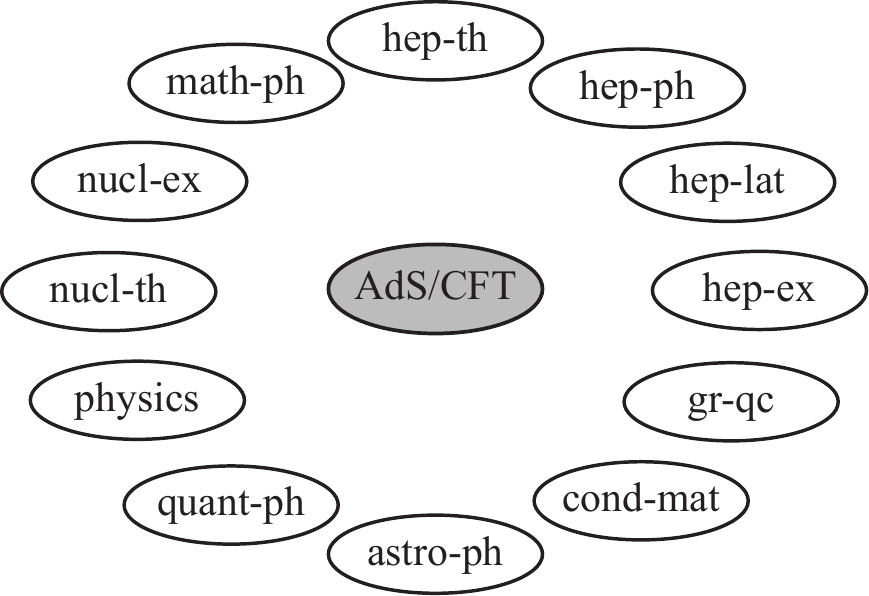} }
\vskip2mm
\caption{The AdS/CFT duality spans all physics arXivs.}
\label{fig:arXiv}
\end{figure}%

The AdS/CFT duality originated from string theory, so it had been discussed in string theory. But the situation is changing in recent years, and AdS/CFT has been discussed beyond theoretical particle physics. This is because AdS/CFT is becoming a powerful tool to analyze the ``real-world." Examples are QCD, nuclear physics, nonequilibrium physics, and condensed-matter physics. In fact, the original AdS/CFT paper \cite{Maldacena:1997re2} has been cited in \textit{all} physics arXivs (\fig{arXiv}). 

For example, the theoretical foundation behind the strong force has been well-understood as QCD. But the perturbation theory often fails because the strong force is literally strong. However, according to AdS/CFT, one can analyze a strongly-coupled gauge theory using the AdS spacetime. So, there are many attempts to analyze the strong force using AdS/CFT%
\footnote{QCD is a $SU(3)$ gauge theory, but AdS/CFT typically considers a $SU(N_c)$ supersymmetric gauge theory, so one should note that AdS/CFT gives only approximate results.}.

One such example is the \keyword{quark-gluon plasma} (QGP). According to QCD, the fundamental degrees of freedom are not protons or neutrons but \keyword{quarks} and \keyword{gluons}. Under normal circumstances, they are confined inside protons and neutrons. But at high enough temperatures, they are \keyword{deconfined} and form the quark-gluon plasma (\sect{QCD_phase}). The QGP experiments are in progress (\sect{heavy_ion}). According to the experiments, QGP behaves like a fluid with a very small \keyword{shear viscosity}. This implies that QGP is strongly coupled, which makes theoretical analysis difficult (Sects.~\ref{sec:unexpected} and \ref{sec:QGP}). However, it turns out that the value of the viscosity implied by the experiments is very close to the value predicted by AdS/CFT using black holes (\chap{QGP}). This triggers the AdS/CFT research beyond string theory community. 

\begin{figure}[tb]
\centering
\includegraphics{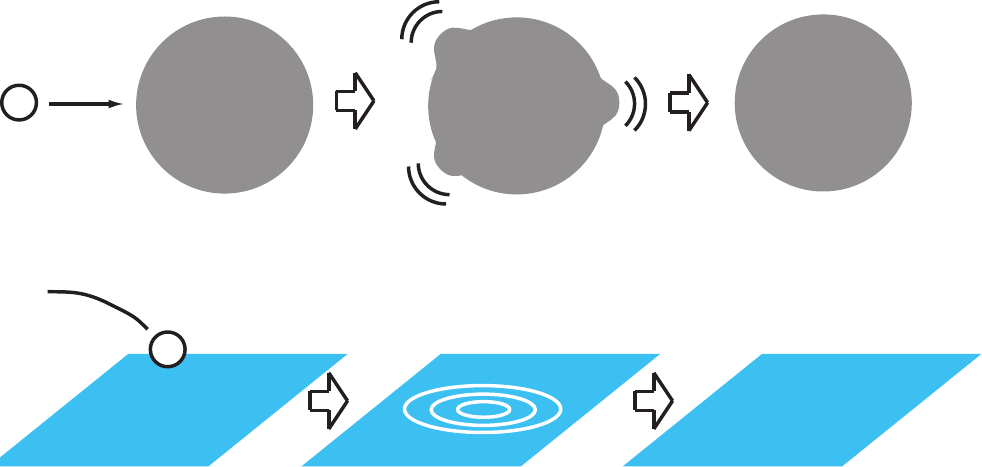}
\vskip2mm
\caption{When one adds perturbations, a black hole behaves like a hydrodynamical system. In hydrodynamics, the dissipation is a consequence of viscosity.}
\label{fig:hydro}
\end{figure}%

How is the black hole related to the viscosity? Here, we give an intuitive explanation (\fig{hydro}). Consider adding a perturbation to a thermal system which is in equilibrium. For example, drop a ball in a water pond. Then, surface waves are generated, but they decay quickly, and the water pond returns to a state of stable equilibrium. This is a \keyword{dissipation} which is a consequence of viscosity. 

This behavior is very similar to a black hole. Again, drop an object to a black hole. Then, the shape of the black hole horizon becomes irregular, but such a perturbation decays quickly, and the black hole returns to the original symmetric shape. If one regards this as a dissipation as well, the dissipation occurs since the perturbation is absorbed by the black hole. Thus, one can consider the notion of viscosity for black holes as well, and the ``viscosity" for black holes should be calculable from the above process. 


Such a phenomenon is in general known as a \keyword{relaxation phenomenon}. In a relaxation phenomenon, one adds a perturbation and sees how it decays. The relaxation phenomenon is the subject of \keyword{nonequilibrium statistical mechanics} or \keyword{hydrodynamics}. The important quantities there are \keyword{transport coefficients}. The viscosity is an example of transport coefficients. A transport coefficient measures how some effect propagates. The correspondence between black holes and hydrodynamics may sound just an analogy, but one can indeed regard that black holes have a very small viscosity; one purpose of this book is to show this.

The AdS/CFT applications are not limited to QCD. Strongly-coupled systems often arise in condensed-matter physics such as high-$T_c$ superconductivity (\sect{Hi-Tc}). Partly inspired by the ``success" of AdS/QGP, researchers try to apply AdS/CFT to condensed-matter physics (\chap{phase}).

As one can see, the applications of AdS/CFT has the ``cross-cultural" character, so researchers in other fields often initiated new applications. For example, the applications of AdS/CFT to the quark-gluon plasma were initiated by nuclear physicists. As another example, the Fermi surface was first discussed in AdS/CFT by a condensed-matter physicist.

\section{Outline}

This book describes applications of AdS/CFT, but it is not our purpose to cover \textit{all} applications of AdS/CFT. This is because so many applications exit and new applications have been proposed very often. We would rather explain the basic idea than cover all applications. Then, as examples, we discuss  following applications:
\begin{itemize}
\item \chap{wilson}: Wilson loops, or quark potentials
\item \chap{QGP}: application to QGP (transport coefficients)
\item \sect{2nd_order}: application to hydrodynamics (second-order hydrodynamics)
\item \sect{H^3}: \HSC
\end{itemize}
But once one gets accustomed to the basic idea, it is not very difficult to apply AdS/CFT to various systems. Essentially what one should do is to repeat a similar exercise.

This book assumes knowledge on elementary general relativity and elementary field theory but does not assume knowledge on black holes and string theory. Also, AdS/CFT has been applied to many different areas of physics, so we explain basics of each area: 
\begin{itemize}
\item Chap.~\ref{chap:BH} and \ref{chap:BH_thermodynamics}: black holes, black hole thermodynamics
\item \chap{QCD}: quantum chromodynamics
\item \chap{string}: superstring theory
\item \chap{hydro}: nonequilibrium statistical mechanics, hydrodynamics
\item \chap{GL}: condensed-matter physics
\end{itemize}
The readers with enough backgrounds may skip some of these chapters.

All chapters end with a list of keywords, and most chapters have a summary. After you read each chapter, you could check your understanding by trying to explain those keywords.
Also, there are exercises in \chap{exercise}. Some are easy, and their aim is to check your understanding. Some are more involved, and their aim is to delve into discussion in the text. Some exercises need knowledge of advanced sections (Sections with ``~\advanced~") as prerequisite, but in such a case, we mention it explicitly. 

This book devotes some pages to explain how one reaches AdS/CFT. But some readers may first want to get accustomed to AdS/CFT through actual computations. In such a case, one would skip \sect{large_N} and \chap{string}.

Sections and footnotes with ``~\advanced~" are somewhat advanced topics and may be omitted in a first reading.

\section{Notation and conventions}\label{sec:notations}

We use the natural units $\hbar=c=1$. We often set the Boltzmann constant $k_\text{B} =1$ for thermodynamic analysis. We restored these constants in some sections though.

We use the metric signature $(-,+, \ldots, +)$, which is standard in general relativity and in string theory. We follow Ref.~\cite{MTW} for the quantities made from the metric, such as the Christoffel symbols and curvature tensors. 
For vector and tensor components, we use Greek indices $\mu, \nu, \ldots$ following the standard convention in general relativity \textit{until \sect{reexamination}.} However, in this book, we have the four-dimensional spacetime where a gauge theory lives and the five-dimensional spacetime where a gravitational theory lives, and we have to distinguish which spacetime dimensions we are talking of. Starting from \sect{reexamination},
\begin{itemize}
\item
Greek indices $\mu, \nu, \ldots$ run though $0,\ldots,3$ and are used for \textit{the four-dimensional spacetime where a gauge theory lives}. We write boundary coordinates in several ways: $x=x^\mu=(t,\bmx)=(t,x,y,z)$.

\item
Capital Latin indices $M, N, \ldots$ run though $0,\ldots,4$ and are used for \textit{the five-dimensional spacetime where a gravitational theory lives}.
\end{itemize}

In field theory, both the Lorentzian formalism and the Euclidean formalism exit. AdS/CFT has both formalisms as well. The Euclidean formalism is useful to discuss an equilibrium state whereas the Lorentzian formalism is useful to discuss a nonequilibrium state, so we use both depending on the context, and we try to be careful so that readers are not confused.

\section{Some useful textbooks and prerequisites}\label{sec:textbooks}

For a review article on AdS/CFT, see Ref.~\cite{Aharony:2000ti}. This is a review written in early days of the AdS/CFT research, but this is still the best review available.

This book explains the minimum amount of string theory. If one would like to learn more details, see Refs.~\cite{Green:1987sp,Polchinski:1998rq,Zwiebach:2004tj}. 
Reference~\cite{Green:1987sp} is the standard textbook till mid 1990s. Reference~\cite{Polchinski:1998rq} is the standard textbook since then. Reference~\cite{Green:1987sp} is a useful supplement to Ref.~\cite{Polchinski:1998rq} however since the former covers materials which are not covered by the latter. Anyhow, these textbooks are written to grow string theory experts. Advanced undergraduate students and researchers in the other fields may find Ref.~\cite{Zwiebach:2004tj} more accessible. 

The other string theory textbooks relatively recently are Refs.~\cite{Becker:2007zj,Kiritsis:2007zz,Johnson:2003gi}.
These textbooks in recent years cover AdS/CFT. 

This book explains basics of black holes but assumes basic concepts of general relativity such as metric, curvature, and the Einstein equation as a prerequisite. But, for completeness, we will review them in \chap{BH}. If this is not enough, read textbooks of general relativity \cite{Schutz:1985jx,Zee:2013dea,Wald:1984rg} (\eg, first 8 chapters of \cite{Schutz:1985jx}).
References~\cite{Schutz:1985jx,Zee:2013dea} are elementary textbooks (but Ref.~\cite{Zee:2013dea} is very modern with many advanced topics in recent years), and Ref.~\cite{Wald:1984rg} is an advanced one.

There are many good textbooks of quantum field theory, but we list only a few here in order of increasing difficulty \cite{Ryder:1985wq,Zee:2003mt,Srednicki:2007qs,Peskin:1995ev,Weinberg:1995mt}. It is advisable for readers to have a rough idea about basic concepts of quantum field theory (such as Feynman rules, renormalization, and $\beta$-function using a simple example like the $\phi^4$-theory) and the Abelian gauge theory.

We will mention textbooks in the other fields in appropriate places. 
Reading them are not prerequisites though. Those textbooks are mentioned if you would like to learn more details.

Finally, we list review articles and textbooks which cover applications of AdS/CFT \cite{Papantonopoulos:2011zz,Horowitz:2012nnc,DeWolfe:2013cua,Hartnoll:2009sz,McGreevy:2009xe,Iqbal:2011ae,CasalderreySolana:2011us,Ammon:2015wua,Zaanen:2015oix}. These cover the materials which are not covered in this book and are useful complements.

\endofsection

\ifx\nameofpaper\undefined 
  \usepackage{macro_natsuume} 
  \def\beginsection{\section*}
  \def\endofsection{\end{document}} 
  \input draft_header.tex
\else 
  \def\beginsection{\chapter}
  \def\endofsection{ } 
\fi

\beginsection{General relativity and black holes
}\label{chap:BH}


\begin{quote}
In this book, black holes frequently appear, so we will describe the simplest black hole, the Schwarzschild black hole and its physics.
\end{quote}

Roughly speaking, a black hole is a region of spacetime where gravity is strong so that even light cannot escape from there. The boundary of a black hole is called the \keyword{horizon}. Even light cannot escape from the horizon, so the horizon represents the boundary between the region which is causally connected to distant observers and the region which is not. 

General relativity is mandatory to understand black holes properly, but a black hole-like object can be imagined in Newtonian gravity. Launch a particle from the surface of a star, but the particle will return if the velocity is too small. In Newtonian gravity, the particle velocity must exceed the escape velocity in order to escape from the star. From the energy conservation, the escape velocity is determined by
\be
\frac{1}{2}v^2=\frac{GM}{r}~.
%
\ee
If the radius $r$ becomes smaller for a fixed star mass $M$, the gravitational potential becomes stronger, so the escape velocity becomes larger. When the radius becomes smaller, eventually the escape velocity reaches the speed of light. Then, no object can escape from the star. Setting $v=c$ in the above equation gives the radius
\begin{equation}
r = \frac{2GM}{c^2}~,
\label{eq:horizon}
\end{equation}
which corresponds to the horizon. For a solar mass black hole, the horizon radius is about 3 km, which is $2.4\times 10^5$ times smaller than the solar radius. 

To be precise, the above argument is false from several reasons:
\begin{enumerate}

\item
First, the speed of light is arbitrary in Newtonian mechanics. As a result, the speed of light decreases as light goes away from the star. But in special relativity the speed of light is the absolute velocity which is independent of observers. 
\item
Newtonian mechanics cannot determine how gravity affects light.
\item
In the above argument, light can temporally leave from the ``horizon." But in general relativity light cannot leave even temporally.
\end{enumerate}
The Newtonian argument has various problems, but the horizon radius \eqref{eq:horizon} itself remains true in general relativity, and we utilize Newtonian arguments again later. 

Below we explain black holes using general relativity, but we first discuss the particle motion in a given spacetime. For the flat spacetime, this is essentially a review of special relativity. We take this approach from the following reasons: (i) We study the particle motion around black holes later in order to understand black hole physics; (ii) The main purpose of this book is not to obtain a new geometry but to study the behavior of a ``probe" such as a particle in a known geometry; (iii) String theory is a natural extension of the particle case below.

\section{Particle action
}

\subsubsection*{Flat spacetime case --- review of special relativity
}\label{sec:particle_action_flat}

First, let us consider the particle motion in the flat spacetime. We denote the particle's coordinates as $x^\mu:=(t,x,y,z)$. According to special relativity, the distance which is invariant relativistically is given by
\be
ds^2 = - dt^2 + dx^2 + dy^2 + dz^2~.
%
\ee
The distance is called timelike when $ds^2<0$, spacelike when $ds^2>0$, and null when $ds^2=0$. For the particle, $ds^2<0$, so one can use the \keyword{proper time} $\tau$ given by
\be
ds^2 = - d\tau^2~.
%
\ee

The proper time gives the relativistically invariant quantity for the particle, so it is natural to use the proper time for the particle action:
\be
\action = - m \int d\tau~.
\label{eq:action_particle1}
\ee
The action takes the familiar form in the nonrelativistic limit. With the velocity $v^i := dx^i/dt$, $d\tau$ is written as $d\tau = dt (1-v^2)^{1/2}$, so
\be
\action = - m \int dt (1-v^2)^{1/2}
\simeq -m \int dt \left(1 - \frac{1}{2}v^2 + \cdots \right), \quad (v \ll 1)~.
%
\ee
In the final expression, the first term represents the particle's rest mass energy, and the second term represents the nonrelativistic kinetic energy. 

\begin{figure}[tb]
\begin{center}
\scalebox{0.75}{ \includegraphics{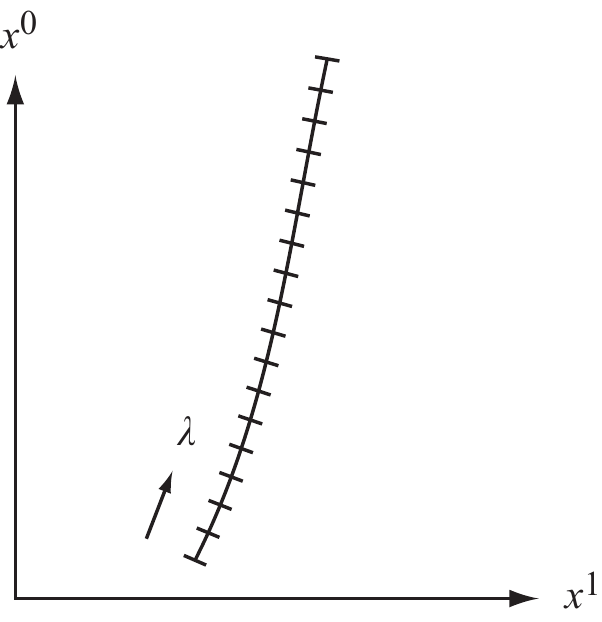} }
\vskip2mm
\caption{A particle draws a world-line in spacetime}
\label{fig:world_line}
\end{center}
\end{figure}

A particle draws a \keyword{world-line} in spacetime (\fig{world_line}). Introducing an arbitrary parametrization $\lambda$ along the world-line, the particle coordinates or the particle motion are described by $x^\mu(\lambda)$. Using the parametrization, 
\be
d\tau^2 
= - \eta_{\mu\nu} dx^\mu dx^\nu 
= - \eta_{\mu\nu} \dotx^\mu \dotx^\nu d\lambda^2 
\quad (\dot{~} := d/d\lambda)~,
%
\ee
so the action is written as
\be
\action = -m \int d\lambda\, \sqrt{ -\eta_{\mu\nu} \dotx^\mu \dotx^\nu } 
= \int d\lambda\, L~.
\label{eq:action_particle2}
\ee
The parametrization $\lambda$ is a redundant variable, so the action should not depend on $\lambda$. In fact, the action is invariant under
\be
\lambda' = \lambda' (\lambda)~.
\label{eq:reparametriz_particle}
\ee

The canonical momentum of the particle is given by
\be
p_\mu = \frac{\del L}{\del \dotx^\mu}
= \frac{ m \dotx_\mu }{ \sqrt{-\dotx^2} }
= m \frac{dx_{\mu}}{d\tau}
%
\ee
($\dotx^2:= \eta_{\mu\nu}\dotx^\mu\dotx^\nu$). Note that the canonical momentum satisfies
\be
p^2 = m^2 \frac{\dotx^2}{-\dotx^2} = -m^2~,
%
\ee
so its components are not independent:
\be
\boxeq{
p^2 = - m^2~.
}
\label{eq:constraint_on_shell}
\ee
The Lagrangian does not contain $x^\mu$ itself but contains only $\dotx^\mu$, so $p_\mu$ is conserved. Thus, $p_\mu = m\, dx_{\mu}/d\tau =\text{(constant)}$, which describes the free motion.

The particle's \keyword{four-velocity}\ $u^\mu$ is defined as
\be
u^\mu := \frac{dx^\mu}{d\tau}~.
%
\ee
In terms of the ordinary velocity $v^i$, 
\be
u^\mu = \frac{dt}{d\tau} \frac{dx^\mu}{dt} = \gamma(1,v^i)~, \quad
\left(\frac{d\tau}{dt}\right)^2 = 1-v^2 := \gamma^{-2}~.
%
\ee
Since $p_\mu = m u_\mu$ and $p^2=-m^2$, $u^\mu$ satisfies $u^2 =-1$.

The action \eqref{eq:action_particle1} is proportional to $m$, and one cannot use it for a massless particle. The action which is also valid for a massless particle is given by
\be
\action = \frac{1}{2}\int d\lambda\, \left\{ e^{-1}  \eta_{\mu\nu} \dotx^\mu \dotx^\nu - e m^2 \right\}~.
\label{eq:action_massless}
\ee
From this action,
\begin{alignat}{2}
&\text{Equation of motion for $e$:} 
& \quad & \dotx^2 + e^2 m^2 = 0~, 
\label{eq:eom_massless1} \\
&\text{Canonical momentum: } 
& \quad & p_\mu = \frac{\del L}{\del \dotx^\mu} = \frac{1}{e} \dotx_\mu 
= \frac{m \dotx_\mu}{ \sqrt{-\dotx^2} }~.
\label{eq:eom_massless2}
%
\end{alignat}
Use \eq{eom_massless1} at the last equality of \eq{eom_massless2}. Using \eq{eom_massless1}, the Lagrangian reduces to the previous one \eqref{eq:action_particle2}:
\be
\frac{1}{2} \left\{ e^{-1} \dotx^2 - e m^2 \right\}
= - m \sqrt{-\dotx^2}~.
%
\ee

This action also has the reparametrization invariance: the action \eqref{eq:action_massless} is invariant under
\begin{align}
\lambda' &= \lambda' (\lambda)~, \\
e' &= \frac{d\lambda}{d\lambda'} e~.
%
\end{align}

\subsubsection*{Particle action (curved spacetime)
}\label{sec:particle_action_curved}

Now, move from special relativity to general relativity. The invariant distance is given by replacing the flat metric $\eta_{\mu\nu}$ with a curved metric $g_{\mu\nu}(x)$:
\be
ds^2 = g_{\mu\nu} dx^\mu dx^\nu~.
\label{eq:line_element}
\ee
Here, we first consider the particle motion in a curved spacetime and postpone the discussion how one determines $g_{\mu\nu}$.

The action is obtained by replacing the flat metric $\eta_{\mu\nu}$ with a curved metric $g_{\mu\nu}$:
\be
\action = -m \int d\tau
= -m \int d\lambda\, \sqrt{ -g_{\mu\nu}(x) \dotx^\mu \dotx^\nu }~.
%
\ee
Just like the flat spacetime, the canonical momentum is given by
\be
p_\mu = m\, \frac{ g_{\mu\nu}(x)\dotx^\nu }{ \sqrt{-\dotx^2} }~, \quad
\dotx^2:= g_{\mu\nu}(x)\dotx^\mu\dotx^\nu~,
%
\ee
and the constraint $p^2 = -m^2$ exists. Also%
\footnote{
What is conserved is $p_\mu$ which may not coincide with $p^\mu$ in general. In the flat spacetime, $p_\mu$ and $p^\mu$ are the same up to the sign, but in the curved spacetime, the functional forms of $p_\mu$ and $p^\mu$ differ by the metric $g_{\mu\nu}(x)$. }, 
\begin{center}
\fbox{
\begin{tabular}{l}
If the metric is independent of  $x^\mu$, its conjugate momentum  $p_\mu$ is conserved.
\end{tabular}
}
\end{center}

The variational principle $\delta \action=0$ gives the world-line which extremizes the action. For the flat spacetime, the particle has the free motion and has the ``straight" world-line. For the curved spacetime, the world-line which extremizes the action is called a \keyword{geodesic}. The variation of the action with respect to $x^\mu$ gives the equation of motion for the particle: 
\be
\frac{d^2x^\mu}{d\tau^2} 
+ \Gamma^\mu_{~\rho\sigma} \frac{dx^\rho}{d\tau}\frac{dx^\sigma}{d\tau} = 0~.
\label{eq:geodesic_eq}
\ee
This is known as the \keyword{geodesic equation}%
\footnote{
Note that we use the proper time $\tau$ here not the arbitrary parametrization $\lambda$. The equation of motion does not take the form of the geodesic equation for a generic $\lambda$. A parameter such as $\tau$ is called an \keyword{affine parameter}. As one can see easily from the geodesic equation, the affine parameter is unique up to the linear transformation $\tau\rightarrow a\tau+b$ ($a, b$: constant). For the massless particle, the proper time cannot be defined, but the affine parameter is possible to define.
}.
Here, $\Gamma^\alpha_{~\mu\nu}$ is the Christoffel symbol: 
\be
\Gamma^\alpha_{~\mu\nu} = \frac{1}{2}\, g^{\alpha\beta} 
(\del_\nu g_{\beta\mu} + \del_\mu g_{\beta\nu} - \del_\beta g_{\mu\nu})~.
%
\ee
The particle motion is determined by solving the geodesic equation. However, black holes considered in this book have enough number of conserved quantities so that one does not need to solve the geodesic equation.

The massless particle action is also obtained by substituting $\eta_{\mu\nu}$ with $g_{\mu\nu}$ in  \eq{action_massless}. The particle action described here can be naturally extended into string and the objects called ``branes" in string theory (\sect{NG}). 

\section{Einstein equation and Schwarzschild metric
}

So far we have not specified the form of the metric, but the metric is determined by the Einstein equation%
\footnote{In \sect{GR}, we summarize the formalism of general relativity for the readers who are not familiar to it.}:
\be
R_{\mu\nu} - \frac{1}{2} g_{\mu\nu} R = 8\pi G\, T_{\mu\nu}~,
%
\ee
where $G$ is the Newton's constant, and $T_{\mu\nu}$ is the energy-momentum tensor of matter fields. The Einstein equation claims that the spacetime curvature is determined by the energy-momentum tensor of matter fields. 

We will encounter various matter fields. Of prime importance in this book is
\be
T_{\mu\nu} = - \frac{\Lambda}{8\pi G} g_{\mu\nu}~,
\label{eq:EM_tensor_cc}
\ee
where $\Lambda$ is called the \keyword{cosmological constant}. In this case, the Einstein equation becomes
\be
R_{\mu\nu} - \frac{1}{2} g_{\mu\nu} R + \Lambda g_{\mu\nu} = 0~.
\label{eq:Einstein_cc}
\ee
From \eq{EM_tensor_cc}, the cosmological constant acts as a constant energy density, and the positive cosmological constant, $\Lambda>0$, has been widely discussed as a dark energy candidate. On the other hand, what appears in AdS/CFT is the negative cosmological constant, $\Lambda<0$. The anti-de~Sitter spacetime used in AdS/CFT is a solution of this case (\chap{AdS}).

For now, let us consider the Einstein equation with no cosmological constant and with no matter fields:
\be
R_{\mu\nu} - \frac{1}{2} g_{\mu\nu} R = 0~.
\label{eq:Einstein_vac}
\ee
The simplest black hole, the \keyword{Schwarzschild black hole}, is the solution of the above equation:
\be
ds^2 = - \left( 1-\frac{2GM}{r} \right) dt^2 + \frac{dr^2}{1-\frac{2GM}{r}}
+ r^2 d\Omega_2^2~.
\label{eq:Schwarzschild}
\ee
Here, $d\Omega_2^2 := d\theta^2 + \sin^2\theta d\varphi^2$ is the line element of the unit $S^2$. We remark several properties of this black hole:
\begin{itemize}

\item
The metric approaches the flat spacetime $ds^2 \rightarrow -dt^2 + dr^2 + r^2d\Omega_2^2$ as $r \rightarrow \infty$.

\item 
As we will see in \sect{geodesics_flat}, $M$ represents the black hole mass. We will also see that the behavior $GM/r$ comes from the four-dimensional Newtonian potential.

\item
The horizon is located at $r = 2GM$ where $g_{00}=0$.

\item
A coordinate invariant quantity such as
\be
R^{\mu\nu\rho\sigma} R_{\mu\nu\rho\sigma} = \frac{48G^2M^2}{r^6} 
\ee
diverges at $r=0$. This location is called a \keyword{spacetime singularity}, where gravity is infinitely strong.

\end{itemize}
We now examine the massive and massless particle motions around the black hole to understand this spacetime more.

\section{Physics of the Schwarzschild black hole
}\label{sec:classic_tests}

\subsection{Gravitational redshift
}\label{sec:red_shift}

\begin{figure}[tb]
\begin{center}
\scalebox{0.75}{ \includegraphics{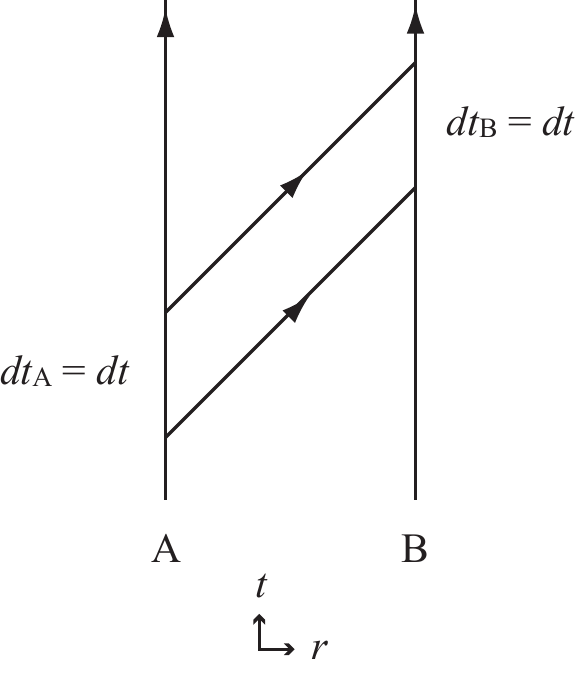} }
\vskip2mm
\caption{Exchange of light between A and B.
}
\label{fig:red_shift}
\end{center}
\end{figure}%

The  \keyword{gravitational redshift} is one of three ``classic tests" of general relativity; the other two are mentioned in \sect{geodesics_flat}.
The discussion here is used to discuss the surface gravity (\sect{surface_grav}) and to discuss the gravitational redshift in the AdS spacetime (\sect{geodesics_AdS}).

Consider two static observers at A and B (\fig{red_shift}). The observer at A sends light, and the observer at B receives light. The light follows the null geodesics $ds^2=0$, so
\begin{align}
ds^2 &= g_{00} dt^2+ g_{rr} dr^2 = 0~, \\
dt^2 &= \frac{g_{rr}}{-g_{00}} dr^2 
\result
\int_A^B dt = \int_A^B \sqrt{ \frac{g_{rr}(r)}{-g_{00}(r)} } dr~.
%
\end{align}
The right-hand side of the final expression does not depend on when light is sent, so the coordinate time until light reaches from A to B is always the same. Thus, if the observer at A emits light for the interval $dt$, the observer at B receives light for the interval $dt$ as well.

However, the proper time for each observer differs since $d\tau^2 = |g_{00}| dt^2$:
\begin{align}
d\tau_A^2 &\simeq |g_{00}(A)| dt^2~, \\
d\tau_B^2 &\simeq |g_{00}(B)| dt^2~.
%
\end{align}
But both observers should agree to the total number of light oscillations, so
\be
\omega_B d\tau_B = \omega_A d\tau_A~.
\ee
The energy of the photon is given by $E=\hbar \omega$, so $ E_B d\tau_B = E_A d\tau_A $, or 
\be
\boxeq{
\frac{E_B}{E_A} =\sqrt{\frac{g_{00}(A)}{g_{00}(B)}}~.
}
\label{eq:red_shift}
\ee

For simplicity, consider the Schwarzschild black hole and set $r_B=\infty$ and $r_A \gg GM$. Then,
\be
%
E_\infty = \sqrt{|g_{00}(A)|} E_A
\simeq E_A - \frac{GM}{r_A} E_A < E_A~.
\label{eq:red_shift_int}
\ee
Here, we used $\sqrt{|g_{00}(A)|} = (1-2GM/r_A)^{1/2} \simeq 1-(GM)/r_A$. Thus, the energy of the photon decreases at infinity. The energy of the photon decreases because the photon has to climb up the gravitational potential. Indeed, the second term of \eq{red_shift_int} takes the form of the Newtonian potential for the photon. Also, suppose that the point A is located at the horizon. Since $g_{00}(A)=0$ at the horizon, $E_\infty\rightarrow0$, namely light gets an infinite redshift.

\subsection{Particle motion
}\label{sec:geodesics}

\subsubsection*{Motion far away
}\label{sec:geodesics_flat}

The particle motion can be determined from the geodesic equation \eqref{eq:geodesic_eq}. However, there are enough number of conserved quantities for a static spherically symmetric solution such as the Schwarzschild black hole, which completely determines the particle motion without solving the geodesic equation.
\begin{itemize}

\item
First, because  of spherical symmetry, the motion is restricted to a single plane, and one can choose the equatorial plane ($\theta=\pi/2$) as the plane without loss of generality. 

\item
Second, as we saw in \sect{particle_action_curved}, when the metric is independent of a coordinate $x^\mu$, its conjugate momentum $p_\mu$ is conserved. For a static spherically symmetric solution, the metric is independent of  $t$ and $\varphi$, so the energy  $p_0$ and the angular momentum  $p_\varphi$ are conserved. 
\end{itemize}
Then, the particle four-momentum is given by
\begin{align}
p_0 &=: -mE~, \\
p_\varphi &=: mL~, \\
p^r &= m \frac{dr}{d\tau}~, \\
p^\theta &=0~.
%
\end{align}
($E$ and $L$ are the energy and the angular momentum per unit rest mass.) Because the four-momentum satisfies the constraint $p^2=-m^2$, 
\be
g^{00}(p_0)^2 + m^2 g_{rr} \left(\frac{dr}{d\tau}\right)^2 + g^{\varphi\varphi}(p_\varphi)^2 = -m^2~.
\label{eq:constraint_on_shell1}
\ee
Substitute the metric of the Schwarzschild black hole. When the angular momentum $L=0$, 
\be
\left(\frac{dr}{d\tau}\right)^2 
= (E^2 - 1) + \frac{2GM}{r}~.
\label{eq:massive_noL}
\ee
Since $(dr/d\tau)^2 \simeq E^2 - 1$ as $r\rightarrow\infty$, $E=1$ represents the energy when the particle is at rest at infinity, namely the rest mass energy of the particle. Differentiating this equation with respect to $\tau$ and using $\tau \simeq t$ in the nonrelativistic limit, one gets
\be
\frac{d^2r}{dt^2} \simeq - \frac{GM}{r^2}~,
\ee
which is nothing but the Newton's law of gravitation. Thus, $M$ in the Schwarzschild black hole \eqref{eq:Schwarzschild} represents the black hole mass.

Similarly, when $L \neq 0$,
\begin{align}
\left(\frac{dr}{d\tau}\right)^2 
&= E^2 - \left( 1-\frac{2GM}{r} \right)\left( 1+\frac{L^2}{r^2} \right) \\
&=  (E^2 - 1) + \frac{2GM}{r} - \frac{L^2}{r^2} + \frac{2GML^2}{r^3}~.
\label{eq:massive_L}
%
\end{align}
The third term in \eq{massive_L} represents the centrifugal force term. On the other hand, the fourth term is a new term characteristic of general relativity. General relativity has ``classic tests" such as 
\begin{itemize}
\item The perihelion shift of Mercury 
\item The light bending 
\end{itemize}
in addition to the gravitational redshift, and both effects come from this fourth term%
\footnote{For the light bending, use the equation for the massless particle instead of \eq{constraint_on_shell1}. }.
The fourth term is comparable to the third term only when the particle approaches $r \simeq 2GM$. This distance corresponds to the horizon radius of the black hole and is about 3km for a solar mass black hole, so the effect of this term is normally very small.

We will generalize the discussion here to a generic static metric in order to discuss the surface gravity in \sect{surface_grav}. We will also examine the particle motion in the AdS spacetime in \sect{geodesics_AdS}.

\subsubsection*{Motion near horizon
}\label{sec:geodesics_horizon}

We now turn to the particle motion near the horizon. How long does it take until the particle reaches the horizon? For simplicity, we assume that the particle is at rest at infinity ($E=1$) and that the particle falls radially ($L=0$). From \eq{massive_noL}, the particle motion for $E=1$ and $L=0$ is given by
\be
\left(\frac{dr}{d\tau}\right)^2 
= \frac{r_0}{r}
%
\ee
($r_0=2GM$). Near the horizon,
\be
\frac{dr}{d\tau} \simeq -1~.
%
\ee
We choose the minus sign since the particle falls inward ($r$ decreases as time passes). So, to go from $r=r_0+R$ to $r=r_0+\epsilon$,
\be
\tau \simeq -\int_{r_0+R}^{r_0+\epsilon} dr = R - \epsilon~.
%
\ee
Namely, the particle reaches the horizon in a \textit{finite proper time.}

However, the story changes from the point of view of the coordinate time $t$. By definition, $p^0 = mdt/d\tau$, and from the conservation law, $p^0 = g^{00}p_0 = m(1-r_0/r)^{-1}$. Then,
\be
\frac{d\tau}{dt} = \frac{r-r_0}{r}~.
%
\ee
Thus,
\be
\left(\frac{dr}{dt}\right)^2 
= \left(\frac{dr}{d\tau}\right)^2 \left(\frac{d\tau}{dt}\right)^2 
= \frac{r_0(r-r_0)^2}{r^3}
%
\ee
or
\be
\frac{dr}{dt} \simeq - \frac{r-r_0}{r_0}
\label{eq:timelike_horizon_t}
\ee
near the horizon, and
\be
t \simeq - r_0 \int_{r_0+R}^{r_0+\epsilon} \frac{dr}{r-r_0} = r_0(\ln R - \ln \epsilon)~,
%
\ee
so $t\rightarrow\infty$ as $\epsilon\rightarrow0$. Namely, it takes an \textit{infinite coordinate time} until the particle reaches the horizon. Incidentally, \eq{timelike_horizon_t} near the horizon takes the same form as the massless case below. Namely, the particle moves with the speed of light near the horizon.

Let us consider the massless case. For the massless particle, $p^2=0$ or $ds^2=0$, so
\begin{align}
& ds^2 = g_{00} dt^2+ g_{rr} dr^2 = 0
\\
&\rightarrow 
\left(\frac{dr}{dt}\right)^2 
= - \frac{ g_{00} }{ g_{rr} }
= \left( 1-\frac{r_0}{r} \right)^2 
\\
&\rightarrow
\frac{dr}{dt} = - \left( 1-\frac{r_0}{r} \right) \simeq - \frac{r-r_0}{r_0}~.
\label{eq:null_horizon_t}
%
\end{align}
Near the horizon, the expression takes the same form as the massive case  \eqref{eq:timelike_horizon_t} as promised. We considered the infalling photon, but one can consider the outgoing photon. In this case, $t\rightarrow\infty$ until the light from the horizon reaches the observer at finite $r$.

There is nothing special to the horizon from the point of view of the infalling particle. But there is a singular behavior from the point of view of the coordinate $t$. This is because the Schwarzschild coordinates $(t,r)$ are not well-behaved near the horizon. Thus, we introduce the coordinate system which is easier to see the infalling particle point of view.

\section{Kruskal coordinates
}

\begin{figure}[tb]
\centering
\scalebox{0.75}{ \includegraphics{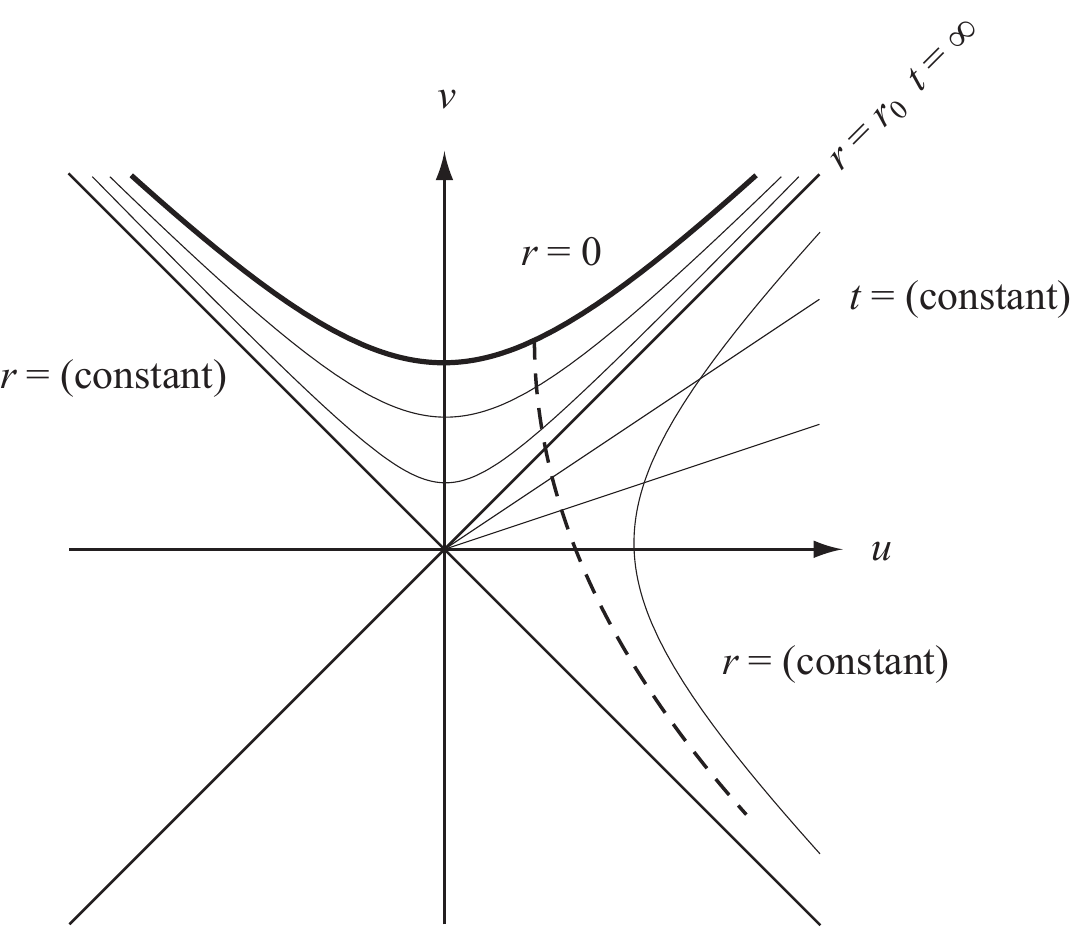} }
\vskip2mm
\caption{Kruskal coordinates. The light-cones are kept at 45$^\circ$, which is convenient to see the causal structure. The dashed line represents an example of the particle path. Once the particle crosses the horizon, it must reach the singularity.
}
\label{fig:kruskal}
\end{figure}%

The particle motion discussed so far can be naturally understood by using a new coordinate system, the \keyword{Kruskal coordinates}. The Kruskal coordinates $(u,v)$ are defined by
\begin{align}
r>r_0 & \left\{
\begin{array}{l}
  u = \left( \frac{r}{r_0}-1 \right)^{1/2} e^{r/(2r_0)} \cosh \left(\frac{t}{2r_0}\right) \\
  v = \left( \frac{r}{r_0}-1 \right)^{1/2} e^{r/(2r_0)} \sinh \left(\frac{t}{2r_0}\right)
\end{array}
\right. 
\label{eq:kruskal} \\
r<r_0 & \left\{
\begin{array}{l}
  u = \left( 1-\frac{r}{r_0} \right)^{1/2} e^{r/(2r_0)} \sinh \left(\frac{t}{2r_0}\right) \\
  v = \left( 1-\frac{r}{r_0} \right)^{1/2} e^{r/(2r_0)} \cosh \left(\frac{t}{2r_0}\right)
\end{array}
\right.
%
\end{align}
By the coordinate transformation, the metric \eqref{eq:Schwarzschild} becomes
\be
ds^2 = \frac{4r_0^3}{r} e^{-r/r_0} ( -dv^2+du^2 ) + r^2 d\Omega_{2}^2~.
\label{eq:metric_kruskal}
\ee
Here, we use not only $(u,v)$ but also use $r$, but $r$ should be regarded as $r=r(u,v)$ and is determined by
\be
\left( \frac{r}{r_0}-1 \right) e^{r/r_0} = u^2-v^2~.
\label{eq:r&uv}
\ee

One can see the following various properties from the coordinate transformation and the metric (see also \fig{kruskal}):
\begin{itemize}

\item
The metric \eqref{eq:metric_kruskal} is not singular at $r=r_0$. There is a singularity at $r=0$. The transformation \eqref{eq:kruskal} is singular at $r=r_0$, but this is not a problem. Because the transformation relates the coordinates which are singular at $r=r_0$ to the coordinates which are not singular at $r=r_0$, the transformation should be singular there. 

\item
The null world-line $ds^2=0$ is given by $dv=\pm du$. In this coordinate system, the lines at 45$^\circ$ give light-cones just like special relativity, which is convenient to see the causal structure of the spacetime. 

\item
The $r=\text{(constant)}$ lines are hyperbolas from \eq{r&uv}. 

\item
In particular, in the limit $r=r_0$, the hyperbola becomes a null line, so \textit{the horizon $r=r_0$ is a null surface}. Namely, the horizon is not really a spatial boundary but is a light-cone. In special relativity, the events inside light-cones cannot influence the events outside light-cones. Similarly, the events inside the horizon cannot influence the events outside the horizon. Then, even light cannot reach from $r<r_0$ to $r>r_0$.

\item
For $r<r_0$, the $r=\text{(constant)}$ lines become spacelike. This means that a particle cannot remain at $r=\text{(constant)}$ because the geodesics of a particle cannot be spacelike. The singularity at $r=0$ is spacelike as well. Namely, the singularity is not a point in spacetime, but rather it is the end of ``time."

\item
The $t=\text{(constant)}$ lines are straight lines. In particular, the $t \rightarrow\infty$ limit is given by $u=v$. One can see that it takes an infinite coordinate time to reach the horizon.

\end{itemize}
To summarize, the particle falling into the horizon cannot escape and necessarily reaches the singularity.

\titlenewterms%

After you read each chapter, try to explain the terms in ``New keywords" by yourself to check your understanding.

\begin{multicols}{2}
\noindent
horizon\\
proper time\\
world-line\\
four-velocity\\
geodesic\\
affine parameter\\
cosmological constant\\
Schwarzschild black hole\\
spacetime singularity\\
gravitational redshift\\
Kruskal coordinates\\
\end{multicols}

\section{Appendix: Review of general relativity
}\label{sec:GR}

Consider a coordinate transformation
\be
x'^\mu = x'^\mu (x)~.
%
\ee
Under a coordinate transformation, a quantity is called a vector if it transforms as
\be
V'^\mu= \frac{\del x'^\mu}{\del x^\nu} V^\nu~,
%
\ee
and as a 1-form if it transforms ``oppositely": 
\be
V'_\mu= \frac{\del x^\nu}{\del x'^\mu} V_\nu~.
%
\ee
The tensors with a multiple number of indices are defined similarly.

In general, the derivative of a tensor such as  $\del_\mu V^\nu$ does not transform as a tensor, but the covariant derivative $\nabla_\mu$ of a tensor transforms as a tensor. The covariant derivatives of the vector and the 1-form are given by
\begin{align}
\nabla_\mu V^\nu &= \del_\mu V^\nu + V^\alpha \Gamma^\nu_{~\alpha\mu}~, \\
\nabla_\mu V_\nu &= \del_\mu V_\nu - V_\alpha \Gamma^\alpha_{~\mu\nu}~.
%
\end{align}
As a useful relation, the covariant divergence of a vector is given by
\be
\nabla_\mu V^\mu = \frac{1}{\sqrt{-g}} \del_\mu \left(\sqrt{-g} V^\mu \right)~,
%
\ee
where $g:= \det g$. This can be shown using a formula for a matrix $M$:
\be
\del_\mu(\det M) = \det M\, \text{tr}(M^{-1}\del_\mu M)~.
\label{eq:matrix_formula}
\ee

For example, $dx^\mu$ transforms as a vector 
\be
dx'^\mu= \frac{\del x'^\mu}{\del x^\nu} dx^\nu
\label{eq:dx_transf}
\ee
and the metric transforms as
\be
g'_{\mu\nu}(x') = \frac{\del x^\rho}{\del x'^\mu}  \frac{\del x^\sigma}{\del x'^\nu}  g_{\rho\sigma}(x)~.
\label{eq:metric_transf}
\ee
Thus, the line element $ds^2 = g_{\mu\nu}dx^\mu dx^\nu$ is invariant under coordinate transformations. Under the infinitesimal transformation $x'^\mu = x^\mu - \xi^\mu(x)$, \eq{metric_transf} is rewritten as
\be
g'_{\mu\nu} (x-\xi) = 
\left(\delta^\rho_{~\mu} + \del_\mu\xi^\rho \right)
\left(\delta^\sigma_{~\nu} + \del_\nu\xi^\sigma \right)g_{\rho\sigma}
%
\ee
or
\begin{align}
g'_{\mu\nu} (x) &= 
g_{\mu\nu}(x) + (\del_\mu\xi^\rho) g_{\rho\nu} + (\del_\nu\xi^\rho) g_{\mu\rho} 
+ \xi^\rho \del_\rho g_{\mu\nu} \\
&= g_{\mu\nu} + \nabla_\mu\xi_\nu +  \nabla_\nu\xi_\mu~.
\label{eq:metric_transf_inf}
%
\end{align}

In general relativity, an action must be a scalar which is invariant under coordinate transformations. From \eq{dx_transf},
\be
d^{4}x' = \left| \frac{\del x'}{\del x} \right| d^{4}x~,
%
\ee
where $| \del x'/\del x |$ is the Jacobian of the transformation. On the other hand, $\sqrt{-g}$ transforms in the opposite manner: 
\be
\sqrt{-g'} = \left| \frac{\del x}{\del x'} \right| \sqrt{-g}~. 
%
\ee
Thus, $d^{4}x\, \sqrt{-g}$ is the volume element which is invariant under coordinate transformations.

The metric is determined by the Einstein-Hilbert action:
\be
\action = \frac{1}{16\pi G} \int d^{4}x\, \sqrt{-g} R~.
%
\ee
Here, $G$ is the Newton's constant, and the Ricci scalar $R$ is defined by the Riemann tensor $R^\alpha_{~\mu\nu\rho}$ and the Ricci tensor $R_{\mu\nu}$ as follows:
\begin{align}
R^\alpha_{~\mu\nu\rho} &=
\del_\nu \Gamma^\alpha_{~\mu\rho} - \del_\rho \Gamma^\alpha_{~\mu\nu}
+ \Gamma^\alpha_{~\sigma\nu}\Gamma^\sigma_{~\mu\rho}
- \Gamma^\alpha_{~\sigma\rho}\Gamma^\sigma_{~\mu\nu}~, \\
R_{\mu\nu} &= R^\alpha_{~\mu\alpha\nu}~, \quad
R = g^{\mu\nu} R_{\mu\nu}~.
%
\end{align}
The variation of the Einstein-Hilbert action gives 
\begin{align}
\delta \action &= \frac{1}{16\pi G} \int d^{4}x\, 
\left\{
\sqrt{-g} R_{\mu\nu} (\delta g^{\mu\nu}) + (\delta\sqrt{-g}) R_{\mu\nu}g^{\mu\nu} \right. 
\nonumber \\
& \left. \hspace{2cm} + \sqrt{-g} (\delta R_{\mu\nu}) g^{\mu\nu} 
\right\}~,
%
\end{align}
where $g^{\mu\nu}$ is the inverse of $g_{\mu\nu}$. The second term can be rewritten by using%
\footnote{Using \eq{matrix_formula} gives $\delta(\sqrt{-g}) = \frac{1}{2} \sqrt{-g} g^{\mu\nu} \delta g_{\mu\nu}$. Then, use another matrix formula $\delta M = -M \delta (M^{-1}) M$ which can be derived from $MM^{-1}=I$. Then, $ \delta g_{\mu\nu} = -g_{\mu\rho} g_{\nu\sigma}\delta g^{\rho\sigma}$. Note that $ \delta g_{\mu\nu} \neq g_{\mu\rho} g_{\nu\sigma}\delta g^{\rho\sigma}$. Namely, we do not use the metric to raise and lower indices of the metric variation.}
\be
\delta \sqrt{-g} = - \frac{1}{2} \sqrt{-g} g_{\mu\nu} \delta g^{\mu\nu}~.
%
\ee
One can show that the third term reduces to a surface term, so it does not contribute to the equation of motion%
\footnote{Care is necessary to the surface term in order to have a well-defined variational principle. This issue will be discussed in Sects.~\ref{sec:SAdS_free} and \ref{sec:action_tensor}.}. 
Therefore,
\be
\delta \action = \frac{1}{16\pi G} \int d^{4}x\, \sqrt{-g} 
\left(R_{\mu\nu} -\frac{1}{2} g_{\mu\nu} R \right) \delta g^{\mu\nu}~,
%
\ee
and by requiring $\delta \action=0$, one gets the vacuum Einstein equation:
\be
R_{\mu\nu} - \frac{1}{2} g_{\mu\nu} R = 0~.
\label{eq:Einstein_vac_again}
\ee
The contraction of \eq{Einstein_vac_again} gives $R_{\mu\nu}=0$.

When one adds the matter action $\action_\text{matter}$, the equation of motion becomes
\be
R_{\mu\nu} - \frac{1}{2} g_{\mu\nu} R = 8\pi G\, T_{\mu\nu}~,
%
\ee
where $T_{\mu\nu}$ is the energy-momentum tensor for matter fields:
\be
T_{\mu\nu} := - \frac{2}{\sqrt{-g}} \frac{\delta \action_\text{matter}}{\delta g^{\mu\nu}}~.
\label{eq:EM_tensor}
\ee
Various matter fields appear in this book, but the simplest term one can add to the Einstein-Hilbert action is given by
\be
\action_{cc} = -\frac{1}{8\pi G} \int d^{4}x\, \sqrt{-g} \Lambda~,
%
\ee
which is the cosmological constant term. From \eq{EM_tensor},
\be
T_{\mu\nu} = - \frac{\Lambda}{8\pi G} g_{\mu\nu}~,
\label{eq:EM_tensor_cc_again}
\ee
so the Einstein equation becomes
\be
R_{\mu\nu} - \frac{1}{2} g_{\mu\nu} R + \Lambda g_{\mu\nu} = 0~.
%
\ee

\endofsection

\ifx\nameofpaper\undefined 
  \usepackage{macro_natsuume} 
  \def\beginsection{\section*}
  \def\endofsection{\end{document}} 
  \input draft_header.tex
\else 
  \def\beginsection{\chapter}
  \def\endofsection{ } 
\fi

\beginsection{Black holes and thermodynamics
}\label{chap:BH_thermodynamics}


\begin{quote}
Quantum mechanically, black holes have thermodynamic properties just like ordinary statistical systems. In this chapter, we explain the relation between black holes and thermodynamics using the example of the Schwarzschild black hole. 
\end{quote}

\section{Black holes and thermodynamics
}

For the Schwarzschild black hole, the horizon radius is given by $r_0=2GM/c^2$. The horizon radius is proportional to the black hole mass, so if matter falls in the black hole, the horizon area increases:
\be
A=4\pi r_0^2=\frac{16\pi G^2M^2}{c^4}~.
\label{eq:horizon_area}
\ee
Also, classically nothing comes out from the black hole, so the area is a non-decreasing quantity%
\footnote{As the black hole evaporates by the Hawking radiation, the horizon area decreases. But the total entropy of the black hole entropy and the radiation entropy always increases (generalized second law).}, 
which reminds us of thermodynamic entropy. Thus, one expects that a black hole has the notion of entropy $\Sbh$: 
\be
\Sbh \propto A ?
\ee
We henceforth call $\Sbh$ as the \keyword{black hole entropy}.

In fact, a black hole obeys not only the second law but also all thermodynamic-like laws as we will see (\fig{thermo}). First of all, a stationary black hole has only a few parameters such as the mass, angular momentum, and charge. This is known as the \keyword{no-hair theorem} (see, \eg, Refs.~\cite{Bekenstein:1996pn,Heusler:1996ft} for reviews). Namely, a black hole does not depend on the properties of the original stars such as the shape and the composition. Conversely, a black hole is constrained only by a few initial conditions, so there are many ways to make a black hole. For example, even if the black hole formed from gravitational collapse is initially asymmetric, it eventually becomes the simple spherically symmetric Schwarzschild black hole (when the angular momentum and the charge vanish). 

\renewcommand{\arraystretch}{1.5}
\begin{figure}[tb]
\begin{center}
\begin{tabular}{|l||c|c|}
\hline
		  & Thermodynamics 		& Black hole	
\\ \hline\hline
Zeroth law & Temperature $T$ is constant	& Surface gravity $\kappa$ is constant \\
		 & at equilibrium			& for a stationary solution
\\ \hline
First law	 & $ dE = T d\Sbh $			& $ dM = \frac{\kappa}{8 \pi G} dA $ 
\\ \hline
Second law & $ dS \geq 0 $			& $ dA \geq 0 $ 
\\ \hline
Third law	   & $S\rightarrow0$ as $T\rightarrow0$ & $S\rightarrow0$ as $T\rightarrow0$? 
\\ \hline
\end{tabular}
\caption{Comparison between laws of thermodynamics and black hole thermodynamics. }
\label{fig:thermo}
\end{center}
\end{figure}
\renewcommand{\arraystretch}{1}

This property of the black hole itself is similar to thermodynamics. Thermodynamics is the theory of many molecules or atoms. According to thermodynamics, one does not have to specify the position and the momentum of each molecule to characterize a thermodynamic system. The system can be characterized only by a few macroscopic variables such as temperature and pressure. The prescription to go from microscopic variables to macroscopic variables is known as the coarse-graining. 

The black hole is described only by a few parameters. This suggests that somehow the black hole is a coarse-grained description. But at present the details of this coarse-graining is not clear. This is because we have not completely established the microscopic theory of the black hole or the quantization of the black hole.

\subsection{Zeroth law}\index{zeroth law of thermodynamics (black hole)}

Let us compare the black hole laws and thermodynamic laws%
\footnote{See \sect{thermodynamics} to refresh your memory of thermodynamics.}. 
A thermodynamic system eventually reaches a thermal equilibrium, and the temperature becomes constant everywhere. This is the zeroth law of thermodynamics. Recall that a black hole eventually becomes the spherically symmetric one even if it is initially asymmetric. Spherical symmetry implies that the gravitational force is constant over the horizon. Then, to rephrase the no-hair theorem, gravity over the horizon eventually becomes constant even if it was not constant initially. This is similar to the zeroth law, and gravity on the horizon corresponds to the temperature. Also, both temperature and gravity are non-negative. A stationary black hole, whose horizon gravity becomes constant, is in a sense a state of equilibrium. 

The gravitational force (per unit mass) or the gravitational acceleration on the horizon is called the \keyword{surface gravity}. In Newtonian gravity, the gravitational acceleration is given by 
\be
a = \frac{GM}{r^2}~,
\ee
so at the horizon $r=r_0$,
\be
\kappa = a(r=r_0) = \frac{c^4}{4GM}~.
\label{eq:surface_gravity_Sch}
\ee

We used Newtonian gravity to derive the surface gravity, but one should not take the argument too seriously. This is just like the Newtonian gravity result for the horizon radius in \chap{BH}. The surface gravity is the force which is necessary to stay at the horizon, but in general relativity, one has to specify who measures it. As discussed below, the surface gravity is the force measured by the asymptotic observer. The infalling observer cannot escape from the horizon, no matter how large the force is. So, the necessary force diverges for the infalling observer himself. But if this force is measured by the asymptotic observer, the force remains finite and coincides with the Newtonian result. Two observers disagree the values of the acceleration because of the gravitational redshift. 

\subsection{Surface gravity \advanced}\label{sec:surface_grav}

Consider a generic static metric of the form
\be
ds^2 = - f(r) dt^2 + \frac{dr^2}{f(r)} + \cdots~.
\label{eq:metric_ansatz}
\ee
Here, $\cdots$ represents the line element along the horizon which is irrelevant to our discussion. Following \sect{geodesics_flat}, the particle motion is determined from 
\be
\left(\frac{dr}{d\tau}\right)^2 = E^2 - f
\result
\frac{d^2r}{d\tau^2} = -\frac{1}{2} f'~, \quad (':=\del_r)~.
\label{eq:surface_gravity_naive}
\ee
For the Schwarzschild black hole, $f=1-2GM/r$, so $d^2r/d\tau^2 =-GM/r^2$. As we saw in \sect{geodesics_flat}, the expression takes the same form as Newtonian gravity. 

In this sense, we used Newtonian gravity to derive the particle's acceleration in the last subsection. But this acceleration is not the covariant acceleration $a^\mu$ but is just the $a^r$ component. It is more suitable to use the proper acceleration which is the magnitude of the covariant acceleration%
\footnote{One can check $a^0=0$ for the particle at rest.}:
\begin{align}
a^2 &:= g_{\mu\nu} a^\mu a^\nu = \frac{f'^2}{4f}~, \\
a &= \frac{f'}{2f^{1/2}}~.
%
\end{align}
The proper acceleration diverges at the horizon since $f(r_0)=0$ at the horizon. This is not surprising since the particle cannot escape from the horizon.

\begin{figure}[tb]
\begin{center}
\scalebox{0.75}{ \includegraphics{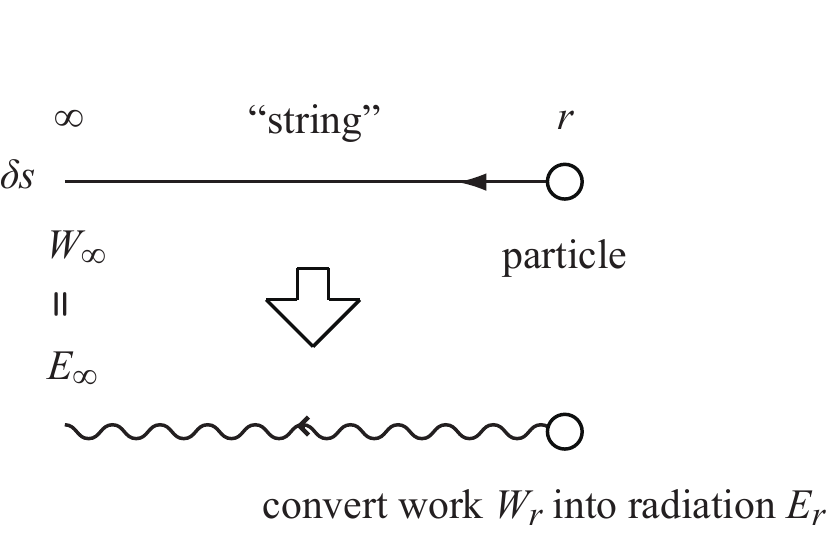} }
\vskip2mm
\caption{
In order to obtain the surface gravity, the asymptotic observer pulls the particle by a massless ``string." The work done to the particle is then converted into radiation. The radiation is subsequently collected at infinity. 
}
\label{fig:surface_gravity}
\end{center}
\end{figure}

The surface gravity is the force (per unit mass) $a_\infty(r_0)$, which is necessary to hold the particle at the horizon by an \textit{asymptotic observer}. In order to obtain the force, suppose that the asymptotic observer pulls the particle at $r$ by a massless ``string" (\fig{surface_gravity}). If the observer pulls the string by the proper distance $\delta s$, the work done is given by
\begin{alignat}{2}
W_\infty &= a_\infty \delta s & \quad & (\text{asymptotic infinity})~, \\
W_r &= a \delta s & \quad & (\text{location }r)~.
%
\end{alignat}
Now, convert the work $W_r$ into radiation with energy $E_r=W_r$ at $r$, and then collect the radiation at infinity. The energy received at infinity, $E_\infty$, gets the gravitational redshift (\sect{red_shift}). \index{gravitational redshift} The redshift formula is \eq{red_shift}:
\be
\frac{E_B}{E_A} =\sqrt{\frac{g_{00}(A)}{g_{00}(B)}}~.
%
\ee
So, $E_\infty$ is given by
\be
E_\infty = \sqrt{ \frac{f(r)}{f(\infty)} } E_r = f(r)^{1/2} a \delta s~,
\label{eq:red_shift_surface}
\ee
where we used $f(\infty)=1$. The energy conservation requires $W_\infty=E_\infty$, so $a_\infty=f^{1/2}a =f'/2$ or
\be
\kappa := a_\infty (r_0) = \frac{f'(r_0)}{2}~,
\label{eq:surface_gravity}
\ee
which coincides with the naive result \eqref{eq:surface_gravity_naive}.

\subsection{First law}

If the black hole mass increases by $dM$, the horizon area increases by $dA$ as well. So, one has a relation
\be
dM \propto dA~.
\ee
To have a precise equation, compare the dimensions of both sides of the equation. First, the left-hand side must be $GdM$ because the Newton's constant and mass appear only in the combination $GM$ in general relativity. The Einstein equation tells how the mass energy curves the spacetime, and $G$ is the dictionary to translate from the mass to the spacetime curvature. 

Then, one can easily see that the right-hand side must have a coefficient which has the dimensions of acceleration. It is natural to use the surface gravity for this acceleration. In fact, the first law $dE=TdS$ tells us that temperature appears as the coefficient of the entropy, and we saw earlier that the surface gravity plays the role of temperature. Thus, we reach
\be
GdM \simeq \kappa dA~.
\ee
By differentiating \eq{horizon_area} with respect to $M$, one can see that
\be
GdM = \frac{\kappa}{8\pi} dA
\label{eq:1st_law}
\ee
including the numerical constant. This is the \textit{first law of black holes.} \index{first law of thermodynamics (black hole)}

We discuss the third law of black holes in \sect{RN}.

\section{From analogy to real thermodynamics}

\subsection{Hawking radiation
}\label{sec:BH_entropy}

We have seen that black hole laws are similar to thermodynamic laws. However, so far this is just an analogy. The same expressions do not mean that they represent the same physics. Indeed, there are several problems to identify black hole laws as thermodynamic laws: 
\begin{enumerate}
\item 
Nothing comes out from a black hole. If the black hole really has the notion of temperature, a black hole should have a thermal radiation. 
\item 
The horizon area and the entropy behave similarly, but they have different dimensions. In the unit $\kB=1$, thermodynamic entropy is dimensionless whereas the area has dimensions. One can make it dimensionless by dividing the area by length squared, but so far we have not encountered an appropriate one.
\end{enumerate}

The black hole is not an isolated object in our universe. For example, matter can make a black hole, and the matter obeys quantum mechanics microscopically. So, consider the quantum effect of matter. If one considers this effect, the black hole indeed emits the black body radiation known as the Hawking radiation, and its temperature is given by
\begin{align}
\kB T &= \frac{\hbar \kappa}{2\pi c} 
\label{eq:temp_vs_surface} \\
&= \frac{\hbar c^3}{8\pi GM}~, \quad \text{(Schwarzschild black hole).}
\end{align}
This is a quantum effect because the temperature is proportional to $\hbar$. We explain later how to derive this temperature, the \keyword{Hawking temperature}. Anyway, once the temperature is determined, we can get the precise relation between the black hole entropy $\Sbh$ and the horizon area $A$. The first law of black hole \eqref{eq:1st_law} can be rewritten as
\begin{align}
d(Mc^2) &= \frac{\kappa c^2}{8\pi G}dA \\
&= \frac{\hbar\kappa}{2\pi \kB c} \frac{\kB c^3}{4G\hbar}dA
= T \frac{\kB c^3}{4G\hbar}dA~.
\end{align}
So, comparing with the first law of thermodynamics $dE=TdS$, one obtains
\be
\boxeq{
\Sbh = \frac{A}{4G\hbar} \kB c^3 = \frac{1}{4}\frac{A}{\lpl^2} \kB~.
}
\label{eq:BH_entropy}
\ee
Here, $\lpl = \sqrt{G\hbar/c^3} \approx 10^{-35}$m is called the \keyword{Planck length}. The Planck length is the length scale where quantum gravity effects become important. Note that one cannot make a quantity whose dimension is the length from the fundamental constants of classical mechanics alone, $G$ and $c$. The black hole entropy now becomes dimensionless because we divide the area by the Planck length squared. Equation~\eqref{eq:BH_entropy} is called the \textit{area law}. \index{area law (black hole entropy)} 

A few remarks are in order. First of all, our discussion here does not ``derive" $\Sbh$ as the real entropy. Microscopically, the entropy is the measure of the degrees of freedom of a system, but we have not counted microscopic states. We are still assuming that classical black hole laws are really thermodynamic laws.

Second, 
\begin{itemize}
\item The black hole entropy is proportional to the ``area." 
\item On the other hand, the statistical entropy is proportional to the ``volume" of a system. 
\end{itemize}
This difference is extremely important for AdS/CFT and gives a clue on the nature of microscopic states of black holes. Because these entropies behave differently, a four-dimensional black hole cannot correspond to a four-dimensional statistical system. But the five-dimensional ``area" is the four-dimensional ``volume." Then, a five-dimensional black hole can correspond to a four-dimensional statistical system. Namely, 
\begin{center}
\fbox{
\begin{tabular}{l}
First clue of AdS/CFT: \\
The black hole entropy suggests that a black hole can be described by \\ 
the usual statistical system whose spatial dimension is one dimension \\ 
lower than the gravitational theory.
\end{tabular}
}
\end{center}
AdS/CFT claims that this statistical system is a gauge theory. Such an idea is called the \keyword{holographic principle} in general \cite{'tHooft:1993gx,Susskind:1994vu}. The holographic principle is the first clue for AdS/CFT, and we will see another clue, the large-$N_c$ gauge theory, in \sect{large_N}. 

Now, our discussion so far uses the Schwarzschild black hole as an example, but the formula \eqref{eq:BH_entropy} itself is generic. The black hole entropy is \textit{always} given by \eq{BH_entropy} as long as 
the gravitational action is written by the Einstein-Hilbert action%
\footnote{For a generic gravitational action, the \bh entropy is given by the Wald formula \index{Wald formula} \cite{Wald:1994yp}.}:
\be
\frac{1}{16\pi G_d}\int d^dx \sqrt{-g} R~.
\ee
Namely, \eq{BH_entropy} is true not only for black holes in general relativity but also for black holes in string theory and in the other gravitational theories.

\subsection{Hawking temperature and Euclidean formalism
}\label{sec:Euclidean}

There are several ways to compute the Hawking temperature. Hawking originally computed it by quantizing matter fields in the black hole background. But the simple way is to require that the Euclidean spacetime be smooth. To do so, one needs a periodic identification in imaginary time, and the temperature is the inverse of this period $\beta$.

By the analytic continuation to Euclidean signature $\tE = it$, the metric \eqref{eq:metric_ansatz} becomes 
\be
ds_\text{E}^2 = + f(r) d\tE^2 + \frac{dr^2}{f(r)} +\cdots~.
%
\ee
Let us focus on the region near the ``horizon" $r \simeq r_0$. Near the horizon, $f(r_0)=0$, so one can approximate%
\footnote{Here, we assume $f'(r_0)\neq 0$. This assumption fails for extreme black holes \index{extreme black hole} where two horizons are degenerate (\sect{RN}). }
$f \simeq f'(r_0) (r-r_0)$. Then,
\begin{align}
ds_\text{E}^2 
&\simeq \frac{dr^2}{f'(r_0) (r-r_0)} + f'(r_0) (r-r_0) d\tE^2 \\
&= d\rho^2 + \rho^2 d\left( \frac{f'(r_0)}{2} \tE \right)^2~,
\label{eq:polar}
\end{align}
where we introduced a new coordinate $\rho := 2\sqrt{ (r - r_0)/f'(r_0) }$. In this coordinate, the metric takes the same form as a plane in polar coordinates if $f'(r_0)\tE/2$ has the period $2\pi$. Otherwise, the metric has a conical singularity at $\rho=0$. The periodicity of $\tE$ is given by $\beta=4\pi/f'(r_0)$ or
\be
\boxeq{
T=\frac{f'(r_0)}{4\pi}~.
}
\label{eq:Hawking_temp}
\ee
If one compares this result with \eq{surface_gravity}, one gets $T=\kappa/(2\pi)$ which is \eq{temp_vs_surface}. For the Schwarzschild black hole, $f(r)=1-r_0/r$, so 
\be
T = \frac{1}{4\pi r_0} = \frac{1}{8\pi GM}~.
%
\ee
Note that the Euclidean metric does not cover inside the horizon $r<r_0$ since the horizon is the origin of polar coordinates.

What is behind the Euclidean formalism is quantum statistical mechanics. In quantum statistical mechanics, the periodicity in the imaginary time is naturally associated with the inverse temperature. We mimic this technique of quantum statistical mechanics and apply to black holes. Then, the ``first-principle" computation of black hole thermodynamic quantities is carried out as follows \cite{Gibbons:1976ue}. (This approach naturally leads to the \keyword{GKP-Witten relation} in AdS/CFT later.) 

The problem of quantum gravity is not completely solved, but it is natural to expect that quantum gravity is defined by a path integral such as
\be
Z = \int {\cal D}g\, e^{i\action_\text{L}[g]}~.
%
\ee
Here, $\action_\text{L}$ is the gravitational action, and the above equation shows the path integral over the metric $g$ schematically. We consider the Euclidean formalism with $\tE=it$, $\action_\text{E}=-i\action_\text{L}$:
\be
Z = \int {\cal D}g\, e^{-\action_\text{E}[g]}~.
\label{eq:Euclidean}
\ee
As we saw above, the Euclidean black hole is periodic in time. Then, it is natural to associate the statistical mechanical partition function \index{partition function} to the black hole and is natural to regard $Z$ as such a partition function. However, it is not clear how to carry out the path integral because it diverges. Also, string theory does not evaluate \eq{Euclidean} itself. In any case, whether one uses string theory or uses the above approach, semiclassical results themselves should agree. Semiclassically, the most dominant contribution to the path integral comes from an extremum of the action, or the classical solution (saddle-point approximation). Under the approximation, 
\be
\boxeq{
Z \simeq e^{-\SosE}~.
}
%
\ee
Here, $\SosE$ is the on-shell action which is obtained by substituting the classical solution to the action. The free energy\index{free energy} $F$ is then given by $\SosE=\beta F$. For example, for the Schwarzschild black hole, one can show that 
\be
F = \frac{r_0}{4G} = \frac{1}{16\pi G T}~.
\label{eq:free_Sch}
\ee
Then, one can get thermodynamic quantities:
\begin{align}
\Sbh &= - \del_T F = \frac{1}{16\pi G T^2} = \frac{\pi r_0^2}{G} = \frac{A}{4G}~, \\
E &= F + T \Sbh = \frac{1}{8\pi G T} = \frac{r_0}{2G} = M~.
%
\end{align}
%
%

We will not carry out the actual computation for the Schwarzschild black hole here but will carry out for the Schwarzschild-AdS$_5$ black holes in \sect{SAdS_free} and \sect{HP}. AdS/CFT identifies the black hole partition function $Z$ as the gauge theory partition function.

\subsection{On the origin of black hole entropy \advanced}

What represents the black hole entropy? It is not easy to answer to the question as we will see below, but let us make some general remarks.

When we are speaking of the entropy of the sun, the meaning is clear. Although a thermodynamic system looks same macroscopically, the system can have different microscopic states. There are number of ways one can distribute, \eg, a given energy to a multiple number of particles. The entropy counts the number of possible quantum states when macroscopic variables such as energy are specified. In general, this number becomes larger when the number of particles becomes larger. So, the entropy is generically proportional to the number of particles.

This suggests that the black hole entropy counts the microscopic states a black hole can have. This number can be estimated as follows \cite{Zurek:1985gd}.
Let us imagine to make a black hole by collecting particles in a small region. The entropy is proportional to the number of particles. To have a larger entropy, we let the total number of particles as many as possible, or we make the black hole from light particles. However, the Compton wavelength of a particle with mass $m$ is about $\lambda \simeq \hbar/m$. We cannot collect particles in a smaller region than the Compton wavelength. But to make a black hole, we must collect them within the Schwarzschild radius. Thus, if the Compton wavelength is larger than the Schwarzschild radius, a black hole is not formed. Consequently, we cannot choose the particle mass freely, and the lightest particle we can use has mass which satisfies $\hbar/m \simeq 2GM$. Then, the total number of particles is given by
\be
N_{max} \simeq \frac{M}{m} \simeq \frac{M}{\frac{\hbar}{2GM}}  \simeq \frac{(GM)^2}{G\hbar}  \simeq \frac{A}{G\hbar}~,
\ee
which is the right order-of-magnitude for the black hole entropy.

This is fine for a rough estimate, but how can one derive the black hole entropy statistical mechanically? The Planck length $\lpl$ appears in the denominator of the area law \eqref{eq:BH_entropy}. This means that the microscopic derivation of the black hole entropy is likely to require quantum gravity. But the formulation of quantum gravity is a difficult problem, so the problem of the black hole entropy is not completely resolved. String theory is the prime candidate of the unified theory including quantum gravity, and string theory can derive the \bh entropy microscopically for some black holes (see, \eg, Refs.~\cite{Horowitz:1996qd,Dabholkar:2012zz} for reviews).

In \sect{Euclidean}, we briefly discussed the semiclassical computation of the black hole entropy. It is natural that such a computation gives the black hole entropy and is not just a coincidence. There should be something behind which justifies the computation. But one can say that it is not satisfactory as the microscopic derivation of the black hole entropy, namely as the statistical derivation of the black hole entropy: 
\begin{itemize}
\item First, \eq{Euclidean} itself is not well-defined and cannot be estimated beyond the semiclassical approximation. 
\item Also, the computation does not tell us what degrees of freedom the black hole entropy corresponds to. 
\end{itemize}

But AdS/CFT gives an interesting interpretation to the semiclassical result. AdS/CFT identifies the partition function computed in this way as the partition function of a gauge theory. AdS/CFT claims the equivalence between two theories, a gauge theory and a gravitational theory, and there are two points of view one can take: 
\begin{itemize}
\item Namely, one can use the gravitational theory to know about the gauge theory. 
\item Or, one can use the gauge theory to know about the gravitational theory. 
\end{itemize}
Because AdS/CFT claims the equivalence, it is hard to say which point of view is more fundamental, namely which theory is more fundamental, but let us take the latter point of view for now. If so, one can regard the origin of the black hole entropy as the entropy of the corresponding gauge theory, namely the number of microscopic states of the gauge theory. Then, the black hole entropy is nothing but the usual statistical entropy. 

However, black holes in AdS/CFT are the ones in AdS spacetime. It is not clear if one can make such an interpretation for any black holes, in particular for the Schwarzschild black hole. The holographic principle seems to suggest that this is somehow possible. 

\section{Other black holes}

\subsection{Higher-dimensional Schwarzschild black holes}
\label{sec:higher}

The four-dimensional Schwarzschild black hole can be generalized to $d$-dimensional spacetime:
\begin{align}
ds_d^2 &= - f(r) dt^2 + \frac{dr^2}{f(r)} 
+ r^2 d\Omega_{d-2}^2~,
\label{eq:Schwarzschild_d_dim} \\
f(r) &= 1-\left(\frac{r_0}{r}\right)^{d-3} = 1- \frac{16 \pi G_d}{(d-2) \Omega_{d-2}} \frac{M}{r^{d-3}}~,
\label{eq:Schwarzschild_d_mass} \\
\Omega_n &= \frac{2\pi^{\frac{n+1}{2}}}{\Gamma(\frac{n+1}{2})}~.
%
\end{align}
Here, $G_d$ is the Newton's constant in $d$-dimensional spacetime, $d\Omega_n^2$ is the line element of the unit $S^n$ (see \sect{higher_AdS} for an explicit construction), and $\Omega_n$ is its area which is written by the Gamma function $\Gamma(n+1/2)$. Some explicit forms of $\Omega_n$ are given by
\begin{gather}
\Omega_0 = 2~, \quad
\Omega_1 = 2 \pi~, \quad
\Omega_2 = 4 \pi~, \quad
\Omega_3 = 2 \pi^2~, \nonumber \\
\Omega_4 = \frac{8}{3} \pi^2~, \quad
\Omega_5 = \pi^3~, \quad
\Omega_6 = \frac{16}{15} \pi^3~, \quad
\Omega_7 = \frac{1}{3} \pi^4~.
\ldots
\end{gather}
The behavior $1/r^{d-3}$ in $f(r)$ comes from the fact that the Newtonian potential behaves as $1/r^{d-3}$ in $d$-dimensional spacetime. 

Using Eqs.~\eqref{eq:BH_entropy} and \eqref{eq:Hawking_temp}, one gets thermodynamic quantities:
\begin{align}
T &= \frac{f'(r_0)}{4\pi} 
= \frac{d-3}{4\pi r_0}~,
\label{eq:temp_Sch_d} \\
\Sbh &= \frac{A}{4G_d} 
= \frac{r_0^{d-2} \Omega_{d-2}}{4G_d}~.
\label{eq:entropy_Sch_d}
%
\end{align}
One can check the first law with these quantities and the mass \eqref{eq:Schwarzschild_d_mass}.

\subsection{Black branes}\label{sec:black_branes}

The black holes in previous subsection are simply the extensions of the four-dimensional black hole, but in higher dimensions, there exist black holes which do not appear in four-dimensions. The Schwarzschild \bh has a spherical horizon, but the horizon can take other topologies in higher dimensions%
\footnote{For four-dimensional asymptotically flat black holes, the topology of black hole horizons must be $S^2$ under appropriate conditions on matter fields. This is known as the \keyword{topology theorem} (see, \eg, Proposition 9.3.2 of Ref.~\cite{HE}) which is part of the no-hair theorem. }.
In particular, the horizon can have an infinite extension. 

A simple example is the five-dimensional ``neutral black string," which is the four-dimensional Schwarzschild \bh times $\mathbb{R}$:
\be
ds_5^2 = - \left(1-\frac{r_0}{r} \right) dt^2 + \frac{dr^2}{1-\frac{r_0}{r}} + r^2 d\Omega_2^2 + dz^2~,
\label{eq:black_string}
\ee
where $z$ represents the $\mathbb{R}$ coordinate. For the five-dimensional Schwarzschild black hole, the horizon, the $r=r_0$ surface has the line element $r_0^2 d\Omega_3^2$ which represents the $S^3$-horizon. Here, the horizon is still given by $r=r_0$, but the horizon has the line element $r_0^2 d\Omega_2^2 + dz^2$. So, the horizon has the topology $S^2\times\mathbb{R}$ and extends indefinitely in $z$-direction. Such a horizon is called a \keyword{planar horizon}. When we say the word ``black hole" in this book, it is mostly a \bh with a planar horizon. A black hole with planar horizon is often called a \keyword{black brane}.

The black branes arise with various dimensionalities. A black brane with $\mathbb{R}^p$ horizon is called a black $p$-brane. Namely, 
\begin{center}
\begin{tabular}{cc}
$ p=0 $		& black hole \\
$ p=1 $		& black string \\
$ p=2 $		& \hspace{5mm} black membrane \\
$ \vdots $	& $ \vdots $
\end{tabular}
\end{center}
To obtain thermodynamic quantities of the neutral black string, notice that the black string is just the four-dimensional Schwarzschild \bh if one compactifies the $z$-direction. Then, the black string and the four-dimensional Schwarzschild \bh have the same thermodynamic quantities. Take the periodicity of $z$ as $V_1$. By the compactification,
\be
\frac{1}{16\pi G_{5}} \int d^{5}x \sqrt{-g_{5}}R_{5} 
=
\frac{V_1}{16\pi G_{5}} \int d^{4}x \sqrt{-g_{4}}\left(R_{4} +\cdots \right)~, 
%
\ee
where the dots denote matter fields which appear by the compactification. Thus, the four-dimensional Newton's constant $G_4$ is written as $1/G_4=V_1/G_5$ from the five-dimensional point of view. Then, thermodynamic quantities are
\be
T = \frac{1}{4\pi r_0}~, \quad
S = \frac{\pi r_0^2}{G_4} = \frac{\pi r_0^2}{G_5}\, V_1~, \quad
M = \frac{r_0}{2G_4} = \frac{r_0}{2G_5}\, V_1~, \quad
P= - \frac{r_0}{4G_5}~,
\label{eq:black_string_thermo}
\ee
where we used the first law $dM=TdS-PdV_1$ to derive the pressure $P$. The entropy of the black string is rewritten as 
\be
S_\text{string} = 4\pi G_5 \frac{M^2}{V_1}~.
\label{eq:black_string_fund}
\ee

\subsubsection*{Gregory-Laflamme instability \advanced}

\begin{figure}[tb]
\centering
\scalebox{0.75}{ \includegraphics{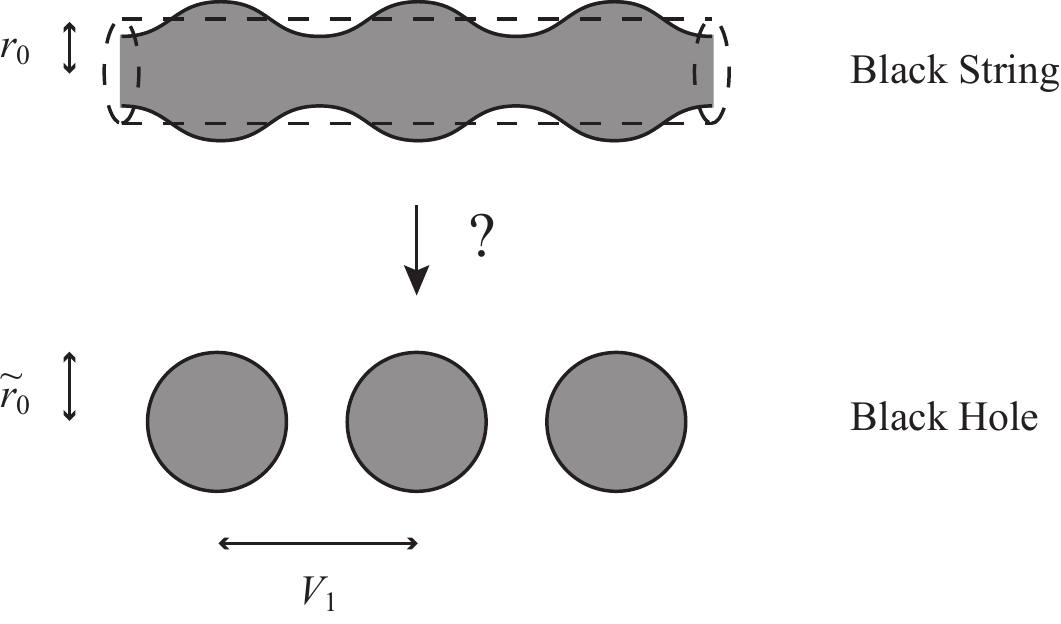} }
\caption{Gregory-Laflamme instability.}
\label{fig:GL_instability} 
\end{figure}

As a side remark, the neutral black string actually has the so-called \keyword{Gregory-Laflamme instability} \cite{Gregory:1993vy} (see, \eg, Refs.~\cite{Horowitz:2012nnc2,Harmark:2007md} for reviews). The black string is not the only neutral solution allowed, but the five-dimensional Schwarzschild \bh is also allowed.
%
%
Even though the $z$-direction is compactified, the metric \eqref{eq:Schwarzschild_d_dim} should be good approximately when the horizon radius $\tilr_0$ satisfies $\tilr_0 \ll V_1$. The thermodynamic quantities behave as
\be
M \simeq \frac{\tilr_0^2}{G_5}~,  \quad
S_\text{BH$_5$} \simeq \frac{\tilr_0^3}{G_5} \simeq \sqrt{G_5}M^{3/2}~.
%
\ee
On the other hand, the entropy of the black string behaves as \eq{black_string_fund}. Consider the microcanonical ensemble with fixed $M$. Then, $S_\text{BH$_5$}>S_\text{string}$ for a large enough $V_1$, so the \bh is favorable entropically. Then, the black string is unstable and would decay to the black hole, but it is still not clear if the transition really occurs (\fig{GL_instability}). Anyway, it has been shown numerically that the neutral black string is unstable under perturbations%
\footnote{It has been argued that the Gregory-Laflamme instability is related to the Rayleigh-Plateau instability in hydrodynamics \cite{Cardoso:2006ks}. \index{Rayleigh-Plateau instability}}.

\subsection{Charged black holes
}\label{sec:RN}

A black hole can have a charge $Q$. The solution is known as the \keyword{Reissner-Nordstr\"{o}m black hole} (RN black hole hereafter):
\begin{align}
ds^2 &= - \left( 1-\frac{2GM}{r}+\frac{GQ^2}{r^2} \right) dt^2 
+ \frac{dr^2}{1-\frac{2GM}{r}+\frac{GQ^2}{r^2}} + r^2 d\Omega_2^2 
\\
&= - \left( 1-\frac{\ro}{r} \right)\left( 1-\frac{\ri}{r} \right) dt^2 
+ \frac{dr^2}{\left( 1-\frac{\ro}{r} \right)\left( 1-\frac{\ri}{r} \right)} + r^2 d\Omega_2^2~,
\label{eq:RN} \\
F_{r0} &= \frac{Q}{r^2}~,
\label{eq:RN_Maxwell}
\end{align}
where $2GM = \ro+\ri$ and $GQ^2 = \ro \ri$. We chose the dimensions of $Q$ as dimensionless. (The charge is dimensionless in the units $c=\hbar=1$.) 

The RN black hole is a solution of the Einstein-Maxwell theory%
\footnote{Note the normalization of the Maxwell action. This choice is standard for the RN black hole.}: 
\be
\action = \int d^{4}x\, \sqrt{-g} \left( \frac{1}{16\pi G} R - \frac{1}{16\pi} F^2 \right)~.
%
\ee
The equations of motion from the above action is given by
\begin{align}
R_{\mu\nu} - \frac{1}{2} g_{\mu\nu} R 
&= 2G\left( F_{\mu\rho} F_\nu^{~\rho} - \frac{1}{4} g_{\mu\nu} F^2 \right)~, \\
\nabla_\nu F^{\mu\nu} &= 0~.
%
\end{align}
Here, we check only the Maxwell equation:
\be
\nabla_\nu F^{\mu\nu} = \frac{1}{\sqrt{-g}} \del_\nu(\sqrt{-g}F^{\mu\nu}) = 0~.
%
\ee
When a black hole has an electric charge, $F_{r0}=E_r \neq 0$, and the only nontrivial equation is $\del_r(\sqrt{-g}F^{r0}) = 0$. For the above metric, $F^{r0}=-F_{r0}$ and $\sqrt{-g} = r^2 \sin^2\theta$, so the solution to the Maxwell equation is given by \eq{RN_Maxwell}.
%
%
The field strength $F_{\mu\nu}$ is written by the gauge potential $A_\mu$ as $ F_{\mu\nu} = \del_\mu A_\nu - \del_\nu A_\mu$. For our electric field, one may choose the gauge potential $A_0$ as
\be
A_0 = - \frac{Q}{r}~.
%
\ee

The black hole \eqref{eq:RN} has two horizons $r=r_\pm$, where 
\be
r_\pm = GM \pm \sqrt{G^2M^2 - GQ^2}~.
\label{eq:outer&inner}
\ee
But note that
\begin{itemize}
\item 
When $\sqrt{G}M>Q$, two horizons exist.
\item
When $\sqrt{G}M<Q$, there is no horizon since \eq{outer&inner} becomes complex.
\item
The limiting case $\sqrt{G}M=Q$ or $\ro=\ri$ is called the \keyword{extreme black hole}.
\end{itemize}
Also, the \bh has a spacetime singularity at $r=0$.

\subsubsection*{Thermodynamic quantities}

Again using Eqs.~\eqref{eq:BH_entropy} and \eqref{eq:Hawking_temp}, one gets
\begin{align}
T 
&= \frac{\ro - \ri}{4\pi \ro^2}
= \frac{1}{2\pi} \frac{ \sqrt{G^2M^2-GQ^2} }{ 2GM \left( GM+\sqrt{G^2M^2-GQ^2} \right) - GQ^2 }~, \label{eq:temp_RN} 
\\
\Sbh 
&= \frac{ 4\pi \ro^2 }{4G}
= \frac{\pi}{G} \left( GM+\sqrt{G^2M^2-GQ^2} \right)^2~.
\label{eq:entropy_RN}
\end{align}
The black holes with $\sqrt{G}M>Q$ emit the Hawking radiation and eventually settle into the extreme black hole. The extreme black hole has zero temperature%
\footnote{
Since the Hawking temperature is proportional to the surface gravity, a zero temperature black hole has zero surface gravity. Then, does the black hole have no gravitational force? As we saw previously, the surface gravity is the gravitational force measured by the asymptotic observer. But the asymptotic observer and the observer near the horizon disagree the value of the force because of the gravitational redshift. Although there is a gravitational force for the observer near the horizon, it vanishes for the asymptotic observer. }.

When a thermodynamic system exchanges particle numbers or charges with the environment, the first law becomes
\be
dM = T d\Sbh + \mu dQ~,
\label{eq:1st_law_charged}
\ee
where $\mu$ is the ``chemical potential." This is just the gauge potential $A_0$ in this case. However, the value of $A_0$ itself is not meaningful due to the gauge invariance:
\be
A_0(x) \to A_0(x) + \del_0 \Lambda(x)~.
\label{eq:bulk_gauge}
\ee
What is physical is the difference of the gauge potential. 

Imagine to add a charge $\Delta Q$ from the asymptotic infinity to the black hole. This requires energy $\Delta E$ proportional to the difference of $A_0$ between the asymptotic infinity and the horizon (for a same sign charge as the black hole). Then, $\mu$ should be given by
\be
\mu = \left. A_0 \right|_{r=\infty} - \left. A_0 \right|_{r=\ro}~,
\label{eq:def_chemical1}
\ee
and $\Delta E = \mu\Delta Q$ which takes the form of the first law. Thus,
\be
\mu = \frac{Q}{\ro} = \sqrt{ \frac{1}{G}\frac{\ri}{\ro} }~.
%
\ee
Using these thermodynamic quantities, one can show the first law \eqref{eq:1st_law_charged}%
\footnote{On the other hand, for the RN black hole (and for the Schwarzschild black hole), the other thermodynamic relations such as the Euler relation and the Gibbs-Duhem relation do not take standard forms (\sect{BH_fund_rels}).}.

\subsubsection*{Third law}

According to the third law of thermodynamics, $S\rightarrow0$ as $T\rightarrow0$ (Nernst-Planck postulate). On the other hand, the entropy of the RN black hole remains finite as $\sqrt{G}M\rightarrow Q$, so the black hole seems to violate the third law.\index{third law of thermodynamics (black hole)} 

This point is still controversial, but a finite entropy at zero temperature means that ground states are degenerate, and there is nothing wrong about degenerate ground states quantum mechanically. However, such a degeneracy is often true only in idealized situations, and the degeneracy is normally lifted by various perturbations. Similarly, the degeneracy of the RN black hole may be lifted by various interactions. 
\section{\titlesummary}

\begin{itemize}
\item 
A \bh behaves as a thermodynamic system and has thermodynamic quantities such as energy (mass), entropy, and temperature.
\item 
The \bh entropy is proportional to the horizon area. 
This suggests that a black hole can be described by 
the usual statistical system whose spatial dimension is one dimension 
lower than the gravitational theory (holographic principle).
\item 
To derive \bh thermodynamic quantities, use the Euclidean path integral formulation and evaluate the path integral by the saddle-point approximation.
\item
These are various black objects (black branes) in higher dimensions.
\end{itemize}

\titlenewterms

\begin{multicols}{2}
\noindent
black hole thermodynamics\\
black hole entropy\\
no-hair theorem\\
surface gravity\\
Hawking temperature\\
Planck length\\
area law\\
holographic principle\\
Euclidean formalism\\
planar horizon\\
black brane\\
Reissner-Nordstr\"{o}m black hole\\
extreme black hole
\end{multicols}

\section{Appendix: Black holes and thermodynamic relations \advanced}\label{sec:BH_fund_rels}

We saw that black holes satisfy the first law. The first law should be always satisfied because the law simply represents the energy conservation. On the other hand, the other thermodynamic relations such as the Euler relation and the Gibbs-Duhem relation do not take standard forms for black holes with \textit{compact horizon}. The story is different for black branes though: these relations take standard forms like standard thermodynamic systems. 

For example, thermodynamic quantities of the Schwarzschild \bh are given by
\be
T = \frac{1}{4\pi r_0}~, \quad
S = \frac{\pi r_0^2}{G_4}~, \quad
E = \frac{r_0}{2G_4}~.
\label{eq:BH_thermo}
\ee
So, the Euler relation does not take the form $E=TS$, but 
\be
E=2TS~,
%
\ee
which is known as the Smarr's formula.\index{Smarr's formula}

For black holes with compact horizon, thermodynamic relations do not take standard forms because these black holes do not satisfy fundamental postulates of thermodynamics (\sect{thermodynamics}). In particular, thermodynamics requires that the entropy is additive over the subsystems. This requirement is unlikely to hold in the presence of a long-range force such as gravity. Another related problem is that the volume $V$ and its conjugate quantity, pressure $P$, do not appear in those black holes (at least as independent variables.)

The postulate in particular implies that the so-called \keyword{fundamental relation} of a standard thermodynamic system 
\be
S = S(E, V)
%
\ee
is a homogeneous \textit{first order} function of extensive variables:
\be
S( \lambda E, \lambda V) = \lambda S(E, V)~.
\ee
Then, the Euler relation and the Gibbs-Duhem relation follow from the fundamental relation. However, if one rewrites \eq{BH_thermo} in the form of the fundamental relation, one gets
\be
S_\text{BH$_4$} = 4\pi G_4 E^2~,
\label{eq:BH_fund}
\ee
which is a homogeneous \textit{second order} function in $E$. Thus, it is no wonder that thermodynamic relations do not take standard forms.

However, the story is different for branes. For branes, one has an additional extensive variable, volume $V$, and the entropy is additive over the subsystems along the brane direction $V=V_A+V_B$. In this sense, the entropy of branes is additive, and branes satisfy the standard postulates of thermodynamics. A simple example is the neutral black string \eqref{eq:black_string}. In \eq{black_string_fund}, we wrote the  entropy of the black string in the form of the fundamental relation:
\be
S_\text{string} = 4\pi G_5 \frac{E^2}{V_1}~.
\label{eq:black_string_fund_again}
\ee
This is indeed a homogeneous first order function in extensive variables. As a result, one can check that the black string does satisfy the standard thermodynamic relations $E = TS-PV_1$ and $SdT-V_1dP=0$. Note that Eqs.~\eqref{eq:BH_fund} and \eqref{eq:black_string_fund_again} are actually the same equation since $V_1/G_5=1/G_4$. 

\endofsection

\ifx\nameofpaper\undefined 
  \usepackage{macro_natsuume} 
  \def\beginsection{\section*}
  \def\endofsection{\end{document}} 
  \input draft_header.tex
\else 
  \def\beginsection{\chapter}
  \def\endofsection{ } 
\fi

\beginsection{Strong interaction and gauge theories}\label{chap:QCD}


\begin{quote}
So far we discussed black holes, but this chapter describes another important element of AdS/CFT,  gauge theory. After we overview the theory of strong interaction, namely QCD, its phase structure, and heavy-ion experiments, we explain the idea of the large-$N_c$ gauge theory. This idea naturally leads us to AdS/CFT.
\end{quote}

\section{Strong interaction and QCD}\label{sec:QCD}

\subsection{Overview of QCD}

Protons and neutrons are made out of more fundamental particles, \keyword{quarks}. Under normal circumstances, quarks are bound together inside protons and neutrons, and one cannot liberate them. The strong interaction is the force which bounds quarks together. 
The theory of the strong interaction is described by a gauge theory called \keyword{quantum chromodynamics}, or QCD%
\footnote{In \sect{YM}, we summarize the formalism of the gauge theory for the readers who are not familiar to it.}. 
QED is a gauge theory based on $U(1)$ gauge symmetry, and QCD is based on $SU(3)$ gauge symmetry. The strong interaction is meditated by \keyword{gluons}, which are the $SU(3)$ gauge fields.

A quark belongs to the \keyword{fundamental representation} of $SU(3)$ and can take 3 states. These degrees of freedom are called \keyword{color}; this is the origin of the word `chromo' in quantum chromodynamics. If we denote the color degrees of freedom by index $i$, a quark can be written as $q_i~(i=1,2,3)$%
\footnote{Quarks have the additional degrees of freedom, \keyword{flavor}. There are 6 flavors, up $u$, down $d$, strange $s$, charm $c$, bottom $b$, and top $t$.}.
An antiquark $\bar{q}^i$ has an anticolor. For the electromagnetic interaction, an electric charge is the source of the interaction; similarly, for the strong interaction, a ``color charge" is the source of the interaction.

A quark has a color, but the quark color typically changes by a quark-gluon interaction. The difference of color is carried off by the gluon. Thus, the gluon carries one color and one anticolor. A gluon behaves as a quark-antiquark pair%
\footnote{This does not mean that a gluon is a quark-antiquark pair. It simply means that a gluon transforms like a quark-antiquark pair under $SU(3)$.} 
$q_i \bar{q}^j$. 
Then, using a $3 \times 3$ matrix on index $(i,j)$, one can write a gluon as 
$(A_\mu)_i^{~j}$. 
Namely, a gluon belongs to the \keyword{adjoint representation} of $SU(3)$.

A gluon, the force carrier itself has color charges, so a gluon has the self-interactions. This point is very different from QED; the force carrier of the electromagnetic force, photon, is electrically neutral. This aspect changes the dynamics of QCD from QED drastically.

The QCD interaction becomes weak at high energy (\keyword{asymptotic freedom}), and one can rely on the perturbative QCD. This was helpful to confirm QCD as the theory of strong interaction since the predictions from the perturbative QCD can be compared with the results from the deep inelastic scattering experiments. On the other hand, the QCD interaction becomes stronger at low energy. Because of this property, quarks and gluons are confined into particles collectively known as \keyword{hadrons}, and only the ``color-neutral" particles appear%
\footnote{Strictly speaking, one should say that only color singlets appear.}.
This is known as the \keyword{color confinement}, but this has not been proven yet.

QCD has the invariance under the quark phase transformation $q \rightarrow e^{i\theta} q$. This is the global $U(1)_B$ symmetry associated with the baryon number conservation.

\subsection{Phase structure of QCD}\label{sec:QCD_phase}

\begin{figure}[tb]
\centering
\scalebox{0.40}{ \includegraphics{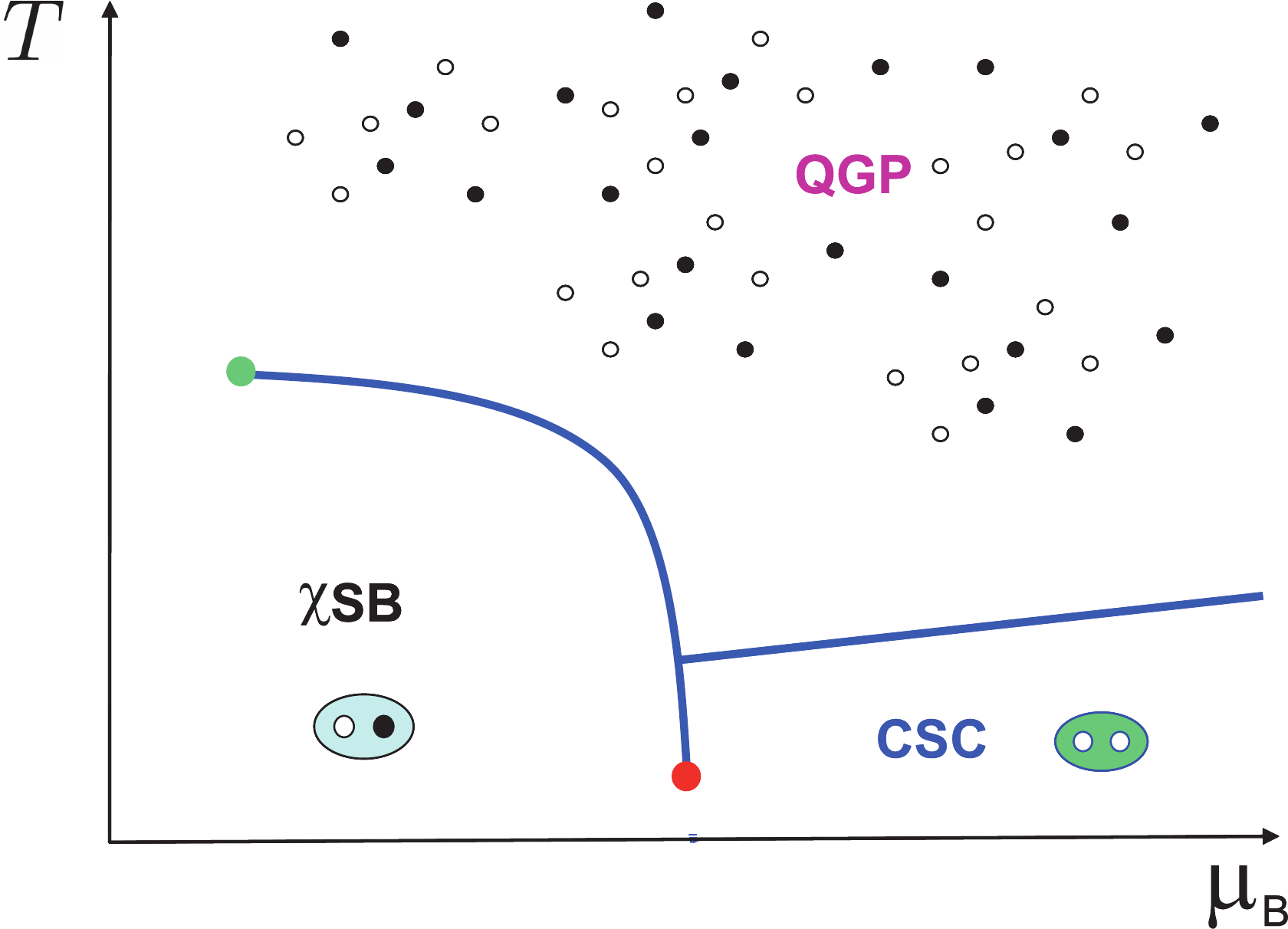} }
\vskip2mm
\caption{QCD phase diagram \cite{Hatsuda:2007rt}. The vertical axis represents temperature, and the horizontal axis represents the so-called ``baryon chemical potential," which is associated with the baryon number conservation (see footnote~\ref{fnote:chemical} in \sect{thermodynamics} for the use of the terminology chemical potential). In the figure, ``$\chi$SB" (which stands for chiral symmetry breaking) represents the hadron phase, and ``CSC" represents the so-called color superconducting phase. This is just a schematic diagram, and the full details of the phase diagram are still unclear; at this moment, it is not clear if there are any other phases in this diagram, and the boundaries of each phases are not clear either. 
}
\label{fig:phase}
\end{figure}

QCD becomes strongly-coupled at low energy, and one cannot rely on the perturbation theory. But this situation suggests that QCD has rich phenomena, richer than QED. In fact, QCD has a rich phase structure, and the full details of the phase structure are still unclear (\fig{phase})%
\footnote{Such a rich phase structure is not limited to QCD and is common to Yang-Mills gauge theories. The large-$N_c$ gauge theories seem to have a rich phase structure in general as we will see gradually.}.
The region ``$\chi$SB" in this figure is the phase where quarks are confined. 

At high temperature, many light particles are excited, and the interaction between quarks is screened by light particles in between. This is the \keyword{Debye screening}, a phenomenon which also occurs in QED%
\footnote{What is screened in QCD is not electric charge but color charges. As a result, the strong interaction becomes weak.}. 
The \keyword{quark-gluon plasma}, or QGP, is the phase where quarks and gluons are deconfined by the screening%
\footnote{See Ref.~\cite{qgp} for the textbook in QGP.}. 

Just like the usual electromagnetic plasma, QGP is made of ``charged ionized" particles in a sense. The differences lie in the meaning of the ``charge" and ``ionization." For QGP, we mean color charges, not electric charges. The quarks of course carry electric charges as well, but it is not important here. Also, what we meant by ionization is the deconfinement for QGP. The transition to QGP is estimated as $T_c \approx 150-200 \text{MeV} \approx 1 \text{fm} \approx 2 \times 10^{12}$K. However, according to lattice simulations, this is not a sharp phase transition but a smooth ``cross-over" at zero baryon chemical potential. (In \fig{phase}, the solid line which represents the phase boundary does not cross the vertical axis, which means the cross-over.)

Naively, one would expect that the perturbation theory is valid in the plasma phase since the deconfinement occurs there. It is true that the plasma should behave like a free gas at high enough temperatures. But the strong interaction does not become weak enough at the temperatures one can realize in current experiments. This is because the QCD coupling constant decreases only logarithmically with energy, $1/\log E$. In fact, there is an experimental indication that QCD is still strongly-coupled in the plasma phase (\sect{unexpected}). But then, even if one would like to discuss plasma properties theoretically, one cannot rely on the perturbative QCD. We need another approach to study QGP.

The lattice simulation is a useful approach at strong coupling, which becomes very powerful in recent years. But this approach is the imaginary-time formalism not the real-time formalism. As a consequence, this is very powerful for equilibrium problems but is not powerful yet for nonequilibrium problems which are the situation at QGP experiments. Also, currently the analysis at finite density is difficult as well.

\subsection{Heavy-ion experiments}\label{sec:heavy_ion}

The quark-gluon plasma is a natural phenomenon from QCD. In fact, theoretically it has been known for a long time. But it was never formed in an experiment, let alone to measure physical properties. But in recent years, there exist \keyword{heavy-ion experiments} whose goals are to form QGP and to measure its physical properties. 

Particle accelerators usually collide $e^+ e^-$, $pp$ or $p\bar{p}$, but the heavy-ion experiments collide heavy nuclei such as gold nuclei. Heavy atoms themselves are electrically neutral. In order to accelerate them, one has to strip electrons and to ionize them. So, such an experiment is called a heavy-ion experiment.

One notable heavy-ion accelerator is the Relativistic Heavy Ion Collider (RHIC\index{RHIC}), which is located at Brookhaven National Laboratory and is in operation since 2000. Another heavy-ion experiment has been performed at CERN since 2010 using the Large Hadron Collider (LHC\index{LHC}). The plasma temperature by these colliders are estimated as about $2T_c$ at RHIC and about $5T_c$ at LHC.

In heavy-ion experiments, one first collides heavy nuclei, and the plasma is formed if the temperature is high enough. But the plasma exists only transiently; the high temperature of the plasma causes it to expand, and then the plasma cools down by the expansion. In the end, the temperature is below the transition temperature, so quarks are confined into hadrons. What is observed is these secondary particles. The above history shows the difficulties of the experiment and its analysis:
\begin{itemize}

\item
The plasma has a very complicated time-evolution so that its analysis is not easy.

\item
What is observed is not QGP itself but only its by-products such as hadrons. One has to infer what happened from the by-products. Thus, it is not an easy job to confirm even the QGP formation.

\item
Many secondary particles are involved in the experiments. The purpose here is the QGP property as a thermal system, so one has to keep track of these particles as many as possible. 

\item
Finally, the perturbative QCD is not very effective for theoretical analysis as discussed below.

\end{itemize}

\subsection{``Unexpected connection between string theory and RHIC collisions"}\label{sec:unexpected}

In 2005, the first detailed RHIC results were announced at the APS annual meeting. This press release has an interesting statement \cite{press}: ``The possibility of a connection between string theory and RHIC collisions is unexpected and exhilarating."%
\footnote{As far as I know, this is the first time string theory has been mentioned in the announcement of a major experiment. }
(Director of the DOE Office of Science) Here, ``string theory" meant the AdS/CFT duality. What is the ``unexpected connection" between string theory and RHIC collisions?

In the RHIC announcement, it was stressed that \textit{quark-gluon-plasma does not behave like a free gas but behaves more like a perfect fluid}%
\footnote{Actually, the announcement stated that QGP behaves like a perfect ``liquid" to avoid the technical term ``fluid," but it was slightly misleading. See footnote~\ref{fnote:liquid} for more details.}.\index{perfect fluid} 
The difference between a free gas and a perfect fluid is that the latter has no viscosity. We will discuss the viscosity and hydrodynamics in \chap{hydro} in details, but let us remind you of freshman physics of viscosity. 

\begin{figure}[tb]
\centering
\includegraphics{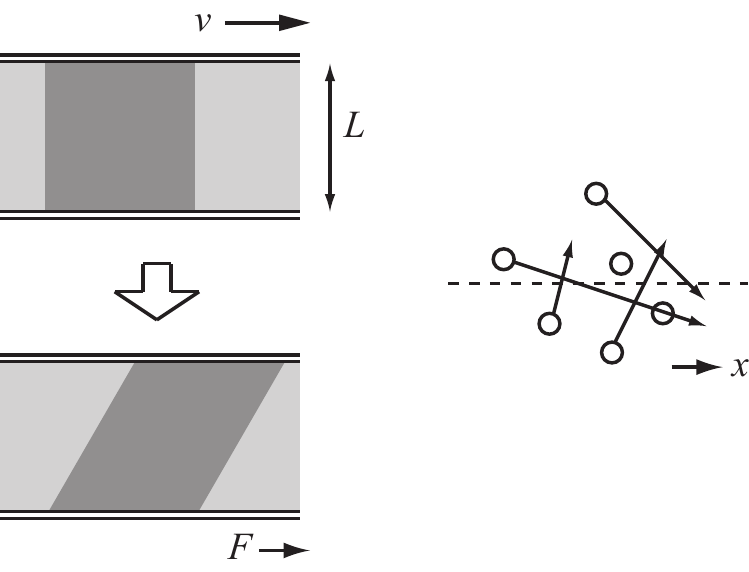}
\vskip2mm
\caption{\textit{Left}: when one moves the upper plate, the fluid is dragged due to the viscosity and the lower plate experiences a force. \textit{Right}: the close-up view of the fluid.
}
\label{fig:viscosity}
\end{figure}%

As a simple example, consider a fluid between two plates and move the upper plate with velocity $v$ (\fig{viscosity}). As the fluid is dragged, the lower plate experiences a force. This force is the manifestation of the viscosity. In this case, the force the lower plate experiences per unit area $F/A$ is given by
\be
\boxeq{
\frac{F}{A}=\eta \frac{v}{L}~.
}
\label{eq:viscosity}
\ee
The proportionality constant $\eta$ is called the \keyword{shear viscosity}.

Microscopically, the viscosity arises due to the momentum transfer between molecules. Figure~\ref{fig:viscosity}(right) shows a close-up view of the fluid and we put an artificial boundary to divide the fluid into two parts. The molecules collide with each other and are exchanged randomly through the boundary. But in the situation where we move the upper plate, the molecules in the upper-half part, on average, have more momentum in the $x$-direction than the ones in the lower-half part. These molecules are exchanged, which means that momentum in the $x$-direction is transported through the boundary. This is the microscopic origin of the viscosity.

The shear viscosity has units of 
$ [\eta] = \pascal\cdot\second = \kilogram\cdot\meter^{-1}\cdot\second^{-1} $
from \eq{viscosity}. Thus, one would expect from dimensional analysis that the shear viscosity is given by
\be
\eta \simeq \rho \bar{v}l_\text{mfp}  \simeq \varepsilon \tau_\text{mfp}
\label{eq:viscosity_micro}
\ee
($\rho$: mass density, $\bar{v}$: mean velocity of particles, $l_{\rm mfp}$: mean-free path, $\varepsilon$: energy density, $\tau_\text{mfp}$: mean-free time).
Now, change the coupling constant. At strong coupling, the mean-free path becomes shorter. The viscosity arises due to the momentum transfer, and the transfer is less effective at strong coupling. So, the viscosity becomes smaller at strong coupling. In particular, a perfect fluid has no viscosity, so a perfect fluid is a strong coupling limit. This is somewhat confusing, but an ideal gas is the free limit, and a perfect fluid is the strong coupling limit%
\footnote{
The argument here has some limitations. First, the fluid must be a Newtonian fluid. [The Newtonian fluids are the fluids which satisfy \eq{viscosity}.] Second, among Newtonian fluids, we consider only the fluids where the momentum transfer occurs due to the mechanism described in the text. For example, ``honey" satisfies neither of these conditions. 

This is an appropriate place to go back to the RHIC announcement of ``perfect liquid." In our discussion, the fluid is always gas-like and the momentum transfer is kinematic (\ie, the momentum is transferred by  particle motion), whereas in a liquid, the momentum transfer by potential energy among particles is more effective (see, \eg, Ref.~\cite{Csernai:2006zz}). What we have in mind is the strong coupling limit of the QGP gas. In this sense, saying that QGP is a perfect liquid is slightly misleading. But in the strong coupling limit, the distinction between a gas and a liquid becomes rather subtle.
\label{fnote:liquid}
}.

A small viscosity implies the strong interaction. According to RHIC, QGP has a small viscosity, so this implies that QGP is strongly-coupled. This is an interesting discovery, but then how can one study such a strong coupling problem theoretically? One cannot rely on the perturbative QCD. This is one main obstacles in the QGP research. String theory may be a good place to study such an issue. It turns out that the prediction of the shear viscosity using AdS/CFT is close to the RHIC result (\chap{QGP}). This is the ``unexpected connection" the RHIC announcement meant.

We will further discuss heavy-ion experiments and their results in \sect{QGP}.

\section{Large-$N_c$ gauge theory
}\label{sec:large_N}

In QCD, the perturbation theory has a limited power, and it is still difficult to solve QCD nonperturbatively. An approximation method, the so-called \keyword{large-$N_c$ gauge theory}, was proposed by 't~Hooft, and this idea is naturally related to string theory \cite{'tHooft:1973jz}%
\footnote{See Chap.~8 of Ref.~\cite{coleman} for the details of the large-$N_c$ gauge theory.}.

There are three colors in QCD, but in the large-$N_c$ gauge theory, one considers a large number of colors. For example, consider a $U(N_c)$ gauge theory. The gauge field is written as an $N_c \times N_c$ matrix,  
$(A_\mu)_i^{~j}$.

Then, the theory has two parameters,
\begin{center}
gauge theory coupling constant $g_\text{YM}$, \qquad number of colors $N_c$~.
\end{center}
Instead, we will use $N_c$ and the \textit{'t~Hooft coupling}\index{t~Hooft coupling@'t~Hooft coupling} $\lambda := g_{\rm YM}^2 N_c$ as independent parameters. The \keyword{large-$N_c$ limit} is given by%
\footnote{If one recovers $g_\text{YM}$, the limit implies $g_\text{YM}\rightarrow0$.}
\begin{center}
$N_c\rightarrow\infty$ while $\lambda$ is kept fixed and large.
\end{center}
This is a nontrivial limit as we will see. Namely, the theory does not reduce to the free one in this limit because the effective coupling is given by $\lambda$. Thus, $\lambda\gg1$ is a strong coupling limit. 
%
%
%
%

The large-$N$ limit of, \eg, $O(N)$ vector model often appears in condensed-matter physics. The large-$N_c$ gauge theory has the same spirit but has important differences. The vector model is solvable in this limit, but the gauge theory is hard to solve even in this limit. Also, the vector model essentially reduces to a classical theory. The gauge theory also reduces to a classical theory, but this classical theory is a \textit{classical gravitational theory} as we will see in a moment. 

\begin{figure}[tb]
\centering
\scalebox{0.75}{ \includegraphics{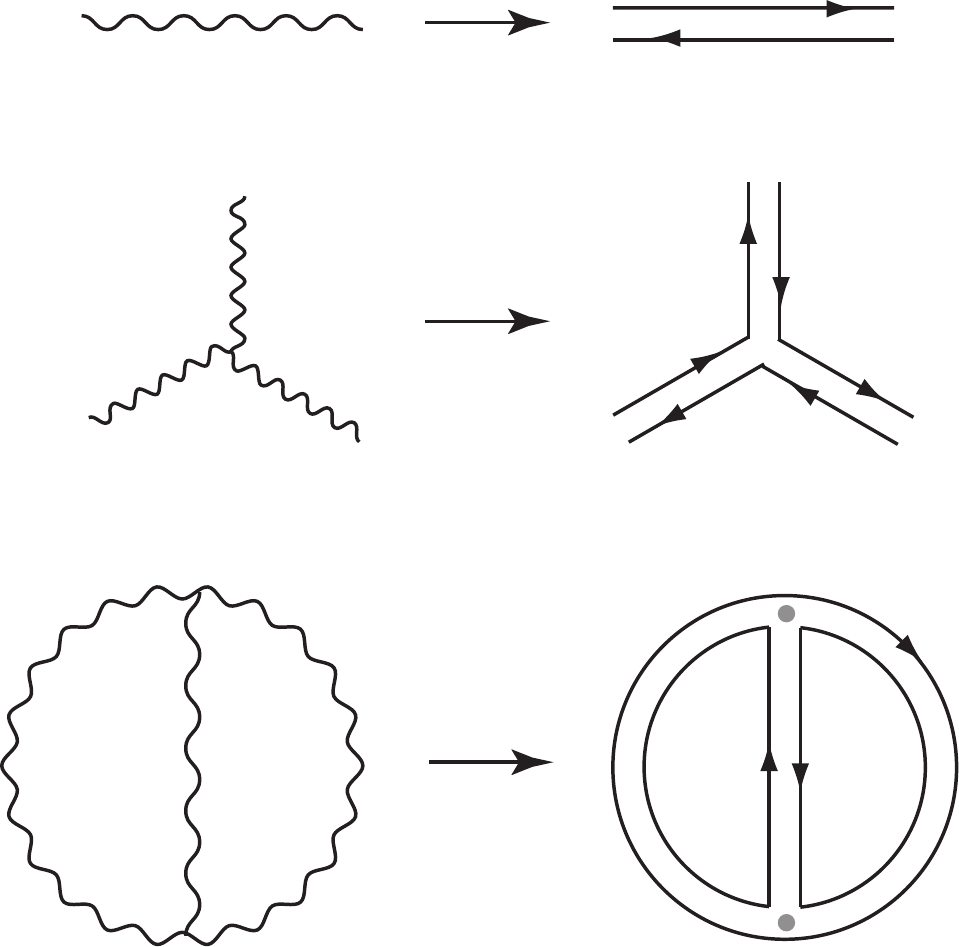} }
\vskip2mm
\caption{The propagator of the gauge field (wavy lines) and the interaction vertex by double-line notations (top, middle). The bottom figure is an example of vacuum amplitudes.
}
\label{fig:double-line-notation}
\end{figure}%

In Feynman diagrams, a gauge theory propagator is represented by a single line. But to keep track of colors, it is convenient to use the ``double-line notation," \index{double-line notation} where the propagator is represented by a double-line with opposite arrows (\fig{double-line-notation}). Then, draw diagrams so that the direction of arrows is preserved. As far as color indices are concerned, the gauge field behaves as a quark-antiquark pair, so one can regard the double-line as the pair. Namely, one line represents the rows of the matrix $(A_\mu)^i_{~j}$, and the other represents the columns. 

Or, in string theory context, a double-line represents the propagation of an open string because an open string represents a gauge field (\chap{string}). A double-line corresponds to the endpoints of a string%
\footnote{We will encounter string theory again in the large-$N_c$ limit below, but this string is not the same as the open string here. String theory appeared below is a closed string theory. }.

\begin{figure*}[tb]
\centering
\scalebox{0.75}{ \includegraphics{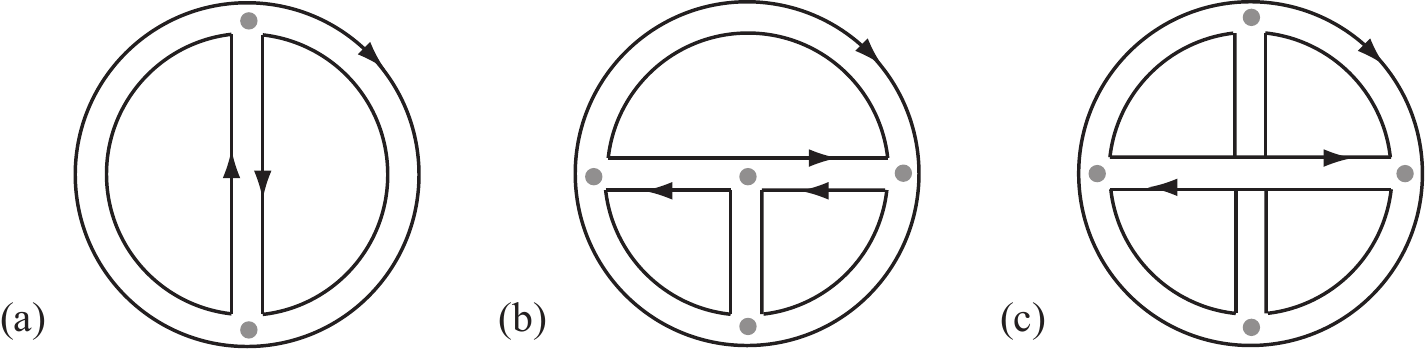} }
\vskip2mm
\caption{Examples of vacuum amplitudes. Fig.~(a) and (b) are planar diagrams, and (c) is a non-planar diagram.
}
\label{fig:diagram}
\end{figure*}

The Lagrangian of a gauge field is schematically written as
\be
{\cal L} = \frac{1}{g_\text{YM}^2} \{ \del A \del A+ A^2 \del A + A^4 \}
= \frac{N_c}{\lambda} \{ \cdots \}~.
%
\ee
From the Lagrangian, one can derive Feynman rules to obtain amplitudes. Here is the summary of rules:
\begin{itemize}


\item
Associate the factor $\lambda/N_c$ for each propagator.

\item
Associate the factor $N_c/\lambda$ for each interaction vertex%
\footnote{In \fig{diagram}, the vertex is represented by a gray dot. Below we will take care only of the factors $N_c$ and $\lambda$ in diagrams, and we will discuss only ``vacuum amplitudes" with no external lines.}.

\item
Associate the factor $N_c$ for each loop (because of the summation over $N_c$ colors).

\end{itemize}
Denote the number of vertices as $V$, the number of propagators as $E$, and the number of loops as $F$. From the Feynman rules, one gets
\be
\left( \frac{N_c}{\lambda} \right)^V
\left( \frac{\lambda}{N_c} \right)^E N_c^F 
= \lambda^{E-V} N_c^{V-E+F}~.
\label{eq:N_counting}
\ee
For example, the diagram~\ref{fig:diagram}(a) has $(V, E, F) = (2,3,3)$, so
\be
\left( \frac{N_c}{\lambda} \right)^2
\left( \frac{\lambda}{N_c} \right)^3 N_c^3
= \lambda N_c^2~.
%
\ee
The diagram~\ref{fig:diagram}(b) has $(V, E, F) = (4,6,4)$, so
\be
\left( \frac{N_c}{\lambda} \right)^4
\left( \frac{\lambda}{N_c} \right)^6 N_c^4
= \lambda^2 N_c^2~.
%
\ee
Thus, these results are summarized schematically as
\be
(1+\lambda+\lambda^2+\cdots) N_c^2 
= f_0(\lambda) N_c^2~.
\label{eq:planar}
\ee

But it is not always the case that diagrams take the form \eqref{eq:planar}. For example, the diagram~\ref{fig:diagram}(c) has $(V, E, F) = (4,6,2)$, so
\be
\left( \frac{N_c}{\lambda} \right)^4
\left( \frac{\lambda}{N_c} \right)^6 N_c^2
= \lambda^2~.
%
\ee
The diagrams such as \fig{diagram}(a), (b) are called \keyword{planar}, and the diagrams such as \fig{diagram}(c) are called \keyword{non-planar}. The word ``planar" (``non-planar")  means that one can (cannot) draw the diagrams on a plane. Including non-planar diagrams, the diagrams are written as
\be
f_0(\lambda) N_c^2 + f_1(\lambda) N_c^0 + f_2(\lambda) \frac{1}{N_c^2} + \cdots~.
\label{eq:planar&nonplanar}
\ee
In the large-$N_c$ limit, the first term of \eq{planar&nonplanar}, the planar diagrams, dominate.

\begin{figure}[tb]
\centering
\scalebox{0.75}{ \includegraphics{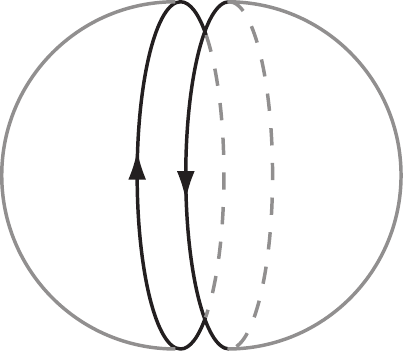} }
\vskip2mm
\caption{Relation between a diagram and a two-dimensional surface.
}
\label{fig:surface}
\end{figure}%

Whether diagrams dominate in the large-$N_c$ limit depends on whether one can draw them on a plane (or a sphere). This strongly suggests the relation between the diagrams and topology. In fact, one can draw the diagram~\ref{fig:diagram}(c) on a torus%
\footnote{This is beyond my ability to draw such an illustration. Anyway, such an illustration is hard to see. I recommend readers to draw such an illustration on their own or to draw diagrams on an object with a handle (such as a doughnut or a mug).}.

\begin{figure}[tb]
\centering
\scalebox{0.75}{ \includegraphics{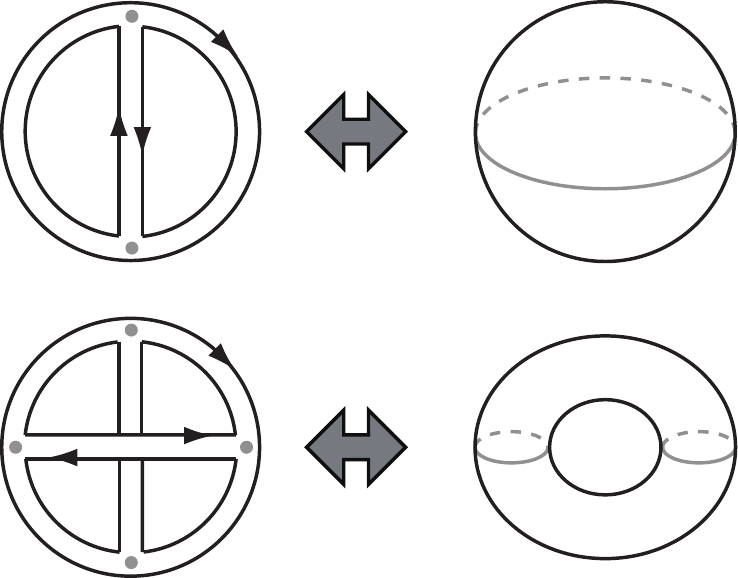} }
\vskip2mm
\caption{Diagrams and topology.}
\label{fig:topology}
\end{figure}

In general, one can understand the relation between these diagrams and topology in the following way. Let us fill in loops with surfaces. One can construct a closed two-dimensional surface (\fig{surface}). Each propagator is an edge of the surface (polygon), and each vertex is a vertex of the polygon. Then, one can regard the number of propagators $E$ as the number of edges, and the number of loops $F$ as the number of faces of the polygons. Now, in this interpretation, the power of $N_c$ in \eq{N_counting} is the Euler characteristic $\chi$, which is a topological invariant: 
\be
\chi = V-E+F~.
%
\ee
For example, a sphere has $\chi=2$, and a torus has $\chi=0$. If we denote the number of handles (genus) as $h$, one can also write $\chi=2-2h$. Namely, the planar diagram \fig{diagram}(a),(b) correspond to the sphere, and the non-planar diagram \fig{diagram}(c) corresponds to the torus (\fig{topology}).

Now, the summation of vacuum diagrams is a partition function, so the partition function is given by
\be
\ln Z_\text{gauge} = \sum_{h=0}^\infty N_c^{\chi} f_h (\lambda)
\label{eq:partition_gauge}
\ee
or
\begin{center}
\fbox{
\begin{tabular}{l}
Second clue of AdS/CFT: \\
The partition function of a large-$\Nc$ theory is given by \\
a summation over the topologies of two-dimensional surfaces.
\end{tabular}
}
\end{center}

As one can see from \fig{diagram}(a), (b), the more propagators we have, the more dominant a diagram is in the large-$N_c$ limit. This limit is hard to evaluate in a field theory. On the other hand, the two-dimensional surface point of view becomes better in this limit. The diagrams become denser in this limit, and they approach a smoother two-dimensional surface.

In the next chapter, we will see that the perturbative expansion of string theory is also given by a summation over two-dimensional topologies (\fig{string_amplitude}). This ``coincidence" leads one to guess that two theories, gauge theory and string theory, are actually equivalent: 
\be
Z_\text{gauge} = Z_\text{string}~.
%
\ee
In particular, a summation of two-dimensional closed surfaces corresponds to the gravitational perturbative expansion. There, the sphere, the planar diagrams, corresponds to classical gravity, and the torus, one of non-planar diagrams, corresponds to gravity at one-loop. In \sect{BH_entropy}, the holographic principle gives the first clue for AdS/CFT, and the large-$N_c$ gauge theory gives us the second clue.  


\section{\titlesummary}

\begin{itemize}
\item 
According to heavy-ion experiments, QGP behaves as a perfect fluid with a very small shear viscosity. This implies that QGP is strongly-coupled.
\item 
It is difficult to compute gauge theories such as QCD at strong coupling. The large-$\Nc$ gauge theory provides an approximation to compute a strong coupling limit.
\item 
The partition function of a large-$\Nc$ theory is given by 
a summation over the topologies of two-dimensional surfaces.
\end{itemize}

\titlenewterms

\begin{multicols}{2}
\noindent
quarks\\
quantum chromodynamics (QCD)\\
gluons\\
fundamental/adjoint representation\\
color\\
flavor\\
asymptotic freedom\\
hadrons\\
color confinement\\
Debye screening\\
quark-gluon plasma (QGP)\\
heavy-ion experiments\\
perfect fluid\\
shear viscosity\\
't~Hooft coupling\\
large-$N_c$ limit\\
double-line notation\\
planar/non-planar\\
topology
\end{multicols}

\section{Appendix: Review of gauge theory}\label{sec:YM}

\newcommand{\Am}{A_\mu}

\head{$U(1)$ gauge theory}

First, recall a $U(1)$ gauge theory. Consider a complex scalar field:
\be
\calL = - \del_\mu \phi^* \del^\mu \phi -m^2 \phi^*\phi~.
%
\ee
The theory is invariant under the global $U(1)$ transformation
\be
\phi \rightarrow U \phi~, \quad (U:= e^{i\alpha})~.
%
\ee
The theory is not invariant under the local $U(1)$ transformation though:
\be
\phi \rightarrow U(x) \phi~.
%
\ee
The local symmetry can be incorporated by introducing a $U(1)$ gauge field $\Am$ and the covariant derivative $D_\mu$:
\be
D_\mu \phi = (\del_\mu - i\Am)\phi~,
%
\ee
where $\Am$ transforms as 
\be
\Am \rightarrow \Am + \del_\mu \alpha = \Am - i(\del_\mu U)U^{-1}~.
%
\ee
The covariant derivative transforms as 
\be
D_\mu \phi \rightarrow U D_\mu \phi~, 
\quad \text{or} \quad
D_\mu \rightarrow U D_\mu U^{-1}~.
%
\ee

One makes the invariant action using covariant (or invariant) quantities such as $D_\mu\phi$ and $F_{\mu\nu} = \del_\mu A_\nu - \del_\nu A_\mu$. The gauge field action is given by
\be
\calL = - \frac{1}{4e^2} F^{\mu\nu}F_{\mu\nu}~.
%
\ee
We normalized the gauge field so that the charge $e$ (coupling constant) appears only as the overall normalization. The usual form is obtained by
\be
\Am \rightarrow e \Am~.
%
\ee
Then, the transformation law and the covariant derivative become
\begin{align}
\Am & \rightarrow \Am + \frac{1}{e}\del_\mu \alpha~,\\
D_\mu \phi &= (\del_\mu - ie\Am)\phi~.
%
\end{align}

\head{$U(\Nc)$ gauge theory}

Now, let us consider a scalar field with $\Nc$ components $\phi_i = (\phi_1, \phi_2, \cdots)$. Consider the Lagrangian given by
\be
\calL = - (\del_\mu \phi)^\dag (\del^\mu \phi) -m^2 \phi^\dag\phi~.
%
\ee
The theory has a global $U(\Nc)$ symmetry:
\be
\phi_i \rightarrow U_i^{~j} \phi_j~,
%
\ee
where $U$ is a unitary $\Nc\times\Nc$ matrix. Again, one can promote the symmetry to a local symmetry by  introducing a gauge field $\Am$ and the covariant derivative $D_\mu$ for $\phi$:
\be
D_\mu \phi = (\del_\mu - i\Am)\phi~,
%
\ee
where $\Am$ is a Hermitian $\Nc\times\Nc$ matrix $(\Am)_i^{~j}$ and transforms as 
\be
\Am \rightarrow U \Am U^{-1} - i(\del_\mu U)U^{-1}~.
\label{eq:A_transf}
\ee
Then, the covariant derivative transforms as 
\begin{align}
D_\mu \phi &\rightarrow (\del_\mu U) \phi + U\del_\mu \phi 
- i\{ U \Am U^{-1} - i(\del_\mu U)U^{-1} \} U \phi \\
&= U (\del_\mu - i\Am) \phi~,
%
\end{align}
so
\be
D_\mu \phi \rightarrow U D_\mu \phi~, 
\quad \text{or} \quad
D_\mu \rightarrow U D_\mu U^{-1}~.
%
\ee

So far, we consider the matter field $\phi_i$, a field in the fundamental representation, but matter fields in the adjoint representation often appear in AdS/CFT. An adjoint scalar $(\phi_A)_i^{~j}$ transforms as
\be
\phi \rightarrow U\phi_A U^{-1}~.
%
\ee
Namely, it transforms as $\phi_i (\phi^\dag)^j$. The gauge field $(\Am)_i^{~j}$ also belongs to the adjoint representation although $\Am$ transformation \eqref{eq:A_transf} itself has an inhomogeneous term $(\del_\mu U)U^{-1}$. The definition of the covariant derivative depends on the fields it acts. Following the same procedure as $\phi_i$, one can show that the covariant derivative of $\phi_A$ is given by
\be
D_\mu \phi_A = \del_\mu \phi_A - i [\Am,\phi_A]~.
%
\ee

One can show
\be
[D_\mu, D_\nu]\phi = \left\{ \del_\mu A_\nu - \del_\nu A_\mu - i [A_\mu, A_\nu] \right\} \phi~.
%
\ee
Note that $D_\mu$ is a differential operator, but its commutator is no longer a differential operator. So, we define the field strength as
\be
F_{\mu\nu} =  -i[D_\mu, D_\nu] = \del_\mu A_\nu - \del_\nu A_\mu - i [A_\mu, A_\nu]~.
%
\ee
Because $D_\mu \rightarrow U D_\mu U^{-1}$, $F_{\mu\nu}$ transforms as
\be
F_{\mu\nu} \rightarrow U F_{\mu\nu} U^{-1}~.
%
\ee
In the $U(1)$ gauge theory, $F_{\mu\nu}$ is invariant, but here $F_{\mu\nu}$ is covariant. We define the gauge theory action as
\be
\calL = - \frac{1}{2g_\text{YM}^2} \text{tr} (F^{\mu\nu}F_{\mu\nu})~.
%
\ee

Write the transformation matrix $U$ as
\begin{align}
U &= e^{iH}~, \\
H &= \alpha \cdot 1 + \alpha^a t^a~. \quad (a=1,\ldots, \Nc^2-1)
%
\end{align}
The parameter $\alpha$ represents $U(1)$ transformation, and the parameters $\alpha^a$ represent $SU(\Nc)$ transformations. The matrices $t^a$ are called $SU(\Nc)$ generators. Because $U$ is a unitary matrix, $t^a$ are traceless Hermitian matrices. In general, the generators satisfy the commutation relation
\be
\left[ t^a, t^b \right] = if_{abc} t^c~,
%
\ee
where $f_{abc}$ is called the structure constant of the gauge group. We normalize the generators so that
\be
\text{tr}\left( t^a t^b \right) = \frac{1}{2} \delta^{ab}~.
%
\ee
Using $t^a$, one can decompose matrix fields such as $\Am$ and $F_{\mu\nu}$, \eg, 
\be
(\Am)_i^{~j} = \Am^a (t^a)_i^{~j}~.
%
\ee
Then, the action becomes
\be
\calL = - \frac{1}{2g_\text{YM}^2} \text{tr} (F^{\mu\nu}F_{\mu\nu})
= - \frac{1}{4g_\text{YM}^2} F^{a\mu\nu}F^a_{\mu\nu}~.
%
\ee

\endofsection

\ifx\nameofpaper\undefined 
  \usepackage{macro_natsuume} 
  \def\beginsection{\section*}
  \def\endofsection{\end{document}} 
  \input draft_header.tex
\else 
  \def\beginsection{\chapter}
  \def\endofsection{ } 
\fi

\beginsection{The road to AdS/CFT
}\label{chap:string}


\begin{quote}
In this chapter, we explain basics of string theory and how one reaches AdS/CFT. The relation between string theory and gauge theory has been discussed for many years, and AdS/CFT solved the ``homework" to some extent. 
\end{quote}

We should give a warning about this chapter. The discussion in this chapter needs string theory to some extent, which may be frustrating to readers. So, we choose to explain string theory intuitively, but, on the other hand, our argument here is not very rigorous. Anyway, \textit{readers do not have to worry about string theory part much if you just want to know the AdS/CFT technique.} This is because we rarely use string theory itself in actual AdS/CFT computations. Nevertheless, we include the material because
%
\begin{enumerate}
\item
AdS/CFT makes a bold claim that a four-dimensional gauge theory is equivalent to a five-dimensional gravitational theory: two theories look different, and even their spacetime dimensions are different. The road to AdS/CFT gives us some other clues why this can be true. 
\item
String theory was originally born as the theory of strong interaction. But there were many problems for the approach at that time. Following the path to AdS/CFT, one can understand how AdS/CFT circumvented these problems. 
\end{enumerate}

\section{String theory: prehistory
}\label{sec:prehistory}

The strong interaction is described by QCD, but string theory was originally born as the theory of strong interaction before QCD.

\begin{figure}[tb]
\centering
\includegraphics{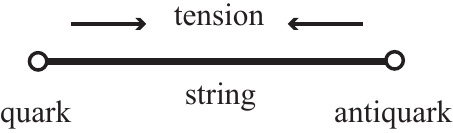}
\vskip2mm
\caption{The string model of a meson.
}
\label{fig:string_model}
\end{figure}

For example, a \keyword{meson} is a quark-antiquark pair, and it is explained well by the ``string" which connects the pair (\fig{string_model}). The string has the tension. In order to separate the pair, one needs energy proportional to the string length $\calR$,  $E \propto \calR$, so quarks cannot be separated, which explains the confinement. In reality, when the string length becomes large enough, the potential energy is large enough to create a new quark-antiquark pair. Then, a string breaks into two strings. But the endpoints of strings are still quarks and antiquarks, and one cannot get individual quarks in isolation. 

Because the string has the tension $\tension$, it would collapse to a point. In order the string not to collapse, one needs to oscillate the string. Or if the string rotates, the string becomes stable by balancing the tension and the centrifugal force. 
But the angular momentum $J$ (or the oscillation) of the string contributes to the energy $M$ of the string, so there is a relation between $J$ and $M$. 

Such a relation can be obtained from the dimensional analysis. We consider the relativistic string and use the unit $c=1$. In this unit, $[J] =\text{M}\text{L}$ and $[\tension] = \text{M}\text{L}^{-1}$. (Recall $\vec{J}=\vec{r}\times\vec{p}$. The mass density has the same dimensions as $\tension$ in these units.) Classically, such a string has only the tension $\tension$ as the dimensionful parameter [see, \eg, \eq{NG}]. Thus, we must have 
\be
J \simeq \frac{1}{\tension} M^2 + \hbar~.
\label{eq:pre_regge}
\ee
Here, we added the second term which is a quantum correction. 

\begin{figure}[tb]
\centering
\scalebox{0.8}{ \includegraphics{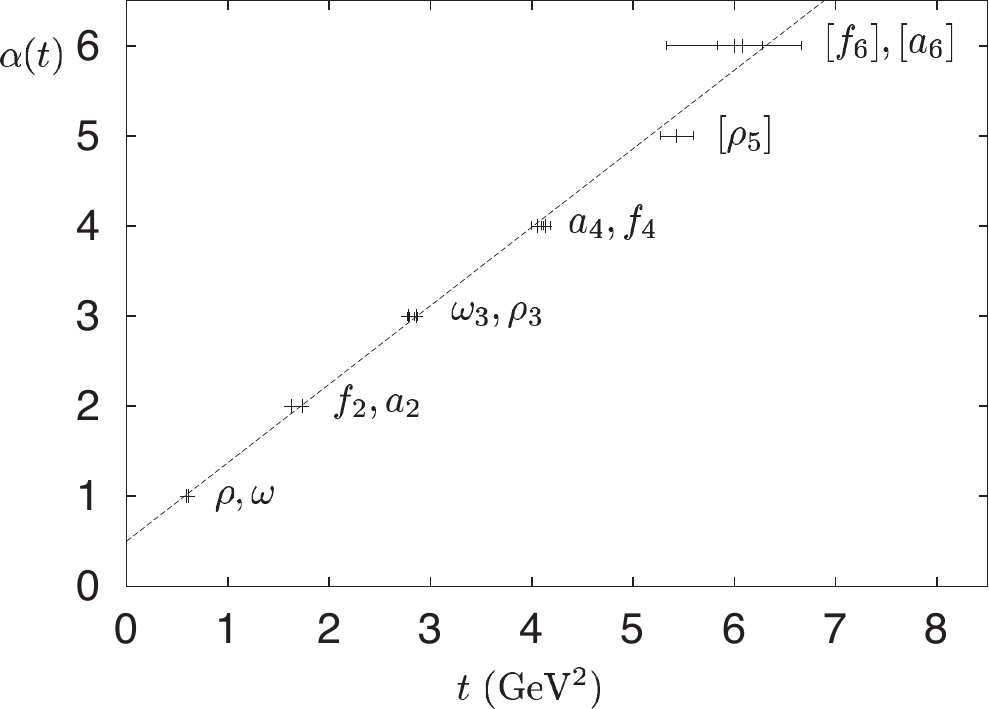} }
\vskip2mm
\caption{An example of Regge trajectories. The horizontal axis represents $M^2$, and the vertical axis represents spins. The mesons in square brackets are not well-known. The straight line is fitted using \eq{regge_trajectory_fit}. Reproduced with permission from Ref.~\cite{Donnachie:2002en}.
}
\label{fig:regge}
\end{figure}

Indeed, the meson spectrum satisfies such a relation. Phenomenologically, the relation \eqref{eq:pre_regge} is fitted by two parameters $\alpha'$ and $\alpha(0)$:
\be
J= \alpha' M^2+\alpha(0)~.
\label{eq:regge}
\ee
From the above discussion, $1/\alpha'$ represents the tension, and $\alpha(0)$ represents the quantum correction. (Note that in the units $c=\hbar=1$, $[\alpha'] =\text{L}^2$ and $\alpha(0)$ is dimensionless.) This relation is known as a \keyword{Regge trajectory}. Figure~\ref{fig:regge} is the Regge trajectory for typical mesons%
\footnote{Such a figure is known as a Chew-Frautschi plot \cite{Chew:1962eu}.}. 

The pions are the most well-known among all mesons. They are made of $u$ and $d$ quarks as well as their antiquarks. The $\rho$ mesons are also made of them. Although the pions have spin 0 (scalar meson), the $\rho$ mesons have spin 1 (vector meson) and heavier mass. Now, there is a series of mesons which are similar to the $\rho$ mesons but have larger spins and heavier masses. These mesons lie on a single line with%
\footnote{To be precise, this line does not represent a single trajectory but represents degenerate trajectories which are distinguished by the other quantum numbers. Incidentally, the $\omega$ mesons contain $s\bar{s}$, but they are similar to the $\rho$ mesons in the sense that they do not carry quantum numbers such as strangeness, charm, bottom, and top.} 
\be
\alpha' \approx 0.9 \text{GeV}^{-2}~, \qquad \alpha(0) \approx 0.5~.
\label{eq:regge_trajectory_fit}
\ee

\begin{figure}[tb]
\centering
\scalebox{0.5}{ \includegraphics{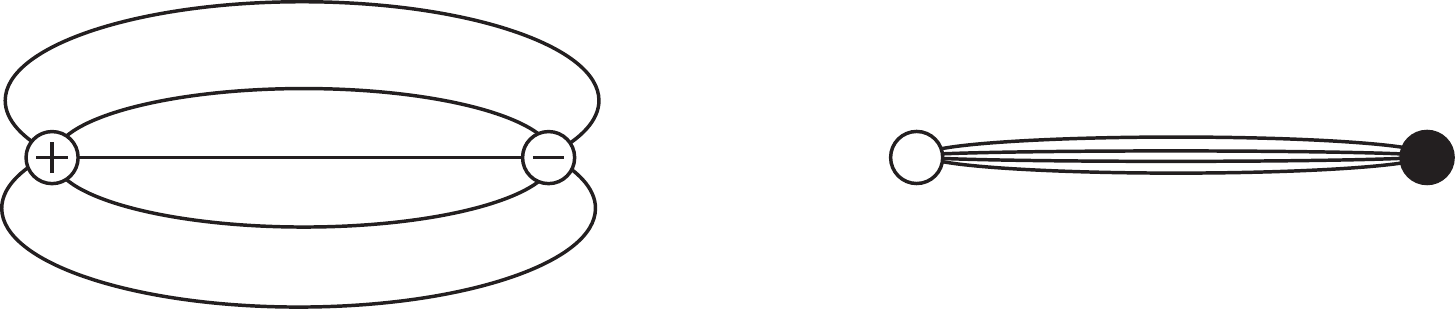} }
\vskip2mm
\caption{The QCD flux lines form a thin flux tube (right) unlike the electromagnetic flux lines (left).
}
\label{fig:flux_tube}
\end{figure}%

The mesons are explained by the string model well. Of course, eventually QCD appeared as the theory of strong interaction, but the string model is not inconsistent with QCD. The flux lines of electrodynamics spread in space. However, color flux lines do not spread but rather form a thin \keyword{flux tube} because gluons have self-interactions (\fig{flux_tube}). It is natural to imagine that mesons are well-described by the string model when the thickness of the flux tube is negligible. 

Thus, string theory has nice properties as a theory of strong interaction, but there are shortcomings as well. In particular, 
\begin{description}

\item[Problem 1:] In order to quantize the string consistently, one needs a higher dimensional spacetime. The bosonic string, which contains only bosons in the spectrum, requires 26-dimensional spacetime, and string theory which also contains fermions requires 10-dimensional spacetime. Otherwise, the theory is not consistent. 
\item[Problem 2:] One can describe the confinement as a string connecting quarks, but what appears in QCD is not only the confining potential. When the quark separation becomes smaller, the thickness of the flux tube is no longer negligible, and the flux lines look more like the usual one. As a consequence, one has the Coulomb-like potential, $E \propto -1/\calR$. The simple string model is unlikely to describe this effect. Also, in the plasma phase, one has the Debye screening by thermally excited light quarks. From the string model point of view, the string becomes easier to break by quark pairs, but this implies that the string description is not very adequate. In a nutshell, the string model looks like a model of confinement only. If string theory really describes QCD, one should be able to discuss the full phase diagram of QCD.

\end{description}
We will reexamine Problem~1 in \sect{reexamination} and Problem~2 in \sect{wilson_intuitive}. 

These problems suggest that something is clearly missing about the correspondence between the strong interaction and string theory. Moreover, QCD was proposed, and QCD was proved to be the right theory of the strong interaction as experimental results accumulated. So, string theory as the strong interaction was once abandoned. Rather, string theory was reincarnated as the unified theory including gravity. We will then quickly look at this aspect of string.

\section{String theory as the unified theory}

\begin{figure}[tb]
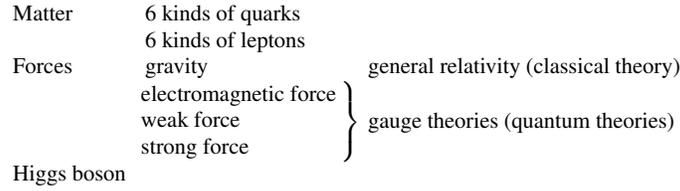

\begin{center}
\begin{tabular}{lll}
Matter	& ~~6 kinds of quarks	& \\
		& ~~6 kinds of leptons	& \\
Forces	& ~~gravity	& general relativity (classical theory) \\
		& $\left. \begin{array}{l}
\text{electromagnetic force} \\
\text{weak force} \\
\text{strong force}
		  \end{array} \right\}$
& gauge theories (quantum theories) \\
Higgs boson &&
\end{tabular}
\caption{The standard model of elementary particles.}
\label{fig:standard_model}
\end{center}
\end{figure}

Figure~\ref{fig:standard_model} shows the contents of the standard model. There are 6 quarks and 6 leptons as matters and 4 forces act on matters. In principle, all phenomena can be understood with these elements, but the standard model has many shortcomings as well. We will not describe these shortcomings, but in brief, the problem is that the standard model is a disjointed framework. 

This is particularly clear in the force sector. Gravity is described by general relativity and this is a classical theory. On the other hand, the other 3 forces are described by gauge theories, and gauge theories can be quantized. So, theoretical foundations and concepts for these forces are completely different. Now, string theory is the theory which unifies all these elements. 

\subsection{String oscillations and elementary particles
}

\begin{figure}[tb]
\centering
\scalebox{0.5}{ \includegraphics{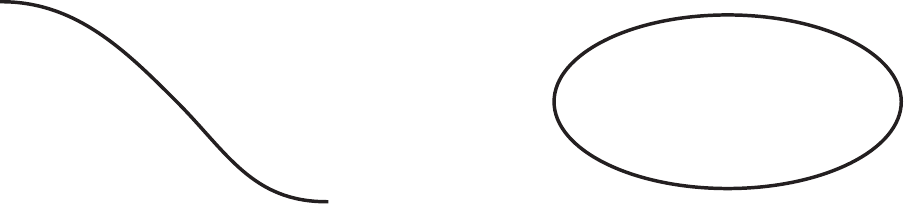} }
\vskip2mm
\caption{Main ingredients of string theory. An open string with endpoints and a closed string without endpoints.}
\label{fig:string}
\end{figure}

The fundamental object in string theory is the very small string (\fig{string}). There are two kinds of string: \keyword{open strings} with endpoints and \keyword{closed strings} with no endpoints. 
The string has only the tension $\tension$ as the dimensionful parameter. Since $[\tension] = \text{L}^{-2}$, we introduce a parameter $l_s$ with the dimension of length%
\footnote{The parameters $\tension$, $\alpha'$, and $l_s$ are related by
\be
\tension = \frac{1}{2\pi\alpha'} =  \frac{1}{2\pi l_s^2}~.
%
\ee
}. 
The parameter $l_s$ is called the \keyword{string length} and represents the characteristic length scale of the string. The string length $l_s$ is the only fundamental length scale in string theory. The string length is about $10^{-34}$ m. But currently the length we can see experimentally is about $10^{-17}$ m. So, macroscopically the string is just a point particle. 

But no matter how small the string is, there is a big difference between strings and point particles; namely, a string can oscillate and a string can oscillate in various ways. This is the advantage of the string. There are many kinds of elementary particles in particle physics. String theory unifies all particles in nature as different oscillations of string. 

\begin{figure}[tb]
\centering
\scalebox{0.5}{ \includegraphics{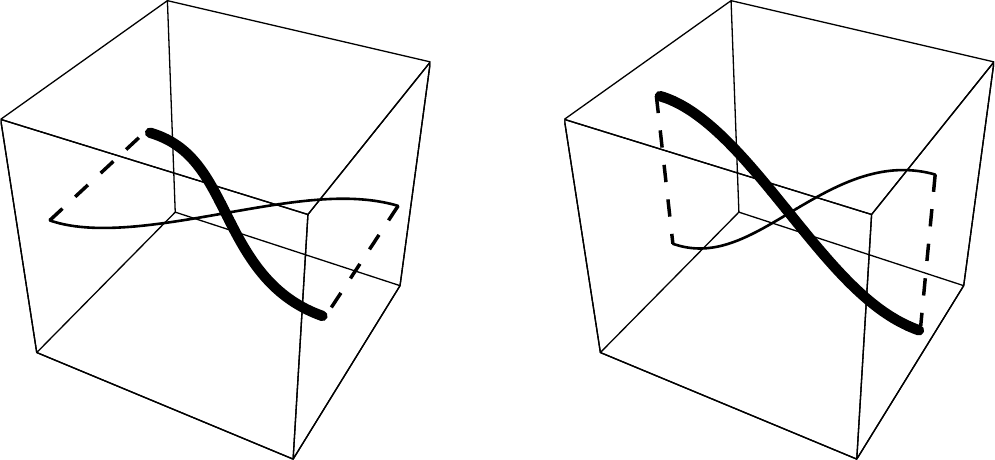} }
\vskip2mm
\caption{The simplest oscillations of an open string. The initial configurations in thick curves become gray curves after the half period. The ``box" here is drawn to see the oscillations easier.}
\label{fig:level1}
\end{figure}

Let us look at string oscillations closely. 
Figure~\ref{fig:level1} shows the simplest open string oscillation (in four dimensions).
As one can see, the string can oscillate in two directions. So, the open string has two degrees of freedom at this level. These degrees of freedom represent two polarizations of a gauge field (photon). In this sense, the open string represents a gauge theory%
\footnote{Classically, such a string oscillation represents a massive particle as one can expect from \eq{regge}. But it is actually massless because of the quantum correction [the second term of \eq{regge}]. This is true only in 26-dimensional spacetime for the bosonic string and in 10-dimensional spacetime for superstrings. This is one reason why higher-dimensional spacetime is necessary in string theory. }. 

\begin{figure}[!tb]
\centering
\scalebox{0.75}{ \includegraphics{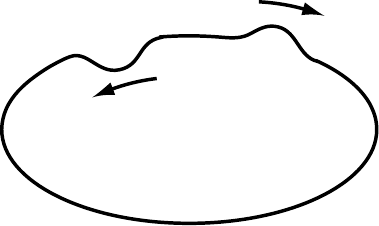} }
\vskip2mm
\caption{A closed string has two independent modes, which corresponds to the spin-2 nature of the graviton.}
\label{fig:closed}
\end{figure}

On the other hand, a closed string represents a graviton. Again, the easiest way to see this is to look at how the string oscillates (\fig{closed}). In general, the oscillations on a string have two modes: the left-moving mode and right-moving mode. For an open string, these modes mix each other at endpoints, but these modes become independent for a closed string. So, one can oscillate the right-moving mode in one direction and the left-moving mode in the other direction. In a sense, a closed string oscillates in two directions simultaneously. This property explains the spin-2 nature of the graviton. In fact, a graviton also oscillates in two directions simultaneously%
\footnote{More precisely, the simplest oscillations of a closed string represents a graviton and two undiscovered scalar particles, the dilaton\index{dilaton} and the axion. Since left and right-moving modes each have two degrees of freedom, a closed string has four degrees of freedom at this level (in four dimensions). The graviton has only two degrees of freedom, and two scalar fields cover the remaining degrees of freedom.}.


\subsection{D-brane}\label{sec:D_brane_primer}

\begin{figure}[tb]
\centering
\subfigure[]{
\scalebox{0.75}{ \includegraphics{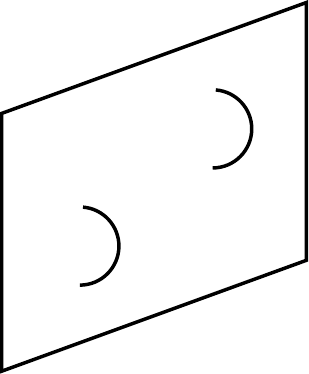} }
\label{fig:Dbrane} } \qquad
\subfigure[]{
\scalebox{0.75}{ \includegraphics{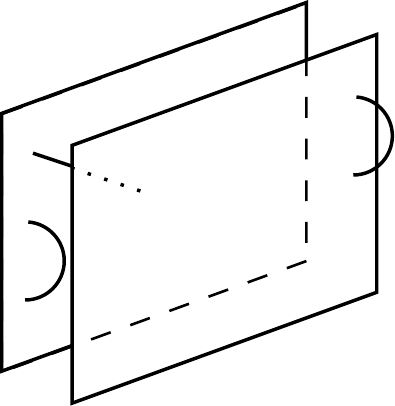} }
\label{fig:Dbrane_SYM} }
\vskip2mm
\caption{
(a) An open string can have endpoints on a D-brane. (b) The $N_c$ coincident D-branes represent a $SU(N_c)$ gauge theory. In the figure, the branes are separated for illustration, but we consider the coincident branes in reality.
}
\label{fig:Dbranes}
\end{figure}

An open string propagates through spacetime just like a closed string, but an open string can also have its endpoints on an object, the so-called \keyword{D-brane} [\fig{Dbrane}] \cite{Dai:1989ua,Polchinski:1995mt}. 
Although these open strings are constrained to have their endpoints on the D-brane, these are open strings, so they still describe gauge theories. But because the endpoints are constrained on the D-brane, the gauge theory described by the D-brane is localized on the D-brane. 

If there are $N_c$ coincident D-branes, open strings can have endpoints in various ways. Open strings can have endpoints on the same brane or can have endpoints on the different branes [two curved lines and one straight line in \fig{Dbrane_SYM}].
These new degrees of freedom correspond to $SU(N_c)$ degrees of freedom%
\footnote{More precisely, the D-branes represent a $U(N_c)$ gauge theory, but the $U(1)$ part describes the center of mass motion of the branes and decouple from the $SU(N_c)$ part.}.

The D-branes arise with various dimensionalities. A D-brane with a $p$-dimensional spatial extension is called the D$p$-brane\index{Dp-brane@D$p$-brane}. Namely, 
\begin{center}
\begin{tabular}{cc}
$ p=0 $		& point-like object \\
$ p=1 $		& string-like object \\
$ p=2 $		&\hspace{5mm} membrane-like object \\
$ \vdots $	& $ \vdots $
\end{tabular}
\end{center}
Thus, the D$p$-brane describes a $(p+1)$-dimensional gauge theory. We are interested in four-dimensional gauge theories, so consider the D3-brane in order to mimic QCD%
\footnote{This is not necessarily the case if one compactifies part of brane directions (\eg, Sakai-Sugimoto model \cite{Sakai:2004cn}). Incidentally, the D1-brane is a string-like object, but it is different from the ``fundamental string" we have discussed so far. 
}.

The D-branes are related to the black branes in \sect{black_branes}. Some of black branes are gravitational descriptions of D-branes.

In \sect{D3_brane}, we discuss more about the D-brane, the gauge theory that the D3-brane represents, and its role in AdS/CFT.

\subsection{Why open and closed strings?}

In string theory, there are two kinds of strings, the open string and the closed string. Why are there two kinds of strings? Or one would say that string theory is not a unified theory since there are two kinds of strings. But actually these are the same object in a sense. To see this, let us consider how strings interact with each other.  

\begin{figure}[tb]
\centering
\scalebox{0.75}{ \includegraphics{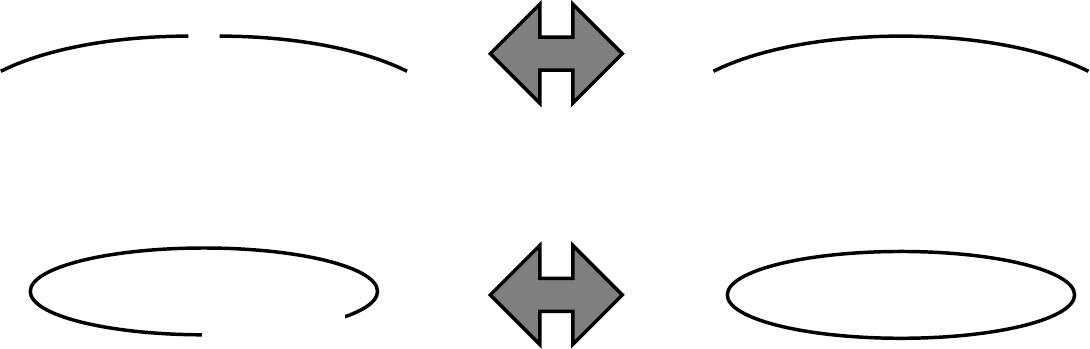} }
\vskip2mm
\caption{\textit{Top}: the simplest interaction for open strings is the joining of two strings into one. \textit{Bottom}: if the endpoints of two open strings can join, the endpoints of a single open string can join as well.}
\label{fig:open_vs_closed1}
\end{figure}

\fig{open_vs_closed1} is the simplest interaction for open strings. Two open strings join into one open string. But if the endpoints of two open strings can join, the endpoints of a single open string can join as well. Otherwise, one has to require a nonlocal constraint on the string dynamics. Thus, the endpoints of a single open string can join to produce a closed string. Namely, if there are open strings, one has to have closed strings. This is the reason why we have two kinds of strings%
\footnote{One can consider string theories with closed string alone, but we will not discuss such theories.}. 

Because an open string represents a gauge theory and a closed string represents gravity, string theory must contain both gauge theory and gravity. This is the unique feature of string theory as the unified theory.

The AdS/CFT duality utilizes this unique feature. The AdS/CFT duality relates a gauge theory to a gravitational theory. This is possible only for the unified theory such as string theory. In conventional field theory, a gauge theory and a gravitational theory are completely separated since theoretical foundations and concepts are very different. As a result, there is no way to get a relation between a gauge theory and a gravitational theory. 

There is a bigger picture behind AdS/CFT. Namely, AdS/CFT requires the existence of a unified theory behind. 

\subsection{String interactions
}\label{sec:string_interaction}

\begin{figure}[tb]
\centering
\subfigure[]{
\scalebox{0.75}{ \includegraphics{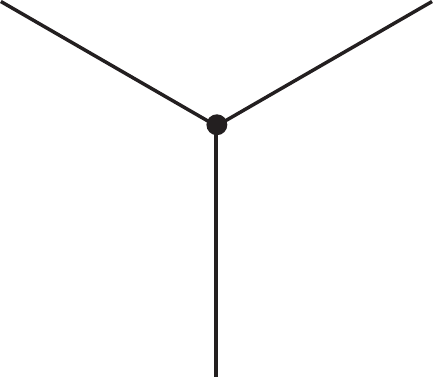} }
} \qquad
\subfigure[]{
\scalebox{0.75}{ \includegraphics{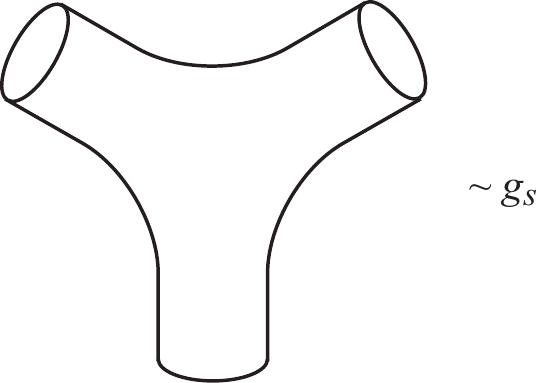} }
}
\vskip2mm
\caption{
(a) A Feynman diagram which represents an interaction vertex. (b) The corresponding interaction for closed strings.}
\label{fig:closed_vertex}
\end{figure}%

\begin{figure}[tb]
\centering
\scalebox{0.75}{ \includegraphics{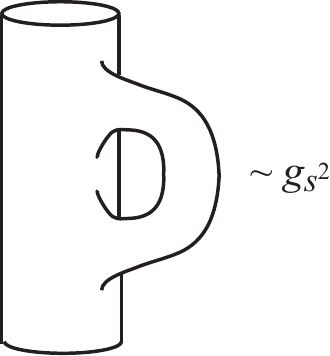} }
\caption{A 1-loop of closed strings.}
\label{fig:closed_loop}
\end{figure}%

\begin{figure}[tb]
\centering
\scalebox{0.75}{ \includegraphics{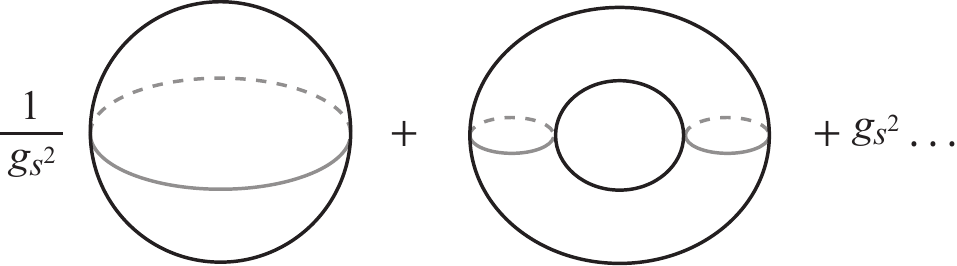} }
\vskip2mm
\caption{The perturbative expansion of string theory is given by the sum over the topologies of world-sheet surfaces.}
\label{fig:string_amplitude}
\end{figure}%

A particle draws a wold-line in spacetime. Similarly, a string sweeps a two-dimensional surface, a \keyword{world-sheet}, in spacetime. In \fig{closed_vertex}(a), we draw a Feynman diagram for a particle, which represents an interaction vertex. For a closed string, such an interaction is replaced by \fig{closed_vertex}(b).

In order to express the strength of the interaction, we assign the \keyword{string coupling constant} $g_s$ for an emission/absorption of a closed string. A closed string 1-loop (\fig{closed_loop}) has an emission and an absorption of a virtual closed string, so it is proportional to $g_s^2$. Figure~\ref{fig:closed_loop} has one ``handle," so we can rephrase the rule to assign $g_s^2$ for each handle. If we have one more loop, we have one more handle, so the diagram has the factor $g_s^4$.

To summarize, string interactions are classified by the world-sheet topologies (\fig{string_amplitude})%
\footnote{Below we often consider only vacuum amplitudes for simplicity, but one can consider scattering amplitudes by adding appropriate external lines. Incidentally, $g_s$ looks like a free parameter of the theory at this stage, but actually it is not a free parameter (\sect{SUGRA}).}. 
If we denote the number of handles (genus) as $h$, a $h$-loop has the factor $g_s^{2h}$. The genus $h$ is related to the Euler characteristic $\chi$ by $\chi=2-2h$, and vacuum amplitudes can be written as $1/g_s^\chi$ using the Euler characteristic%
\footnote{If one adds three external lines (factor $g_s^3$) to the sphere $1/g_s^2$, the diagram is proportional to $g_s$, which is consistent with the interaction vertex in \fig{closed_vertex}(b).}. 

\begin{figure}[tb]
\centering
\subfigure[]{
\scalebox{0.75}{ \includegraphics{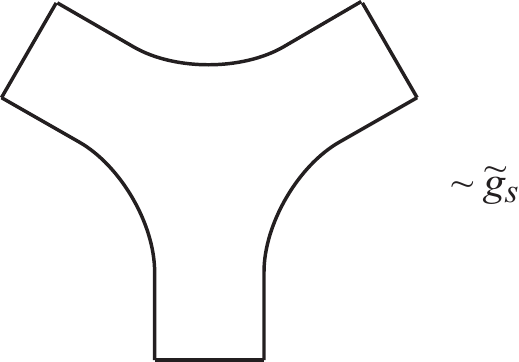} }
} \qquad
\subfigure[]{
\scalebox{0.75}{ \includegraphics{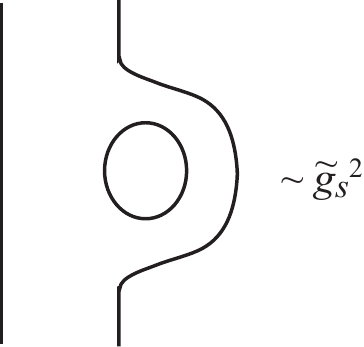} }
}
\vskip2mm
\caption{(a) An open string interaction. (b) A 1-loop of open strings. }
\label{fig:open_vertex}
\end{figure}

\begin{figure}[tb]
\centering
\scalebox{0.75}{ \includegraphics{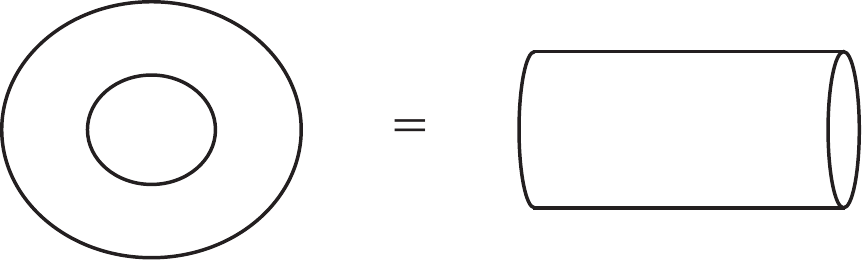} }
\vskip2mm
\caption{An open string 1-loop can be viewed as an emission of a closed string.}
\label{fig:open_vs_closed2}
\end{figure}

The open string case is similar. Assign the open string coupling constant $\tilgs$ for an emission of an open string. Then, the open string interactions are classified by the topologies of world-sheets with holes (\fig{open_vertex}). However, the open string naturally accompanies the closed string as we saw earlier,  and open string interactions are related to closed string interactions. In fact, an open string 1-loop can be viewed as an emission of a closed string (\fig{open_vs_closed2}), so we obtain 
\be
g_s \simeq \tilgs^2~.
%
\ee

\subsection{Supergravity: classical gravity approximation of string theory
}\label{sec:SUGRA}

We have seen that the simplest oscillations of strings correspond to gauge theories and gravity. But a string has an infinite number of harmonics, so string theory predicts an infinite number of elementary particles. Since string oscillations cost energy, these infinite number of harmonics correspond to massive particles whose masses are determined by energy of oscillations. This mass is typically $O(1/l_s)$, which is extremely heavy as elementary particles. Thus, what normally matters is a set of the lowest oscillations only; they correspond to known particles.

String theory describes a finite number of known particles when one can ignore the string length $l_s$, so string theory should be written by standard field theories. Moreover, when $g_s \ll 1$, only the lowest order of topology expansion, tree diagrams, dominates. In such a case, classical field theories should be enough to describe string theory. Such a theory is known as \keyword{supergravity}%
\footnote{See Ref.~\cite{Freedman:2012zz} for a textbook of supergravity. The name, supergravity, comes from the fact that the theory has local supersymmetry, but we will not discuss the terms with fermions below.}.

The form of supergravity action can be determined from general principles%
\footnote{More properly, one calculates string Feynman diagrams as in \sect{string_interaction} and writes down a field theory action which reproduces these scattering amplitudes.}.
For example, one can show that string theory graviton has general covariance, so the gravity part should be given by
\be
\action = \frac{1}{16\pi G_{10}} \int d^{10}x\, \sqrt{-g} R
\label{eq:sugra_leading}
\ee
just like general relativity. We write it as a 10-dimensional action since superstring theory can be quantized consistently only in 10-dimensional spacetime. $G_{10}$ is the Newton's constant in 10-dimensional spacetime. Similarly, one can show that gauge fields in string theory have gauge invariance, so the gauge theory part should be given by the usual gauge theory action.

In string theory, the fundamental coupling constant is the string coupling constant $g_s$. In supergravity, coupling constants are $G_{10}$ and the gauge field coupling constant $g_\text{YM}$. Then, one can relate $g_s$ to supergravity coupling constants. By expanding $g_{\mu\nu} = \eta_{\mu\nu} + h_{\mu\nu}$, the graviton action is schematically written as
\be
\action = \frac{1}{16\pi G_{10}}  \int d^{10}x\,  \{ \del h \del h + h \del h\del h + h^2 \del h\del h + \cdots \}~.
\label{eq:action_graviton}
\ee
This implies that a graviton emission is proportional to $G_{10}^{1/2}$. Since a closed string emission is proportional to $g_s$, one gets $G_{10} \propto g_s^2$.

Similarly, the gauge field action is schematically written as%
\footnote{
We write the $(p+1)$-dimensional action instead of a 10-dimensional action because one can consider $(p+1)$-dimensional gauge theories using D-branes.
}
\be
\action =  \frac{1}{g_\text{YM}^2} \int d^{p+1}x\, \{ \del A \del A+ A^2 \del A + A^4 \}~.
\label{eq:action_gauge}
\ee
This implies that a gauge field emission is proportional to $g_\text{YM}$. Since an open string emission is proportional to $\tilgs \simeq g_s^{1/2}$, 
\be
g_\text{YM}^2 \propto g_s~.
%
\ee

The dimensional analysis further constrains the form of field theory coupling constants. Since we usually take the metric as dimensionless, the Newton's constant has dimensions $[G_{10}] = \text{L}^{8}$ from \eq{action_graviton}. In string theory, the fundamental string scale is only the string length $l_s$ which represents the characteristic length scale of the string. Thus, $G_{10} \simeq g_s^2 l_s^8$. 
Similarly, the gauge field has dimensions $[A] =\text{L}^{-1}$ from \eq{action_gauge}. In order for the action to be dimensionless, $[g_\text{YM}^2] = \text{L}^{p-3}$. Thus, $g_\text{YM}^2 \simeq g_s l_s^{p-3}$. To summarize, 
\be
\boxeq{
G_{10} \simeq g_s^2 l_s^8~, \quad
g_\text{YM}^2 \simeq g_s l_s^{p-3}~.
}
\label{eq:string_vs_qft}
\ee

The action \eqref{eq:sugra_leading} is the approximation when one can ignore the string length $l_s$. When the string length can no longer be ignored, the action has an infinite number of terms with powers of curvature tensors such as
\be
\action = \frac{1}{16\pi G_{10}} \int d^{10}x \sqrt{-g}\, (R + l_s^2 R^2 + \cdots)~.
\label{eq:sugra_alpha'}
\ee
Such corrections are known as \textit{$\alpha'$-corrections}\index{alpha'-corrections@$\alpha'$-corrections}%
\footnote{Recall that the parameter $\alpha'$ is related to the string length $l_s$ by $\alpha'=l_s^2$.}.
From the field theory point of view, these terms correspond to higher derivative terms in an effective field theory%
\footnote{See Ref.~\cite{Polchinski:1992ed} for a review of effective field theories.}, 
and they originated from integrating out massive fields with mass $O(1/l_s)$. Roughly speaking, these terms can be ignored when $l_s^2 R \ll 1$, or the length scale of the curvature is much larger than the string length. 

In the topology expansion of string theory, these terms correspond to the lowest order diagrams, the sphere. Note that the classical gravity approximation of string theory corresponds to \eq{sugra_alpha'}, not \eq{sugra_leading}. However, one often considers the case where the spacetime curvature is small. In such a case, the action \eqref{eq:sugra_leading} is enough, or the action \eqref{eq:sugra_alpha'} with the lowest $\alpha'$-corrections is enough.

Finally, to be more precise, supergravity typically takes the form
\be
\action = \frac{1}{16\pi G_{10}} \int d^{10}x\, \sqrt{-g}\, e^{-2\phi} \left\{R + 4(\nabla\phi)^2 \right\} + \cdots
\label{eq:sugra_dilaton}
\ee
instead of the action \eqref{eq:sugra_leading}. Here, $\phi$ is a scalar field called the \keyword{dilaton}. One characteristic feature of \eq{sugra_dilaton} is that the Ricci scalar part does not take the simple Einstein-Hilbert form but has the dilaton factor $e^{-2\phi}$. This means that the Newton's constant or the string coupling constant actually correspond to the constant part of the dilaton $e^{-2\phi_0}$. Readers do not have to worry this fact much since the dilaton is constant in the simplest AdS/CFT (the D3-brane case in the language of D-branes in
\sect{D_brane_primer}). But the dilaton may have a nontrivial behavior in a generic AdS/CFT. A nontrivial dilaton means that the Newton's constant is not really a constant but it can vary. Similarly, the gauge field coupling constants can vary because the gauge theory actions can have a dilaton factor. 

\section{Reexamine string theory as the theory of strong interaction
}\label{sec:reexamination}

\subsection{Comparison of partition functions
}


We have seen in \sect{large_N} that the vacuum amplitudes of large-$N_c$ gauge theories are given by a summation over the topologies of two-dimensional surfaces. This structure is the same as the closed string amplitudes in \sect{string_interaction}. Since the closed string represents gravity, the closed string expansion corresponds to the gravitational perturbative expansion. In particular, the sphere corresponds to classical gravity, and the torus corresponds to gravity at 1-loop. 

This implies that \textit{a large-$N_c$ gauge theory is represented by string theory and in particular is represented by classical gravitational theory of string theory in the large-$N_c$ limit.} This is an amazing conclusion. The original theory is a gauge theory which does not include gravity, but it is represented by a gravitational theory in the special limit. In this way, the relation between the gauge theory and string theory, which was once abandoned (\sect{prehistory}), again emerges.

More explicitly, we saw in \sect{large_N} that the gauge theory partition function $Z_\text{gauge}$ is given by
\be
\ln Z_\text{gauge} = \sum_{h=0}^\infty N_c^{\chi} f_h (\lambda)~.
\label{eq:partition_gauge_again}
\ee
On the other hand, from the discussion in this chapter, the string theory partition function $Z_\text{string}$ is given by
\be
\ln Z_\text{string} = \sum_{h=0}^\infty \left(\frac{1}{g_s}\right)^{\chi} \tilde{f}_h (l_s)~.
\label{eq:partition_string}
\ee
Since string theory has $\alpha'$-corrections, We write the dependence as  $\tilde{f}_h (l_s)$. Equations~\eqref{eq:partition_gauge_again} and \eqref{eq:partition_string} have the same structure, so we arrive at the relation
\be
%
Z_\text{gauge} = Z_\text{string}~,
\label{eq:pre_GKPW1}
\ee
and the parameters between two theories are related by%
\footnote{The parameters $N_c$ and $\lambda$ are dimensionless whereas $G$ and $l_s$ are dimensionful, but we make the dimensions to come out right in \sect{towards_ads}.}
\be
N_c^2 \propto \frac{1}{g_s^2} \propto \frac{1}{G}~, \quad
\lambda \leftrightarrow l_s~,
\label{eq:pre_dictionary}
\ee
where we also used the relation $G \propto g_s^2$.

The analysis of large-$N_c$ gauge theories leads us to string theory as the theory of strong interaction. However, as we mentioned before, one needs higher-dimensional spacetime in order to quantize string theory consistently. If one assumes the $d$-dimensional \poincare\ invariance%
\footnote{The \poincare\ invariance is also called the inhomogeneous Lorentz invariance. $ISO(1,3)=\mathbb{R}^{1,3} \times SO(1,3)$.} 
$ISO(1,d-1)$, $d=26$ for the bosonic string which contains only bosons in the spectrum, and $d=10$ for string theory which also contains fermions in the spectrum. String theory in four-dimensional spacetime is not consistent. Namely,
\begin{itemize}

\item
According to gauge theories, a large-$N_c$ gauge theory is represented by a string theory.
\item
On the other hand, string theory is not valid as a four-dimensional theory.

\end{itemize}
These two statements sound inconsistent. 

These two statements are actually not inconsistent. To realize this, one should recall that string theory is a theory with gravity as well. Namely, string theory should admit not only the flat spacetime as the solution but also curved spacetimes as the solutions. One may use such a curved spacetime to represent a large-$N_c$ gauge theory. The curved spacetime itself can be a higher-dimensional one. The resulting theory can be interpreted as a flat four-dimensional theory if the curved spacetime has only the $ISO(1,3)$ invariance instead of $ISO(1,d-1)$. For this purpose, the spacetime dimensionality must be at least five-dimensions, and the fifth direction is curved.

The appearance of the five-dimensional curved spacetime is also natural from the black hole entropy. A five-dimensional black hole has entropy which is proportional to the five-dimensional ``area." But an ``area" in five dimensions is a ``volume" in four dimensions: This is appropriate as the entropy of a four-dimensional statistical system.

To summarize%
\footnote{I am not aware of who first made this claim explicitly, but see, \eg, Refs.~\cite{Polyakov:1981rd,Natsuume:1992ky} for early attempts.},
\begin{center}
\fbox{
\begin{tabular}{l}
A large-$N_c$ gauge theory is represented by a five-dimensional curved \\ spacetime with four-dimensional \poincare\ invariance $ISO(1,3)$.
\end{tabular}
}
\end{center}
Such a metric is written as%
\footnote{
The factor $\Omega$ does not depend on $(t, \bmx_3)$ because of the translational invariance on $(t, \bmx_3)$. Incidentally, one can consider the metric of the form
\be
ds_{5}^2 = \Omega_1(w')^2 (-dt^2 + d\bmx_3^2) + \Omega_2(w')^2 dw'^2~,
%
\ee
but the metric reduces to \eq{ansatz} by redefining $w'$.}
\be
ds_{5}^2 = \Omega(w)^2 (-dt^2 + d\bmx_3^2) + dw^2~.
\label{eq:ansatz}
\ee
%
%
Here, $x^\mu=(t, \bmx_3)=(t,x,y,z)$.
The metric contains the combination $-dt^2+d\bmx_3^2$, so it has the $ISO(1,3)$ invariance on $(t, \bmx_3)$. Thus, $\bmx_3$-directions correspond to the spatial directions of the large-$N_c$ gauge theory.

\danger{Hereafter, Greek indices $\mu, \nu, \ldots$ run though $0,\ldots,3$ and are used for the four-dimensional ``boundary" spacetime where a gauge theory lives. On the other hand, capital Latin indices $M, N, \ldots$ are used for the five-dimensional ``bulk" spacetime where a gravitational theory lives.}

\subsection{Scale invariance and its consequences
}\label{sec:scale_inv}

Without further information, we cannot go any further into the relation between the large-$N_c$ gauge theory and the curved spacetime. First, it is not clear what kind of gauge theory is represented by the curved spacetime. Namely, it is not clear if such a gauge theory is close to QCD or is the one which behaves differently from QCD. Second, it is not clear what kind of curved spacetime is represented by the gauge theory. One cannot determine the factor $\Omega(w)$. We will impose one more condition to identify both the gauge theory and the curved spacetime. As we argue below, the appropriate condition is the \keyword{scale invariance}. 


\head{Scale invariant gauge theory (classical)}

When a theory has no scale or no dimensionful parameter, it is natural to expect that physics does not change under the scale transformation:
\be
x^\mu \rightarrow a x^\mu~.
\label{eq:scale_transf1}
\ee
The (classical) scale invariance is an important property of four-dimensional pure gauge theories. For example, consider the Maxwell theory:
\be
\action = - \frac{1}{4e^2} \int d^4x\, F^{\mu\nu}F_{\mu\nu}~.
%
\ee
The action has no dimensionful parameter and is invariant under the scaling with
\be
A_\mu \rightarrow \frac{1}{a} A_\mu~.
%
\ee
If a field $\Phi$ transforms as
\be
\Phi \rightarrow a^{-\Delta}\Phi~,
\label{eq:conf_dim}
\ee
under the scaling \eqref{eq:scale_transf1}, we call that the field has \keyword{scaling dimension}%
\footnote{The scaling dimension is also called the conformal weight or the conformal dimension.} \index{conformal weight}\index{conformal dimension}
$\Delta$. The Maxwell field has scaling dimension 1. Roughly speaking, the scaling dimension coincides with the mass dimension (the Maxwell field has mass dimension~1) although these two dimensions are different notions. 

The energy-momentum tensor of the Maxwell theory is given by
\be
T_{\mu\nu} = \frac{1}{e^2}\left(F_{\mu\rho} F_\nu^{~\rho} - \frac{1}{4} \eta_{\mu\nu} F^2\right)~.
%
\ee
A convenient way to compute $T_{\mu\nu}$ is to couple gravity to a field theory, use 
\be
T_{\mu\nu} = -\frac{2}{\sqrt{-g}} \frac{\delta \action}{\delta g^{\mu\nu}}~,
%
\ee
and then set $g_{\mu\nu}=\eta_{\mu\nu}$. For the Maxwell theory, the energy-momentum tensor is traceless:
\be
\boxeq{
T^\mu_{~\mu}=0~.
}
%
\ee
This is related to the scale invariance. But the relation is rather subtle, so we leave it to appendix (\sect{weyl_inv}).

\head{Scale invariant gauge theory (quantum)}

Pure gauge theories such as the Maxwell theory are classically scale invariant in four dimensions. So, it is natural to impose the scale invariance in our discussion. However, gauge theories are not scale invariant quantum mechanically. Renormalization introduces a renormalization scale which breaks the scale invariance. As a result, the coupling constants change with the energy scale, and we can have phenomena such as confinement.  For example, the 1-loop $\beta$-function for the $SU(N_c)$ gauge theory is given by
\begin{align}
\beta(g_\text{YM}) 
  &= \mu \frac{dg_\text{YM}}{d\mu} \\
  &= - \frac{11}{48\pi^2} g_\text{YM}^3 N_c~,
\label{eq:1loop_beta_pure}
\end{align}
where $\mu$ is a renormalization scale. The $\beta$-function is negative, which means that the theory is asymptotically free.

However, there is a special class of gauge theories which keep scale invariance even quantum mechanically. For example, one would make the right-hand side of \eq{1loop_beta_pure} vanish by adding appropriate matter fields. We focus on such gauge theories. In particular, the so-called \textit{${\cal N}=4$ super-Yang-Mills theory}\index{N=4 super-Yang-Mills theory@${\cal N}=4$ super-Yang-Mills theory} (SYM)
has the largest symmetries among such theories. Here, ${\cal N}=4$ means that the theory has 4 supercharges which are the maximum number of supercharges for a four-dimensional gauge theory. 

The field contents of the ${\cal N}=4$ SYM are given by 
\begin{itemize}
\item gauge field $A_\mu$, 
\item 6 scalar fields $\phi_i$, 
\item 4 Weyl fermions $\lambda_I$
\end{itemize}
(The color indices are suppressed for simplicity. We will also suppress the spinor indices.)  All these fields transform as adjoint representations of a gauge group. 

The action for the ${\cal N}=4$ SYM is given by%
\footnote{Here, $F_{\mu\nu} := F_{\mu\nu}^a t^a$. We normalize the gauge group generators $t^a$ by $\text{tr}(t^a t^b) = \frac{1}{2} \delta^{ab}$.}
\be
{\cal L} = \frac{1}{g_\text{YM}^2}
\text{tr} \left[ -\frac{1}{2} F_{\mu\nu}^2 - (D_{\mu}\phi_i)^2 
- i \bar{\lambda}_I \gamma^{\mu}D_{\mu}\lambda^I
+ O(\phi^4) + O(\lambda\lambda\phi) \right]~.
\nonumber
\ee
First three terms are standard kinetic terms for the gauge field, the scalar fields, and the fermions. In addition, the action contains interaction terms which are written only schematically; $\phi^4$ interactions and Yukawa-like interactions. The theory has no dimensionful parameter and the theory is scale invariant classically. This can be seen as follows. The gauge field and the scalars have mass dimension 1 and the fermions have mass dimension $3/2$, so all terms in the action have mass dimension 4, which means that all parameters must be dimensionless.

The ${\cal N}=4$ SYM is also scale invariant quantum mechanically. We will not show this, but let us check it at 1-loop from the $\beta$-function. When there are adjoint fermions and adjoint scalars, the 1-loop $\beta$-function for the $SU(N_c)$ gauge theory is given by
\be
\beta(g_\text{YM}) = - \frac{g_\text{YM}^3}{48\pi^2} N_c \left(11 - 2n_f - \frac{1}{2}n_s \right)~,
%
\ee
where $n_f$ and $n_s$ are the numbers of Weyl fermions and real scalars, respectively%
\footnote{See, \eg, Sect.~78 of Ref.~\cite{Srednicki:2007qs2} for the derivation.}. 
The $\beta$-function indeed vanishes for the ${\cal N}=4$ SYM since $n_f=4$ and $n_s=6$.


\head{Curved spacetime and scale invariance}

Now, consider how the four-dimensional scale invariance \eqref{eq:scale_transf1} constrains the five-dimensional curved spacetime \eqref{eq:ansatz}:
\be
ds_{5}^2 = \Omega(w)^2 (-dt^2 + d\bmx_3^2) + dw^2~.
\label{eq:ansatz2}
\ee
Under the scale transformation $x^\mu \rightarrow a x^\mu$, the metric \eqref{eq:ansatz2} is invariant if $\Omega(w)$ transforms as
\be
\Omega(w)^2 \rightarrow a^{-2} \Omega(w)^2~.
\label{eq:scale_Omega}
\ee
For this to be possible, $w$ must be transformed in a nontrivial way, but the line element along $w$ is $dw^2$. So, only the translation is allowed on $w$. Since $w$ has the dimension of length, let us introduce a parameter $L$ which has the dimension of length, and write the translation as%
\footnote{We can set this without loss of generality. We can always bring the translation into this form by redefining the parameter $L$.}
\be
w \rightarrow w + L \ln a~.
\label{eq:scale_transf2}
\ee
Under the scale transformation and \eq{scale_transf2}, $\Omega(w)$ must transform as \eq{scale_Omega}. This determines the form of $\Omega(w)$ uniquely:
\begin{align}
ds_{5}^2 
& = e^{-2w/L} (-dt^2 + d\bmx_3^2) + dw^2 \\
&= \left(\frac{r}{L}\right)^2 (- dt^2 + d\bmx_3^2) + L^2 \frac{dr^2}{r^2}~,
%
\end{align}
where $r=Le^{-w/L}$. This is known as the five-dimensional \keyword{anti-de~Sitter spacetime}, or the AdS$_5$ spacetime (\chap{AdS}). The length scale $L$ is known as the \keyword{AdS radius}. This is the spacetime which corresponds to the scale-invariant large-$N_c$ gauge theory.

The AdS$_5$ spacetime is the solution of the Einstein equation with the negative cosmological constant (\chap{AdS}):
\be
\action_5 = \frac{1}{16\pi G_{5}} \int d^{5}x\, \sqrt{-g_5} (R_5-2\Lambda)~,
\label{eq:gravity_side}
\ee
where $2\Lambda := -12/L^2$.

\danger{Readers may be puzzled that we introduce a scale $L$ in a ``scale-invariant theory." But the gravitational theory already has a scale $G_5$ or $l_s$. From the four-dimensional gauge theory point of view, we will see that $L$ appears only as the combination of dimensionless quantities $L^3/G_5$ and $L/l_s$.}

So far we focused only on the \poincare\ invariance and the scale invariance. Actually, both the ${\cal N}=4$ SYM and the AdS$_5$ spacetime have a larger symmetry. It is often the case in a relativistic field theory that the \poincare\ invariance $ISO(1,3)$ and the scale invariance combine into a larger symmetry, the \keyword{conformal invariance} $SO(2,4)$ \cite{Polchinski:1987dy}. The AdS$_5$ spacetime also has the $SO(2,4)$ invariance (\chap{AdS}).

To summarize, by imposing a further condition, the scale invariance, we arrive at
\begin{itemize}
\item Scale-invariant gauge theories such as the ${\cal N}=4$ SYM, and 
\item The AdS$_5$ spacetime.
\end{itemize}

\subsection{AdS/CFT, finally}\label{sec:towards_ads}

We now have enough information to carry out actual computations. First of all, \eq{pre_GKPW1} is replaced by
\be
\boxeq{
Z_\text{CFT} = Z_\text{AdS$_5$}~.
}
\label{eq:pre_GKPW2}
\ee
The left-hand side is the partition function of a gauge theory with scale invariance (conformal invariance). The right-hand side is the partition function of string theory on the AdS$_5$ spacetime. Such a relation is known as the \keyword{GKP-Witten relation}%
\footnote{The word ``GKP" are the initials of Gubser, Klebanov, and Polyakov who proposed such a relation independently from Witten \cite{Gubser:1998bc,Witten:1998qj}.}.
However, we do not really need to evaluate the full string partition function as we argue below.

\head{Relation of parameters}
Let us revisit the relation of parameters between two theories \eqref{eq:pre_dictionary}:
\be
N_c^2 \propto \frac{1}{g_s^2} \propto \frac{1}{G}~, \quad
\lambda \leftrightarrow l_s~.
\label{eq:pre_dictionary_again}
\ee
We did not make the dimensions to come out right at that time, but the AdS$_5$ spacetime introduces one more length scale $L$, so it is natural to expect 
\be
N_c^2 \simeq \frac{L^3}{G_5}~.
\label{eq:pre_dictionary1}
\ee
The 't~Hooft coupling $\lambda$ is related to the $\alpha'$-corrections. In the $\alpha'$-corrections, the combination $l_s^2 R$ appears (\sect{SUGRA}). The curvature of the AdS$_5$ spacetime is $R \simeq 1/L^2$. Thus, 
\be
\lambda \simeq \left(\frac{L}{l_s}\right)^\#~,
\label{eq:pre_dictionary2}
\ee
where \# is some number, which cannot be determined without additional information. The results here can be justified using the D-brane (\sect{D3_brane}). The D-brane can fix the number in \eq{pre_dictionary2} and the numerical coefficients in Eqs.~\eqref{eq:pre_dictionary1} and \eqref{eq:pre_dictionary2} as%
\footnote{The overall coefficient for $\lambda = (L/l_s)^4$ depends on the normalization of gauge group generators $t^a$. See, \eg, Sect.~13.4 of Ref.~\cite{Kiritsis:2007zz2}. We use $\text{tr}(t^a t^b) = \frac{1}{2} \delta^{ab}$.}
\be
\boxeq{
N_c^2 = \frac{\pi}{2} \frac{L^3}{G_5}~, \quad
\lambda = \left(\frac{L}{l_s}\right)^4~.
}
\label{eq:dictionary_preview}
\ee
Such relations are known as the \keyword{AdS/CFT dictionary}.

\head{The GKP-Witten relation in the large-$\Nc$ limit}
Going back to \eq{pre_GKPW2}, one does not really need to evaluate the full string partition function on the right-hand side. In the large-$N_c$ limit, it is enough to use the classical gravitational theory on the right-hand side. In such a case, one can use the saddle-point approximation just like when we argued black hole thermodynamics (\sect{Euclidean}):
\be
Z_\text{CFT ($N_c\gg1$)} = e^{-\SosE+O\left(l_s^2\right)}~.
\label{eq:pre_GKPW3}
\ee
Here, $\action_\text{E}$ is the Euclidean action of action \eqref{eq:gravity_side}, and $\SosE$  is the on-shell action which is obtained by substituting the classical solution of the metric (AdS$_5$ spacetime) to the action. 

In \eq{pre_GKPW3}, $O(l_s^2)$ represent the $\alpha'$-corrections. One can ignore the corrections when the spacetime curvature is small. From \eq{dictionary_preview}, ignoring the $\alpha'$-corrections corresponds to taking $\lambda \gg 1$ in the gauge theory side. Therefore, we arrive at our final relation: 
\be
\boxeq{
Z_\text{CFT ($N_c \gg \lambda\gg1$) } = e^{-\SosE}~.
}
\label{eq:GKPW_large_lambda}
\ee
This limit is hard to evaluate in a field theory even in the large-$N_c$ limit. However, the right-hand side tells us that this limit can be evaluated just using general relativity. We evaluate this relation in various examples below.

Now, the AdS$_5$ spacetime is not the only solution of \eq{gravity_side}. A black hole can exist in the AdS$_5$ spacetime (\chap{AdS_BH}). As we saw in \chap{BH_thermodynamics}, a black hole is a thermodynamic system, so the following correspondence is natural:
\begin{itemize}
\item Gauge theory at zero temperature $\leftrightarrow$ AdS$_5$ spacetime
\item Gauge theory at finite temperature $\leftrightarrow$ AdS$_5$ black hole
\end{itemize}
The discussion in this section is heuristic. But one can make a more systematic argument using the D-brane in string theory. This is because the D-brane can ``connect" the ${\cal N}=4$ SYM and the AdS$_5$ spacetime.

\section{\titlesummary}

\begin{itemize}
\item
In string theory, open strings represent a gauge theory, and closed strings represent a gravitational theory.
\item 
The partition function of a closed string theory is given by a summation over the world-sheet topologies. This reminds us of large-$\Nc$ gauge theories.
\item 
This clue of the partition function and the holographic principle suggest that a large-$\Nc$ gauge theory is represented by a five-dimensional curved spacetime with four-dimensional \poincare\ invariance $ISO(1,3)$.
\item 
The scale invariance is the useful guideline to determine the large-$\Nc$ gauge theory and the curved spacetime in question.
\item
Classically, a four-dimensional pure gauge theory is scale invariant. Quantum mechanically, it is not. However, there is a special gauge theory which keeps scale invariance even quantum mechanically. This is the \Nfour\ SYM. The \Nfour\ SYM actually has a larger symmetry, the conformal invariance $SO(2,4)$. 
\item
On the other hand, the \poincare\ invariance and the scale invariance in four-dimensions determine the curved spacetime counterpart as the AdS$_5$ spacetime. The AdS$_5$ spacetime actually has a larger symmetry $SO(2,4)$, which is the same as the \Nfour\ SYM.
\item
Superstring theory has extended objects, D-branes, other than strings. Some of black branes are the gravitational descriptions of D-branes. The D-branes are useful to establish the precise relation between the gauge theory and the curved spacetime.
\end{itemize}
In this chapter, the AdS spacetime and the AdS black hole appear, and we study these spacetimes in details in the following chapters.

\titlenewterms

Keywords in square brackets appear in appendices.

\begin{multicols}{2}
\noindent
meson\\
Regge trajectory\\
superstring theory\\
open/closed string\\
string length\\
D-brane\\
world-sheet\\
string coupling constant\\
supergravity\\
$\alpha'$-corrections\\
dilaton\\
scale invariance\\
conformal invariance\\
\newtermapp{Weyl invariance} \\
conformal field theory (CFT)\\
scaling dimension\\
${\cal N}=4$ super-Yang-Mills theory\\
anti-de~Sitter (AdS) spacetime\\
AdS radius\\
GKP-Witten relation\\
AdS/CFT dictionary\\
\newtermapp{R-symmetry}\\
\newtermapp{decoupling limit}\\
\newtermapp{near-horizon limit}\\
\newtermapp{gauged supergravity}
\end{multicols}

\section{Appendix: Scale, conformal, and Weyl invariance \advanced}\label{sec:weyl_inv}

In this appendix, we discuss three closely related symmetries, the scale invariance, the conformal invariance, and (local and global) Weyl invariances. 

The scale transformation $x^\mu \rightarrow a x^\mu$ implies $ds^2 = \eta_{\mu\nu} dx^\mu dx^\nu \rightarrow a^2\eta_{\mu\nu} dx^\mu dx^\nu$. So, it is convenient to consider the scale transformation of the metric $g_{\mu\nu} \rightarrow a^2 g_{\mu\nu}$, which is called the Weyl transformation. 

\head{The local Weyl invariance}
We start with the local Weyl transformation $g_{\mu\nu} \rightarrow a(x)^2 g_{\mu\nu}$ which is strongest among the symmetries we discuss here. The Maxwell theory in the curved spacetime
\be
\action = - \frac{1}{4e^2} \int d^4x\, \sqrt{-g} g^{\mu\nu} g^{\rho\sigma} F_{\mu\rho}F_{\nu\sigma}
%
\ee
is local Weyl invariant with $A_\mu \rightarrow A_\mu$. \index{Weyl invariance} 

If a theory is local Weyl invariant under $\delta g_{\mu\nu} = \epsilon(x) g_{\mu\nu}$ or $\delta g^{\mu\nu} = -\epsilon(x) g^{\mu\nu}$, 
\begin{align}
0 &= \delta \action = \int d^4x\, \frac{\delta \action}{\delta g^{\mu\nu}} \delta g^{\mu\nu} 
\nonumber \\
&= \frac{1}{2} \int d^4x\, \sqrt{-g}\, T_{\mu\nu} \left(\epsilon(x) g^{\mu\nu}\right)
= \frac{1}{2} \int d^4x\, \sqrt{-g}\, \epsilon(x) T^\mu_{~\mu}~.
%
\end{align}
In order for this to be true for any $\epsilon(x)$, the energy-momentum tensor must be traceless:
\be
T^\mu_{~\mu} = 0~.
%
\ee
When $g_{\mu\nu}=\eta_{\mu\nu}$, the local Weyl invariance reduces to the conformal invariance. 

Note that $A_\mu \rightarrow a^{-1} A_\mu$ under the scale transformation $x^\mu \rightarrow a x^\mu$ whereas $A_\mu \rightarrow  A_\mu$ under the local Weyl transformation. Thus, the scaling dimension under the local Weyl transformation in general differs from the scaling dimension under the scale transformation and from the naive mass dimension. As a related issue, for the local Weyl transformation, one has to assign a scaling dimension to the metric since it transforms nontrivially. Then, for a tensor, the scaling dimension depends on index positions (\ie, either lower or upper). See, \eg, App.~D of Ref.~\cite{Wald:1984rg2} and Ref.~\cite{Baier:2007ix} for more details. 

\head{The global Weyl invariance}
The global Weyl invariance $\delta g^{\mu\nu} = -\epsilon g^{\mu\nu}$ requires
\be
0 = \delta \action = \frac{\epsilon}{2} \int d^4x\, \sqrt{-g}\, T^\mu_{~\mu}~,
%
\ee
so the trace does not have to vanish but vanishes up to a total derivative. When $g_{\mu\nu}=\eta_{\mu\nu}$,
\be
T^\mu_{~\mu} = -\del_\mu K^\mu
\label{eq:virial}
\ee
for some $K^\mu$. The global Weyl invariance reduces to the scale invariance in the flat spacetime.

\head{The conformal invariance}

The conformal invariance is the flat spacetime limit of the local Weyl invariance $\eta_{\mu\nu} \rightarrow a(x)^2 \eta_{\mu\nu}$. For a scale-invariant theory, $T^\mu_{~\mu}$ vanishes up to a total derivative. A theorem states \cite{Polchinski:1987dy} that for a conformal invariant theory, 
\be
K^\mu = -\del_\nu L^{\nu\mu} 
\result 
T^\mu_{~\mu}=\del_\mu\del_\nu L^{\nu\mu}
%
\ee
for some $L^{\nu\mu}$ and that there exists an ``improved" energy-momentum tensor $\tilde{T}_{\mu\nu}$ which is traceless. (In the flat spacetime, the energy-momentum tensor is not unique. We discuss the related issue in the curved spacetime below.) For many four-dimensional relativistic field theories, there is no nontrivial candidate for $K^\mu$ which is not a divergence. Then, the scale invariance implies the conformal invariance. 

\head{Scalar field example}
As another example, consider a scalar field:
\be
\action = -\frac{1}{2} \int d^4x\, \{ (\del_\mu\phi)^2+m^2\phi^2 \}~.
%
\ee
When $m=0$, the theory is scale invariant with $\phi \rightarrow a^{-1}\phi$. The scalar field has scaling dimension 1. But when $m\neq0$, the theory is not scale invariant since we do not scale $m$. 

The energy-momentum tensor of the massless scalar field is given by
\be
T_{\mu\nu} = \del_\mu\phi \del_\nu\phi - \frac{1}{2}\eta_{\mu\nu} (\del\phi)^2~,
\label{eq:EM_scalar}
\ee
so $T_{\mu\nu}$ is not traceless:
\be
T^\mu_{~\mu} = - (\del\phi)^2 = -\frac{1}{2} \del^2\phi^2~,
\label{eq:trace_scalar}
\ee
where we used the equation of motion $\del^2\phi=0$ in the last equality. Comparing with \eq{virial}, $K^\mu = -\del^\mu\phi^2/2$. However, $K^\mu$ is a divergence, so the ``improved" energy-momentum tensor $\tilde{T}_{\mu\nu}$ exists:
\be
\tilde{T}_{\mu\nu} = T_{\mu\nu} + \frac{1}{6} (\eta_{\mu\nu}\del^2-\del_\mu\del_\nu)\phi^2~,
\label{eq:EM_scalar_improved}
\ee
which is traceless using the equation of motion. Then, the massless scalar theory is conformally invariant. 

Now, consider the Weyl invariance. In the curved spacetime, a simple extension of the scalar theory (so-called ``comma-goes-to-semicolon rule") is
\be
\action = -\frac{1}{2} \int d^4x\, \sqrt{-g} g^{\mu\nu} \del_\mu\phi \del_\nu\phi~.
\label{eq:scalar_curved1}
\ee
As one can see easily, the theory is global Weyl invariant with $\phi\rightarrow a^{-1}\phi$. However, it is not local Weyl invariant. As a result, the energy-momentum tensor is not traceless as in \eq{trace_scalar}. 

However, given a flat spacetime action, the curved spacetime extension is not unique. In the flat spacetime, $T_{\mu\nu}$ is not unique, and this is its curved-spacetime counterpart. There is a local Weyl invariant scalar theory:
\be
\action = -\frac{1}{2} \int d^4x\, \sqrt{-g} \{ (\nabla\phi)^2+ \xi R \phi^2 \}~,
\label{eq:scalar_curved2}
%
\ee
where $\xi=1/6$. The theory is local Weyl invariant with $\phi\rightarrow a(x)^{-1}\phi$. The energy-momentum tensor is given by
\be
T_{\mu \nu} = \del_{\mu} \phi \del_{\nu} \phi - \frac{1}{2} g_{\mu \nu} (\del \phi)^2 
+ \xi \left\{ g_{\mu \nu} \nabla^2 - \nabla_{\mu} \nabla_{\nu} + R_{\mu \nu} - \frac{1}{2} g_{\mu \nu} R \right\} \phi^2.
\label{eq:EM_scalar_conf}
\ee
One can show that it is traceless using the equation of motion. In the flat spacetime, \eq{EM_scalar_conf} reduces to
\be
T_{\mu\nu} = \del_\mu\phi \del_\nu\phi - \frac{1}{2}\eta_{\mu\nu} (\del\phi)^2 
 + \xi (\eta_{\mu\nu}\del^2-\del_\mu\del_\nu)\phi^2~,
%
\ee
which is the curved-spacetime counterpart of $\tilde{T}_{\mu\nu}$ \eqref{eq:EM_scalar_improved}.

\section{Appendix: D-brane and AdS/CFT \advanced
}\label{sec:D3_brane}

\danger{It is not necessary for beginners to understand this appendix completely. It is enough to go back to this section after you get accustomed to AdS/CFT computations. }

The D3-brane in \sect{D_brane_primer} has two descriptions as 
\begin{itemize}
\item gauge theory,
\item black brane in supergravity.
\end{itemize}
As we see below, these two descriptions give the \Nfour\ SYM and the AdS$_5$ spacetime in appropriate limits.

\subsection{D-brane and gauge theory
}\label{sec:gauge}

In order to know more about the gauge theory the D-brane describes, let us look at open string oscillations more carefully.

The open strings on a D-brane are bounded on the D-brane, so the D3-brane represents a four-dimensional gauge theory, but superstring theory actually requires 10-dimensional spacetime for consistency. Then, the open strings on the D-brane still oscillate in the full 10-dimensional spacetime. Thus, the simplest open string oscillations have 8 degrees of freedom instead of 2. What are these degrees of freedom? In other words, what kind of gauge theory the D3-brane represents?

To see this, note that there are two types of string oscillations. First are the oscillations in the brane and the other are the oscillations out of the brane. From the four-dimensional point of view [in terms of $SO(1,3)$ representations], the former represents a gauge field, and the latter represents scalars. The spatial dimension is 9 and the brane dimension is 3, so there are 6 scalar fields. Thus, the gauge theory represented by the D3-brane inevitably comes with scalar fields. In addition, there are fermions due to supersymmetry (which comes from the supersymmetry in superstring but we omit the details.) The Lorentz transformation properties are different, but they all come from similar string oscillations, which means that all these fields transform as the adjoint representation of $SU(\Nc)$. Namely, the theory has no fundamental representation such as quarks%
\footnote{We describe one simple way to include fundamental representations in \sect{wilson_intuitive}.}. 
These properties coincide with the \Nfour\ SYM. 

The D-brane consideration provides us one more important information about the \Nfour\ SYM. If we have only D3-branes, as in the present case, the directions transverse to the brane are all isotropic. These directions correspond to the scalar fields $\phi_i$, so the isotropy means that there is a global $SO(6)$ symmetry for $\phi_i$. Such a global symmetry is known as 
\keyword{R-symmetry}.
Thus,  combining with the conformal invariance $SO(2,4)$, 
\begin{center}
\fbox{
\begin{tabular}{l}
The ${\cal N}=4$ SYM has the global $SO(2,4) \times SO(6)_R$ symmetry. 
\end{tabular}
}
\end{center}

We get the \Nfour\ SYM from the D-brane, 
but it is not yet clear if the D-brane is simply described by a gauge theory. This is because string theory is more than a gauge theory.
\begin{itemize}
\item In particular, string theory contains graviton since string theory is the unified theory.
\item The simplest oscillations (harmonics) of an open string corresponds to a gauge theory as we saw before, but the string has higher harmonics which correspond to massive particles. 
\end{itemize}
At this point, it is not clear if these effects can be ignored. 

First, let us consider gravity. According to general relativity, any energy-momentum tensor curves spacetime. The D-brane has some energy, so how does the D-brane curve spacetime? Since gravity is described by the Newtonian potential
\be
\phi_\text{Newton} \simeq \frac{GM}{r}~,
\ee
one can measure the effect of curvature by $GM$. However, this is the case for a point particle in three-dimensional space. We have to use the 10-dimensional Newton's constant $G_{10}$ instead of the four-dimensional Newton's constant $G_4$. Also, the brane has the spatial extension. So, we have to use the mass density of the brane, $\tension_3$, instead of $M$. We also have to take into account that the number of the spatial dimensions transverse to the brane is six. Then, 
\be
\phi_\text{Newton} \simeq \frac{G_{10} \tension_3}{r^4}~.
\ee
According to string theory, the D-brane mass density is given by
\be
\tension_3 \simeq \frac{N_c}{g_s}\frac{1}{l_s^4}~.
\label{eq:D3_tension}
\ee
Since $G_{10} \simeq g_s^2 l_s^8$ from \eq{string_vs_qft}, 
\be
\phi_\text{Newton} \simeq \frac{g_s N_c l_s^4}{r^4}~.
\label{eq:Newton_D3}
\ee

On the other hand, the effective coupling constant of the gauge theory is the 't~Hooft coupling constant $\lambda := g_\text{YM}^2 N_c \simeq g_s N_c$.  To summarize,
\begin{alignat}{2}
&\text{Gravity by brane: }  & \quad G_{10} \tension_3 &\simeq g_s N_c l_s^4~, \\
&\text{effective coupling of gauge theory: }  & \quad \lambda &\simeq g_s N_c~.
%
\end{alignat}
If one takes the limit $l_s \rightarrow 0$ (decoupling limit), one can leave nontrivial dynamics of the gauge theory and at the same time one can ignore the effect of gravity.

Second, let us consider the massive particles. They have the mass $M \simeq 1/l_s$. Thus, they also decouple from the theory in the limit $l_s \rightarrow 0$.

In conclusions, what we get from the D3-brane is
\begin{center}
(\Nfour\ SYM) \\ + (supergravity in the 10-dimensional flat spacetime)
\end{center}

\subsection{D-brane and curved spacetime
}

\begin{figure}[tb]
\centering
\scalebox{0.75}{ \includegraphics{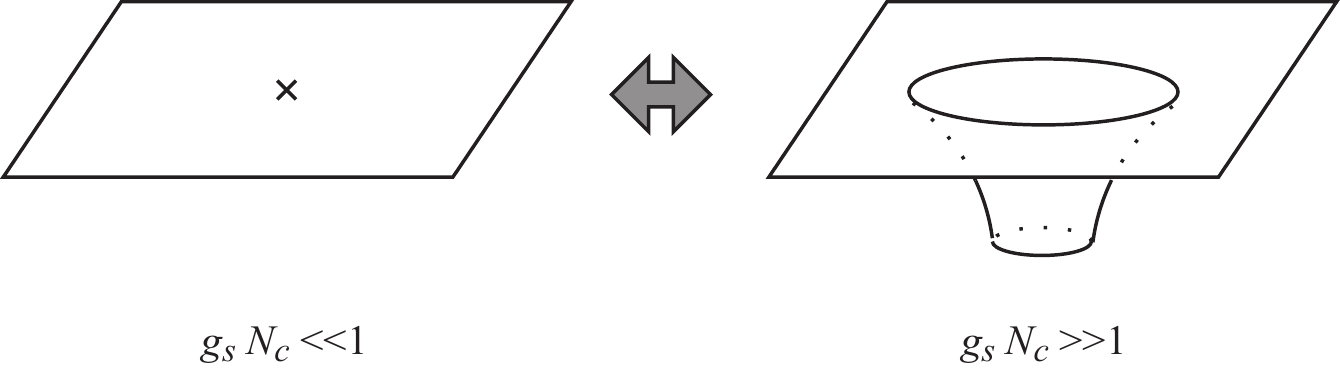} }
\vskip2mm
\caption{When $g_s N_c \ll 1$, the curved spacetime region by the D-brane is small so that one can approximate the geometry by a flat spacetime with a source (left). However, when $g_s N_c \gg 1$, a macroscopic curved spacetime appears (right).}
\label{fig:near_horizon}
\end{figure}

The D3-brane describes the \Nfour\ SYM in the flat spacetime. But this conclusion is not valid when $g_s N_c \gg 1$ because the D-brane starts to curve the spacetime. Going back to the Newtonian potential argument, we essentially took the $GM \rightarrow 0$ limit in order to ignore the effect of gravity. But one cannot ignore gravity near the origin $r\rightarrow 0$ even in this limit. As a consequence, a curved spacetime appears near the origin $r^4 \ll g_s N_c l_s^4$ (\fig{near_horizon}). 

In the $g_s N_c \ll 1$ limit, the curved region is small. Then, one can consider that there exists a source represented by the D-brane in the flat spacetime. This is the case we considered in the above subsection. But a macroscopic curved spacetime appears in the $g_s N_c \gg 1$ limit.

The supergravity description is appropriate in the $g_s N_c \gg 1$ limit. The black D3-brane is given by \cite{Horowitz:1991cd}
\begin{align}
ds_{10}^2 &= Z^{-1/2} (- dt^2 + d\bmx_3^2) + Z^{1/2} (dr^2+r^2 d\Omega_5^2)~,
\label{eq:D3_extreme} \\
Z &= 1+\left(\frac{L}{r}\right)^4~, \quad
L^4 \simeq g_s N_c l_s^4~,
\label{eq:D3_extreme2}
%
\end{align}
where $\bmx_3=(x,y,z)$ represents the directions of the spatial extension of the brane. The behavior of the factor $Z$ comes from the Newtonian potential \eqref{eq:Newton_D3}. We are interested in how the D-brane curves the spacetime near the origin. So, taking the limit $r \ll L$ (\keyword{near-horizon limit}), one gets
\be
ds_{10}^2 \rightarrow \left(\frac{r}{L}\right)^2 (- dt^2 + d\bmx_3^2)
+ L^2 \frac{dr^2}{r^2}+L^2 d\Omega_5^2~.
\label{eq:D3_near_horizon}
\ee
The part $L^2 d\Omega_5^2$ represents $S^5$ with radius $L$. The remaining five-dimensional part is the AdS$_5$ spacetime. This near-horizon limit corresponds to the decoupling limit from the gauge theory point of view%
\footnote{
\advanced To see the relation, we need to discuss the decoupling limit more carefully. Denote the typical scale of the gauge theory as $l_\text{obs}$. One can ignore the string length scale $l_s$ when
\be
l_s \ll l_\text{obs}~.
\label{eq:decoupling1}
\ee
As $l_\text{obs}$, let us take the ``W-boson" mass scale. Here, the ``W-boson" is the gauge boson which arises by breaking the gauge group $SU(N_c) \rightarrow SU(N_c-1) \times U(1)$. This corresponds to separating a D-brane from $N_c$ D-branes. The gauge boson corresponds to an open string, but the open string between the D-brane and $(N_c-1)$ D-branes is stretched, and the open string has a tension. So, the corresponding W-boson becomes massive. The mass is given by
\be
\frac{1}{l_\text{obs}} \simeq \frac{r}{l_s^2} 
\label{eq:gauge_boson}
\ee
[The tension of the open string is $O(1/l_s^2)$]. From Eqs.~\eqref{eq:decoupling1} and \eqref{eq:gauge_boson}, one gets
\be
r \ll l_s~.
\label{eq:decoupling2}
\ee
This is the decoupling limit. 

When $g_s N_c \gg 1$, the $r \ll l_s$ limit implies $r \ll L$, which is the near-horizon limit. 
}.

Therefore, let us roughly divide the spacetime made by the D3-brane into AdS$_5 \times S^5$ near the origin and the nearly flat spacetime around:
\begin{center}
(supergravity on AdS$_5 \times S^5$) \\ + (supergravity in the 10-dimensional flat spacetime)
\end{center}

As we will see in \chap{AdS}, the AdS$_5$ spacetime has the $SO(2,4)$ invariance. In addition, the full geometry \eqref{eq:D3_near_horizon} involves $S^5$ which has the $SO(6)$ invariance. This is the same symmetry as the ${\cal N}=4$ R-symmetry. Thus,
\begin{center}
\fbox{
\begin{tabular}{l}
The gravity side also has the global $SO(2,4) \times SO(6)$ symmetry.
\end{tabular}
}
\end{center}

The gauge theory description fails when $g_s N_c \gg 1$. On the other hand, the supergravity description fails when $g_s N_c \ll 1$. This is because the curvature of the metric \eqref{eq:D3_near_horizon} behaves as
\be
R^{MNPQ} R_{MNPQ} \propto \frac{1}{g_s N_c l_s^4}~.
%
\ee
When $g_s N_c \ll 1$, the curvature becomes large. The $\alpha'$-corrections \index{alpha'-corrections@$\alpha'$-corrections} become important, which can change the metric \eqref{eq:D3_near_horizon} in general. The metric such as \eq{D3_near_horizon} is trustable when $g_s N_c \gg 1$.

\begin{figure}[tb]
\centering
\subfigure[${\cal N}=4$ SYM]{
\scalebox{0.75}{ \includegraphics{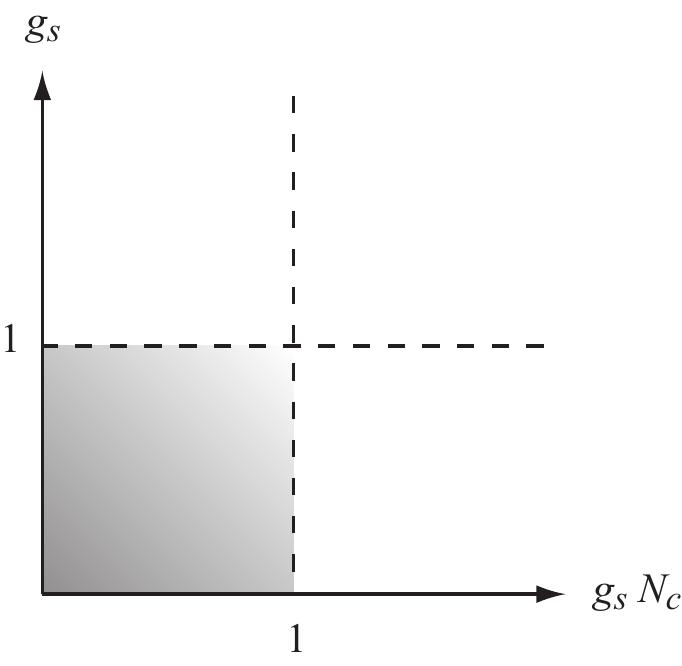} }
} 
\subfigure[Supergravity on AdS$_5 \times S^5$]{
\scalebox{0.75}{ \includegraphics{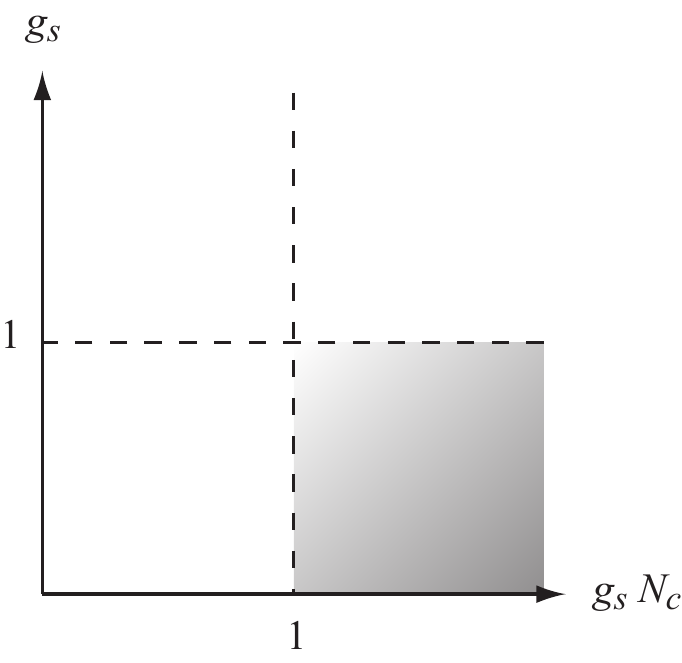} }
}
\vskip2mm
\caption{
We obtained two systems from the D3-brane, the \Nfour\ SYM and supergravity on  AdS$_5 \times S^5$. These two theories are complementary to each other. Each theory has a region where analysis is relatively easy (shaded region), but they do not overlap. AdS/CFT claims that two systems are equivalent.
}
\label{fig:ads_cft}
\end{figure}%

\subsection{Gauge theory and curved spacetime
}

To summarize our discussion so far, we obtained two systems from the D3-brane, the \Nfour\ SYM and supergravity on AdS$_5 \times S^5$. We also obtained a common system which is decoupled from the rest, supergravity in the 10-dimensional flat spacetime, in both cases. Since this part is common, one can forget it. Then, the \Nfour\ SYM corresponds to supergravity on AdS$_5 \times S^5$. 

These two descriptions of the brane, gauge theory and supergravity, are complementary to each other (\fig{ads_cft}). 
The former description is valid when $g_s \ll 1$ and $g_s N_c \ll 1$, whereas the latter description is valid when $g_s \ll 1$ and $g_s N_c \gg 1$. 
%
%

However, in principle, both theories exist for all $g_s$ and $g_s N_c$. Taking \sect{reexamination} discussion into account, it is natural to imagine that both theories are equivalent for all $g_s$ and $g_s N_c$. If this is true, we can make computations using one theory even when the other theory is hard to compute. 

We should stress that this is not a logical consequence. We obtained two systems from the D3-brane. But these two systems are two different limits of the brane. In order to justify the above claim, one has to compare results in both theories computed at the same $g_s$ and $g_s N_c$. There are many circumstantial evidences when one can rely on supersymmetry, but we will not discuss such evidences in this book. Rather, we go through computations related to actual applications, and we will see that gravity results can indeed be interpreted as physical quantities of gauge theories. We also compare the AdS/CFT results with experiments and with the other theoretical tools such as the lattice gauge theory in order to see that AdS/CFT is likely to be true.

\subsection{What D-brane tells us
}\label{sec:definition}

The D-brane taught us two things. First, the GKP-Witten relation \eqref{eq:pre_GKPW2} becomes more precise:
\be
%
Z_\text{\Nfour} = Z_\text{AdS$_5 \times S^5$}~.
%
\ee
The left-hand side is the partition function of the \Nfour\ SYM, and the right-hand side is the partition function of string theory on AdS$_5 \times S^5$. 

We mainly focused on the conformal invariance. But, in retrospect, we are able to identify the gauge theory because we take the \Nfour\ R-symmetry into account. One would say that we added $S^5$ on the gravity side to reflect the R-symmetry on the gauge theory side. Thus, when we try to find a gravity dual, it is in general important to take care of the symmetries both theories have.

Second, from the D-brane, we are able to obtain the AdS/CFT dictionary \eqref{eq:dictionary_preview}. This is possible by combining various expressions we encountered:
\begin{align}
L^4 &\simeq g_s N_c l_s^4~, \\
G_{10} 
&\simeq g_s^2 l_s^8~, \\
g_s &\simeq g_\text{YM}^2~.
%
\end{align}
Eliminating $g_s$ from these expressions give
\be
N_c^2 \simeq \frac{L^8}{G_{10}}~, \quad \lambda \simeq \left(\frac{L}{l_s} \right)^4~.
\label{eq:dictionary_preliminary}
\ee

Now, $S^5$ corresponds to the R-symmetry from the point of view of the \Nfour\ SYM and is an important part of the theory. But one often compactifies $S^5$ and consider the resulting five-dimensional gravitational theory. The theory obtained in this way is called \keyword{gauged supergravity}.

The actual procedure of the $S^5$ compactification is rather complicated, and the full gauged supergravity action is complicated as well. But in our case, we do not have to go through such a computation, and we can infer the action. The five-dimensional part is the AdS$_5$ spacetime, so the five-dimensional action should be given by \eq{gravity_side}. Since $S^5$ has radius $L$, the compactification gives
\be
\frac{1}{16\pi G_{10}} \int d^{10}x \sqrt{-g_{10}}R_{10} 
=
\frac{L^5 \Omega_5}{16\pi G_{10}} \int d^{5}x \sqrt{-g_{5}}\left(R_{5} 
+\cdots \right)~.  
\nonumber
%
\ee
Thus, the five-dimensional Newton's constant is given by  $G_5 \simeq G_{10}/L^5$. Then, the AdS/CFT dictionary \eqref{eq:dictionary_preliminary} can be rewritten by $G_5$. If one works out numerical coefficients, the results are
\be
\boxeq{
N_c^2 = \frac{\pi}{2} \frac{L^3}{G_5}~, \quad
\lambda = \left(\frac{L}{l_s}\right)^4~.
}
\label{eq:dictionary}
\ee

\subsubsection*{Finite-temperature case
}

The metric \eqref{eq:D3_extreme} is the zero-temperature solution of the D3-brane, or the extreme black hole \index{extreme black hole} solution, which corresponds to the \Nfour\ SYM at zero temperature. In order to discuss the finite temperature gauge theory, we use the finite temperature solution of the D3-brane:
\begin{align}
ds_{10}^2 &= Z^{-1/2} (- h dt^2 + d\bmx_3^2) + Z^{1/2} (h^{-1}dr^2+r^2 d\Omega_5^2)~,
\\
Z &= 1+\left(\frac{L}{r}\right)^4~, \\
h &= 1-\left(\frac{r_0}{r}\right)^4~. 
\label{eq:D3_non_extreme}
%
\end{align}
The horizon is located at $r=r_0$. Again take the near-horizon limit $r \ll L$. In order to remain outside the horizon $r > r_0$ even in the near-horizon limit, take $r_0 \ll L$. Then, the resulting geometry is
\be
ds_{10}^2 \rightarrow \left(\frac{r}{L}\right)^2 (- h dt^2 + d\bmx_3^2)
+ L^2 \frac{dr^2}{hr^2}+L^2 d\Omega_5^2~.
\label{eq:D3_near_horizon2}
\ee
This is known as the \textit{Schwarzschild-AdS$_5$ black hole}\index{Schwarzschild-AdS black hole (planar horizon)} (\chap{AdS_BH}). When $h=1$, the geometry reduces to the AdS$_5$ spacetime. Also, the factor similar to $h$ appeared in the higher-dimensional Schwarzschild black holes.



\endofsection

\ifx\nameofpaper\undefined 
  \usepackage{macro_natsuume} 
  \def\beginsection{\section*}
  \def\endofsection{\end{document}} 
  \input draft_header.tex
\else 
  \def\beginsection{\chapter}
  \def\endofsection{ } 
\fi

\beginsection{The AdS spacetime}\label{chap:AdS}


\begin{quote}
The AdS spacetime is one of spacetimes with constant curvature. To be familiar with the AdS spacetime, we first discuss spaces with constant curvature such as the sphere and then discuss spacetimes with constant curvature.
\end{quote}

\section{Spacetimes with constant curvature}

\subsection{Spaces with constant curvature}

Before we discuss spacetimes with constant curvature such as the AdS spacetime, let us consider a simple example, the sphere $S^2$. Consider the three-dimensional Euclidean space with metric
\be
ds^2 = dX^2 + dY^2 + dZ^2~.
\label{eq:embed_s2}
\ee
The sphere is defined by the surface which satisfies the constraint
\be
X^2 + Y^2 + Z^2 = L^2~.
\label{eq:surface_s2}
\ee
This constraint can be solved by the familiar spherical coordinates:
\be
X = L \sin\theta \cos\varphi~, \quad Y = L \sin\theta \sin\varphi~, \quad Z = L \cos\theta~.
\ee
In this coordinate system, the metric then becomes 
\be
ds^2 = L^2(d\theta^2 + \sin^2\theta\, d\varphi^2)~.
\label{eq:metric_s2}
\ee
The sphere $S^2$ has the $SO(3)$ invariance. This is because the surface \eqref{eq:surface_s2} respects the $SO(3)$ invariance of the ``ambient space," the three-dimensional Euclidean space \eqref{eq:embed_s2}. By an $SO(3)$ transformation, any point on $S^2$ can be mapped to the other points. In this sense, $S^2$ is homogeneous. The sphere has a constant positive curvature and the Ricci scalar is given by
\be
R = \frac{2}{L^2}~.
\ee

\begin{figure}[tb]
\centering
\scalebox{0.75}{  \includegraphics{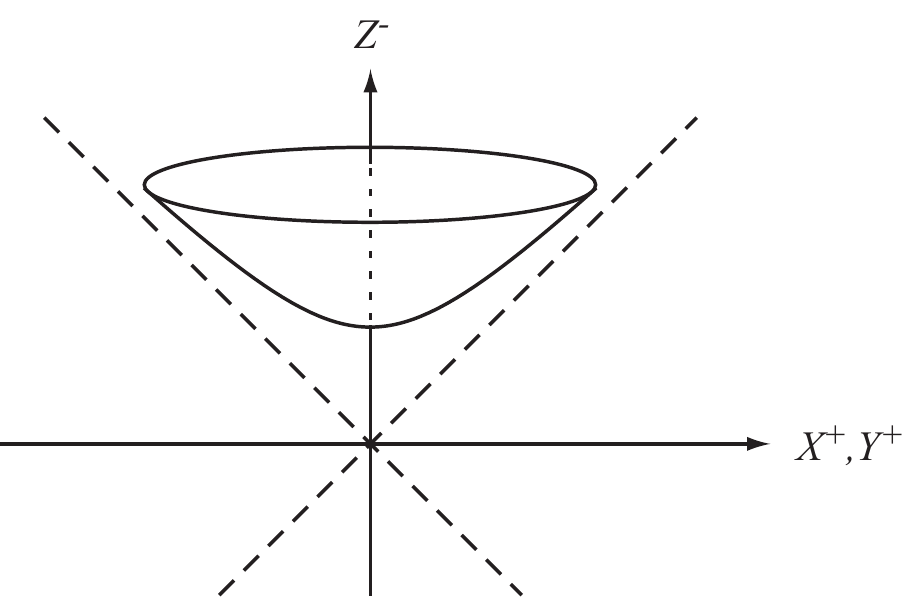} }
\vskip2mm
\caption{The embedding of $H^2$ into $\mathbb{R}^{1,2}$. The superscript signs refer to the signature of the ambient spacetime ($-$: timelike, $+$: spacelike).}
\label{fig:H2}
\end{figure}%

The space with constant negative curvature is known as the \keyword{hyperbolic space} $H^2$. It is much harder to visualize the space with constant negative curvature. It is partly because the hyperbolic space cannot be embedded into the three-dimensional Euclidean space unlike the sphere. But the hyperbolic space can be embedded into the three-dimensional Minkowski space. The hyperbolic space is defined by
\begin{gather}
ds^2 = - dZ^2 + dX^2 + dY^2~,
\label{eq:embed_h2} \\
-Z^2 + X^2 + Y^2 = - L^2~,
\label{eq:surface_h2}
\end{gather}
(\fig{H2}). The hyperbolic space is homogeneous in the sense that it respects the $SO(1,2)$ invariance of the ambient Minkowski spacetime. Namely, any point on the surface can be mapped to the other points by an $SO(1,2)$ ``Lorentz" transformation. Note that the hyperbolic space is \textit{not} the familiar hyperboloid in the three-dimensional Euclidean space:
\begin{gather}
ds^2 = dZ^2 + dX^2 + dY^2~, \\
-Z^2 + X^2 + Y^2 = - L^2~.
\end{gather}
The hyperboloid is embedded into the three-dimensional Euclidean space. Thus, the hyperboloid does not respect the $SO(3)$ invariance of the ambient space, and it is not homogeneous.

In order to solve the constraint \eqref{eq:surface_h2}, take a coordinate system which is slightly different from the sphere:
\be
X = L \sinh\rho \cos\varphi~, \quad Y = L \sinh\rho \sin\varphi~, \quad Z = L \cosh\rho~.
\ee
Then, the metric of the hyperbolic space is given by
\be
ds^2 = L^2 (d\rho^2 + \sinh^2\rho\, d\varphi^2)~.
\label{eq:metric_h2}
\ee
As one can see from this metric, the hyperbolic space itself does not have a timelike direction although we embed it into the three-dimensional Minkowski spacetime. Namely, the hyperbolic space is a space not a spacetime. The curvature is constant negative:
\be
R = - \frac{2}{L^2}~.
\ee

It is not easy to visualize the hyperbolic space, but there is a familiar example. The mass-shell condition \eqref{eq:constraint_on_shell} in special relativity is nothing but the hyperbolic space. For the particle with mass $m$, the canonical momentum $p_\mu=(p_0,p_1,p_2)$ satisfies
\be
p^2 = - (p_0)^2 + (p_1)^2 + (p_2)^2 = - m^2~.
\ee
This takes the same from as the embedding equation \eqref{eq:surface_h2} for $H^2$.

\subsection{Spacetimes with constant curvature
}

\begin{figure}[tb]
\centering
\scalebox{0.75}{ \includegraphics{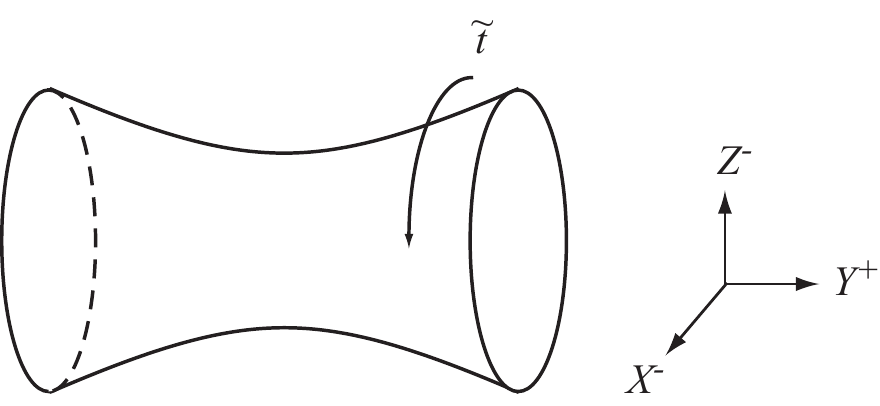} }
\vskip2mm
\caption{The embedding of AdS$_2$ into $\mathbb{R}^{2,1}$. The timelike direction $\tg$ is periodic, so we consider the covering space. }
\label{fig:AdS2}
\end{figure}%

So far we discussed spaces with constant curvature, but now consider \textit{spacetimes} with constant curvature. The AdS$_2$ spacetime\index{AdS spacetime} can be embedded into a flat spacetime with \textit{two} timelike directions (\fig{AdS2}):
\begin{gather}
ds^2 = - dZ^2 - dX^2 + dY^2~,
\label{eq:embed_ads2} \\
-Z^2 - X^2 + Y^2 = - L^2~.
\label{eq:surface_ads2}
\end{gather}
The parameter $L$ is called the \keyword{AdS radius}. The AdS$_2$ spacetime has the $SO(2,1)$ invariance. Just like $S^2$ and $H^2$, take a coordinate system
\be
Z = L \cosh\rho \cos\tg~, \quad 
X = L \cosh\rho \sin\tg~, \quad 
Y = L \sinh\rho~.
\label{eq:def_global}
\ee
Then, the metric becomes
\be
ds^2 = L^2 (- \cosh^2\rho\, d\tg^2 + d\rho^2)~.
\label{eq:metric_ads2}
\ee
This coordinate system $(\tg, \rho)$ is called the \keyword{global coordinates}. Although we embed the AdS spacetime into a flat spacetime with two timelike directions $X$ and $Y$, the AdS spacetime itself has only one timelike direction.

From \eq{def_global}, the coordinate $\tg$ has the periodicity $2\pi$, so the timelike direction is periodic. This is problematic causally%
\footnote{One can have closed timelike curves, where causal curves are closed. }, 
so one usually unwraps the timelike direction and considers the covering space of the AdS$_2$ spacetime, where $-\infty<\tg<\infty$. The AdS spacetime in AdS/CFT is this covering space. The AdS$_2$ spacetime has a constant negative curvature $R=-2/L^2$.

\begin{figure}[tb]
\centering
\scalebox{0.75}{ \includegraphics{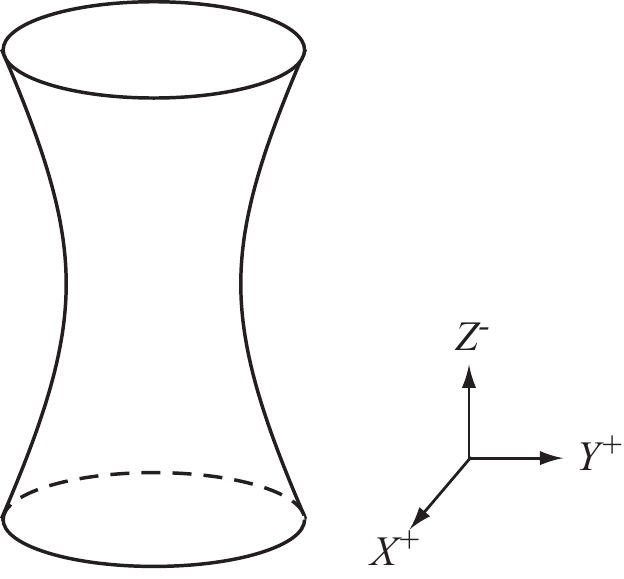} }
\vskip2mm
\caption{The embedding of dS$_2$ into $\mathbb{R}^{1,2}$. The figure looks the same as \fig{AdS2}, but the signature of the ambient spacetime differs from \fig{AdS2}.}
\label{fig:dS2}
\end{figure}%

The \keyword{de~Sitter spacetime} is another spacetime with constant curvature, but this time constant positive curvature. It does not frequently appear in AdS/CFT, but the spacetime itself has been widely discussed in connection with the dark energy in cosmology. The two-dimensional de~Sitter spacetime, or the dS$_2$ spacetime is defined by
\begin{gather}
ds^2 = - dZ^2 + dX^2 + dY^2~,
\label{eq:embed_ds2} \\
-Z^2 + X^2 + Y^2 = + L^2~,
\label{eq:surface_ds2}
\end{gather}
(\fig{dS2}). The dS$_2$ spacetime has the $SO(1,2)$ invariance. In the coordinates 
\be
X = L \cosh\tg \cos\theta~, \quad Y = L \cosh\tg \sin\theta~, \quad Z = L \sinh\tg~,
\ee
the metric becomes 
\be
ds^2 = L^2 (- d\tg^2 + \cosh^2\tg\, d\theta^2)~.
\label{eq:metric_ds2}
\ee
The dS$_2$ spacetime has a constant positive curvature $R=2/L^2$. One often considers the AdS$_5$ spacetime for applications to AdS/CFT, but for the dS spacetime, one often considers the dS$_4$ spacetime for applications to cosmology.

\subsection{Relation with constant curvature spaces}

We saw various spaces and spacetimes, but the spacetimes with constant curvature are related to the spaces with constant curvature. Take a constant curvature spacetime. The Euclidean rotation of the timelike direction gives a constant curvature space:
\be
\boxeq{
\begin{aligned}
\text{AdS}_2 &\xrightarrow{\text{Euclidean}}{} \text{H}^2 \\
\text{dS}_2 &\xrightarrow{\text{Euclidean}}{} S^2
\end{aligned}
}
\ee
The dS$_2$ spacetime and $S^2$ are defined by
\begin{align}
\text{dS}_2: &&  ds^2 &= - dZ^2 \!+\! dX^2 \!+\! dY^2~, & -Z^2 \!+\! X^2 \!+\! Y^2 &= L^2~.
\\
S^2: && ds^2 &= dZ^2 \!+\! dX^2 \!+\! dY^2~, & Z^2 \!+\! X^2 \!+\! Y^2 &= L^2~.
\end{align}
The dS$_2$ spacetime becomes $S^2$ by $Z_\text{E} = iZ$. The AdS$_2$ spacetime and $H^2$ are defined by
\begin{align}
\text{AdS}_2: && ds^2 &= -dZ^2 \!-\! dX^2 \!+\! dY^2~, & -Z^2 \!-\! X^2 \!+\! Y^2 &= -L^2~.
\\
H^2: && ds^2 &= - dZ^2 \!+\! dX^2 \!+\! dY^2~, & -Z^2 \!+\! X^2 \!+\! Y^2 &= -L^2~.
\end{align}
The AdS$_2$ spacetime becomes $H^2$ by $X_\text{E} = iX$. The Euclidean spacetime is often used in AdS/CFT.

\subsection{Various coordinate systems of AdS spacetime
}\label{sec:AdS_coordinates}

So far we discussed the AdS spacetime using the global coordinates $(\tg, \rho)$:
\be
\frac{ds^2}{L^2} = - \cosh^2\rho\, d\tg^2 + d\rho^2~.
\ee
But various other coordinate systems appear in the literature.

\subsubsection*{Static coordinates
$(\tg, \rs)$}
\index{static coordinates}

The coordinate $\rs$ is defined by $\rs := \sinh\rho$. The metric becomes
\be
\frac{ds^2}{L^2} = - (\rs^2+1)\, d\tg^2 + \frac{d\rs^2}{\rs^2+1}~.
\label{eq:ads2_static}
\ee
This coordinate system is useful to compare with the AdS black hole (\sect{spherical_SAdS}).

\subsubsection*{Conformal coordinates
$(\tg, \theta)$}
\index{conformal coordinates}

\begin{figure}[tb]
\centering
\scalebox{0.75}{ \includegraphics{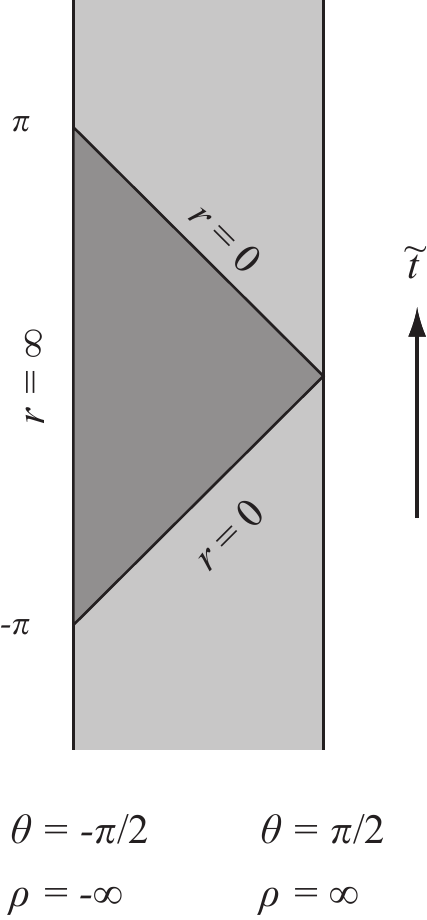} }
\vskip2mm
\caption{The AdS$_2$ spacetime in conformal coordinates. The \poincare\ coordinates cover only part of the full AdS spacetime which is shown in the dark shaded region (\poincare\ patch).}
\label{fig:AdS2_Penrose}
\end{figure}

The coordinate $\theta$ is defined by $\tan \theta := \sinh\rho$ ($\theta: -\pi/2 \rightarrow \pi/2$).
The metric becomes flat up to an overfall factor (conformally flat)\index{conformally flat}%
\footnote{The name ``conformal coordinates" is not a standard one.}:
\be
\frac{ds^2}{L^2} = \frac{1}{\cos^2\theta} (- d\tg^2 + d\theta^2 )~.
\label{eq:ads2_conf}
\ee
The AdS spacetime is represented as \fig{AdS2_Penrose} in this coordinate system. What is important is the existence of the spatial ``boundary" at $\theta=\pm\pi/2$. This boundary is called the \keyword{AdS boundary}. The AdS boundary is located at $\rs\rightarrow\infty$ in static coordinates and is located at $\rP\rightarrow\infty$ in \poincare\ coordinates below. The existence of the boundary means that one should specify the boundary condition on the AdS boundary in order to solve initial-value problems. From the AdS/CFT point of view, this boundary condition corresponds to specifying external sources one adds in the gauge theory side (\chap{GKPW}).

\subsubsection*{\poincare\ coordinates 
$(t, \rP)$}
\index{Poincar\'{e} coordinates}

This coordinate system is defined by
\begin{align}
Z &= \frac{L\rP}{2} \left( -t^2 + \frac{1}{\rP^2} + 1 \right)~, \\
X &= L \,\rP\, t~, \\
Y &= \frac{L\rP}{2} \left( -t^2 + \frac{1}{\rP^2} - 1 \right)~, 
%
\end{align}
($\rP>0$, $t:-\infty\rightarrow\infty$). The metric becomes
\be
\boxeq{
\frac{ds^2}{L^2} = - \rP^2 dt^2 + \frac{d\rP^2}{\rP^2}~.
}
\label{eq:ads2_poincare}
\ee
This is the most often used coordinate system in AdS/CFT. This coordinate system is also useful to compare with the AdS black hole (\sect{SAdS}).

\subsection{Higher-dimensional cases
}\label{sec:higher_AdS}

The spaces/spacetimes with constant curvature discussed so far can be easily generalized to the higher-dimensional case. In all examples, adding $p$ spatial directions to ambient spacetimes gives the $(p+2)$-dimensional cases.

\subsubsection*{The sphere $S^{p+2}$}

The unit $S^{p+2}$ has the $SO(p+3)$ invariance and is defined by
\begin{gather}
ds_{p+3}^2 = d\omega_1^2 + \cdots + d\omega_{p+3}^2~, \\
\omega_1^2 + \cdots + \omega_{p+3}^2 = + 1~.
\end{gather}
The $S^{p+2}$ symmetry contains the $SO(p+2)$ invariance as a subgroup, which can be used to write the metric. For example, the unit $S^3$ coordinates can be chosen using $S^2$ coordinates $(\theta_2, \theta_3)$ as 
\be
\omega_2 = r \cos\theta_2~, \quad 
\omega_3 = r \sin\theta_2 \cos\theta_3~, \quad
\omega_4 = r \sin\theta_2 \sin\theta_3~,
%
\ee
and $\omega_1^2=1-r^2$. Then,
\be
ds^2 = d\omega_1^2 + dr^2 + r^2 d\Omega_2^2 
= \frac{dr^2}{1-r^2} + r^2 d\Omega_2^2~,
%
\ee
where $d\Omega_2^2 = d\theta_2^2 + \sin^2\theta_2\,d\theta_3^2$.
In the coordinate $r^2 = \sin^2\theta_1$ $(0 \leq \theta_1 <\pi)$, 
\be
d\Omega_3^2 = d\theta_1^2 + \sin^2\theta_1 d\Omega_2^2~.
%
\ee
The similar construction can be done for $S^n$. So, one can construct the $S^n$ metric iteratively:
\begin{align}
d\Omega_n^2 
&= d\theta_1^2 + \sin^2\theta_1 d\Omega_{n-1}^2 \\
&= d\theta_1^2 
+ \sin^2\theta_1 d\theta_2^2 
+ \cdots 
+ \sin^2\theta_1 \cdots \sin^2\theta_{n-1} d\theta_n^2~,
%
\end{align}
where $0 \leq \theta_i <\pi$ $(1 \leq i \leq n-1)$ and $0 \leq \theta_n <2\pi$.

\subsubsection*{The hyperbolic space $H^{p+2}$}

$H^{p+2}$ has the $SO(1,p+2)$ invariance and is defined by
\begin{gather}
ds_{p+3}^2 = - dX_0^2 + dX_1^2 + \cdots + dX_{p+2}^2~, \\
-X_0^2 + X_1^2 + \cdots + X_{p+2}^2 = - L^2~.
\end{gather}
For $p=0$, we set $X_0=Z$, $X_1=X$, and $X_2=Y$. 
The $H^{p+2}$ symmetry contains the $SO(p+2)$ invariance as a subgroup, so one can utilize the unit $S^{p+1}$ coordinates $\omega_i$ $(i =1, \ldots, p+2)$ to write the $H^{p+2}$ metric:
\begin{gather}
X_0 = L \cosh\rho, \quad X_i = L\sinh\rho\, \omega_i~. 
%
\end{gather}
Then, the metric becomes
\be
\frac{ds^2}{L^2} = d\rho^2 + \sinh^2\rho\, d\Omega_{p+1}^2~.
\ee

\subsubsection*{The AdS$_{p+2}$ spacetime%
}

The AdS$_{p+2}$ spacetime has the $SO(2,p+1)$ invariance and is defined by
\begin{gather}
ds_{p+3}^2 = - dX_0^2 - dX_{p+2}^2 + dX_1^2 + \cdots + dX_{p+1}^2~, \\
-X_0^2 - X_{p+2}^2 + X_1^2+ \cdots + X_{p+1}^2 = - L^2~.
\end{gather}
For $p=0$, we set $X_0=Z$, $X_{p+2}=X$, and $X_{p+1}=Y$. 
Just like AdS$_2$, the AdS$_{p+2}$ spacetime becomes  $H^{p+2}$ by $X_\text{E} = iX_{p+2}$.

The AdS$_{p+2}$ symmetry contains the $SO(p+1)$ invariance as a subgroup, and one can utilize the unit $S^{p}$ coordinates $\omega_i$ $(i =1, \ldots, p+1)$ to write the AdS$_{p+2}$ metric:
\begin{gather}
X_0 = L \cosh\rho \cos\tg~, \quad X_{p+2} = L \cosh\rho \sin\tg~, \quad
X_i = L \sinh\rho\, \omega_i~. 
%
\end{gather}
This is the global coordinates for AdS$_{p+2}$. The metric becomes
\be
\frac{ds^2}{L^2} = - \cosh^2\rho\, d\tg^2 + d\rho^2 + \sinh^2\rho\, d\Omega_{p}^2~.
\ee
One can define the other coordinate systems just like AdS$_2$. In static coordinates where $\rs := \sinh\rho$, the metric becomes
\be
\frac{ds^2}{L^2} = - (\rs^2+1)\, d\tg^2 + \frac{d\rs^2}{\rs^2+1} + \rs^2 d\Omega_{p}^2~.
\ee
In conformal coordinates where $\tan \theta := \sinh\rho$, the metric becomes
\be
\frac{ds^2}{L^2} = \frac{1}{\cos^2\theta} (- d\tg^2 + d\theta^2 + \sin^2\theta\, d\Omega_{p}^2)~.
\ee
For AdS$_{p+2}$, the \poincare\ coordinates are defined by
\begin{align}
X_0 &= \frac{L\rP}{2} \left( x_i^2 - t^2 + \frac{1}{\rP^2} + 1 \right)~, \\
X_{p+2} &= L \,\rP\, t~, \\
X_i &= L\, \rP\, x_i \quad (i = 1,\ldots, p)~, \\
X_{p+1} &= \frac{L\rP}{2} \left( x_i^2 - t^2 + \frac{1}{\rP^2} - 1 \right)~.
%
\end{align}
The metric becomes
\be
\boxeq{
\frac{ds^2}{L^2} = \rP^2 (-dt^2+d\bmx_p^2) + \frac{d\rP^2}{\rP^2}~.
}
\ee
When $p=3$, the metric coincides with the near-horizon limit of the D3-brane \eqref{eq:D3_near_horizon}. 

%
%
%
%
%
%
%
%

\subsubsection*{The dS$_{p+2}$ spacetime
}

The dS$_{p+2}$ spacetime has the $SO(1,p+2)$ invariance and is defined by
\begin{gather}
ds_{p+3}^2 = - dX_0^2 + dX_1^2 + \cdots + dX_{p+2}^2~, \\
- X_0^2 + X_1^2 + \cdots + X_{p+2}^2 = + L^2~.
\end{gather}
For $p=0$, we set $X_0=Z$, $X_1=X$, and $X_2=Y$. 
Just like dS$_2$, the dS$_{p+2}$ spacetime becomes $S^{p+2}$ by $X_\text{E} = iX_0$. 

The dS$_{p+2}$ symmetry contains the $SO(p+2)$ invariance as a subgroup, and one can utilize the unit $S^{p+1}$ coordinates $\omega_i$ $(i =1, \ldots, p+2)$ to write the dS$_{p+2}$ metric:
\begin{gather}
X_0 = L \sinh\tg, \quad X_i = L \cosh\tg\, \omega_i~. 
%
\end{gather}
Then, the metric becomes
\be
\frac{ds^2}{L^2} = - d\tg^2 + \cosh^2\tg\, d\Omega_{p+1}^2~.
\ee

\subsubsection*{Maximally symmetric spacetimes}

We saw that these spacetimes have a large number of symmetries like $S^2$. In fact, they are called \keyword{maximally symmetric spacetimes} which admit the maximum number of symmetry generators. As a familiar example, the Minkowski spacetime is also a spacetime with constant curvature (namely $R=0$) and is a maximally symmetric space. The $(p+2)$-dimensional Minkowski spacetime has the $ISO(1,p+1)$ \poincare\ invariance. The number of symmetry generators is $(p+1)(p+2)/2$ for $SO(1,p+1)$ and $(p+2)$ for translations, so $(p+2)(p+3)/2$ in total. 
%
%
%
%
%
%
This is the maximum number of generators. Figure~\ref{fig:sym} summarizes the symmetries of spaces/spacetimes with constant curvature. They all have the same maximum number of generators.

As a consequence of a maximally symmetric spacetime, it is known that the Riemann tensor is written as 
\be
R_{ABCD} = \pm \frac{1}{L^2} (g_{AC}g_{BD} - g_{AD}g_{BC})
%
\ee
($+$ for positive curvature and $-$ for negative curvature). Then, the Ricci tensor and the Ricci scalar are 
\be
R_{MN} = \pm \frac{p+1}{L^2} g_{MN}~, \quad
R = \pm \frac{(p+1)(p+2)}{L^2}~.
%
\ee
When $p=0$, we recover $R=\pm 2/L^2$. Then, a maximally symmetric spacetime is a solution of the Einstein equation with an appropriately chosen cosmological constant $\Lambda$:
\be
R_{MN} - \frac{1}{2} g_{MN} R = \mp \frac{p(p+1)}{2L^2} g_{MN}~,
\label{eq:Einstein_cc2}
\ee
which fixes $\Lambda=\pm p(p+1)/(2L^2)$.

\begin{figure}[tb]
\begin{center}
\begin{tabular}{|c|c||c|c|}
\hline
Space & Symmetry & Spacetime & Symmetry \\
\hline
$S^{p+2}$	& $SO(p+3)$	& dS$_{p+2}$	& $SO(1,p+2)$ \\
$H^{p+2}$	& $SO(1,p+2)$	& AdS$_{p+2}$	& $SO(2,p+1)$ \\
\hline
\end{tabular}
\caption{Symmetries of spaces/spacetimes with constant curvature.}
\label{fig:sym}
\end{center}
\end{figure}

\section{Particle motion in AdS spacetime \advanced
}\label{sec:geodesics_AdS}

Readers may skip this section in a first reading because the AdS physics is not intuitively very clear. 

\subsubsection*{Gravitational redshift
}

Let us consider the gravitational redshift in static coordinates $(\tg,\rs)$. \index{static coordinates}
%
%
Consider two static observers, the observer A at the origin $\rs=0$ and the observer B at $\rs=\rs_B \gg 1$. The observer A sends a photon. From the redshift formula \eqref{eq:red_shift}, the photon energy received by B is given by
\be
\frac{E_B}{E_A} =\sqrt{\frac{g_{00}(A)}{g_{00}(B)}}
\result
E_B \simeq \left(\frac{1}{\rs_B}\right) E_A~.
\label{eq:red_shift_AdS}
\ee
Thus, the photon energy decreases at B. In particular, when $\rs_B\rightarrow\infty$, $E_B\rightarrow0$, so the photon gets an infinite redshift. The gravitational redshift comes from the gravitational potential, so this implies that $\rs=0$ is the ``bottom" of the gravitational potential well. The fact that the photon gets an infinite redshift is similar to a black hole, but the AdS spacetime is not a black hole.

\subsubsection*{Photon motion
}

Let us examine the photon and particle motions in the AdS spacetime. Such analysis can be done just like the particle motion analysis in the Schwarzschild black hole (\sect{geodesics_horizon}), but the photon and the particle have strange behaviors in the AdS spacetime. Take the static coordinates and consider a particle which starts from the bottom of the gravitational potential $\rs=0$ to $\rs=\infty$. We set $L=1$ below for simplicity.

First, consider the photon motion $ds^2=0$ from the point of view of the coordinate time $\tg$. This is the simplest in conformal coordinates:
\be
ds^2 = \frac{1}{\cos^2\theta} (- d\tg^2 + d\theta^2 ) = 0
\result 
\frac{d\theta}{d\tg} = 1~.
%
\ee
Since $\rs=\tan\theta$, $\rs:0\rightarrow\infty$ corresponds to $\theta:0\rightarrow\pi/2$, and
\be
\tg = \int_0^{\pi/2} d\theta = \frac{\pi}{2}~.
%
\ee
Namely, the photon reaches the AdS boundary in a \textit{finite} coordinate time. We thus need a boundary condition at the AdS boundary: how the photon behaves at the AdS boundary depends on the boundary condition at the AdS boundary (\eg, it reflects back to the origin).

But consider the motion from the point of view of the ``proper time" $\tau$ (one cannot define the proper time for the photon, but one can define an affine parameter, so $\tau$ is an affine parameter\index{affine parameter}). From $p^2=0$, 
\be
g^{00} E^2 + g_{\rs\rs} \left(\frac{d\rs}{d\tau}\right)^2 = 0~.
\ee
We define the energy $E$ for the photon as $p_0=:-E$. Then,
\be
- \frac{E^2}{\rs^2+1} + \frac{1}{\rs^2+1} \left( \frac{d\rs}{d\tau} \right)^2 = 0
\result
\frac{d\rs}{d\tau} = E~.
\label{eq:null_dtau}
\ee
Thus, $\rs=E\tau$, namely it takes an \textit{infinite} affine parameter time until the photon reaches the AdS boundary.

\subsubsection*{Particle motion
}

\begin{figure}[tb]
\centering
\scalebox{0.75}{ \includegraphics{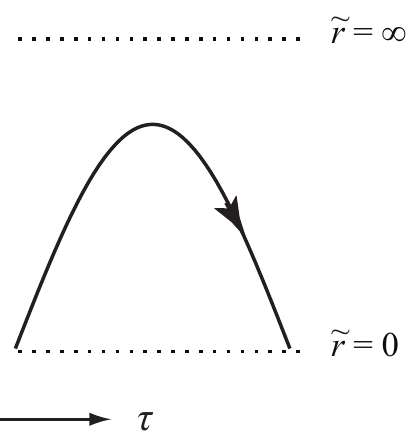} }
\caption{In the AdS spacetime, a particle cannot reach the boundary $\rs\rightarrow\infty$ and comes back to the origin in a finite proper time and coordinate time. }
\label{fig:potential_well} 
\end{figure}%

Now, consider the particle motion from the point of view of the proper time $\tau$. From $p^2=-m^2$, 
\begin{align}
g^{00} E^2 + g_{\rs\rs} \left(\frac{d\rs}{d\tau}\right)^2 = -1 
&\result
- \frac{E^2}{\rs^2+1} + \frac{1}{\rs^2+1} \left( \frac{d\rs}{d\tau} \right)^2 = -1 \\
&\result
\left( \frac{d\rs}{d\tau} \right)^2 = (E^2-1) - \rs^2~.
\label{eq:timelike_dtau}
%
\end{align}
Take $E>1$ so that the right-hand side of \eq{timelike_dtau} is positive at $\rs=0$. But the right-hand side becomes negative as $\rs\rightarrow\infty$, so the particle cannot reach the AdS boundary $\rs\rightarrow\infty$. This is the effect of the gravitational potential well in the AdS spacetime (\fig{potential_well}). The turning point $\rs_*$ is given by $\rs_* = \sqrt{E^2-1}$. Then,
\be
\tau = \int_0^{\rs_*} \frac{ d\rs }{ \sqrt{(E^2\!-\!1) \!-\! \rs^2} }
= \int_0^{\pi/2} d\varphi 
= \frac{\pi}{2}~,
\label{eq:timelike_tau}
\ee
where $\rs =: \sqrt{E^2-1} \sin\varphi$.

How about the motion in the coordinate time $\tg$? By definition $p^0 = md\tg/d\tau$, and from the conservation law, $p^0 = g^{00}p_0 = -mg^{00}E$. Thus,
\be
\frac{d\tg}{d\tau} = \frac{E}{-g_{00}}~.
%
\ee
Using \eq{timelike_tau}, one gets
\be
\tg = E \! \int_0^{\rs_*} \! \frac{ d\rs }{ (\rs^2\!+\!1)\sqrt{(E^2\!-\!1) \!-\! \rs^2} }
= E\! \int_0^{\pi/2} \! \frac{d\varphi }{ (E^2\!-\!1)\sin^2\varphi \!+\! 1 }
\!= \frac{\pi}{2}~.
%
\ee
Namely, the particle returns to $\rs=0$ with the same amount of the coordinate time $\tg=\pi$ as the photon. The amount of time is the same in the particle proper time $\tau$. Also, the time for the particle to return does not depend on energy $E$. The particle goes further as you increase $E$, but it always returns with the same amount of time. These results are summarized in \fig{particle_motion}.

\begin{figure}[tb]
\begin{center}
\begin{tabular}{|c||c|c|}
\hline
					& Photon & Particle \\
\hline
affine time $\tau$		& $\infty\ (\rs: 0\rightarrow\infty)$	& $\pi/2\ (\rs: 0\rightarrow \rs_*)$  \\
coordinate time $\tg$		& $\pi/2\ (\rs: 0\rightarrow\infty)$	& $\pi/2\ (\rs: 0\rightarrow \rs_*)$  \\
\hline
\end{tabular}
\caption{The affine time and the static coordinate time $\tg$ for the particle and the photon to reach the AdS boundary.}
\label{fig:particle_motion}
\end{center}
\end{figure}

\subsubsection*{\poincare\ coordinates and \poincare\ patch \index{Poincar\'{e} coordinates}
}

We examined the motion using the static coordinate time $\tg$, but let us examine the same question using the \poincare\ coordinate time $t$. For simplicity, we consider only the photon motion. Consider the photon motion from $\rP=R$ to $\rP=0$. From  
\be
ds^2 = -\rP^2dt^2 + \frac{d\rP^2}{\rP^2} = 0
\result 
\frac{d\rP}{dt} = -\rP^2~,
%
\ee
one gets
\be
t = -\int_R^{\epsilon} \frac{ d\rP }{ \rP^2 }
= \left.\frac{1}{\rP} \right|_R^\epsilon 
= \frac{1}{\epsilon}-\frac{1}{R}
\rightarrow\infty~.
%
\ee
Thus, it takes an infinite coordinate time $t$ until the photon reaches $\rP=0$.

But this is not the case in the affine parameter $\tau$. Just like \eq{null_dtau}, 
\be
\frac{d\rP}{d\tau} = -E~,
%
\ee
so it takes a finite affine time from $\rP=R$ to $\rP=0$%
\footnote{As in \eq{null_dtau}, it takes an infinite affine time from $\rP=\infty$ to $\rP=0$, but it takes a finite amount of time from finite $R$ to $\rP=0$.}.

The behavior of the \poincare\ coordinates is similar to the Schwarzschild coordinates. Because the photon reaches $\rP=0$ in a finite affine parameter, the \poincare\ coordinates cover only part of the full AdS spacetime%
\footnote{Otherwise, the spacetime would be \keyword{geodesically incomplete} which signals a spacetime singularity. The geodesic incompleteness means that there is at least one geodesic which is inextensible in a finite ``proper time" (in a finite affine parameter). This is the definition of a spacetime singularity.}. 
From the point of view of the full AdS geometry, $t\rightarrow\infty$ as $\rP\rightarrow0$ because of the ill-behaved coordinates just like the Schwarzschild coordinates. The region covered by \poincare\ coordinates is called the \textit{\poincare\ patch} (\fig{AdS2_Penrose}).\index{Poincar\'{e} patch}

In this sense, the location $\rP=0$ is similar to the horizon but has a difference from the horizon. The region inside the horizon cannot influence the region outside. But the AdS spacetime does not have such a region. In conformal coordinates, clearly there is no region which cannot influence the AdS boundary. The \poincare\ patch is rather similar to the \keyword{Rindler spacetime}:
\be
ds_{p+2}^2 = -r\, dt^2 + \frac{dr^2}{r} + d\bmx_p^2~.
\label{eq:rindler}
\ee
The Rindler spacetime is just the Minkowski spacetime although it does not look so (\probset{rindler}). The difference is that the Rindler spacetime covers only part of the Minkowski spacetime. Physically, the Rindler coordinates represent the observer who experiences a constant acceleration and there is a horizon, $r=0$, for the observer. But the horizon is observer-dependent, and the Minkowski spacetime as a whole has no horizon.

Similarly, there is a horizon at $\rP=0$ for the observer inside the \poincare\ patch, but the AdS spacetime as a whole has no horizon. Whether $\rP=0$ is a horizon or not depends on the point of view%
\footnote{For example, in the near-horizon limit, the extreme RN-AdS$_4$ \bh becomes the AdS$_2$ spacetime in \poincare\ coordinates (\sect{others_HSC}). In this case, $\rP=0$ corresponds to the true horizon of the black hole. So, whether $\rP=0$ is a true horizon or not depends on the context.}. 
As a finial remark, the Rindler spacetime has a Hawking temperature,
but the AdS spacetime in \poincare\ coordinates has zero temperature. The Rindler spacetime has the following thermodynamic quantities:
\be
T = \frac{1}{4\pi}~, \quad
s = \frac{1}{4G}~, \quad
\varepsilon = 0~, \quad
P= \frac{1}{16\pi G}~.
%
\ee

\section{Remarks on AdS/CFT interpretations
}\label{sec:interpretation}

\subsubsection*{The symmetries of the AdS spacetime
}

Let us rewrite the AdS$_5$ spacetime in \poincare\ coordinates%
\footnote{So far we took dimensionless coordinates, but we take dimensionful ones below. }:
\be
ds_5^2 = \left( \frac{r}{L} \right)^2 (-dt^2+dx^2+dy^2+dz^2) 
+ L^2\frac{dr^2}{r^2}~.
\label{eq:ads_poincare}
\ee
The spacetime has the $SO(2,4)$ invariance, part of which can be seen easily in \poincare\ coordinates and are physically important:
\begin{enumerate}
\item
\textit{Four-dimensional \poincare\ invariance}: The metric has the \poincare\ invariance $ISO(1,3)$ on $x^\mu=(t,x,y,z)$. This corresponds to the \poincare\ invariance of the dual gauge theory in four-dimensional spacetime, so $x^\mu$ is interpreted as the spacetime coordinates of the gauge theory. Similarly, for the AdS$_{p+2}$ spacetime, $p$ represents the number of the spatial dimensions of the dual gauge theory.
\item
\textit{Four-dimensional scale invariance}: The metric is invariant under
\be
\boxeq{
x^\mu \rightarrow a x^\mu, \quad r \rightarrow \frac{1}{a}\, r~.
}
\label{eq:scale_inv}
\ee
\end{enumerate}

Under the scaling, $r$ transforms as energy which is conjugate to $t$. This is one reason why the gauge theory is four-dimensional whereas the gravitational theory is five-dimensional. Namely, the $r$-coordinate has the interpretation as the gauge theory energy scale. The ${\cal N}=4$ SYM has the scale invariance, but the invariance is realized geometrically in the dual gravitational theory.

Let us examine more about the dual gauge theory energy. In AdS/CFT, the gauge theory time corresponds to the coordinate $t$ not to the proper time $\tau$. The proper time $\tau_r$ for the static observer at $r$ is related to the coordinate $t$ as
\be
d\tau_r^2 = |g_{00}(r)| dt^2~.
\label{eq:tau_vs_t}
\ee
Then, the proper energy for the observer at $r$, $E(r)$, is related to the gauge theory energy, $E_t$,  as%
\footnote{This equation looks similar to the gravitational redshift \eqref{eq:red_shift_AdS}, but the interpretations are different. The gravitational redshift compares proper energies at different radial positions. The UV/IR relation compares the proper energy and the energy conjugate to the coordinate $t$ at one radial position. }
\be
E_t =\sqrt{|g_{00}(r)|} E(r)
\simeq  \left(\frac{r}{L}\right) E(r)~.
\label{eq:UV/IR}
\ee
Consider an excitation in the AdS spacetime with a given proper energy (for example, a string). The gauge theory energy depends on where the excitation is located. The gauge theory energy is larger if the excitation is nearer the AdS boundary. The relation \eqref{eq:UV/IR} is called the \keyword{UV/IR relation}.

In general relativity, one does not put emphasis on a coordinate time since it depends on coordinate systems. Rather one puts emphasis on the proper time. However, in AdS/CFT, the coordinate time plays the special role as the gauge theory time. Thus, we use $t$ for a \textit{gauge theory interpretation.}

\subsubsection*{The Hawking temperature and the proper temperature \advanced
}

Similar remarks also apply to thermodynamic quantities of the AdS black holes in \chap{AdS_BH}. The Hawking temperature is determined from the periodicity of the imaginary time $t_\text{E}$. This means that the Hawking temperature is the temperature measured by the coordinate $t$. 

The Hawking temperature $T$ differs from the temperature as seen by the observer at $r$. The proper time $\tau_r$ and $t$ are related to each other by \eq{tau_vs_t}, so the \keyword{proper temperature} $T(r)$ is given by
\be
T(r)=\frac{1}{\sqrt{|g_{00}(r)|}} T~.
\label{eq:proper_temp}
\ee
For asymptotically flat black holes, the proper temperature for the asymptotic observer coincides with the Hawking temperature since $|g_{00}|\rightarrow 1$ as $r\rightarrow\infty$. But for AdS black holes, $|g_{00}|\propto r^2$ as $r\rightarrow\infty$, so the proper temperature for the asymptotic observer vanishes%
\footnote{This should coincide with the surface gravity (for the asymptotic observer). When we discussed the surface gravity, we assumed that the spacetime is asymptotically flat [in the last expression of \eq{red_shift_surface}, we used $f(\infty)=1$]. For AdS black holes, the surface gravity vanishes by taking $f(\infty)\rightarrow\infty$ into account. The expression $f'(r_0)/2$ corresponds to the ``surface gravity" in the $t$-coordinate.}.

In AdS/CFT, the coordinate $t$ has the special interpretation as the gauge theory time. Therefore, when we discuss temperatures, we will use the Hawking temperature (gauge theory temperature) not the proper temperature.

\section{\titlesummary}

\begin{itemize}
\item 
The AdS spacetime is the spacetime of constant negative curvature.
\item 
The AdS$_5$ spacetime has the $(3\!+\!1)$-dimensional timelike boundary known as the AdS boundary.
\item 
The AdS$_5$ spacetime has the $SO(2,4)$ invariance. This symmetry is the same as the $(3\!+\!1)$-dimensional conformal invariance of the \Nfour\ SYM.
\item
The AdS$_5$ spacetime in \poincare\ coordinates coincides with the near-horizon limit of the D3-brane. 
\item
In AdS/CFT, the gauge theory time is the coordinate time not the proper time. Accordingly, the AdS radial coordinate has the interpretation as the gauge theory energy scale. Given a proper energy of an excitation, the gauge theory energy is larger if the excitation is nearer the AdS boundary. 
\end{itemize}

\titlenewterms

\begin{multicols}{2}
\noindent
hyperbolic space\\
global coordinates\\
static coordinates\\
conformal coordinates\\
Poincar\'{e} coordinates\\
Poincar\'{e} patch\\
AdS boundary\\
de~Sitter spacetime\\
maximally symmetric spacetimes\\
UV/IR relation\\
proper temperature
\end{multicols}

\endofsection
\ifx\nameofpaper\undefined 
  \usepackage{macro_natsuume} 
  \def\beginsection{\section*}
  \def\endofsection{\end{document}} 
  \input draft_header.tex
\else 
  \def\beginsection{\chapter}
  \def\endofsection{ } 
\fi

\beginsection{AdS/CFT - equilibrium
}\label{chap:AdS_BH}


\begin{quote}
In this chapter, we compute thermodynamic quantities for the Schwarzschild-AdS black hole, from which one can get thermodynamic quantities of the strongly-coupled ${\cal N}=4$ super-Yang-Mills theory. We also compare the results with the free gas result.
\end{quote}

\section{The AdS black hole
}\label{sec:SAdS}

Black holes can exist in the AdS spacetime. The simplest AdS black hole is known as the \textit{Schwarzschild-AdS black hole}\index{Schwarzschild-AdS black hole (planar horizon)} (SAdS black hole hereafter). Just like the Schwarzschild black hole, one can consider AdS black holes with spherical horizon (\sect{spherical_SAdS}), but we consider AdS black holes with planar horizon or AdS black branes for the time being.

The SAdS$_5$ black hole is a solution of the Einstein equation with a negative cosmological constant \eqref{eq:Einstein_cc2} like the AdS$_5$ spacetime. The metric is given by
\begin{align}
ds_5^2& = - \left(\frac{r}{L}\right)^2 h(r) dt^2 
+ \frac{dr^2}{ \left(\frac{r}{L}\right)^2 h(r) } 
+ \left(\frac{r}{L}\right)^2 (dx^2 \!+\! dy^2 \!+\! dz^2)~,
\label{eq:sads5} \\
h(r) &= 1 - \left( \frac{r_0}{r} \right)^4~.
\end{align}
The horizon is located at $r=r_0$. When $r_0=0$, the metric reduces to the AdS$_5$ spacetime in \poincare\index{Poincar\'{e} coordinates} coordinates \eqref{eq:ads_poincare}. The $g_{00}$ component contains the factor $r_0^4/(L^2 r^2)$. The $O(r^{-2})$
 behavior comes from the Newtonian potential which behaves as $r^{-2}$ in the five-dimensional spacetime. 

The coordinates $(x,y,z)$ represent $\mathbb{R}^3$ coordinates. In the Schwarzschild black hole, this part was $r^2 d\Omega^2$ which represents a spherical horizon, but here the $r=r_0$ horizon extends indefinitely in $(x,y,z)$-directions. 

The AdS spacetime is a spacetime with constant curvature, but the SAdS$_5$ black hole is not. For example, there is a curvature singularity at $r=0$.

The AdS spacetime is invariant under the scaling $x^\mu \rightarrow a x^\mu$, $r \rightarrow r/a$. Under the scaling, the SAdS \bh metric becomes
\be
ds_5^2 \rightarrow - \left(\frac{r}{L}\right)^2 \left\{ 1 - \left( \frac{a r_0}{r} \right)^4 \right\} dt^2 
+ \cdots~,
%
\ee
so one can always scale the horizon radius. Namely, black holes with different horizon radii are all equivalent. As we will see, the temperature of the black hole is $T \propto r_0$, so one can change the temperature by the scaling. This means that all temperatures are equivalent and the physics is the same (except at zero temperature). Thus, there is no characteristic temperature such as a phase transition.
From the gauge theory point of view, this is because the \Nfour\ SYM is scale invariant and there is no dimensionful quantity except the temperature. 

Another way of saying this is that the metric is invariant under
\be
x^\mu \rightarrow a x^\mu, \quad r \rightarrow \frac{1}{a}\, r,  \quad r_0 \rightarrow \frac{1}{a}\, r_0~.
\label{eq:scale_inv2}
\ee
Using this scaling, one can obtain the functional form of the \bh temperature. The temperature $T$ scales as the inverse time, so $T \rightarrow T/a$ under the scaling. Comparing the scaling of $r_0$, one gets $T \propto r_0$. The dimensional analysis then fixes $T \propto r_0/L^2$. See \eq{temp_SAdS5} for the numerical coefficient.


\section{Thermodynamic quantities of AdS black hole
}\label{sec:sads_thermo}

\subsection{Thermodynamic quantities
}

Here, we compute thermodynamic quantities of the SAdS$_5$ black hole. In AdS/CFT, they are interpreted as thermodynamic quantities of the dual \Nfour\ SYM at strong coupling. In order to rewrite black hole results as gauge theory results, one needs the relation of the parameters between two theories. This is given by the AdS/CFT dictionary\index{AdS/CFT dictionary} in \sect{towards_ads}:
\be
\boxeq{
N_c^2 = \frac{\pi}{2} \frac{L^3}{G_5}~, \quad
\lambda = \left(\frac{L}{l_s}\right)^4~.
}
\label{eq:dictionary_again}
\ee
On the left-hand side, we have gauge theory parameters which are written in terms of gravity parameters on the right-hand side.

First, the temperature is given by
\begin{align}
T &= \frac{f'(r_0)}{4\pi} \\
&= \frac{1}{4\pi}\frac{1}{L^2} \left. \left( 2r + \frac{2r_0^4}{r^3} \right) \right|_{r=r_0} \\
&= \frac{1}{\pi}\frac{r_0}{L^2}~,
\label{eq:temp_SAdS5}
\end{align}
where we used \eq{Hawking_temp}.


For this black hole, the horizon has an infinite extension, and the entropy itself diverges, so it is more appropriate to use the entropy density $s$. Let the spatial extension of the \bh as $0 \leq x,y,z \leq L_x, L_y, L_z$. (This is just an infrared cutoff to avoid divergent expressions.) The gauge theory coordinates are $(x,y,z)$, so the gauge theory volume is $V_3:=L_x L_y L_z$. This is different from the horizon ``area" since the 
line element is $(r/L)^2(dx^2+dy^2+dz^2)$. Then, from the area law \eqref{eq:BH_entropy},
\be
S = \frac{A}{4G_5} = \frac{1}{4 G_5}\left(\frac{r_0}{L}\right)^3 V_3
%
\ee
or
\be
s = \frac{S}{V_3} = \frac{1}{4 G_5}\left(\frac{r_0}{L}\right)^3~.
\label{eq:entropy_SAdS5}
\ee
One would write the area law as
\be
s = \frac{a}{4G_5}~,
\label{eq:BH_entropy_density}
\ee
where $a:=A/V_3$ is the ``horizon area density."
%
%
%
%
Using the temperature \eqref{eq:temp_SAdS5} and the AdS/CFT dictionary, one gets
\be
\boxeq{
s = \frac{\pi^2}{2} N_c^2 T^3~.
}
\label{eq:entropy_N=4}
\ee
The rest of thermodynamic quantities can be determined using thermodynamic relations. The first law $d\varepsilon=Tds$ can determine the energy density $\varepsilon$%
\footnote{An integration constant is discarded. It is not allowed since we have only $T$ as the dimensionful quantity. This can be checked explicitly from the \bh partition function, but there are examples with nonvanishing integration constants when there are other dimensionful quantities (see, \eg, \sect{HP}). }:
\be
\varepsilon =\frac{3}{8}\pi^2N_c^2T^4~.
\label{eq:energy_N=4}
\ee
The Euler relation $\varepsilon=Ts-P$ then determines the pressure $P$:
\be
P =\frac{1}{8}\pi^2N_c^2T^4
\result
\boxeq{
P=\frac{1}{3}\varepsilon~.
}
\label{eq:pressure_N=4}
\ee
Here, we obtained thermodynamic quantities by the area law and thermodynamic relations, but one can obtain these quantities by evaluating the black hole free energy $F(T,V_3)$.
In thermodynamics, the free energy is defined by $F(T,V_3):=E-TS = -P(T)V_3$, so the result should be
\be
F = - \frac{V_3}{16\pi G_5} \frac{r_0^4}{L^5} 
= - \frac{V_3 L^3}{16 \pi G_5} \pi^4 T^4
= - \frac{1}{8}\pi^2N_c^2T^4 V_3
%
\ee
from \eq{pressure_N=4}, which is indeed confirmed in \sect{SAdS_free}.

This is a simple exercise, but one can learn many from the results as described below.

\subsubsection*{The temperature dependence
}

The temperature dependence $\varepsilon \propto T^4$ represents the \keyword{Stefan-Boltzmann law}. The \Nfour\ SYM is scale invariant, and there is no dimensionful quantity except the temperature. Then, it is obvious that we have the Stefan-Boltzmann law from the field theory point of view. What is nontrivial is the proportionality constant. We will compare the coefficient with the free gas case. 

However, from the black hole point of view, it is highly nontrivial that a black hole obeys the Stefan-Boltzmann law. For example, for the five-dimensional Schwarz\-schild black hole, $M \propto 1/T^2$, so the black hole takes a completely different form from the Stefan-Boltzmann form.

Moreover, the Schwarzschild black hole has a negative heat capacity $C:=dM/dT<0$, and there is no stable equilibrium. 
Namely, the black hole emits the Hawking radiation, and the black hole loses its mass as the result of the radiation. This raises the temperature of the black hole, which makes the Hawking radiation more active. In this way, the temperature becomes high indefinitely, and the black hole does not reach an equilibrium.

The absence of the stable equilibrium is true not only for the Schwarzschild black hole but also for a gravitational system in general. 
In a gravitational system, once the system deviates from the equilibrium, the fluctuations do not decay but grow in time, and the system deviates further from the equilibrium%
\footnote{This is because the gravitational force is a long-range force. The Schwarzschild black hole can be stabilized if one confines it in a small enough ``box" \cite{York:1986it}. In the AdS black hole, there is a potential barrier (\sect{geodesics_AdS}) which gives a natural notion of the box.}. 
Of course, this is one reason why our universe is not uniform 
and is the driving force to produce astronomical objects such as the solar system. Thus, usual black holes cannot correspond to the standard statistical system such as the gauge theory. The AdS black hole is very special in this sense.

Let us compare the Schwarzschild black hole and the AdS black hole more carefully:
\begin{alignat}{2}
&\text{(5-dimensional) Schwarzschild:} & M &\simeq \frac{1}{G_5 T^2}~, \\
&\text{(5-dimensional) Schwarzschild-AdS:}\quad & \varepsilon &= \frac{(\pi L)^3}{4G_5} T^4~.
\end{alignat}
For the Schwarzschild black hole, there is no dimensionful quantities other than $G_5$ and $T$. On the other hand, for the AdS black hole, there is another dimensionful quantity, the AdS radius $L$. As one can see from the AdS/CFT dictionary \eqref{eq:dictionary_again}, $L^3$ and $G_5$ combine to give only a dimensionless quantity $\Nc$. In this way, the Stefan-Boltzmann law can appear for the AdS black hole. On the other hand, for the Schwarzschild black hole, one cannot eliminate the Newton's constant $G_5$, so the Stefan-Boltzmann law cannot appear.

Also, note that the \bh entropy can be interpreted as a four-dimensional statistical entropy because we consider a \bh in the five-dimensional spacetime. Unlike the usual statistical systems, the \bh entropy is not proportional to the volume of the system but is proportional to the area of the horizon. That is why the five-dimensional \bh entropy can be naturally interpreted as a $(3+1)$-dimensional quantity. 

\subsubsection*{The $N_c^2$ dependence}

The entropy density is proportional to $O(N_c^2)$. This implies that the dual gauge theory is in the unconfined plasma phase. The entropy counts the degrees of freedom of a system. An $SU(N_c)$ gauge theory has $O(N_c^2)$ color degrees of freedom. In the unconfined phase, these contribute to the entropy. In the confined phase, only the $SU(N_c)$ gauge singlets contribute to the entropy, and the entropy is not proportional to $O(N_c^2)$. The \Nfour\ SYM is scale invariant, and only the plasma phase exists.

\subsubsection*{Traceless energy-momentum tensor
}

The energy density and pressure satisfies $\varepsilon=3P$, so the energy-momentum tensor is traceless $T^\mu_{~\mu}=0$. This also comes from the scale invariance of the \Nfour\ SYM. The \Nfour\ SYM has only the temperature as a dimensionful quantity. When thermodynamic quantities are the only dimensionful quantities, the dimensional analysis (Stefan-Boltzmann law) and thermodynamic relations immediately imply the traceless%
\footnote{Form the field theory point of view, the tracelessness $T^\mu_{~\mu}=0$ is an operator statement and is valid irrespective of states (either the vacuum state or a thermal state) as long as there is no trace anomaly. }.
%
%
%
%

\subsubsection*{Comparison with free gas result
}

Let us compute the entropy density for the \Nfour\ SYM in the free gas limit in order to compare with the AdS/CFT result. First, for the photon gas, the entropy density is given by 
\be
s_\text{photon} 
= \frac{2\pi^2}{45} T^3 \times 2
%
\ee
(\sect{weak_coupling}). The last factor 2 comes from photon's polarizations. The entropy density for the \Nfour\ SYM is given just by replacing the factor 2 with the \Nfour\ degrees of freedom $N_\text{dof}$. Let us count them. The theory has the gauge field and 6 adjoint scalars, so the bosonic degrees of freedom are $ N_\text{boson} = (2+6) \times (N_c^2-1) $. The theory has the adjoint fermions as well, but the fermionic degrees of freedom are the same as the bosonic ones from supersymmetry: $ N_\text{fermion} = N_\text{boson} $. Also, the fermions contribute 7/8 of the bosons to the entropy (\sect{weak_coupling}). So, $ N_\text{dof} = N_\text{boson} + (7/8) N_\text{fermion} = 15 (N_c^2-1) $. Thus, the entropy density in the free gas limit is given by
\be
s_\text{free} 
= \frac{2\pi^2}{45} N_\text{dof} T^3 
= \frac{2\pi^2}{3} (N_c^2-1) T^3
\simeq \frac{2\pi^2}{3} N_c^2 T^3~.
\label{eq:pYM}
\ee
In the last equality, we took the large-$N_c$ limit. 

Equations~\eqref{eq:entropy_N=4} and \eqref{eq:pYM} have the same functional form, but the coefficients differ \cite{Gubser:1996de}. They are related to each other by 
\be
\boxeq{
s_\text{BH} = \frac{3}{4} s_\text{free}~.
}
\label{eq:comparison_entropy}
%
\ee
Why the 3/4 ``discrepancy"? The black hole computation or the AdS/CFT computation correspond to the strong coupling result (large-$N_c$ limit) not to the free gas result. Namely, AdS/CFT predicts that the entropy of the \Nfour\ SYM at strong coupling becomes 3/4 of the free gas result. It is difficult to check the prediction since technical tools to compute strongly-coupled gauge theories are rather limited, but a similar behavior is obtained in lattice simulations (\chap{QGP}).

\subsection{Free gas computation
}\label{sec:weak_coupling}

Consider the partition function of free particles. Denote the energy level as $\omega_i$, and denote the particle number occupying $\omega_i$ as $n_i$. The total energy of a microscopic state is specified by $n_1,n_2,\dots$ and is given by $E(n_1,n_2,\dots) = \sum_i n_i \omega_i$. The partition function then becomes
\be
Z = \sum e^{-\beta E(n_1,n_2,\dots)}
= \sum_{n_1}\sum_{n_2}\cdots e^{ -\beta (n_1 \omega_1 + n_2 \omega_2 + \cdots)}~,
%
\ee
where $\beta:=1/T$. For bosons, $n_i$ takes the value from 0 to $\infty$, so
\be
Z_\text{B} 
= \left( \sum_{n_1=0}^\infty e^{-\beta n_1 \omega_1} \right) \times \cdots
= \prod_{i=1}^\infty \frac{1}{ 1-e^{\beta\omega_i} }~.
%
\ee
Then,
\begin{align}
\ln Z_\text{B} 
&= - \sum_{i=1}^\infty \ln (1-e^{-\beta \omega_i}) \\
&\rightarrow -V \int  \frac{d^3q}{(2\pi)^3} \ln(1-e^{-\beta \omega}) 
\label{eq:line2} \\
&= -\frac{V}{(2\pi)^3} 4\pi \int_0^\infty dq\, q^2 \ln(1-e^{-\beta q}) 
\label{eq:line3} \\
&= -\frac{V}{2\pi^2} \frac{1}{\beta^3} \int_0^\infty dx\, x^2 \ln(1-e^{-x})~, \quad(x:=\beta q)
\label{eq:line4} \\
&= V \frac{\pi^2}{90\beta^3}~.
%
\end{align}
Here, we replaced the sum by the integral in \eq{line2} by taking the $V\rightarrow\infty$ limit. We also used the dispersion relation $\omega=|q|$ for massless particles in \eq{line3}. The integral in \eq{line4} (as well as the corresponding integral for fermions below) is evaluated as
\be
\int_0^\infty dx\, x^2 \ln(1-e^{-x}) = - \frac{\pi^4}{45}~, \quad
\int_0^\infty dx\, x^2 \ln(1+e^{-x}) = \frac{7\pi^4}{360}~.
%
\ee
Therefore, thermodynamic quantities are given by
\begin{align}
\varepsilon_\text{B} &= -\frac{1}{V} \frac{ \del \ln Z_\text{B} }{\del\beta}
=\frac{\pi^2}{30\beta^4}~, \\
s_\text{B} &= \frac{1}{V} \ln Z_\text{B} + \beta \varepsilon_\text{B} 
=\frac{2\pi^2}{45\beta^3}~.
%
\end{align}

Similarly, for fermions, $n_i$ takes the value either 0 or 1, so
\begin{align}
Z_\text{F} &= (1+e^{\beta\omega_1}) \times \cdots 
= \prod_{i=1}^\infty (1+e^{\beta\omega_i})~, \\
\ln Z_\text{F} 
&= \sum_{i=1}^\infty \ln (1+e^{-\beta \omega_i}) \\
&\rightarrow V \int  \frac{d^3q}{(2\pi)^3} \ln(1+e^{-\beta\omega}) \\
&= \frac{V}{(2\pi)^3} 4\pi \int_0^\infty dq\, q^2 \ln(1+e^{-\beta q}) \\
&= V \frac{7\pi^2}{720\beta^3}~.
%
\end{align}
Thus,
\be
\ln Z_\text{F} = \frac{7}{8} \ln Z_\text{B}~. 
%
\ee

\section{The AdS \bh with spherical horizon
}\label{sec:spherical_SAdS}

The AdS \bh with spherical horizon \index{Schwarzschild-AdS black hole (spherical horizon)} is given by 
\be
ds_5^2 =  - \left( \frac{r^2}{L^2} + 1 - \frac{r_0^4}{L^2r^2} \right) dt^2 
+ \frac{dr^2}{ \frac{r^2}{L^2} + 1 - \frac{r_0^4}{L^2r^2} } 
+ r^2 d\Omega_3^2~.
\label{eq:spherical_sads5}
%
\ee
When $r_0=0$, the metric reduces to the AdS$_5$ metric in static coordinates. The horizon is located at $r=r_+$, where
\be
\frac{r_+^2}{L^2} + 1 - \frac{r_0^4}{L^2r_+^2} = 0 
\result
r_0^4 = r_+^4 + L^2r_+^2~.
\label{eq:horizon_spherical}
\ee

Unlike the planar horizon case, black holes with different horizon radii are inequivalent. In particular, the large \bh limit corresponds to the planar horizon case \eqref{eq:sads5}. To see this, consider the scale transformation $t \rightarrow a t$, $r \rightarrow r/a$ and take the $a\rightarrow0$ limit. We want to consider the large \bh limit, but we keep the coordinate value $r_+$ of the horizon fixed under the scale transformation. To do so, transform $r_+ \rightarrow r_+/a$ or $r_0 \rightarrow r_0/a$ [we consider the large \bh limit, so take only the $O(r_+^4)$ term in \eq{horizon_spherical} into account]. Then,
\be
ds_5^2 \rightarrow - \left( \frac{r^2}{L^2} \!+\! a^2 \!-\! \frac{r_0^4}{L^2r^2} \right) dt^2 
+ \frac{dr^2}{ \frac{r^2}{L^2} \!+\! a^2 \!-\! \frac{r_0^4}{L^2r^2} } 
+ \left( \frac{r}{a} \right)^2 d\Omega_3^2~.
%
\ee
The metric does not reduce to \eq{spherical_sads5} under the scaling. The radius of the $S^3$ horizon grows as $r/a$. In the $a\rightarrow0$ limit, the horizon becomes flat, and one can approximate $(r/a)^2 d\Omega_3^2 \simeq (r/L)^2 d\bmx_3^2$ using $\mathbb{R}^3$ coordinates. The resulting geometry is the planar horizon one.

Black holes with different horizon radii are inequivalent, so not all temperatures are equivalent. In fact, we will see in \sect{HP} that a phase transition can occur for the black hole.

\head{Remark}
Readers do not have to worry much at this point, but we mainly focus on the planar horizon in this book from the following reasons.
\begin{itemize}

\item As we saw in \sect{D3_brane}, the near-horizon limit of the D3-brane gives the \bh with planar horizon.

\item The coordinates $(x,y,z)$ correspond to the gauge theory spatial coordinates, so a \bh with planar horizon corresponds to a standard gauge theory on $\mathbb{R}^3$. A \bh with compact horizon corresponds to a gauge theory on compact space. It is interesting in its own right because it has rich physics such as phase transition. To put differently, such a \bh is more complicated, so we postpone the discussion. 

\item In later chapters, we add perturbations to black holes and take the low-energy $\omega\rightarrow0$, long-wavelength limit $q\rightarrow0$ for hydrodynamic analysis of large-$N_c$ plasmas. But for a black hole with compact horizon, the wavelength of the perturbation necessarily becomes the \bh scale. 

To be more precise, let us consider the perturbation of a field $\phi$ on \bh backgrounds. For the planar horizon, we can decompose the perturbation as the plane wave, $\phi \propto e^{-i\omega t+iqz}$. In this case, one can take the $\omega\rightarrow0, q\rightarrow0$ limits. For the spherical horizon, we decompose the perturbation using spherical harmonics. For $S^2$ horizon,  $\phi \propto Y_{lm}(\theta,\varphi)$. As a result, the spectrum is discrete so that we cannot take the limit.

\end{itemize}

\section{\titlesummary}

\begin{itemize}
\item
In the AdS$_5$ spacetime, a \bh with planar horizon exists. We consider black holes with planar horizon from various reasons. 
\item
The SAdS$_5$ \bh (with planar horizon) obeys the $(3\!+\!1)$-dimensional Stefan-Boltzmann law like a standard statistical system.
\item 
The entropy of the $(4\!+\!1)$-dimensional \bh is interpreted as a $(3\!+\!1)$-dimensional quantity because of the area law of the \bh entropy.
\item 
The entropy of the SAdS$_5$ \bh is 3/4 of the entropy of the \Nfour\ SYM in the free gas limit. This implies that the entropy of the \Nfour\ SYM at strong coupling becomes 3/4 of the free gas entropy.
\end{itemize}

\titlenewterms

\begin{multicols}{2}
\noindent
Schwarzschild-AdS black hole \\
Stefan-Boltzmann law\\
\newtermapp{Gibbons-Hawking action}\\
\newtermapp{extrinsic/intrinsic curvature}\\
\newtermapp{counterterm action}\\
\newtermapp{holographic renormalization}
\end{multicols}

\section{Appendix: AdS \bh partition function
\advanced}\label{sec:SAdS_free}

In this section, we evaluate the semiclassical partition function for the SAdS$_5$ black hole. The partition function gives thermodynamic quantities in \sect{sads_thermo}%
\footnote{\advanced
There are actually several ways to compute thermodynamic quantities. For example, one can use the Brown-York tensor \cite{Balasubramanian:1999re}. Or one can read them directly from the ``fast falloff" of the metric (\chap{GKPW}). These methods are simpler but are essentially equivalent to the method here. In this book, we always start from the GKP-Witten relation if possible. You would learn these alternative methods once you get accustomed to AdS/CFT computations. 
}.

First of all, the gravitational action is schematically written as
\be
\action_\text{E} = \action_\text{bulk}+ \action_\text{GH} + \action_\text{CT}~.
%
\ee
Here, $\action_\text{bulk}$, $\action_\text{GH}$, and $\action_\text{CT}$ are called the bulk action, the Gibbons-Hawking action, and the counterterm action, respectively. We explain these actions below and evaluate on-shell actions for the SAdS$_5$ black hole. We use the following coordinates:
\begin{align}
ds_5^2 &= \left(\frac{r}{L}\right)^2 (hd\tE^2+d\bmx_3^2)+L^2\frac{dr^2}{h r^2}~, 
& h &= 1- \left(\frac{r_0}{r}\right)^4~, \\
&= \left(\frac{r_0}{L}\right)^2\frac{1}{u^2} (hd\tE^2+d\bmx_3^2)+L^2\frac{du^2}{h u^2}~, 
& h &= 1- u^4~, 
\label{eq:sads5_u}
\end{align}
($u:=r_0/r$). Note that we consider Euclidean actions and the Euclidean metric here%
\footnote{
The Lorentzian action $\action_\text{L}$ and the Euclidean action $\action_\text{E}$ are related by $i \action_\text{L} = i\int dt L = -\int d\tE (-L) = - \action_\text{E}$ from $\tE = it$．
}. 
In the coordinate $r$, the \bh ``horizon" is located at $r=r_0$ and the AdS boundary is located at $r=\infty$. In the coordinate $u$, the \bh ``horizon" is located at $u=1$ and the AdS boundary is located at $u=0$. 

\subsubsection*{The bulk action
}\label{sec:bulk_free}

The bulk action is the five-dimensional action. For the SAdS$_{p+2}$ black hole, it is the standard Einstein-Hilbert action:
\begin{align}
\action_\text{bulk} &= - \frac{1}{16\pi G_{p+2}} \int d^{p+2}x\, \sqrt{g} \left( R -2 \Lambda \right)~, \\
2\Lambda &= - \frac{p(p+1)}{L^2}~.
%
\end{align}
The SAdS$_5$ \bh corresponds to $p=3$. The contraction of the Einstein equation
\be
R_{MN} - \frac{1}{2} g_{MN} R + \Lambda g_{MN} = 0
%
\ee
gives
\be
R = - \frac{(p+1)(p+2)}{L^2}~.
%
\ee
Thus, the on-shell bulk action%
\footnote{We write on-shell actions $\Sos$ as $\action$ below for simplicity.}
is  
\be
\action_\text{bulk} = \frac{p+1}{8\pi G_{p+2} L^2} \int d^{p+2}x \sqrt{g}~,
\label{eq:bulk_free}
\ee
\ie, it is proportional to the spacetime volume. For the SAdS$_5$ black hole,
\begin{align}
\action_\text{bulk} 
&= \frac{1}{2\pi G_{5}}\frac{r_0^4}{L^5} \int_0^\beta dt \int d\bmx \int_u^1 du \frac{1}{u^5}
\\
&\xrightarrow{u\rightarrow0}{}
\left. \frac{\beta V_3}{8\pi G_5} \frac{r_0^4}{L^5} \left(\frac{1}{u^4}-1 \right) \right|_{u=0} 
\label{eq:bulk_free_bdy} \\
&=:  \frac{\beta V_3}{16\pi G_5} \frac{r_0^4}{L^5} \hatS_\text{bulk}~,
%
\end{align}
where $V_3$ is the volume of $\bmx$-directions.
One can check that the factor in front of $\hatS_\text{bulk}$ is common to the other on-shell actions, so we will set $\beta=V_3=r_0=L=16\pi G_5=1$ below. 

The bulk action diverges as $u\rightarrow 0$ because it is proportional to the spacetime volume. 

\subsubsection*{The Gibbons-Hawking action
}\label{sec:GH_free}

It is necessary to add the following ``surface term" to the Einstein-Hilbert action in order to have a well-defined variational problem (\sect{GH_tensor}):
\be
\action_\text{GH} = - \frac{2}{16\pi G_{p+2}} \int d^{p+1}x\, \sqrt{\gamma}\, K~.
%
\ee
The action is known as the \keyword{Gibbons-Hawking action}. This is a surface term evaluated at $u=0$. Here, $\gamma_{\mu\nu}$ is the $(p+1)$-dimensional metric at the surface, and $K$ is the trace of the \keyword{extrinsic curvature} of the surface as described below. 

\danger{This is a surface term, so it does not affect the equation of motion (Einstein equation), but it does affect the value of the on-shell action. Elementary general relativity textbooks rarely put emphasis on the Gibbons-Hawking action since the equation of motion is fundamentally important for the dynamics of curved spacetimes. But for thermodynamic properties, one is interested in the \textit{whole} value of the on-shell action in the Euclidean formalism, so this action is equally important. \\
\indent
Moreover, in some cases, it is the only contribution to the on-shell action. For the Schwarzschild black hole, $R=0$, so the bulk action makes no contribution to the on-shell action. The contribution to the on-shell action entirely comes from the Gibbons-Hawking action.}
For simplicity, consider a diagonal metric such as the SAdS black hole. The $(p+1)$-dimensional metric $\gamma_{\mu\nu}$ is given by decomposing the metric as (``The ADM decomposition")
\be
ds_{p+2}^2 = g_{uu} du^2 + \gamma_{\mu\nu} dx^\mu dx^\nu~.
%
\ee
The unit normal to the $u=\text{(constant)}$ surface is given by%
\footnote{The vector $n^M$ is ``outward-pointing" pointing in the direction of decreasing $u$ or increasing $r$. }
\be
g_{MN}n^M n^N = 1 
\result
n^u = - \frac{1}{ \sqrt{g_{uu}} }~.
%
\ee
Then, $K$ is given by%
\footnote{For a diagonal metric, the extrinsic curvature itself is given by
\be
K_{\mu\nu} = \frac{1}{2} n^u \del_u \gamma_{\mu\nu}~,
%
\ee
and the trace is defined by $K:=\gamma^{\mu\nu}K_{\mu\nu}$. Use the matrix formula \eqref{eq:matrix_formula} $\del_\mu(\det M) = \det M\, \text{tr}(M^{-1}\del_\mu M)$ to derive \eq{trace_K} from $K_{\mu\nu}$.}
\be
K = n^u \frac{ \partial_u\sqrt{\gamma} }{ \sqrt{\gamma} }~.
\label{eq:trace_K}
\ee


%
%


There are two kinds of curvature: intrinsic and extrinsic. The Riemann tensor is the former. The extrinsic curvature relies on the notion of a higher-dimensional spacetime. Namely, the extrinsic curvature of a spacetime relies on how the spacetime is embedded into a higher-dimensional spacetime. The Riemann tensor does not have to assume the existence of such a higher-dimensional spacetime. 

\begin{figure}[tb]
\centering
\scalebox{0.75}{ \includegraphics{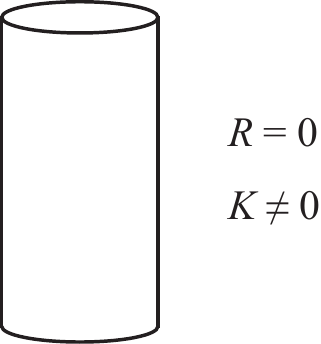} }
\caption{For the cylinder, the intrinsic curvature vanishes ($R=0$), but the extrinsic curvature can be nonvanishing ($K\neq0$).}
\label{fig:cylinder} 
\end{figure}%

For example, consider a cylinder (\fig{cylinder}). The cylinder has no intrinsic curvature $R=0$ since the cylinder is just a flat space with a periodic boundary condition. But the cylinder is ``curved." This is the curvature which arises by embedding it into the three-dimensional space. This is the extrinsic curvature. In cylindrical coordinates
\be
ds^2 = dz^2+ dr^2+ r^2 d\theta^2~,
%
\ee
the cylinder is the $r=\text{(constant)}$ surface. Then, $n^r=1$, $\gamma_{\mu\nu}=\text{diag}(1,r^2)$, and $\sqrt{\gamma}=r$, so the extrinsic curvature is indeed nonvanishing $K=1/r$.

In general relativity, one considers the extrinsic curvature of a surface when one embeds the surface into a spacetime or when one discusses the canonical formalism (one slices the spacetime into a series of spacelike hypersurfaces). In our case, we are talking of the extrinsic curvature of the four-dimensional timelike hypersurface sliced at $u=\text{(constant)}$ of the five-dimensional AdS black hole. This hypersurface is the spacetime where the boundary gauge theory lives.

For the SAdS$_5$ black hole, $n^u=-uh^{1/2}$ and $\sqrt{\gamma} = u^{-4}h^{1/2}$, so
\be
\action_\text{GH} =
 \left. 2uh^{1/2}\left[ u^{-4}h^{1/2} \right]'  \right|_{u=0} 
\xrightarrow{u\rightarrow0}{}
 \left. -\frac{8}{u^4} + 4 \right|_{u=0}~.
\label{eq:GH_free_bdy}
\ee
Note that the Gibbons-Hawking action also diverges as $u\rightarrow0$.

\subsubsection*{The counterterm action
}\label{sec:counter_free}

As we saw above, the bulk action and the Gibbons-Hawking action diverge as $u\rightarrow0$. To cancel the divergence, we add to the action another surface term, the \keyword{counterterm action}. 

In AdS/CFT, this divergence is interpreted as the ultraviolet divergence of the dual field theory. In the dual field theory, it is clear how to handle the divergence. One carries out the renormalization and adds a finite number of local counterterms to the bare action. Following this field theory prescription, we add a counterterm action in order to have a finite gravitational partition function. This procedure is known as the \keyword{holographic renormalization}%
\footnote{See, \eg, Ref.~\cite{Skenderis:2002wp} for a review of the holographic renormalization. The counterterm action contains the AdS scale $L$, so one cannot regulate the divergence if the spacetime is not asymptotically AdS. In particular, one cannot use this prescription for asymptotically flat black holes such as the Schwarzschild black hole. See \sect{HP} for a prescription of such a case.}. 

Thus, the counterterm action is chosen so that
\begin{itemize}

\item It is written only in terms of boundary quantities ($\gamma_{\mu\nu}$ and the quantities made of $\gamma_{\mu\nu}$ such as the Ricci scalar%
\footnote{Readers should not confuse $\cR$ with $R$, the Ricci scalar in the bulk spacetime. $R_{MN}$ is a tensor in $(p+2)$-dimensions, and $\cR_{\mu\nu}$ is a tensor in the $(p+1)$-dimensions.}  
$\cR$). 

\item It consists of only a finite number of terms.

\item The coefficients of these terms are chosen once and for all in order to cancel divergences.

\end{itemize}
When $p \leq 5$, the counterterm action is given by
\begin{align}
\action_\text{CT} &= \frac{1}{16\pi G_{p+2}} \int  d^{p+1}x\, \sqrt{\gamma} 
  \left\{ 
  \frac{2p}{L} 
  + \frac{L}{p-1}\cR \right. 
\nonumber \\
& 
 \left. - \frac{L^3}{(p-3)(p-1)^2} \left(\cR^{\mu\nu}\cR_{\mu\nu} - \frac{p+1}{4p}\cR^2\right)
+ \cdots
\right\}.
%
\end{align}
The above terms are enough to cancel power-law divergences for $p \leq 5$, but one may need the other terms to cancel log divergences. For black holes with planar horizon, $\cR_{\mu\nu}=0$, so only the first term contributes to thermodynamic quantities:
\be
\action_\text{CT} = \left. 6u^{-4}h^{1/2} \right|_{u=0} 
\xrightarrow{u\rightarrow0}{}
 \left. \frac{6}{u^4} - 3 \right|_{u=0}~.
\label{eq:counter_free_bdy}
\ee

\subsubsection*{The partition function and thermodynamic quantities
}

To summarize our results \eqref{eq:bulk_free_bdy}, \eqref{eq:GH_free_bdy}, and \eqref{eq:counter_free_bdy},
\begin{align}
\action_\text{bulk} &= \left. \frac{2}{u^4} - 2 \right|_{u=0}~, \\
\action_\text{GH} &=  \left.  -\frac{8}{u^4} + 4 \right|_{u=0}~, \\
\action_\text{CT} &=  \left.  \frac{6}{u^4} - 3 \right|_{u=0}~.
%
\end{align}
Combining these results, we get a simple result $\action_\text{bulk} + \action_\text{BH} + \action_\text{CT} = -1$ after all these computations. Recovering dimensionful quantities, we get
\be
\SosE =  - \frac{\beta V_3}{16\pi G_5} \frac{r_0^4}{L^5}~.
%
\ee
The on-shell action is related to the partition function $Z$ and the free energy \index{free energy} $F$ as
\be
Z = e^{-\SosE}~, \quad
\SosE = \beta F~.
%
\ee
Then, the free energy becomes
\be
F =  - \frac{V_3}{16\pi G_5} \frac{r_0^4}{L^5} 
= - \frac{(\pi L T)^4}{16 \pi G_5 L} V_3 
= -\frac{1}{8}\pi^2 N_c^2 T^4 V_3~,
%
\ee
where we used $T=r_0/(\pi L^2)$. Thermodynamic quantities are then derived from the free energy as
\begin{align}
s &= - \frac{1}{V_3}\del_T F 
= \frac{(\pi L T)^3}{4G_5} = \frac{1}{4G_5} \left(\frac{r_0}{L}\right)^3
= \frac{1}{2}\pi^2 N_c^2 T^3~, \\
P &= - \del_{V_3} F
= \frac{(\pi L T)^4}{16\pi G_5 L} = \frac{1}{16\pi G_5 L} \left(\frac{r_0}{L}\right)^4
= \frac{1}{8}\pi^2 N_c^2 T^4~, 
\label{eq:pressure_sads} \\
\varepsilon &= \frac{F}{V_3} + T s
= \frac{3(\pi L T)^4}{16\pi G_5 L} = \frac{3}{16\pi G_5 L} \left(\frac{r_0}{L}\right)^4
= \frac{3}{8}\pi^2 N_c^2 T^4~.
%
\end{align}

\endofsection

\ifx\nameofpaper\undefined 
  \usepackage{macro_natsuume} 
  \def\beginsection{\section*}
  \def\endofsection{\end{document}} 
  \input draft_header.tex
\else 
  \def\beginsection{\chapter}
  \def\endofsection{ } 
\fi

\beginsection{AdS/CFT - adding probes
}\label{chap:wilson}


\begin{quote}
In real experiments, one often adds ``probes" to a system to examine its response. Or one adds impurities to a system to see how they change the properties of the system. In this chapter, we discuss how to add probes in AdS/CFT. As a typical example, we add ``quarks" to gauge theories as probes and see the behavior of quark potentials.
\end{quote}

Coupling new degrees of freedom to the original system often arises new phenomena. Adding some new degrees of freedom to AdS/CFT should be also interesting. This is practically important as well. The \Nfour\ SYM is clearly insufficient to mimic real worlds completely since, \eg, it does not have quarks. 

In string theory, there are various fields and branes, so one may would like to add them. The resulting geometries or solutions have been known for some cases, but it is in general very difficult to solve the Einstein equation when there are multiple number of fields and branes. 

So, one often adds them as ``probes." This is just like the particle motion analysis in curved spacetime (Sects.~\ref{sec:classic_tests} and \ref{sec:geodesics_AdS}). One fixes the background geometry and considers the case where the backreaction of the probe onto the geometry is negligible. 

In this chapter, as a typical example, we add ``quarks" to large-$\Nc$ gauge theories as a probe and analyze quark potentials. In \sect{H^3}, we see another example of a probe system, holographic superconductors.

\section{Basics of Wilson loop
}

\begin{figure}[tb]
\centering
\subfigure[]{
\scalebox{0.75}{ \includegraphics{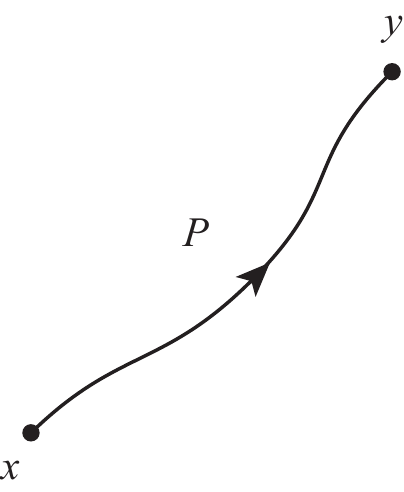} }
\label{fig:path}} \qquad
\subfigure[]{
\scalebox{0.70}{ \includegraphics{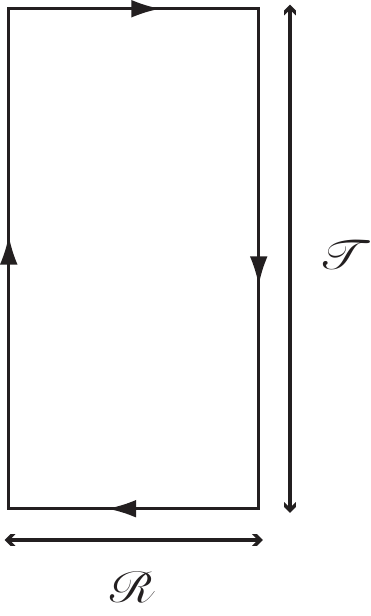} }
\label{fig:wilson}}
\vskip2mm
\caption{(a) Path $P$. (b) The Wilson loop represents a quark-antiquark pair.}
\end{figure}

The \keyword{Wilson loop} is an important observable in gauge theory, and it represents the quark-antiquark potential physically. As an example, consider a $U(1)$ gauge theory with gauge transformation given by
\begin{align}
\phi(x) &\rightarrow e^{i\alpha(x)} \phi(x)~, \\
A_\mu(x) &\rightarrow A_\mu(x) + \del_\mu \alpha(x)~.
%
\end{align}
A nonlocal operator such as $\phi(x)\phi^*(y)$ is not gauge invariant in general and is not an observable. But the following quantity is gauge-invariant:
\be
\phi(x) e^{i \!\int_P dx^\mu A_\mu } \phi^*(y)~,
%
\ee
where $P$ is an arbitrary path from point $x$ to $y$ [\fig{path}]. It transforms as
\begin{align}
\phi(x) e^{i \!\int_P dx^\mu A_\mu } \phi^*(y)
& \rightarrow
\phi(x) e^{i\alpha(x)} e^{i \!\int_P dx^\mu (A_\mu+\del_\mu \alpha) } e^{-i\alpha(y)} \phi^*(y)\\
& = \phi(x) e^{i \!\int_P dx^\mu A_\mu } \phi^*(y)~.
%
\end{align}
Or if one takes a closed path $P$, $W_P$ itself is gauge invariant. Thus, we define the following operator:
\begin{alignat}{2}
W_P(x,y) &= e^{i \int_P dx^\mu A_\mu }  \quad & (\text{Wilson line})~, \\
W_P(x,x) &= e^{i \oint dx^\mu A_\mu }  \quad & (\text{Wilson loop})~.
%
\end{alignat}

The Wilson loop represents the coupling of the gauge field to a test charge. Consider a charged particle with world-line $y^\mu(\lambda)$. The current is given by
\be
J^\mu(x) = \oint d\lambda\, \frac{dy^\mu}{d\lambda} \delta\left( x^\mu-y^\mu(\lambda) \right)~.
%
\ee
The sign of the charge depends on the sign of $dy/d\lambda$. Here, we take $dy/d\lambda>0$ for a positive charge. For a given closed path, $dy/d\lambda$ can be both positive and negative, so we have both a positive charge and a negative charge [\fig{wilson}]. Namely, the closed path describes the process of creating a ``quark-antiquark pair" from the vacuum, pulling them a distance $\calR$ apart, interacting for time $\calT$, and annihilating them. If one uses $J^\mu$, the coupling of the gauge field to  the point particle action is written as $\delta\action = \int d^4x\, A_\mu J^\mu$. This perturbed action $\delta\action$ can be rewritten as the exponent of the Wilson loop:
\be
\delta \action
= \!\int d^4x\, A_\mu(x) J^\mu(x) 
= \!\oint d\lambda\, \frac{dy^\mu}{d\lambda} A_\mu(y(\lambda))
= \!\oint dy^\mu A_\mu(y)~.
%
\ee
Therefore, the Wilson loop represents a partition function in the presence of a test charge:
\be
\bra W_P \ket = \frac{Z[J]}{Z[0]}~.
%
\ee

Such a partition function gives the quark-antiquark potential. Let us write the Euclidean partition function formally as
\be
Z = \bras{f} e^{-H\calT} \kets{i}
%
\ee
($\kets{i}$ and $\kets{f}$ are the initial state and the final state, respectively). If one uses a complete set of energy eigenstates $H \kets{n} = E_n \kets{n}$,
\be
Z = \sum_n e^{-E_n \calT} \norm{f}{n}\norm{n}{i} 
 \xrightarrow{\calT\rightarrow\infty}{} e^{-E_0 \calT}~.
%
\ee
Thus, in the $\calT\rightarrow\infty$ limit, the Euclidean partition function is dominated by the ground state and gives the ground state energy. When the kinetic energy is negligible, it gives the quark-antiquark potential energy%
\footnote{From $\tE = it$, the Lorentzian action $\action_L$, the Euclidean action $\action_E$, and the potential $V$ are related to each other by $i \action_L = i\int dt (-V) = -\int d\tE V = - \action_E$.}.
Consequently,
\be
\bra W_P \ket \simeq e^{-V(\calR)\calT}~.
%
\ee
One can show that the horizontal parts of \fig{wilson} are negligible in the large $\calT\rightarrow\infty$ limit.

When the quark is confined like QCD, the potential grows with the separation $\calR$, so $V(\calR) \simeq \sigma \calR (\calR \gg 1)$, where $\sigma$ is called the string tension. Then,
\be
\bra W_P \ket \simeq e^{-O(\calR\calT)} = e^{-\sigma A}~.
%
\ee
The exponent is proportional to the area of the Wilson loop $A = \calR\calT$. This behavior is known as the \textit{area law}\index{area law (Wilson loop)}. An unconfined potential behaves differently. The Coulomb potential decays with the separation, and one can show that 
\be
\bra W_P \ket \simeq e^{-O(\calR)}
\quad (\text{when } \calR=\calT \gg 1)~.
%
\ee
This is known as the perimeter law. In this way, the Wilson loop provides a criterion for the confinement. 

Here, we consider only the $U(1)$ gauge theory, but a similar discussion can be done for a Yang-Mills theory.

\section{Wilson loops in AdS/CFT: intuitive approach
}\label{sec:wilson_intuitive}



\begin{figure}[tb]
\centering
\subfigure[]{
\scalebox{0.75}{ \includegraphics{Ch5_SYM.pdf} }
\label{fig:Dbrane_SYM_again} } \qquad
\subfigure[]{
\scalebox{0.5}{ \includegraphics{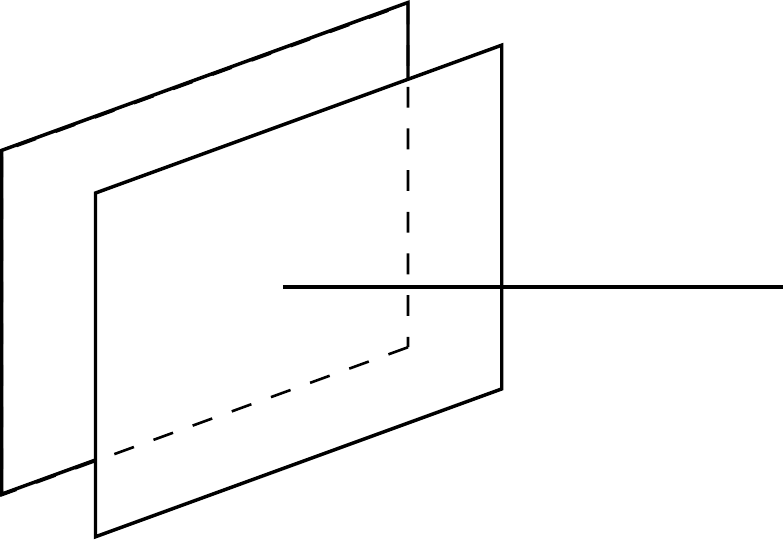} }
\label{fig:quark} }
\vskip2mm
\caption{
(a) An open string can have endpoints on a D-brane. The $N_c$ coincident D-branes represent a $SU(N_c)$ gauge theory. (b) A long string represents a massive ``quark.''
}
\end{figure}%

Let us consider the Wilson loop in AdS/CFT. The Wilson loop in AdS/CFT gives a typical example of adding a probe system to the original system. The AdS/CFT results can be understood intuitively. So, before we go through an actual computation, we first explain what kind of results one can expect in various situations. Then, we confirm our intuitive explanation via an actual computation.

The matter fields in the ${\cal N}=4$ SYM are all in the adjoint representation. So, one first has to understand how to realize the fundamental representation\index{fundamental representation} such as  a quark in AdS/CFT. Below, we describe one simple way to add such matter. 

To do so, recall how the adjoint representation appeared for the D-brane [\fig{Dbrane_SYM_again}]. 
The open strings can have endpoints on a D-brane, but when there are multiple number of D-branes, an open string can have endpoints in various ways; there are $N_c^2$ possibilities. This means that the string transforms as the adjoint representation of $SU(N_c)$ gauge theory. 

Now, consider an infinitely long string [\fig{quark}]. In this case, the string can have endpoints in $N_c$ different ways. This means that the string transforms as the fundamental representation of $SU(N_c)$ gauge theory. In this sense, such a long string represents a ``quark." Such a string has an extension and tension, so the string has a large mass, which means that the long string represents a heavy quark. We discuss the Wilson loop in AdS/CFT using such a string.

We saw earlier that the string model of QCD does not describe potentials other than the confining potential (Problem 2 of \sect{prehistory}). However, one can avoid this problem in AdS/CFT, and one can get the Coulomb potential which appears at short distances in QCD. The AdS/CFT result differs from the simple string model one essentially because of the curved spacetime effect as discussed below. Note that we avoided Problem 1 of \sect{prehistory} by the same trick.

First, we discuss the simplest case, the pure AdS case, to understand the basic idea of the AdS/CFT quark potential. In this case, one gets only the Coulomb potential. We then consider a more generic AdS spacetime and get a confining potential as well. Also, if we consider a black hole, we can recover behaviors in plasma phase.

\subsubsection*{The pure AdS spacetime
}

\begin{figure}[tb]
\centering
\scalebox{0.75}{ \includegraphics{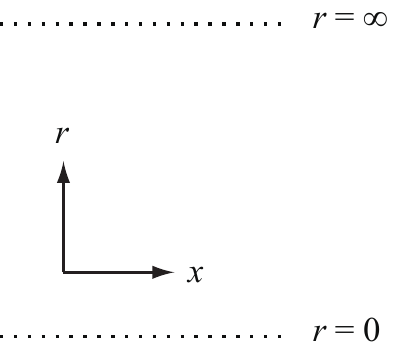} }
\vskip2mm
\caption{Schematically drawn AdS spacetime. The horizontal direction represents one of three-dimensional space the gauge theory lives. The vertical direction represents the AdS radial coordinate. The radial coordinate extends from $r=0$ to $r=\infty$, but we draw in a compact region for illustration.}
\label{fig:potential1}
\end{figure}%

The AdS metric in \poincare\ coordinates is written as
\be
ds^2 = \left( \frac{r}{L} \right)^2(-dt^2+dx^2+\cdots)~.
\label{eq:metric_AdS}
\ee
The line element has the factor $r^2$. We measure the gauge theory time and distance using $t$ and $x$, but they differ from the proper time and distance of the AdS spacetime. This is the important point, and the qualitative behavior of the quark potential can be understood using this fact.

\begin{figure}[tb]
\centering
\scalebox{0.75}{ \includegraphics{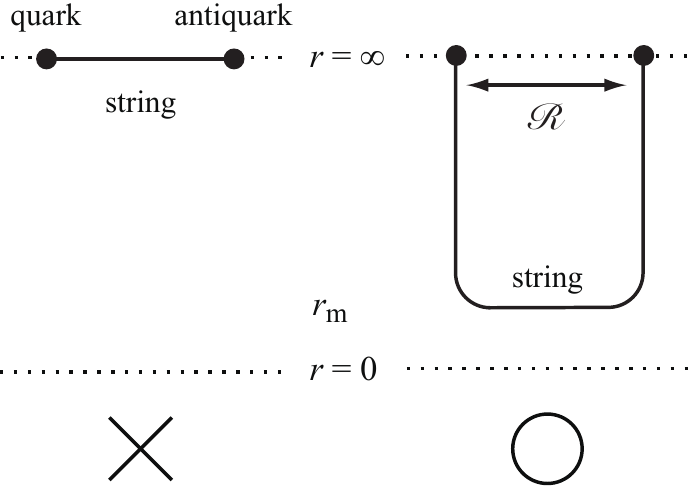} }
\vskip2mm
\caption{The straight string is not the lowest energy state (left), and the string which goes inside the AdS spacetime is the lowest energy state (right).}
\label{fig:potential2}
\end{figure}

Figure~\ref{fig:potential1} shows the AdS spacetime schematically. Denote the quark-antiquark separation as $\Delta x=\calR$. The quark-antiquark pair is represented by a string which connects the pair. The string has the tension, so the tension tends to minimizes the string length. At first glance, one would connect the pair by a straight string at  $r=\infty$ (\fig{potential2}). But this does not minimize the string length. This is because the coordinate distance does not represent a true distance (proper distance) in a curved spacetime. The figure does not show the proper length properly, so one needs a care. For the AdS spacetime, the proper length of the string actually gets shorter if the string goes inside the AdS spacetime ($r\neq\infty$). The line element has the factor $r^2$, so the proper length $r\Delta x$ gets shorter near the origin.

\begin{figure}[tb]
\centering
\scalebox{0.75}{ \includegraphics{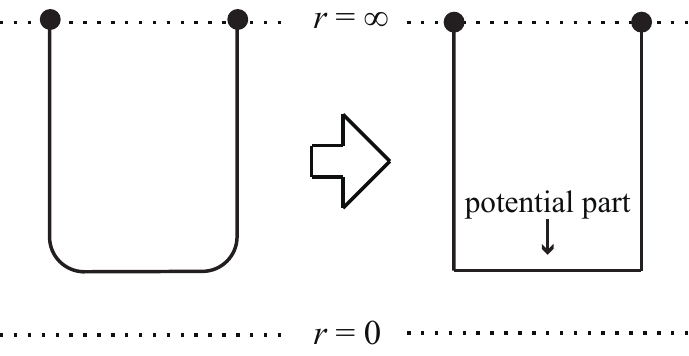} }
\vskip2mm
\caption{The string which connects the quark pair (left) is approximated by a rectangular string. The energy of the horizontal string gives the quark potential. }
\label{fig:potential3}
\end{figure}%

According to the analysis of \sect{wilson_details}, this string is roughly divided into two parts: the part the string extends vertically, and the part the string extends horizontally. So, for simplicity, let us approximate the configuration by a rectangular string (\fig{potential3}). Only the horizontal string contributes to the quark potential. This part varies as we vary the quark separation $\calR$. On the other hand, the vertical string does not vary much. This part simply describes the quark mass. 

\begin{figure}[!tb]
\centering
\scalebox{0.75}{ \includegraphics{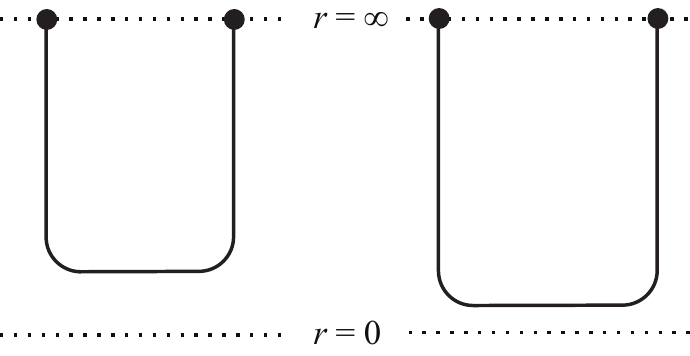} }
\vskip2mm
\caption{The behavior of the string as we vary the quark separation. The larger $\calR$ lowers the string turning point $\rmin$ as $\rmin \propto 1/\calR$.}
\label{fig:potential4}
\end{figure}%

We need a little more information to compute the potential. The explicit computation shows that the string turning point $r=\rmin$ behaves as
\be
\rmin \propto L^2/\calR
\label{eq:r_min}
\ee
(\fig{potential4}). Also, the line element \eqref{eq:metric_AdS} gives two consequences. First, the proper length of the horizontal string is $(r/L)\calR$, so the string energy $E(r)$ is given by
\be
E(r) \propto \left( \frac{r}{L} \right) \calR~.
\label{eq:proper_energy}
\ee
Second, this energy is the proper energy and not the gauge theory energy. The timelike direction also has the factor $r^2$ as in \eq{metric_AdS}. The gauge theory time is the coordinate time $t$ not the proper time. As a result, the gauge theory energy differs from the proper energy $E(r)$. From \eq{metric_AdS}, the proper time $\tau_r$ is related to the gauge theory time by $\tau_r=(r/L) t$, so the proper energy is related to the gauge theory energy $E_t$ by
\be
E_t = \left( \frac{r}{L} \right) E(r)~.
\label{eq:UV/IR2}
\ee
This is the UV/IR relation\index{UV/IR relation} in \sect{interpretation} \cite{Susskind:1998dq,Peet:1998wn}. Thus, the potential is given by
\begin{align}
E_t &=  \left( \frac{\rmin}{L} \right) E(r)
\propto \left( \frac{\rmin}{L} \right)^2 \calR \\
&\propto \frac{L^2}{\calR}~,
\label{eq:coulomb}
%
\end{align}
where we also used \eq{r_min}. This result \cite{Rey:1998ik,Maldacena:1998im} has two important points:
\begin{enumerate}
\item
First, we obtained the Coulomb potential $E \propto 1/\calR$ not the confining potential $E \propto \calR$. Namely, the string connecting the quark-antiquark pair does not necessarily implies a confining potential, but it can describe an unconfining potential using the curved spacetime. \textit{In this way, we resolved Problem 2 of the string model in \sect{prehistory}.} But then, how can we describe the confining potential in AdS/CFT? We will discuss this point below. 
\item
Second, the potential is proportional to $L^2$. According to the AdS/CFT dictionary, $L^2 \propto \lambda^{1/2}$, so the potential is proportional to $(g_\text{YM}^2 N_c)^{1/2}$. But perturbatively, the potential is proportional to $g_\text{YM}^2 N_c$. This is because the AdS/CFT result corresponds to the large-$\Nc$ limit and represents a nonperturbative effect%
\footnote{For the \Nfour\ SYM at zero temperature, the potential is evaluated nonperturbatively from the field theory point of view, and it indeed behaves as $\lambda^{1/2}$ at strong coupling \cite{Erickson:2000af,Drukker:2000rr}.}.
\end{enumerate}

Let us evaluate the potential for a generic metric for later use. By repeating the above argument, the potential energy becomes
\be
\boxeq{
E_t =  \sqrt{-g_{00}}|_{r_m} E(r) = \frac{1}{2\pi l_s^2} \sqrt{-g_{00}g_{xx}}|_{r_m} \, \calR~,
}
\label{eq:formula}
\ee
where the metric is evaluated at $r=\rmin$. We also included the factor of the string tension $T=1/(2\pi l_s^2)$ which we ignored in \eq{proper_energy}.

\subsubsection*{The confining phase
}

\begin{figure}[tb]
\centering
\scalebox{0.75}{ \includegraphics{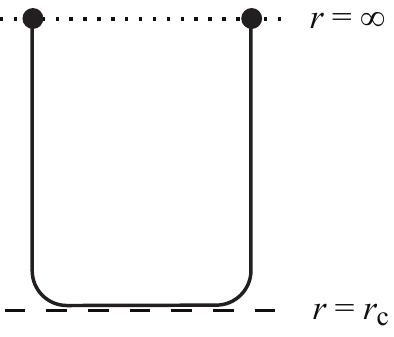} }
\vskip2mm
\caption{In the cutoff AdS spacetime, the string reaches the end of the space, $r=r_c$, when the quark separation is large enough.}
\label{fig:potential5}
\end{figure}%

AdS/CFT can also describe the confining potential which the old string model can describe qualitatively well. The pure AdS spacetime corresponds to the \Nfour\ SYM not to QCD. The \Nfour\ SYM is scale invariant and the confining phase does not exist even at zero temperature. We need to modify the simple AdS geometry to describe a theory which is closer to QCD.

Many examples are known about how the AdS spacetime is deformed if one deforms the \Nfour\ SYM. But we use a simple model to simplify our analysis here \cite{Polchinski:2001tt}.\index{cutoff AdS spacetime} The AdS spacetime extends from $r=\infty$ to $r=0$, but in this model, we cut off the AdS spacetime at $r=r_c$ (\fig{potential5}). Let us suppose that the confinement happens at  a low-energy scale $\Lambda$. In AdS/CFT, the $r$-coordinate has the interpretation as the gauge theory energy scale. So, the confinement means that the AdS spacetime is modified deep inside the AdS spacetime $r \propto \Lambda$. The cutoff AdS roughly represents this effect%
\footnote{The cutoff AdS is a toy model for the confinement, but we discuss an explicit example in \sect{ads_soliton}.
}. 

Even though we modify the spacetime, there is little difference if the string is far enough from the cutoff $r=r_c$. One gets the Coulomb potential like the pure AdS spacetime. But if the quark separation $\calR$ is large enough, there is a new effect.

In the AdS spacetime, the turning point of the string behaves as $\rmin \propto 1/\calR$. But in the cutoff AdS spacetime, the string reaches at $r=r_c$ for a large enough $\calR$. Once the string reaches there, the string cannot go further. Thus, from \eq{formula}, the energy of the horizontal string is given by
\be
E_t \propto r_c^2 \calR \simeq O(\calR)~,
\ee
which is indeed the confining potential.

After all, what contributes to the potential energy is the string at the cutoff $r=r_c$, so the AdS/CFT computation \textit{essentially reduces to the old string model one.} AdS/CFT takes the advantage of the old string model and at the same time overcomes the difficulty of the model.

\subsubsection*{The plasma phase
}

\begin{figure}[tb]
\centering
\scalebox{0.75}{ \includegraphics{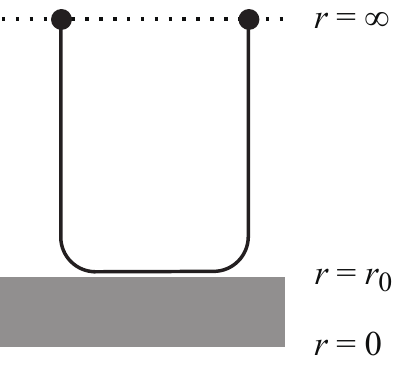} }
\vskip2mm
\caption{The plasma phase case. The shaded region represents the black hole.}
\label{fig:finite_temp}
\end{figure}%

We now consider the finite temperature case or the plasma phase. According to AdS/CFT, the \Nfour\ SYM at finite temperature corresponds to the AdS \bh (\fig{finite_temp}).

At finite temperature, there is a \bh horizon at $r=r_0$. But if the string is far enough from the black hole, the geometry is approximately the AdS spacetime, so one approximately has the Coulomb potential. But if the string reaches the horizon, there is a new effect.

For a black hole, the line element in the timelike direction has the unique behavior, and the relation \eqref{eq:UV/IR2} between $E(r)$ and $E_t$ is modified. For the Schwarzschild-AdS$_5$ (SAdS$_5$) black hole, the line element is given by
\be
ds^2 = - \left(\frac{r}{L}\right)^2 \left\{ 1 - \left( \frac{r_0}{r} \right)^4 \right\} dt^2 + \cdots~,
\ee
so $g_{00}=0$ at the horizon $r=r_0$. Thus, \eq{formula} gives
\be
E_t = 0~.
\label{eq:wilson_debye}
\ee
Namely, the horizontal string has no contribution to the energy. Thus, there is no force 
when the quark separation is large enough. This is the Debye screening\index{Debye screening} in AdS/CFT \cite{Rey:1998bq,Brandhuber:1998bs,Bak:2007fk} (\probset{wilson_debye}).

\subsubsection*{Return of Wilson loops
}

The Wilson loop argument here was proposed in less than two weeks after the systematic AdS/CFT researches started in 1998. Various extensions were made within a month. But people started to come back to such simple analysis since 2006.

What changed the situation? In the past, such a computation was made to find circumstantial evidences of AdS/CFT. Namely, one would like to check whether AdS/CFT correctly reproduces the behavior of gauge theories or not. People do not really have real applications in mind. This is understandable since supersymmetric gauge theories are different from QCD, so probably one was reluctant to apply them to the ``real world." But in recent years, people revisits such analysis and compute various effects by taking into account the real experimental situations.

As discussed in \sect{QCD_phase}, the perturbative QCD is not very effective even in the plasma phase. Thus, heavy-ion physicists try to identify the typical ``fingerprints" of QGP. Some of the fingerprints discussed to date are
\begin{enumerate}
\item Small shear viscosity (\chap{QCD}, \ref{chap:QGP}) 
\item Jet quenching
\item $J/\Psi$-suppression
\end{enumerate}

\begin{figure}[tb]
\centering
\scalebox{0.5}{ \includegraphics{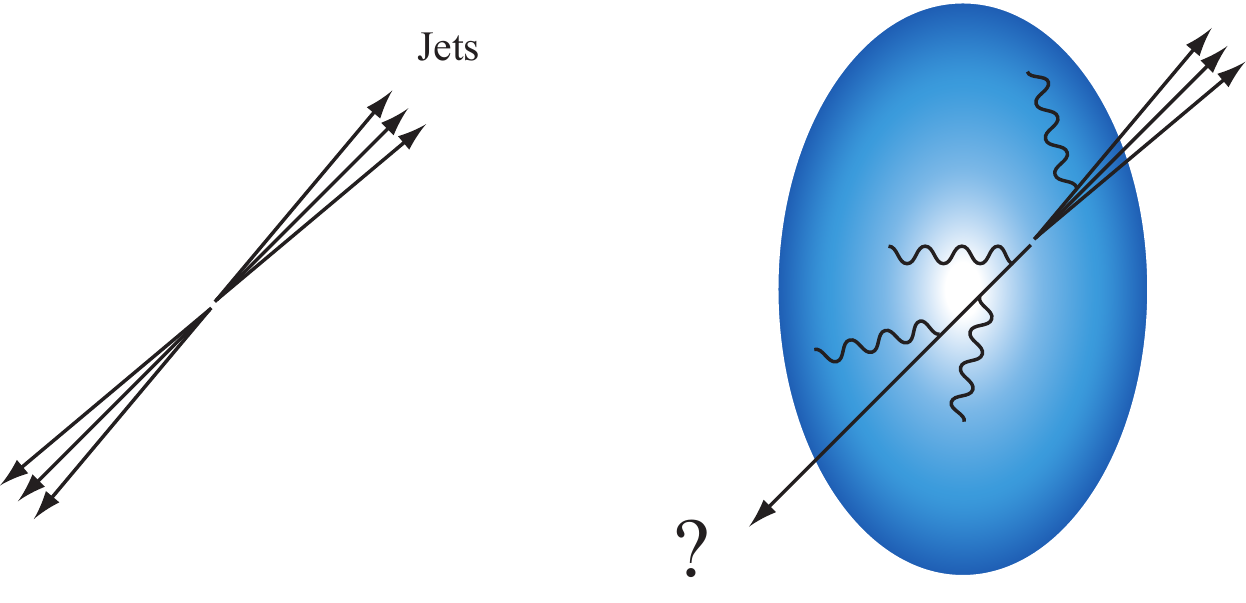} }
\vskip2mm
\caption{\textit{Left}: the collision of nuclei in vacuum. \textit{Right}: jet quenching in the plasma. The ellipsoid represents the plasma.}
\label{fig:jet_quenching}
\end{figure}%

\begin{figure}[tb]
\centering
\scalebox{0.75}{ \includegraphics{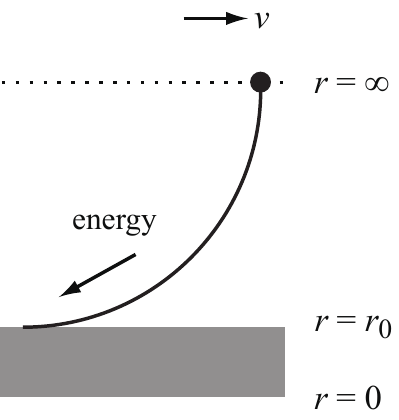} }
\vskip2mm
\caption{Jet quenching in AdS/CFT.}
\label{fig:ads_jet_quenching}
\end{figure}%

In the parton hard-scattering, jets are often formed. A jets is a collection of hadrons which travel roughly in the same direction. If jets are formed in the plasma medium, the energy of the jets are absorbed by the medium, so the number of observed hadrons are suppressed. This is the \keyword{jet quenching} (\fig{jet_quenching}). 

Another fingerprint is the \textit{$J/\Psi$-suppression}\index{J/Psi-suppression@$J/\Psi$-suppression} \cite{Matsui:1986dk}. $J/\Psi$ is a ``charmonium" which consists of $c\bar{c}$. Since a charm quark is heavy ($\approx 4.2$ GeV), the charm pair production occurs only at the early stage of heavy-ion collisions. Now, if the production occurs in the plasma medium, the interaction between $c\bar{c}$ is screened by the light quarks and gluons in between, which is the Debye screening. Then, the charm quark is more likely to bind with the plasma constituents rather than the charm antiquark. The result is the suppression of $J/\Psi$ production. 

These phenomena have been discussed in AdS/CFT. For example, consider the jet quenching  \cite{Liu:2006ug,Herzog:2006gh,CasalderreySolana:2006rq,Gubser:2006bz,Gubser:2008as,Chesler:2008wd}. So far, we considered the static quark to obtain the potential. But in this case, one is interested in how the quark loses its energy. So, move the quark (string) with velocity $v$ along the $x$-direction. Then, the string is dragged as in \fig{ads_jet_quenching}. The string is dragged because the energy of the string flows towards the horizon. This energy loss is interpreted as the energy loss of the quark in the plasma medium.


\section{String action
}\label{sec:NG}

\begin{figure}[tb]
\centering
\scalebox{0.75}{ \includegraphics{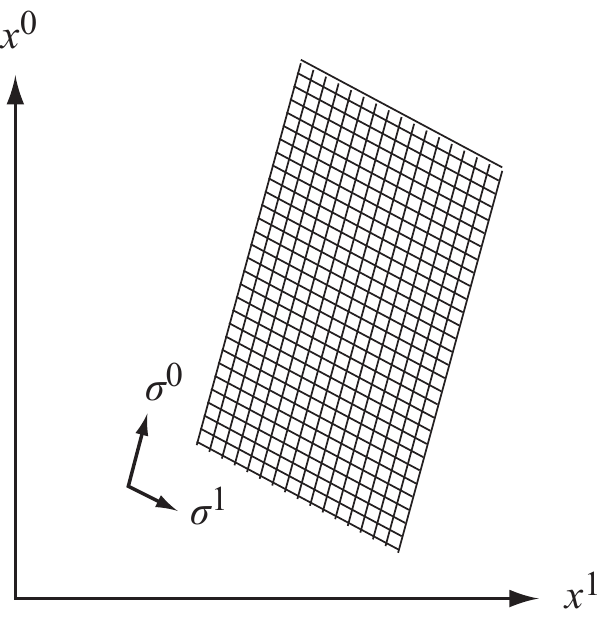} }
\vskip2mm
\caption{A string sweeps a world-sheet in spacetime.}
\label{fig:world_sheet}
\end{figure}%

In order to confirm the intuitive explanation in the last section, let us first consider the string action. The string action is obtained using the similar argument as the particle action in \sect{particle_action_flat}. A particle draws a world-line in spacetime. Similarly, a string sweeps a two-dimensional surface, a world-sheet, \index{world-sheet} in spacetime (\fig{world_sheet}). We write the particle action by the proper length of the world-line. Similarly, it is natural to write the string action by the area $A$ of the world-sheet:
\be
\action = -\tension \int dA~.
\label{eq:action_string}
\ee
The parameter $\tension$ has the dimensions $[\tension] =\text{L}^{-2}$, which makes the action dimensionless. Physically, it represents the string tension. It is convenient to introduce a parameter $l_s$ with the dimension of length and to write the tension as
\be
\tension = \frac{1}{2\pi l_s^2}~.
\ee
The parameter $l_s$ represents the characteristic length scale of the string (string length\index{string length}). 

Just as the particle action, introduce coordinates $\sigma^a=(\sigma^0,\sigma^1)$ on the world-sheet. Then, the world-sheet is described by $x^M(\sigma^a)$. Using the world-sheet coordinates $\sigma^a$, the spacetime metric is written as
\begin{align}
ds^2 &= \eta_{MN} dx^M dx^N
 = \eta_{MN} \frac{\del x^M}{\del \sigma^a} \frac{\del x^N}{\del \sigma^b} d\sigma^a d\sigma^b 
 \label{eq:induced_def} \\
& =: h_{ab}\, d\sigma^a d\sigma^b~,
%
\end{align}
where $h_{ab}$ is known as the \keyword{induced metric}. What we are doing here is essentially the same as the embedding of a hypersurface into a higher-dimensional spacetime in \chap{AdS}. For example, embed $S^2$ into $\mathbb{R}^3$:
\be
ds^2 = dX^2 + dY^2 + dZ^2
 = d\theta^2 + \sin^2\theta\, d\varphi^2 
%
\ee
In this case, we take $S^2$ coordinates as $\sigma^a =(\theta,\varphi)$, and the induced metric is given by
\be
h_{ab} = \left(
\begin{array}{cc}
1 & 0 \\
0 & \sin^2\theta
\end{array}
\right)~.
\label{eq:induced_s2}
\ee

Using the world-sheet coordinates, one can write the area element as
\be
dA = d^2\sigma \sqrt{ -\det h_{ab} }~.
\label{eq:surface_element}
\ee
For $S^2$, $dA = \sin\theta\, d\theta\, d\varphi$, which is the familiar area element for $S^2$. The coordinates $\sigma^a$ are just parametrizations on the world-sheet, so the area element is invariant under 
\be
\sigma^{a'} = \sigma^{a'} (\sigma^b)~.
\label{eq:reparametriz_string}
\ee
The situation is similar to general relativity. In general relativity, one writes the volume element as $d^dx \sqrt{-g}$, and the volume element is invariant under coordinate transformations. The only difference is whether one considers a spacetime or a world-sheet. In general relativity, one considers the volume element in spacetime and the coordinate transformation in spacetime, whereas  \eq{surface_element} is the area element on the world-sheet and \eq{reparametriz_string} is the coordinate transformation on the world-sheet.

Using \eq{surface_element}, one gets the \keyword{Nambu-Goto action}:
\be
\boxeq{
\action_\text{NG} = - \tension \int d^2\sigma \sqrt{ -\det h_{ab}}~.
}
\label{eq:NG}
\ee
From \eq{induced_def}, the induced metric is written as
\be
h_{ab} = \left(
\begin{array}{cc}
\dotx \cdot \dotx & \dotx \cdot x' \\
\dotx \cdot x' & x' \cdot x'
\end{array}
\right)
\quad (\dot{~}:= \del_{\sigma^0},~': = \del_{\sigma^1})~.
%
\ee
Just as in \eq{induced_s2}, this is a matrix on $(a,b)$ indices. Using this, we can write the Nambu-Goto action as
\be
\action_\text{NG} = - \tension \int d^2\sigma \sqrt{ (\dotx \cdot x')^2 -\dotx^2 x'^2 }~.
%
\ee

One can consider a few extensions of the action:
\begin{enumerate}

\item
Here, we used the Minkowski spacetime as the ambient spacetime. But one can get the curved spacetime case by replacing $\eta_{MN}$ by $g_{MN}(x)$ like the particle action case in \sect{particle_action_curved}.

\item
A brane action is obtained similarly. For the D$p$-brane, with the $(p+1)$-dimensional induced metric $h_{ab}$, one writes the action as%
\footnote{Note the factor of the dilation $e^{-\phi}$. The dilaton $\phi$ and the string coupling constant $g_s$ are related by $g_s \simeq e^\phi$, so this factor means that the mass density of the D-brane is proportional to $1/g_s$ [\eq{D3_tension}].}
\be
\action_\text{D$p$} = - \tension_p \int d^{p+1}\sigma\, e^{-\phi} \sqrt{ -\det h_{ab}}~.
%
\ee
Such a brane can be added as a probe just like the string.
\end{enumerate}

\section{Wilson loops in AdS/CFT: actual computation
}\label{sec:wilson_details}

\begin{figure}[tb]
\centering
\scalebox{0.75}{ \includegraphics{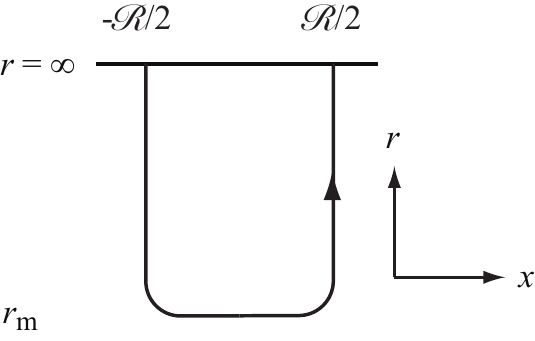} }
\vskip2mm
\caption{The configuration to compute the Wilson loop}
\label{fig:gauge}
\end{figure}%

In this section, we confirm our intuitive explanation in \sect{wilson_intuitive} by an actual computation. As an example, we compute the Wilson loop in the pure AdS$_5$ spacetime. 

The quark potential is given by the energy of the string in AdS/CFT. So, the starting point is the Nambu-Goto action \eqref{eq:NG}. The action has the reparametrization invariance on the world-sheet, so we can choose convenient world-sheet coordinates by coordinate transformations (gauge fixing). Here, we take the \keyword{static gauge}%
\footnote{The string has the turning point at $r=\rmin$, so our gauge is not well-defined in reality. But this is no problem because it is enough to consider only the half of the string by symmetry. One normally takes the gauge $\sigma^0 = t$, $\sigma^1 = x$, and $r=r(x)$ instead of \eq{static_gauge}. The computation is slightly easier in our gauge.
} (\fig{gauge}):
\be
\sigma^0 = t~, \quad
\sigma^1 = r~, \quad
x=x(r)~.
\label{eq:static_gauge}
\ee
The induced metric on the AdS$_5$ spacetime is given by
\begin{align}
%
ds_5^2 &= \left( \frac{r}{L} \right)^2 (-dt^2+d\bmx_3^2) + L^2 \frac{dr^2}{r^2} \\
&= - \left( \frac{r}{L} \right)^2 dt^2 +
\left\{ \left( \frac{L}{r} \right)^2 + \left( \frac{r}{L} \right)^2 x'^2 \right\} dr^2
\quad (' := \del_r)~,
%
\end{align}
so the determinant of the induced metric becomes
\be
-\det h_{ab} = 1 + \left( \frac{r}{L} \right)^4 x'^2~.
%
\ee
Then, the action is given by
\be
%
\action = - \frac{1}{2\pi l_s^2} \int d^2\sigma \sqrt{ -\det h_{ab} }
= - \frac{\calT}{2\pi l_s^2} \int dr\, \sqrt{ 1 + \left( \frac{r}{L} \right)^4 x'^2 }~,
\label{eq:action1}
\ee
where $\calT$ is the time duration in $t$. The Lagrangian does not contain $x$, so there is a conserved momentum $p_x$ which is conjugate to $x$:
\be
%
\frac{\del L}{\del x'} 
\propto \frac{ \left( \frac{r}{L} \right)^4 x' }{ \sqrt{ 1 + \left( \frac{r}{L} \right)^4 x'^2 } }
= \text{(constant)}~.
\label{eq:cyclic_mom}
\ee
Let us determine the constant. The string has the turning point at $r=\rmin$. At the turning point, $\del_r x|_{r=\rmin} = \infty$, so the constant is given by
\be
\left.
\text{(constant)} 
= \frac{ \left( \frac{r}{L} \right)^4 x' }{ \sqrt{ 1 + \left( \frac{r}{L} \right)^4 x'^2 } } 
\right|_{r=\rmin}
= \left( \frac{\rmin}{L} \right)^2~.
\label{eq:const}
\ee
Solving \eq{cyclic_mom} in terms of $x'$, one gets
\be
x'^2 = \left( \frac{L}{r} \right)^4 \frac{1}{ \left( \frac{r}{\rmin} \right)^4 - 1 }~.
\label{eq:eom_sol}
\ee

One can determine the string configuration $x(r)$ by solving \eq{eom_sol}. We take $x=0$ at $r=\rmin$, so $x(r)$ is given by the integral
\be
\int_0^x dx 
= \int_{\rmin}^{r} 
\left( \frac{L}{r} \right)^2 \frac{dr}{ \sqrt{ \left(\frac{r}{\rmin} \right)^4 - 1} }~.
\label{eq:eom_int}
\ee
In particular, $x=\calR/2$ at $r\rightarrow\infty$, so \eq{eom_int} gives
\begin{align}
\frac{\calR}{2}
&= \frac{L^2}{\rmin}
\int_1^\infty \frac{dy}{ y^2 \sqrt{y^4-1} } \quad (y:=r/\rmin) \\
&= \frac{L^2}{\rmin} \frac{ \sqrt{2} \pi^{3/2} }{ \Gamma(\frac{1}{4})^2}~.
\label{eq:rm_vs_R}
%
\end{align}
From \eq{rm_vs_R},
\be
\rmin \simeq \frac{L^2}{\calR}~,
\label{eq:r_min_AdS}
\ee
which justifies \eq{r_min}. Also, when $r \gg \rmin$, \eq{eom_int} gives
\be
\frac{\calR}{2} - x 
= \frac{L^2}{\rmin}
\int_{r/\rmin}^\infty \frac{dy}{ y^2 \sqrt{y^4-1} } 
\simeq r^{-3}~.
%
\ee
The string quickly approaches $x=\calR/2$ for large $r$, which confirms the \fig{potential3} behavior.

We determined the string configuration. We now evaluate the action \eqref{eq:action1} to compute the quark potential. Substituting \eq{eom_sol} into \eq{action1}, one gets
\be
%
\action = - \frac{\calT}{2\pi l_s^2} \int_{\rmin}^\infty dr\, \frac{ \left( \frac{r}{\rmin} \right)^2 }{ \sqrt{\left( \frac{r}{\rmin} \right)^4 - 1} }~.
%
\ee
Then, the potential energy is given by
\be
%
E = - \frac{2\action}{\calT}
= \frac{2}{2\pi l_s^2} \rmin \int_1^\infty \frac{y^2 dy}{\sqrt{y^4-1}}~.
%
\ee
The integral actually diverges, but this reflects the fact that the quark is infinitely heavy. We must subtract the quark mass contribution%
\footnote{See Ref.~\cite{Drukker:1999zq} for a more appropriate procedure.}.
The isolated string configuration is given by $x'=0$. By substituting $x'=0$ into \eq{action1}, one obtains the quark mass contribution:
\begin{align}
%
\action_0 &= - \frac{\calT}{2\pi l_s^2} \int_0^\infty dr~, \\
E_0 &= \frac{2}{2\pi l_s^2} \int_0^\infty dr~.
%
\end{align}
Thus,
\be
%
E-E_0 = \frac{2}{2\pi l_s^2} \rmin 
\left\{ 
\int_1^\infty \left(\frac{y^2}{\sqrt{y^4-1}} -1 \right)dy - 1 
\right\}~.
\label{eq:potential_ads}
\ee
The expression is proportional to $\rmin$, but $ \rmin \simeq L^2/\calR $ from \eq{r_min_AdS}, so we get the Coulomb potential $E \simeq 1/\calR$. The evaluation of the integral in \eq{potential_ads} gives
\be
E-E_0 = - \frac{4\pi^2}{\Gamma(\frac{1}{4})^4} \frac{\lambda^{1/2}}{\calR}~,
%
\ee
which agrees with our intuitive explanation \eqref{eq:coulomb}.

\section{\titlesummary}

\begin{itemize}
\item 
Adding probes to the original system is a simple but useful way to explore the system further. 
\item 
As an example, we add Wilson loops to various asymptotically AdS spacetimes. The Wilson loop is an important nonlocal observable in a gauge theory, and it represents the quark-antiquark potential.
\item 
In AdS/CFT, the Wilson loop corresponds to adding an infinitely long string extending from the AdS boundary.
\item
In the pure AdS spacetime, the holographic Wilson loop gives the Coulomb potential which is a curved spacetime effect. The potential is proportional to $(g_\text{YM}^2 N_c)^{1/2}$ which represents a strong coupling effect.
\item
If one changes background geometries, one gets various quark potentials such as the confining potential and the Debye screening. Also, if one considers dynamical strings, one can discuss dynamical problems such as the jet quenching in the plasma phase. 
\end{itemize}

\titlenewterms

\begin{multicols}{2}
\noindent
Wilson loop\\
cutoff AdS spacetime\\
jet quenching\\
$J/\Psi$-suppression\\
induced metric\\
Nambu-Goto action\\
static gauge\\
\newtermapp{AdS soliton}
\end{multicols}

\section{Appendix: A simple example of the confining phase
}\label{sec:ads_soliton}

In the text, we discussed the cutoff AdS spacetime as a toy model of the confining phase. Here, as an explicit example, we discuss the $S^1$-compactified \Nfour\ SYM and its dual geometry. 

\subsubsection*{AdS soliton}

The SAdS$_5$ black hole is given by
\begin{align}
ds_5^2 &= \left(\frac{r}{L}\right)^2 (-hdt^2+dx^2+dy^2+dz^2)+L^2\frac{dr^2}{h r^2}~, 
\label{eq:SAdS_app} \\
h &= 1- \left(\frac{r_0}{r}\right)^4~.
\end{align}
We now compactify the $z$-direction as $0 \leq z <l$.

However, the compacfified SAdS$_5$ black hole is not the only solution whose asymptotic geometry is $\mathbb{R}^{1,2} \times S^1$. The ``double Wick rotation" 
\be
z' = it~, \quad z = it'
%
\ee
of the black hole gives the metric
\be
ds_5^2 = \left(\frac{r}{L}\right)^2 (-dt'^2+dx^2+dy^2+hdz'^2)+L^2\frac{dr^2}{h r^2}~,
\label{eq:AdS_soliton}
\ee
which has the same asymptotic structure $\mathbb{R}^{1,2} \times S^1$. The geometry \eqref{eq:AdS_soliton} is known as the \keyword{AdS soliton} \cite{Horowitz:1998ha}.

As Euclidean geometries, they are the same, but they have different Lorentzian interpretations. The AdS soliton is not a black hole. Rather, because of the factor $h$ in front of $dz'^2$, the spacetime ends at $r=r_0$ just like the Euclidean black hole. From the discussion in the text, this geometry describes a confining phase. 

For the SAdS black hole, the imaginary time direction has the periodicity $\beta = \pi L^2/r_0$ to avoid a conical singularity. Similarly, for the AdS soliton, $z'$ has the periodicity $l$ given by
\be
l = \frac{\pi L^2}{r_0}~.
%
\ee

\subsubsection*{Wilson loop}

Let us consider the quark potential in this geometry. Take the quark separation as $\calR$ in the $x$-direction. This corresponds to a Wilson loop on the $t'$-$x$ plane. Since the geometry ends at $r=r_0$, the formula \eqref{eq:formula} gives
\be
E_t \propto \sqrt{-g_{t't'}g_{xx}}|_{r_0} \, \calR 
= \left(\frac{r_0}{L}\right)^2 \calR~,
\label{eq:wilson_ads_soliton}
\ee
which is a confining potential (\probset{wilson_ads_soliton}).

In \sect{wilson_intuitive}, we considered the Wilson loop in the SAdS black hole and discussed the Debye screening. Here, we consider a Wilson loop in the same Euclidean geometry, but the Wilson loop here is different from the one in \sect{wilson_intuitive}:
\begin{itemize}

\item For the AdS soliton, we consider the Wilson loop on the $t'$-$x$ plane (temporal Wilson loop), but as the black hole, this is a Wilson loop on the $z$-$x$ plane or a spatial Wilson loop.
\item For the black hole, we considered the temporal Wilson loop on the $t$-$x$ plane, but as the AdS soliton, this is a spatial Wilson loop on the $z'$-$x$ plane.

\end{itemize}

At high temperature $Tl>1$, the AdS soliton undergoes a first-order phase transition to the SAdS black hole (\sect{phase_example}).

\endofsection

\ifx\nameofpaper\undefined 
  \usepackage{macro_natsuume} 
  \def\beginsection{\section*}
  \def\endofsection{\end{document}} 
  \input draft_header.tex
\else 
  \def\beginsection{\chapter}
  \def\endofsection{ } 
\fi

\beginsection{Basics of nonequilibrium physics
}\label{chap:hydro}


\begin{quote}
So far we discussed equilibrium physics. In order to apply AdS/CFT to nonequilibrium physics, we explain the basics of nonequilibrium physics. We explain it both from the microscopic point of view (linear response theory) and from the macroscopic point of view (hydrodynamics).
\end{quote}

\section{Linear response theory
}

\subsection{Ensemble average and density matrix
}

In statistical mechanics, one considers an ensemble average. An ensemble average of an operator $\calO$ is defined by
\be
\bra \calO \ket := \sum_i w_i  \bras{\alpha_i}  \calO \kets{\alpha_i}~.
%
\ee
The states $\kets{\alpha_i}$ do not have to form a complete set, but they are normalized. The coefficient $w_i$ represents the statistical weight and satisfies $\sum_i w_i=1$ and $w_i \geq 0$. Inserting a complete set $\kets{b}$ twice, one gets
\be
\bra \calO \ket 
= \sum_{b',b''} \left( \sum_i w_i  \norm{b''}{\alpha_i} \norm{\alpha_i}{b'} \right)
 \bras{b'} \calO \kets{b''}~.
%
\ee
Introducing the \keyword{density matrix} $\rho$ defined by
\be
\rho := \sum_i w_i \kets{\alpha_i} \bras{\alpha_i}~,
%
\ee
one can rewrite
\begin{align}
\bra \calO \ket 
&= \bras{b''}\rho\kets{b'} \bras{b'}\calO\kets{b''} \\
&= \text{tr} \left[ \rho \calO \right]~.
%
\end{align}
The condition $\sum_i w_i=1$ is rewritten as $\text{tr} (\rho) = 1$.

As an example, consider spin-$1/2$ systems:
\be
S_z \kets{\pm} = \pm\frac{1}{2}\kets{\pm}~.
\ee
With the $S_z$ basis, the pure ensemble%
\footnote{Traditionally, pure ensemble and mixed ensemble are called as pure states and mixed states.}
 which contains $\kets{+}$ only is
\be
\rho = \kets{+}\bras{+} 
= \left(
 \begin{array}{c}
  1 \\
  0
 \end{array}
\right)
\left(
 \begin{array}{cc}
  1, & 0\\
 \end{array}
\right)
=  \begin{pmatrix}
  1&0 \\
  0&0
 \end{pmatrix}~.
%
\ee
One can easily check $\rho_\text{pure}^2 = \rho_\text{pure}$, $\text{tr} (\rho_\text{pure}^2) = 1$ for a pure ensemble. The mixed ensemble which contains $\kets{+}$ and $\kets{-}$ with equal weights is
\be
\rho = \frac{1}{2} 
 \begin{pmatrix}
  1&0 \\
  0&1
 \end{pmatrix}~.
%
\ee

In a canonical ensemble, $\rho$ takes the form
\be
\rho = \frac{1}{Z} e^{-\beta H}~,
\label{eq:rho_canonical}
\ee
where $Z:= \text{tr}[e^{-\beta H}]$ is the partition function, $\beta:=1/T$, and $H$ is the Hamiltonian.

In the Schr\"{o}dinger picture, a state evolves in time. Denoting the time-evolution operator as $U(t,t_0)$, the time evolution is written as $\kets{\alpha, t} = U(t, t_0) \kets{\alpha, t_0}$. Then, the density matrix evolves as
\be
\rho(t) = \sum_i w_i \kets{\alpha_i ,t} \bras{\alpha_i, t}
= U(t,t_0) \rho(t_0) U^{-1}(t,t_0)~,
\label{eq:density_matrix:evolution}
\ee
or
\be
i\del_t \rho = -[\rho, H]~.
%
\ee

\subsection{Linear response theory
}\label{sec:linear_response}

In real experiments, one often adds an external source and sees the \keyword{response} to the operator $\calO$, $\delta\bra\calO\ket$, which couples to the external source. Some examples are%
\footnote{Some explanation is probably necessary for the fluid. Of course, in real experiments, one does not curve our spacetime. We regard the spacetime fluctuation as an external source here because we would like to discuss the fluid on the same footing as the other systems such as the magnetic system. (We explain this in \sect{diffusion}, but this is convenient to derive the so-called Kubo formula.) In any case, there is no problem to consider such a source and a response in principle since spacetime fluctuations bring up fluctuations in the energy-momentum tensor according to general relativity.}
\begin{center}
\begin{tabular}{llcl}
 & external source $\source$ & $\rightarrow$ & response to $\calO$ \\
 \\
magnetic system: & magnetic field $H$ && magnetization $m$ \\
charged system:  & gauge potential $\mu$ && charge density $\rho$  \\
conductor:  & vector potential $A^{(0)}_i$ && current $J^i$ \\
fluid: & spacetime fluctuation $h^{(0)}_{\mu\nu}$ && energy-momentum tensor $T^{\mu\nu}$
\end{tabular}
\end{center}
The purpose of the \keyword{linear response theory} is to examine such a response. The word ``linear" means that the response is studied at linear order in external source%
\footnote{In this book, we consider only at the linear level, but nonlinear cases are interesting both in statistical mechanics and in AdS/CFT.}.

The response can be determined using the time-dependent perturbation theory in quantum mechanics. We start with the Schr\"{o}dinger picture, where the operator $\calO$ is independent of time. Write the full Hamiltonian as
\be
H=H_0+\delta H(t)~,
%
\ee
where $H_0$ is the Hamiltonian when the external source is absent. Denote the perturbed Hamiltonian $\delta H$ or the perturbed action $\delta \action$ as
\begin{align}
\delta H(t) &= - \int d^3x\,  \source(t,\bmx) \calO(\bmx)~, \\
\delta \action &= \int d^4x\,  \source(t,\bmx) \calO(\bmx)~.
\label{eq:perturb}
\end{align}

%
%

The ensemble average of $\calO$ under the perturbation is written as
\be
\bra \calO(t,\bmx) \ketj = \text{tr} \left[ \rho(t) \calO(\bmx) \right]~.
%
\ee
The subscript ``$s$" represents the average in the presence of the external source. We assume that the unperturbed system is in equilibrium at $t=t_0\rightarrow-\infty$. Thus, $\rho(t_0)$ is given by an equilibrium density matrix, \eg, \eq{rho_canonical} for a canonical ensemble. We then turn on the source $\source$ at $t=t_0$. 

From the evolution of the density matrix \eqref{eq:density_matrix:evolution},
\be
\bra \calO(t,\bmx) \ketj 
= \text{tr} \left[ \rho(t_0) U^{-1}(t,t_0) \calO(\bmx) U(t,t_0) \right]~.
\label{eq:evol_Heisenberg}
%
\ee
In the Heisenberg picture where the operator evolves in time, one defines $\calO_H = U^{-1} \calO U$. But we use the interaction picture here. Namely, the operator evolves in time with the unperturbed Hamiltonian $H_0$. The operator in the interaction picture $\calO_I (t,t_0)$ is then written as $\calO_I = U_0^{-1} \calO U_0$, where $U_0$ is the time-evolution operator using $H_0$, or $U_0=e^{-iH_0(t-t_0)}$. From the Schr\"{o}dinger equation, time-evolution operators satisfy
\begin{align}
i\del_t U &= HU~, \\
i\del_t U_0 &= H_0U_0~.
%
\end{align}
Then, $U$ is written as
\begin{align}
U(t,t_0) &= U_0(t,t_0) U_1(t,t_0)~, \\
i\del_t U_1 &= \delta H_I U_1~,
%
\end{align}
where $\delta H_I := U_0^{-1} \delta H U_0$ is the perturbed Hamiltonian \textit{in the interaction picture}. The equation for $U_1$ is formally solved as
\begin{align}
U_1(t,t_0) 
&=1 - i \int_{t_0}^t dt' \delta H_I(t') \nonumber \\
& + (-i)^2 \int_{t_0}^t dt' \int_{t_0}^{t'} dt'' \delta H_I(t') \delta H_I(t'') + \cdots 
\label{eq:dyson_series} \\
&=: T \exp\left[ -i \int _{t_0}^t dt' \delta H_I(t')  \right]~,
%
\end{align}
where $T$ is the time-ordered product. In the interaction picture, \eq{evol_Heisenberg} is written as 
\be
\bra \calO(t, \bmx) \ketj 
=\text{tr} \left[ \rho(t_0) U_1^{-1}(t,t_0) \calO_I(t,\bmx) U_1(t,t_0) \right]~. 
\ee
Write this equation explicitly using \eq{dyson_series} but taking only up to first order in $\source$:
\begin{align}
\bra \calO(t, \bmx) \ketj 
&= \text{tr} \left[ \rho(t_0) 
\left( 1 + i \int_{t_0}^t dt' \delta H_I(t') + \cdots \right) 
 \calO_I(t, \bmx) \right. \nonumber\\
& \left.\hspace{1cm} \times
 \left( 1 - i \int_{t_0}^t dt' \delta H_I(t') + \cdots \right)
\right] \\
&= \text{tr} \left[ \rho(t_0) \calO_I(t, \bmx) \right] \nonumber \\
& - i\, \text{tr} \left[ \rho(t_0) \int_{t_0}^t dt' 
[\calO_I(t,\bmx), \delta H_I(t') ] \right] + \cdots~.
%
\end{align}
Consider the first term. $\rho(t_0)$ is the equilibrium density matrix $\rho_\text{eq}$ with respect to $H_0$. Since $\del_t \rho_\text{eq}=0$, $\rho_\text{eq}$ and $H_0$ commute, and 
\be
\text{tr} \left[ \rho(t_0) \calO_I(t, \bmx) \right] = \text{tr} \left[ \rho_\text{eq} \calO(\bmx) \right].
%
\ee
This is just the ensemble average when the external source is absent, namely $\bra \calO \ketnoj$. 
Taking $t_0 \rightarrow -\infty$, we get
\begin{align}
\delta \bra \calO(t, \bmx) \ket 
&:= \bra \calO(t, \bmx) \ketj - \bra \calO \ketnoj \\
&= i \int^{t}_{-\infty} dt' \int^{\infty}_{-\infty} d^3x'\, 
\left\bra [\calO(t,\bmx), \calO(t',\bmx') ] \right\ketnoj \source(t',\bmx') \\
&= i \int^{\infty}_{-\infty} d^4x'\, 
\theta(t-t') \left\bra [\calO(t,\bmx), \calO(t',\bmx') ] \right\ketnoj \source(t',\bmx')~,
\label{eq:linear_response}
\end{align}
where $\theta(t-t')$ is the step function. Equation~\eqref{eq:linear_response} tells us that the response is determined by the ensemble average with the \textit{equilibrium} density matrix $\rho(t_0)$. Note that $\calO(t,\bmx)$ evolves in time with $H_0$.
This is the great advantage as well as the limitation of the linear response theory. Namely, we need only the equilibrium density matrix to compute the response, which simplifies our analysis. At the same time, we circumvent the question of what the nature of nonequilibrium statistical mechanics is. 

We define the \keyword{retarded Green's function} $G_R^{\calO\calO}$ as
\be
G_R^{\calO\calO}(t-t', \bmx-\bmx') := -i \theta(t-t')
\left\bra [\calO(t,\bmx), \calO(t',\bmx') ] \right\ketnoj~.
\ee
Then, \eq{linear_response} becomes%
\footnote{Various conventions are found in the literature about the sign of the perturbed Hamiltonian $\delta H$ (the sign of the perturbed action $\delta \action$) and the sign of the retarded Green's function $G_R$. Accordingly, the linear response relation \eqref{eq:linear_response:real} or \eqref{eq:linear_response:mom} may have a sign difference.}
\be
\delta \bra \calO(t, \bmx) \ket 
= - \int^{\infty}_{-\infty} d^4x'\, 
G_R^{\calO\calO}(t-t', \bmx-\bmx') \source(t',\bmx')~.
\label{eq:linear_response:real}
\ee

The Fourier transformation of this equation gives
\be
\boxeq{
\delta \bra \calO(k) \ket = - \,G_R^{\calO\calO}(k)\source(k)~,
}
\label{eq:linear_response:mom}
\ee
where $k_\mu = (\omega, \bmq)$, and $G_R^{\calO\calO}(k)$ is the retarded Green's function in momentum space:
\be
G_R^{\calO\calO}(k) = -i \int^{\infty}_{-\infty} d^4x\, e^\ikx 
\theta(t)  \left\bra [\calO(t,\bmx), \calO(0,\bm{0}) ] \right\ketnoj~.
\ee
AdS/CFT can determine the Green's function $G_R$. In condensed-matter physics, $\chi:=-G_R$ is called the \keyword{response function}%
\footnote{The response function differs from the Green's function by a minus sign traditionally since it is natural to define the response function as $\chi = \delta\bra\calO\ketj/\delta\source$. }.

Let us look at several examples:
\begin{itemize}
\item For fluids, a perturbed Lagrangian is given, \eg, by $\delta{\cal L} = \sourceG(t) T^{xy}(x)$, and 
\begin{align}
\delta \bra T^{xy} \ket &= -G_R^{xy,xy} \sourceG~, 
\label{eq:response_emtensor} \\
G_R^{xy,xy} &= -i \int^{\infty}_{-\infty} d^4x\, e^\ikx 
\theta(t)  \left\bra [T^{xy}(t,\bmx), T^{xy}(0,\bm{0}) ] \right\ketnoj~.
\end{align}
\item The case where a conserved current $J^\mu$ exists:
\begin{itemize}
\item For a charged system, $\delta{\cal L} = A^{(0)}_{0}(t) J^0 (x)$, and 
\begin{align}
\delta \bra \rho \ket &= -G_R^{\rho\rho} \mu~, 
\label{eq:response_rho} \\
G_R^{\rho\rho} &= -i \int^{\infty}_{-\infty} d^4x\, e^\ikx
\theta(t)  \left\bra [\rho(t,\bmx), \rho(0,\bm{0}) ] \right\ketnoj~,
\end{align}
where $\rho=J^0$ is the charge density, and $\mu= \sourceA{0}$ is the gauge potential conjugate to  $\rho$.
\item For a conductor, $\delta{\cal L} = A^{(0)}_{x}(t) J^x (x)$, and
\begin{align}
\delta \bra J^x \ket &= -G_R^{xx} A^{(0)}_x~, 
\label{eq:response_current_micro} \\
G_R^{xx} &= -i \int^{\infty}_{-\infty} d^4x\, e^\ikx
\theta(t)  \left\bra [J^x(t,\bmx), J^x(0,\bm{0}) ] \right\ketnoj~.
\end{align}
\end{itemize}
\end{itemize}

\subsection{Transport coefficient: an example
}\label{sec:conductivity}

The Green's functions are related to \keyword{transport coefficients}. As a simple example, let us see this for the \keyword{conductivity} $\sigma$.

The conductivity is given by Ohm's law $\delta \bra J^x \ket = \sigma \sourceE$. In the gauge $\sourceA{0}=0$, the external electric field is  $\sourceE = -\del_t \sourceA{x} \stackrel{FT}{\rightarrow} i\omega \sourceA{x}$ (``$\stackrel{FT}{\rightarrow}$" means Fourier-transformed quantities), so
\be
\delta \bra J^x \ket = i\omega\sigma \sourceA{x}~.
\label{eq:response_current}
\ee
But this is the linear response relation for the current \eqref{eq:response_current_micro}:
\be
\delta \bra J^x \ket = -G_R^{xx} A^{(0)}_x~.
\nonumber
\ee
Thus,%
\footnote{This is the AC conductivity or the frequency-dependent conductivity. The DC conductivity is given by the $\omega\rightarrow0$ limit of \eq{conductivity_ac_kubo}:
\be
\sigma(\omega\rightarrow0) = - \lim_{\omega\rightarrow 0} \frac{1}{\omega} \text{Im}\, G_R^{xx}(\omega,\bmq=0)~.
\label{eq:conductivity_kubo}
\ee}
\be
\sigma(\omega) = -\frac{G_R^{xx}(\omega,\bmq=0)}{i\omega}~.
\label{eq:conductivity_ac_kubo}
\ee
Such a relation, a relation between a transport coefficient and a Green's function, is called a \keyword{Kubo formula}.





Below we consider fluids and relate $G_R^{xy,xy}$ to the viscosity while we explain the basics of hydrodynamics.

\section{Thermodynamics
}\label{sec:thermodynamics}

Below we summarize basic thermodynamics. See Ref.~\cite{callen} for the details; we follow the axiomatic approach of this reference.


\head{Fundamental postulates of thermodynamics}
\begin{itemize}

\item Postulate 1: The equilibrium state is completely described by a set of \keyword{extensive variables} (energy $E$, volume $V$, and particle number%
\footnote{We consider a system of one particle species for simplicity.}
$Q$). These extensive variables are independent variables of a thermodynamic system. 

\item Postulate 2: There exists the entropy $S$ at equilibrium. The entropy is additive over the subsystems, continuous, differentiable, and a monotonically increasing function of the energy. The equilibrium state of a composite system is the state which maximizes the entropy, \ie, the state where the sum of subsystem entropies is maximized.

\end{itemize}

\head{The fundamental relation} 
From the postulates, the entropy is a function of the extensive variables and is written as
\be
S = S(E, V, Q)~.
\label{eq:fund_entropy}
\ee
This relation is called the \keyword{fundamental relation}. Since the entropy is additive, the fundamental relation is a homogeneous first order function of the extensive variables%
\footnote{This does not hold for black holes with compact horizon. See \sect{BH_fund_rels}.}:
\be
S( \lambda E, \lambda V, \lambda Q) = \lambda S(E, V, Q)~.
\ee
The monotonic property implies
\be
\left. \frac{\del S}{\del E} \right|_{V, Q} >0~.
\label{eq:monotonic}
\ee
The continuity, differentiability, and monotonic property imply that the fundamental relation is invertible with the entropy:
\be
E=E(S,V,Q)~,
\label{eq:fund_energy}
\ee
with
\be
E( \lambda S, \lambda V, \lambda Q) = \lambda E(S, V, Q)~.
\label{eq:homo_1st_order}
\ee
Equation~\eqref{eq:fund_entropy} is called the entropy representation, and \eq{fund_energy} is called the energy representation. In general, a ``representation" specifies a set of independent variables. The independent variables are $(E,V,Q)$ in the entropy representation. Below we utilize the energy representation.

The fundamental relation has all thermodynamic information about a system, and all thermodynamic relations below are derived from the fundamental relation.

As an example, consider the photon gas. The fundamental relation is given by%
\footnote{This relation can be derived from statistical mechanics using the results of \sect{weak_coupling}. Rewrite $\varepsilon_\text{B}$ in terms of $s_\text{B}$ (by taking into account that the photon has two degrees of freedom).}
\be
E(S,V) = \frac{3}{4} c^{1/3} S^{4/3}V^{-1/3}~, \quad c=\frac{45}{4\pi^2}~.
\label{eq:fund_rel_photon}
\ee
This equation is indeed a homogeneous first order function of the extensive variables.

\head{The intensive variables}\index{intensive variables} 
The temperature $T$, the pressure $P$, and the ``chemical potential"%
\footnote{Thermodynamics is traditionally applied to systems with various chemical components, so $\mu$ is called the chemical potential. But the use of $\mu$ is not limited to a chemical system. For example, $\mu$ can be a gauge potential for a charged particle or can be a potential which is conjugate to non-chemical particle number (such as the baryon number). We use the same terminology, chemical potential, even for those cases. In any case, $\mu$ represents the potential energy for a matter flow. \label{fnote:chemical}}
$\mu$ are defined by partial derivatives of the fundamental relation:
\begin{align}
T &= \left(\frac{\del E}{\del S} \right)_{V, Q} = T(S,V,Q)~, \\
P &= -\left(\frac{\del E}{\del V} \right)_{S, Q} = P(S,V,Q)~, \\
\mu &= \left(\frac{\del E}{\del Q} \right)_{S,V} = \mu(S,V,Q)~.
%
\end{align}
Because the fundamental relation is a first order homogeneous function of the extensive variables, the intensive variables are zeroth order homogeneous functions as they should be. These equations which express intensive variables by extensive variables are called the \keyword{equations of state}. From \eq{monotonic}, one gets $T>0$.

\head{The first law of thermodynamics} \index{first law of thermodynamics}
The fundamental relation and the definition of intensive variables give
\be
dE = T dS - P dV + \mu dQ~. 
\ee

\head{The Euler relation} \index{Euler relation}
Differentiating \eq{homo_1st_order} with respect to $\lambda$
\be
\frac{ \del E(\lambda S, \ldots) }{ \del(\lambda S) } \frac{ \del(\lambda S) }{\del \lambda} + \cdots 
= E(S,V,Q)~,
%
\ee
and setting $\lambda=1$ gives
\be
E = TS - PV + \mu Q~. 
\ee
The Euler relation is not a fundamental relation because it involves intensive variables.

\head{The Gibbs-Duhem relation}\index{Gibbs-Duhem relation}
The relation among intensive variables. From the Euler relation and the first law, one gets
\be
S dT - V dP + Q d\mu = 0~.
\ee

\head{Thermodynamic potentials} 
So far, we used representations by extensive variables $(E,V,Q)$ or $(S,V,Q)$. But in real experiments, one can control intensive variables, \eg, temperature, more easily than extensive variables. In such a case, it is more convenient to use some of intensive variables as independent variables. For that purpose, we use \keyword{thermodynamic potentials} which are obtained by the partial Legendre transformation of the fundamental relation. The equilibrium is given by the state which minimizes thermodynamic potentials. Thermodynamic potentials have all thermodynamic information about a system as well%
\footnote{In this sense, a thermodynamic potential should be called a fundamental relation in a particular representation.}.

For example, the \keyword{free energy} $F=F(T,V,Q)$ is defined by
\be
F(T,V,Q) := E - TS~.
%
\ee
From
\be
dF = - SdT - PdV + \mu dQ~,
%
\ee
one gets
\be
S = - \left(\frac{\del F}{\del T} \right)_{V, Q}~.
\ee

As an example, consider the free energy for the photon gas. From the fundamental relation \eqref{eq:fund_rel_photon}, the temperature is given by $T(S,V)=\del E/\del S = (c S/V)^{1/3}$. Solving this equation in terms of $S$, one gets $S(T,V)=T^3 V/c$. Thus, 
\be
F = E-TS = -\frac{1}{4} c^{1/3} S^{4/3} V^{-1/3} = -\frac{1}{4c} T^4 V~.
\ee

The \keyword{grand canonical potential} $\Omega=\Omega(T,V,\mu)$ is defined by
\begin{align}
\Omega(T,V,\mu) &:= E - TS - \mu Q  \\
&= -P(T,\mu)V~.
%
\end{align}
We used the Euler relation on the second line. From
\be
d\Omega=- SdT - PdV - Q d\mu~, 
%
\ee
one gets
\be
S = - \left(\frac{\del \Omega}{\del T} \right)_{V, \mu}~, \quad
Q = - \left(\frac{\del \Omega}{\del \mu} \right)_{T, V}~.
\ee
Thermodynamic potentials are related to the \keyword{partition function} $Z$ in statistical mechanics. For example,
\be
Z = e^{-\beta\Omega}~.
\ee

\head{Spatially homogeneous system}
In such a case, it is convenient to use the energy density $\varepsilon:= E/V$, the entropy density $s:= S/V$, and the number density $\rho:=Q/V$. Then, the first law becomes
\be
d\varepsilon = T ds + \mu d\rho~. 
\ee


\section{Hydrodynamics
}\label{sec:hydro}

\subsection{Overview of hydrodynamics
}

We have seen that the retarded Green's function represents the response of a system from the microscopic point of view. But it is a different issue whether one can actually compute the Green's function microscopically. AdS/CFT can compute the Green's function. On the other hand, one can narrow down the necessary information from the Green's function using macroscopic considerations such as conservation laws and the low-energy effective theory. Namely, we do not have to know the complete form of the Green's function. This formalism is \keyword{hydrodynamics}%
\footnote{Landau-Lifshitz \cite{fluid_text} is rather old, but it is still a good textbook on hydrodynamics.}.

Hydrodynamics describes the macroscopic behavior of a system. Of primary interest is conserved quantities. This is because they are guaranteed to survive in the low-energy $\omega\rightarrow0$, long-wavelength limit $q\rightarrow0$ (hydrodynamic limit\index{hydrodynamic limit}). For example, in the diffusion problem below, the current conservation takes the form
\be
\del_0 \rho + \del_i J^i = 0~.
\ee
When Fourier transformed [\eq{diffusion_dispersion}], this equation implies a mode with $\omega\rightarrow0$ as $q\rightarrow0$.

Typical macroscopic variables other than conserved quantities are
\begin{itemize}

\item \keyword{Nambu-Goldstone mode} (if there is a continuous symmetry breaking)~,
\item \textit{Order parameter}\index{order parameter} (if there is a phase transition).

\end{itemize}
Hydrodynamics refers to \textit{dynamics of these macroscopic variables in any system}. In particular, note that hydrodynamics is not limited to the literal fluids such as water. For example, a spin system in condensed-matter physics is a hydrodynamic system.

From the field theory point of view, hydrodynamics is an effective theory\index{effective theory}. In an effective theory, one writes down an action with all terms consistent with symmetry, but the coefficients of terms depend on the details of a microscopic theory, so one cannot determine them in the formalism of effective theory alone. For hydrodynamics, these coefficients are called \keyword{transport coefficients}. These coefficients are necessary information for us to know responses. We do not need the complete Green's functions but need part of Green's functions which are represented by transport coefficients as we will see below%
\footnote{For example, in order to extract the conductivity, it is enough to know the $O(\omega)$ part of the Green's function $G_R^{xx}$ in the $\bmq\rightarrow0$ limit. See \eq{conductivity_kubo}.
}. 
AdS/CFT can carry out computations of microscopic theories using gravitational theories, so AdS/CFT can determine these coefficients for particular theories.

\subsection{Example: diffusion problem
}\label{sec:diffusion}

The problem of fluids, in particular the viscous fluids, is rather complicated. So, let us first consider a simple example, the diffusion problem of a charge. The various issues we encounter in the diffusion problem are common to the fluid problem.

Consider a current $J^\mu$, where $\rho := J^0$ is a conserved charge density. Here, we do not specify the current explicitly:
\begin{itemize}

\item In condensed-matter applications, the current may be the usual electromagnetic current, but it is not limited to the  electromagnetic current.

\item For example, in QCD, there is the $U(1)_B$ current associated with the baryon number conservation. 

\item In the \Nfour\ SYM, there are R-currents \index{R-currents} associated with the R-symmetry.\index{R-symmetry}

\end{itemize}
In any case, $\rho$ is a ``number density" or a ``charge density" in the Noether's theorem sense associated with a global $U(1)$ symmetry.

For the diffusion problem, the variables and the conservation law are
\begin{alignat}{2}
&\text{variables:} && J^\mu = (\rho, J^i)~, \\
&\text{conservation law:}\quad && \del_\mu J^\mu =0~.
\label{eq:conservation}
\end{alignat}
In $(3+1)$-dimensions, there are four variables whereas the conservation law gives only one equation, so the equation of motion is not closed. In order to close the equation of motion, we introduce the \keyword{constitutive equation} which is a phenomenological equation. For the diffusion problem, the constitutive equation is known as \keyword{Fick's law}:
\be
J^i= -D \del^i \rho~.
\label{eq:constitutive}
\ee
Fick's law tells that a charge gradient produces a current; this is natural physically. The proportionality constant $D$ is the \keyword{diffusion constant}. If one combines the conservation law with Fick's law, $\rho$ and $J_i$ decouple, and one gets the equation for $\rho$ only:
\be
0 = \del_\mu J^\mu = \del_0 \rho + \del_i J^i = \del_0 \rho - D \del_i^2 \rho~.
\label{eq:diffusion}
\ee
This is the \keyword{diffusion equation}.

Let us solve the diffusion equation in $(1+1)$-dimensions. We take the initial condition $\rho(t=0,x)=\delta(x)$. Make a Fourier transformation in space and a Laplace transformation in time:
\begin{align}
\rho (t, q) &= \int_{-\infty}^{\infty} dx\, e^{-iqx} \rho(t, x)~, \\
\tilde{\rho} (\omega, q) &= \int_{0}^{\infty} dt\, e^{i\omega t} \rho(t, q)~.
%
\end{align}
Then, \eq{diffusion} is solved as
\be
\tilde{\rho} (\omega, q) = \frac{1}{-i\omega+Dq^2} \rho(t=0,q) = \frac{1}{-i\omega+Dq^2}~.
\ee
Namely, $\tilde{\rho} (\omega, q)$ has a pole on the negative imaginary axis in the complex $\omega$-plane:
\be
\boxeq{
\omega = -iD q^2~.
}
\label{eq:diffusion_dispersion}
\ee
We will encounter such a dispersion relation over and over again.

\begin{figure}[tb]
\centering
\scalebox{0.5}{ \includegraphics{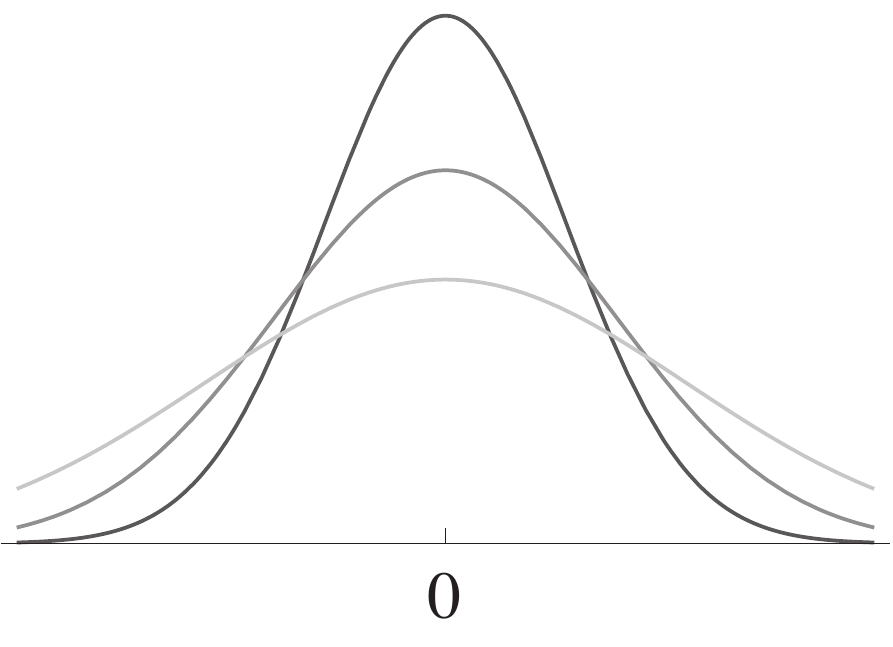} }
\vskip2mm
\caption{The charge diffusion.}
\label{fig:diffusion}
\end{figure}

Make the inverse transformation to rewrite the momentum space solution by the real space solution. By the inverse transformation in $t$, the contour integral picks up the pole, so $\rho(t,q)$ decays exponentially in time: $\rho(t,q) \propto e^{-Dq^2t}$. Then, the time scale of the diffusion, the \keyword{relaxation time}, is $\tau\simeq1/(Dq^2)$. The inverse transformation in $x$ gives
\be
\rho(t,x) = \frac{1}{\sqrt{4\pi Dt}}\exp\left(-\frac{x^2}{4Dt}\right)~, \quad (t\geq 0)~.
\label{eq:diffusion_sol}
\ee
The solution shows that the delta-function distribution of the charge is spread as time passes (\fig{diffusion}).

The constitutive equation \eqref{eq:constitutive} is physically natural, but in general it is determined to satisfy the second law of thermodynamics. As a simple example, consider the heat diffusion (or heat conduction). In this case, the conservation law and the first law of thermodynamics are given by
\begin{align}
\del_0 \varepsilon + \del_i q^i &= 0~, 
\label{eq:heat_conserv} \\
d\varepsilon &= T ds~,
\end{align}
($q^i$: heat flow). Combining them, we get
\begin{align}
0 &= \del_0 s + \frac{1}{T} \del_i q^i \\
&= \del_0 s + \del_i \left(\frac{1}{T} q^i \right) + \frac{\del_i T}{T^2} q^i~.
%
\end{align}
Then, the evolution of the entropy is given by
\be
\del_0 S = \int d^3x\, \del_0 s = - \int d^3x\, \frac{\del_i T}{T^2} q^i~,
\ee
where we discarded a surface integral. If we require
\be
q^i = -\kappa\, \del^i T
\label{eq:Fourier}
\ee
(and $\kappa \geq 0$), the entropy satisfies the second law:
\be
\del_0 S = \kappa \int d^3x\, \left(\frac{\del_i T}{T} \right)^2 \geq 0~.
\ee
The coefficient $\kappa$ is called the \keyword{heat conductivity}, which is a transport coefficient. Equation~\eqref{eq:Fourier} is similar to Fick's law, but for the heat diffusion, it is called Fourier's law%
\footnote{In order to obtain a closed expression from \eq{heat_conserv} and \eq{Fourier}, use the heat capacity $C=\del\varepsilon/\del T$. Then, one gets $\del_0\varepsilon = (\kappa/C) \del_i^2\varepsilon$. The heat diffusion constant $D_T$ satisfies $\kappa = D_T C$, which is the heat version of \eq{diffusion_vs_conductivity}. }.

As we saw earlier, a transport coefficient can be obtained from the linear response theory microscopically, but one needs a little trick to obtain the Kubo formula for the diffusion constant. Our discussion of the linear response so far considers the case
\be
\text{external source}~\phi_0 \rightarrow \text{response}~\delta\bra\calO\ket~.
\ee
But, in the diffusion problem, we are interested in the behavior of charge fluctuation $\delta\bra\rho\ket$. Such a problem is statistical in nature and cannot be expressed as a perturbed Hamiltonian. Rather, it should arise from the nonequilibrium density matrix. But then we lose the advantage of the linear response theory, where a response can be computed just using the equilibrium density matrix. What should we do? The heat diffusion $\delta\varepsilon$ and the fluid case $\delta \bra T^{\mu\nu} \ket$ share the same problem. They are the case
\be
\text{thermal internal force} \rightarrow \text{response}~\delta\bra\calO\ket~.
\ee

In this case, one can imagine the inhomogeneity of $\rho$ as coming from an external source. Then, one can apply the linear response theory. We apply an inhomogeneous chemical potential $\mu$ and produce an inhomogeneous $\rho$:
\be
J^i = - D \del^i \rho 
= - D \left( \frac{\del\rho}{\del\mu} \right) \del^i \mu~.
\ee
Now, $\chi_T := \del\rho/\del\mu$ is the thermodynamic susceptibility\index{thermodynamic susceptibility}, and $E^i := -\del^i \mu$ is the ``electric field" for the current. Then,
\be
J^i = (D\chi_T) E^i~.
\label{eq:diffusion_as_conduction}
\ee
But this is nothing but Ohm's law $J^i = \sigma E^i$, where $\sigma$ is the conductivity. Then, there is a simple relation among the diffusion constant, the thermodynamic susceptibility, and the conductivity:
\be
\sigma = D \chi_T~.
\label{eq:diffusion_vs_conductivity}
\ee
Because we have a Kubo formula for $\sigma$ (\sect{conductivity}), we can determine $D$ from the Kubo formula and \eq{diffusion_vs_conductivity}:
\be
D = - \frac{1}{\chi_T} \lim_{\omega\rightarrow 0} \frac{1}{\omega} \text{Im}\, G_R^{xx}(\omega,\bmq=0)~.
\label{eq:diffusion_kubo}
\ee
The lesson here is to rewrite a thermal internal force problem in terms of an external source problem. We will do the same thing for fluids. When we will derive the Kubo formula for the viscosity, we apply a ``fictitious" gravitational field and see the response $\delta \bra T^{\mu\nu} \ket$ (\sect{viscosity_kubo})%
\footnote{The philosophy is somewhat similar to the derivation of the energy-momentum tensor in \sect{scale_inv}. Even if one is not really interested in field theory in the curved spacetime, the coupling with gravity is a convenient way to derive the energy-momentum tensor.}.
But in AdS/CFT, this fictitious gravitational field will have a new interpretation: it is part of the ``real" five-dimensional gravitational field.

There is a different method to obtain $D$. The linear response relation for $\rho$ reads as
\be
\delta \bra \rho \ket = -G_R^{\rho\rho} \mu~.
\label{eq:diffusion_pole}
\ee
Since $\rho$ has a pole at $\omega=-iDq^2$, $G_R^{\rho\rho}$ should have the pole, and from the pole we can determine $D$. Note that we use a Green's function different from \eq{diffusion_kubo}. Also, we are interested in the \textit{value} of the Green's function $G_R^{xx}$, whereas we are interested in the \textit{pole} of the Green's function $G_R^{\rho\rho}$.

To summarize our discussion so far,
\begin{itemize}

\item Hydrodynamics describes dynamics of macroscopic variables. Typical macroscopic variables are conserved quantities (such as charge density, energy density, and momentum density), Nambu-Goldstone modes, and order parameters.

\item For conserved quantities, we obtain a closed form expression from the conservation law and the constitutive equation.

\item The constitutive equation is introduced phenomenologically, and it is determined to satisfy the second law of thermodynamics. The constitutive equation introduces a transport coefficient.

\item Local thermal equilibrium: we consider a nonequilibrium situation, but it is close to an equilibrium. Namely, we assume that thermodynamic equilibrium holds in smaller scales than the characteristic length scale of the problem and that time and position-dependent thermodynamic quantities make sense. For example, the temperature is originally defined at equilibrium and is constant everywhere, but we consider $T(t,x)$ for the heat diffusion problem.

\item We would like to determine responses. The linear response theory narrows down to retarded Green's functions, and hydrodynamics further narrows down to transport coefficients. AdS/CFT can determine transport coefficients from the microscopic point of view. However, AdS/CFT does not have to assume the formalism of hydrodynamics here. Namely, we can cross-check the formalism of hydrodynamics itself (\sect{2nd_order}).

\item There are two ways to determine a transport coefficient:
\begin{itemize}
\item The method from the $O(\omega)$ coefficient of $G_R^{xx}$ (Kubo formula).
\item The method from the pole of $G_R^{\rho\rho}$ and the dispersion relation.
\end{itemize}

\end{itemize}
We have more to say about the diffusion problem: we will come back to the problem in \sect{2nd_order}.

\subsection{Perfect fluid
}

As a simple fluid, we first consider the case where macroscopic variable is $T^{\mu\nu}$ only:
\begin{alignat}{2}
&\text{variables:} && T^{\mu\nu}~, \\
&\text{conservation law:}\quad && \del_\mu T^{\mu\nu} = 0~.
\label{eq:EM_conservation}
\end{alignat}
$T^{\mu\nu}$ is a rank-two symmetric tensor, so in $(3+1)$-dimensions, it has 10 components whereas the conservation law provides only four equations, so the equations of motion are not closed like the diffusion problem. Again, one can construct the constitutive equation which expresses currents in terms of conserved quantities (in this case, the energy density and the momentum density). But it is more convenient to introduce a new set of variables to close the equations of motion. We choose these variables as
\begin{center}
energy density $\varepsilon(x)$~，four-velocity field of fluid $u^\mu(x)$~.
\end{center}
The \keyword{velocity field} specifies the fluid velocity at each spacetime point. There are in total four variables as we wish (one variable for $\varepsilon$ and three for $u^\mu$. There are only three independent components for $u^\mu$ since $u^2=-1$.)

Below we determine the form of $T^{\mu\nu}$ via derivative expansion. The zeroth order of the derivative expansion corresponds to the \keyword{perfect fluid}. If one is not allowed to use $\del_\mu$, the only symmetric combinations we can make are $u^\mu u^\nu$ and $\eta^{\mu\nu}$. We thus write $T^{\mu\nu}$ as
\be
T^{\mu\nu} = (\varepsilon+P)u^\mu u^\nu + P \eta^{\mu\nu}~.
\label{eq:EM_constitutive}
\ee
This is the constitutive equation for the perfect fluid. We explain below how we choose the coefficients $(\varepsilon+P)$ and $P$. Here, we introduced the pressure $P$, but it is not an independent variable as we saw in \sect{thermodynamics}. Consider the (local) \keyword{rest frame}, where $u^\mu\stackrel{RF}{=}(1,0,0,0)$ (``$\stackrel{RF}{=}$" denotes an expression which is valid only in the rest frame). In the rest frame, $T^{\mu\nu}$ takes the form 
\be
T^{\mu\nu} \stackrel{RF}{=} 
	\begin{pmatrix}
	  \varepsilon & 0 & 0 & 0 \\
	  0 & P & 0 & 0 \\
	  0 & 0 & P & 0 \\
	  0 & 0 & 0 & P \\
	\end{pmatrix}~.
\ee
Each component of $T^{\mu\nu}$ has the following physical meaning:
\begin{itemize}

\item 
$T^{00}$ represents the energy density. $T^{00}\stackrel{RF}{=}\varepsilon$, so $\varepsilon$ is the energy density in the rest frame. (In hydrodynamics, all scalar quantities are defined in the rest frame.)

\item
$T^{0i}$ represents the momentum density.
There is no fluid motion in the rest frame, so it vanishes in the frame. Note that the momentum density $T^{0i}=(\varepsilon+P)u^0u^i$ has the contribution not only from the energy density but also from the pressure as well in a relativistic fluid%
\footnote{The appearance of $(\varepsilon+P)$ in a relativistic fluid plays a vital role in various phenomena. The gravitational collapse of a star to a black hole comes from this effect. For a star to be stable, a pressure is necessary to overcome gravity. But for a massive enough star, once nuclear reactions are over, no pressure can hold the star. In general relativity, the whole $T^{\mu\nu}$ is the source of the gravitational field, so a large pressure is self-destructive; it only increases the source. As a result, a black hole is born.}.

\item
Roughly speaking, $T^{ji}$ represents forces between adjacent fluid elements. 
For the perfect fluid, $T^{ij} \stackrel{RF}{=} P \delta^{ij}$. The pressure is the same in all directions (isotropy) and is perpendicular to the surface. This is known as Pascal's law. In general, the force does not have to be perpendicular to the surface (Recall \fig{viscosity}). We have the isotropy because we construct $T^{\mu\nu}$ from $u^\mu$ and $\eta^{\mu\nu}$ alone. For an anisotropic fluid, one has a new vector, say $V^\mu$, which specifies a particular spatial direction. Using $V^\mu$, one can add a new symmetric term to $T^{\mu\nu}$ which is absent in \eq{EM_constitutive}.

\end{itemize}

We now write down closed expressions using the constitutive equation as in the diffusion problem. Each components of \eq{EM_conservation} correspond to hydrodynamic equations as 
%
%
\begin{align}
\del_\mu T^{\mu 0} &= 0 \rightarrow \text{continuity equation,} 
\label{eq:EM_conservation_0} \\
\del_\mu T^{\mu i} &= 0 \rightarrow \text{Euler equation.}
\label{eq:EM_conservation_i}
\end{align}
First, consider \eq{EM_conservation_0}: 
\be
\del_0 T^{00} + \del_i T^{i0} = 0~.
\label{eq:EM_conservation_0c}
\ee
In order to see the physical meaning of the equation, integrate it over some volume $V$ and use Gauss' law:
\be
\del_0 \int_{V} d^3x\, T^{00} + \int_{\del V} dS\, n_i T^{i0} = 0~,
%
\ee
where $n^i$ is the outward unit normal to the surface $\del V$. The first term represents the energy change in the volume, so the second term represents energy flowing out through the surface in unit time. Or $T^{i0}$ represents the energy flux across the $i$ surface, the $x^i=\text{(constant)}$ surface. The energy flux is equivalent to the $i$ momentum density $T^{0i}$. 

Contract \eq{EM_conservation} with $u_\nu$. In the rest frame, this corresponds to \eq{EM_conservation_0}:
\be
u_\nu \del_\mu \left\{ (\varepsilon+P)u^\mu u^\nu + P \eta^{\mu\nu} \right\} = 0~.
\label{eq:continuity_tmp}
\ee
Since $u^\mu$ is normalized as $u^2=-1$,
\be
0 = \del_\mu u^2 = 2 u_\nu \del_\mu u^\nu~.
%
\ee
Then, \eq{continuity_tmp} reduces to
\begin{align}
0 &= u^\mu \del_\mu \varepsilon + (\varepsilon+P)\del_\mu u^\mu \\
&= \frac{d\varepsilon}{d\tau} + (\varepsilon+P)\del_\mu u^\mu~,
\label{eq:continuity}
%
\end{align}
where $d/d\tau := u^\mu \del_\mu$. This is the relativistic \keyword{continuity equation} (see below).

Second, consider \eq{EM_conservation_i}: 
\be
\del_0 T^{0i} + \del_j T^{ji} = 0~.
\label{eq:EM_conservation_ic}
\ee
Again, integrating it over some volume gives
\be
\del_0 \int_{V} d^3x\, T^{0i} + \int_{\del V} dS\, n_j T^{ji} = 0~.
%
\ee
The first term represents the $i$ momentum change in the volume, so the second term represents the  momentum flowing out through the surface in unit time. Or $T^{ji}$ represents the $i$ momentum flux across the $j$ surface. The momentum flux is force per unit area, so $T^{ji}$ represents the force between adjacent fluid elements. Then, \eq{EM_conservation_ic} is essentially Newton's second law, $F=ma$. 

Evaluating $\del_\mu T^{\mu i} = 0$, we get
\begin{align}
0 &=
 \del_\mu \left\{ (\varepsilon+P)u^\mu u^i + P \eta^{\mu i} \right\} \\
&\stackrel{RF}{=} (\varepsilon+P)u^\mu \del_\mu u^i + \del^i P~.
\label{eq:Euler_rest}
\end{align}
In the rest frame, $u^i$ must be differentiated; otherwise, it vanishes. The derivative of $u^i$ is nonvanishing even in the rest frame. In an arbitrary frame, one can show that \eq{Euler_rest} is replaced by
\be
(\varepsilon+P) \frac{du^i}{d\tau} + u^i\frac{dP}{d\tau} + \del^i P = 0~,
\label{eq:Euler}
\ee
which is the relativistic \keyword{Euler equation} (see below).

The continuity equation, the first law $d\varepsilon = T ds$, and the Euler relation $\varepsilon + P = T s$ lead to
\be
0 = u^\mu \del_\mu \varepsilon + (\varepsilon+P)\del_\mu u^\mu
= T \del_\mu (s u^\mu)~.
\ee
The entropy is conserved, and the fluid is called adiabatic. Namely, there is no dissipation (no viscosity, no heat conduction) in a perfect fluid. 


\subsubsection*{Non-relativistic limit
}

The non-relativistic limit corresponds to scaling
\be
x^i \rightarrow a^{-1} x^i~, \quad
t \rightarrow a^{-2} t~, \quad
v \rightarrow a v~, \quad
P \rightarrow a^2 P~, \quad
%
\ee
and taking the $a\rightarrow0$ limit. The scaling implies $v \ll 1$ and $\varepsilon \gg P$. The  scaling also implies  $L/T \ll 1$, where $L$ and $T$ are the characteristic length and time scales for changes in the fluid. Under the scaling, 
\be
u^\mu = \gamma(1,v^i) \rightarrow (1, av^i)~, \quad
u^\mu \del_\mu = \gamma(\del_0+v^i\del_i) \rightarrow a^2(\del_0+v^i\del_i)~.
%
\ee
Then, at lowest order in $a$, \eq{continuity} becomes
\be
\frac{d\varepsilon}{d\tau} + (\varepsilon+P)\del_\mu u^\mu 
\rightarrow
a^2 \{ \del_0\varepsilon + v^i \del_i\varepsilon + \varepsilon\del_i v^i \} + O(a^4)~,
%
\ee
so we get the non-relativistic continuity equation:
\be
\del_0 \varepsilon + \del_i (\varepsilon v^i) = 0~.
%
\ee
Similarly, \eq{Euler} becomes
\be
(\varepsilon+P) \frac{du^i}{d\tau} + u^i\frac{dP}{d\tau} + \del^i P
\rightarrow 
a^3 \{ \varepsilon (\del_0+ v^j \del_j) v^i + \del^iP \} +O(a^5)~,
%
\ee
so we get the non-relativistic Euler equation:
\be
\varepsilon (\del_0 + v^j \del_j) v^i + \del^iP = 0~.
%
\ee

\subsection{Viscous fluid
}

Let us proceed to the next order in the derivative expansion. At next order, one can include the effect of the dissipation which is not included in the perfect fluid. We write new terms appearing at this order as $\tau^{\mu\nu}$:
\be
T^{\mu\nu} = (\varepsilon+P)u^\mu u^\nu + P \eta^{\mu\nu} +\tau^{\mu\nu}~.
\label{eq:EM_constitutive_dissipative}
\ee
In the rest frame, $T^{0i} = 0$ since the momentum density must vanish in this frame. Also, we define $\varepsilon:=T^{00}$ in the rest frame. Then, the only nonzero components are $\tau^{ij}$ in the rest frame. Assuming the isotropy, one gets two independent transport coefficients:
\begin{alignat}{3}
\tau_{ij}
& \RF -\eta &\left(\del_i u_j + \del_j u_i - \frac{2}{3}\delta_{ij}\del_k u^k \right) 
 -  \zeta & \delta_{ij}\del_k u^k~. &
\label{eq:dissipative} \\
&& \text{traceless} \qquad & \text{trace} &
\nonumber
\end{alignat}
The coefficient $\eta$ in the traceless part is the \keyword{shear viscosity}, and the coefficient $\zeta$ in the trace part is the \keyword{bulk viscosity}. One can notice that the shear viscosity part is the differential form of the fluid example between two plates (\sect{unexpected}). 

As in the diffusion problem, the constitutive equation \eqref{eq:dissipative} is justified from the second law of thermodynamics. We saw that the time evolution of the entropy follows from the continuity equation, so let us consider the continuity equation for the viscous fluid. As in the perfect fluid,
\begin{align}
0 &= u_\nu \del_\mu T^{\mu\nu} \\
&= - u^\mu \del_\mu \varepsilon - (\varepsilon+P)\del_\mu u^\mu + u_\nu \del_\mu \tau^{\mu\nu} \\
&= - T \del_\mu(s u^\mu) - \tau^{\mu\nu}\del_\mu u_\nu~.
\label{eq:continuity_viscous1}
\end{align}
Here, we used $\tau^{\mu\nu}u_\nu = 0$ which implies $\tau^{00}=\tau^{0i} \RF 0$ in the rest frame. Unlike the perfect fluid, the entropy is not conserved due to the second term of \eq{continuity_viscous1}. Now, a symmetric tensor $A^{ij}$ can be written as
\begin{align}
A^{ij} &= A^{\bra ij \ket} + A \delta^{ij}~, \\
\text{where~~} A^{\bra ij \ket} &:= \frac{1}{2}(A^{ij} + A^{ji} - 2A\delta^{ij})~, \quad
A := A^i_{~i}/3~.
\end{align}
Using this expression, \eq{continuity_viscous1} can be rewritten as 
\begin{align}
T \del_\mu(s u^\mu) &\stackrel{RF}{=} - \tau^{\bra ij \ket} \sigma_{ij} + \tau\theta~,
\label{eq:continuity_viscous2} \\
\sigma_{ij} &:= \del_{\bra i} u_{j \ket}~, \quad
\theta := \del_k u^k~,
%
\end{align}
in the rest frame. Then, if we require
\be
\tau^{\bra ij \ket} = - 2 \eta \sigma^{ij}~, \quad
\tau = - \zeta \theta~,
\label{eq:requirement_viscous}
\ee
(and $\eta, \zeta \geq 0$), the right-hand side of \eq{continuity_viscous2} is non-negative:
\be
T \del_\mu(s u^\mu) \stackrel{RF}{=} 2\eta \sigma^{ij}\sigma_{ij} + \zeta \theta^2 \geq 0~.
%
\ee
Equation~\eqref{eq:requirement_viscous} leads to \eq{dissipative}.

We considered the continuity equation. The rest of the conservation equation, $\del_\mu T^{\mu i} = 0$, gives the equation of motion for the viscous fluid. However, in this book, we do not use the hydrodynamic equations directly, and we go back to the conservation law each time, so here we give only the form in the rest frame:
\be
(\varepsilon+P) \frac{du^i}{d\tau} + \del^i P - \eta \del_j^2 u^i - \left(\zeta + \frac{1}{3} \eta \right) \del^i\del_j u^j \stackrel{RF}{=} 0~.
%
\ee
This corresponds to the non-relativistic hydrodynamic equation for the viscous fluid:
\be
%
\varepsilon (\del_0 + v^j \del_j) v^i + \del^iP 
- \eta \del_j^2 v^i - \left(\zeta + \frac{1}{3} \eta \right) \del^i\del_j v^j = 0~.
\label{eq:generalized_NS}
\ee
In the non-relativistic case, one often considers the \keyword{Navier-Stokes equation}:
\be
%
\varepsilon (\del_0 + v^j \del_j) v^i + \del^iP - \eta \del_j^2 v^i = 0~.
%
\ee
In the Navier-Stokes equation, one assumes \keyword{incompressibility} or the constant energy density. Then, the continuity equation gives $\del_i v^i=0$, so the last term of \eq{generalized_NS} vanishes. However, the assumption is problematic in special relativity because the speed of sound $c_s^2 := \del P/\del\varepsilon$ diverges. Thus, we do not impose incompressibility for a relativistic fluid.

We construct the constitutive equation via derivative expansion. We stop at first order in this chapter, but one can proceed the derivative expansion at higher orders (\sect{2nd_order}).

\subsection{When a current exists
\advanced}

\subsubsection*{Perfect fluid with current
}

Let us go back to the leading order in the derivative expansion, and consider the case where a current $J^\mu$ also exits as macroscopic variables.
\begin{alignat}{2}
&\text{variables:}\quad && T^{\mu\nu}~, \quad J^\mu~, \\
&\text{conservation laws:}\quad && \del_\mu T^{\mu\nu} = 0~, \quad \del_\mu J^\mu = 0~.
%
\end{alignat}
We choose the number density $\rho(t,x)$ instead of  $J^\mu$ to close the equations of motion. At this order, the current must be proportional to $u^\mu$: 
\begin{align}
T^{\mu\nu} &= (\varepsilon+P)u^\mu u^\nu + P \eta^{\mu\nu}~, \\
J^\mu &= \rho u^\mu~.
%
\end{align}
In the diffusion problem, the constitutive equation is written as $J^i = -D\del^i \rho$, but this equation contains a derivative. Such a term arises at next order. The term $\rho u^\mu$ represents the fluid motion or the \keyword{convection} which is not included in the diffusion problem. 

The continuity equation is the same as the perfect fluid. One can show the entropy conservation equation $\del_\mu (s u^\mu) = 0$ from the continuity equation.

\subsubsection*{Viscous fluid with current
}\label{sec:viscous_fluid&current}

Let us proceed to the next order in the derivative expansion:
\begin{align}
T^{\mu\nu} &= (\varepsilon+P)u^\mu u^\nu + P \eta^{\mu\nu} +\tau^{\mu\nu}~, 
\label{eq:EM_constitutive_viscous_current} \\
J^\mu &= \rho u^\mu + \nu^\mu~.
%
\end{align}
As in the viscous fluid, we define  $\varepsilon:=T^{00}$ and $\rho := J^0$ in the rest frame. Then, only nonzero components are $\tau^{ij}$ and $\nu^i$.

We have spatial flows, but in this case, there are two currents (current and momentum density). These two currents do not have to coincide with each other. Then, the notion of the ``rest frame" is ambiguous; the rest frame depends on which current we choose. In the rest frame, $u^i=0$, so the ambiguity implies that the definition of $u^\mu$ depends on the rest frame we choose. 

The definition here follows Landau and Lifshitz, and $u^\mu$ represents the momentum density (energy flux). Thus, the momentum density $T^{0i}$ vanishes in the rest frame. On the other hand, in this definition, the particle flux $J^i$ does not vanish in the rest frame due to $\nu^i$. 

The particle flux vanishes when $J^i = \rho u^i+\nu^i = 0$, so $u^i\neq0$. The particle flux vanishes, but the energy flux is nonvanishing in this case. Then, this energy flux represents the heat conduction, and $\nu^i \propto \kappa$, where $\kappa$ is the heat conductivity\index{heat conductivity}. For a perfect fluid, the particle flux coincides with the energy flux, so there is no heat conduction. Thus, each term of $J^\mu$ has the physical meaning as
\begin{alignat}{2}
%
J^\mu = \rho u^\mu	&	& +	& \nu^\mu~.  \\
\text{convection} 	&	&	& \text{conduction/diffusion} \nonumber
\end{alignat}
The explicit form of $\nu^i$ can be determined like $\tau^{ij}$, but we omit the discussion here.

We choose the energy flux as $u^\mu$, which is known as the \keyword{Landau-Lifshitz frame}. Alternatively, one can choose the particle flux as $u^\mu$, which is known as the \keyword{Eckart frame}%
\footnote{The Eckart frame is also called as the Particle frame and the N-frame. The corresponding names for the Landau-Lifshitz frame are the Energy frame and the E-frame. (The E-frame \textit{does not} stand for Eckart, which is rather confusing.)}: 
\begin{alignat}{2}
\text{energy flux } & \text{for } u^\mu & ~\rightarrow~ & \text{Landau-Lifshitz frame} \nonumber \\
\text{particle flux } & \text{for } u^\mu & ~\rightarrow~ & \text{Eckart frame} \nonumber
%
\end{alignat}
In the Eckart frame, $u^\mu$ represents the particle flux, so $\nu^\mu=0$, whereas $T^{\mu\nu}$ needs an additional term $q^{(\mu} u^{\nu)}$, where $q^\mu$ is the heat current.

In principle, the choice of $u^\mu$ should be just a choice of frames, but actually it is not; there is a subtle issue (\sect{2nd_order}).

\subsection{Kubo formula for viscosity
}\label{sec:viscosity_kubo}

A transport coefficient is related to a retarded Green's function. We now have enough knowledge of hydrodynamics to derive the Kubo formula for the viscosity.

As in the diffusion problem, the shear viscosity arises as the response under a thermal internal force. A quick way to derive the Kubo formula is to couple fictitious gravity to the fluid and see the responses of $T^{\mu\nu}$ under gravitational perturbations $h^{(0)}_{\mu\nu}$. Of course, one does not curve our spacetime for real fluid experiments. This is just a convenient way to derive the Kubo formula. 

However, according to general relativity, spacetime fluctuations bring up fluctuations in the energy-momentum tensor. Also, if one considers hydrodynamics in astrophysics, one really needs the effect of spacetime curvature. Moreover, the derivation here has a natural interpretation in AdS/CFT (\chap{GKPW}).

\danger{Note that we consider the four-dimensional curved spacetime $g^{(0)}_{\mu\nu}$ where the gauge theory lives, not the five-dimensional curved spacetime like AdS/CFT. In AdS/CFT, $g^{(0)}_{\mu\nu}$ has the five-dimensional origin, but the argument here itself is independent of AdS/CFT.}

Thus, following the philosophy of the linear response theory,
\begin{enumerate}
\item Add a gravitational perturbation in the four-dimensional spacetime. As in the diffusion problem, the perturbation specifies hydrodynamic variables $u^\mu$. Use hydrodynamics to write the response $\delta \bra T^{\mu\nu} \ket$.
\item Comparing the expression with the linear response theory, one relates the shear viscosity to a retarded Green's function.
 %
\end{enumerate}
In order to consider a gravitational perturbation, the constitutive equation \eqref{eq:EM_constitutive_dissipative} with \eqref{eq:dissipative} is not appropriate since it is defined in the flat spacetime. We first extend the constitutive equation to the curved spacetime. The extension is given by
\begin{align}
T^{\mu\nu} &= (\varepsilon+P)u^\mu u^\nu + P g^{(0)\mu\nu} 
+ \tau^{\mu\nu}~, 
\label{eq:constitutive_curved} 
\\
\tau^{\mu\nu} &= 
- P^{\mu\alpha} P^{\nu\beta} \left\{ 
\eta \left( \nabla_\alpha u_\beta \!+\! \nabla_\beta u_\alpha 
- \frac{2}{3} g^{(0)}_{\alpha\beta} \nabla\!\cdot\! u \right) 
+ \zeta g^{(0)}_{\alpha\beta} \nabla\!\cdot\! u 
\right\}~.
\label{eq:dissipative_curved}
%
\end{align}
Here,
\begin{itemize}

\item $\nabla_\mu$ represents the covariant derivative in the curved spacetime.

\item The tensor $P^{\mu\nu} := g^{(0)\mu\nu} + u^\mu u^\nu$ is called the \keyword{projection tensor} along spatial directions. The projection tensor enables us to write the constitutive equation in a covariant manner. (This tensor is also necessary in the flat spacetime to get the Lorentz-invariant expression.) In the flat spacetime, the rest frame takes $u^\mu \stackrel{RF}{=} (1,0,0,0)$, so 
\be
P^{\mu\nu} \stackrel{RF}{=} \text{diag}(0,1,1,1)~.
\label{eq:projection_LRF}
\ee
Thus, $P^{\mu\nu}$ clearly acts as the projection along spatial directions.

\end{itemize}
%
%





We write the coordinate system as $x^\mu = (t, x, y, z)$. One can consider a generic perturbation, 
but the following form is enough to evaluate the shear viscosity:
\be
g^{(0)}_{\mu\nu} = 
\begin{pmatrix}
  -1 & 0 & 0 & 0 \\
  0 & 1 & h^{(0)}_{xy}(t) & 0 \\
  0 & h^{(0)}_{xy}(t) & 1 & 0 \\
  0 & 0 & 0 & 1 
\end{pmatrix}~.
\label{eq:metric_kubo}
\ee
We use \eq{dissipative_curved} and compute $\tau^{xy}$ to linear order in the perturbation. (The symbol ``$\sim$" below denotes terms up to the linear order.)

This is  a spatially homogeneous perturbation, so even if the fluid has a motion, the fluid motion must be homogeneous $u_i=u_i(t)$. But a motion in a particular direction is forbidden from the parity invariance. Consequently, there is no fluid motion, and $u^\mu = (1,0,0,0)$ or $u_\mu = (-1,0,0,0)$%
\footnote{In this case, the simple physical consideration was enough to determine $u^\mu$. In general, given an external perturbation, the conservation equation determines hydrodynamic variables $u^\mu$ and $\varepsilon$ in terms of the perturbation.}.
Then, the nonzero contribution in the covariant derivative comes from the Christoffel symbol only:
\be
\nabla_x u_y 
= \del_x u_y - \Gamma^\alpha_{xy} u_\alpha
= -\Gamma^0_{xy} u_0
=\Gamma^0_{xy}~.
%
\ee
The component $\nabla_y u_x$ is obtained similarly. The other components vanish. The component $\nabla_x u_y$ is already linear in perturbation, so $(\nabla \cdot u)$ is second order. Thus, only the first two terms in the curly brackets in \eq{dissipative_curved} contribute to $\tau^{xy}$:
\be
\delta \bra \tau^{xy}  \ket 
\sim -\eta (\nabla_x u_y + \nabla_y u_x)~.
\ee
For the projection tensor, it is enough to use \eq{projection_LRF} since the quantities inside the curly brackets are already first order. Evaluating the Christoffel symbol, we get
\be
\Gamma^0_{xy} = \frac{1}{2}g^{(0)00}(\del_y g^{(0)}_{0x} + \del_x g^{(0)}_{0y} - \del_0 g^{(0)}_{xy})
= \frac{1}{2} \del_0 h^{(0)}_{xy}~.
\ee
Thus,
\be
\delta \bra \tau^{xy}  \ket 
= -2 \eta \Gamma^0_{xy}  
= - \eta \del_0 h^{(0)}_{xy}~.
\ee
After the Fourier transformation,
\be
%
\boxeq{
\delta \bra \tau^{xy}(\omega,\bmq=0)  \ket = i\omega \eta \sourceG~.
}
\label{eq:Txy_vs_eta}
\ee
This is the desired result. The result takes the same form as the linear response relation  \eqref{eq:response_emtensor}. Comparing \eqref{eq:response_emtensor} and \eqref{eq:Txy_vs_eta} gives the Kubo formula for $\eta$:
\be
\eta = - \lim_{\omega\rightarrow 0} \frac{1}{\omega} \text{Im}\, G_R^{xy,xy}(\omega,\bmq=0)~.
\label{eq:viscosity_kubo}
\ee

Using a similar argument, one can get the Kubo formula for the other transport coefficients such as $\zeta$%
\footnote{To derive this result, consider the perturbation of the form
\be
g^{(0)}_{\mu\nu} = \text{diag} \left(-1,1+h^{(0)}(t), 1+h^{(0)}(t),1+h^{(0)}(t) \right)~.
\ee
}:
\begin{align}
\zeta &= - \frac{1}{9} \lim_{\omega\rightarrow 0} \frac{1}{\omega} \text{Im}\, G_R^\text{tr}(\omega,\bmq=0)~, \\
G_R^\text{tr} &= -i \int^{\infty}_{-\infty} d^4x\, e^\ikx
\theta(t)  \left\bra [T^{i}_{~i} (t,\bmx), T^{j}_{~j}(0,\bm{0}) ] \right\ket~.
%
\end{align}

\subsection{Linearized hydrodynamic equations and their poles
}\label{sec:linearized}

In the diffusion problem, there are two ways to determine the diffusion constant: the Kubo formula \eqref{eq:diffusion_kubo} and the pole of the response $\delta \bra \rho \ket$ \eqref{eq:diffusion_pole}. Similarly, the other components of $\delta \bra T^{\mu\nu} \ket$ have poles from which one can determine the shear viscosity.


\head{Linearization}
In order to obtain the diffusion pole, it was enough to solve the diffusion equation. Similarly, it is not necessary to couple to the external source, gravitational fields, to obtain the poles. So, we return to the flat spacetime and simply linearize hydrodynamic equations.

We define
\be
\varepsilon (t,\bmx) = \delta\varepsilon(t,\bmx) + \barepsilon~, \quad
v^i (t,\bmx) = \delta v^i(t,\bmx)~, 
%
\ee
and so on. Here, ``~\={ }~" denotes an equilibrium value, and $\delta$ denotes the deviation from the equilibrium. We keep only the linear order in $\delta$. One can set $\bar{v}^i=0$ without loss of generality. The four-velocity $u^\mu$ and the velocity $v^i$ are related by $u^0 = 1/\sqrt{1-v^2} \sim 1$, $u^i = v^i/\sqrt{1-v^2} \sim v^i$.

\head{Tensor decomposition}
We are interested in dispersion relations, so we consider inhomogeneous perturbations. One can take the perturbations of the form
\be
u^i = u^i(t,z)~, \quad
P = P(t,z)~.
%
\ee
Then, there is a little group\index{little group} $SO(2)$ acting on $x^a = (x,y)$. We classify $T^{\mu\nu}$ components by their transformation properties under $SO(2)$. The tensor decomposition is convenient because each modes transform differently and their equations of motion are decoupled with each other. 

\begin{itemize}

\item As a warm-up exercise, first classify $J^\mu$ components:
\begin{alignat}{2}
&\text{vector mode: }&& J^{a}~, \\
&\text{scalar mode: }&& J^{0}, J^{z}~.
%
\end{alignat}
The scalar mode components do not transform under $SO(2)$, and the vector mode components transform as vectors. 
Then, the diffusion problem in \sect{diffusion} is summarized in the language of the tensor decomposition as follows:
\begin{itemize}
\item The charge density $\rho$ belongs to the scalar mode, so the scalar mode has a pole at $\omega=-iDq^2$ from the diffusion equation.
\item The current $J^x$ belongs to the vector mode, so the vector mode gives the conductivity $\sigma$ from Ohm's law.
\end{itemize}

\item Similarly, we classify $T^{\mu\nu}$ components as follows:
\begin{alignat}{2}
&\text{tensor mode: }&& T^{xy}, T^{xx}=-T^{yy}~, \\
&\text{vector mode: }&& T^{0a}, T^{za}~, \\
&\text{scalar mode: }&& T^{00}, T^{0z}, T^{zz}, T^{xx}=T^{yy}~.
%
\end{alignat}
The derivation of the Kubo formula in \sect{viscosity_kubo} utilized the tensor mode.

\end{itemize}
Because each mode is decoupled, we consider the linearized equation of motion for each mode. One can start with hydrodynamic equations such as \eq{generalized_NS}, but here we start with the original conservation law $\del_\mu T^{\mu\nu} = 0$ and the constitutive equation. 

\subsubsection*{Vector mode}

Let us write down the vector mode components:
\begin{align}
T^{za} &= (\varepsilon+P) u^z u^a - \eta (\del^z u^a+\del^a u^z) \sim - \eta \del^z u^a~, \\
T^{0a} &= (\varepsilon+P) u^0 u^a \sim (\barepsilon+\barP) u^a~.
%
\end{align}
For the perturbation of the form $u^a \sim e^{-i\omega t+iqz}$, the $a$-component of the conservation law becomes
\be
0 = \del_0 T^{0a} + \del_z T^{za} 
\sim (\barepsilon+\barP) \del_0 u^a - \eta \del_z^2 u^a 
\propto -i\omega (\barepsilon+\barP) + \eta q^2~.
\ee
Thus, the dispersion relation is given by
\begin{align}
\omega &= - i \frac{\eta}{\barepsilon+\barP} q^2 \\
&= - i \frac{\eta}{\barT\bars} q^2 \quad (\text{for } \mu=0)~.
\label{eq:dispersion_vector}
\end{align}
Comparing with the diffusion problem \eqref{eq:diffusion_dispersion}, one can see that $D_{\eta} := \eta/(\barepsilon+\barP)$ plays the role of a diffusion constant. The corresponding quantity in the non-relativistic limit ($\barP \ll \barepsilon$), $\nu := \eta/\barepsilon$, is called the \keyword{kinematic viscosity}.

As we will see in \chap{QGP}, the combination $\eta/\bars$ is a \textit{particularly interesting quantity in AdS/CFT}%
\footnote{In later chapters, we omit ``~\={ }~" for simplicity, so we will write $\eta/s$  instead of $\eta/\bars$.}.
When there is a chemical potential, $\mu \neq 0$, $\eta/\bars \neq \barT D_{\eta}$, but what is important is $\eta/\bars$ not $\barT D_{\eta}$. This is because AdS/CFT predicts a universal result for $\eta/\bars$.

\subsubsection*{Scalar mode}

Similarly, the scalar mode components are
\begin{align}
T^{00} &= (\varepsilon+P) u^0 u^0 - P \sim \varepsilon~, \\
T^{0z} &= (\varepsilon+P) u^0 u^z \sim (\barepsilon+\barP) u^z~, \\
T^{zz} &= (\varepsilon+P) u^z u^z + P 
  - \eta \left( 2 - \frac{2}{3} \right) \del_z u^z - \zeta \del_z u^z \\
  &\sim P - \left( \frac{4}{3}\eta+\zeta \right) \del_z u^z~.
%
\end{align}
The $t$ and $z$ components of the conservation law become
\begin{align}
0 &=  \del_0 T^{00} + \del_z T^{z0}
\sim \del_0 \varepsilon+(\barepsilon+\barP) \del_z u^z~,
\label{eq:conservation_0} \\
0 &= \del_0 T^{0z} + \del_z T^{zz}~.
\label{eq:conservation_z}
\end{align}Subtracting the time derivative of \eq{conservation_0} from the $z$-derivative of \eq{conservation_z} gives
\begin{align}
0 &= - \del_0^2 T^{00} + \del_z^2 T^{zz} \\
&= -\del_0^2 \varepsilon + \del_z^2 P - \left( \frac{4}{3}\eta+\zeta \right) \del_z^3 u^z \\
&= -\del_0^2 \varepsilon + \del_z^2 P 
+ \frac{\frac{4}{3}\eta+\zeta}{\barepsilon+\barP} \del_0 \del_z^2 \varepsilon~,
%
\end{align}
where we used \eq{conservation_0} in the last expression. Using the speed of sound\index{speed of sound} $c_s$, $\del_z^2 P$ is written as $\del_z^2 P =(\del P/\del \varepsilon) \del_z^2 \varepsilon =: c_s^2 \del_z^2 \varepsilon$. Then, in momentum space, one gets
\begin{align}
&\omega^2+i \Gamma_\text{s} \omega q^2 -c_s^2 q^2 = 0~,
\label{eq:sound_FT} \\
&\Gamma_\text{s} := \frac{1}{\barepsilon+\barP} \left( \frac{4}{3}\eta+\zeta \right)~.
\end{align}
Rewrite \eq{sound_FT} in the form of the dispersion relation $\omega=\ldots$. In hydrodynamics, we consider the derivative expansion, and the equation of motion for the viscous fluid contains at most two derivatives. So, we keep terms $O(q^2)$ in the dispersion relation:
\be
\omega = \pm c_s q - \frac{i}{2} \Gamma_\text{s} q^2 + O(q^3)~.
\ee
The first term represents a pair of sound waves which propagate with velocity $c_s$. The second term takes the same form as \eq{diffusion_dispersion} and represents the damping of sound waves. The coefficient is called the \keyword{sound attenuation constant}. Unlike the vector mode, the sound wave damping depends both on $\eta$ and $\zeta$.

Figure~\ref{fig:hydro_mode_summary} summarizes transport coefficients one can derive from each mode. This table does not exhaust all possibilities to determine transport coefficients though. (For example, as mentioned in \sect{viscosity_kubo}, there is a Kubo formula for $\zeta$ which is a scalar mode computation.)

\begin{figure}[tb]
\begin{center}
\begin{tabular}{|l|l|l|}
\hline
		& $T^{\mu\nu}$							& $J^\mu$ \\
\hline
tensor	& $\eta$ (Kubo, \sect{viscosity_kubo})		& $-$ \\
vector	& $\eta$ (pole, \sect{linearized})			& $\sigma$ (Kubo, \sect{diffusion}) \\
scalar	& $\eta, \zeta, c_s$ (pole, \sect{linearized})	& $D$ (pole,  \sect{diffusion}) \\
\hline
\end{tabular}
\caption{Tensor decomposition of conserved quantities and transport coefficients one can derive from each mode. ``Kubo" and ``pole" indicate the methods used to derive transport coefficients.}
\label{fig:hydro_mode_summary}
\end{center}
\end{figure}

\section{\titlesummary}

\begin{itemize}
\item 
The linear response theory provides the microscopic description of nonequilibrium physics (at linear level). According to the linear response theory, a response of a system is represented by a retarded Green's function.
\item 
On the other hand, hydrodynamics provides the macroscopic description of nonequilibrium physics. According to hydrodynamics, we do not have to know the full Green's function in order to know the response. It is enough to know transport coefficients.
\item 
The shear viscosity arises in the derivative expansion of the fluid energy-momentum tensor $T^{\mu\nu}$. In order to know the shear viscosity, it is convenient to couple gravity $h^{(0)}_{\mu\nu}$ to the fluid.
\item
See also the summary of the diffusion problem in \sect{diffusion}.
\end{itemize}
In later chapters, we obtain the transport coefficients of strongly-coupled gauge theories using knowledge 
we obtained here. But before we do so, we now discuss AdS/CFT in nonequilibrium situations. The philosophy of the linear response theory will be useful there.

\titlenewterms

\begin{multicols}{2}
\noindent
density matrix\\
linear response theory\\
retarded Green's function\\
response function\\
transport coefficients\\
conductivity\\
Kubo formula\\
hydrodynamics\\
constitutive equation\\
Fick's law\\
diffusion constant\\
relaxation time\\
rest frame\\
velocity field\\
continuity equation\\
Euler equation\\
shear/bulk viscosity\\
Navier-Stokes equation\\
Landau-Lifshitz/Eckart frame\\
projection tensor\\
tensor decomposition\\
kinematic viscosity\\
speed of sound\\
sound attenuation constant
\end{multicols}

\endofsection

\ifx\nameofpaper\undefined 
  \usepackage{macro_natsuume} 
  \def\beginsection{\section*}
  \def\endofsection{\end{document}} 
  \input draft_header.tex
\else 
  \def\beginsection{\chapter}
  \def\endofsection{ } 
\fi

\beginsection{AdS/CFT - non-equilibrium
}\label{chap:GKPW}


\begin{quote}
The GKP-Witten relation is the most important equation to apply AdS/CFT to nonequilibrium physics. We explain the relation using simple examples. 
\end{quote}

\section{GKP-Witten relation
}

\begin{figure}[tb]
\centering
\scalebox{0.75}{ \includegraphics{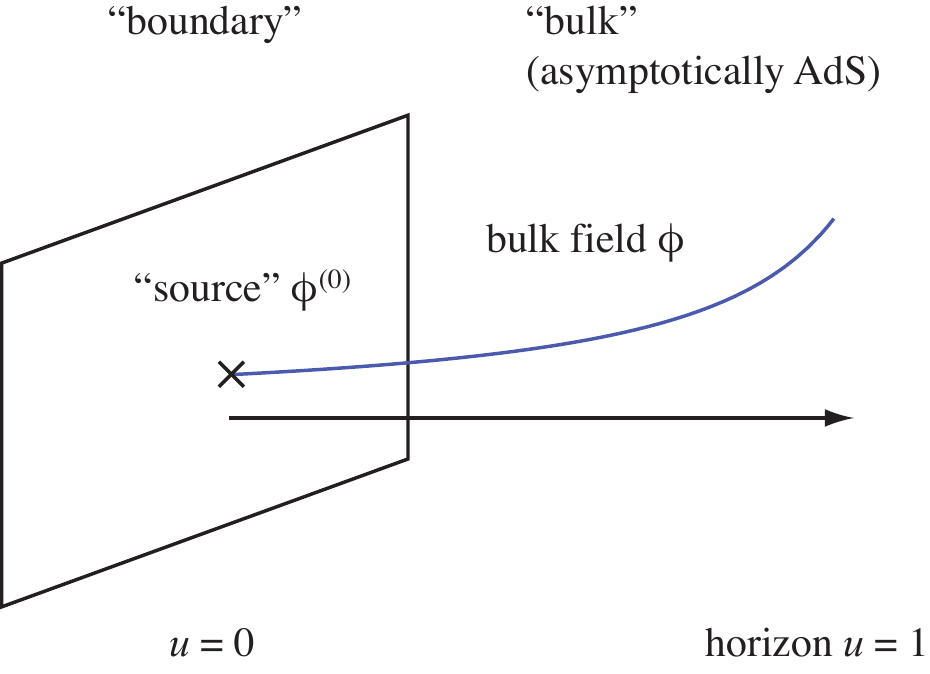} }
\caption{This figure illustrates some of the conventions used for the GKP-Witten relation.}
\label{fig:GKP-Witten}
\end{figure}

AdS/CFT claims the equivalence
\be
Z_\text{gauge} = Z_\text{AdS}~.
%
\ee
In nonequilibrium situations, one can use the following relation (\keyword{GKP-Witten relation} in a narrow sense) \cite{Gubser:1998bc2,Witten:1998qj2}%
\footnote{The GKP-Witten relation is actually formulated as the Euclidean relation. Here, we are interested in dynamics, so we use the Lorentzian GKP-Witten relation. There exist several important differences between the Euclidean relation and the Lorentzian relation. For example, there is a difference in the boundary condition at the black hole horizon (\sect{scalar}). Also, one should not take the Lorentzian relation too literally (\sect{Lorentzian_prescription}).}:
\be
\boxeq{
\left\bra \exp \left(i\int
\source \calO \right) \right\ket  
= e^{i\Sos[\phi|_{u=0}=\source]}~.
}
\label{eq:GKP-W}
\ee
Here,
\begin{itemize}

\item As in previous chapters, the left-hand side represents a four-dimensional gauge theory (\keyword{boundary theory}), and the right-hand side represents a five-dimensional gravitational theory (\keyword{bulk theory}).

\item $\phi$ represents a particular field in the gravitational theory, and $\calO$ represents a particular operator in the gauge theory. $\phi$ and $\calO$ are written only schematically; some explicit examples are discussed below.

\item The AdS boundary (\sect{AdS_coordinates}) is located at $u=0$ in an appropriate coordinate system.

\end{itemize}
We explain the other conventions below (see also \fig{GKP-Witten}). Let us look at both sides of the relation carefully to understand the relation.

\subsubsection*{Left-hand side (gauge theory)
}\label{sec:GKP-Witten_left}

The left-hand side of the GKP-Witten relation is
\be
%
\text{(LHS)} = \left\bra \exp \left(i\int \source \calO \right) \right\ket~,
\label{eq:lhs}
\ee
where $\bra \cdots \ket$ denotes an ensemble average. This is the generating functional of a four-dimensional field theory when an external source $\source$ is added. This takes the same form as \eq{perturb} in the linear response theory. As we saw in the linear response theory, if one can compute this left-hand side, one can know transport coefficients, but the actual computation is difficult in general at strong coupling. In AdS/CFT, we evaluate this using the right-hand side of the GKP-Witten relation, a gravitational theory.

\subsubsection*{Right-hand side (gravitational theory)
}

The right-hand side of the GKP-Witten relation,
\be
%
\text{(RHS)} = e^{i\Sos[\phi|_{u=0}=\source]}
\ee
is the generating functional of a five-dimensional gravitational theory. We use the saddle-point approximation and approximate the generating functional by the classical solution of the gravitational theory. The procedure is similar to the computation of black hole thermodynamic quantities in \sect{Euclidean}. From the gauge theory point of view, the saddle-point approximation corresponds to the large-$\Nc$ limit (\chap{string}). 

Thus, as in the computation of thermodynamic quantities,
\begin{itemize}

\item $\Sos$ represents the on-shell action. Namely, one solves the equation of motion for the bulk field%
\footnote{The bulk field $\phi$ in the GKP-Witten relation is not just a scalar field like \eq{matter_action} but represents bulk fields in the five-dimensional gravitational theory collectively. Incidentally, if $\phi$ is not constant as $u\rightarrow 0$ but behaves as $\phi \sim u^{\Delta_-}$, one defines $\source$ as $\phi|_{u=0} = \source u^{\Delta_-}$ (see the massive scalar field example below).}
$\phi$ under the boundary condition at the AdS boundary $\phi|_{u=0}=\source$. $\Sos$ is obtained by substituting the solution to the action. 

\item The bulk field $\phi$ satisfies the equation of motion, so the on-shell action reduces to a surface term on the AdS boundary. In this way, we obtain a four-dimensional quantity from the five-dimensional quantity. This surface term is identified as the generating functional of the gauge theory.

\end{itemize}
In the above sense, the generating functional of the gauge theory is defined at the AdS boundary. One loosely calls it that the gauge theory ``lives" on the boundary.

From the four-dimensional point of view, $\source$ is an external source, but from the five-dimensional point of view, $\phi$ is a field propagating in the five-dimensional spacetime. Namely, AdS/CFT claims that an external source of the field theory can have a five-dimensional origin, which is an important point of the GKP-Witten relation. Or one could say that 
\begin{center}
\fbox{
\begin{tabular}{l}
A bulk field acts as an external source of a boundary operator.\end{tabular}
}
\end{center}

We would like to obtain the generating functional of a gauge theory, and the GKP-Witten relation claims that it can be obtained by evaluating the classical action of a gravitational theory. Then, for example, the standard one-point function is obtained from the GKP-Witten relation as
\be
\bra \calO  \ketnoj = \left.\frac{\delta \Sos[\source]}{\delta \source}\right|_{\source=0}~.
%
\ee
But in the linear response theory, we are also interested in the response of a system. In such a case, our interest is the one-point function in the presence of the external source: 
\be
\bra \calO  \ketj = \frac{\delta \Sos[\source]}{\delta \source}~.
\label{eq:GKPW_1pt}
\ee
Below we compute this quantity in various examples.

\subsubsection*{The form of the gravitational theory}

From the discussion in \chap{string}, we use the five-dimensional general relativity with the negative cosmological constant to evaluate the on-shell action%
\footnote{\advanced In addition to the bulk action \eqref{eq:EH}, one generally needs to take into account appropriate boundary actions such as the Gibbons-Hawking action and the counterterm actions (both for gravity and for matter fields) as we saw in \sect{SAdS_free}. 
}:
\be
\action = \frac{1}{16\pi G_5} \int d^{5}x\, \sqrt{-g} \left( R - 2\Lambda \right) + \action_\text{matter}~,
\label{eq:EH}
\ee
where the matter action $\action_\text{matter}$ typically includes the Maxwell field and the scalar field%
\footnote{We choose the dimensions of matter fields as $[A_M] = \text{L}^{-1} = \text{M}$ and $[\phi] = \text{L}^0 = \text{M}^0$ so that the dimensions here coincide with the scaling dimensions which appear later.}
\be
\action_\text{matter} =  \frac{1}{16\pi G_5}\int d^5x\, \sqrt{-g} \left\{ - \frac{L^2}{4} F_{MN}^2 - \frac{1}{2}(\nabla_M\phi)^2 - V(\phi)  \right\}~.
\label{eq:matter_action}
\ee
What kind of five-dimensional bulk matter fields actually appear depends on the four-dimensional boundary gauge theory we consider. In order to know the details, one generally needs the knowledge of string theory and D-branes (\chap{string}), but we see several examples in \chap{others}.


There are various asymptotically AdS spacetimes as solutions to such an action. In particular, we often have in mind the Schwarzschild-AdS or SAdS black hole:
\begin{align}
ds_5^2 &= \left(\frac{r_0}{L}\right)^2\frac{1}{u^2} (-hdt^2+d\bmx_3^2)+L^2\frac{du^2}{h u^2}~, \quad
h = 1- u^4~, 
\label{eq:sads5_u_ch10} \\
\text{(matter)}&= 0~.
\end{align}
The AdS boundary is located at $u=0$, and the horizon is located at $u=1$. But for now, we focus only on the asymptotic behavior:
\be
ds_5^2 \sim \frac{1}{u^2} (-dt^2+d\bmx_3^2+du^2)~, 
\quad (u\rightarrow0)~.
\label{eq:asymptotic_ads2}
\ee
For simplicity, we set $r_0=L=16\pi G_5=1$ here and for the rest of this chapter. The symbol ``$\sim$" denotes an equality which is valid asymptotically. Then, the computation goes as follows:
%
%
\begin{enumerate}

\item Consider such a spacetime as the background and add a perturbation.
\item The perturbation can be gravitational perturbations or matter perturbations. Choose a perturbation of a bulk field $\phi$ corresponding to the boundary operator $\calO$ which we would like to compute.
\item Once we obtained the classical solution of the perturbation, obtain the on-shell action by substituting the solution into the action.

\end{enumerate}
We now consider a few simple examples.

\section{Example: scalar field}\label{sec:scalar}



\danger{This section is the typical example of the AdS/CFT computation. After all, computations in AdS/CFT are more or less similar to this example.}

\subsubsection*{Correspondence between a bulk field and a boundary operator
}

As the simplest example of a bulk field, let us consider the massless scalar field%
\footnote{It is enough to consider the scalar action only for the discussion below because the Einstein-Hilbert action is independent of the scalar field. 
}:
\be
\action = -\frac{1}{2} \int d^5x \sqrt{-g} (\nabla_M\phi)^2~.
\label{eq:scalar_action}
\ee
This is just an illustrative example, so we do not explicitly specify the boundary operator which couples to the scalar. We simply denote the operator as $\calO$. 

We evaluate the on-shell action of \eq{scalar_action}, but here we focus on the asymptotic behavior at the AdS boundary $u\rightarrow 0$. So, it is enough to use the metric \eqref{eq:asymptotic_ads2}. Also, for simplicity, we consider a static homogeneous solution $\phi=\phi(u)$ along the boundary directions. Using $\sqrt{-g} \sim u^{-5}$, $g^{uu} \sim u^2$, and integrating \eq{scalar_action} by parts, we get
\begin{align}
\action &\sim  \int d^4x\, du\,-\frac{1}{2u^3} \phi'^2 
\qquad (':= \partial_u)
\\
&= \int d^4x\, \int_0^1 du\, \left( -\frac{1}{2u^3} \phi\phi' \right)' 
   + \left( \frac{1}{2u^3} \phi' \right)' \phi 
\\
&= \int d^4x\, \left. \frac{1}{2u^3}\phi\phi'\right|_{u=0} 
   + \int d^4x\, du\, \left( \frac{1}{2u^3} \phi' \right)' \phi~.
\label{eq:scalar_action2}
%
\end{align}
The second term of \eq{scalar_action2} is simply the equation of motion:
\be
\left( \frac{1}{2u^3} \phi' \right)' \sim 0~.
\label{eq:scalar_eom_asymptotic}
\ee
The equation of motion is second order in derivatives, so there are two independent solutions. Their asymptotic forms are 
\be
\phi \sim \source\left(1+\vev u^4\right)~,
\quad (u\rightarrow0)~.
\label{eq:massless_scalar_falloff}
\ee
One can check this easily by substituting the solution into \eq{scalar_eom_asymptotic}.

Using the equation of motion, the action \eqref{eq:scalar_action2} reduces to a surface term on the AdS boundary:
\be
\Sos \sim  \int d^4x\, \left. \frac{1}{2u^3}\phi\phi'\right|_{u=0}~.
\ee
Thus, substituting the asymptotic form of the solution \eqref{eq:massless_scalar_falloff}, we get
\be
\Sos[\source] = \int d^4x\, 2\phi^{(0)\,2}\vev~.
\ee
Then, the one-point function is obtained from the GKP-Witten relation as
\be
\bra \calO  \ketj = \frac{\delta \Sos[\source]}{\delta \source} = 4\vev \source~.
\label{eq:kw}
\ee

To summarize our result,
\be
%
\phi \sim \source + \frac{1}{4} \bra \calO \ketj\, u^4~,
\quad (u\rightarrow0)
\label{eq:faster_op}
\ee
or%
\footnote{\advanced More precisely, the fast falloff means a normalizable mode. A normalizable mode can be quantized, whereas a non-normalizable mode cannot be quantized and should be regarded as an external source. Then, there are cases where even the slow falloff is a normalizable mode. This indeed happens, and one can exchange the role of the external source and the operator in such a case (\sect{massive_more}). One should take this into account for the field/operator correspondence to really work.}
%
%
\be
\fbox{$
\begin{aligned}
\text{slow falloff of bulk field} & \rightarrow & \text{source} \\
\text{fast falloff of bulk field} & \rightarrow & \text{response}
\end{aligned}
$}
\label{eq:bulk_bdy_mapping}
\ee
This is the scalar field example, but a similar relation holds for the Maxwell field and the gravitational field. The procedure is more complicated but is the same%
\footnote{\advanced ``The fast falloff as the response" is a useful phrase to remember, but  it is not always true.
In some cases, additional terms in the action may modify the relation. In such a case, one should go back to the GKP-Witten relation (see, \eg, Sects.~\ref{sec:r-charged} and \ref{sec:action_tensor}).}.

At this stage, the $O(1)$ and $O(u^4)$ terms are independent solutions. But they should be related by a linear response relation. In AdS/CFT, this arises by imposing a boundary condition inside the bulk spacetime (on the black hole horizon).
We discuss the boundary condition below, but after we impose a boundary condition, $\vev$ is uniquely determined, or  the $O(1)$ and $O(u^4)$ terms are uniquely determined up to an overall coefficient $\source$ for a linear perturbation.  This is the reason why we write the asymptotic form as \eq{massless_scalar_falloff}. 

Then, \eq{kw} is nothing but the linear response relation \eqref{eq:linear_response:real}:
\be
\delta \bra \calO(t, \bmx) \ket 
= - \int^{\infty}_{-\infty} d^4x'\, 
G_R^{\calO\calO}(t-t', \bmx-\bmx') \source(t',\bmx')~.
\label{eq:linear_response:real_again}
\ee
Comparing Eqs.~\eqref{eq:kw} and \eqref{eq:linear_response:real_again}, one can see that AdS/CFT determines the retarded Green's function $G_R$ as%
\footnote{In many examples below, $\bra\calO\ketnoj = 0$, so $\delta\bra\calO\ket = \bra\calO\ketj - \bra\calO\ketnoj = \bra\calO\ketj$.}
\be
G_R^{\calO\calO}(k=0) = -4\vev~.
%
\ee
This is just an example, so we do not really compute $\vev$ here%
\footnote{\advanced This problem is actually trivial in the sense that $\vev=0$ for the static homogeneous perturbation from the conformal invariance. One can show this by computing $\vev$ in the SAdS black hole background with the regularity condition at the horizon.},
but we will essentially compute it for the time-dependent case in \sect{tensor_mode_sol}. 
Note that we need to find a solution throughout the bulk spacetime. What we need is a quantity on the AdS boundary, but we need to solve the equation of motion in the entire bulk spacetime since we have to impose a boundary condition inside the bulk spacetime. This is typical in AdS/CFT.

\subsubsection*{Scaling dimension
}

The AdS spacetime has the scale invariance under
\be
x^\mu \rightarrow a x^\mu, \quad 
u \rightarrow a u~. 
\label{eq:scale_inv_again}
\ee
If a quantity $\Phi$ transforms as
\be
\Phi \rightarrow a^{-\Delta}\Phi~,
%
\ee
under the scaling, we call that the quantity has \keyword{scaling dimension} $\Delta$. 
From the boundary point of view, \eq{scale_inv_again} is the scale transformation \eqref{eq:scale_transf1} of a four-dimensional field theory, and the definition of the scaling dimension here coincides with the field theory one \eqref{eq:conf_dim}. 

From the bulk point of view, the scale transformation is just a coordinate transformation, so the bulk field $\phi$ is invariant under the scale transformation. Then, from \eq{faster_op},
%
%
\be
\boxeq{
\begin{aligned}
\source &\rightarrow \source: & \source \text{ has scaling dimension 0~,} \\
\bra \calO \ketj &\rightarrow a^{-4} \bra \calO \ketj: & \bra \calO \ketj \text{ has scaling dimension 4~.}
\end{aligned}
}
\label{eq:conf_dim_massless}
\ee
Then, the perturbed action of the boundary theory is invariant under the scale transformation:
\begin{alignat}{2}
\delta \action_\text{QFT} = \int & d^4x\, & \source & \calO \\
		& \downarrow & \downarrow\hspace{2mm}&  \downarrow \nonumber \\
		& a^{4}	& a^{0}\,			& a^{-4} \nonumber 
\end{alignat}
The gravitational field $h^{(0)}_{\mu\nu}$ and the energy-momentum tensor $T^{\mu\nu}$ have such scaling dimensions (\sect{gravity}).

\subsubsection*{The field/operator correspondence: explicit examples
}\label{sec:dictionary}\index{field/operator correspondence}

What kind of boundary operators actually correspond to the bulk fields? In order to know the correspondence, one in general needs the knowledge of string theory. But here let us focus on the ``universal sector" of the theory that is common to any theory. 

In hydrodynamics, one is primarily interested in conserved quantities such as $T^{\mu\nu}$ and a current $J^\mu$. If a theory has the associated symmetries, any theory has these operators. From the discussion in \chap{hydro}, these operators are the responses under the external sources $h_{\mu\nu}^{(0)}$ and $A_{\mu}^{(0)}$. The corresponding bulk fields are obtained by promoting the external sources into the five-dimensional propagating fields. Therefore, 
\begin{center}
\fbox{
\begin{tabular}{ccccc}
Boundary operators & & external sources & & Bulk fields \\
$T^{\mu\nu}$  & $\leftrightarrow$ & $h_{\mu\nu}^{(0)}$ & $\rightarrow$ & gravitational field $h_{MN}$ \\
$J^\mu$ & $\leftrightarrow$ & $A_{\mu}^{(0)}$ & $\rightarrow$ & Maxwell field $A_M$
\end{tabular}
}
\end{center}

\noindent
The explicit forms of the field/operator correspondence are described below. 

\danger{The Maxwell field $A_M$ in the above table is a five-dimensional bulk $U(1)$ field and is different from a four-dimensional boundary $SU(N_c)$ gauge field $(A_\mu)^i_{~j}$. For example, this Maxwell field may be the external electromagnetic field which couples to the \Nfour\ SYM.}

\subsubsection*{Boundary condition at horizon}

We impose a boundary condition inside the bulk spacetime. For a time-independent perturbation, we impose the regularity condition at the horizon. We add a perturbation in a background geometry, so the effect of the perturbation should remain small and should not affect the geometry. As a result, the energy-momentum tensor $T_{MN}$ of the perturbation must be finite. (It is more appropriate to use coordinate-invariant quantities such as the trace.)

For a time-dependent perturbation, we impose the so-called \textit{``incoming-wave" boundary condition}\index{incoming-wave boundary condition} at the horizon or perturbations are only absorbed by the horizon. The boundary condition is physically natural since nothing comes out from the horizon%
\footnote{\advanced This boundary condition is a difference from the Euclidean formalism. In the Euclidean formalism, there is no region inside the ``horizon," so one does not impose the incoming-wave boundary condition.}. 

This boundary condition is the origin of the dissipation. General relativity itself is invariant under the time-reversal, but a dissipation is not a time-reversal process. What breaks the time-reversal invariance is this boundary condition.
See \sect{tensor_mode_sol} for an explicit computation.

\section{Other examples}\label{sec:other_examples}

\subsection{Maxwell field}\label{sec:current}



As another example, let us consider the boundary theory with a $U(1)$ current $J^\mu$. At this stage, we do not specify the current explicitly. The current may be the usual electromagnetic current, but it is not restricted to the electromagnetic current (\sect{diffusion}). In any case, $\rho :=\bra J^0 \ket$ is a number density or a charge density. Let $\mu$ as the chemical potential which is conjugate to $\rho$. 

From the above discussion, such a case corresponds to adding a bulk Maxwell field $A_0$. Then, we now evaluate the action
\be
\action =  -\frac{1}{4} \int d^5x\, \sqrt{-g} F_{MN}^2~.
\ee
For simplicity, consider $A_0=A_0(u)$. The Lagrangian becomes $\sqrt{-g}g^{00}g^{uu} (F_{u0})^2 \sim -u^{-5}u^4 A_0'^2 = -u^{-1} A_0'^2$, so the equation of motion is
\be
(u^{-1} A_0')' \sim 0~.
%
\ee
(For the SAdS$_5$ black hole, this equation of motion is actually valid for all $u$.) Then, the asymptotic behavior is given by
\be
A_0 \sim \sourceA{0}\left(1+\vevA{0} u^2\right)~,
\quad (u\rightarrow0) 
\label{eq:gauge_falloff} 
\ee
instead of \eq{massless_scalar_falloff}. The physical meaning of the behavior is clear. The slow falloff is constant because of the gauge invariance. (The gauge potential appears with a derivative in $F_{MN}$.) The fast falloff represents Coulomb's law in the bulk five-dimensional spacetime. 

As in the scalar field example, the evaluation of the on-shell Maxwell action gives the relation such as 
\begin{align}
\mu &= \sourceA{0}~, \\
\bra J^0 \ketj &= -\cA \vevA{0} \sourceA{0}~,
\label{eq:response_rho_ads}
\end{align}
where $\cA$ is an appropriate factor which can be obtained by actually evaluating the on-shell action (see \probset{fieldop_Maxwell} and \sect{SAdS5_diff} for the explicit expression for the \Nfour\ SYM). The latter equation is the linear response relation \eqref{eq:response_rho} for $J^0$. 


A similar relation holds for $A_x$:
\be
\bra J^x \ketj = \cA \vevA{x} \sourceA{x}~,
\label{eq:response_current_ads}
\ee
which corresponds to \eq{response_current}. But we saw in \sect{conductivity} that the conductivity $\sigma$ is given by Ohm's law:
\be
\bra J^x \ketj = \sigma \sourceE = i\omega \sigma \sourceA{x}~.
%
\ee
Thus,
\be
\sigma(\omega) = \cA \frac{\vevA{x}}{i\omega}~.
\ee
Then, the conductivity is obtained by solving the bulk $A_x$ equation of motion and extracting $\vevA{x}$. For the holographic superconductor (\sect{H^3}), one shows the divergence of the conductivity from such a computation.

\head{Scaling dimension}
The Maxwell field is a one-form, so under the scale transformation \eqref{eq:scale_inv_again}, it transforms as $A_M(x,u) \rightarrow A_M/a$. Namely, the Maxwell field has scaling dimension 1. Then, following the scalar field example, one gets
\be
\begin{split}
\sourceA{\mu}(x) &\rightarrow a^{-1} \sourceA{\mu}~, \\
\bra J^\mu(x) \ketj &\rightarrow a^{-3} \bra J^\mu \ketj~.
\end{split}
\ee
Namely, the current has scaling dimension 3. This scaling dimension coincides with the naive mass dimension of a current. (In the units $c=\hbar=1$, the charge is dimensionless.)

\subsubsection*{Remark on chemical potential}

The chemical potential is given by $\mu = \left. A_0 \right|_{u=0}$. Of course, the constant of $A_0$ itself is not meaningful due to the gauge invariance. What is physical is the difference of the gauge potential (\sect{RN}). Or the relation itself is not gauge invariant. In AdS/CFT, one normally chooses the gauge $A_0=0$ at the \bh horizon $u=1$. Then, 
\be
\mu = \left. A_0 \right|_{u=0} - \left. A_0 \right|_{u=1}
%
\ee
reduces to the above relation.

\subsection{Massive scalar field}\label{sec:massive_scalar}


For a massive scalar field,
\be
\action_\text{bulk} = -\frac{1}{2} \int d^{p+2}x\, \sqrt{-g} \{ (\nabla_M\phi)^2 + m^2 \phi^2 \}~.
\ee
A scale-invariant four-dimensional boundary theory cannot have a mass term, but it is perfectly fine for the dual five-dimensional bulk field to have a mass term. A five-dimensional mass corresponds to adding an operator with a different scaling dimension in the four-dimensional boundary theory.

The equation of motion is given by
\be
\frac{1}{\sqrt{-g}} \del_M \left(\sqrt{-g} g^{MN}\del_N \phi \right) - m^2 \phi = 0~.
%
\ee
Solve the equation asymptotically to obtain the asymptotic behavior. Instead of \eq{massless_scalar_falloff}, one obtains
\be
\phi \sim u^\Delta~, \qquad
\Delta(\Delta-p-1)=m^2~,
\label{eq:scalar_falloff}
%
\ee
or
\be
%
\Delta_\pm = \frac{p+1}{2} \pm \sqrt{\frac{(p+1)^2}{4}+m^2}~.
\label{eq:conf_weight_scalar}
\ee
Equation~\eqref{eq:conf_weight_scalar} implies that the theory is stable even for $m^2<0$ if 
\be
m^2 \geq -\frac{(p+1)^2}{4}~, 
\label{eq:BF_bound}
\ee
which is known as the \keyword{Breitenlohner-Freedman bound} \cite{Breitenlohner:1982bm}.

The slow falloff behaves as $\phi \sim u^{\Delta_-}$. So, we define $\source$ as 
\be
\phi \sim \source u^{\Delta_-} + \cphi \bra\calO\ketj\, u^{\Delta_+}~,
\quad (u\rightarrow0)
\label{eq:massive_scalar_falloff}
\ee
where $\cphi$ is an appropriate factor which can be obtained by actually evaluating the on-shell scalar action (see \sect{massive_more} for more details). As in the massless scalar,
\be
\begin{split}
\source &\rightarrow a^{-\Delta_-} \source~, \\
\bra \calO \ketj &\rightarrow a^{-\Delta_+} \bra \calO \ketj~. 
\end{split}
\ee
From \eq{conf_weight_scalar}, $\Delta_++\Delta_-=p+1$, so the boundary action $\delta \action_\text{QFT} = \int d^{p+1}x\, \source \calO$ is again scale invariant. When $p=3$ and $m=0$, $(\Delta_-, \Delta_+) = (0,4)$, which reduces to the massless scalar example.

When $m^2>0$, the slow falloff diverges as $\phi \sim u^{\Delta_-}$ at the AdS boundary. Such a perturbation has a diverging energy-momentum tensor $T_{MN}$ and affects the background geometry, so it cannot be added as a perturbation. From the field theory point of view, $\calO$ is an irrelevant operator whose dimension is $\Delta_+ > p+1$, and its effect grows in the UV.

\subsection{Gravitational field}\label{sec:gravity}

\head{Boundary metric}
The AdS metric takes the form
\be
ds_5^2 = \frac{1}{u^2}(\eta_{\mu\nu} dx^\mu dx^\nu + du^2)~.
%
\ee
Note that the metric in the boundary theory is $\eta_{\mu\nu}$. This is not the bulk metric itself, and they are related to each other by the factor $u^{-2}$. Then, when the bulk metric takes a more general form
\be
ds_5^2 = g_{\mu\nu} dx^\mu dx^\nu + \frac{du^2}{u^2}~,
%
\ee
it is natural to define the boundary metric $g^{(0)}_{\mu\nu}$ as $g_{\mu\nu}|_{u=0}=g^{(0)}_{\mu\nu}\, u^{-2}$. This is analogous to the massive scalar case. Since $\phi$ behaves as $\phi \sim u^{\Delta_-}$, we define $\source$ as $\phi|_{u=0} = \source u^{\Delta_-}$.

For the SAdS black hole \eqref{eq:sads5_u_ch10}, $g_{00}$ differs from $g_{ii}$ because of the factor $h$, but the boundary metric remains flat from the above definition of the boundary metric. The subleading term essentially represents the boundary energy-momentum tensor $\bra T^{\mu\nu}\ket$ at equilibrium as described below. 

\head{Fefferman-Graham coordinate}
In order to write down the field/operator correspondence for the gravitational field, it is convenient to use the \keyword{Fefferman-Graham coordinate} instead of the SAdS coordinate. In the Fefferman-Graham coordinate, $g_{\tilu\tilu}=1/\tilu^2$. In the SAdS coordinate, $g_{uu} \neq 1/u^2$, so we rewrite the solution in $\zfg$. The transformation is given by
\be
u^2 = \frac{ 2\zfg^2/\zfg_0^2 }{ 1+\zfg^4/\zfg_0^4 }~.
%
\ee
Then, the metric becomes
\be
ds_5^2 = \frac{1}{\zfg^2} \left[ 
- \frac{ ( 1-\zfg^4/\zfg_0^4 )^2 }{ 1+\zfg^4/\zfg_0^4 } dt^2 
+ (1+ \zfg^4/\zfg_0^4 ) d\bmx_3^2 + d\zfg^2 \right]~,
%
\ee
where the horizon is located at $\zfg=\zfg_0=\sqrt{2}/(\pi T)$. The metric asymptotically behaves as
\be
ds_5^2 \sim \frac{1}{\zfg^2} \left[ 
(-1 + 3 \zfg^4/\zfg_0^4) dt^2 
+ (1+ \zfg^4/\zfg_0^4 ) d\bmx_3^2 + d\zfg^2 \right]~, \quad
(\zfg\rightarrow0)~.
\label{eq:sads5_FG_asymp}
\ee

\head{The field/operator correspondence}
In this subsection, we use the Fefferman-Graham coordinate because the field/operator correspondence for the gravitational field takes a particularly simple form in this coordinate:
\be
g_{\mu\nu} \sim \frac{1}{\tilu^2} \left( \eta^{(0)}_{\mu\nu} + 4\pi G_5 \bra T^{\mu\nu}\ketnoj\, \tilu^4 \right)~, 
\quad (\tilu\rightarrow0)~,
\label{eq:field_op_grav}
\ee
where we recovered $G_5$. 

In the SAdS coordinate $u$, the field/operator correspondence is modified. The coordinates $\zfg$ and $u$ are related by $\zfg = u \{ 1+O(u^4) \}$. They differ only at subleading order, so the definition of the slow falloff is not affected, but the subleading term, the fast falloff, is affected.

The bulk metric $g_{\mu\nu}$ is a tensor with 2 lower indices, so it transforms as $g_{\mu\nu} \rightarrow g_{\mu\nu}/a^2$ under the scale transformation \eqref{eq:scale_inv_again}. 
Since the factor $\tilu^{-2}$ scales in the same way, 
\be
\begin{split}
\eta^{(0)}_{\mu\nu} &\rightarrow \eta^{(0)}_{\mu\nu}~, \\
\bra T^{\mu\nu} \ketnoj &\rightarrow a^{-4}  \bra T^{\mu\nu} \ketnoj~.
\end{split}
\ee
Namely, $\bra T^{\mu\nu} \ketnoj$ has scaling dimension 4, which coincides with the naive mass dimension of the energy-momentum tensor in $(3+1)$-dimensions. 

We computed thermodynamics quantities in \chap{AdS_BH}, but the field/operator correspondence \eqref{eq:field_op_grav} gives a simple way to evaluate thermodynamic quantities. From the asymptotic behavior \eqref{eq:sads5_FG_asymp}, 
\be
\bra T^{\mu\nu} \ketnoj 
= \frac{1}{4\pi G_5}\frac{1}{\zfg_0^4}\, \text{diag} (3,1,1,1) 
= \frac{\pi^2}{8} \Nc^2 T^4 \text{diag} (3,1,1,1)~,
%
\ee
which agrees with the result in \chap{AdS_BH}. 

So far, we considered the case where the boundary metric remains flat. Similar results hold when we add a boundary metric perturbation. For example, add the tensor mode $h^{(0)}_{xy}$. The asymptotic behavior of the bulk perturbation $h_{xy}$ is given by
\be
h_{xy} \sim \frac{1}{\tilu^2} \left( h^{(0)}_{xy} + 4\pi G_5 \bra T^{xy}\ketj\, \tilu^4 \right)~,
\quad (\tilu\rightarrow0)~,
%
\ee
in the Fefferman-Graham coordinate. In \chap{QGP}, we will essentially derive this result in the SAdS coordinate $u$.

\section{On the Lorentzian prescription
\advanced}\label{sec:Lorentzian_prescription}

The GKP-Witten relation is actually formulated as the Euclidean relation. One encounters troubles if one takes the Lorentzian relation too literally. Here, we discuss these problems and gives the prescription to get correct results. This issue is not problematic for a static perturbation like \sect{scalar} but is problematic for a time-dependent perturbation.

Again, consider the massless scalar field as an example:
\be
\action = -\frac{1}{2} \int d^5x \sqrt{-g} (\nabla_M\phi)^2~.
%
\ee
We consider the scalar in the SAdS$_5$ black hole but this time not only asymptotically.
%
%
We consider an inhomogeneous perturbation along the boundary directions, and it is convenient to make the Fourier transformation for the boundary directions $x^\mu=(t,\bmx)$:
\be
\phi (t, \bmx, u)
  = \int (dk)\,
  e^{-i \omega t + i \bmq \cdot \bmx} \phip(u)~, 
\quad (dk):=\frac{d^4k}{(2\pi)^4}~.
%
\ee
A computation similar to \sect{scalar} leads to
\begin{align}
%
\action &= -\frac{1}{2} \int d^4x\, du\,\sqrt{-g}\{ g^{\mu\nu} \del_\mu\phi\del_\nu\phi + g^{uu}\phi'^2 \}
\\
&= -\frac{1}{2} \int d^4x\, du \int (dk)(dk')\, u^{-5} \{ g^{\mu\nu}(ik'_\mu)(ik_\nu)\phi_{k'}\phip 
+ hu^2  \phi_{k'}'\phip' \} \nonumber \\
& \times e^{i(k+k')\cdot x}
\\
&= \int (dk) \int_0^1 du\,  \left\{ - \frac{h}{2u^3}  \phim'\phip' - \frac{g^{\mu\nu}k_\mu k_\nu}{2u^5}  \phim\phip \right\}
\\
&= \int (dk) \int_0^1 du\,  \left\{ \left(- \frac{h}{2u^3}  \phim\phip'\right)' + \frac{1}{2} (\text{EOM}) \phim \right\}~.
\label{eq:scalar_action_k}
\end{align}
The equation of motion is given by
\be
\left(\frac{h}{u^3} \phip' \right)' - \frac{g^{\mu\nu}k_\mu k_\nu}{u^5}  \phip = 0~.
%
\ee
Asymptotically, it behaves as
\be
u^3 \left(\frac{1}{u^3} \phip' \right)' - k^2  \phip \sim 0~,
%
\ee
where $k^2 = \eta^{\mu\nu} k_\mu k_\nu$. Again, set $\phip \sim u^\Delta$. Then,
\be
\Delta (\Delta-4) u^{\Delta-2} -k^2 u^\Delta \sim 0~.
%
\ee
In this case, there is a $k^2$-dependence, but it is subleading in $u$ and can be ignored. Thus, the asymptotic form \eqref{eq:massless_scalar_falloff} is unchanged.

From \eq{scalar_action_k}, we obtain the on-shell action:
\be
%
\Sos = \int (dk)  \left\{ 
\left. \frac{h}{2u^3}  \phim\phip' \right|_{u=0} 
- \left. \frac{h}{2u^3}  \phim\phip' \right|_{u=1} 
\right\}~.
%
\ee
Here, we take into account the contribution from the horizon since the horizon provides a boundary in the Lorentzian formalism unlike the Euclidean formalism. We discuss this contribution further below.
Write the solution as $\phi_k = \source_k f_k(u)$. By definition, $f_k$ satisfies $f_k(u\rightarrow0)=1$. Then,
\begin{align}
%
\Sos[\source] &= \int (dk)\, \phim^{(0)} \left\{
\left. \Fp \right|_{u=0} - \left.\Fp \right|_{u=1} 
\right\} \phip^{(0)}~,
\label{eq:bdy_bc} \\
\Fp(u) &:= \frac{h}{2u^3} f_{-k}f_{k}'~.
\label{eq:def_Fk}
%
\end{align}

According to the GKP-Witten relation, the one-point function is obtained by
\be
\bra \calO_k  \ketj = \frac{\delta \Sos[\source]}{\delta \source_{-k}}~.
%
\ee
Taking the derivative of \eq{bdy_bc} with respect to $\source_{-k}$ gives
\be
%
\bra \calO_k  \ketj \stackrel{?}{=} \left\{
\left. (\Fp+\Fm) \right|_{u=0}
- \left. (\Fp+\Fm) \right|_{u=1} \right\} \source_k~.
\label{eq:Lorentzian_prescription_naive}
\ee
But this is inappropriate as a one-point function. The Fourier-transformed one-point function should be complex in general, but the above expression has no imaginary part:
\begin{enumerate}

\item We take into account the contribution from the horizon since the horizon provides a boundary in the Lorentzian formalism. But one can show that $\left.\Fp \right|_{u=0}^{u=1}$ has no imaginary part since the imaginary part at the horizon cancels out the imaginary part at the AdS boundary%
\footnote{For example, this can be checked by substituting the $O(\omega)$ solution \eqref{eq:perturbative_sol_app1}  of $f_k$ into \eq{def_Fk}. }.

\item Even if one throws away the contribution from the horizon, we still have a problem. Since the field $\phi(t,\bmx,u)$ is real, the Fourier component satisfies $\phip(u)=\phim^*(u)$. Then, $\Fm \simeq f_{k}f_{-k}'=f_{-k}^* f_{k}'^* \simeq \Fp^*$, so $\Fp+\Fm = \Fp+\Fp^* = 2 \text{Re}\, \Fp$, and the one-point function still lacks the imaginary part.

\end{enumerate}
These problems come from the fact that we use the Lorentzian GKP-Witten relation too literally which was originally defined as the Euclidean relation. In order to obtain the correct one-point function, use the following prescription \cite{Son:2002sd}:
\be
\boxeq{
\bra \calO_k  \ketj = \left. 2\Fp \right|_{u=0} \source_k~.
}
\label{eq:Lorentzian_prescription}
\ee
The prescription is justified using the real-time like formalism of finite-temperature field theory \cite{Herzog:2002pc}. We discussed only the one-point function, but the Lorentzian prescription of the $n$-point function has been known as well \cite{Skenderis:2008dh}.

\section{\titlesummary}

\begin{itemize}
\item 
The GKP-Witten relation is the most important relation in AdS/CFT. The relation relates the generating functional of a four-dimensional gauge theory and the generating functional of a five-dimensional gravitational theory.
\item
A boundary operator has a corresponding bulk field. For example, any boundary theory with appropriate symmetries has $T^{\mu\nu}$ and $J^\mu$. The corresponding bulk fields are the gravitational field $h_{MN}$ and the Maxwell field $A_M$.
\item 
A bulk field $\phi$ has a slow falloff and a fast falloff as independent solutions. The slow falloff represents $\source$, the external source of an operator $\calO$ in the boundary theory, and the fast falloff represents $\bra\calO\ketj$, its response. 
\item
The falloffs also determine the scaling dimensions of the external source and the operator.
\end{itemize}

\titlenewterms

\begin{multicols}{2}
\noindent
bulk/boundary theory\\
slow/fast falloff\\
field/operator correspondence\\
incoming-wave boundary condition\\
Breitenlohner-Freedman (BF) bound\\
\newtermapp{standard/alternative quantization}
\end{multicols}

\section{Appendix: More about massive scalars \advanced}\label{sec:massive_more}

\head{Holographic renormalization}
A holographic renormalization is necessary to derive the field/operator correspondence for a massive scalar field. As an example, we consider the case $p=2$ and $m^2=-2$, or $(\Delta_-, \Delta_+)=(1,2)$. 

The counterterm action is given by
\be
\action_\text{CT} = - \frac{\Delta_-}{2} \int d^{p+1}x\, \sqrt{-\gamma}\, \phi^2~. 
%
\ee
The on-shell action becomes
\be
\Sos 
= \underline{\action_\text{bulk}} + \underline{\action_\text{CT}}
= \left. \int d^3x\, \left(\frac{1}{2u^2}\phi\phi' - \frac{1}{2u^3}\phi^2\right) \right|_{u=0}~. 
%
\ee
The first term is the bulk contribution. Substituting the asymptotic form of the scalar
\be
\phi \sim \source\left(u + \vev u^2 \right)~,
\quad (u\rightarrow0)~,
%
\ee
one gets a finite result:
\be
\Sos = \int d^3x\, \frac{1}{2} \phi^{(0)\,2}\vev~.
%
\ee
Then, the one-point function is given by
\be
\bra \calO  \ketj 
= \frac{\delta \Sos[\source]}{\delta \source} 
= \vev \source~.
%
\ee
The general formula is 
\begin{align}
\bra \calO  \ketj 
&= (\Delta_+ - \Delta_-) \vev \source \\
&= \{ 2\Delta_+ - (p+1) \} \vev \source~.
%
\end{align}
When $p=3$ and $m=0$, one reproduces \eq{kw}, 
$\bra \calO  \ketj = 4\vev \source$.

\begin{figure}[tb]
\centering
\scalebox{0.65}{ \includegraphics{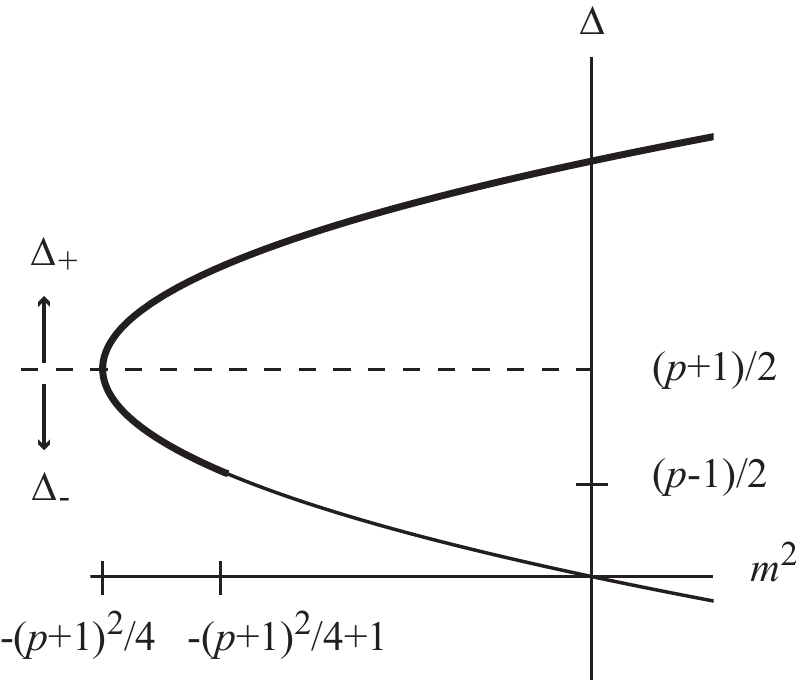} }
\caption{Scaling dimension $\Delta$ and $m^2$ from \eq{conf_weight_scalar}. On the thick curve, the falloffs can be interpreted as operators.}
\label{fig:scaling_dim}
\end{figure}

\head{``Alternative quantization"}
From the scaling dimension \eqref{eq:conf_weight_scalar}, the above procedure gives an operator with dimension $\Delta_+ \geq (p+1)/2$. However, it is known that unitarity allows a local scalar operator with dimension $\Delta \geq (p-1)/2$. The missing operators $(p-1)/2 \leq \Delta < (p+1)/2$ are provided by the slow falloff. When
\be
%
- \frac{(p+1)^2}{4} \leq m^2 \leq - \frac{(p+1)^2}{4}+1~,
%
\ee
the slow falloff is also normalizable so that one can exchange the role of the external source and the operator \cite{Klebanov:1999tb}. The procedure we described so far is known as the ``standard quantization" \index{standard quantization} whereas this procedure is known as the ``alternative quantization" (\fig{scaling_dim}). \index{alternative quantization}

For the alternative quantization, the counterterm action is given by
\be
\action_\text{CT} = \int d^{p+1}x\, \sqrt{-\gamma}\,
\left(\phi n^M \nabla_M \phi + \frac{\Delta_-}{2} \phi^2 \right)~,
%
\ee
where $n^M$ is the unit normal to the boundary, so $g_{MN} n^{M}n^{N}=1$ and $n^u = -1/\sqrt{g_{uu}}$.

Substituting the asymptotic form of the scalar
\be
\phi \sim \vev\left(\source u + u^2 \right)~,
\quad (u\rightarrow0)~,
%
\ee
the on-shell action becomes
\begin{align}
\Sos 
= \underline{\action_\text{bulk}} + \underline{\action_\text{CT}}
&= \left. \int d^3x\, \left(\frac{1}{2u^2} \phi\phi'- \frac{1}{u^2} \phi\phi' +\frac{1}{2u^3} \phi^2 \right) \right|_{u=0} \\
&= -\int d^3x\, \frac{1}{2} \phi^{(1)\,2}\source~.
%
\end{align}
The one-point function is given by
\begin{align}
\bra \calO  \ketj 
&= \frac{\delta \Sos[\vev]}{\delta \vev} \\
& = - \source \vev 
 = -(\Delta_+ - \Delta_-) \source \vev~.
%
\end{align}
The result is just the exchange of the external source and the operator ($\source\leftrightarrow\vev$ and $\Delta_+\leftrightarrow\Delta_-$). 

It is often important to generalize the boundary conditions at the AdS boundary, and this provides one example.

\endofsection

\ifx\nameofpaper\undefined 
  \usepackage{macro_natsuume} 
  \def\beginsection{\section*}
  \def\endofsection{\end{document}} 
  \input draft_header.tex
\else 
  \def\beginsection{\chapter}
  \def\endofsection{ } 
\fi

\beginsection{Other AdS spacetimes}\label{chap:others}


\begin{quote}
We discuss various spacetimes which often appear in AdS/CFT. We describe charged AdS black holes, the Schwarzschild-AdS black hole in the other dimensions, various branes (M-branes and D$p$-branes), and some other examples. Many of them are asymptotically AdS spacetimes, but some are not.
\end{quote}

\section{Overview of other AdS spacetimes
}\label{sec:others}

So far we discussed the AdS$_5$ spacetime and the Schwarzschild-AdS$_5$ (SAdS$_5$) black hole, but various other asymptotically AdS spacetimes appear in the literature. Here, we briefly discuss them. We discuss some of the geometries further in \sect{others_details}. Currently, one cannot directly compute QCD in AdS/CFT. So, one analyzes, \eg, the black hole which is dual to the \Nfour\ SYM, but the results are not the ones for QCD. Therefore, when one computes a physical quantity, the result may be theory-dependent. It is desirable to compute a physical quantity in various theories in order to know the theory dependence.  

There are various approaches below, but in all cases, the symmetry of both the gauge theory and the gravitational theory plays an important role. 

\subsection*{Charged AdS black holes
}

As in the asymptotically flat spacetime, one can consider charged black holes in AdS spacetime. The Einstein-Maxwell theory gives a simple example: 
\be
\action =  \frac{1}{16\pi G_5} \int d^{5}x\, \sqrt{-g} \left(R - 2\Lambda - \frac{L^2}{4}F_{MN}^2 \right)~.
%
\ee
The bulk Maxwell field $A_M$ corresponds to the boundary current $J^\mu$. So, the electric charge case, $A_0 \neq 0$, corresponds to a finite chemical potential or a finite charge density. The solution is known as the \keyword{Reissner-Nordstr\"{o}m AdS black hole} (RN-AdS black hole hereafter).

One generalization of the charged AdS black hole is the \keyword{R-charged black hole} \cite{Behrndt:1998jd,Kraus:1998hv,Cvetic:1999xp}. The R-charged black hole is dual to the \Nfour\ SYM at a finite chemical potential. In this case, the bulk gravitational action is complicated, but it schematically takes the form
\be
{\cal L}_5 =  \text{(Gravity)} + \text{(three $U(1)$ gauge fields)} + \text{(two real scalars)}~.
%
\ee
The \Nfour\ SYM is dual to the AdS$_5 \times S^5$ spacetime. An R-charged black hole is obtained by adding an angular momentum along $S^5$. The sphere $S^5$ has the $SO(6)$ symmetry which is rank three, so one can add at most three independent angular momenta.

If one compactifies $S^5$, one gets Kaluza-Klein gauge fields \index{Kaluza-Klein gauge field} from the compactified metric%
\footnote{See \sect{vector} for the simple $S^1$ compactification.}.
The angular momenta along $S^5$ become the charges under the Kaluza-Klein gauge fields from the five-dimensional point of view. The symmetry of $S^5$ corresponds to the internal symmetry of the dual gauge theory, R-symmetry, so such a charge is called a R-charge\index{R-charge}.

When all three charges are equal, the R-charged black hole reduces to the RN-AdS black hole above. However, the R-charged black hole in general differs from the RN-AdS black hole. This is because scalar fields are nonvanishing for the R-charged black hole. For asymptotically flat black holes, a scalar must vanish, and the existence of a scalar does not affect a black hole solution \cite{Bekenstein:1971hc}. This is because of the no-hair theorem\index{no-hair theorem}. But this is not the case for the R-charged black hole. Namely, the no-hair theorem is often violated in the AdS spacetime. 

The R-charged black hole undergoes a second-order phase transition \cite{Cai:1998ji,Cvetic:1999ne}.

\subsection*{Schwarzschild-AdS black holes in arbitrary dimensions
}

From the gravitational theory point of view, one may consider the SAdS$_{p+2}$ black hole as a simple extension of the SAdS$_5$ black hole. The dual gauge theory should be a $(p+1)$-dimensional conformal theory. The details of the dual gauge theory are unclear though because we do not go through the brane argument such as \sect{D3_brane}. But when $p=2$ and 5, there are brane realizations, the M2-brane and the M5-brane below.


\subsection*{Other branes (M-branes)
}

In string theory, there appear various branes, and the AdS/CFT duality based on these branes has been discussed (see, \eg, Ref.~\cite{Marolf:2011zs} for a review of these branes).

The \keyword{M-branes} appear in the 11-dimensional theory known as the \keyword{M-theory}. The M-theory does not yet have a microscopic formulation unlike string theory, but we know several clues about the theory:
\begin{itemize}

\item First, the M-theory is not a theory of strings. For a superstring theory to be consistent, the spacetime dimensionality must be 10.

\item At low energy, the M-theory is described by the 11-dimensional supergravity. This is just as  superstring theories are described by 10-dimensional supergravities at low energy. Just as the D-brane is a solution of a supergravity [\cf, \eq{D3_extreme}], the M-brane is a solution of the 11-dimensional supergravity. What we know about M-branes mostly comes from the 11-dimensional supergravity.

\item Although the M-theory is not a string theory, it is related to string theory. The $S^1$-compactification of the M-theory becomes the so-called \keyword{type IIA superstring theory}. As a result, the M-branes and the D-branes are related to each other.

\end{itemize}

There are two kinds of M-branes: the M2-brane and the M5-brane. The M2-brane reduces to the AdS$_4 \times S^7$ spacetime in the near-horizon limit. From the symmetry of the geometry, the dual gauge theory must be a $(2+1)$-dimensional conformal theory with R-symmetry $SO(8)$. Also, the M2-brane at finite temperature reduces to the SAdS$_4 \times S^7$ spacetime. But the full details of the dual gauge theory are still unclear. Unlike string theory, the M-theory does not yet have a microscopic formulation, so we do not have a microscopic tool such as the D-brane. 

Moreover, AdS/CFT predicts a characteristic behavior of thermodynamic quantities. For example,  the energy density of the M2-brane behaves as $\varepsilon \propto \Nc^{3/2} T^3$. The temperature dependence $T^3$ is natural and comes from the Stefan-Boltzmann law \index{Stefan-Boltzmann law} in $(2+1)$-dimensions. But the $\Nc$-dependence is not easy to understand. For gauge theories, it is natural to have the dependence $\Nc^2$ like the D3-brane, but the M2-brane does not have such a dependence. The power 3/2 implies that the M2-brane is not described by a simple gauge theory but is described by a highly nontrivial gauge theory. The dual gauge theory is not completely understood, but there are important progresses in recent years, and the power 3/2 has been understood to some extent \cite{Aharony:2008ug,Drukker:2010nc}.

On the other hand, the M5-brane reduces to the AdS$_7 \times S^4$ spacetime in the near-horizon limit. Then, the dual gauge theory must be a $(5+1)$-dimensional conformal theory with R-symmetry $SO(5)$. Also, the M5-brane at finite temperature reduces to the SAdS$_7 \times S^4$ spacetime. Again, thermodynamic quantities have a characteristic behavior in $\Nc$, \eg, $\varepsilon \propto \Nc^3 T^6$.

\subsection*{Other branes (D$p$-branes)
}

One can consider the D$p$-brane \index{D$p$-brane} with $p \neq 3$ \cite{Itzhaki:1998dd}. As in the D3-brane case, the D$p$-brane is dual to the $(p+1)$-dimensional SYM. The geometry has the \poincare\ invariance $ISO(1,p)$ and R-symmetry $SO(9-p)$. But the geometry does not reduce to the AdS spacetime in the near-horizon limit, so it does not have a conformal invariance. In particular, there is no scale invariance%
\footnote{The theory is scale invariant up to an overall scaling of the metric. An extension of such a geometry is known as the \keyword{hyperscaling violating geometry} \cite{Huijse:2011ef}.} 
because the dilaton is nontrivial (\sect{SUGRA}). From the gauge theory point of view, the $(p+1)$-dimensional gauge theory does not have a scale invariance because the coupling constant is dimensionful as $[g_\text{YM}^2] = \text{L}^{p-3}$. 

The energy density behaves as%
\footnote{One normally considers the $p<5$ case because the heat capacity diverges for $p=5$ and becomes negative for $p>5$.}
\be
\varepsilon \propto (g_\text{YM}^2 N_c)^{(p-3)/(5-p)} N_c^2 T^{2(7-p)/(5-p)}~.
\label{eq:energy_Dp}
\ee
In this case, we do not have the $(p+1)$-dimensional Stefan-Boltzmann law because we have a dimensionful quantity $g_\text{YM}$ other than the temperature. Also, thermodynamic quantities depend on the \thooft\ coupling $\lambda$ unlike the D3-brane and the M-branes.

However, there are exceptions. For $p=1$, $\varepsilon \propto \lambda^{-1/2} N_c^2 T^3 \propto N_c^{3/2} T^3$, which is the same behavior as the M2-brane. Similarly, for $p=4$, $\varepsilon \propto \lambda N_c^2 T^3 \propto N_c^{3} T^6$, which is the same behavior as the M5-brane. These behaviors are not just coincidences. They arise because some D-branes have their origins in the 11-dimensional M-branes although we do not discuss the details here.

\subsection*{Extensions from the gauge theory point of view
}

Starting from the \Nfour\ SYM, one can obtain new gauge theories by adding deformations. \index{deformations} For example, the \Nfour\ SYM has various matter fields in the adjoint representation, and one can add deformations to make these fields massive. The resulting theories have less supersymmetry compared to the \Nfour\ SYM. Also, the \Nfour\ SYM is scale invariant, so the theory does not have the confining phase but has only the plasma phase. The deformations of the \Nfour\ SYM are generally not scale invariant and some have a dynamical confinement like QCD. Some examples of dual geometries are 
\begin{itemize}
\item Pilch-Warner geometry (${\cal N}=2^*$ theory) \index{Pilch-Warner geometry} \cite{Pilch:2000ue,Buchel:2003ah},
\item Polchinski-Strassler geometry \index{Polchinski-Strassler geometry}  (${\cal N}=1^*$ theory) \cite{Polchinski:2000uf}, 
\item Klebanov-Strassler geometry \index{Klebanov-Strassler geometry} \cite{Klebanov:2000hb,Gubser:2001ri}.
\end{itemize}
This approach has both the advantage and the disadvantage:
\begin{itemize}
\item Advantage: the dual gauge theory is clear. 
\item Disadvantage: the resulting gravitational theory and the metric are often very complicated and difficult to analyze.
\end{itemize}



\subsection*{Extensions from the gravitational theory point of view
}

When one adds a deformation to the \Nfour\ SYM, one starts with the gauge theory point of view. In contrast, one could start with the gravitational theory point of view. There are at least two approaches:

\begin{enumerate}

\item \textit{Start with a gravitational theory which is easy to handle.} When one adds a deformation, the resulting gravitational action is often highly complicated and is hard to analyze. To avoid the problem, one could choose a theory which is natural or simple from the gravitational theory point of view. The SAdS$_{p+2}$ or RN-AdS black holes are the examples. The \textit{holographic superconductor}\index{holographic superconductors} is another typical example (\sect{H^3}). The holographic superconductor has the action 
\begin{align}
& \action = \int d^{p+2}x \sqrt{-g} \left[ R - 2\Lambda - \frac{1}{4} F_{MN}^2  
- |D_M\Psi|^2 -V(|\Psi|) \right]~, \\
& D_M := \nabla_M - i \qm A_M~,
%
\end{align}
where $\Psi$ is a complex scalar field.

\item \textit{Start with the symmetry of the metric.} In this case, one first writes down the metric with the desired symmetry and then tries to find a gravitational theory which admits such a metric as a solution. A typical example is the \keyword{Lifshitz geometry} \cite{Kachru:2008yh}. The metric 
\be
\frac{ds^2}{L^2} = -r^{2z}dt^2 + r^2 d\bmx_p^2 + \frac{dr^2}{r^2}
\label{eq:Lifshitz}
\ee
is invariant under the anisotropic scaling 
\be
t \rightarrow a^z t~, \bmx \rightarrow a \bmx~, r \rightarrow r/a
%
\ee
instead of the scaling discussed so far which corresponds to $z=1$. Such an anisotropic scaling is partly motivated by the dynamic critical phenomena in \sect{exponents}.

This approach starts with the metric, so the first questions one would ask are what kind of gravitational theory admits such a solution and whether such a gravitational theory is natural from the string theory point of view. Also, the metric \eqref{eq:Lifshitz} has a spacetime singularity at $r=0$ (when $z \neq 1$)%
\footnote{Unlike the Schwarzschild black hole, a curvature invariant such as $R^{MNPQ}R_{MNPQ}$ does not diverge at this singularity, but the tidal force diverges there. Some branes have similar spacetime singularities. A spacetime singularity is called a s.p. (scalar polynomial) singularity \index{scalar polynomial singularity} if a curvature invariant has a divergence and is called a p.p. (parallelly propagated) singularity \index{parallelly propagated singularity} if the tidal force diverges. There are various kinds of spacetime singularities and one cannot completely classify them. Thus, in general relativity, a spacetime singularity is defined operationally as geodesic incompleteness. \index{geodesically incomplete} }. 
The curvature becomes strong at the origin, and $\alpha'$-corrections could become important, so physical quantities one computes may not be reliable. 


\end{enumerate}
These approaches also have both the advantage and the disadvantage: 
\begin{itemize}
\item Advantage: one can choose a simple gravitational theory. One can use various gravitational theories. 
\item Disadvantage: the dual field theory is not very clear. In particular, one lacks the precise AdS/CFT dictionary which relates the parameters of the gravitational theory and the field theory, so quantitative predictions may be difficult.
\end{itemize}


\subsection*{Other holographic theories}

Finally, various generalizations of AdS/CFT have been proposed in the literature. They are holographic theories in a broad sense. Examples include
\begin{itemize}
\item dS/CFT correspondence (see Ref.~\cite{Spradlin:2001pw} for a review).
\item Kerr/CFT correspondence (see Ref.~\cite{Bredberg:2011hp} for a review).
\item Correspondence between the Rindler space \eqref{eq:rindler} and an incompressible fluid \cite{Bredberg:2010ky}.
\item Higher-spin holography \cite{Klebanov:2002ja}.
\end{itemize}

\titlenewterms

\begin{multicols}{2}
\noindent
Reissner-Nordstr\"{o}m AdS black hole\\
R-charged black hole\\
Kaluza-Klein gauge field\\
M-branes\\
M-theory\\
D$p$-brane\\
deformations
\end{multicols}

\endofsection

\ifx\nameofpaper\undefined 
  \usepackage{macro_natsuume} 
  \def\beginsection{\section*}
  \def\endofsection{\end{document}} 
  \input draft_header.tex
\else 
  \def\beginsection{\chapter}
  \def\endofsection{ } 
\fi

\section{Appendix: Explicit form of other AdS spacetimes \advanced
}\label{sec:others_details}

This appendix will be included in the published version.

\subsection{SAdS$_{p+2}$ black hole}\label{sec:SAdSp+2}

\subsection{M-branes}\label{sec:M_brane}

\subsection{D$p$-brane}\label{sec:Dp_brane}

\subsection{RN-AdS black holes}\label{sec:RNAdS}

\subsection{$d=5$ R-charged black hole and holographic current anomaly}\label{sec:r-charged}

\endofsection

\ifx\nameofpaper\undefined 
  \usepackage{macro_natsuume} 
  \def\beginsection{\section*}
  \def\endofsection{\end{document}} 
  \input draft_header.tex
\else 
  \def\beginsection{\chapter}
  \def\endofsection{ } 
\fi

\beginsection{Applications to quark-gluon plasma
}\label{chap:QGP}


\begin{quote}
For strongly-coupled Yang-Mills plasmas, AdS/CFT predicts that the ratio of the shear viscosity to the entropy density takes a universal small value. This is true for all known examples in the strong coupling limit. This prediction turns out to be close to the value of the real quark-gluon plasma (QGP).
\end{quote}

\section{Viscosity of large-$N_c$ plasmas
}\label{sec:large_N_viscosity}

\subsection{Transport coefficients of \Nfour\ plasma
}\label{sec:tensor_mode}

We now compute transport coefficients for the \Nfour\ plasma. To summarize our discussion so far,
\begin{itemize}

\item The \Nfour\ SYM is dual to the five-dimensional pure Einstein gravity with a negative cosmological constant (\chap{string}). The solutions are the AdS$_5$ spacetime at zero temperature and the Schwarzschild-AdS$_5$ (SAdS$_5$) \bh at finite temperature (\chap{AdS_BH}).

\item We will apply AdS/CFT to the quark-gluon plasma or QGP, so we will consider the finite temperature case. We add perturbations to the \bh and solve perturbation equations. From the gauge theory point of view, this gives the response to the operator which couples to the perturbation (\chap{GKPW}).
\item One can compute various transport coefficients of the \Nfour\ SYM depending on which bulk perturbations one considers.

\end{itemize}
For later convenience, we repeat some of the equations in previous chapters. The SAdS$_5$ \bh metric is given by
\be
ds_\text{SAdS$_5$}^2 = \left(\frac{r_0}{L}\right)^2\frac{1}{u^2} (-hdt^2+d\bmx_3^2)+L^2\frac{du^2}{h u^2}~, \quad
h = 1- u^4~, 
\label{eq:sads5_u_again}
\ee
where the horizon is located at $u=1$ and the AdS boundary is located at $u=0$. The temperature and the entropy density of the \bh are computed in Eqs.~\eqref{eq:temp_SAdS5} and \eqref{eq:entropy_SAdS5}:
\be
T = \frac{1}{\pi}\frac{r_0}{L^2}~, \quad
s = \frac{1}{4 G_5}\left(\frac{r_0}{L}\right)^3~.
\label{eq:thermo_SAdS5}
\ee
Also, the AdS/CFT dictionary is \eq{dictionary_preview}:
\be
%
N_c^2 = 8 \pi^2 \frac{L^3}{16\pi G_5}~, \quad
\lambda = \left(\frac{L}{l_s}\right)^4~.
\label{eq:dictionary_Ch12}
\ee

To begin with, the shear viscosity is the only nontrivial transport coefficient for the \Nfour\ SYM as a fluid. In a scale invariant theory, the energy-momentum tensor is traceless (\sect{sads_thermo}):
\be
T^\mu_{~\mu}=0~.
%
\ee
This immediately fixes
\be
\text{bulk viscosity}\quad \zeta = 0~, \quad 
\text{speed of sound}\quad c_s^2:=\frac{\del P}{\del\varepsilon} = \frac{1}{3}~.
\ee
Of course, one can check if $\zeta$ and $c_s$ really take these values, which is an important consistency check.

From the discussion in \sect{viscosity_kubo}, if one adds the perturbation $\sourceG$ in the boundary theory, the response $\delta \bra T^{xy} \ket$ is given by
\be
\delta \bra \tau^{xy}  \ket = i\omega \eta \sourceG~,
\label{eq:Txy_vs_eta_again}
\ee
where $\tau^{xy}$ is the dissipative part of $T^{xy}$. Thus, one can determine $\eta$ by computing this response in AdS/CFT. So, we consider the bulk perturbation $h_{xy}$:
\be
ds_5^2 = ds_\text{SAdS$_5$}^2 + 2h_{xy}\, dx dy~.
%
\ee
%
%

One can show that $(g^{xx} h_{xy})$ obeys the equation of motion for a massless scalar field, so we can apply the results in \sect{scalar}. Namely, the asymptotic behavior is given by
\be
\left(g^{xx} h_{xy} \right) 
\sim \sourceG \left\{ 1+ \vevG u^4 \right\}~,
\quad (u\rightarrow0)~.
\ee
Also, the fast falloff represents the response $\delta \bra \tau^{xy} \ket$. 
The evaluation of the on-shell gravitational action in \sect{action_tensor} shows the field/operator correspondence:
\be
\delta \bra \tau^{xy} \ket = \frac{1}{16\pi G_5} \frac{r_0^4}{L^5} 4\vevG \sourceG~.
\label{eq:response_hydro_ads}
\ee
In the units $r_0=L=16\pi G_5=1$, \eq{response_hydro_ads} reduces to \eq{kw}. Comparing Eqs.~\eqref{eq:Txy_vs_eta_again} and \eqref{eq:response_hydro_ads}, and using the temperature and the entropy density \eqref{eq:thermo_SAdS5}, one gets
\be
i \nw \frac{\eta}{s} = \vevG~, \qquad \nw := \frac{\omega}{T}~.
\label{eq:vev_G}
\ee

To determine $\vevG$, one has to actually solve the perturbation equation.  This is done in \sect{tensor_mode_sol}. 
%
%
In this example, the perturbation equation reduces to the massless scalar, so it may sound easy to solve but it is not even in this case. This is because the perturbation equation has more than three regular singular points, so the analytic solution is not expected in general. This is a typical problem in \bh physics, and as a result, the perturbation problem in a \bh background typically requires a numerical computation.

Fortunately, our primary interest here is the hydrodynamic limit where  $\omega\rightarrow 0$ as $q\rightarrow 0$. Then, one can expand the perturbation equation as a double series in $(\omega, q)$, and it is enough to solve the low orders in the expansion. In this limit, perturbation equations can often be solved analytically.

In this way, we solve the  the perturbation equation in \sect{tensor_mode_sol}, and the result \eqref{eq:tensor_sol_asymptotic} is
\be
\vevG = \frac{1}{4\pi} i \nw + O(\omega^2, q^2)~.
\label{eq:perturbative_sol}
\ee
Comparing Eqs.~\eqref{eq:vev_G} and \eqref{eq:perturbative_sol}, we obtain
\be
\boxeq{
\frac{\eta}{s} = \frac{1}{4\pi}~.
}
\label{eq:universality}
\ee
The value of $\eta$ itself is obtained by using the AdS/CFT dictionary  \eqref{eq:dictionary_Ch12} and the entropy density \eqref{eq:thermo_SAdS5}:
\be
\eta = \frac{1}{16\pi G_5}\left(\frac{r_0}{L}\right)^3 
= \frac{\pi}{8} N_c^2 T^3~.
\ee
The entropy density $s = a/(4G_5)$ is proportional to the  the ``horizon area density" $a$,
and $\eta/s$ is constant, so $\eta$ is also proportional to $a$. This is true not only for the \Nfour\ SYM but also for the other large-$\Nc$ plasmas. We will see this in \sect{universality}.

\head{Remarks}
As we saw in \sect{hydro}, there are two methods to obtain $\eta/s$:
\begin{itemize}

\item The method from the $O(\omega)$ coefficient of the tensor mode $\delta \bra T^{xy} \ket$ (``Kubo formula method"). This is the method discussed here. 

\item The method from the pole of the vector mode and the dispersion relation

\end{itemize}
One can check that $\eta/s$ from the dispersion relation gives the same result (Sects.~\ref{sec:background} and \ref{sec:vector}).

The \Nfour\ SYM has R-symmetries and associated $U(1)$ currents $J^\mu$ (\sect{gauge}). The gravitational dual actually has Maxwell fields (\sect{r-charged}). One can add Maxwell fields as probes and compute the diffusion constant $D$ or the conductivity $\sigma$. The computations are very similar to the shear viscosity but are somewhat easier (\sect{SAdS5_diff}). 

\subsection{Implications}\label{sec:implications}

In order to understand the AdS/CFT result, let us pause here to consider the implications of the result. 

\head{Viscosity of ordinary materials}
First, let us compare the AdS/CFT result with the viscosity of ordinary materials. Recovering dimensionful quantities and using $\hbar \approx 1.05 \times 10^{-34}~\joule\cdot\second$ and $\kB \approx 1.38 \times 10^{-23}~\joule\cdot\kelvin^{-1}$, the AdS/CFT prediction is
\be
\frac{\eta}{s} = \frac{\hbar}{4\pi \kB}
\approx 6.1 \times 10^{-13}~\kelvin\cdot\second~.
\ee
As an example of ordinary materials, take the nitrogen gas. According to Ref.~\cite{nist}, the nitrogen has
\be
\eta \approx 16.6~\micro\pascal\cdot\second~, \quad
s \approx 189~\joule\cdot\kelvin^{-1}\cdot\mole^{-1}
\ee
at $273.15~\kelvin$ (0 \textcelsius) and atmospheric pressure (1 atm $\approx 0.1 \mega\pascal = 1 \bbar$). 
The ideal gas under the above condition occupies $22.4~\liter$ ($=22.4 \times 10^{-3} \meter^3$) per mole, so
\be
s \approx 189 \times \frac{1}{22.4 \times 10^{-3}} 
\approx 8.4 \times 10^3~\joule\cdot\kelvin^{-1}\cdot\meter^{-3}~.
\ee
Thus, $\eta/s$ for the nitrogen is
\begin{align}
\frac{\eta}{s} \approx \frac{16.6 \times 10^{-6}}{8.4 \times 10^3}
&\approx 2.0 \times 10^{-9}~\kelvin\cdot\second \\
&\approx 3.3 \times 10^3~\frac{\hbar}{4\pi \kB}~,
%
\end{align}
which is $3\times10^3$ times larger than the AdS/CFT result. Namely, 
\begin{center}
\fbox{
\begin{tabular}{l}
The strongly-coupled large-$\Nc$ plasma has an extremely small $\eta/s$ \\
compared with ordinary materials.
\end{tabular}
}
\end{center}

\head{Typical behavior of $\eta/s$}
Let us also consider the typical behavior of $\eta/s$. 
As an example, consider the $\coup \phi^4$-theory, and estimate physical quantities at tree-level. The dimensional analysis and tree diagrams determine the behavior of various quantities as follows%
\footnote{For perturbative estimates, see, \eg, Ref.~\cite{Jeon:1994if,Jeon:1995zm} ($\coup \phi^4$-theory), Ref.~\cite{Arnold:2000dr} (QCD), and Ref.~\cite{Huot:2006ys} (\Nfour\ SYM). }:
\begin{alignat}{2}
&\text{number density: } && n \propto T^3~, 
\label{eq:phi4_estimate1} \\
\displaybreak
&\text{cross section: } && \sigma \propto \frac{\coup^2}{T^2}~, \\
&\text{mean-free path: } && l_\text{mfp} \simeq \frac{1}{n \sigma} \propto \frac{1}{\coup^2 T}~, \\
&\text{energy density: } && \varepsilon \propto T^4~, \\
&\text{shear viscosity: } && \eta \simeq \varepsilon l_\text{mfp} \propto \frac{T^3}{\coup^2}~.
\label{eq:phi4_estimate5}
%
\end{alignat}
Thus%
\footnote{The entropy density depends on coupling constants only weakly as we saw for the \Nfour\ SYM (\sect{sads_thermo}).}, 
\be
\frac{\eta}{s} \simeq \frac{1}{\coup^2}~.
\label{eq:eta/s_perturbative}
\ee

The comparison between this perturbative estimate and the AdS/CFT result $\eta/s=1/(4\pi)$ gives the following implications:
\begin{itemize}

\item 
In the AdS/CFT result, there is no coupling dependence $\lambda$. This is because the AdS/CFT result is the strong coupling limit $\lambda\rightarrow\infty$.

\item 
In the $\coup \phi^4$-theory, the naive extrapolation of the perturbative estimate to the $\coup\rightarrow\infty$ limit gives the perfect fluid $\eta/s \rightarrow 0$. But, in AdS/CFT, $\eta/s$ remains finite even in the strong coupling limit. AdS/CFT suggests that $\eta/s$ cannot be small indefinitely and is saturated. 

\item 
At finite coupling, one expects that the AdS/CFT result becomes
\be
\frac{\eta}{s} \geq \frac{1}{4\pi}~.
%
\ee
We discuss more about the corrections in \sect{coupling}.

\end{itemize}

\head{Does QGP have a small viscosity?}
The analysis of the $\coup \phi^4$-theory also clarifies an often confusing issue. We loosely mentioned ``QGP has a very small viscosity," but more precisely what we meant is that \textit{$\eta/s$ is small}. 

This point is important because the viscosity of QGP itself is not small. \textit{The QGP viscosity itself is very large} because we consider plasmas at high temperature and $\eta \propto T^3$.

To see this, let us go back to Eqs.~\eqref{eq:phi4_estimate1}-\eqref{eq:phi4_estimate5} 
and consider the physical interpretations. At high temperature, many particles are created and the number density becomes larger, so $l_\text{mfp}$ becomes shorter. However, this does not imply that the shear viscosity itself becomes smaller. This is because the shear viscosity is the energy-momentum transfer. At high temperature, each particle has a higher energy on average and many particles are created. Therefore, the energy transfer becomes more effective at high temperature, which results in a larger shear viscosity.

However, note the coupling constant dependence of the shear viscosity. Although $\eta \propto T^3$, a strongly-coupled fluid has a smaller viscosity than a weakly-coupled fluid (if one fixes temperature). It is in this sense that the shear viscosity of QGP is small.

But it is rather confusing to take into account both the temperature dependence and the coupling constant dependence. A better way is to make a dimensionless quantity (in natural units) by combining with another physical quantity such as \eq{eta/s_perturbative}.
%
%
Then, only the coupling constant dependence remains, and the meaning of the ``small viscosity" at strong coupling becomes clear. 

Also, what appears in hydrodynamic equations is this combination $\eta/s$ [see, \eg, \eq{dispersion_vector}], so this is the appropriate quantity to consider.

\subsection{Universality of $\eta/s$
}\label{sec:universality}


We obtained the shear viscosity in the simplest example, the SAdS$_5$ \bh which is dual to the \Nfour\ SYM. But as we discuss below, one can compute $\eta/s$ for various large-$\Nc$ gauge theories, and the result is $\eta/s=1/(4\pi)$ in all known examples (in the large-$\Nc$ limit)%
\footnote{It has been pointed out that the universality argument below does not hold for anisotropic plasmas \cite{Natsuume:2010ky,Erdmenger:2010xm}.}. Therefore, one concludes 
\begin{center}
\fbox{
\begin{tabular}{l}
The strongly-coupled large-$\Nc$ plasmas have the universal small value of \\
$\eta/s=1/(4\pi)$.
\end{tabular}
}
\end{center}
If large-$\Nc$ gauge theories are good approximations to QCD, one would imagine that the result equally applies to QGP. In fact, according to RHIC experiments, QGP behaves as a fluid with a very small $\eta/s$, and the value is close to $1/(4\pi)$ (\sect{QGP}).



We should stress that the ratio $\eta/s$ is universal, but the values of $\eta$ and $s$ themselves have no universality, and they depend on the theory one considers. 

The entropy density counts the degrees of freedom of a system, so it clearly depends on the theory. Also, the entropy density of the \Nfour\ SYM is $s \propto T^3$ from the dimensional analysis, but if a system has the other scales, the entropy density depends on these scales. For example, it depends on the chemical potential $\mu$ at finite density. 

What the universality implies is that $\eta$ and $s$ have the same functional form even in such cases, so by taking the ratio $\eta/s$, the details of the system cancel, and only the fundamental constants ($\hbar$ and $\kB$) appear. Of course, we discard many information about the system by taking the ratio, but we obtain a robust prediction.
\subsubsection*{Major examples of universality
}

Major examples which satisfy the universality are
\begin{enumerate}

\item[(i)] \textit{Conformal theories} \cite{Policastro:2001yc,Policastro:2002se} (All examples below are the computations in plasma phases.)

\item[(ii)] \textit{Non-conformal theories} \cite{Kovtun:2003wp,Buchel:2003tz}: 
The \Nfour\ SYM is scale invariant, but the universality holds for theories with explicit scales, non-conformal theories. Some examples are D$p$-branes with $p\neq3$ and the Klebanov-Strassler geometry at finite temperature (\chap{others}). The latter theory has a dynamical confinement scale like QCD.

\item[(iii)] \textit{Theories in the other spacetime dimensions} \cite{Kovtun:2003wp,Herzog:2002fn}: 
The universality holds in the $(p+1)$-dimensional boundary spacetimes. Some examples are SAdS$_{p+2}$ black holes and D$p$-branes.

\item[(iv)]  \textit{Theories at finite chemical potential} \cite{Mas:2006dy,Son:2006em,Saremi:2006ep,Maeda:2006by,Benincasa:2006fu}:
In the QCD phase diagram, the chemical potential is equally as important as the temperature. The universality holds at a finite chemical potential. For the \Nfour\ SYM, one can add chemical potentials associated with global R-symmetries. The dual geometries are charged AdS black holes (\chap{others}). Historically, this example was important since there was a conjecture that the universality no longer holds at a finite chemical potential. Thus, the result was a surprise, which motivated many works to study $\eta/s$ under various circumstances.

\item[(v)] \textit{Theories with the fundamental representation such as quarks} \cite{Mateos:2006yd}: 
The theories described so far have matter fields in the adjoint representation and not in the fundamental representation such as quarks. However, the universality holds even in the presence of the fundamental matter. One way to include the fundamental matter is to include D7-branes in addition to D3-branes. We discussed one simple way to realize the fundamental matter in \chap{wilson}, and the D3-D7 system is an extension of the method. This D3-D7 system is known as Karch-Katz model \cite{Karch:2002sh}. The D7-brane changes the metric, but one can evaluate the effect of its backreaction perturbatively if one adds D7-branes as a probe.  

\item[(vi)]  \textit{Time-dependent \Nfour\ plasma} \cite{Janik:2006ft}:
Large-$N_c$ plasmas described so far are stationary ones. The real plasma at heavy-ion experiments is of course a rapidly changing system, and it is desirable to study such a plasma as well. In the gravity side, this corresponds to a time-dependent black hole. The universality has been shown for such a case.

\end{enumerate}

How can one show the universality? In order to show the cases (i)-(iv), consider a gravitational theory with various matter fields such as Maxwell fields and scalar fields. One can solve the perturbation equation just like the SAdS$_5$ case and obtain \eq{universality} \cite{Benincasa:2006fu}. The technical reasons for the universality are two-folds:
\begin{itemize}

\item First, the perturbation $h_{xy}$ belongs to the tensor mode. We take the boundary coordinates as $x^\mu=(t,x,y,z)$ and consider the perturbation propagating in the $z$-direction or $k_\mu=(\omega, 0,0,q)$. Then, there is a little group $SO(2)$ \index{little group} acting on $(x,y)$ which keeps $k_\mu$ invariant. We can classify the energy-momentum tensor and the gravitational perturbations by their transformation properties under the little group. The components are classified as the scalar mode, the vector mode, and the tensor mode (Sects.~\ref{sec:linearized} and \ref{sec:tensor}). The tensor decomposition is convenient because each modes transform differently and their equations of motion are decoupled with each other. 

Now, the perturbation $h_{xy}$ belongs to the tensor mode in this classification. Even if a gravitational theory has Maxwell fields and scalar fields, they do not transform as tensors, so they are not coupled with the tensor mode.

One can show the the perturbation equation for the tensor mode takes the same form as the massless scalar in general. This equation can be solved for an arbitrary \bh background in the hydrodynamic limit, and one can show that $\eta$ is proportional to the horizon area density $a$.

\item  Second, as mentioned in \sect{BH_entropy}, the entropy density is also proportional to $a$ as long as the gravitational action takes the form of the Einstein-Hilbert action.

\item Thus, if one takes the ratio $\eta/s$, $a$ cancels out and only the numerical coefficient remains in the ratio which is \eq{universality}.

\end{itemize}

\head{Power of the duality}
This universality tells us the {\it power of the duality}. In principle, one can reach the universality if one develops strong coupling computations such as lattice simulation powerful enough. Then, one would wonder if we really need the AdS/CFT duality. But the universality would not be very transparent in such an approach. In AdS/CFT, the universality is transparent: it is tied to the universal nature of black holes, \eg, the area law of \bh entropy. Namely, 
\begin{center}
\fbox{
\begin{tabular}{l}
Some results which are not very transparent in original variables (gauge theory) \\
can be transparent in different variables (gravitational theory).
\end{tabular}
}
\end{center}
In a duality, one can look at the same physics from a different point of view: this is the power of a duality.

\subsection{How to solve perturbation equation
}\label{sec:tensor_mode_sol}

Here, we actually solve the perturbation equation for the tensor mode and derive \eq{perturbative_sol}. Setting $\phi := g^{xx} h_{xy}$, one can show that $\phi$ obeys the equation of motion for a massless scalar:
\be
\frac{1}{\sqrt{-g}} \del_\mu ( \sqrt{-g} g^{\mu\nu} \del_\nu \phi ) = 0
%
\ee
(see \sect{action_tensor} for the derivation). Consider the perturbation of the form  $\phi = \phip(u) e^{-i\omega t +iqz}$. In the SAdS$_5$ \bh background, the perturbation equation becomes
\be
\frac{u^3}{h} \left( \frac{h}{u^3} \phip' \right)' 
+ \frac{ \nw^2-\nq^2 h }{ \pi^2 h^2 } \phip = 0~,
\label{eq:scalar_eom}
\ee
where
\be
':=\del_u~, \qquad \nw := \frac{\omega}{T}~, \qquad \nq := \frac{q}{T}~.
\ee
Below we solve this differential equation with $\nq=0$. We impose the following boundary conditions:
\begin{itemize}
\item The horizon $u\rightarrow1$: the ``incoming-wave" boundary condition,
\item The AdS boundary $u\rightarrow0$: the Dirichlet condition $\phip(u\rightarrow0) = \source_k$. 
\end{itemize}

When we have a differential equation, the first question we should ask is the structure of singular points; how many singular points the equation has and whether they are regular or not. Equation~\eqref{eq:scalar_eom} has more than 3 regular singular points, so the analytic solution cannot be expected in general.  The equation has 4 regular singular points at 
\be
u=0~, \pm1~, \infty~.
\ee
However, in this computation, we are interested only in the hydrodynamic limit $\omega\rightarrow 0$. In this limit, the equation can be solved analytically. 

\subsubsection*{Incorporating boundary condition at horizon
}

We first solve \eq{scalar_eom} near the horizon $u \simeq 1$ in order to incorporate the ``incoming-wave" boundary condition.\index{incoming-wave boundary condition} Near the horizon, $h = 1-u^4 \simeq 4(1-u)$, and the perturbation equation becomes
\be
\phip'' - \frac{1}{1-u} \phip' + \left( \frac{\nw}{4\pi} \right)^2 \frac{1}{(1-u)^2} \phip \simeq 0~.
\label{eq:scalar_eom_horizon}
\ee
Setting $\phip \simeq (1-u)^\lambda$, one gets
\be
(1-u)^{\lambda-2} \left\{ \lambda(\lambda-1) + \lambda + \left( \frac{\nw}{4\pi} \right)^2 \right\} = 0~.
\label{eq:determining_eq}
\ee
Thus, the solutions are
\be
\phip \propto (1-u)^{\pm i \nw/(4\pi)}~, \qquad (u\rightarrow 1)~.
\label{eq:in&out}
\ee
One way to solve a differential equation is to utilize a power series expansion about some $u$. In such a case, \eq{determining_eq} is the indicial equation to determine the lowest power of $(1-u)$. Proceeding further, one can obtain the solution by a power series expansion%
\footnote{We want a solution which is applicable from $u=1$ to $u=0$, so the radius of convergence of the power series solution must reach $u=0$. In general, the radius of convergence is at least as large as the distance to the nearest singular point in the complex $u$-plane. For our problem, the nearest singular point of $u=1$ is $u=0$, so the radius of convergence reaches $u=0$.}, 
but we we do not need to take the approach here. See \sect{HSC_more}.

The solutions \eqref{eq:in&out} may be unfamiliar, but there is a natural interpretation in the \keyword{tortoise coordinate}. The tortoise coordinate $r_*$ is defined such that the metric in the $(t,r_*)$-directions becomes conformally flat:
\be
ds^2 = -f(r) dt^2+ \frac{dr^2}{g(r)} = f(r) (-dt^2+dr_*^2)~.
\ee
This coordinate is useful for physical interpretations near the horizon since the Laplacian takes the standard Minkowski form, $\nabla^2 \propto -\del_t^2+\del_{r_*}^2$. The SAdS$_5$ \bh takes
\be
ds^2 = \left(\frac{r_0}{L}\right)^2 \frac{h}{u^2} \left\{ -dt^2 + \frac{L^4}{r_0^2} \frac{du^2}{h^2} \right\} +\cdots~,
\ee
so the tortoise coordinate is given by
\begin{align}
%
u_* &= -\frac{L^2}{r_0} \int_0^u \frac{du}{h} \\
&\sim \frac{1}{4\pi T} \ln(1-u) ~, \qquad (u\rightarrow 1)~.
%
\end{align}
The horizon $u=1$ corresponds to $u_*\rightarrow-\infty$. As discussed in \sect{geodesics_horizon}, if one uses the Schwarzschild time coordinate $t$, it takes an infinite amount of coordinate time to reach the horizon. In the tortoise coordinate, this familiar effect is interpreted that the horizon is infinitely far away. Since $ds^2 \propto -dt^2+du_*^2$, the infinite amount of the coordinate time is possible if the horizon is located at infinity $u_*\rightarrow-\infty$.

Using the tortoise coordinate and combining the time-dependence, the near-horizon solutions are written as
\be
\phi \propto e^{-i\omega t}(1-u)^{\pm i \nw/(4\pi)} \simeq e^{-i \omega (t \mp u_*)}~.
\ee
The solutions take the standard plane-wave form, and two solutions represent either ``incoming" to the horizon (the lower sign in the double sign expression) or ``outgoing" from the horizon. We impose the ``incoming-wave" boundary condition, so we choose the solution
\begin{alignat}{2}
\phip &\propto (1-u)^{- i \nw/(4\pi)}~, & \qquad & (u\rightarrow 1)
\\
&\simeq (1-u^4)^{- i \nw/(4\pi)}~, & \qquad & (u\rightarrow 1)~.
\label{eq:incoming}
\end{alignat}

Below we solve the perturbation equation as a power series in $\nw$ and compare with the near-horizon solution. So, expand the near-horizon solution \eqref{eq:incoming} in terms of $\nw$ as well. From $a^x \sim 1 + x \ln a + \frac{1}{2} x^2(\ln a)^2 + \cdots$,
\be
\phip \propto 1 - \frac{i\nw}{4\pi} \ln(1-u^4) + \cdots
~, \qquad (u\rightarrow 1)~.
\label{eq:tensor_sol_horizon}
\ee
The solution \eqref{eq:incoming} is valid near the horizon $1-u \ll 1$ or $\ln(1-u) \gg 1$. On the other hand, the above $\nw$-expansion is valid when $\nw \ln(1-u) \ll 1$. These two conditions are satisfied for a small enough $\nw$.

\subsubsection*{The $\nw$-expansion}

We obtained the near-horizon solution, but we now obtain the solution for all $u$ as a power series in $\nw$:
\be
\phip = F_0(u) + \nw F_1(u) + \cdots~.
%
\ee
Then, \eq{scalar_eom} becomes
\be
\left( \frac{h}{u^3} F_i' \right)' = 0~,
\label{eq:scalar_eom0}
\ee
where $i=0,1$. Equation~\eqref{eq:scalar_eom0} is easily solved as
\be
F_i = A_i + B_i \ln (1-u^4) ~,
%
\ee
where $A_i$ and $B_i$ are integration constant. Thus,
\be
\phip = (A_0 + \nw A_1) + (B_0 + \nw B_1) \ln (1-u^4) + O(\nw^2)~.
%
\ee
Finally, we impose boundary conditions. The boundary condition at the AdS boundary $u\rightarrow0$ is given by $\phip(u\rightarrow0) = \source_k$, so
\be
A_0 + \nw A_1 = \source_k~.
%
\ee
The boundary condition at the horizon reduces to \eq{tensor_sol_horizon}, which determines the rest of integration constant:
\be
B_0 + \nw B_1 = -\frac{i\nw}{4\pi} \source_k~.
%
\ee

To summarize, the solution is given by
\begin{align}
\phip 
& = \source_k \left\{ 1 - \frac{i\nw}{4\pi} \ln (1-u^4) + O(\nw^2) \right\}~, 
&& (0\leq u\leq 1)~. 
\label{eq:perturbative_sol_app1} \\
&\sim \source_k \left\{ 1+ \frac{1}{4\pi} i \nw u^4 + \cdots \right\}~, 
&& (u\rightarrow 0)~.
\label{eq:tensor_sol_asymptotic}
\end{align}
The $O(\nw)$ coefficient $1/(4\pi)$ gives the value of $\eta/s$.

\section{Comparison with QGP
}\label{sec:QGP}

\subsection{How can one see viscosity in the experiment?
}\label{sec:elliptic_flow}

\begin{figure}[tb]
\centering
\scalebox{0.7}{ \includegraphics{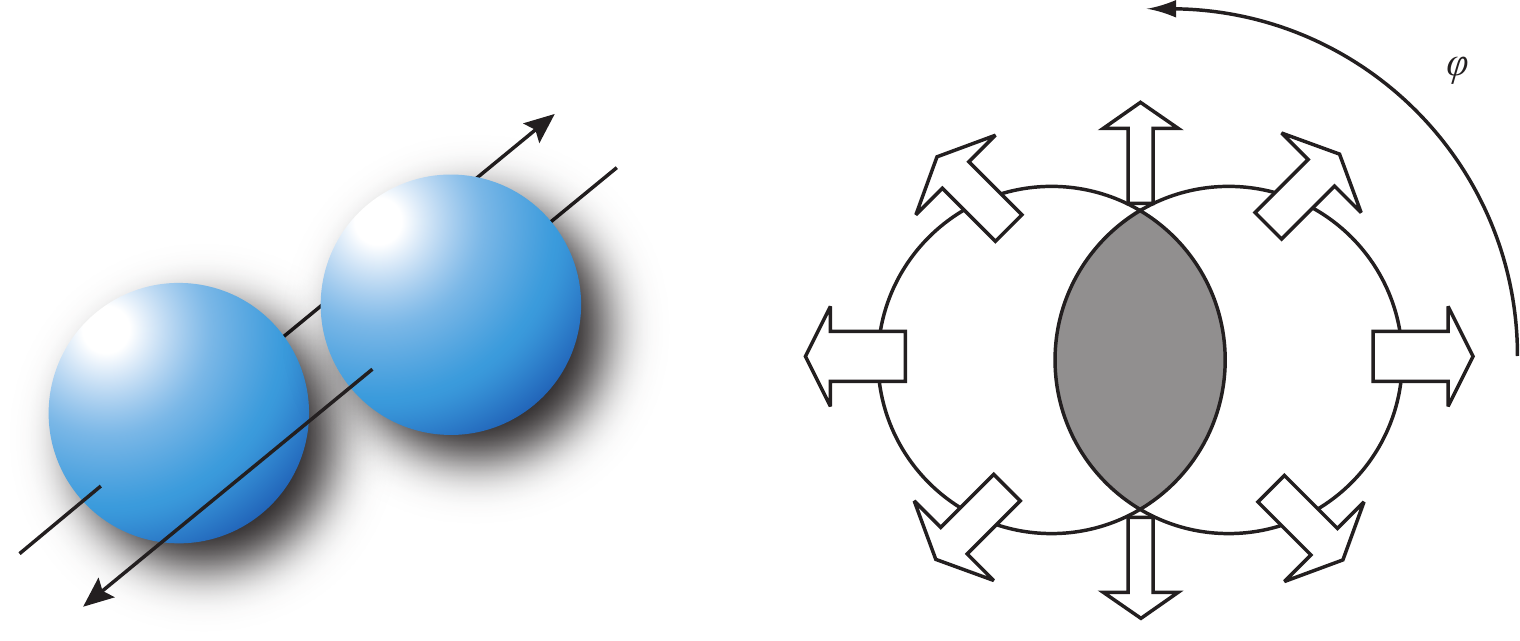} }
\vskip2mm
\caption{Elliptic flow. The nuclei overlap in the shaded region. The width of arrows represents the produced particle number $N$.}
\label{fig:elliptic_flow}
\end{figure}

Experimentally, the shear viscosity of QGP can be measured through the phenomenon called the \keyword{elliptic flow}. In heavy-ion experiments, one collides heavy nuclei. The elliptic flow is a phenomenon when nuclei undergo off-axis collisions instead of head-on collisions (\fig{elliptic_flow}). In such a collision, the nuclei overlap in the almond-shaped region in the figure. The strong interaction is a short-range force, so the plasma formation occurs in this almond-shaped region. 

Let us consider particles which are created near the center of the almond:
\begin{itemize}
\item Suppose that the particles initially have the momenta in the longer direction of the almond. In this direction, the particles have to travel longer until they get out. So, there are more chances to collide with the other particles. As a result, the distribution of the finial momentum directions becomes more or less isotropic due to the scatterings.
\item On the other hand, if the particles initially have the momenta in the shorter direction, the particles have less chances for collisions. Thus, the particles tend to preserve the initial momentum directions.
\end{itemize}
Then, the overall particle distribution is not isotropic, and more particles should be observed along the shorter direction. The important point is that the anisotropy comes from interactions of particles, namely from the viscosity. For an ideal gas, the particle distribution should be isotropic.

To represent such anisotropy, expand the produced particle number $N$ as a Fourier series:
\be
\frac{dN}{d\varphi} = N \left\{ 1+ 2v_2 \cos(2\varphi) + \cdots \right\}~,
\ee
where $\varphi$ represents the angle from the shorter axis (reaction plane). The coefficient $v_2$ is called the elliptic flow.

\begin{figure}[tb]
\centering
\scalebox{0.45}{ \includegraphics{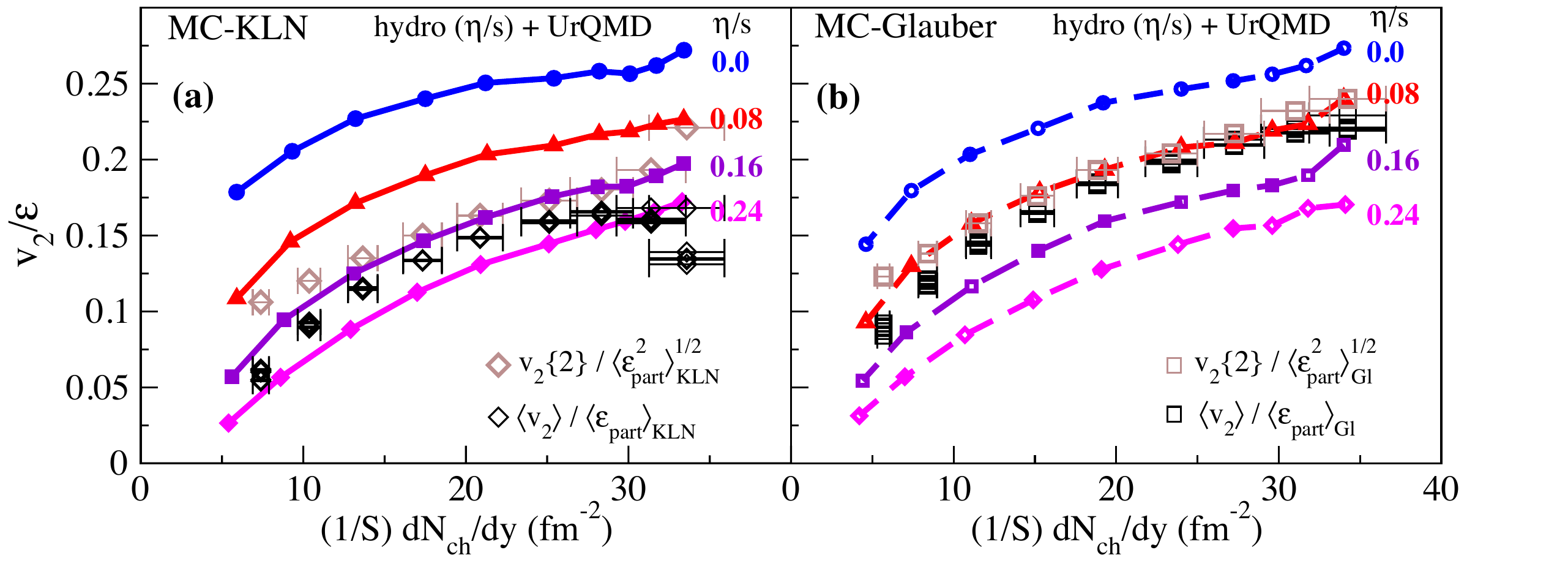} }
\vskip2mm
\caption{Comparison between RHIC results and hydrodynamic simulations \cite{Song:2010mg}. Figures~(a) and (b) use different sets of initial conditions. (``KLN" on the left and ``Glauber" on the right indicate sets of initial conditions.) The figures compare experimental results (plots with error bars) and hydrodynamic simulations with various values of the viscosity (solid lines on the left and dashed lines on the right.) The horizontal axis represents particle multiplicity per overlap area, and the vertical axis represents the elliptic flow $v_2$ (normalized by eccentricities $\epsilon$ of the overlap region.) }
\label{fig:song}
\end{figure}

Figures~\ref{fig:song} compare experimental results with hydrodynamic analysis \cite{Song:2010mg}. The vertical axis represents $v_2$. A larger $v_2$ implies stronger interaction. For hydrodynamic analysis, each figure plots four curves with different values of $\eta/s$. The elliptic flow is indeed largest for the perfect fluid case which is the naive strong coupling limit. From this analysis, they conclude that 
\be
\frac{1}{4\pi} < \frac{\eta}{s} < 2.5 \times \frac{1}{4\pi}~, 
\qquad (\Tc < T \lesssim 2\Tc)~.
%
\ee
Main uncertainties come from different sets of initial conditions. (See also Ref.~\cite{Song:2012ua} for the status of hydrodynamic simulations as of 2012.)
%
%

In such analysis, one numerically simulates the fluid using hydrodynamic equations. Since one does not know the values of transport coefficients such as the viscosity, one tries simulations using various values of transport coefficients, fits with experimental results, and determine the values of transport coefficients. However, QGP has a complicated time-evolution, so hydrodynamic simulations alone cannot determine the viscosity. There are various uncertainties:
\begin{itemize}

\item First, one has to specify the initial conditions as the fluid (such as initial velocity fields and energy density). They depend on dynamics of the system before one can use hydrodynamics, namely before QGP achieves local equilibrium. One specifies these conditions using QCD, but this is a difficult problem by itself. If one does not know this dynamics well, one has to try various initial conditions as well.
\item Also, QGP expands and cools down by the expansion. Eventually, the temperature is below the transition temperature, so quarks are confined into hadrons (hadronization). But this process is not completely understood.
\item To determine $\eta$, one needs at least $(2+1)$-dimensional simulations which require more computational powers than the $(1+1)$-dimensional simulations. Such simulations become possible only in recent years.
\item The standard relativistic viscous hydrodynamics actually has unphysical instabilities, and numerical simulations of relativistic viscous fluids are not possible. One can remove such instabilities, but this requires to introduce new transport coefficients. But the results then depend on the values of these new transport coefficients. (So, the simulation of \fig{song} is not the one of the standard relativistic hydrodynamics.) We briefly discuss this issue in \sect{2nd_order}.
\end{itemize}

\subsection{Comparison with lattice simulation
}

\begin{figure}[tb]
\centering
\scalebox{0.8}{ \includegraphics{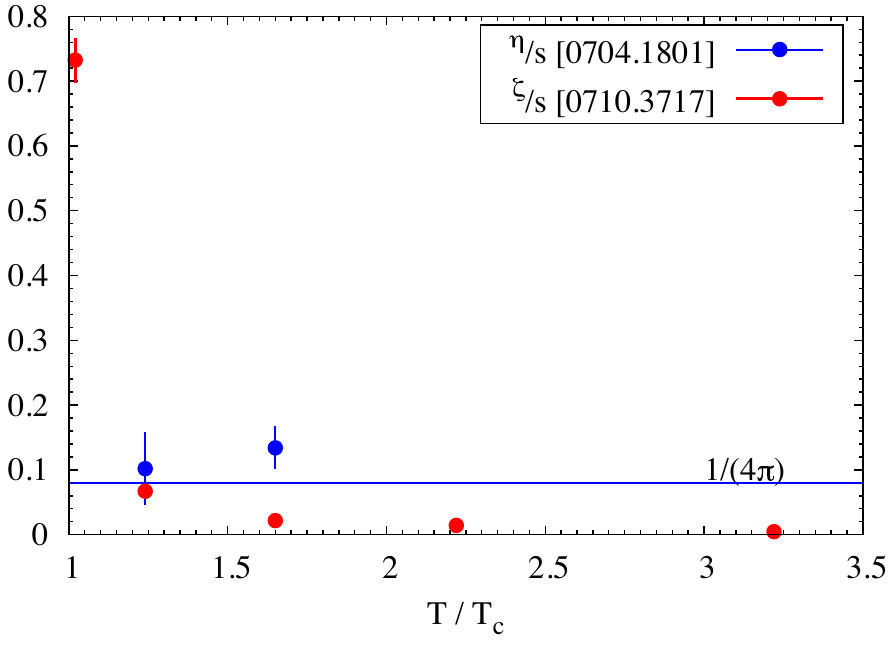} }
\vskip2mm
\caption{A lattice simulation for $\eta/s$ \cite{Meyer:2007ic}. This is the result for pure $SU(3)$ gauge theory. The horizontal line corresponds to the AdS/CFT result. Adapted from slides at the  conference NFQCD2008.}
\label{fig:lattice_eta}
\end{figure}

Hydrodynamic simulations above determine the value of viscosity phenomenologically by comparing with experiments. But, in principle, one can determine the value using QCD. Namely, if one evaluates the Kubo formula in QCD, one can determine the viscosity microscopically. Of course, such a computation is difficult analytically at strong coupling (this is why we use AdS/CFT), but lattice simulations become possible, and the results are close to \eq{universality}. According to Ref.~\cite{Meyer:2007ic}, pure $SU(3)$ gauge theory has
\begin{alignat}{2}
\frac{\eta}{s} &= 0.134(33)~, 
&\qquad& (T = 1.65 \Tc)~,\\
\text{\textit{cf.}~} \frac{\eta}{s} &= \frac{1}{4\pi} \approx 0.08~,
&\qquad& \text{(AdS/CFT)~.} 
%
\end{alignat}
(\fig{lattice_eta}. See Ref.~\cite{Nakamura:2004sy} for an early work.) This approach has both advantage and disadvantage:
\begin{itemize}

\item Advantage: Hydrodynamic simulations have various uncertainties which come from the complicated QGP time-evolution, but this approach is free from such uncertainties in principle. It should be able to answer the question if strongly-coupled gauge theories really have a very small $\eta/s$.
\item Disadvantage: It does not answer the question if what one sees in heavy-ion experiments are really the effects caused by a small $\eta/s$ or not. Also, this computation in \fig{lattice_eta} itself partly uses AdS/CFT results, so it is not completely an independent result from AdS/CFT.
\end{itemize}
Note that this is the result for pure $SU(3)$ gauge theory. The AdS/CFT result applies to supersymmetric gauge theories in the large-$\Nc$ limit. But this result \cite{Meyer:2007ic} implies that the universality holds in a good approximation even for gauge theories which do not have supersymmetry and are not in the large-$\Nc$ limit.

Both the hydrodynamic analysis and the lattice simulation suggest that $\eta/s$ is close to $1/(4\pi)$ but is larger than $1/(4\pi)$. This is natural since the AdS/CFT result is the strong coupling result and one expects that $\eta/s \geq 1/(4\pi)$ at finite coupling as mentioned in \sect{implications}.

\subsection{Why study supersymmetric gauge theories?
}

One can study various gauge theories using AdS/CFT, but they are supersymmetric gauge theories. This is because we do not yet have the gravitational dual to QCD. So, in AdS/CFT, we use supersymmetric gauge theories such as the \Nfour\ SYM. Then, one would ask if supersymmetric gauge theories have anything to do with real QCD. There could be at least two possible answers:

\begin{itemize}
\item The first answer is ``universality." Here, the word universality means the observables which have universal behaviors among gauge theories. If one can find such an observable and can compute it, one would expect that the result equally applies to QCD. The ratio $\eta/s$ is one example.
\item The first answer applies only to restricted class of observables. Although we lose the rigor, the second answer is that ``\Nfour\ SYM may not be far from QCD in some cases." 
\end{itemize}

\begin{figure}[tb]
\centering
\scalebox{0.65}{ \includegraphics{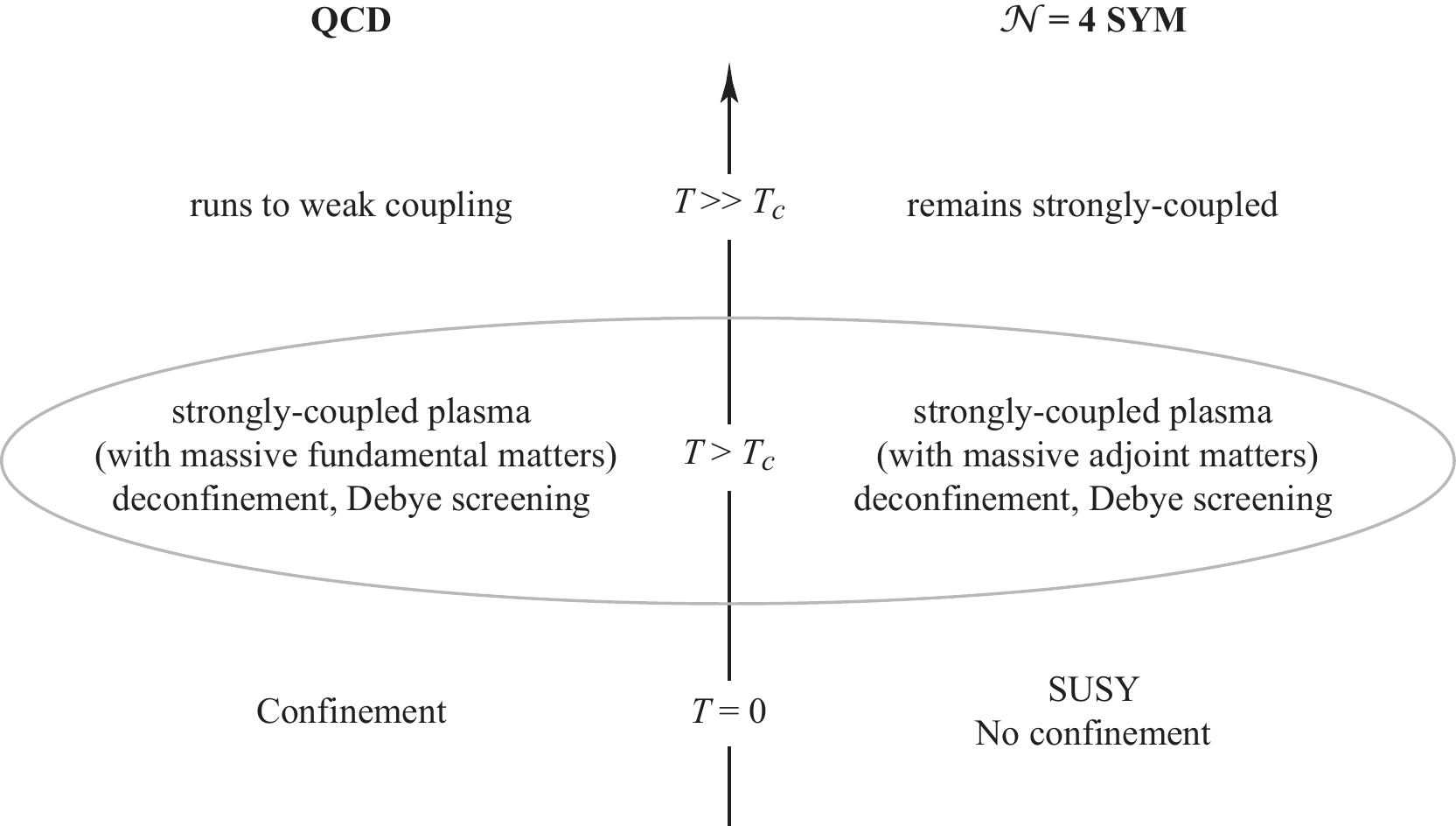} }
\vskip2mm
\caption{Comparison between QCD and \Nfour\ SYM in three temperature regions.}
\label{fig:comparison}
\end{figure}

To be clear, we are not saying that the \Nfour\ SYM is always close to QCD. To understand this point, let us compare the \Nfour\ SYM with QCD as we change the temperature (\fig{comparison}). 
In general, these two theories are very different. At high temperature, QCD runs to weak coupling, but the \Nfour\ SYM is scale invariant; if it is strongly-coupled at one temperature, it remains strongly-coupled. At zero temperature, these two theories are again different. QCD is confining whereas the \Nfour\ SYM is not confining due to supersymmetry.

However, at intermediate temperature, both theories are strongly-coupled plasmas, and they show similar behaviors such as the deconfinement and the Debye screening. Compared with QCD, the \Nfour\ SYM has extra matter fields in the adjoint representation. However, at finite temperature, both adjoint fermions and scalars have masses $m^2 \simeq O(\lambda T^2)$. This effect breaks supersymmetry at finite temperature%
\footnote{We loosely mentioned ``supersymmetric gauge theories" even at finite temperature, but this means gauge theories which have supersymmetry at zero temperature. }. 
Then, one expects that the \Nfour\ SYM at finite temperature is close to a pure Yang-Mills theory. \textit{It is this region in which one expects that supersymmetric gauge theories may be close to QCD.}

\begin{figure}[tb]
\centering
\scalebox{0.6}{ \includegraphics{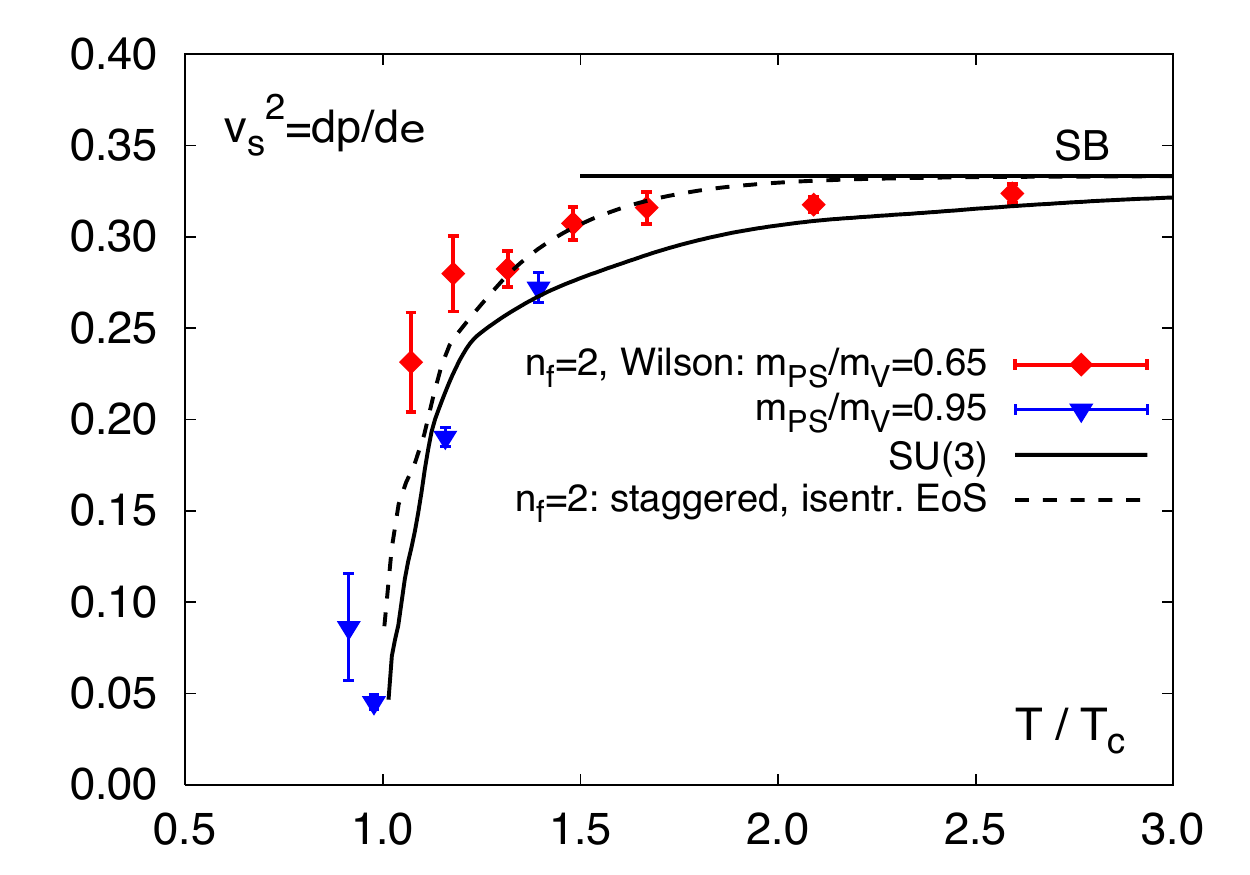} }
\vskip2mm
\caption{The speed of sound by lattice simulations \cite{Karsch:2006sf}. The results of several groups by different methods are shown (\eg, different lattice fermion actions and different masses), but they all have the similar tendency. ``SB" represents the scale-invariant value $1/3$.}
\label{fig:lattice_sound}
\end{figure}

For example, compare the speed of sound. In a scale-invariant theory, the energy density $\varepsilon$ and the pressure $P$ are related by $P=3\varepsilon$, 
so the speed of sound $c_s$ is given by
\be
c_s^2 = \frac{dP}{d\varepsilon}=\frac{1}{3}~.
\ee
According to lattice simulations (\fig{lattice_sound}), the speed of sound for QCD deviates from $1/\sqrt{3}$ significantly near the transition temperature $\Tc$, but it quickly approaches $1/\sqrt{3}$ around $T > 2\Tc$. Thus, the \Nfour\ SYM may be a good approximation to QCD in this temperature region.

\begin{figure}[tb]
\centering
\scalebox{0.6}{ \includegraphics{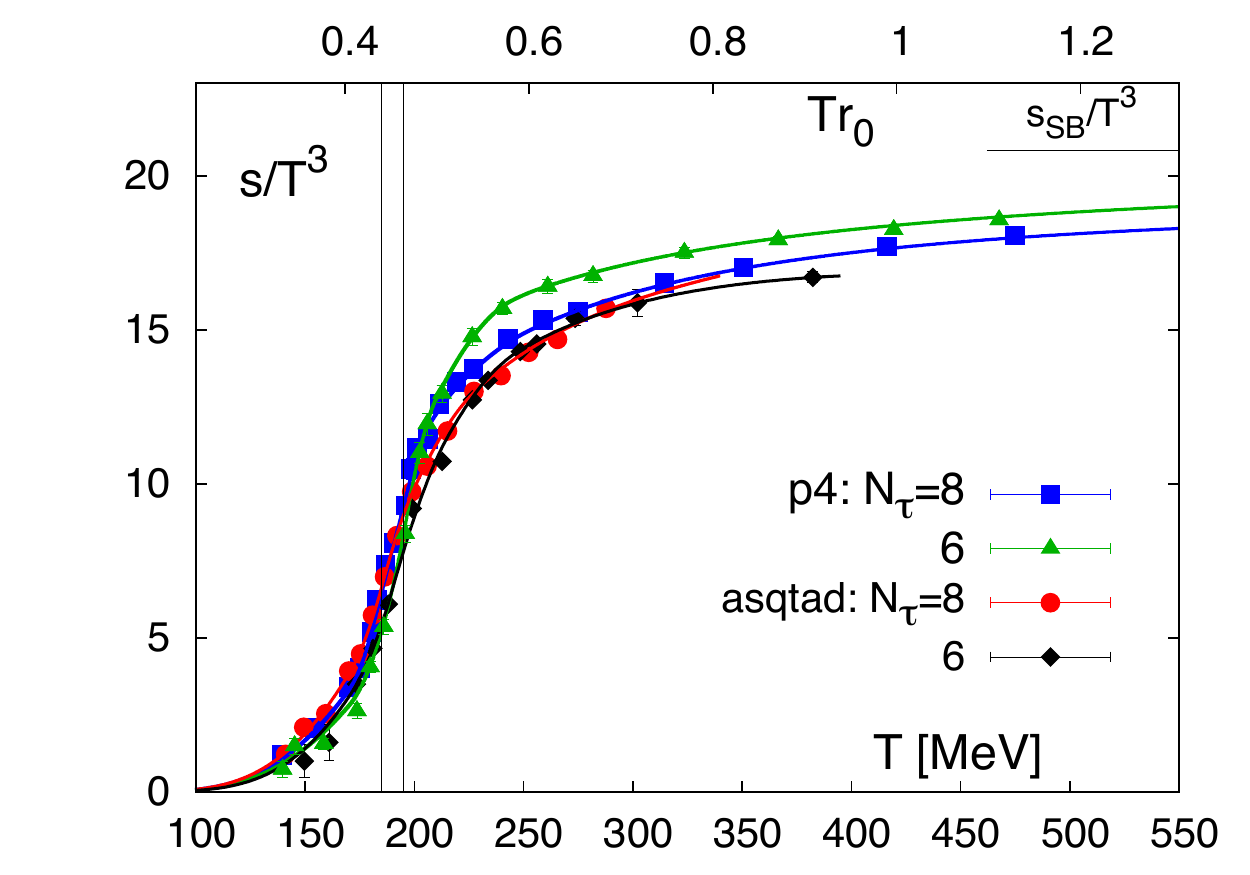} }
\vskip2mm
\caption{The entropy density by lattice simulations \cite{Bazavov:2009zn}. `SB" represents the scale-invariant free gas one.}
\label{fig:lattice_entropy}
\end{figure}%

As lattice simulation results, let us look at the entropy density as well. According to AdS/CFT, the entropy density of the \Nfour\ SYM at strong coupling becomes $3/4$ of the free gas result (\sect{sads_thermo}). A similar behavior has been seen in lattice simulations (\fig{lattice_entropy}). Just as the speed of sound, one should not compare the \Nfour\ SYM with QCD near the transition temperature. The numerical coefficient of the entropy density decreases slightly from the free gas result%
\footnote{This decrease does not seem to agree with $3/4$ of the AdS/CFT result. But (1) the number $3/4$ is likely to be theory-dependent \cite{Nishioka:2007zz}; (2) the AdS/CFT result is the strong coupling result. One has to take these issues into account for comparison.}. 

\section{Other issues
}

\subsection{Viscosity bound
}

So far, we considered relativistic Yang-Mills plasmas. \index{viscosity bound} For those plasmas, AdS/CFT gives $\eta/s=1/(4\pi)$ in the strong coupling limit. At finite coupling, one expects
\be
\frac{\eta}{s} \geq \frac{\hbar}{4\pi \kB}
\label{eq:bound}
\ee
as mentioned in \sect{implications}. 

\begin{figure}[tb]
\centering
\scalebox{0.6}{ \includegraphics{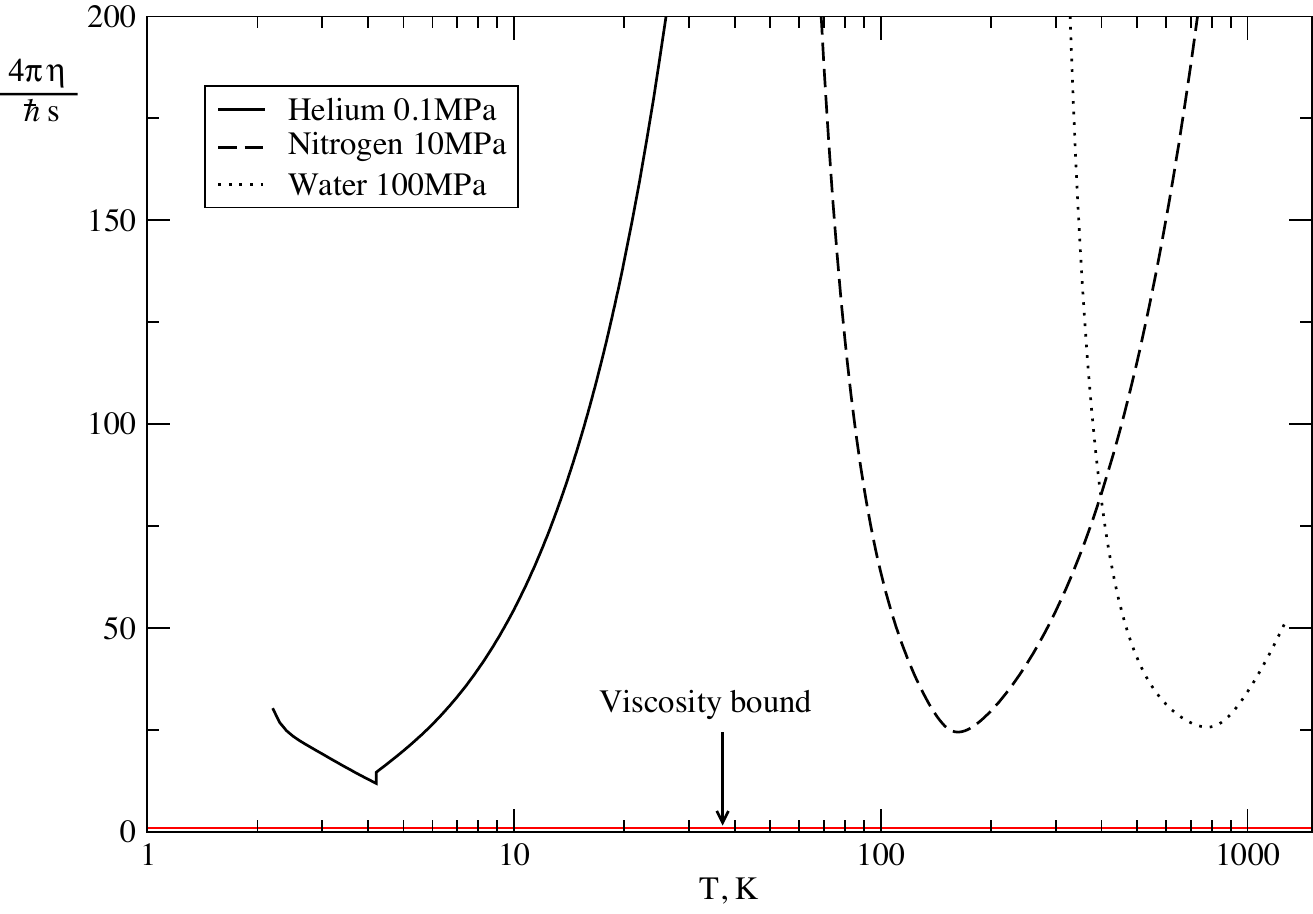} }
\vskip2mm
\caption{$\eta/s$ for helium, nitrogen, and water (under certain pressures) \cite{Kovtun:2004de}. There is a horizontal line which almost overlaps with the horizontal temperature axis: this is the AdS/CFT value.
}
\label{fig:bound}
\end{figure}

But there is a conjecture \cite{Kovtun:2004de} which claims that \textit{any} fluid satisfies \eq{bound}%
\footnote{See, \eg, Ref.~\cite{Cremonini:2011iq} for the current status of the conjecture though.}. 
Figure~\ref{fig:bound} shows $\eta/s$ for several fluids. These fluids certainly satisfy the relation.

A simple argument suggests the existence of the bound. In general, the shear viscosity and the entropy density of a fluid are given by
\be
\eta \simeq \rho \bar{v}l_\text{mfp},\qquad s \simeq \frac{\rho}{m}\kB
\ee
($\rho$: mass density, $\bar{v}$: mean velocity of particles of the fluid, $l_{\rm mfp}$: mean-free path, $m$: particle mass). Then, the viscosity bound $\eta/s \gtrsim \hbar/\kB$ implies $l_\text{mfp} \gtrsim \hbar/(m\bar{v})$, namely the mean-free path should be longer than the de~Broglie wavelength. For usual fluids, the particle picture is valid, so the relation is expected to hold.

Care is necessary for superfluids. A superfluid such as the liquid helium $^4$He has a nonzero viscosity (\fig{bound}). It is true that no viscous resistance is observed when the liquid helium goes through a narrow pipe. On the other hand, a viscous drag is observed when a test body is moved in the liquid, which indicates that the liquid helium has a component with a nonzero viscosity. 

According to the \keyword{two-fluid model}, a superfluid consists of the superfluid component and the normal component. The superfluid component has no viscosity, but the normal component has a nonzero viscosity, so a superfluid has a nonzero viscosity as a whole. The viscous drag in the liquid is caused by the normal component. The normal component represents the effect of thermal fluctuation, and it always exists at finite temperatures. The particle description is valid for the normal component. 




\subsection{Corrections to the large-$\Nc$ limit
\advanced}\label{sec:coupling}

The AdS/CFT result $\eta/s=1/(4\pi)$ is the strong coupling limit or the large-$\Nc$ limit. In order to compare the AdS/CFT result with QGP more accurately, one may need to take into account the corrections to the large-$\Nc$ limit. 

AdS/CFT has two independent parameters, $\lambda$ and $\Nc$. Their values for QGP differ from AdS/CFT:
\begin{itemize}

\item QGP: $\lambda=O(10)$, $N_c=3$ ($g_\text{YM}=O(1)$) \label{item:qgp}
\item AdS/CFT: $\lambda\rightarrow\infty$, $N_c\rightarrow\infty$ ($g_\text{YM}\rightarrow 0$) \label{item:ads}

\end{itemize}
So, there are two kinds of corrections, $1/\lambda$-corrections and $1/\Nc$-corrections. Here, we briefly discuss these corrections.



\subsubsection*{$1/\lambda$-corrections}

The $1/\lambda$-corrections correspond to the $\alpha'$-corrections \index{alpha'-corrections@$\alpha'$-corrections} from the gravitational point of view (\sect{towards_ads}). For the \Nfour\ SYM, the $\alpha'$-corrections in the dual supergravity take the form
\be
\action = \frac{1}{16\pi G_{10}}\int d^{10}x \sqrt{-g} 
\left\{ R + l_s^6 O(R^4) + \cdots \right\}~.
\label{eq:alpha'_N=4}
\ee
From the AdS/CFT dictionary \eqref{eq:dictionary_Ch12}, $l_s^6$ means $O(\lambda^{-3/2})$ corrections. If one takes the corrections into account, $\eta/s$ becomes \cite{Buchel:2004di,Buchel:2008sh}
\be
\frac{\eta}{s} = \frac{1}{4\pi}
\left( 1+\frac{120}{8}\zeta(3)\frac{1}{\lambda^{3/2}}+\cdots \right)~,
\label{eq:lambda_correction}
\ee
where $\zeta(3)=1.20206\ldots$. To estimate the magnitude of the correction, use $\alpha_\text{QCD}:=g_\text{YM}^2/(4\pi)=1/2$ or $\lambda=6\pi$, which is often used for QGP analysis. The correction is about 22\% increase if we use $\lambda=6\pi$.

Although the universality holds for many different black holes, why does $\eta/s$ deviate from $1/(4\pi)$ with corrections?
\begin{itemize}

\item First, the area law $s=a/(4G_5)$ holds \textit{as long as the gravitational action takes the form of the Einstein-Hilbert action} (\sect{BH_entropy}). But the corrections \eqref{eq:alpha'_N=4} do not satisfy the assumption. So, the area law does not hold.
\item Second, when we solve the tensor mode perturbation, we use the fact that the perturbation equation takes the form of the massless scalar field. But the $1/\lambda$-corrections are higher derivative terms, so the perturbation equation no longer takes the form of the massless scalar field. 

\end{itemize}
The deviation from $1/(4\pi)$ arises from these effects.


However, unlike the $\lambda\rightarrow\infty$ result, there is no universality for the corrections  \eqref{eq:lambda_correction}. Namely, the result is specific to the \Nfour\ SYM and is not universally true. This is because the explicit form of $\alpha'$-corrections depend on the gauge theory or the dual gravitational theory one considers. Moreover, generally many terms contribute to the $\alpha'$-corrections and not all terms are known. So, it would be nice if one could compute the $1/\lambda$-corrections in the other gauge theories (\eg, D$p$-branes with $p\neq3$), but it is not possible in general. Thus, it is not clear either if the above estimate of 22\% increase is realistic or not. 

Instead , one would simply choose a gravitational theory with some $\alpha'$-corrections one can handle and study the $1/\lambda$-corrections of the theory \cite{Brigante:2007nu}. In this case, the dual gauge theory is not clear though. 

\subsubsection*{$1/\Nc$-corrections}

The $1/\Nc$-corrections correspond to string loop corrections or quantum gravity corrections from the gravitational point of view. String theory is the theory of quantum gravity, but in general it is difficult to evaluate quantum gravity effects. Thus, the $1/\Nc$-corrections are difficult to compute as well, and the corrections have not been evaluated for the \Nfour\ SYM. 

However, the $1/\Nc$-corrections have been studied from different motivations. First of all, there are phenomena which are never visible in the large-$\Nc$ limit. Such a phenomenon is essentially tied to fluctuations, and they are particularly important in low spatial dimensions. For example, the fluctuations become large in low spatial dimensions, and there is no spontaneously symmetry breaking. But the fluctuations are suppressed in the large-$\Nc$ limit so that one has spontaneously symmetry breaking even in low spatial dimensions (Sects.~\ref{sec:exponents} and \ref{sec:H^3}).

In principle, such a phenomenon can be explored by taking the $1/\Nc$-corrections into account. In practice, quantitative computations are difficult because they are quantum gravity effects. But for those applications, one is mainly interested in qualitative behaviors of the phenomena, so the discussion is possible to some extent \cite{Kovtun:2003vj,CaronHuot:2009iq,Anninos:2010sq,Natsuume:2010bs}. 

To conclude, the $1/\lambda$-corrections have been evaluated for the \Nfour\ SYM and may be small enough, but it is not clear for the other gauge theories. On the other hand, the $1/\Nc$-corrections have not been evaluated for the \Nfour\ SYM. Therefore, at this moment, it is not clear if the strong coupling expansion in AdS/CFT is good enough to describe the real strongly-coupled QGP.

\subsection{Revisiting diffusion problem: hydrodynamic application
\advanced}\label{sec:2nd_order}

When we solve the perturbation equation for the tensor mode, we solved it as a power series expansion in $\omega$. As one can see from \eq{perturbative_sol}, there should be $O(\omega^2, q^2)$-corrections to $\vevG$. The corrections must be the ones for the response $\delta \bra T^{\mu\nu} \ket$, but how can we interpret them in the boundary theory?

We utilized hydrodynamics for the boundary interpretation. From the field theory point of view, ``standard" hydrodynamics corresponds to low orders in effective theory expansion. Namely, 
\begin{itemize}
\item The zeroth order: the perfect fluid.
\item The first order: the viscous fluid. 
\end{itemize}
In \chap{hydro}, we stopped at the first order, but one can continue the derivative expansion. The resulting hydrodynamics is called \keyword{causal hydrodynamics} or \keyword{second-order hydrodynamics}. In particular, the \keyword{Israel-Stewart theory} \cite{muller,Israel:1976tn,Israel:1979wp} is well-known and has been widely studied in the literature.  

In such a theory, new parameters or new transport coefficients appear from the higher order terms, but little is known about them. Just like hydrodynamics, second-order hydrodynamics is a framework: it does not tell the values of these parameters. However, AdS/CFT can determine these new transport coefficients if one solves the bulk perturbation equation up to higher orders \cite{Baier:2007ix2,Bhattacharyya:2008jc,Natsuume:2007ty}. The results obtained in AdS/CFT have been immediately applied to QGP hydrodynamic simulations. 

\subsubsection*{The diffusion case
}

To understand the physical meaning of causal hydrodynamics or second-order hydrodynamics, let us go back to the diffusion problem. In \sect{diffusion}, we saw that the diffusion equation has the following solution:
\be
\rho(t,x) = \frac{1}{\sqrt{4\pi Dt}}\exp\left(-\frac{x^2}{4Dt}\right)~.
\ee
This solution starts with the delta-function distribution, and it is smeared as the time passes. The point is that the solution is nonzero everywhere although it is exponential. In particular, it is nonvanishing even outside the light-cone $x>ct$. This implies \textit{an infinite velocity for the signal propagation, so it does not satisfy causality.} 

Mathematically, the diffusion equation is parabolic (\fig{diff_eqs}): it has first derivative in time but has second derivatives in space. Such an asymmetry is important for the physics of diffusion which is asymmetric in time, but this means that it does not satisfy causality. 

So, what is wrong? We used two equations to derive the diffusion equation, the conservation equation and the constitutive equation (Fick's law\index{Fick's law}). The conservation equation has the fundamental importance, but Fick's law is a phenomenological equation, so what is wrong is Fick's law:
\be
J_i= -D \del_i \rho~.
%
\ee 
Suppose that the charge gradient vanishes at some instance, $\del_i \rho=0$ for $t=0$. Then, Fick's law  tells that the current vanishes immediately, \ie, $J_i(t \geq 0)=0$. But this sounds unnatural; in reality, the current should decay in some finite time period. So, modify Fick's law as follows: 
\be
\tR \del_0 J_i + J_i= -D \del_i \rho~.
\label{eq:modified}
\ee 
The first term is a new term added, and $\tR$ is a new transport coefficient. When $\tR=0$, it gives the original Fick's law. Equation~\eqref{eq:modified} has the solution where the current decays exponentially: $J_i(t \geq 0)=J_i(0) e^{-t/\tR}$. Namely, $\tR$ is the relaxation time for the charge current%
\footnote{
One often calls it just as the ``relaxation time" in literature, but one had better specify what is relaxing. For example, the relaxation time for the charge density is given by $\tau_{\rho} \simeq 1/(Dq^2)$. The relaxation time here is the one for the current. }.

\begin{figure}[tb]
\begin{center}
\begin{tabular}{lrl}
Parabolic: &
 $ - \del_0 \rho + D \del_i^2 \rho = 0 $ & diffusion equation \\
Hyperbolic: &
 $ - \del_0^2 \phi + \del_i^2 \phi = 0 $ & wave equation  \\
Elliptic: &
 $ \del_i^2 \phi = 0 $ & Laplace equation 
\end{tabular}
\caption{The classification of differential equations and their typical examples in physics.}
\label{fig:diff_eqs}
\end{center}
\end{figure}

Now, combine the conservation equation with the modified law \eqref{eq:modified}: in this case, one gets the \keyword{telegrapher's equation}:
\be
\tR \del_0^2\rho + \del_0 \rho - D \del_i^2 \rho = 0~.
\label{eq:telegrapher}
\ee
This is a \textit{hyperbolic} equation. The new term is second derivative in time, whereas the original term is first derivative in time, so the new term becomes important for rapid evolution. 

A combination of $D$ and $\tR$ gives a quantity with a dimension of speed: 
\be
v_\text{\,front}^2 :=D/\tR~. 
\label{eq:v_front}
\ee
This gives the characteristic velocity for the signal propagation. 
So, if $v_\text{\,front} < c$, causality is fine. This is the origin of the name, causal hydrodynamics. Let us rewrite the telegrapher's equation in momentum space. \eq{telegrapher} is written as 
\be
- \tR \omega^2 - i \omega+ Dq^2 = 0~,
\ee
so the dispersion relation becomes
\be
\omega = - i D q^2 - i D^2 \tR q^4 + O(q^6)
\label{eq:telegrapher_dispersion}
\ee
for the hydrodynamic mode. The first term is the familiar dispersion relation \eqref{eq:diffusion_dispersion} we saw earlier, and the second term is the correction due to causal hydrodynamics. 

What we have done so far is just an effective theory expansion in higher orders, and $\tR$ is a new transport coefficient associated with the higher order expansion. The original diffusion problem corresponds to the case where the current relaxation happens very fast so that we can approximate $\tR=0$. Namely, the time scale $\tau$ of the problem is given by $\tau \gg \tR$.

From the point of view of an effective theory, one had better include $O(\omega^3, q^3)$ terms further at higher energy and shorter wavelength which we ignored. In order to check causality, one actually needs a dispersion relation which is valid for all energy. So, the issue of causality can be answered only if we sum all terms in the effective theory expansion. Equation~\eqref{eq:v_front} is just a rough estimate. This raises the question if second-order hydrodynamics is really useful. Actually, standard ``first-order" hydrodynamics such as the diffusion equation has the other difficulties, and second-order hydrodynamics is useful to solve these problems (see below). 

\subsubsection*{Fluid case
}

Now, move from the diffusion problem to the fluid problem. The problem of fluids is rather complicated, so we make only some general remarks. For fluids, we again have two equations:
\begin{itemize}
\item 
The conservation equation: $\nabla_\mu T^{\mu\nu} = 0$,
\item 
The constitutive equation:
\begin{align}
T^{\mu\nu} &= (\varepsilon+P)u^\mu u^\nu + P g^{(0)\mu\nu} 
+ \tau^{\mu\nu}~,
\\
\tau^{\mu\nu} &= 
- P^{\mu\alpha} P^{\nu\beta} \left\{ 
\eta \left( \nabla_\alpha u_\beta \!+\! \nabla_\beta u_\alpha 
- \frac{2}{3} g^{(0)}_{\alpha\beta} \nabla\!\cdot\! u \right) 
+ \zeta g^{(0)}_{\alpha\beta} \nabla\!\cdot\! u 
\right\}~.
\label{eq:dissipative_curved_again}
%
\end{align}

\end{itemize}
Once again, the constitutive equation is imposed as a phenomenological equation. Higher orders in effective theory correspond to adding higher order terms in the derivative expansion. As the result, the $O(\omega^2,q^2)$-corrections appear in \eq{dissipative_curved_again}. For example, the response $\delta\bra \tau^{xy} \ket$ is modified from \eq{Txy_vs_eta_again} as
\be
\delta \bra \tau^{xy}  \ket 
= \left[ 
i\omega \eta -\eta \tau_\pi \omega^2+\frac{\kappa}{2}\{(p-2) \omega^2+q^2 \}
\right]
\sourceG~,
\label{eq:kubo_causal}
\ee
(for $p\geq3$ conformal fluids) if we include $O(\omega^2,q^2)$-corrections \cite{Baier:2007ix2}. Here, $\tau_\pi$ and $\kappa$ are two of new transport coefficients which appear in second-order hydrodynamics. The coefficient $\tR$ is the relaxation time for the current. Similarly, $\tau_\pi$ has the physical meaning as the relaxation time of the momentum flux. From the bulk point of view, these new terms correspond to the $O(\omega^2,q^2)$-corrections to $\vevG$.
The bulk computation determines 
\be
\tau_\pi = \frac{2-\ln2}{2\pi T}~, \quad
\kappa = \frac{\eta}{\pi T}~,
\label{eq:transport_causal}
\ee
for the \Nfour\ SYM in the large-$\Nc$ limit (\probset{2nd}).

AdS/CFT determines these transport coefficients, but AdS/CFT tells us more. The formalism of second-order hydrodynamics must be consistent with AdS/CFT results. Second-order hydrodynamics has a long history: it has been discussed more than 100 years if we include nonrelativistic cases. But no unique formalism has been obtained. Namely, the Israel-Stewart theory may not be the most general effective theory. 

In fact, in order to interpret AdS/CFT results consistently, there should exist new higher order terms which were missing in the Israel-Stewart theory. These new terms introduce a further set of transport coefficients at second order which had not been discussed. The coefficient $\kappa$ in \eq{kubo_causal} is one example. Namely, 
\begin{center}
\fbox{
\begin{tabular}{l}
One can cross-check the hydrodynamic framework itself using AdS/CFT results.
\end{tabular}
}
\end{center}


As we mentioned above, first-order hydrodynamics has difficulties other than causality. Second-order hydrodynamics is useful to solve these problems:
\begin{enumerate}

\item \textit{Unphysical instabilities:}
Standard relativistic first-oder hydrodynamics has unphysical instabilities \cite{hiscock_lindblom2,hiscock_lindblom3}. Second-order hydrodynamics is free from this problem (at least for linear perturbations). From a practical point of view, the instability implies that we have no control on numerical simulations as soon as viscosity is introduced. Numerical simulation for first-order hydrodynamics simply does not exist. For a numerical simulation, we are forced to consider second-order hydrodynamics. That is the reason why the AdS/CFT results for second-order hydrodynamics was immediately applied to QGP numerical simulations.

\item \textit{Frame dependence:}
We have discussed the Landau-Lifshitz frame\index{Landau-Lifshitz frame} and the Eckart frame\index{Eckart frame} (\sect{viscous_fluid&current}). For a nonequilibrium state, one in general has various flows associated with different currents, so the notion of the ``fluid rest frame" is ambiguous: a different flow defines a different ``fluid rest frame." Two commonly used choices are the Landau-Lifshitz frame and the Eckart frame. 

In principle, these are just a choice of rest frames. However, they are not just a choice of frame in first-order theories, and they are actually different theories. This is because the transformation of these ``frames" is impossible within the framework of first-order theories. Namely, first-order theories are frame-dependent. 

In order to guarantee the frame-independence, one needs to take into account some second-order terms. But then one had better consider the full second-order theory. 
This problem is clear in the above instability problem. 
In the Eckart frame, the instability is more severe than the Landau-Lifshitz frame.

\end{enumerate}

\section{\titlesummary}

\begin{itemize}
\item 
According to AdS/CFT, a strongly-coupled large-$\Nc$ plasma has the universal value of $\eta/s=1/(4\pi)$. This is true for all known examples (in the large-$\Nc$ limit). The value is extremely small compared with ordinary materials.
\item 
This universality is tied to the universal nature of black holes, \eg, the area law of \bh entropy. 
\item 
QGP has a small $\eta/s$ whose value is close to the AdS/CFT prediction. This is supported by heavy-ion experiments and lattice simulations.
\end{itemize}

\titlenewterms

\begin{multicols}{2}
\noindent
tortoise coordinate\\
elliptic flow\\
viscosity bound\\
causal/second-order hydrodynamics\\
\newtermapp{quasinormal mode}
\end{multicols}

\endofsection

\ifx\nameofpaper\undefined 
  \usepackage{macro_natsuume} 
  \def\beginsection{\section*}
  \def\endofsection{\end{document}} 
  \input draft_header.tex
\else 
  \def\beginsection{\chapter}
  \def\endofsection{ } 
\fi

\section{Appendix: Tensor mode action
\advanced}\label{sec:action_tensor}

In this appendix, we derive the action for the tensor mode and evaluate the on-shell action. The on-shell action then gives the field/operator correspondence \eqref{eq:response_hydro_ads} for the gravitational perturbation.

As we saw in \sect{SAdS_free}, the gravitational action consists of the bulk action $\action_\text{bulk}$, the Gibbons-Hawking action $\action_\text{GH}$, and the counterterm action $\action_\text{CT}$%
\footnote{We consider the Euclidean action in \sect{SAdS_free} but we consider the Lorentzian action here.}:
\be
\action = \action_\text{bulk}+ \action_\text{GH} + \action_\text{CT}~.
%
\ee
Below we evaluate each action for the tensor mode perturbation. We use the coordinate system \eqref{eq:sads5_u_again}. We also set $\phi = g^{xx} h_{xy}$.
%
%
We make the Fourier transformation in the boundary spacetime directions $x^\mu=(t,\bmx)=(t,x,y,z)$: 
\be
\phi (t, \bmx, u)
  = \int (dk)\,
  e^{-i \omega t + i \bmq \cdot \bmx} \phip(u)~, 
\quad (dk):=\frac{d^4k}{(2\pi)^4}~.
%
\ee

\subsubsection*{The bulk action
}\label{sec:bulk_tensor}

The bulk action is the standard Einstein-Hilbert action:
\begin{align}
\action_\text{bulk} &= \frac{1}{16\pi G_5} \int d^{5}x\, \sqrt{-g} \left( R -2 \Lambda \right) 
= \frac{r_0^4}{16\pi G_5 L^5} \hatS_\text{bulk}~, \\
2\Lambda &= - \frac{p(p+1)}{L^2}~.
%
\end{align}
One can check that the factor in front of $\hatS_\text{bulk}$ is common to the other on-shell actions, so we will set $r_0=L=16\pi G_5=1$. For the tensor mode perturbation,
\be
\action_\text{bulk} \sim \action_0 + \action_2~,
%
\ee
where ``$\sim$" denotes terms up to $O(\phi^2)$, and one can show%
\footnote{This can be derived analytically, but one may use \textit{Mathematica}. }
\begin{align}
\action_0 &= -\int_0^1 du\, \frac{8V_4}{u^5}
  \\
\action_2 &= \int_0^1 du\, \left[
 \frac{h}{u^3}\left( \frac{3}{2}\phim'\cdot\phip'+2\phim\cdot\phip'' \right) 
- \frac{8}{u^4} \phim\cdot\phip' \right.
 \nonumber \\
& \qquad\left. + \left( \frac{\nw^2-\nq^2h}{2u^3h} + \frac{4}{u^5} \right) \phim\cdot\phip \right]~,
\label{eq:bulk_tensor}
\end{align}
where $V_4$ is the four-dimensional ``volume" in the $(t, \bmx)$-directions, and
\be
f_{-k} \cdot g_{k} :=\int(dk)\, f_{-k}g_{k}~, \quad
':=\del_u~, \quad
\nw := \frac{\omega}{T}~, \quad 
\nq := \frac{q}{T}~.
\nonumber
\ee
Note that the action contains the term with $\phi''$ unlike the usual scalar field. 

Let us consider a generic action of the form
\be
\action = \int du\, \calL(\phi, \phi', \phi'')~.
%
\ee
The variation of the action is given by
\begin{align}
\delta \action 
&= \int du\, 
\left( \frac{\del\calL}{\del\phi} \right) \delta\phi 
+ \left( \frac{\del\calL}{\del\phi'} \right) \delta\phi'
+ \left( \frac{\del\calL}{\del\phi''} \right) \delta\phi''
\label{eq:action_variation1} \\
&=  \left. \left\{ \frac{\del\calL}{\del\phi'} - \left( \frac{\del\calL}{\del\phi''} \right)' \right\} \delta\phi 
+ \frac{\del\calL}{\del\phi''} \delta\phi' \right|_\text{bdy} 
\nonumber \\
& + \int du\, \left\{ \left( \frac{\del\calL}{\del\phi''} \right)'' - \left( \frac{\del\calL}{\del\phi'} \right)' + \frac{\del\calL}{\del\phi} \right\} \delta\phi~.
\label{eq:action_variation2}
\end{align}
The second line of \eq{action_variation2} is the equation of motion. For the tensor mode, one can check that it coincides with the massless scalar one:
\be
\frac{u^3}{h} \left( \frac{h}{u^3} \phip' \right)' 
+ \frac{ \nw^2-\nq^2 h }{ \pi^2 h^2 } \phip = 0~.
\label{eq:scalar_eom_app}
\ee

Now, a quadratic action can be written as
\be
2 \action \sim \int du\, 
\left( \frac{\del\calL}{\del\phi} \right)\phi 
+ \left( \frac{\del\calL}{\del\phi'} \right)\phi'
+ \left( \frac{\del\calL}{\del\phi''} \right)\phi''~.
%
\ee
Following the similar steps as Eqs.~\eqref{eq:action_variation1}-\eqref{eq:action_variation2}, the on-shell action becomes%
\footnote{We write the on-shell action $\Sos$ as $\action$ for simplicity.}
\be
2 \Sos \sim 
\left. \left\{ \frac{\del\calL}{\del\phi'} - \left( \frac{\del\calL}{\del\phi''} \right)' \right\} \phi 
+ \frac{\del\calL}{\del\phi''} \phi' \right|_\text{bdy}~.
%
\ee
Using the explicit form of the action \eqref{eq:bulk_tensor}, one gets
\be
\hatS_\text{bulk} 
\! \xrightarrow{u\rightarrow0}{} \!
 \left.V_4 \left(\!-\frac{2}{u^4} \!+\!2 \right) 
 \!+\! \left( \frac{1}{u^4} \!-\!1 \right)\phim\cdot\phip \!-\! \frac{3}{2u^3}\phim\cdot\phip' \right|_{u=0}~.
\label{eq:bulk_tensor_bdy}
\ee

\subsubsection*{The Gibbons-Hawking action
}\label{sec:GH_tensor}

Upon variation of an action, surface terms arise. As usual in the variational problem, setting $\left. \delta\phi \right|_\text{bdy} = 0$ removes the terms which are proportional to $\left. \delta\phi \right|_\text{bdy}$. But the gravitational action \eqref{eq:bulk_tensor} contains the term with $\phi''$ unlike the standard field theory action. Then, the variation gives the terms which are proportional to $\left.\delta\phi'\right|_\text{bdy}$ as one can see on the first line of \eq{action_variation2}. The Gibbons-Hawking action \index{Gibbons-Hawking action} cancels the terms in order to have a well-defined variational problem:
\be
\action_\text{GH} = \frac{2}{16\pi G_{p+2}} \int d^{p+1}x\, \sqrt{-\gamma}\, K~.
%
\ee
See \sect{GH_free} for notations. For the SAdS$_5$ black hole, $n^u=-uh^{1/2}$ and $\sqrt{-\gamma} = u^{-4}h^{1/2}(1-\phi^2)^{1/2}$, so
\begin{align}
{\cal L}_\text{GH} 
&=  2\sqrt{-\gamma} K = 2 n^u \partial_u\sqrt{-\gamma} \\
&= -2uh^{1/2}\left[ u^{-4}h^{1/2}(1-\phi^2)^{1/2} \right]' \\
&\sim 2u^{-3} h \phi\phi'-2uh^{1/2}(u^{-4}h^{1/2})' \left(1-\frac{1}{2}\phi^2 \right)~.
\label{eq:GH_tensor}
\end{align}
Thus,
\be
\action_\text{GH} 
\xrightarrow{u\rightarrow0}{}
\left. V_4 \left(\frac{8}{u^4} \!-\! 4\right) 
\!+\! \left( -\frac{4}{u^4} \!+\! 2 \right)\phim\cdot\phip \!+\! \frac{2}{u^3}\phim\cdot\phip' \right|_{u=0}~.
\label{eq:GH_tensor_bdy}
\ee

The $\left. \delta\phi' \right|_{u=0}$ term which arises from the Gibbons-Hawking action cancels with the $\left. \delta\phi' \right|_{u=0}$ term which arises from the bulk action:
\begin{align}
\delta \calL_\text{bulk}
&= \left. \frac{\del\calL_\text{bulk}}{\del\phip''} \delta\phip' + \cdots \right|_{u=0}^{u=1} 
= \left. -\frac{2}{u^3}h\phim\cdot\delta\phip'+ \cdots \right|_{u=0}~, \\
\delta \calL_\text{GH}
&= \left. \frac{\del\calL_\text{GH}}{\del\phip'} \delta\phip' +\cdots \right|_{u=0}
= \left. + \frac{2}{u^3}h\phim\cdot\delta\phip' + \cdots \right|_{u=0}~,
%
\end{align}
where $\cdots$ represents the terms which are proportional to $\delta\phi$.

\subsubsection*{The counterterm action
}\label{sec:counter_tensor}

The counterterm action \index{counterterm action} is given by
\begin{align}
\action_\text{CT} &= - \frac{1}{16\pi G_{p+2}} \int  d^{p+1}x\, \sqrt{-\gamma} 
  \left\{ 
  \frac{2p}{L} 
  + \frac{L}{p-1}\cR \right. 
\nonumber \\
& 
  \left. - \frac{L^3}{(p-3)(p-1)^2} \left(\cR^{\mu\nu}\cR_{\mu\nu} - \frac{p+1}{4p}\cR^2\right)
+ \cdots
\right\}.
%
\end{align}
See \sect{counter_free} for notations. According to the Kubo formula, the shear viscosity arises at $O(\omega)$ term in the response, and one can ignore the $O(\cR)$ terms which contain at least two derivatives. Evaluating only the first term gives
\begin{align}
{\cal L}_\text{CT} 
&= -6 \sqrt{-\gamma} \\
&= -6u^{-4}h^{1/2}(1-\phi^2)^{1/2} \\
&\sim -6u^{-4}h^{1/2}\left(1-\frac{1}{2}\phi^2\right)~.
%
\end{align}
Thus, 
\be
\action_\text{CT} 
\xrightarrow{u\rightarrow0}{}
\left. V_4\left(-\frac{6}{u^4}+3\right) 
+ \left( \frac{3}{u^4} - \frac{3}{2} \right)\phim\cdot\phip \right|_{u=0}~.
\label{eq:counter_tensor_bdy}
\ee
\subsubsection*{The full on-shell action
}

To summarize our results \eqref{eq:bulk_tensor_bdy}, \eqref{eq:GH_tensor_bdy}, and \eqref{eq:counter_tensor_bdy},
\begin{align}
\action_\text{bulk} &= \left. V_4\left( -\frac{2}{u^4} + 2 \right) 
 + \left( \frac{1}{u^4 }-1 \right)\phim\cdot\phip - \frac{3}{2u^3}\phim\cdot\phip' \right|_{u=0}~,
 \nonumber \\
\action_\text{GH} &= \left. V_4\left( \frac{8}{u^4} - 4 \right) 
 + \left( -\frac{4}{u^4} + 2 \right)\phim\cdot\phip +\frac{2}{u^3}\phim\cdot\phip' \right|_{u=0}~,
 \nonumber \\
\action_\text{CT} &= \left. V_4\left( -\frac{6}{u^4}+3 \right) 
 + \left( \frac{3}{u^4} - \frac{3}{2} \right)\phim\cdot\phip \right|_{u=0}~.
 \nonumber
%
\end{align}
Combining these results, we get
\be
\Sos = \left. V_4-\frac{1}{2}\phim\cdot\phip+\frac{1}{2u^3}\phim\cdot\phip'  \right|_{u=0}~.
\label{eq:tensor_on_shell1}
\ee

\subsubsection*{Field/operator correspondence
}

If we recover dimensionful quantities, the on-shell action becomes
\be
\Sos = \frac{r_0^4}{16\pi G_5 L^5} 
\left\{ V_4 
+ \int (dk) 
\left. \left[ -\frac{1}{2}\phim\phip+\frac{1}{2u^3}\phim\phip'  \right] \right|_{u=0}
\right\}~.
\label{eq:tensor_bdy}
\ee
The first term which does not depend on the fluctuation is the Lorentzian version of the free energy computed in \sect{SAdS_free}. The third term takes the same form as the massless scalar case in \sect{scalar}, but we have an additional term, the second term. We discuss the implication of the second term below.

Since $\phip$ satisfies the equation of motion for the massless scalar, the asymptotic behavior is given by \eq{massless_scalar_falloff}:
\be
\phip \sim \phip^{(0)} \left(1+\phip^{(1)} u^4 \right)~.
%
\ee
Thus, the on-shell action becomes
\be
\Sos = \frac{r_0^4}{16\pi G_5 L^5} 
\left\{ V_4 
+ \int (dk)\,
\phim^{(0)} \left( - \frac{1}{2} + 2\phip^{(1)} \right) \phip^{(0)} \right\}~.
\label{eq:tensor_bdy_bc}
\ee

We now evaluate $\bra T^{xy}_k \ketj$. The perturbed action in the boundary theory is 
\be
\delta\action = \int d^4x\,  \frac{1}{2} h^{(0)}_{\mu\nu} T^{\mu\nu} =  \int d^4x\,  \sourceG T^{xy}
%
\ee
from the definition of the energy-momentum tensor. Then, the GKP-Witten relation gives the one-point function as
\be
\bra T^{xy} \ketj
= \frac{\delta \Sos}{\delta \sourceG} = \frac{\delta \Sos}{\delta \source}~.
%
\ee
Note that $\sourceG=\source$. But by taking into account the Lorentzian prescription in \sect{Lorentzian_prescription}, one obtains
\be
\bra T^{xy}_k \ketj =
\frac{r_0^4}{16\pi G_5 L^5} \left( - 1 + 4\phip^{(1)} \right) \phip^{(0)}~.
\label{eq:response_ads_full}
\ee
This is the desired result. The second term coincides with \eq{response_hydro_ads}.

Then, how about the first term of \eq{response_ads_full}? This comes from the action \eqref{eq:tensor_bdy} which is absent in the scalar field action. Because the on-shell action has an additional term, the field/operator correspondence \eqref{eq:kw} is modified from the scalar case as \eq{response_ads_full}. Namely, the rule ``the fast falloff as the response" is modified, 
and
$\phip^{(1)}$ does not represent the full response. 

The existence of the additional term has a natural hydrodynamic interpretation. We discussed the constitutive equation in the curved spacetime in \sect{viscosity_kubo}. In order to derive the Kubo formula, it is enough to consider only the dissipative part, $\tau^{\mu\nu}$. But note that the complete constitutive equation is given by
\be
T^{\mu\nu} = (\varepsilon+P)u^\mu u^\nu + P g^{(0)\mu\nu} 
+ \tau^{\mu\nu}
\label{eq:constitutive_curved_again}
\ee
in the curved spacetime. Then, when one adds gravitational perturbations, $\delta\bra T^{\mu\nu} \ket$ has an extra term to $\delta\bra\tau^{\mu\nu}\ket$: the second term of \eq{constitutive_curved_again} gives a term  which is proportional to pressure. If we expand%
\footnote{Note $g^{(0)\mu\nu} = \eta^{(0)\mu\nu} - h^{(0)\mu\nu}$.}
 $g^{(0)}_{\mu\nu} = \eta^{(0)}_{\mu\nu} + h^{(0)}_{\mu\nu}$, 
\begin{align}
\delta \bra T^{xy} \ket
&= P g^{(0)xy} -2 \eta \Gamma^0_{xy}  \\
&= - P {h}^{(0)}_{xy} - \eta \del_0h^{(0)}_{xy}  \\
&\xrightarrow{FT}{} (- P + i\omega \eta) \sourceG~.
\label{eq:Txy_vs_eta_full}
\end{align}
The first term of \eq{response_ads_full} represents this pressure term. Comparing Eqs.~\eqref{eq:response_ads_full} and \eqref{eq:Txy_vs_eta_full}, one gets
\be
P = \frac{1}{16\pi G_5 L} \left(\frac{r_0}{L}\right)^4
= \frac{\pi^2}{8} N_c^2 T^4~.
\ee
This agrees with \eq{pressure_sads} which is obtained from the free energy.

\section{Appendix: Some other background materials \advanced}\label{sec:background}

\subsection{Tensor decomposition}\label{sec:tensor}

\head{Maxwell perturbations}
Maxwell field has the gauge invariance
\be
A_M \rightarrow A_M+\del_M\Lambda~,
%
\ee
so we fix the gauge. We choose the gauge
\be
\text{gauge-fixing condition:}\quad A_u=0~.
%
\ee

The rest of components can be classified by their transformation properties under the little group $SO(2)$ just as the tensor decomposition of $J^\mu$ in \sect{linearized}. We make the Fourier transformation in the boundary spacetime directions $x^\mu=(t,\bmx)=(t,x,y,z)$
\be
A_{\mu} (t, \bmx, u)
  = \int (dk)\,
  e^{-i \omega t + i \bmq \cdot \bmx} A_{\mu}(\omega,\bmq, u)~,
\quad (dk):=\frac{d^4k}{(2\pi)^4}~,
%
\ee
and take $k_\mu=(\omega, 0,0, q)$. Then, 
\begin{alignat}{2}
&\text{vector: }&& A_{a}~, \\
&\text{scalar: }&& A_{0}, A_{z}~,
%
\end{alignat}
where $x^a = (x,y)$. The scalar mode components do not transform under $SO(2)$, and the vector mode components transform as vectors.

\head{Gravitational perturbations}
Similarly, one can make the tensor decomposition for gravitational perturbations. The gravitational perturbations have the gauge invariance, and they transform as  \eq{metric_transf_inf} under the coordinate transformation:
\be
h_{MN} \rightarrow
h_{MN} + \nabla_M\xi_N +  \nabla_N\xi_M~,
%
\ee
so we fix the gauge. In the five-dimensional spacetime, the coordinate transformation has five degrees of freedom $\xi^M$, so one needs five gauge-fixing conditions. We choose the gauge
\be
\text{gauge-fixing condition:}\quad h_{uM}=0~.
%
\ee
Again, the other components can be classified by the little group $SO(2)$:
\begin{alignat}{2}
&\text{tensor: }&& h_{xy}, h_{xx}=-h_{yy}~, \\
&\text{vector: }&& h_{0a}, h_{za}~, \\
&\text{scalar: }&& h_{00}, h_{0z}, h_{zz}, h_{xx}=h_{yy}~.
%
\end{alignat}

Figure~\ref{fig:mode_summary} summarizes transport coefficients one can derive from each mode.

\begin{figure}[tb]
\begin{center}
\begin{tabular}{|l|l|l|}
\hline
		& Gravitational field			& Maxwell field \\
\hline
tensor	& $\eta$ (Kubo, \sect{tensor_mode_sol})	& $-$ \\
vector	& $\eta$ (pole, \sect{vector})			& $\sigma$ (Kubo, \sect{SAdS5_diff}) \\
scalar	& $\eta, \zeta, c_s$ (pole)		& $D$ (pole,  \sect{SAdS5_diff}) \\
\hline
\end{tabular}
\caption{Tensor decomposition of bulk fields and transport coefficients one can derive from each mode. ``Kubo" and ``pole" indicate the methods used to derive transport coefficients. This book does not cover the gravitational scalar mode computation.}
\label{fig:mode_summary}
\end{center}
\end{figure}

\subsection{How to locate a pole}\label{sec:QNM}

In the text, we solved the tensor mode to derive $\eta/s$, and $\eta/s$ is the $O(\omega)$-coefficient of the tensor mode $\delta \bra T^{xy} \ket$ or $\vevG$. One can derive transport coefficients from other modes. One could compute the responses such as $\delta \bra T^{\mu\nu} \ket$ and $\delta \bra J^\mu \ket$ just like $\delta \bra T^{xy} \ket$, but it is often the case that our primary interest is hydrodynamic poles. 
For example, from the discussion in \sect{linearized}, the gravitational vector mode should have a pole at
\be
\omega = - i\frac{\eta}{Ts} q^2~.
%
\ee
One can determine $\eta/s$ by obtaining this dispersion relation. In such a case, we can simplify the problem slightly. 

As an example, consider the massless scalar field in \sect{scalar}. The bulk field has the asymptotic behavior
\be
\phi \sim \source\left(1+\vev u^4\right) = A + B u^4~,
\quad (u\rightarrow0)~.
%
\ee
Then, the retarded Green's function is given by
\be
G_R \propto \vev = \frac{B}{A}~.
%
\ee
Our purpose here is to find a pole which corresponds to $A=0$. Thus, it is enough to solve the perturbation equation under the boundary condition $A=0$ at the AdS boundary. As we will see below, this is possible if $\omega$ and $q$ satisfies a relation which is the dispersion relation. Namely, \textit{a vanishing slow falloff problem determines the location of poles.}

In general relativity, such a computation is called \keyword{quasinormal mode} computation%
\footnote{Quasinormal mode computations are traditionally carried out for asymptotically flat black holes, and one imposes the ``outgoing-wave" boundary condition at $r\rightarrow\infty$. But it is natural to impose the Dirichlet condition for an asymptotically AdS spacetime \cite{Horowitz:1999jd} since the spacetime has a potential barrier by the cosmological constant (\sect{geodesics_AdS}).}. 
The prefix ``quasi" indicates that the poles are located on the complex $\omega$-plane unlike normal modes. From the \bh point of view, the complex frequency represents the absorption by the \bh and comes from the ``incoming-wave" boundary condition on the horizon. From the hydrodynamic point of view, this represents the dissipation, \eg, by the viscosity. 

\section{Appendix: Gravitational vector mode computation
\advanced}\label{sec:vector}

In the text, we have seen the tensor mode computation, but we can determine transport coefficients from the vector mode and the scalar mode as well. For example, the vector mode has a pole at
\be
\omega = - i\frac{\eta}{Ts} q^2~.
\label{eq:vector_pole}
\ee
Here, we compute $\eta/s$ from the vector mode and check that the result agrees with the tensor mode computation. Since we are interested in a pole, we solve the vector mode under the vanishing slow falloff condition (\sect{QNM}). 

\subsubsection*{Perturbation equation
}

We would like to find the pole of the vector mode of $T^{\mu\nu}$, so we consider the vector mode perturbation. 
The vector mode perturbations are $h_{0a}$ and $h_{za}$. The $a=x$ and $a=y$ cases are the same from symmetry, so we consider the $a=x$ case. We first write down the perturbation equation for the vector mode. One can of course derive it from the Einstein equation, but it is easier to derive it if one utilizes its symmetry.

The vector mode satisfies a (modified) Maxwell equation. This is because there is a translational invariance along $x$. The translational invariance allows us to consider a ``fictitious" $S^1$-compactification. Let the index $\alpha$ run through only noncompact directions (directions except $x$). By the $S^1$-compactification, the five-dimensional metric $g_{MN}$ is decomposed as 
\be
g_{MN} \rightarrow  
g_{xx}~, g_{\alpha x}~, \text{and } g_{\alpha\beta}~.
%
\ee
%
%
From the ``four-dimensional" point of view%
\footnote{Note that four-dimensions here refer to noncompact directions $(t,y,z,u)$ not to the four-dimensional boundary directions $(t,x,y,z)$.}, one cannot ``see" the $x$-direction, so these components behave as a scalar, a vector, and a tensor, respectively. In particular, $g_{\alpha x}=h_{\alpha x}$ behaves as a Maxwell field which is known as a Kaluza-Klein gauge field. Let us write the metric as
\be
ds_5^2 = g_{MN} dx^M dx^N = e^{2\sigma} (dx+A_\alpha dx^\alpha)^2 + g_{\alpha\beta} dx^\alpha dx^\beta~,
\label{eq:metric_KK}
\ee
where $e^{2\sigma}:=g_{xx}$. In our case, only nonzero vector components are $h_{0x}=g_{xx} A_0$ and $h_{zx}=g_{xx} A_z$. 
Upon the compactification, the five-dimensional action becomes%
\footnote{\advanced See the other textbooks for the derivation of this action, but the form of the action is determined by symmetry. The metric \eqref{eq:metric_KK} is invariant under
\be
x \rightarrow c x~, \quad
A_\alpha \rightarrow c A_\alpha~, \quad
e^\sigma \rightarrow c^{-1} e^\sigma~, \quad
g_{\alpha\beta} \rightarrow g_{\alpha\beta}~.
\label{eq:scale_transf_KK}
\ee
The action \eqref{eq:action_KK} should keep the invariance, which determines the $\sigma$-dependence. Symmetry also requires that the components $A_\alpha$ behave as the Maxwell field. The five-dimensional coordinate transformation contains the transformation
\be
x \rightarrow x - \Lambda(x^\alpha)~.
%
\ee
From the four-dimensional point of view, the coordinate transformation is nothing but the gauge transformation:
\be
A_\alpha \rightarrow A_\alpha + \del_\alpha \Lambda~.
%
\ee
}
\begin{align}
\action_5 &= \frac{1}{16\pi G_5} \int d^5x\, \sqrt{-g_5} R_5 
\\
&= \frac{1}{16\pi G_5} \int dx \int d^4x\, \sqrt{-g_4} e^\sigma \left( R_4 -\frac{1}{4}e^{2\sigma} F_{\alpha\beta}^2 \right)~.
\label{eq:action_KK}
\end{align}
%
%
%
Note that the Kaluza-Klein gauge field action has the extra factor $e^{3\sigma}=g_{xx}^{3/2}$. Thus, the $A_\alpha$ field equation is given by
\be
\del_\beta(e^{3\sigma}\sqrt{-g_4} F^{\alpha\beta})=0~.
%
\ee

The $\alpha=u$, 0, and $z$ components of the field equation are
\begin{align}
%
g^{00} \omega A_0' - q g^{zz} A_z' &=0~,
\label{eq:maxwell1} \\
\partial_u \left( \sqrt{-\tilg} g^{00} g^{uu} A_0'\right)
 -  \sqrt{-\tilg} g^{00} g^{zz} \left( \omega q A_z + q^2 A_0 \right) &=0~,
\label{eq:maxwell2} \\
\partial_u \left( \sqrt{-\tilg} g^{zz} g^{uu} A_z'\right) 
 -  \sqrt{-\tilg} g^{00} g^{zz} \left( \omega q A_0 + \omega^2 A_z \right) &=0~,
 \label{eq:maxwell3}
\end{align}
where $~{}' := \partial_u$ and $\sqrt{-\tilg}:=e^{3\sigma}\sqrt{-g_4}=e^{2\sigma}\sqrt{-g_5}$. From Eqs.~\eqref{eq:maxwell1} and \eqref{eq:maxwell2}, one gets a decoupled equation for $A_0'$ only:
\be
\frac{d}{du} \left[ 
\frac{\partial_u ( \sqrt{-\tilg} g^{00} g^{uu} A_0')} {\sqrt{-\tilg} g^{00} g^{zz}}
\right] 
+ \left(-\frac{g^{00}}{g^{zz}} \omega^2 - q^2\right) A_0' =0\,.
\label{eq:master1}
\ee
This is the perturbation equation for the vector mode.
For the SAdS$_5$ black hole, \eq{master1} reduces to
\begin{align}
\frac{1}{hu^3} \left( hu^3 F' \right)'+ \frac{\nw^2 - \nq^2 h}{\pi^2 h^2} F &= 0~,
\label{eq:vector_eom} \\
F(u) &:= u^{-3} A_0' = u^{-3} \left(\frac{h_{0x}}{g_{xx}}\right)'~.
%
\end{align}
Unlike the tensor mode, the vector mode equation does not take the form of the massless scalar one \eqref{eq:scalar_eom}. Solving the equation asymptotically, one obtains the asymptotic behavior of $F$:
\be
F \sim \frac{A}{u^2} + B~,
\quad (u\rightarrow0)~.
\label{eq:vector_asymptotic}
\ee

\head{The near-horizon solution}
We again solve the perturbation equation \eqref{eq:vector_eom} as in \sect{tensor_mode_sol}. First, solve the equation near the horizon. Near the horizon, \eq{scalar_eom_horizon} becomes
\be
F'' - \frac{1}{1-u} F' + \left( \frac{\nw}{4\pi} \right)^2 \frac{1}{(1-u)^2} F \simeq 0~,
%
\ee
which is the same as \eq{scalar_eom_horizon}. Then, the solution with the ``incoming-wave" boundary condition is given by
\begin{align}
F &\propto (1-u)^{-i \nw/(4\pi)}~, \qquad (u\rightarrow 1) \\
&\simeq 1 - \frac{i\nw}{4\pi} \ln(1-u) + O(\nw^2)~.
\label{eq:vector_sol_horizon}
\end{align}

\head{The $(\nw, \nq)$-expansion}
Second, solve \eq{vector_eom} as a double-series expansion in $(\nw, \nq)$:
\be
F = F_0(u) + \nw F_{10}(u) + \nq^2 F_{01}(u) + \cdots~.
%
\ee
From \eq{vector_eom}, each variables obey the following equations:
\begin{align}
\hat{\cal L} F_0 &= 0~,
 \label{eq:vector_eom0} \\
\hat{\cal L} F_{10} &= 0~,
 \label{eq:vector_eom10} \\
\hat{\cal L} F_{01} &= j_{01}~,
 \label{eq:vector_eom01} 
%
\end{align}
where
\be
\hat{\cal L} \varphi := \left(p(u) \varphi' \right)'~, \quad
p(u) := hu^3~, \quad
j_{01}(u) := \frac{F_0}{\pi^2} u^3~.
\ee
We solve Eqs.~\eqref{eq:vector_eom0}-\eqref{eq:vector_eom01} by imposing the boundary conditions at the horizon $u\rightarrow1$ and at the AdS boundary $u\rightarrow0$. The boundary condition at the horizon reduces to \eq{vector_sol_horizon}, so the boundary conditions for each variables are 
\begin{align}
F_0(u\rightarrow1) &= C~, 
 \label{eq:bc_F0} \\
F_{10}(u\rightarrow1)  &= - \frac{iC}{4\pi} \ln(1-u)~, 
 \label{eq:bc_F10} \\
F_{01}(u\rightarrow1) &= 0~.
 \label{eq:bc_F01}
%
\end{align}
In order to solve Eqs.~\eqref{eq:vector_eom0}-\eqref{eq:vector_eom01},
\begin{itemize}

\item First, solve the homogeneous equation $\hat{\cal L} \varphi = 0$. Denote two independent solutions of the homogeneous equation as $\varphi_1$ and $\varphi_2$.

\item Then, the solution of the inhomogeneous equation $\hat{\cal L} \varphi = j$ is also written in terms of $\varphi_1$ and $\varphi_2$. The solution is given by
\be
\varphi (u) =
  \varphi_1(u) \int^1_u du'\, \frac{j(u') \varphi_2(u')}{p(u')W(u')}
 - \varphi_2(u) \int^1_u du'\, \frac{j(u') \varphi_1(u')}{p(u')W(u')}
\label{eq:inhomogeneous_sol} 
\ee
under the boundary condition \eqref{eq:bc_F01}, where $W$ is the Wronskian: 
$W(u) := \varphi_1 \varphi_2' - \varphi_1' \varphi_2 $.
%
%

\end{itemize}

\head{Boundary conditions}
So, first solve $\hat{\cal L} \varphi = 0$, and one obtains $\varphi_1$ and $\varphi_2$ as
\be
\varphi = C_1 \varphi_1 + C_2 \varphi_2 
= C_1 + C_2 \left\{ \frac{1}{2} \ln\frac{1+u^2}{1-u^2} - \frac{1}{u^2} \right\}~,
\label{eq:homogeneous_sol} 
\ee
where $C_1$ and $C_2$ are integration constants. Near the horizon, $\varphi$ behaves as
\be
\varphi \sim C_1 + C_2 \left\{ - \frac{1}{2} \ln(1-u) - 1 \right\}~, \qquad (u\rightarrow 1)~.
%
\ee
The integration constants for $F_0$ and $F_{10}$ are determined by comparing this near-horizon behavior with the boundary conditions at the horizon 
\eqref{eq:bc_F0}-\eqref{eq:bc_F01}. Then, use \eq{inhomogeneous_sol} to obtain $F_{01}$. The results are
\begin{align}
F_0 &= C~, \\
F_{10} &= \frac{iC}{2\pi} \left\{ \frac{1}{2} \ln\frac{1+u^2}{1-u^2} - \frac{1}{u^2}  + 1 \right\}~, \\
F_{01} &= \frac{C}{8\pi^2} \left(\frac{1}{u^2}-1 \right)~.
%
\end{align}

\head{Results}
To summarize, the solution is given by
\be
\frac{F}{C} = 1+ \frac{i}{2\pi} \nw \left\{ \frac{1}{2} \ln\frac{1+u^2}{1-u^2} - \frac{1}{u^2} + 1 \right\}
 + \frac{1}{8\pi^2} \nq^2 \left(\frac{1}{u^2}-1 \right) + \cdots~.
%
\ee
The solution asymptotically behaves as \eq{vector_asymptotic}:
\be
\frac{F}{C} \xrightarrow{u\rightarrow0}{} \frac{\nq^2-4\pi i\nw}{8\pi^2u^2} + O(u^0)~.
%
\ee
We impose the vanishing slow falloff condition, so the $O(u^{-2})$ term must vanish. This is possible if $\nw$ and $\nq$ satisfy
\be
\nw = -\frac{i}{4\pi} \nq^2 
\result
\omega = -\frac{i}{4\pi T} q^2~.
\ee
Comparing this with the dispersion relation \eqref{eq:vector_pole}, one obtains
\be
\frac{\eta}{s} = \frac{1}{4\pi}~.
\ee
This agrees with the tensor mode result \eqref{eq:universality}.

\section{Appendix: \Nfour\ diffusion constant and conductivity \advanced}\label{sec:SAdS5_diff}

We consider 
\be
\action =  -\frac{1}{4} \alpha\int d^5x\, \sqrt{-g} F_{MN}^2
\label{eq:action_Maxwell}
\ee
in the SAdS$_5$ background, where $\alpha$ is an appropriate normalization factor. We choose the gauge $A_u=0$. 
We again consider the perturbation of the form $A_\mu =A_\mu(u)e^{-i\omega t+iqz}$. 
The Maxwell field has the vector mode and the scalar mode (\sect{tensor}), and both determine the diffusion constant $D$ or the conductivity $\sigma$ (\sect{diffusion}).

\subsection{Conductivity from Maxwell vector mode}

Recall how to derive the conductivity. Following \sect{current}, the vector mode $A_x$ behaves as
\be
A_x \sim \sourceA{x}\left(1+\vevA{x} u^2\right)~,
\quad (u\rightarrow0) 
%
\ee
and the fast falloff is the current $J^x$:
\be
\bra J^x \ketj = \cA \vevA{x} \sourceA{x}~,
\label{eq:response_current_ads5}
\ee
where $\cA$ is an appropriate factor. On the other hand, from Ohm's law,
\be
\bra J^x \ketj = \sigma \sourceE = i\omega \sigma \sourceA{x}~.
%
\ee
Thus,
\be
i\omega\sigma = \cA\vevA{x}~.
\label{eq:fast_vs_cond_ads5}
\ee
The conductivity $\sigma$ is proportional to the $O(\omega)$-coefficient of the vector mode $\vevA{x}$. Then, the computation is similar to the tensor mode one in \sect{tensor_mode_sol}.

From the Maxwell equation, the $A_x$ equation (with $q=0$) becomes
\be
\frac{u}{h}\left(\frac{h}{u} A_x'\right)' + \frac{\nw^2}{\pi^2 h^2} A_x = 0~.
\label{eq:Ax_eom}
\ee

\head{The near-horizon solution}
Fist, solve \eq{Ax_eom} near the horizon. Near the horizon, \eq{Ax_eom} reduces to the same form as \eq{scalar_eom_horizon}. Then, the solution with the ``incoming-wave" boundary condition is given by
\begin{alignat}{2}
A_x &\propto (1-u)^{- i \nw/(4\pi)}~, & \qquad & (u\rightarrow 1)
\\
&\simeq 1 - \frac{i\nw}{4\pi} \ln(1-u) + O(\nw^2)~, & \qquad & (u\rightarrow 1)~.
\label{eq:Ax_sol_horizon}
\end{alignat}

\head{The $\nw$-expansion}
Second, solve \eq{Ax_eom} in the $\nw$-expansion:
\be
A_x = F_0(u) + \nw F_1(u) + \cdots~.
%
\ee
Then, \eq{Ax_eom} becomes 
\be
\left( \frac{h}{u} F_i' \right)' = 0~, \quad (i=0, 1)~,
%
\ee
which is solved as
\be
A_x = (A_0 + \nw A_1) + (B_0 + \nw B_1) \ln\left( \frac{1+u^2}{1-u^2} \right) + O(\nw^2)~,
%
\ee
where $A_i$ and $B_i$ are integration constant.

\head{Boundary conditions}
Finally, impose boundary conditions. The boundary condition at the AdS boundary fixes
$A_0 + \nw A_1 = \sourceA{x}$. Imposing the boundary condition at the horizon \eqref{eq:Ax_sol_horizon}, the solution is given by
\begin{align}
A_x
& = \sourceA{x} \left\{ 1 + \frac{i\nw}{4\pi} \ln\left( \frac{1+u^2}{1-u^2} \right) + O(\nw^2) \right\}~, 
&& (0\leq u\leq 1)~. 
\\
&\sim \sourceA{x} \left\{ 1+ \frac{i\nw}{2\pi} u^2 + \cdots \right\}~, 
&& (u\rightarrow 0)~.
\label{eq:Ax_sol_asymptotic}
\end{align}

\head{Results}
The solution \eqref{eq:Ax_sol_asymptotic} determines $\vevA{x} = i\omega/(2\pi T)$, so
\be
\sigma = \frac{\cA}{2\pi T}
\label{eq:cond_ads5}
\ee
from \eq{fast_vs_cond_ads5}. The factor $\cA$ can be obtained by actually evaluating the on-shell Maxwell action. The computation gives $\cA=2\alpha\, r_0^2/L^3$ and $\alpha = L^2/(32\pi G_5)$ for the \Nfour\ SYM%
\footnote{We change the normalization of the Maxwell field in \sect{r-charged} as $A_M \rightarrow A_M/\sqrt{2}$, which is rather conventional. See, \eg, Ref.~\cite{Policastro:2002se}.}. 
Thus,
\begin{align}
\cA &= \frac{L^3}{32\pi G_5}2\pi^2T^2 = \frac{1}{8}\Nc^2 T^2~,\\
\sigma &= \frac{L^3}{32\pi G_5}\pi T = \frac{1}{16\pi}\Nc^2 T~.
%
\end{align}

\subsection{Diffusion constant from Maxwell vector mode}


In \sect{diffusion}, we saw that $\sigma$ and $D$ are related by $\sigma=D\chi_T$, where $\chi_T$ is the thermodynamic susceptibility. Thus, once one gets $\sigma$, one can get $D$ by computing $\chi_T$.

Following \sect{current}, $A_0$ behaves as
\be
A_0 \sim \sourceA{0}\left(1+\vevA{0} u^2\right)~,
\quad (u\rightarrow0)~.
%
\ee
The slow falloff is the chemical potential $\mu$, and the fast falloff is the charge density $\rho$:
\be
\bra \rho \ketj = -\cA \vevA{0} \mu~,
%
\ee
where the factor $\cA$ is common to \eq{response_current_ads5}. Then, the thermodynamic susceptibility is given by
\be
\chi_T = \frac{\del\bra\rho\ketj}{\del\mu} = -\cA \vevA{0}~.
\label{eq:suscep_sol}
\ee
The static solution for $A_0$ is given by%
\footnote{The $A_u=0$ condition does not completely fix the gauge. The gauge transformation $A_\mu(x,u) \rightarrow A_\mu(x,u)+\del_\mu \Lambda(x)$ is still allowed. As discussed in \sect{current}, we partly fix the gauge by requiring $\left. A_0 \right|_{u=1}=0$. }
$A_0(u)=\mu(1-u^2)$ or $\vevA{0}=-1$. Then, 
\be
\chi_T= \cA = \frac{L^3}{32\pi G_5}2\pi^2T^2 = \frac{1}{8}\Nc^2 T^2~.
%
\ee
Using the formula  $\sigma=D\chi_T$ and the conductivity \eqref{eq:cond_ads5}, one gets
\be
D= \frac{1}{2\pi T}~.
\label{eq:diffusion_kubo_ads5}
\ee
Equation~\eqref{eq:fast_vs_cond_ads5} can be rewritten as
\be
\vevA{x} = i\omega \frac{\sigma}{\cA}
= i\omega \frac{\sigma}{\chi_T}
= i\omega D~.
%
\ee
In our conventions, the $O(i\omega)$ coefficient of $\vevA{x}$ directly gives the diffusion constant.

\subsection{Diffusion constant from Maxwell scalar mode}

From the diffusion equation, $\delta\bra\rho\ket$ has a pole at 
\be
\omega = - iDq^2~.
%
\ee
Here, we compute $D$ from the Maxwell scalar mode and check that the result agrees with the Maxwell vector mode. Since we are interested in a pole, we solve the scalar mode under the vanishing slow falloff condition (\sect{QNM}). The computation is similar to the gravitational vector mode one in \sect{vector}. 

The scalar mode components are $A_0$ and $A_z$. The decoupled equation is already given in \eq{master1} (replace $\sqrt{-\tilg}$ by $\sqrt{-g}$). For the SAdS$_5$ black hole, \eq{master1} becomes
\be
\frac{1}{hu} \left( hu F' \right)'+ \frac{\nw^2 - \nq^2 h}{\pi^2 h^2} F = 0~,
\quad
F(u) := u^{-1} A_0'~.
\label{eq:A0_eom} 
\ee
The asymptotic behavior of $F$ is given by
\be
F \sim A \ln u + B~,
\quad (u\rightarrow0)~.
\label{eq:scalar_asymptotic}
\ee

\head{The near-horizon solution}
First, solve the equation near the horizon. The solution with the ``incoming-wave" boundary condition is given by
\be
F \propto 1 - \frac{i\nw}{4\pi} \ln(1-u) + O(\nw^2)~.
\label{eq:A0_sol_horizon}
\ee

\head{The $(\nw, \nq)$-expansion}
Second, solve \eq{A0_eom} as a doble-series expansion in $(\nw, \nq)$:
\be
F = F_0(u) + \nw F_{10}(u) + \nq^2 F_{01}(u) + \cdots~.
%
\ee
Each variables obey the equations
\begin{align}
\hat{\cal L} F_0 &= 0~,
 \label{eq:Maxwell_scalar_eom0} \\
\hat{\cal L} F_{10} &= 0~,
 \label{eq:Maxwell_scalar_eom10} \\
\hat{\cal L} F_{01} &= j_{01}~,
 \label{eq:Maxwell_scalar_eom01} 
%
\end{align}
where
\be
\hat{\cal L} \varphi := \left(p(u) \varphi' \right)'~, \quad
p(u) := hu~, \quad
j_{01}(u) := \frac{F_0}{\pi^2} u~.
\ee
From \eq{A0_sol_horizon}, the boundary conditions at the horizon for each variables are
\begin{align}
F_0(u\rightarrow1) &= C~, 
 \label{eq:A0_bc_F0} \\
F_{10}(u\rightarrow1)  &= - \frac{iC}{4\pi} \ln(1-u)~, 
 \label{eq:A0_bc_F10} \\
F_{01}(u\rightarrow1) &= 0~.
 \label{eq:A0_bc_F01}
%
\end{align}

\head{Boundary conditions}
The solution of the homogeneous equation $\hat{\cal L} \varphi=0$ is given by
\be
\varphi = C_1 \varphi_1 + C_2 \varphi_2 
= C_1 + C_2 \ln\frac{4u^4}{1-u^4}~,
\label{eq:A0_homogeneous_sol} 
\ee
where $C_1$ and $C_2$ are integration constants. 
The integration constants for $F_0$ and $F_{10}$ are determined by comparing \eq{A0_homogeneous_sol} with the boundary conditions at the horizon 
\eqref{eq:A0_bc_F0}-\eqref{eq:A0_bc_F01}. Use \eq{inhomogeneous_sol} to obtain $F_{01}$. The results are
\be
F_0 = C~, \quad
F_{10} = \frac{iC}{4\pi} \ln\frac{4u^4}{1-u^4}~, \quad
F_{01} = -\frac{C}{4\pi^2} \ln\frac{2u^2}{1+u^2}~.
%
\ee

\head{Results}
To summarize, the solution is given by
\be
\frac{F}{C} = 1+ \frac{i}{\pi} \nw \ln\frac{4u^4}{1-u^4} - \frac{\nq^2}{4\pi^2} \ln\frac{2u^2}{1+u^2}+ \cdots~.
%
\ee
The solution asymptotically behaves as \eq{scalar_asymptotic}:
\be
\frac{F}{C} \xrightarrow{u\rightarrow0}{} \frac{\ln u}{\pi} \left( i\nw - \frac{\nq^2}{2\pi} \right) + O(u^0)~.
\ee
We impose the vanishing slow falloff condition, so the $O(\ln u)$ term must vanish, which determines the dispersion relation:
\be
\nw = -\frac{i}{2\pi} \nq^2 
\result 
\omega = -\frac{i}{2\pi T} q^2~.
\ee
Comparing this with the dispersion relation $\omega=-iDq^2$, we obtain $D=1/(2\pi T)$, which agrees with the Maxwell vector mode result \eqref{eq:diffusion_kubo_ads5}.

\endofsection

\ifx\nameofpaper\undefined 
  \usepackage{macro_natsuume} 
  \def\beginsection{\section*}
  \def\endofsection{\end{document}} 
  \input draft_header.tex
\else 
  \def\beginsection{\chapter}
  \def\endofsection{ } 
\fi

\beginsection{Basics of phase transition
}\label{chap:GL}


\begin{quote}
In this chapter, we explain the basics of phase transitions and related phenomena (critical phenomena and superconductivity) mainly using mean-field theories.
\end{quote}

\section{Phase transition
}


When one changes control parameters such as temperature in a thermodynamic system, the system may undergo a transition to a macroscopically different state which is more stable. This is a \keyword{phase transition}. In a phase transition, a thermodynamic potential such as free energy becomes non-analytic. In the $n$th order phase transition, analyticity is broken in the $n$th derivative of a thermodynamic potential. Namely,
\begin{itemize}

\item First-order phase transition: $F$ is continuous, but $F'$ is discontinuous%
\footnote{Here, $F'$ means a derivative with respect to any independent variables of the thermodynamic potential.}.

\item Second-order phase transition: $F$ and $F'$ are continuous, but $F''$ is discontinuous (or diverges).

\end{itemize}
A second-order phase transition often appears as the end point of a first-order phase transition. 

\begin{figure}[tb]
\centering
\subfigure[]{
\scalebox{0.7}{ \includegraphics{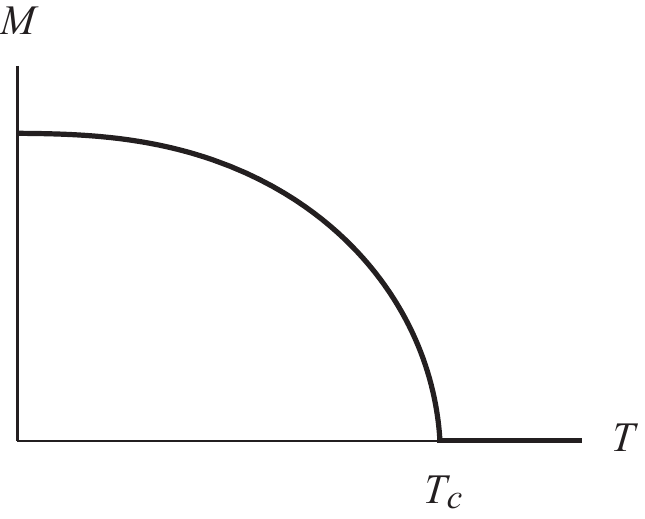} }
\label{fig:order}} \qquad
\subfigure[]{
\scalebox{0.7}{ \includegraphics{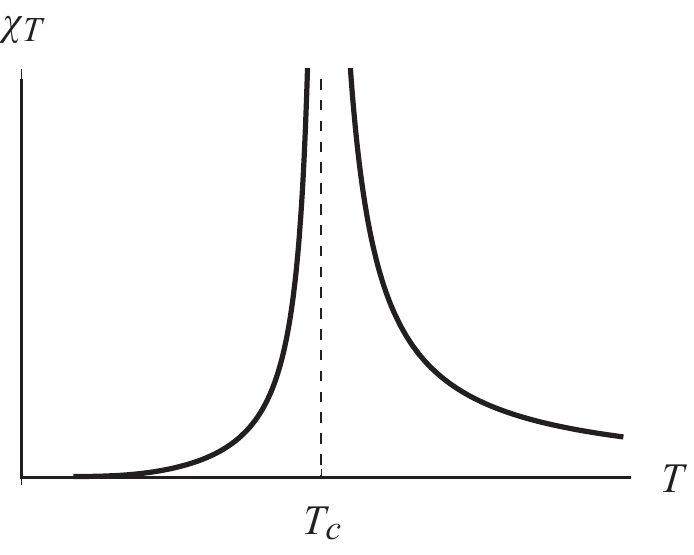} }
\label{fig:chiT}}
\vskip2mm
\caption{(a) Temperature-dependence of the order parameter. (b) The divergence of the magnetic susceptibility.}
\label{fig:condensate}
\end{figure}

As an example, a ferromagnet has a spontaneous magnetization $M$ below the transition temperature $\Tc$ and the magnetization vanishes at $T=\Tc$ [\fig{order}]. A macroscopic variable such as $M$ which characterizes two phases is called the \keyword{order parameter}.

As in the other thermodynamic systems, one can consider the free energy $F=F(T,M)$ for a magnetic system. However, in real experiments, one usually uses the magnetic field $H$ as the control parameter, so we use the Gibbs free energy%
\footnote{Originally, the Gibbs free energy chooses pressure $P$ as an independent variable, but one often calls such a potential the Gibbs free energy as well.}
which is the Legendre transformation of the free energy:
\be
G(T,H)=F-MH~.
%
\ee
Then,
\be
M = - \left(\frac{\del G}{\del H} \right)_{T}~.
\ee
Below we use the mean-field theory to determine $F$ itself. For a ferromagnet, $M$ is continuous at the transition point (when $H$=0), which implies a second-order phase transition.

In the second-order phase transition, the spin \keyword{correlation length} $\xi$ diverges $\xi\rightarrow\infty$ at the transition point, and the system has a macroscopic correlation length. As the consequence, various physical quantities diverge. For example, the magnetic susceptibility behaves as $\chi_T=\del m/\del H \rightarrow\infty$ [\fig{chiT}]. To parametrize such a divergence, one introduces the \keyword{critical exponent} $\gamma$ and writes $\chi_T \propto |T- \Tc|^{-\gamma}$. Critical exponents depend on the symmetry, the spatial dimensionality, and so on but do not depend on the details of microscopic physics. This is called the \keyword{universality}. If two systems have the same set of critical exponents, they are said to belong to the same universality class. Different physical systems may belong to the same universality class. For example, the 3d Ising model and the liquid-gas phase transition belong to the same universality class%
\footnote{See, \eg, Refs.~\cite{critical_text1,critical_text2} for textbooks on critical phenomena.}.

\subsection{Second-order phase transition
}\label{sec:phase_2nd}

Below we mainly use the \keyword{mean-field theory} to discuss phase transitions. In statistical mechanics, the partition function $Z$ may be written using the order parameter $m$ as
\be
Z = \int {\cal D}m\, e^{-\beta \action[m;T,H]}~.
%
\ee
We call $\action[m;T,H]$ the ``pseudo free energy"%
\footnote{We call pseudo free energy because $\action$ is not the free energy itself. (The functional integral over $m$ is not carried out.) The free energy is the on-shell action $\Sos$ below from the mean-field point of view.}.
The order parameter $m$ does not have to be an elementary field of the microscopic theory. In the mean-field theory, one ignores thermal fluctuations of the order parameter $m$ and evaluates $Z$ in the saddle-point approximation:
\be
Z \simeq e^{-\beta\Sos(T,H)}~,
%
\ee
where $\Sos$ represents the ``on-shell action" which is obtained by substituting the solution of $m$. Then, the Gibbs free energy is given by $G(T,H)=\Sos(T,H)$. The mean-field theory is the same technique as the one we used to evaluate the partition function of a gravitational theory.

\begin{figure}[tb]
\centering
\scalebox{0.75}{ \includegraphics{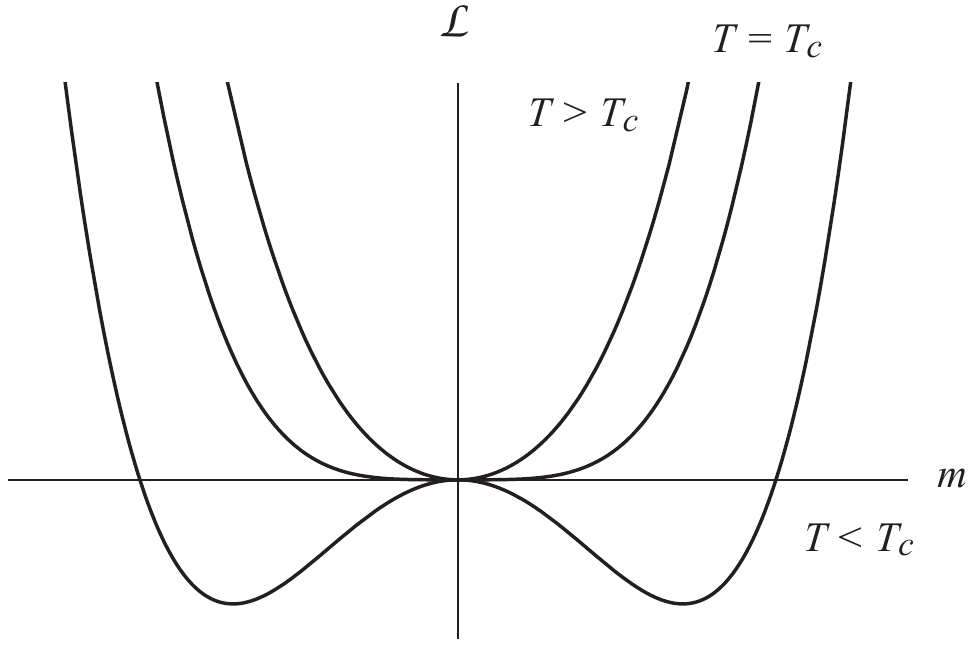} }
\vskip2mm
\caption{The pseudo free energy for a second-order phase transition.}
\label{fig:GL_2nd}
\end{figure}

As an example, let us consider the mean-field theory of the Ising model. For simplicity, we first consider the spatially homogeneous case. Instead of $G$ and $\action$, it is convenient to use the free energy density $g$ where $G=:gV$ and the pseudo free energy density $\calL$ where $\action=:\calL V$. The magnetization density $m$ is defined by $m :=\sum_i \bra S_i\ket/V$. For the Ising model, the pseudo free energy is determined from the following properties:
\begin{enumerate}

\item The order parameter $m$ is small near $T=\Tc$, so one can expand $\calL$ as a power series in $m$. 

\item When the external field is absent, it is natural to impose the spin-reversal symmetry%
\footnote{In fact, the Ising model Hamiltonian is given by
\be
H =-J \sum_{\bra i,j\ket} S_iS_j~,
\ee
where the sum is taken over the nearest-neighbor pairs, so it is invariant under $S_i\rightarrow-S_i$.}
$S_i\rightarrow-S_i$ for spin variables $S_i$, so $\calL$ is an even function of $m$. 
%
%

\end{enumerate}
Then, the pseudo free energy density $\calL$ can be written as
\be
\calL[m; T, H] = \calL_0 + \frac{1}{2} a m^2 + \frac{1}{4} b m^4 + \cdots -m H~.
\label{eq:GL_ferro1}
\ee 
Such a theory is called the \keyword{Ginzburg-Landau theory} (GL theory hereafter). The GL theory has a Higgs-like potential in particle physics (\fig{GL_2nd}). When $a>0$, the potential takes the minimum at the origin $m=0$, which is the behavior at high temperature. On the other hand, when $a<0$, the spontaneous symmetry breaking occurs, and the system has a nonzero magnetization $m\neq0$, which is the behavior at low temperature. So, we take the form
\be
a = a_0 (T- \Tc) + \cdots \quad (a_0>0)~, \qquad
b = b_0 + \cdots \quad (b_0>0)~.
\ee

The saddle point is given by $\del_m \calL |_{H=0} = am+bm^3 = 0$, which determines the spontaneous magnetization:
\be
m = \sqrt{-\frac{a}{b}} \propto ( \Tc-T)^{1/2}~.
%
\ee
Then, the free energy density $g=\Los$ for $T<\Tc$ is given by
\be
g(T) = \calL_0 + m^2\left( \frac{a}{2}+\frac{b}{4}m^2 \right) = \calL_0 - \frac{a^2}{4b}~,
%
\ee
so the specific heat is given by
\be
C_H = -T\frac{\del^2 g}{\del T^2} = \frac{a_0^2}{2b_0} T~,
\qquad (H=0, T< \Tc)~.
%
\ee

When one adds the magnetic field, the saddle point is given by
\be
\del_m \calL = am+bm^3 - H = 0~,
\label{eq:eom_w/H}
\ee
so at the critical point $a=0$,
\be
m^3 \propto H~, \qquad (T= \Tc)~.
%
\ee
Differentiating \eq{eom_w/H} with respect to $H$ gives $(a+3bm^2)\del m/\del H-1=0$. Then, the magnetic susceptibility%
\footnote{It is also known as the thermodynamic susceptibility.\index{thermodynamic susceptibility} } 
$\chi_T$ is 
\be
\chi_T:= \frac{\del m}{\del H} = \frac{1}{a+3bm^2} = \frac{1}{-2a} \propto \frac{1}{|\eT|}~,
\qquad (T< \Tc)~.
\label{eq:thermodynamic_susceptibility}
\ee

If we define four critical exponents $(\alpha, \beta, \gamma, \delta)$ as
\begin{align}
&\text{specific heat: } & C_H & \propto |\eT|^{-\alpha}~, & \\
&\text{spontaneous magnetization: } & m & \propto |\eT|^{\beta}~, & (T< \Tc), \\
&\text{magnetic susceptibility: } & \chi_T & \propto |\eT|^{-\gamma}~, & \\ 
&\text{critical isotherm: } & m & \propto |H|^{1/\delta}~, & (T= \Tc),
\end{align}
our results are summarized as
\be
(\alpha, \beta, \gamma, \delta) = \left(0, \frac{1}{2}, 1, 3\right)~.
\ee

Here, we obtained the exponents $(\alpha,\gamma)$ in the low-temperature phase $T<\Tc$, but they take the same values in the high-temperature phase $T>\Tc$. When $T>\Tc$, $m=0$, so $\Los = \calL_0$ and $\alpha=0$. Also, $\chi_T = 1/a$ from \eq{thermodynamic_susceptibility}, so $\gamma=1$.

\subsection{First-order phase transition
}\label{sec:phase_1st}

If a system does not have the symmetry $m \rightarrow -m$, the $O(m^3)$ term can exist, which leads to a first-oder phase transition:
\begin{align}
\calL[m; T, H] &= \frac{1}{2} a m^2 - \frac{1}{3} c m^3 + \frac{1}{4} b m^4 + \cdots -m H~,
\label{eq:GL_1st} \\
a &= a_0(T-T_0)+\cdots~.
%
\end{align}
Below we set $H=0$. In this case, $T=T_0$ will differ from the transition temperature $\Tc$.

\begin{figure}[tb]
\centering
\scalebox{0.75}{ \includegraphics{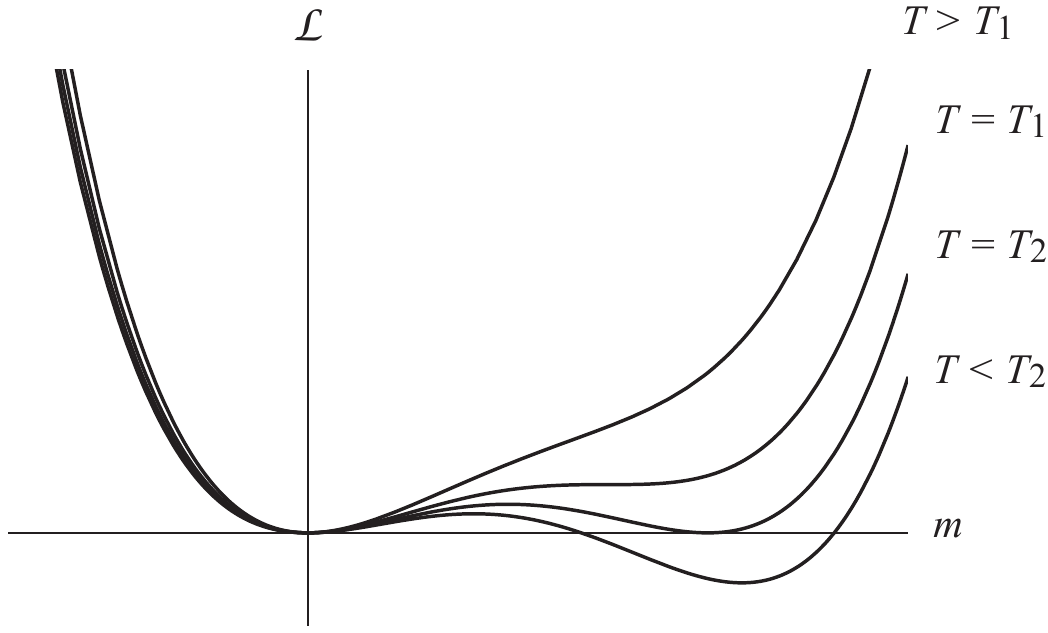} }
\vskip2mm
\caption{The pseudo free energy for a first-order phase transition.}
\label{fig:GL_1st}
\end{figure}

At high enough temperature $a \gg 0$, $m=0$ as in the last subsection. When $a<a_1$ or $T<T_1$, the pseudo free energy develops a saddle point at $m\neq0$ (\fig{GL_1st}). But the saddle point $m\neq0$ has a higher free energy than the state $m=0$, so the saddle point is a metastable local minimum. The saddle points are determined by $\del_m \calL = m(a-cm+bm^2) = 0$, so
\be
m=0, \quad m = \frac{c \pm \sqrt{c^2-4ab}}{2b}~.
%
\ee
The temperature $T_1$ is determined from the condition that the $m\neq0$ solution is real:
\be
a_1 = \frac{c^2}{4b}~.
%
\ee

Decrease temperature further. When $a<a_2$ or $T<T_2$, the $m\neq0$ state has a lower free energy than the  $m=0$ state. Then, $m$ changes from $m=0$ to $m\neq0$ discontinuously. So, this is a first-order phase transition (in the second-order phase transition, $m$ changes continuously at $T=\Tc$), and this temperature $T_2$ is the transition temperature. At $T=T_2$, we have degenerate minima with $\calL=0$. The solutions of $\calL = m^2(a/2-cm/3+bm^2/4) = 0$ are given by
\be
m=0, \quad m = \frac{2c \pm \sqrt{2(2c^2-9ab)}}{3b}~.
%
\ee
The temperature $T_2$ is determined from the condition that the $m\neq0$ solution is real:
\be
a_2 = \frac{2c^2}{9b}~.
%
\ee
In the second-order phase transition, $m$ can be as small as one wises if one approaches the critical point, which justifies the power series expansion \eqref{eq:GL_ferro1}. But in the first-order phase transition, $m$ does not need to be small, so one should be careful if the power series expansion \eqref{eq:GL_1st} is really valid. 

\subsection{Inhomogeneous case
}\label{sec:phase_nonuniform}

Go back to the second-order phase transition and consider the spatially inhomogeneous case. We add the $(\del_i m)^2$ term to guarantee the homogeneity of the equilibrium state:
\be
\action[m; T, H] = \int d^{\ds} x\,\left[ \frac{1}{2}\,(\del_i m)^2 \!+ \frac{a}{2}\,m^2 \!+ \frac{b}{4}\, m^4 \!+\!\cdots \!-\! m H \right]~,
\label{eq:GL_ferro2}
\ee 
where $\ds$ denotes the number of \textit{spatial dimensions}. The $am^2$ term plays the role of the ``mass term" for the order parameter field $m$, and $m$ becomes ``massless" at the critical point. Then, the characteristic length or the correlation length $\xi$ diverge, and $m$ has a long-range correlation:
\be
\xi = \frac{1}{ \text{(mass)} } = a^{-1/2} \propto (\eT)^{-1/2}~.
%
\ee 

For simplicity, we focus on the high-temperature phase. Recall the homogeneous case. To obtain the critical exponents $(\alpha,\gamma)$ in the high-temperature phase, the quartic term $m^4$ was not essential and can be ignored. Similarly, we consider only up to quadratic terms here. Then, the saddle point is determined by
\be
0 = \frac{\delta \action}{\delta m} = -\del_i^2 m + a m - H = 0~.
%
\ee
As in the homogeneous case, we define the ``susceptibility" as%
\footnote{It is also known as the static susceptibility \index{static susceptibility} or the static response function. Incidentally, the correlation function $G(\bmx-\bmx')$ is defined by
\be
G(\bmx-\bmx') 
:= \bra m(\bmx) m(\bmx') \ket - \bra m(\bmx) \ket \bra m(\bmx') \ket 
= \frac{1}{\beta^2} \frac{\delta^2 \ln Z}{\delta H(\bmx)\delta H(\bmx')}~,
%
\ee
and the magnetization is defined by
\be
m(\bmx) = \frac{1}{\beta} \frac{\delta \ln Z}{\delta H(\bmx)}~,
%
\ee
so the correlation function and the susceptibility are related by
\be
G(\bmx-\bmx') = \frac{1}{\beta} \frac{\delta m(\bmx)}{\delta H(\bmx')} := T\chi(\bmx-\bmx')~.
%
\ee
}
\be
\chi(\bmx-\bmx') = \frac{\delta m(\bmx)}{\delta H(\bmx')}~.
%
\ee
Then,
\be
(-\del_\bmx^2+a)\chi(\bmx-\bmx') = \delta(\bmx-\bmx')~.
%
\ee
After the Fourier transformation, one gets
\be
(\bmq^2+a)\chi_\bmq = 1~.
%
\ee
Since $\chi_T =1/a$, $\chi_T=\chi_{\bmq\rightarrow0}$. When $T \neq \Tc$, $\chi(\bmr)$ takes the Yukawa form:
\be
\chi(\bmr) \propto r^{-(\ds-1)/2} e^{-r/\xi}~,
%
\ee
so the correlation is lost beyond the distance $r \simeq \xi$. When $T = \Tc$, $\chi(\bmr)$ takes the Coulomb form:
\be
\chi(\bmr) \propto r^{-\ds+2} ~,
%
\ee
so the correlation is long-ranged.

If we define critical exponents $(\nu,\eta)$ which appear in the inhomogeneous case as
\begin{align}
&\text{static susceptibility:  } & \chi(\bmr\,) & \propto e^{-r/\xi}~, & (T \neq  \Tc), \\
&                    && \propto r^{-\ds+2-\eta}~, & (T= \Tc), \\
&\text{correlation length: } & \xi & \propto |\eT|^{-\nu}~, &
\label{eq:nu-def} 
\end{align}
we obtained
\be
(\nu, \eta) = \left(\frac{1}{2}, 0\right)~.
\ee

\subsection{Critical phenomena
}\label{sec:exponents}

\subsubsection*{Critical exponents and scaling relations
}

In critical phenomena, there appear six critical exponents traditionally. For the GL theory, 
\be
(\alpha, \beta, \gamma, \delta, \nu, \eta) = \left(0, \frac{1}{2}, 1, 3,  \frac{1}{2}, 0\right)
\label{eq:GL_exponents}
\ee
from the results in Sects.~\ref{sec:phase_2nd} and \ref{sec:phase_nonuniform}. There are six critical exponents, but not all are independent, and they satisfy \keyword{scaling relations}:
\begin{align}
\alpha + 2\beta+\gamma =2~, 
\label{eq:scaling1} \\
\gamma = \beta(\delta-1)~, 
\label{eq:scaling2} \\
\gamma = \nu (2-\eta)~, 
\label{eq:scaling3} \\
2-\alpha = \nu \ds~.
\label{eq:scaling4} 
%
\end{align}
A relation such as \eq{scaling4} which depends on the spatial dimensionality $\ds$  is known as a \keyword{hyperscaling relation}. One can check that the GL theory satisfies the scaling relations except the hyperscaling relation. We will come back to the breakdown of the hyperscaling relation. 

The scaling relations themselves are valid independent of the details of our theory (\eg, GL theory). Then, when one studies critical phenomena in AdS/CFT, one had better pay attention to the issue whether these relations hold or not (although we will encounter only simple GL exponents in this book).

There are four scaling relations, so only two exponents are independent among six exponents, which suggests that there is some structure behind these relations. In fact, these relations can be derived from the \keyword{scaling law} for the free energy $g(\epT, H)$:
\be
g(\epT, H) = b^{-\ds} g(b^{\yt} \epT, b^{\yh} H)~,
\label{eq:static_law_energy}
\ee
where $\epT:= (T- \Tc)/ \Tc$. If one chooses $b^{\yt} \epT=1$, \eq{static_law_energy} becomes
\be
g (\epT, H) = \epT^{\ds/\yt} g(1, \epT^{-\yh/\yt} H)
 =: \epT^{\ds/\yt} \tilde{g} (\epT^{-\yh/\yt} H)~.
\label{eq:static_law_energy2}
\ee
 
%
%
The scaling law determines six critical exponents in terms of two parameters $(\yt, \yh)$:
\begin{align}
C_H(\epT,0) &\propto \frac{\del^2 g(\epT,0)}{\del \epT^2} \propto \epT^{\ds/\yt-2}~, 
\\
m(\epT,0) &\propto \left.\frac{\del g(\epT,H)}{\del H}\right|_{H=0} \propto \epT^{(\ds-\yh)/\yt}~, 
\\
\chi_T(\epT,0) &\propto \left.\frac{\del^2 g(\epT,H)}{\del H^2}\right|_{H=0} \propto \epT^{(\ds-2\yh)/\yt}~, 
\\
m(0,H) &\propto \frac{\del g(0,H)}{\del H} \propto b^{-\ds+\yh} g_2(0,b^{\yh}H) = H^{(\ds-\yh)/\yh}~.
%
\end{align}
Here, $g_2$ is the partial derivative of $g$ with respect to the second argument (magnetic field), and we chooses $b^{\yh}H=1$. Similarly, there is a scaling law for the response function which determines $(\nu,\eta)$. In this way, one obtains
\begin{align}
\alpha &= 2-\ds/\yt~, \label{eq:alpha} \\
\beta &= \frac{\ds-\yh}{\yt}~, \label{eq:beta} \\
\gamma &= \frac{2\yh-\ds}{\yt}~, \label{eq:gamma} \\
\delta &= \frac{\yh}{\ds-\yh}~, \label{eq:delta} \\
\nu &= 1/\yt~, \label{eq:nu} \\
\eta &= \ds - 2\yh + 2~. \label{eq:eta} 
%
\end{align}
Equations~(\ref{eq:alpha})-(\ref{eq:eta}) satisfy the scaling relations (\ref{eq:scaling1})-(\ref{eq:scaling4}). 

%
%

The scaling law is justified from the renormalization group, but one can easily check that the GL theory  \eqref{eq:GL_ferro2} satisfies the scaling law \eqref{eq:static_law_energy}. The GL theory is rewritten as
\begin{align}
\calL &\simeq a_0 t\, m^2 + b_0 m^4 - m H \\
&\simeq t^2 \{ a_0 \tilm^2 + b_0 \tilm^4 - \tilm\tilH \}~,
%
\end{align}
where we rescaled $m = \tilm t^{1/2}$ and $H = \tilH t^{3/2}$ to isolate the $t$-dependence. The solution of the equation of motion takes the form $\tilm=f(\tilH)$, so the free energy density becomes
\be
%
\Los \simeq t^2 \{ a_0 f(\tilH)^2 + b_0 f(\tilH)^4 - \tilH f(\tilH) \}~.
%
\ee
This takes the form of the scaling law with $d_s/\yt=2$ and $\yh/\yt=3/2$.

\subsubsection*{Critical exponents and spatial dimensionality
}

\begin{figure}[tb]
\centering
\scalebox{0.65}{ \includegraphics{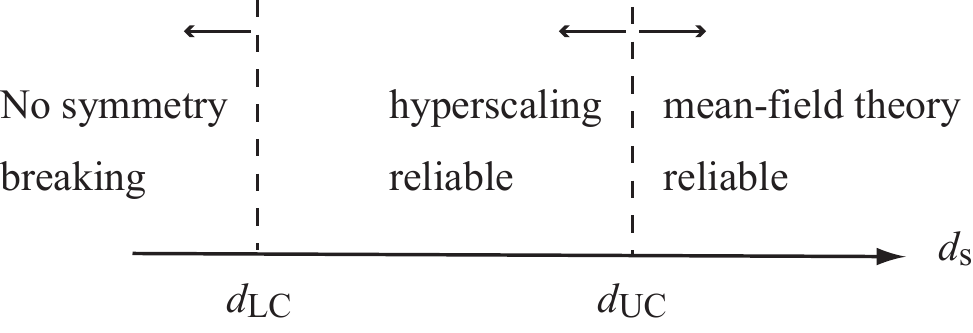} }
\caption{Critical exponents and spatial dimensionality.}
\label{fig:exponents} 
\end{figure}

As is clear from the discussion so far, mean-field results do not depend on the spatial dimensionality $\ds$. This is because the mean-field theory ignores the statistical fluctuations of the order parameter. In general, the effect of fluctuations becomes more important in low spatial dimensions, so mean-field exponents, which do not depend on the dimensionality, may be modified in low dimensions. On the other hand, the effect is less important in high spatial dimensions. There are two important dimensionalities (\fig{exponents}): 
\begin{itemize}

\item The \keyword{upper critical dimension} $d_\text{UC}$: For high enough dimensions $\ds \geq d_\text{UC}$, the fluctuations are not important, and mean-field exponents are reliable. The value of $d_\text{UC}$ can be estimated using the so-called Ginzburg criterion (\probset{ginzburg}):
\be
d_\text{UC} = \frac{2\beta+\gamma}{\nu} = \frac{2-\alpha}{\nu}~.
\label{eq:critical_dim}
\ee
For the GL theory, $d_\text{UC}=4$. Thus, mean-field exponents are reliable for $\ds \geq 4$ but are modified for the real $\ds=3$.

\item The \keyword{lower critical dimension} $d_\text{LC}$: For low enough dimensions $\ds \leq d_\text{LC}$, the fluctuations are too large so that there is no phase transition and no symmetry breaking at finite temperature (Coleman-Mermin-Wagner theorem\index{Coleman-Mermin-Wagner theorem} \cite{Coleman:1973ci,Mermin:1966fe}). For a system with a discrete symmetry such as the Ising model, $d_\text{LC}=1$. For a system with a continuous symmetry, $d_\text{LC}=2$.

\end{itemize}

The hyperscaling relation \index{hyperscaling relation} \eqref{eq:scaling4} depends on the dimensionality, but mean-field results do not. Thus, the hyperscaling relation is not always valid in the mean-field theory. However, mean-field results are reliable for $\ds \geq d_\text{UC}$. This implies that the hyperscaling relation is not reliable above the upper critical dimension. This is because the so-called dangerously irrelevant operators \index{dangerously irrelevant operators} may exist in free energy and break the scaling law of the free energy. For the mean-field theory, the hyperscaling relation can be satisfied only for $\ds = d_\text{UC}$. The GL theory actually satisfies the hyperscaling relation when $\ds=4$.

\subsubsection*{Static and dynamic critical phenomena
}

Roughly speaking, there are two kinds of critical phenomena: 
\begin{itemize}

\item Static critical phenomena: This is the critical phenomena discussed so far. The critical phenomena normally implies this static critical phenomena. In the static critical phenomena, various thermodynamic quantities have singular behaviors and one parametrizes singularities by critical exponents.

\item Dynamic critical phenomena \cite{hohenberg_halperin}: Near the critical point, dynamic quantities also have singular behaviors. In the dynamic case, one interesting quantity is the relaxation time $\tau$. The relaxation time measures how some disturbance decays in time. In the dynamic critical phenomena, the relaxation time also diverges
\footnote{There are various modes in a system, and their relaxation times all differ in general, so we had better specify which mode has the singular relaxation time. In a phase transition, singular behaviors occur in the quantities related to the order parameter. Thus, $\tau$ here is the relaxation time of the order parameter. However, the order parameter may couple to the other modes in the system, and the other modes may have singular behaviors as well (mode-mode coupling). \index{mode-mode coupling} As a consequence, there appear multiple number of dynamic critical exponents in general.}:
\be
\tau \simeq \xi^z \rightarrow \infty~.
\ee
\end{itemize}
The exponent $z$ is called a dynamic critical exponent. The diverging relaxation time means that the relaxation of the system slows down. So, such a phenomenon is also known as the \keyword{critical slowing down}.

The details of the dynamic exponent depend on dynamic universality classes. The dynamic universality class depends on additional properties of the system which do not affect the static universality class. In particular, conservation laws play an important role to determine dynamic universality class. A conservation law forces the relaxation to proceed more slowly. As a consequence, even if two systems belong to the same static universality class, they may not belong to the same dynamic universality class. To study the dynamic critical phenomena, one uses the time-dependent extension of the GL theory, the time-dependent GL equation \index{time-dependent GL equation} or the TDGL equation. Both static and dynamic critical phenomena have been discussed in the context of AdS/CFT \cite{Maeda:2008hn}.

\section{Superconductivity
}\label{sec:SC}

\subsection{Ginzburg-Landau theory of superconductivity}

We now consider superconductivity/superfluidity as a phenomenon associated with a phase transition. Phenomenologically, superconductivity is described by a simple extension of the GL theory:
\begin{align}
\calL[\psi; T, A_i]  &= \frac{\hbar^2}{2\ms} | D_i\psi |^2 +a |\psi|^2 +\frac{b}{2} |\psi|^4 + \frac{1}{4} F_{ij}^2~,
\label{eq:GL_super} \\
D_i &:= \del_i - i \frac{\es}{\hbar} A_i~, 
\end{align}
where we take the gauge $A_0=0$. New ingredients compared with previous cases are two-folds:
\begin{enumerate}

\item The order parameter $\psi$ is a complex field.
\item The system is coupled with the gauge field $A_i$. When the gauge field is not coupled, the GL theory describes superfluids%
\footnote{The corresponding theory for superfluids is known as the Gross-Pitaevskii theory.\index{Gross-Pitaevskii theory}}.

\end{enumerate}
The complex order parameter comes from the fact that superconductivity/superfluidity are macroscopic quantum phenomena. Microscopically, superconductivity is described by the \keyword{BCS theory}. According to the BCS theory, the motion of an electron causes a distortion of the lattice whose effect is mediated to another electron. As a result, there is an attractive interaction between electrons (with opposite spin and momenta) mediated by \keyword{phonons} (lattice vibrations), and the electron pair forms the \keyword{Cooper pair}. The condensation of the Cooper pair causes superconductivity. The order parameter $\psi$ corresponds to the wave function of the Cooper pair and is called the \keyword{macroscopic wave function}
\footnote{The order parameter is the Cooper pair which consists of two electrons, so $\ms=2m$ and $\es=2e$.}.

Superconductors are coupled with the gauge field. Characteristic features from the coupling are
\begin{itemize}
\item Zero resistivity or the diverging DC conductivity.
\item The \keyword{Meissner effect} which expels a magnetic field (perfect diamagnetism).
\end{itemize}
Among these effects, what is unique to superconductivity is the Meissner effect as discussed below.

We discuss superconductivity using the pseudo free energy $\calL$. First, consider the spatially homogeneous case. When the gauge field is not coupled, the discussion is the same as the magnetic system, and
\be
|\psi|^2 = -\frac{a}{b} \propto (\Tc-T)
\label{eq:sol_condensate}
\ee
in the low-temperature phase. Also, the order parameter has an effective mass term $a|\psi|^2$.

In the inhomogeneous case, decompose $\psi$ as $\psi(x) =|\psi(x)|e^{i\es\phase(x)/\hbar}$. Then, \eq{GL_super} becomes 
\be
\calL
= \frac{\hbar^2}{2\ms} (\del_i |\psi|)^2 + \frac{\es^2}{2\ms} |\psi|^2 (\del_i \phase - A_i)^2 +a |\psi|^2 + \cdots~.
%
\ee 
Thus,
\begin{enumerate}

\item When the gauge field is not coupled, the phase $\phase$ is massless unlike the amplitude $|\psi|$. The phase $\phase$ is the \textit{Nambu-Goldstone boson}\index{Nambu-Goldstone mode} associated with the spontaneous symmetry breaking of the global $U(1)$ symmetry.
\item When the gauge field is coupled, one can absorb $\phase$ into the gauge field by a gauge transformation:
\begin{align}
\psi(x) &\rightarrow e^{i\es\alpha(x)/\hbar} \psi(x)~, \\
A_i(x) &\rightarrow A_i(x) + \del_i \alpha(x)~,
%
\end{align}
and the Nambu-Goldstone boson does not appear. But the gauge field becomes massive because of the term $\es^2|\psi|^2A_i^2/(2\ms)$, which is the \keyword{Higgs mechanism}.
\end{enumerate}

\subsubsection*{Characteristic length scales of superconductors
}

To summarize our discussion so far, both the order parameter and the gauge field have effective masses. So, a \SC\ has two characteristic length scales:
\begin{enumerate}
\item The correlation length%
\footnote{In the \SC\ literature, it is often called the GL coherence length.\index{GL coherence length} } 
$\xi$ from the order parameter mass:
\be
\xi^2 = \frac{\hbar^2}{2\ms |a|}~.
%
\ee
This represents the characteristic length scale of the spatial variation for the order parameter. The order parameter $|\psi|$ is approximately constant inside the \SC\ beyond the distance $\xi$ (when there is no magnetic field). In particle physics, the mass is analogous to the Higgs mass.

\item The magnetic penetration length $\lambda$ from the gauge field mass:
\be
\lambda^2 = \frac{\ms}{\es^2|\psi|^2} = \frac{\ms}{\es^2}\frac{b}{|a|}~.
%
\ee
Since the gauge field becomes massive, the gauge field can enter into the \SC\ only up to the length $\lambda$. In particle physics, the mass is analogous to the W-boson mass.

\end{enumerate}
Then, the GL theory is characterized by a dimensionless parameter $\kappa$, the GL parameter, \index{GL parameter} using $\lambda$ and $\xi$:
\be
\kappa:= \frac{\lambda}{\xi} = \frac{\ms}{\es\hbar} \sqrt{2b}~.
%
\ee
The \SCs\ are classified by the value of $\kappa$ as
\begin{itemize}
\item \keyword{Type I \SCs}: $\kappa<1/\sqrt{2}$ or $\xi>\sqrt{2} \lambda$.
\item \keyword{Type II \SCs}: $\kappa>1/\sqrt{2}$ or $\xi<\sqrt{2} \lambda$.
\end{itemize}
High-$\Tc$ \SCs\ are type II \SCs.

In type II \SCs, the penetration length is larger than the correlation length. So, the magnetic field can enter the \SC\ keeping the superconducting state on the whole. The magnetic field penetration arises by forming vortices. 

\begin{figure}[tb]
\centering
\scalebox{0.75}{ \includegraphics{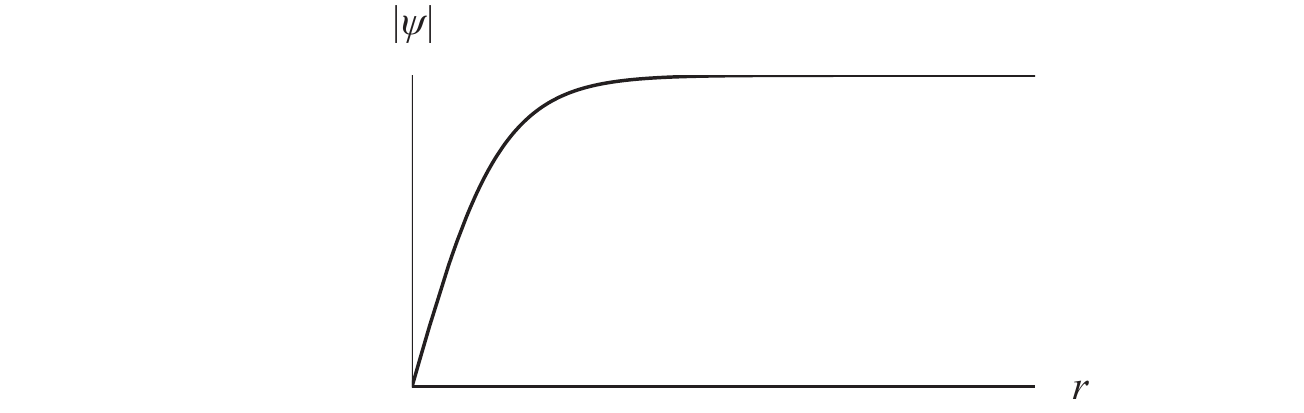} }
\caption{An isolated vortex solution.}
\label{fig:SC_vortex} 
\end{figure}

The \keyword{vortex} solution takes the form 
\be
\psi = |\psi(r)| e^{i\theta}~,
%
\ee
where we take the cylindrical coordinates $ds^2=dz^2+dr^2+r^2d\theta^2$. At the center of the vortex, $\psi=0$, so the superconducting state is broken there (\fig{SC_vortex}). The phase changes by $2\pi$ (winding number one) as we go around the vortex. This implies that the vortex has a quantized magnetic flux $\Phi=h/\es$.

As one increases the magnetic field, the magnetic field begins to penetrate into the \SC\ and vortices appear at $B_{c1}=B_c$ known as the \keyword{lower critical magnetic field}. As one increases the magnetic field further, more and more vortices are created and vortices overlap. Eventually, the superconducting state is completely broken at the \keyword{upper critical magnetic field} $B_{c2}$. The vortices form a lattice. In the GL theory, the triangular lattice is the most favorable configuration. 

\subsubsection*{London equation and its consequences
}

From the pseudo free energy $\calL$, the field equation for $A_i$ is given by
%
%
\be
\del_j F^{ij} = J^i~,
\label{eq:gauge_eom}
\ee
where
\begin{align}
J_i &:= - \frac{\del \calL}{\del A^i}  
= -\frac{i\hbar}{2\ms}\es\{ \psi^*D_i\psi - \psi(D_i\psi)^* \} \\
&= -\frac{i\hbar}{2\ms}\es( \psi^*\del_i\psi - \psi\del_i\psi^*) - \frac{\es^2}{\ms}|\psi|^2 A_i \\
&= \frac{\es^2}{\ms}|\psi|^2 (\del_i\phase-A_i)~.
\label{eq:def_current}
\end{align}
In the gauge where the Nambu-Goldstone boson $\phase$ is eliminated, \eq{def_current} becomes
\be
J_i = -\frac{\es^2}{\ms}|\psi|^2 A_i = -\frac{1}{\lambda^2}A_i~.
\label{eq:London}
\ee
This is known as the \keyword{London equation}. The London equation is another example of a linear response relation. The generic linear response relation is a nonlocal expression (\sect{linear_response}) whereas the London equation is a local expression. This is because the London equation is a phenomenological equation, so the long-wavelength limit is implicitly assumed. The nonlocal extension of the London equation is known as the Pippard equation.

\begin{figure}[tb]
\centering
\scalebox{0.75}{ \includegraphics{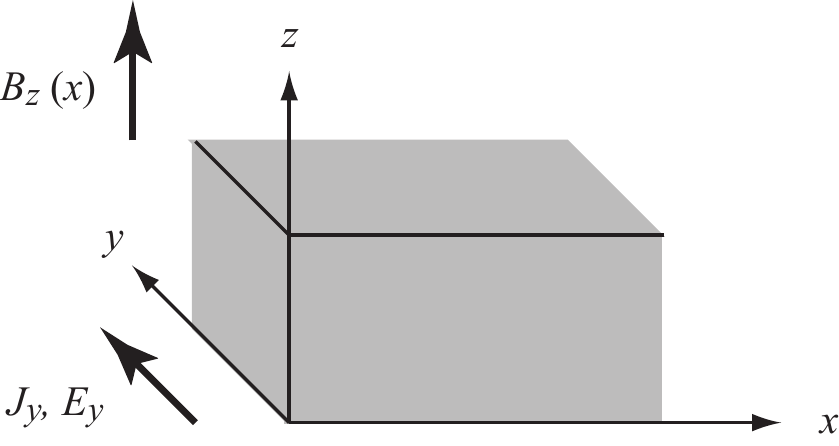} }
\caption{Superconductor under a magnetic field.}
\label{fig:SC_magnetic} 
\end{figure}

From the London equation, one gets the Meissner effect and the diverging DC conductivity.
\begin{enumerate}

\item The Meissner effect: If we choose $A_i = (0, A_y(x), 0)$ for simplicity, $F_{xy}=\del_x A_y = B_z(x)$ (\fig{SC_magnetic}). Combining the London equation and the Maxwell equation  \eqref{eq:gauge_eom}, one gets
\be
\del_x F^{yx} = J^y 
\result 
-\del_x^2 A_y = - \frac{1}{\lambda^2} A_y~.
\label{eq:Higgs}
\ee
Thus,
\be
B_z(x) = B_0 e^{-x/\lambda}~.
%
\ee
Namely, the magnetic field decays exponentially inside a \SC.

\item The diverging DC conductivity: The time-derivative of the London equation gives 
\be
\del_t J_i = \frac{1}{\lambda^2} E_i~,
\label{eq:SC_electric}
\ee
since $E_i = -\del_t A_{i}$, or in Fourier components, 
\be
J_i = \frac{i}{\omega}\frac{1}{\lambda^2} E_i (\omega, \bmq=0)~.
\label{eq:pole}
\ee
The electric field grows the current in \eq{SC_electric}, or $\text{Im}[\sigma(\omega)]$ has a $1/\omega$ pole in \eq{pole}. This implies a diverging DC conductivity as discussed below.

\end{enumerate}

%
%
%
%

\subsection{Normal, perfect, and superconductors}\label{sec:drude}

Characteristic features of \SCs\ are the diverging DC conductivity and the Meissner effect. But its essence is in the Meissner effect. A diverging conductivity also appears in a perfect conductor, but the Meissner effect $B=0$ is unique to a \SC. A perfect conductor can only explain $\del B/\del t = 0$ as discussed below. 

Here, we consider a simple model of conductivity, the \keyword{Drude model}, and study its electromagnetic responses. The model is helpful to distinguish a normal conductor, perfect conductor, and \SC. It is also helpful to get a rough idea of the AC conductivity for a \SC.

The Drude model uses the classical mechanics for the electron motion:
\be
m \frac{dv}{dt} = eE - m\frac{v}{\tau}~,
\label{eq:drude}
\ee
where $v$ is the mean-velocity of the electron and $\tau$ is the relaxation time due to the scattering of the electron. 

\head{Normal conductor}
In a normal conductor, the stationary solution of \eq{drude} is given by $v=eE\tau/m$. Then, the current is
\be
J = nev = \frac{ne^2\tau}{m} E = \sigma E
\label{eq:NC_electric}
\ee
($n$: number density of electrons). This is Ohm's law with $\sigma = ne^2\tau/m$. The dissipation is necessary for a finite conductivity, and the electric field is necessary to keep $v$ and $J$ constant.

Microscopically, a conductor is described as a \keyword{Fermi liquid}%
\footnote{See Ref.~\cite{Polchinski:1992ed2} for a review.}. 
In a conductor, many electrons are interacting with each other via Coulomb interactions. In principle, a conductor is a highly-complicated interacting problem. Yet, according to the Fermi liquid theory, one can treat a conductor as a collection of weakly-interacting ``electrons" (more precisely, quasiparticles which are ``dressed electrons.") The BCS theory is also based on the weakly-coupled Fermi liquid description. 

\head{Perfect conductor}
A perfect conductor is the limit $\tau\rightarrow\infty$ in \eq{drude}:
\be
\frac{\del J}{\del t} = \frac{ne^2}{m} E~.
\label{eq:PC_electric}
\ee
Ohm's law is replaced by \eq{PC_electric}: the electric field accelerates the electron or one can have a steady current with no electric field. Note that the electric response of a \SC\ \eqref{eq:SC_electric} takes the same form as \eq{PC_electric}. 

In order to see the diverging DC conductivity, consider the AC conductivity for the normal conductor. Using $E(t)=E e^{-i\omega t}$ and $J(t)=J e^{-i\omega t}$, \eq{drude} becomes
\be
J(t) = \frac{ne^2\tau}{m} \frac{1}{1-i\omega\tau} E(t)~.
%
\ee
Then, the complex conductivity $\sigma(\omega)$ and its $\tau\rightarrow\infty$ limit are
\begin{align}
\text{Re}[\sigma(\omega)] &= \frac{ne^2\tau}{m} \frac{1}{1+\omega^2\tau^2} 
\xrightarrow{\tau\rightarrow\infty}{} \frac{ne^2}{m} \pi\delta(\omega)~,
\\
\text{Im}[\sigma(\omega)] &= \frac{ne^2\tau}{m} \frac{\omega\tau}{1+\omega^2\tau^2}
\xrightarrow{\tau\rightarrow\infty}{} \frac{ne^2}{m} \frac{1}{\omega}~.
%
\end{align}
Thus, $\text{Re}[\sigma]$ has the diverging DC conductivity, and at the same time $\text{Im}[\sigma]$ has the $1/\omega$ pole.

More generally, the real part and the imaginary part of $\sigma(\omega)$ are related by the \keyword{Kramers-Kronig relation}:
\be
\text{Im}[\sigma(\omega)] = 
- \frac{1}{\pi} {\cal P} \int_{-\infty}^{\infty} 
\frac{\text{Re}[\sigma(\omega')] d\omega'}{\omega'-\omega}~,
\label{eq:kramers_kronig}
\ee
where ${\cal P}$ denotes the principal value. From the formula, a delta function in $\text{Re} [\sigma]$ is reflected into the $1/\omega$ pole in $\text{Im}[\sigma]$.

\head{Perfect conductors versus superconductors}
The conductivity takes the same form both in the \SC\ \eqref{eq:SC_electric} and in the Drude model \eqref{eq:PC_electric}, so clearly the diverging DC conductivity is inadequate to distinguish them. Then, how does one distinguish them? Equation~\eqref{eq:PC_electric} may be written as
\be
\del_t J_i = - \frac{1}{\lambda_D^2} \del_t A_i~, \quad
\lambda_D^2 = \frac{m}{ne^2}~.
\label{eq:London_like}
\ee
Compared with \eq{London}, this is the time derivative of the London equation. Thus, the Drude form follows from the London equation, but the converse is not true. Or \textit{the London equation makes a stronger claim than the Drude model, which distinguishes \SCs\ from perfect conductors.}

This difference arises in magnetic responses. Because $\bm{B}=\bm{\nabla} \times \bm{A}$, the London equation implies 
\be
\bm{\nabla} \times \bm{J} + \frac{1}{\lambda^2} \bm{B} = 0~,
\label{eq:SC_magnetic}
\ee
whereas the Drude form only implies 
\be
\frac{\del}{\del t} \left(\bm{\nabla} \times \bm{J} + \frac{1}{\lambda^2} \bm{B} \right) = 0~.
\label{eq:PC_magnetic}
\ee
In a perfect conductor, \eq{PC_magnetic} tends to $\del B/\del t=0$ instead of $B=0$. Namely, the \textit{exclusion} of a magnetic field from entering a sample can be explained by a perfect conductivity. On the other hand, a magnetic field in an originally normal state is also \textit{expelled} as it is cooled below $T_c$. This cannot be explained by perfect conductivity since it tends to trap flux in. 

\begin{figure}[tb]
\centering
\scalebox{0.5}{\includegraphics{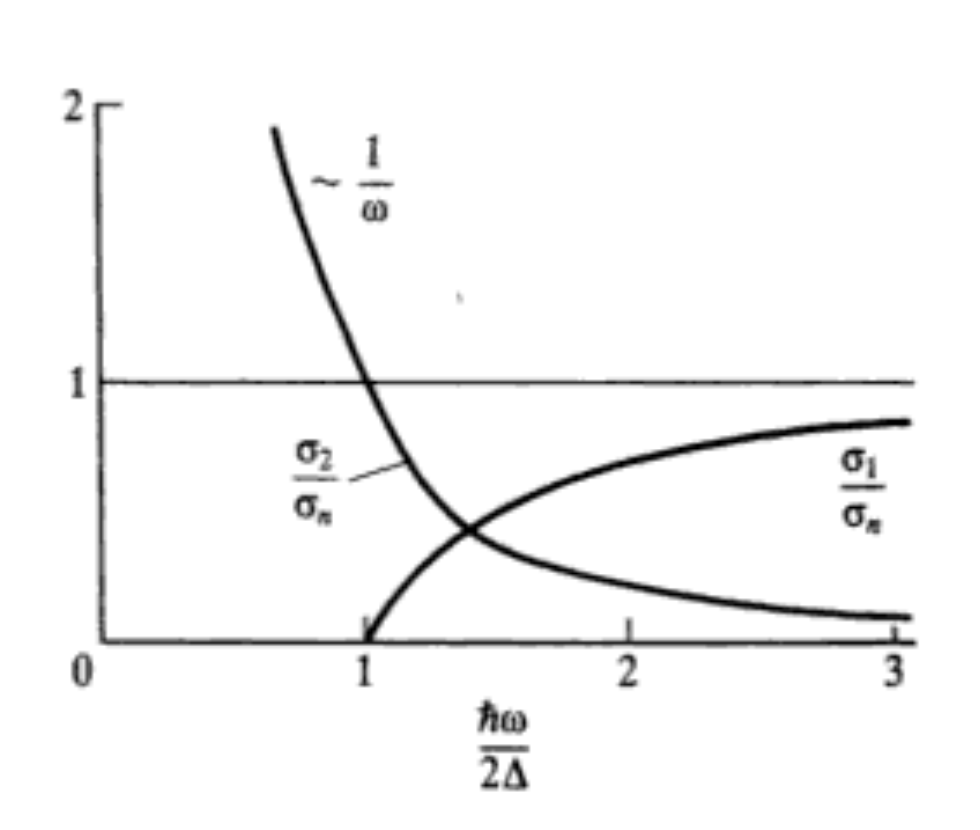} }
\caption{Complex conductivity for a typical superconductor at zero temperature \cite{tinkham}. Here, $\sigma_1$ and $\sigma_2$ represent the real and imaginary parts, respectively.}
\label{fig:tinkham} 
\end{figure}

\head{AC conductivity of \SCs}
Let us go back to the conductivity and briefly remark the other aspects of the AC conductivity for a \SC. At low-$\omega$, the AC conductivity $\text{Re}[\sigma(\omega)]$ vanishes at zero temperature (\fig{tinkham}), but there is a small AC conductivity at finite temperature. 

According to the two-fluid model, a superfluid consists of the superfluid component and the normal component. The superfluid component has no dissipation, but the normal part has a dissipation. The normal components are thermally excited electrons. If one increases temperature, more and more electrons are excited, which blocks the Cooper pair formation. Eventually, this process destroys superconductivity at $T_c$. 

One would describe the AC conductivity of the two-fluid model \index{two-fluid model} using the Drude model as
\be
\sigma(\omega) \simeq 
\frac{n_s \es^2}{\ms} \pi\delta(\omega) 
+ \frac{n_n \es^2\tau_n}{\ms} 
+ i \frac{n_s \es^2}{\ms} \frac{1}{\omega}~,
\label{eq:2_fluid_electric}
\ee
where $n_s$ is the number density of the superfluid component with $\tau_s\rightarrow\infty$, and $n_n$ is the number density of the normal component with a finite $\tau_n$. Comparing the last term of \eq{2_fluid_electric} with \eq{pole}, one gets $n_s=|\psi|^2$. 

Equation~\eqref{eq:2_fluid_electric} is valid for low-$\omega$. For $\hbar\omega>2\Delta$, where $\Delta$ is called the energy gap, a photon can break up a Cooper pair and create thermally excited electrons. These additional electrons increase the AC conductivity for $\hbar\omega>2\Delta$ (\fig{tinkham}).

\subsection{High-$\Tc$ superconductors}\label{sec:Hi-Tc}

\begin{figure}[tb]
\centering
\scalebox{0.4}{ \includegraphics{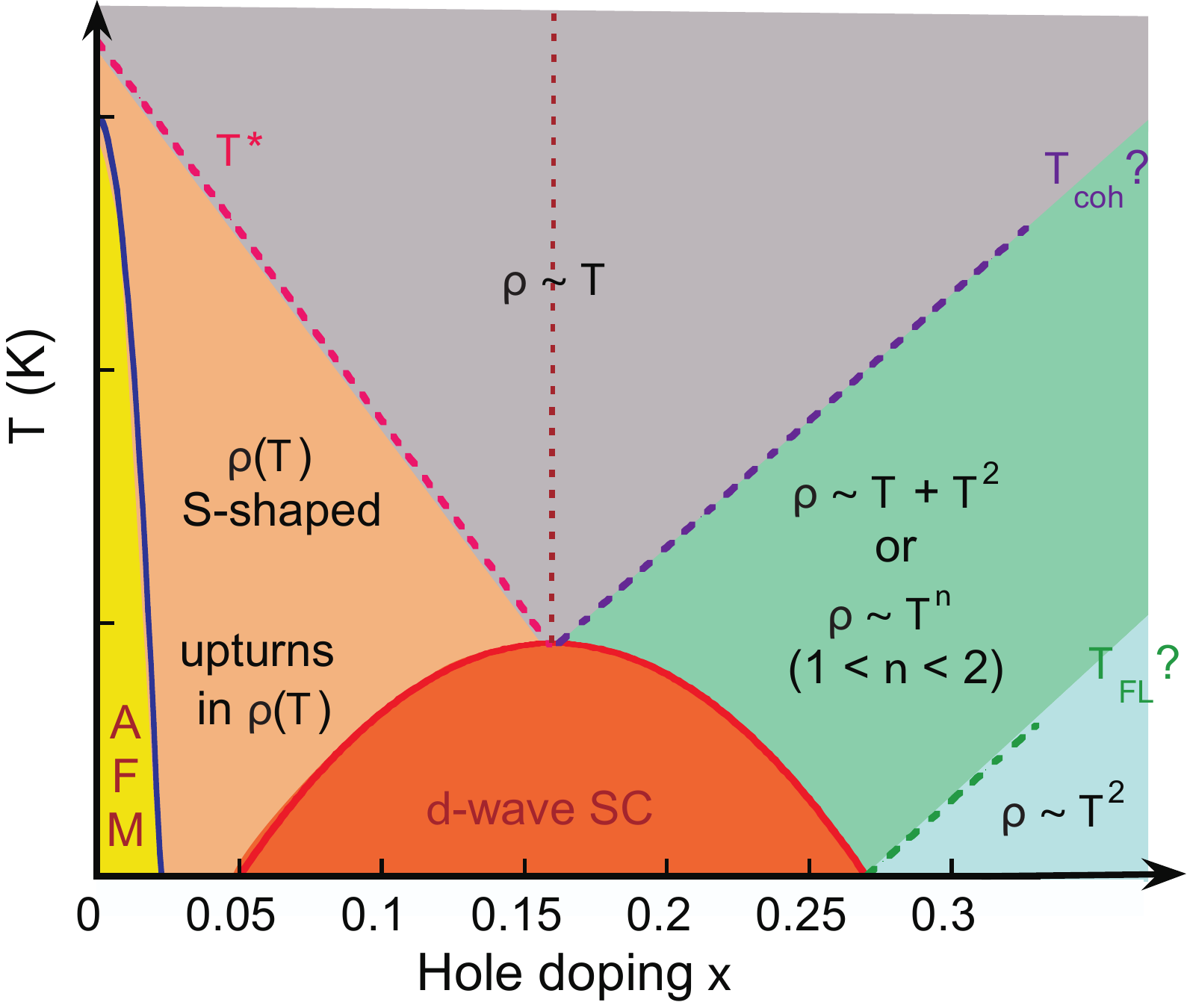} }
\caption{Typical phase diagram of high-$\Tc$ \SCs\ with the resistivity parallel to Cu-O planes \cite{hussey}. 
The dome-shaped region is the $d$-wave superconducting phase. The other regions are (from left to right) antiferromagnetic phase (AFM), pseudo gap region, non-Fermi liquid phase, and the Fermi liquid phase.}
\label{fig:Hi-Tc} 
\end{figure}%

After the discovery of superconductivity in 1911 (Hg, 4K), the highest $\Tc$ had been 23K in Nb$_3$Ge. In 1986, \textit{high-$\Tc$ \SCs}\index{high-Tc superconductors@high-$\Tc$ \SCs} were found. These are copper oxide compounds or \keyword{cuprates}. 
Among cuprates, some well-known systems are
\begin{itemize}
\item Y-Ba-Cu-O system (``YBCO")
\item Bi-Sr-Ca-Cu-O system (``BSCCO")
\end{itemize}
Currently, HgBa$_2$Ca$_2$Cu$_3$O$_8$ has the highest $\Tc$ ($\sim 135$K). Under high pressures, $\Tc$ of the material rises to about 165K. 

The subject of \hightc\ \SCs\ is diverse, and it is not our purpose here to give a detailed overview of \hightc\ superconductivity. Here, we explain some of basic terms and properties. 

The most important property of \hightc\ materials is that $\Tc$ seems too high from the traditional BCS picture. In the traditional theory, highest $\Tc$ is expected to be 30-40K. The BCS theory is based on the weakly-coupled Fermi liquid description, so the description seems insufficient to describe \hightc\ materials. This suggests that one needs a strong coupling description. We see several other indications of this below.

From the crystal point of view, another important property is that their crystal structures share the two-dimensional Cu-O planes. The electrons or holes on the planes are responsible for the \SC.

Figure~\ref{fig:Hi-Tc} shows a typical phase diagram of \hightc\ materials. The solid lines are phase boundaries while the dashed lines indicate crossovers. The horizontal axis represents the doping. For example, consider the originally discovered \hightc\ material, La$_{2-x}$Sr$_x$CuO$_4$. For the material, one replaces La$^{3+}$ by Sr$^{2+}$. The material must be charge neutral under the replacement, which results in adding positively charged \keyword{holes}. This is called the \keyword{hole doping}. These mobile holes are responsible for conduction.

When $x=0$, the system is an anti-ferromagnetic insulator. This suggests a strong electron-electron interaction. As one increases $x$, the material becomes conducting. It becomes superconducting at some point, and $\Tc$ keeps increasing. However, if one further increases $x$, $\Tc$ starts to decrease, and the superconducting phase is eventually gone. The transition temperature $\Tc$ becomes highest at the \keyword{optimal doping}. A smaller $x$ is called \keyword{underdoping}, and a larger $x$ is called \keyword{overdoping}. 

In the superconducting phase, the electrons form the Cooper pair, but the pairing mechanism is not well-understood. The Cooper pair wave function has a different symmetry. The wave function of an electron in an atom is classified by the orbital angular momentum, \eg, $s$ ($l=0$), $p$ ($l=1$), and $d$ ($l=2$). Similarly, the Cooper pair wave function may be the $s$-wave, $p$-wave, and $d$-wave. The conventional \SCs\ are $s$-wave \SCs, but \hightc\ \SCs\ are $d$-wave \SCs.

A \hightc\ material has a very rich phase structure, and what is interesting is not only the superconducting phase, but the whole phase diagram is interesting. 
\begin{itemize}

\item First, the overdoped region shows the Fermi liquid behavior. For a Fermi liquid, the resistivity is proportional to $T^2$. 

\item But the normal state immediately above the superconducting dome shows \textit{linear resistivity}. The region is called the \keyword{non-Fermi liquid} or the \keyword{strange metal}. Thus, the weakly-coupled description of the Fermi liquid is certainly insufficient to describe \hightc\ materials.

\item Region left to $T^*$ is called the \keyword{pseudo gap region} which is also mysterious. It is not a phase since there seems no long-range order, but there seem indications of an energy gap.

\end{itemize}

\section{\titlesummary}

\begin{itemize}
\item 
The mean-field theory or the GL theory describe phase transitions. In second-order phase transitions, one encounters the critical phenomena where various quantities have power-law behaviors and these powers (critical exponents) obey the universality.
\item 
A simple extension of the GL theory describes superconductivity/superfluidity.
\item 
High-$\Tc$ superconductivity is not completely understood, but it is likely to involve strong-coupling physics.
\end{itemize}

\titlenewterms

\begin{multicols}{2}
\noindent
phase transition\\
order parameter\\
correlation length\\
critical exponent\\
universality\\
mean-field theory\\
Ginzburg-Landau (GL) theory\\
thermodynamic susceptibility\\
static susceptibility\\
scaling relations\\
hyperscaling relation\\
scaling law\\
upper/lower critical dimension\\
Coleman-Mermin-Wagner theorem\\
static/dynamic critical phenomena\\
BCS theory\\
phonons\\
Cooper pair\\
macroscopic wave function\\
Meissner effect\\
Nambu-Goldstone boson\\
Higgs mechanism\\
GL parameter\\
Type I/Type II superconductors\\
vortex\\
lower/upper critical magnetic field\\
London equation\\
Drude model\\
Fermi/non-Fermi liquid\\
Kramers-Kronig relation\\
two-fluid model\\
high-$\Tc$ \SCs\\
cuprates\\
hole doping\\
optimal doping/underdoping/overdoping\\
strange metal\\
pseudo gap region
\end{multicols}

\endofsection

\ifx\nameofpaper\undefined 
  \usepackage{macro_natsuume} 
  \def\beginsection{\section*}
  \def\endofsection{\end{document}} 
  \input draft_header.tex
\else 
  \def\beginsection{\chapter}
  \def\endofsection{ } 
\fi

\beginsection{AdS/CFT - phase transition
}\label{chap:phase}


\begin{quote}
There are systems with phase transitions in AdS/CFT. We discuss typical examples of the first-order phase transition and of the second-order phase transition. 
\end{quote}

\section{Why study phase transitions in AdS/CFT?
}

The \Nfour\ SYM has only the plasma phase, but in this chapter, we consider theories and black holes which undergo phase transitions. The reasons are two-folds:
\begin{enumerate}

\item First, it is interesting as gauge theories. QCD has a rich phase structure. The phase structure and its related phenomena have been widely investigated both theoretically and experimentally. Condensed-matter tools such as \chap{GL} may be enough to understand QCD phase transitions theoretically. But AdS/CFT may bring new insights into the problem.
\item Second, one often encounters strongly-coupled systems in condensed-matter physics as well. Of particular interest is high-$T_c$ superconductivity. It would be nice if one could have a dual gravity description. Understanding its rich phase structure is likely to be the key to solve the mysteries of the system. 

\end{enumerate}

But there is an immediate problem in order to apply AdS/CFT to condensed-matter systems. The simplest AdS/CFT uses the AdS$_5$ spacetime, which is dual to the ${\cal N}=4$ SYM. It is different from QCD but is not very ``far" in the sense that they are both Yang-Mills theories. However, it is not clear if any large-$N_c$ gauge theory is behind in condensed-matter systems. 
So, one approach one takes is to 
\begin{itemize}
\item Put aside the correspondence with real condensed-matter systems for the time being. 
\item Simply realize interesting condensed-matter behaviors in the context of large-$N_c$ theory. 
\end{itemize}
Our hope is that the  AdS/CFT duality will give some insights even to real condensed-matter systems. 

Now, what kind of phenomena should we focus on? AdS/CFT may predict a variety of exotic phenomena, but they may not be very relevant unless one can see them in laboratories. Also, we cannot cover all possible phenomena, so in this book, we focus on robust phenomena in low-energy physics such as superconductivity and critical phenomena.

\section{First-order phase transition
}\label{sec:HP}

\subsection{Simple example}\label{sec:phase_example}

We now study the holographic phase transition. We have been evaluating \bh partition function by a saddle-point approximation:
\be
Z_\text{gauge} \simeq e^{-\SosE}~,
%
\ee
where $\Sos$ represents the on-shell action which is obtained by substituting the classical solution $g$ to the action. However, the classical solution may not be unique. For example, if we have two saddle points, we must sum over the saddle points:
\be
Z_\text{gauge} \simeq e^{-\SosE} + e^{-\SosE'}~.
%
\ee
Here, $\SosE$ and $\SosE'$ are the on-shell actions obtained by the classical solutions $g$ and $g'$, respectively. The dominant contribution comes from the solution with lower free energy.

As a simple example, let us consider the $S^1$-compactified \Nfour\ SYM with periodicity $l$. The \Nfour\ SYM on $\mathbb{R}^3$ is scale invariant and has no dimensionful parameter except the temperature. One can always change the temperature by a scaling, and all temperatures are equivalent. Introducing a scale $l$ in the gauge theory changes physics drastically. One can no longer change the temperature by a scaling, and the theory is parametrized by a dimensionless parameter $Tl$. As we will see, the theory has a first-order phase transition at $Tl=1$.

As discussed in \sect{ads_soliton}, there are two possible dual geometries which approach $\mathbb{R}^{1,2} \times S^1$ asymptotically. The first one is the Schwarzschild-SAdS$_5$ (SAdS$_5$) black hole:
\begin{align}
ds_5^2 &= \left(\frac{r}{L}\right)^2 (-hdt^2+dx^2+dy^2+dz^2)+L^2\frac{dr^2}{h r^2}~, 
\\
h &= 1- \left(\frac{r_0}{r}\right)^4~, \quad 0<z\leq l~,
\end{align}
and the second one is the AdS soliton: \index{AdS soliton}
\begin{align}
ds_5^2 &= \left(\frac{r}{L}\right)^2 (-dt'^2+dx^2+dy^2+hdz'^2)+L^2\frac{dr^2}{h r^2}~,\\
l &= \frac{\pi L^2}{r_0}~,
%
\end{align}
which is obtained from the black hole by the ``double Wick rotation" $z' = it~, z = it'$. 

The SAdS \bh describes the plasma phase whereas the AdS soliton describes the confining phase. For the AdS soliton, the Wilson loop shows the linear potential (\sect{ads_soliton}). Also, the AdS soliton does not have entropy since it is not a black hole (more precisely, its entropy is suppressed by $1/N_c^2$ compared to the black hole), which is appropriate as the confining phase. 

At high temperature, the AdS soliton undergoes a first-order phase transition to the SAdS black hole. Then, the phase transition describes a confinement/deconfinement transition in the dual gauge theory%
\footnote{This is an example of a confinement/deconfinement transition, but this does not explain the QCD  confinement/deconfinement transition. First, the theory here is the \Nfour\ SYM and is not QCD. Second, this is a phenomenon which happens in large-$N_c$ gauge theories on compact spaces.}.

Thus, evaluate the free energy difference between the SAdS \bh and the AdS soliton. In \sect{SAdS_free}, we computed the free energy for the SAdS black hole:
\be
F_\text{BH} =  - \frac{V_3}{16\pi G_5} \frac{r_0^4}{L^5} 
= - \frac{V_3 L^3}{16 \pi G_5} \pi^4 T^4~.
%
\ee
The AdS soliton has the same Euclidean geometry, so the free energy for the AdS soliton takes the same form (if expressed in terms of $r_0$). But for the AdS soliton, $r_0$ is not related to the temperature $T$ but is related to the $S^1$ periodicity $l$:
\be
F_\text{soliton} =  - \frac{V_3}{16\pi G_5} \frac{r_0^4}{L^5} 
= - \frac{V_3 L^3}{16 \pi G_5} \frac{\pi^4}{l^4}~.
%
\ee
Then, the free energy difference is given by
\be
\Delta F = F_\text{BH} - F_\text{soliton} 
= - \frac{V_3 L^3}{16 \pi G_5} \pi^4 \left(T^4 - \frac{1}{l^4}\right)~.
%
\ee
Thus,
\begin{itemize}

\item At low temperature $T<1/l$, the stable solution is the AdS soliton which describes the confining phase.
\item At high temperature $T>1/l$, the stable solution is the black hole which describes the unconfining phase.

\end{itemize}
Because one forms a \bh which has entropy, the entropy is discontinuous at $Tl=1$. Since the first derivative of free energy, $S = -\del_T F$, is discontinuous there, this is a first-order phase transition. 

\subsection{Hawking-Page transition}\label{sec:HP_S3}

The phase transition we saw above is generally called the \keyword{Hawking-Page transition} \cite{Hawking:1982dh}. We now discuss the original Hawking-Page transition which uses a SAdS \bh with spherical horizon. The SAdS$_5$ black hole with $S^3$ horizon \index{Schwarzschild-AdS black hole (spherical horizon)} (\sect{spherical_SAdS}) is given by
\be
ds_5^2 =  - \left( \frac{r^2}{L^2} + 1 - \frac{r_0^4}{L^2r^2} \right) dt^2 
+ \frac{dr^2}{ \frac{r^2}{L^2} + 1 - \frac{r_0^4}{L^2r^2} } 
+ r^2 d\Omega_3^2~.
%
\ee
As $r\rightarrow\infty$, the metric along the AdS boundary approaches
\be
ds_4^2 \sim \left( \frac{r}{L} \right)^2 (-dt^2 + L^2 d\Omega_3^2)~,
\quad (r\rightarrow\infty)~,
%
\ee
so the dual gauge theory is the \Nfour\ SYM on $S^3$ with radius $L$.

As in the previous example, we introduce the scale $L$ in the gauge theory, so the theory is parametrized by a dimensionless parameter $TL$, and it has a first-order phase transition at an appropriate $TL$, but the details are slightly more complicated%
\footnote{In the large-$\Nc$ limit, a phase transition can occur even in finite volume.}.

\begin{figure}[tb]
\centering
\scalebox{0.75}{ \includegraphics{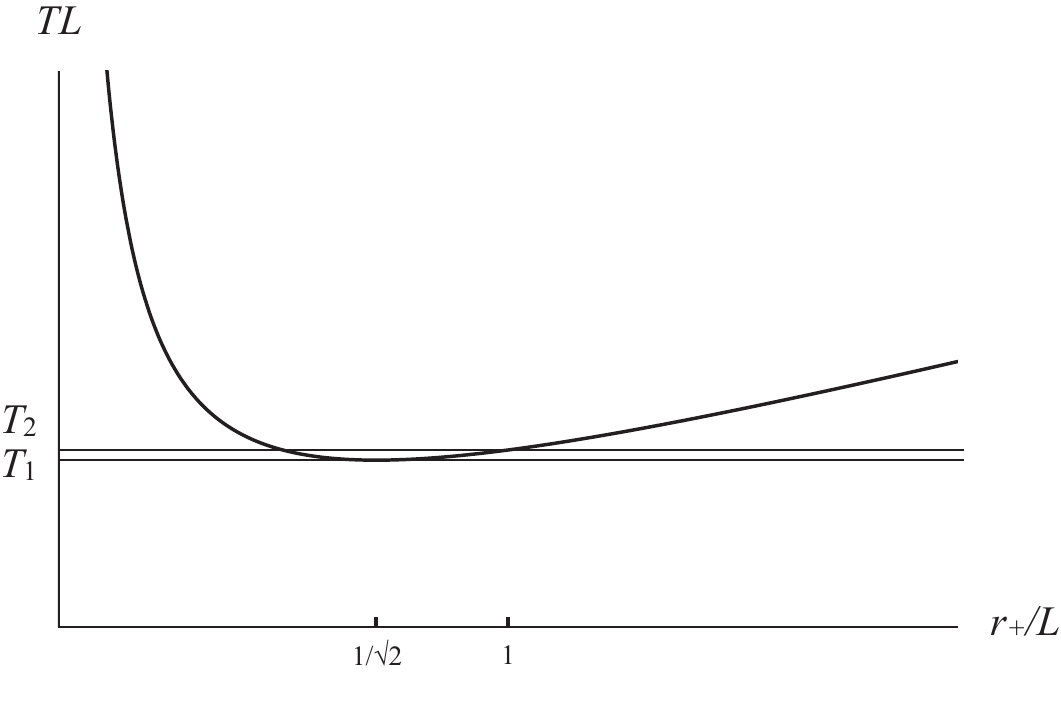} }
\vskip2mm
\caption{The horizon radius $\ro$ versus the temperature for the SAdS$_5$ \bh with spherical horizon. For a given temperature, there are two values of $\ro$. The \bh has the lowest temperature at $T=T_1$, and the Hawking-Page transition occurs at $T=T_2$.}
\label{fig:HP_temp}
\end{figure}

 
The horizon is located at $r=\ro$, where
\be
\frac{\ro^2}{L^2} + 1 - \frac{r_0^4}{L^2\ro^2} = 0 
\result
r_0^4 = \ro^4 + L^2\ro^2~.
%
\ee
The temperature is given by
\be
T = \frac{ 2\ro^2 + L^2}{2\pi \ro L^2}~.
%
\ee
Figure~\ref{fig:HP_temp} shows the horizon radius $\ro$ versus the temperature. For a given temperature, there are two values of $\ro$. The temperature has the minimum $T_1:=\sqrt{2}/(\pi L)$ when $\ro = L/\sqrt{2}$. The solution with $\ro < L/\sqrt{2}$ is called the ``small black hole." The small \bh is small enough compared with the AdS scale $L$, and the effect of the cosmological constant is negligible. Then, the behavior of the small \bh is similar to the asymptotically flat Schwarzschild black hole. In fact, when $\ro \ll L$,
\be
T \simeq \frac{1}{2\pi \ro}~,
%
\ee
which is the behavior of the five-dimensional Schwarzschild solution \eqref{eq:temp_Sch_d}.

On the other hand, the solution with $\ro > L/\sqrt{2}$ is called the ``large black hole." The behavior of the large \bh is similar to the SAdS$_5$ solution with planar horizon. In fact, when $\ro \gg L$,
\be
T \simeq \frac{\ro}{\pi L^2}~.
%
\ee
 
The \bh solution does not exist when $T<T_1$. In this case, the solution is the ``thermal AdS spacetime," which is the AdS spacetime with Euclidean time periodicity $\beta_0$. As we increase temperature, there is a phase transition from the thermal AdS spacetime to the SAdS$_5$ black hole. Again, this is a confinement/deconfinement transition \cite{Witten:1998zw}.
 

%
%
%
%

We again evaluate the free energy difference between the SAdS$_5$ \bh and the thermal AdS$_5$ spacetime. The free energy is obtained by repeating a computation similar to \sect{SAdS_free}. In general, in order to obtain the free energy, one needs to evaluate not only the bulk action, but also the Gibbons-Hawking action, and the counterterm action. But, in this case, it is enough to evaluate only the bulk action if one is interested in the free energy difference.

Namely, one obtains the finite free energy using the spacetime without \bh which is called the \keyword{reference spacetime}. This prescription is often used in general relativity, and one uses the reference spacetime method for asymptotically flat black holes such as the Schwarzschild black hole.

In \sect{SAdS_free}, we saw that the on-shell bulk action becomes the spacetime volume \eqref{eq:bulk_free}. Evaluating the volume both for the SAdS$_5$ \bh and for the thermal AdS$_5$ spacetime, one obtains
\begin{align}
\action_\text{SAdS} 
&= \frac{1}{2\pi G_{5} L^2} \int_0^\beta dt \int d\Omega_3 \int_{\ro}^{r} dr \, r^3
= \left. \frac{\beta \Omega_3}{8\pi G_5 L^2} (r^4-\ro^4) \right|_{r=\infty},
\\
\action_\text{AdS} 
&= \frac{1}{2\pi G_{5} L^2} \int_0^{\beta_0} dt \int d\Omega_3 \int_{0}^{r} dr \, r^3
= \left. \frac{\beta_0 \Omega_3}{8\pi G_5 L^2} r^4 \right|_{r=\infty}.
%
\end{align}

Care is necessary for the thermal AdS$_5$ temperature. The temperature of a \bh is determined by requiring that the Euclidean geometry does not have a conical singularity. The periodicity $\beta_0$ for the thermal AdS$_5$ spacetime is arbitrary, but we are interested in the free energy difference, so we must match the thermal AdS$_5$ temperature with the SAdS$_5$ one. The Hawking temperature of the SAdS$_5$ \bh is $T$, but this differs from the proper temperature $T(r)$ at radius $r$ [\eq{proper_temp}]:
\be
T(r)=\frac{1}{\sqrt{|g_{00}(r)|}} T~.
%
\ee
The correct procedure is to match this proper temperature in the reference spacetime method%
\footnote{The difference between $\beta_0(r)$ and $\beta$ is subleading in $r$, so this prescription matters when the on-shell action diverges such as the reference spacetime regularization. If one employs the counterterm regularization, the free energy remains finite for both spacetimes, so the difference does not matter.}.
Thus,
\be
\beta f^{1/2} =  \beta_0(r) f_0^{1/2}~,
%
\ee
where $f$ and $f_0$ are the $g_{00}$ components for the SAdS$_5$ \bh and the thermal AdS$_5$ spacetime. Then, the free energy difference is given by
\begin{align}
\Delta F 
&:= \frac{1}{\beta}(\action_\text{SAdS} - \action_\text{AdS}) \\
&=
\left. \frac{\Omega_3}{8\pi G_5 L^2} 
\left\{ 
 r^4 - \ro^4 - r^4 
 \sqrt{\frac{ \left(\frac{r}{L} \right)^2 + 1-  \frac{r_0^4}{L^2r^2} }{\left(\frac{r}{L} \right)^2+1}} 
\right\} \right|_{r=\infty} \\
& \xrightarrow{r\rightarrow\infty}{}
\frac{\Omega_3}{8\pi G_5 L^2} 
\left\{ 
 r^4 - \ro^4 - r^4 \left( 1 - \frac{r_0^4}{2r^4} \right) 
\right\} \\
&=  - \frac{\pi \ro^2}{8G_5 L^2} (\ro^2-L^2)~.
%
\end{align}
Therefore, $\Delta F<0$ when $\ro>L$ or $T>T_2:=3/(2\pi L)$. The small \bh never satisfies the condition and is not allowed as a stable equilibrium.

\begin{figure}[tb]
\centering
\scalebox{0.75}{ \includegraphics{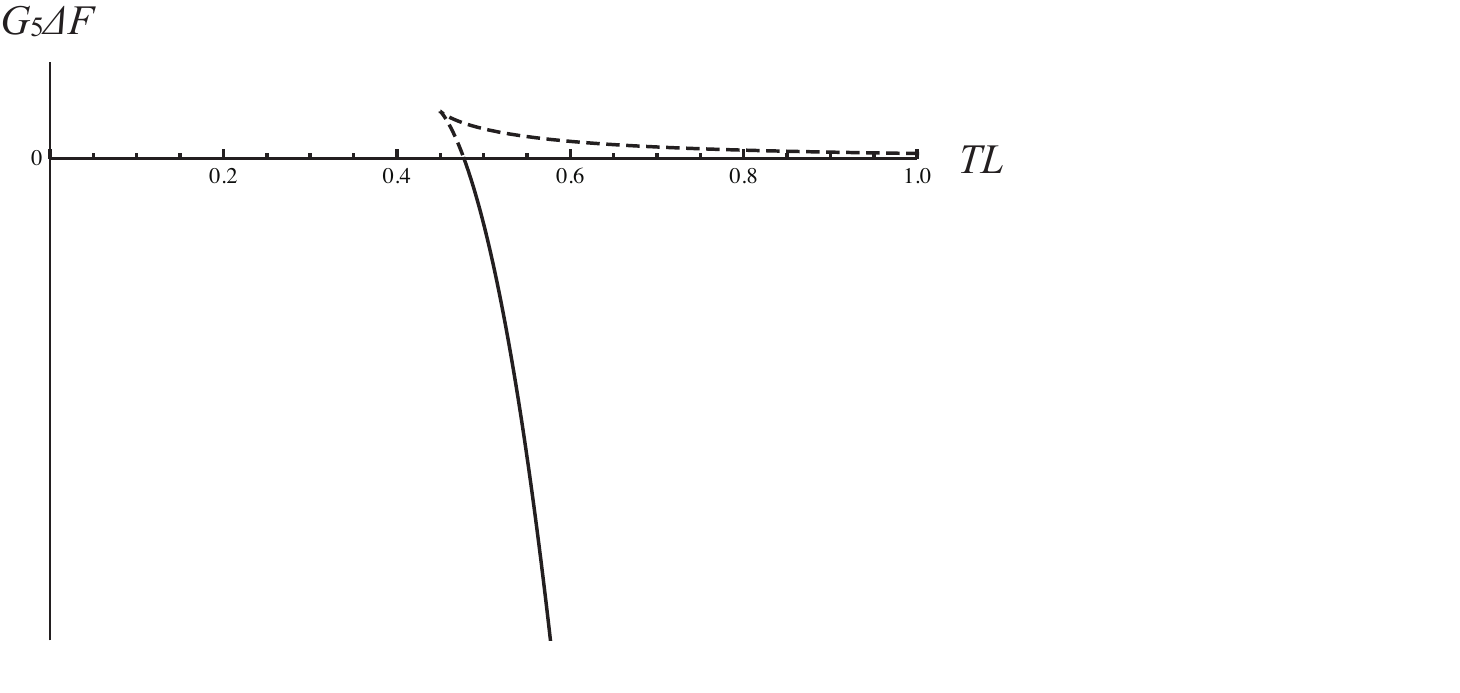} }
\vskip2mm
\caption{The free energy $\Delta F(T)$ for the SAdS$_5$ \bh with spherical horizon. The region with $\Delta F>0$ is plotted with a dashed line because the \bh not stable in the region and the thermal AdS$_5$ spacetime is stable.}
\label{fig:HP_free}
\end{figure}

We plot $\Delta F(T)$ in \fig{HP_free}. As one increases the temperature, the plot shows the following behavior:
\begin{itemize}

\item $T<T_1$: A \bh cannot exist, and the thermal AdS spacetime is the stable state.
\item
$T_1 \leq T<T_2$: A \bh can exist but has $\Delta F>0$, so again the thermal AdS spacetime is the stable state. A \bh decays to the thermal AdS spacetime via the Hawking radiation.
\item
$T \geq T_2$: A large \bh has $\Delta F<0$, so there is a discontinuous change from the thermal AdS spacetime to the large black hole. The transition temperature is $T=T_2$.

\end{itemize}
Because we form a \bh at $T=T_2$, this is again a first-order phase transition as in \sect{phase_example}. One would compare with the mean-field theory in \sect{phase_1st}. When $T_1 \leq T<T_2$, the pseudo free energy develops a saddle point of a metastable state which corresponds to a black hole. But the state is not globally stable, so it decays to the globally stable state, the thermal AdS spacetime.

One can obtain thermodynamic quantities from $\Delta F$:
\begin{align}
S &= - \del_T \Delta F 
= \frac{\pi^2 \ro^3}{2 G_5} 
= \frac{\Omega_3 \ro^3}{4 G_5}~,
\\
E &= \Delta F + TS 
= \frac{3\pi \ro^2 (\ro^2 + L^2)}{8G_5 L^2}
= \frac{3 \pi r_0^4}{8G_5L^2}~,
\\
C &= \frac{\pi^2}{2G_5} \frac{ 3\ro^3(2\ro^2+L^2) }{ 2\ro^2-L^2 }~.
%
\end{align}


If one repeats the computation in \sect{SAdS_free}, one can obtain the free energy itself:
\be
F_\text{SAdS} =  - \frac{\pi}{8G_5 L^2} \left(\ro^4 - L^2\ro^2 -\frac{3}{4} L^4 \right)~,
%
\ee
which gives
\be
S = \frac{\pi^2 \ro^3}{2G_5}~, \quad
E = \frac{3 \pi r_0^4}{8 G_5L^2} + \frac{3 \pi L^2}{32 G_5}~.
\ee
Note that the energy has a constant term. From the AdS/CFT dictionary, $L^3/G_5=2N_c^2/\pi$, so the constant is rewritten as
\be
E_0 = \frac{3N_c^2}{16L}~.
\label{eq:Casimir_ads}
\ee
This $E_0$ can be interpreted as the \keyword{Casimir energy} from the boundary point of view%
\footnote{The Casimir energy is the zero-point energy of a field theory with a boundary condition.}
\cite{Balasubramanian:1999re}. For free fields on $S^3\times \mathbb{R}$ with radius $L$, the Casimir energy is given by
\be
E_\text{Casmir} =\frac{1}{960L} (4n_0 + 17n_{1/2} + 88n_1)~,
\label{eq:Casimir_qft}
\ee
where $n_i$ represents the number of field species \cite{BD}. For the \Nfour\ SYM,
\begin{alignat}{2}
&\text{real scalars: }~& n_0 &= 6(N_c^2-1)~, \\ 
&\text{Weyl fermions:}~& n_{1/2} &= 4(N_c^2-1)~, \\ 
&\text{vector fields:}~& n_1 &= N_c^2-1~. 
%
\end{alignat}
Substituting these into \eq{Casimir_qft}, one obtains
\be
E_\text{Casmir} =\frac{3(N_c^2-1)}{16L}~,
%
\ee
which agrees with \eq{Casimir_ads} in the large-$\Nc$ limit.

\section{Second-order phase transition: Holographic superconductors
}\label{sec:H^3}


\subsection{Overview}

We now turn to second-order phase transitions. First-order phase transitions were relatively easy in the sense that introducing another scale was enough to achieve. But these examples are transitions from a non-black hole geometry to a \bh geometry, and they are necessarily first-order transitions. In a second-order transition, the entropy must be continuous, so we need a transition from a \bh to another black hole.

However, recall the no-hair theorem. \index{no-hair theorem} According to the theorem, a black hole often has a few parameters, mass, angular momentum, and charge. Given these quantities, the black hole solution is unique. But in order to have a phase transition, we need a multiple number of saddle points.

Thus, one would expect that a second-order transition is more subtle to achieve. Is there any? Let us consider gravity systems starting from the simplest one to more complicated ones. We focus on black holes with planar horizon which has an infinite extension and not black holes with compact horizon:
\begin{itemize}

\item The simplest gravity system is pure gravity, and the solution is the SAdS black hole.\index{Schwarzschild-AdS black hole (planar horizon)} The dual gauge theory is the ${\cal N}=4$ SYM. The ${\cal N}=4$ SYM has no dimensionful quantity other than temperature, so all temperatures are equivalent. Thus, there is no phase transition, and we need a more complicated system. 

\item The next simple system is the Einstein-Maxwell system, and the solution is the charged black hole or the Reissner-Nordstr\"{o}m-AdS (RN-AdS) black hole \index{Reissner-Nordstr\"{o}m AdS black hole} (\sect{others}). The \bh has an additional dimensionful quantity set by the chemical potential $\mu$ or has an additional dimensionless parameter $\mu/T$. Thus, not all temperatures are equivalent in this case. But the system still has no phase transition. Adding a charge is not enough. It is likely that this is related to the no-hair theorem. 

\item What we can do is to add a scalar. Usually (for asymptotically flat solutions), the existence of a scalar does not affect a black hole solution from the no-hair theorem. But the theorem is not entirely true for higher-dimensional spacetime or for the AdS spacetime. So, adding a scalar can give a black hole with nonzero scalar. 

\end{itemize}
Thus, we arrive at an Einstein-Maxwell-scalar system. Below we consider such a system to discuss superconductivity and critical phenomena.





We asked what system is necessary from the gravity side, but let us ask a slightly different question from the field theory side. If one is interested in superconductivity as a second-order transition system, what kind of ingredients are necessary to realize a \SC\ in AdS/CFT?
\begin{itemize}

\item First, we will use a field theory, so the theory has a conserved energy-momentum tensor. 
\item One unique characteristic of a \SC\ is the zero resistivity or the diverging conductivity. In order to see this, we need a $U(1)$ current. 
\item In addition, a \SC\ is a phenomenon associated with a phase transition. For a \SC, the order parameter is the charged scalar operator $\bra \calO \ket$. The operator $\bra \calO \ket$ corresponds to the ``macroscopic wave function" in the Ginzburg-Landau (GL) theory.

\end{itemize}
Given these field theory ingredients, the AdS/CFT dictionary (\sect{dictionary}) tells necessary bulk fields. The boundary operators $T^{\mu\nu}$, $J^{\mu}$, and $\bra \calO \ket$ correspond to the bulk metric $g_{MN}$, Maxwell field  $A_M$, and complex scalar $\Psi$, respectively. So, we again end up with an Einstein-Maxwell-complex scalar system. Thus,  there are many kinds of \keyword{holographic superconductors}, but one typically considers an Einstein-Maxwell-complex scalar system \cite{Gubser:2008px,Hartnoll:2008vx}:
\begin{align}
& \action = \int d^{p+2}x \sqrt{-g} \left[ R - 2\Lambda - \frac{1}{4} F_{MN}^2  
- |D_M\Psi|^2 -V(|\Psi|) \right]~,
\label{eq:H^3_action} \\
& D_M := \nabla_M - i\qm A_M~, \\
& V(|\Psi|) = m^2 |\Psi|^2~.
\end{align}
This system looks similar to the GL theory \eqref{eq:GL_ferro2} with two differences. First, in this model, gravity is coupled. Second, the order parameter has only the mass term and has no $O( |\Psi|^4)$ term. One could include nonlinear terms, but {\it nonlinear terms are not necessary to achieve a symmetry breaking.}

This system has the RN-AdS black hole with $\Psi=0$ as a solution (\sect{others}).\index{Reissner-Nordstr\"{o}m AdS black hole} This corresponds to the normal state. But at low enough temperature $T<T_c$, the solution becomes unstable and undergoes a second-order phase transition. As a result, the solution is replaced by a \bh with $\Psi\neq0$. Thus, $\Psi$ characterizes the phase transition, and the corresponding operator $\bra \calO \ket$ indeed can be interpreted as the order parameter in the field theory side.

The detail of the dual field theory is unclear though because we do not go through the brane argument such as \sect{D3_brane} unlike the \Nfour\ SYM. Rather, we just collect the minimum ingredients which realize superconductivity. But the low-temperature phase corresponds to a certain kind of superconductivity or superfluid. The justifications come from the diverging conductivity, the London equation, and the existence of an energy gap as we see below.

\subsection{Probe limit}

In this system, gravity is coupled with a multiple number of matter fields. Analyzing such a system is difficult in the low-temperature phase (superconducting phase). So, one often carries out numerical computations or employs approximations. We mainly consider the ``probe limit," \index{probe limit} where we add matter fields as probes. In the probe limit, one redefines matter fields as $\Psi \rightarrow \Psi/\qm$ and $A_M \rightarrow A_M/\qm$ and takes the scalar charge $\qm \rightarrow \infty$:
\be
\action = \int d^{p+2}x \sqrt{-g} \left[ R - 2\Lambda + \frac{1}{\qm^2} 
\left\{ -\frac{1}{4} F_{MN}^2  - |D_M\Psi|^2 - V(|\Psi|) \right\} \right]~.
\label{eq:probe_action}
\ee
As one can see from the action, the Maxwell field and the scalar field decouple from gravity in this limit. One can ignore the backreaction of matter fields onto the geometry, so the \bh solution is simply a pure gravity one, the SAdS black hole, \index{Schwarzschild-AdS black hole (planar horizon)} and it is enough to solve matter equations on the \bh geometry. For $p=2$, the metric is given by
\be
ds_4^2 = \left(\frac{r_0}{L}\right)^2\frac{1}{u^2} (-hdt^2+dx^2+dy^2)+L^2\frac{du^2}{h u^2}~, \quad
h = 1- u^3~.
%
\ee

Of course, the SAdS \bh itself has no dimensionful quantity other than temperature, so one has to introduce another scale to have a phase transition. We introduce a chemical potential. Then, the system is parametrized by $\mu/T$. 

When $\Psi=0$, 
the solution $A_0$ takes the same form as \eq{gauge_falloff}:
\be
A_0 = \mu(1-u)~, \quad A_i = A_u = 0, \quad \Psi = 0~,
\label{eq:background_matter}
\ee
where we take the gauge $A_u=0$ and $\left. A_0\right|_{u=1}=0$ (\sect{current}). This is the solution in the high-temperature phase.

Like the GL theory, the system becomes unstable at $T_c$ because $\Psi$ becomes tachyonic. However, unlike the GL theory, the holographic \SC\ does not have the Higgs-like potential, so the instability comes from a different origin. The instability comes from the coupling of the scalar field with the Maxwell field. The complex scalar has the effective mass given by
\be
m_\text{eff}^2 = m^2 - (-g^{00}) A_0^2~,
\label{eq:effective_mass}
\ee
from the action \eqref{eq:probe_action}. Thus, $\Psi$ becomes sufficiently tachyonic for a large enough $\mu/T$ or for low enough temperature. 

\subsubsection*{Critical phenomena of holographic \SC}

Typically, one chooses the spatial dimensionality $p$ of the boundary theory and the scalar mass squared as $p=2$ and $m^2=-2/L^2$. The choice $p=2$ would remind readers that in high-$\Tc$ materials the two-dimensional Cu-O plane plays an important role. This is certainly part of the reason, but the reason here is mainly technical: the equations of motion are often easiest to solve (with the choice of scalar mass). 

The choice $m^2=-2/L^2$ means that $\Psi$ is actually tachyonic from the beginning, but a certain range of tachyonic mass is allowed in the AdS spacetime if the mass satisfies the Breitenlohner-Freedman (BF) bound \eqref{eq:BF_bound}: \index{Breitenlohner-Freedman bound}
\be
m^2 \geq -\frac{(p+1)^2}{4L^2} = -\frac{9}{4L^2}~,
\label{eq:BF_bound_again}
\ee
so the choice satisfies the BF bound. 

We consider the solution of the form $\Psi=\Psi(u)$ and $A_0=A_0(u)$. Then, matter equations are
\begin{align}
& A_0'' - \frac{2|\Psi|^2}{u^2h} A_0 = 0~, 
\label{eq:HSC_matter_explicit1} \\
& \frac{u^2}{h}\left(\frac{h}{u^2} \Psi'\right)' +\left( \frac{A_0^2}{\Ts^2 h^2}-\frac{L^2m^2}{u^2 h} \right) \Psi = 0~,
\label{eq:HSC_matter_explicit2}
\end{align}
where $':=\del_u$ and $\Ts:=4\pi T/3=r_0/L^2$.

First, determine the critical temperature $\Tc$. Near $\Tc$, the scalar field remains small, and one can ignore the backreaction of $\Psi$ onto the Maxwell field. In this region, one can use \eq{background_matter} for $A_0$, and it is enough to solve the $\Psi$ equation only. 

Following the discussion of scalar fields (\sect{massive_scalar}), $\Psi$ behaves as
\begin{alignat}{2}
\Psi &\sim \Psi^{(0)} u^{\Delta_-} + \cpsi \bra\calO\ketj\, u^{\Delta_+}
&\quad& (u\rightarrow0)~,
\label{eq:GL_scalar_falloff} \\
\Delta_\pm &= \frac{p+1}{2} \pm \sqrt{\frac{(p+1)^2}{4}+L^2m^2}~,&&
\end{alignat}
where $\bra \calO \ketj$ is the order parameter in the presence of the external source $\Psi^{(0)}$, 
and $\cpsi$ is an appropriate factor. For our choice, $(\Delta_-, \Delta_+)=(1,2)$.

At $\Tc$, the system has the spontaneous condensation without an external source, so solve the $\Psi(u)$ equation by imposing the boundary condition $\Psi^{(0)}=0$. A regular solution with a nonvanishing $\bra \calO \ket$ can exist only for $T\leq\Tc$, which determines $\Tc$ (\sect{HSC_more}).

\begin{figure}[tb]
\centering
\subfigure[]{ 
\scalebox{0.3}{ \includegraphics{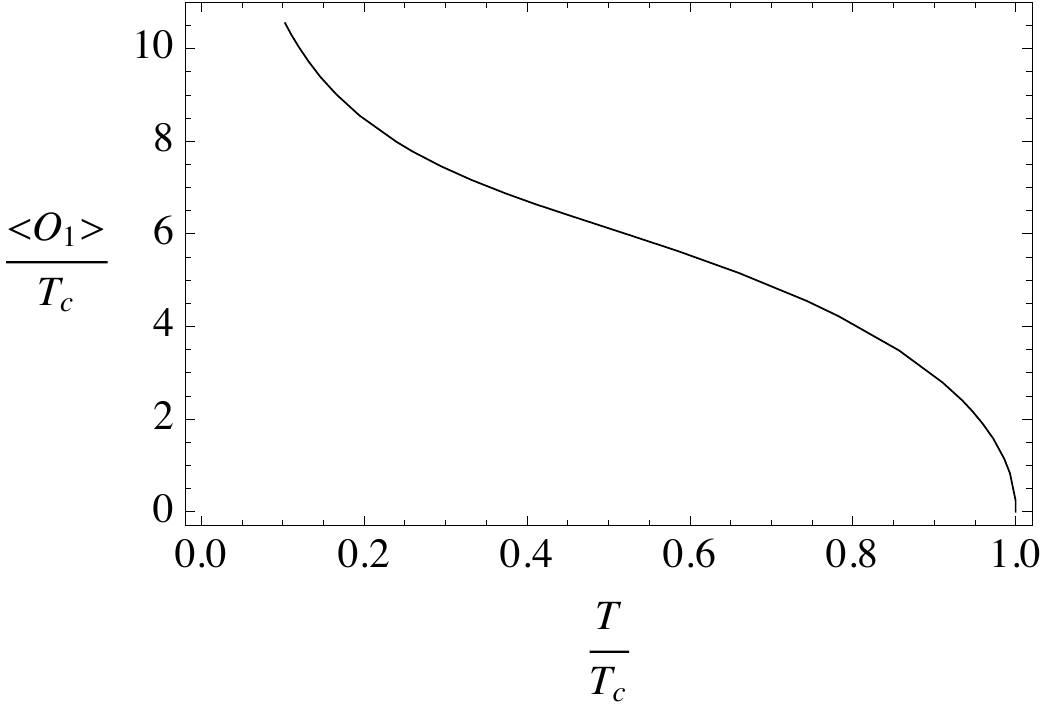} } 
} 
\subfigure[]{
\scalebox{0.25}{ \includegraphics{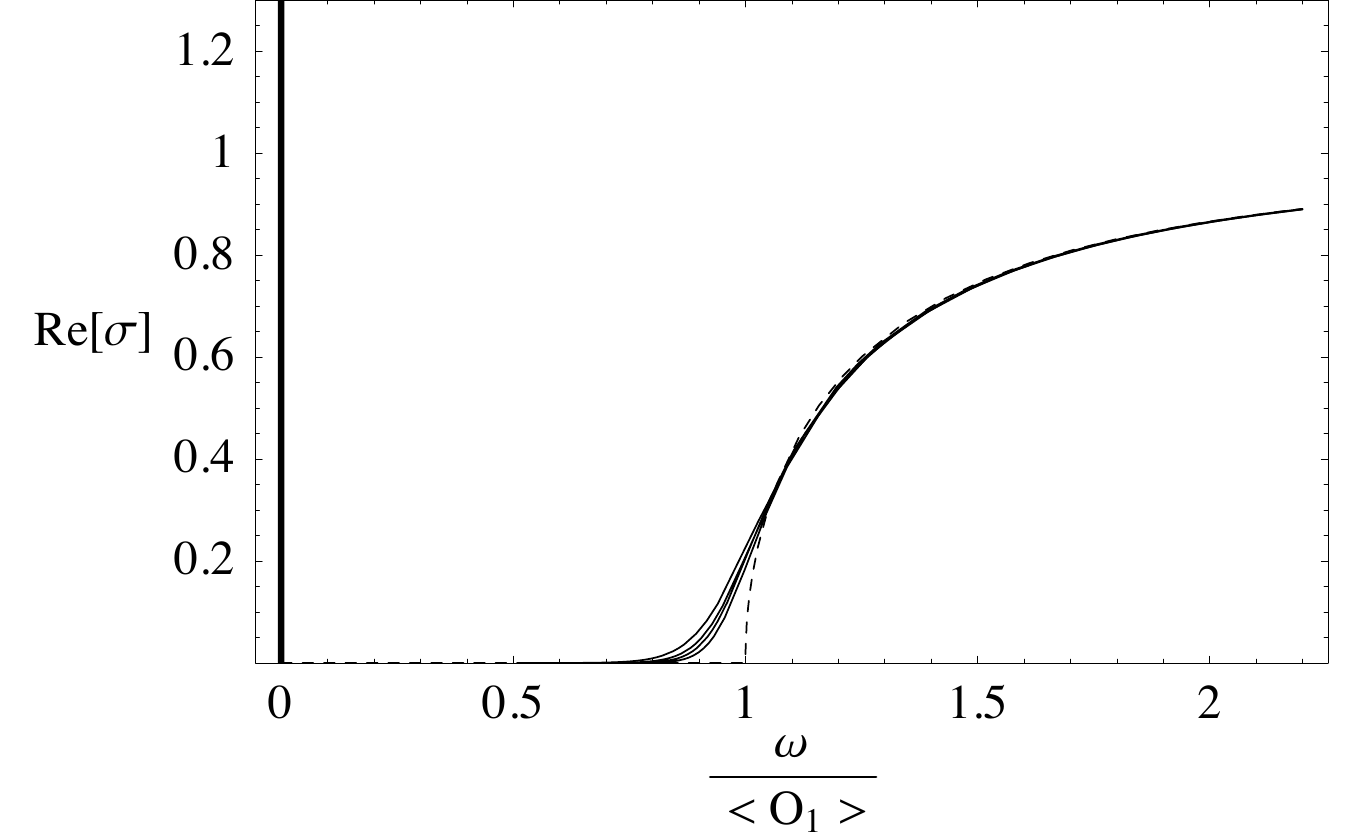} } 
} 
\subfigure[]{ 
\scalebox{0.25}{ \includegraphics{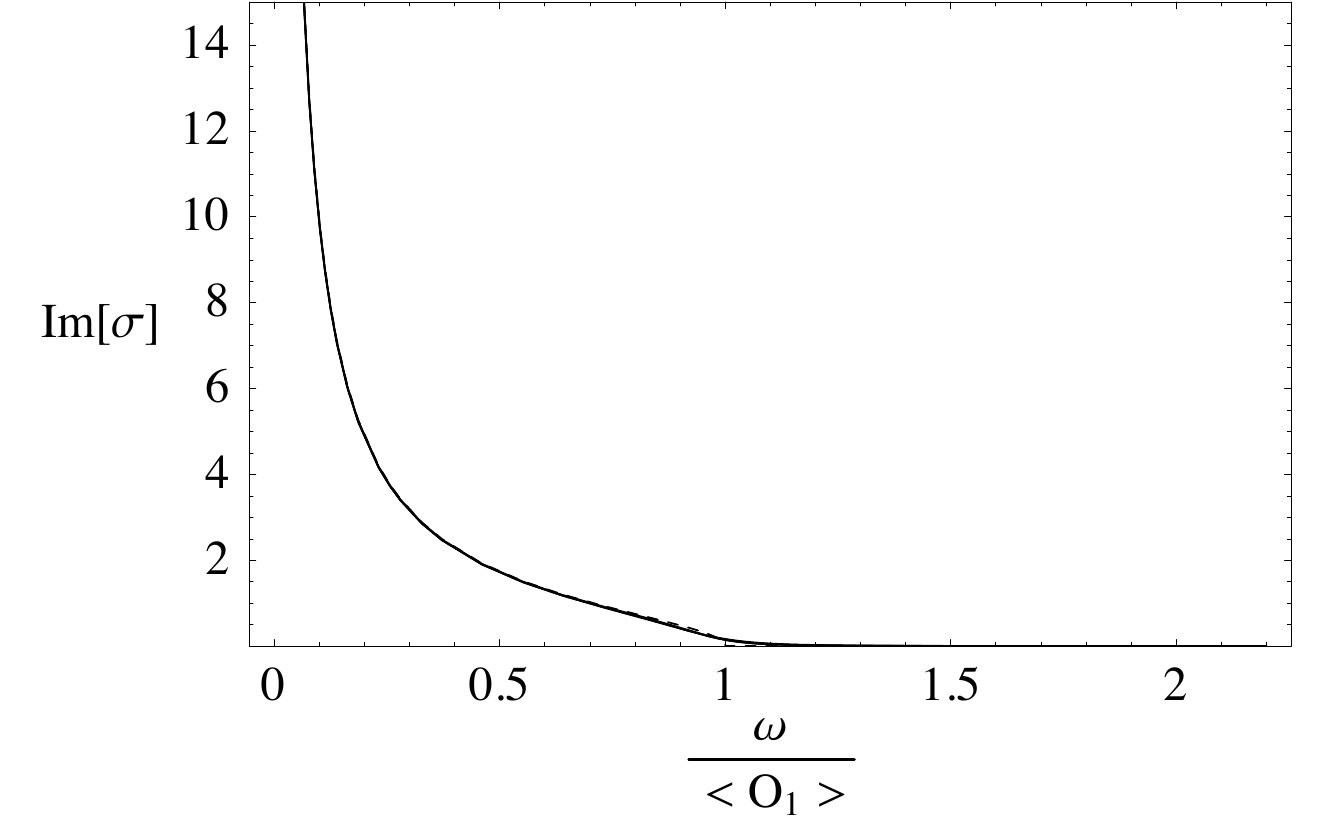} } 
}
\vskip2mm
\caption{Numerical results of (a) the order parameter $\bra\calO\ketnoj$, (b) $\text{Re}[\sigma(\omega)]$, and (c) $\text{Im}[\sigma(\omega)]$ (for the choice $p=2$ and $m^2 =-2/L^2$) \cite{Hartnoll:2008vx}.}
\label{fig:H^3} 
\end{figure}

Then, solve the matter equations in the low-temperature phase. Figure~\ref{fig:H^3}(a) shows a numerical result%
\footnote{A \textit{Mathematica} code is available from Chris Herzog's website \cite{herzog}.}.
Near the critical point $\Tc$, 
\be
\bra \calO \ketnoj \propto (T_c-T)^{1/2}~,
\label{eq:condensate}
\ee
which agrees with the GL theory of the second-order transition, $\beta=1/2$. 

If the holographic \SC\ indeed shows the critical phenomena in \chap{GL}, and if it is indeed represented by the GL theory, the other critical exponents should agree as well. First, in the probe limit, the spacetime geometry is just the SAdS$_5$ black hole, so the singular behavior is not reflected into the geometry. Thus, the heat capacity remains constant, which implies $\alpha=0$. Also, one can solve the $\Psi$ equation in $(\omega,q)$-expansion, and \eq{GL_scalar_falloff} gives the ``order parameter response function"
\be
\chi_k = \frac{\del \bra \calO_k \ketj}{\del \Psi^{(0)}_k}~,
%
\ee
from which one can extract the exponents $(\gamma, \nu, \eta, z)$. The results agree with the GL theory \eqref{eq:GL_exponents} \cite{Maeda:2009wv}. Because the values of these exponents are the GL ones, they obviously satisfy scaling relations \eqref{eq:scaling1}-\eqref{eq:scaling4}
except the hyperscaling relation. 

The GL theory is a mean-field theory. The holographic \SC\ agrees with the mean-field theory because we consider the large-$\Nc$ limit. In this limit, fluctuations are suppressed so that mean-field results are exact (\probset{factorization}). The critical exponents of the holographic \SC\ are independent of the spatial dimensionality, which is another indication of mean-field results%
\footnote{In particular, a phase transition occurs even when the boundary theory is $(2+1)$-dimensional, but this does not contradict with the Coleman-Mermin-Wagner's theorem \cite{Coleman:1973ci2,Mermin:1966fe2}.}.

As a condensed-matter system, it is natural that critical exponents agree with the GL theory, but this is not entirely obvious as a gravitational theory. First, we are talking of critical phenomena in a black hole background, not a usual statistical system. In the presence of gravity, the usual notion of statistical mechanics does not hold in general. For example, the Schwarzschild \bh has a negative heat capacity, so there is no stable equilibrium. Thus, it is not clear either if the standard theory of critical phenomena is valid for black holes. (Of course, AdS/CFT claims that AdS black holes are equivalent to usual statistical systems such as gauge theories, so the standard theory had better hold in AdS/CFT.)

Second, in this case, we have critical exponents of the standard GL theory, but it is not always the case. If one changes the system, one still gets mean-field exponents but nonstandard exponents. Such a system should have nonstandard mean-field free energy, \ie, not like \eq{GL_ferro2}. In this sense, second-order phase transitions in AdS/CFT seem to have rich phenomena even in the framework of mean-field theory.

\subsubsection*{Conductivity of holographic \SC
}

We now consider electromagnetic responses. As we saw in \sect{SC}, the London equation \eqref{eq:London} is crucial to the electromagnetic responses of a \SC. One can show that the London equation holds for the holographic \SC, so the electromagnetic responses of the holographic \SC\ are the ones of a \SC, but let us look at the details of the responses. 

We first consider the conductivity, the AC conductivity $\sigma(\omega)$. The diverging DC conductivity $\text{Re}[\sigma(\omega\rightarrow0)] \propto \delta(\omega)$ cannot be seen in a numerical computation but can be seen from the $1/\omega$ pole in $\text{Im}[\sigma]$ (\sect{drude}). 
%
%

Following \sect{current}, we add $A_x=A_x(u) e^{-i\omega t}$. The vector potential $A_x$ behaves as
\be
A_x \sim \sourceA{x}\left(1+\vevA{x} u\right)~,
\quad (u\rightarrow0) 
%
\ee
and the fast falloff is the current $J^x$:
\be
\bra J^x \ketj = \cA \vevA{x} \sourceA{x}~,
\label{eq:response_current_ads4}
\ee
where $\cA$ is an appropriate factor. On the other hand, from Ohm's law,
\be
\bra J^x \ketj = \sigma \sourceE = i\omega \sigma \sourceA{x}~,
%
\ee
so
\be
i\omega\sigma(\omega) = \cA\vevA{x}~.
\label{eq:fast_vs_cond_ads4}
\ee

The $A_x$ equation is written as 
\be
\frac{1}{h}(h A_x')' +\left( \frac{\omega^2}{\Ts^2 h^2}-\frac{2\Psi^2}{u^2 h} \right) A_x = 0~.
\label{eq:Ax_eom_4}
\ee
The important term is the last term proportional to $\Psi^2$. This term is essential to the diverging DC conductivity. In the high-temperature phase, $\Psi=0$, so the term vanishes. Then, the problem is similar to the SAdS$_5$ conductivity computed in \sect{SAdS5_diff}. (Of course, we considered the $p=3$ case there, but the structure is the same. See \probset{SAdS4_diff}.) We solved the conductivity in the $\omega$-expansion, and the result \eqref{eq:Ax_sol_asymptotic} was
\be
\vevA{x} = \frac{i\omega}{2\pi T} + \cdots. \quad (\text{for SAdS$_5$ with $\Psi=0$})~.
%
\ee
Namely, the solution starts with the $O(i\omega)$ term and there is no $\omega$-independent term. From \eq{fast_vs_cond_ads4}, the DC conductivity $\text{Re}[\sigma(\omega\rightarrow0)]$ is then finite, and $\text{Im}[\sigma(\omega)]$ has no $1/\omega$ pole. 

On the other hand, in the low-temperature phase, $\Psi \neq 0$, and the $\Psi^2$ term in the $A_x$ equation is nonvanishing. In this case, an $\omega$-independent term can exist in $\vevA{x}$ (\sect{HSC_more}). Figures~\ref{fig:H^3}(b) and (c) show numerical results of the conductivity in the low-temperature phase. The conductivity has the following two properties just as usual \SCs\ (Fig.~\ref{fig:tinkham}):
\begin{enumerate}

\item The diverging DC conductivity $\text{Re}[\sigma(\omega\rightarrow0)]$. $\text{Im}[\sigma(\omega)]$ indeed has a $1/\omega$ pole.
%
%

\item The AC conductivity $\text{Re}[\sigma(\omega)]$ almost vanishes for $\omega<\omega_g$, which indicates the existence of the energy gap $\omega_g$.

\end{enumerate}
In addition, the AC conductivity approaches constant for large-$\omega$. This simply reflects the $(2+1)$-dimensional nature of the problem. In $(2+1)$-dimensions, $J(x)$ has mass dimension 2, and $E(x)$ has mass dimension 2, so the conductivity is dimensionless.

\subsubsection*{Magnetic response
}

For \SCs, magnetic responses are interesting and important as well. In particular, the magnetic field is expelled because of the Meissner effect.\index{Meissner effect}
For the holographic \SC, the boundary theory has the $U(1)$ gauge field $\sourceA{\mu}$, but the gauge field is added as an external source. The gauge field is not dynamical in the boundary theory. Thus, the Meissner effect does not arise, and a magnetic field can penetrate the holographic \SC. When we discussed the Meissner effect, we showed that the gauge field becomes massive by combining two equations:
\begin{itemize}
\item London equation \eqref{eq:London}, 
\item Maxwell equation. 
\end{itemize}
But for the holographic \SC, there is no Maxwell equation for the boundary $U(1)$ gauge field. This is the reason why there is no Meissner effect.

Even though the Meissner effect does not arise, the London equation must hold. The London equation is just the response of the current under the external source. Whether photon is dynamical or not should be irrelevant to the response itself. One can show that the London equation holds for the holographic \SC%
\footnote{From the field theory point of view, the London equation is a consequence of the spontaneous symmetry breaking of the gauge symmetry \cite{Weinberg:1986cq}. The argument holds even when the gauge field is nondynamical.}.

Near $\Tc$, the scalar field remains small, and the bulk Maxwell equation $\del_x(\sqrt{-g}F^{yx})=\del_y(\sqrt{-g}F^{xy})=0$ reduces to $\del_x F_{yx}=\del_y F_{xy}=0$. Then, one can apply a constant magnetic field into the holographic \SC:
\be
B = F_{xy} = \del_x A_y~, \quad
A_y = Bx~.
%
\ee
The magnetic field is perpendicular to the boundary spatial direction.

A large enough magnetic field destroys the superconducting state. This is true for the holographic \SC\ as well \cite{Nakano:2008xc,Albash:2008eh,Hartnoll:2008kx}. Let us go back to the effective mass argument. The effective mass argument is useful to understand the instability of the system, but this time consider with a vector potential $A_i(\bmx)$:
\be
m_\text{eff}^2 = m^2 + \left\{ - (-g^{00})A_0^2 + g^{ii}A_i(\bmx)^2 \right\}~.
%
\ee
As we saw earlier, $A_0$ contributes with minus sign, so it tends to {\it destabilize} the normal state, which leads to the superconductivity. On the other hand, $A_i(\bmx)$ contributes with plus sign, so it  tends to {\it stabilize} the normal state. Thus, the superconducting state becomes unstable under a large enough magnetic field.

Superconductors are classified as type I and type II \SCs\ depending on the magnetic response. For the holographic \SC, a magnetic field can can penetrate \SCs. In this sense, the holographic \SC\ is the  ``extreme type II" \SC. (The superfluid is often called so as well.) In type II \SCs, the penetration of the magnetic field arises by forming vortices. The vortex solutions and the vortex lattice have been constructed for the holographic \SC\ \cite{Albash:2009ix,Montull:2009fe,Maeda:2009vf}. For the vortex lattice, the triangular lattice is the most favorable configuration just as the GL theory. 

The holographic \SC\ does not have a dynamical $U(1)$ gauge field. One would regard the system as a superfluid rather than a \SC. The superfluid property of the system has been investigated as well (see, \eg, Ref.~\cite{Herzog:2008he}). 

\subsection{Other issues \advanced}\label{sec:others_HSC}

\subsubsection*{Beyond the probe limit}

One can actually solve the full Einstein-Maxwell-scalar problem without the probe limit. We do not discuss the details, but two remarks are worth to mention here. 

First, in the full problem, the DC conductivity diverges even in the high-temperature phase, \ie, for the RN-AdS$_4$ black hole. This is not superconductivity but perfect conductivity. 

One can boost the whole system in the $x$-direction. In a charged system, this produces a steady current without an electric field, which signals perfect conductivity. Since we have a nontrivial $A_0(u)$, the boost produces $A_x(u)$. Then, the current $J^x$ is nonvanishing even though there is no electric field ($\sourceA{x}=$constant). In the probe limit, the boost is not possible since we fixed the \bh background.

Second, in the full problem, even the neutral scalar ($\qm=0$) becomes unstable. The stability analysis is particularly simple at $T=0$, so consider the extreme RN-AdS$_4$ black hole. There are actually two mechanisms for the instability of the scalar \cite{Hartnoll:2008kx,Gubser:2008pf,Denef:2009tp}:
\begin{itemize}

\item First, as we saw above, the effective mass becomes more tachyonic because of the Maxwell field. 
\item Second, the extreme black hole effectively becomes AdS$_2$ near the horizon, but the BF bound becomes stringent for a smaller $p$: 
\be
m^2 \geq -\frac{(p+1)^2}{4L^2}~.
%
\ee
Namely, even if the scalar mass is above the BF bound for AdS$_4$ asymptotically, the scalar may be below the BF bound for AdS$_2$, so the scalar becomes unstable near the horizon.

\end{itemize}

\begin{figure}[tb]
\centering
\scalebox{0.65}{ \includegraphics{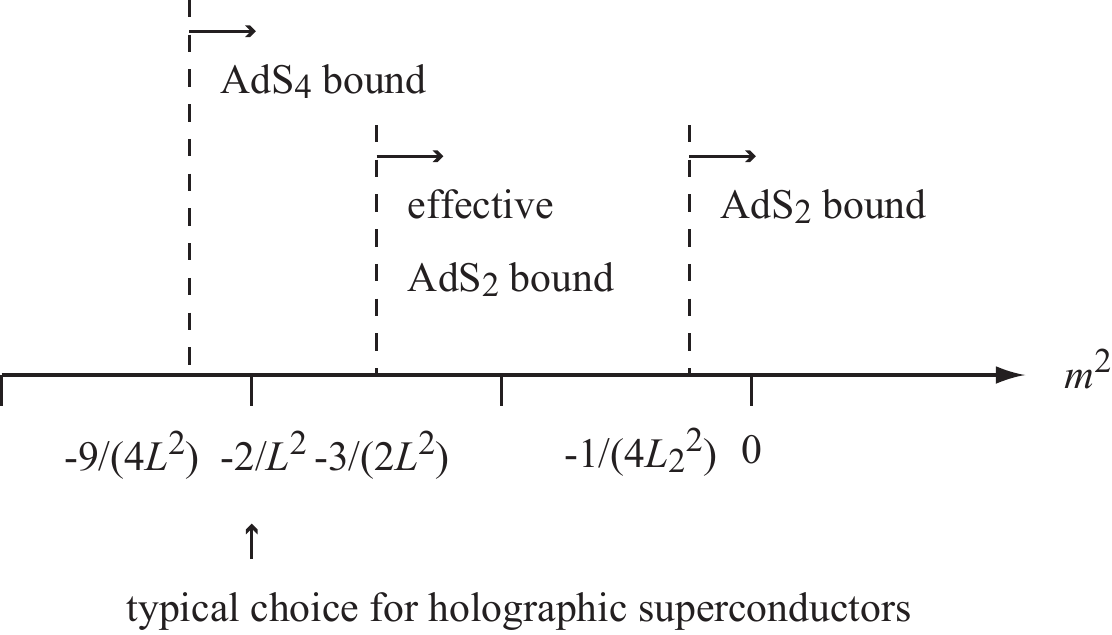} }
\caption{Various masses appeared in the text. Even if the scalar mass is above the BF bound for AdS$_4$ asymptotically, the scalar may be below the BF bound for the effective AdS$_2$.}
\label{fig:BF_bound} 
\end{figure}

See \sect{RNAdS} for the explicit form of the RN-AdS$_4$ black hole. The extreme limit is given by $\ro=\ri=r_0$ or $\alpha=1$. In the extreme limit, the near-horizon geometry becomes%
\footnote{We change the normalization of the Maxwell field in \sect{RNAdS} as $A_M \rightarrow L A_M$ to take into account the standard normalization convention for the \HSC\ \eqref{eq:H^3_action}.}
\begin{align}
ds_4^2 &\rightarrow - f dt^2 + \frac{d\tilr^2}{f} + \left(\frac{r_0}{L}\right)^2 d\bmx_2^2~,\\
f &\rightarrow 6\left(\frac{\tilr}{L}\right)^2~,\\
A_0 &\rightarrow \frac{2\sqrt{3}}{L}\tilr~,
%
\end{align}
where $\tilr:=r-r_0$. The geometry reduces to AdS$_2 \times \mathbb{R}^2$ with the effective AdS$_2$ radius $L_2^2:=L^2/6$. Thus, even if the scalar mass $m^2=-2/L^2$ is above the BF bound asymptotically, \ie,
\be
m^2 \geq -\frac{9}{4L^2}~, \quad
\text{(BF bound for $p=2$)}
%
\ee
it is below the BF bound for AdS$_2$:
\be
m^2 \geq -\frac{1}{4L_2^2} = -\frac{3}{2L^2}~,  \quad
\text{(effective BF bound for $p=0$)}
\label{eq:instability_AdS2}
\ee
The effective AdS$_2$ radius is actually smaller than the AdS$_4$ radius, $L_2^2=L^2/6$, but the scalar can be unstable even taking this into account. See also \fig{BF_bound}.

Now, consider the charged scalar case. In the near-horizon limit, the effective mass becomes
\be
%
m_\text{eff}^2 = m^2 - (-g^{00}) \qm^2 A_0^2 \rightarrow m^2 - 2\qm^2~,
%
\ee
so the BF bound \eqref{eq:instability_AdS2} is rewritten as 
\be
%
L_2^2 m_\text{eff}^2 = \frac{L^2}{6} (m^2-2\qm^2) \geq -\frac{1}{4}~.
\label{eq:instability_zero}
\ee
The instability occurs if the scalar does not satisfy the condition. 

\subsubsection*{Other models}

Previously, we arrive at an Einstein-Maxwell-scalar system to evade the no-hair theorem, but this is not the only possibility. A Yang-Mills hair is possible even in asymptotically flat spacetimes, so an Einstein-Yang-Mills system is another option. In such a system, the order parameter is a vector (Yang-Mills field), so it is a $p$-wave holographic \SC\ \cite{Gubser:2008zu,Gubser:2008wv}. In the Einstein-Maxwell-scalar system, the order parameter is a scalar, so it is a $s$-wave holographic \SC. 

The $d$-wave holographic \SCs\ are potentially interesting since high-$T_c$ materials are $d$-wave \SCs. The order parameter should be a massive spin-2 field, but finding an action for a charged spin-2 field in curved spacetime is an unsolved issue \cite{Benini:2010pr}. 

\subsubsection*{Limitations of the current model
}

We have used a particular model \eqref{eq:H^3_action} to discuss holographic superconductivity, but this phenomenon does not depend on the details of the system. We saw the phase transition even in the probe limit, so the key property of this phenomenon is the instability of matter fields. Holographic superconductivity should arise in a broad range of theories which satisfy the stability argument \eqref{eq:effective_mass}. Namely, holographic superconductivity is a robust phenomenon. This is naturally expected since superconductivity is a robust phenomenon at low temperature.

Unfortunately, the holographic \SC\ lacks the microscopic picture of the order parameter $\calO$. One starts with the gravitational theory, so the details of the dual field theory, in particular the nature of $\calO$, is unclear. The model does not tell if $\calO$ is a composite operator like the Cooper pair and does not tell about the paring mechanism at strong coupling. Of course, such a problem is also common to the GL theory. So, the holographic \SC\ is the holographic GL theory rather than the holographic BCS theory. 

A quarter of a century has passed after the discovery of high-$T_c$ superconductivity, but its theoretical understanding is still insufficient. One feature of high-$T_c$ superconductivity is its rich phase structure: these materials typically have non-Fermi liquid phase, pseudogap region, anti-ferromagnetic insulator phase in addition to the superconducting phase and the Fermi liquid phase. These phases are likely to be tied very closely so that it is probably mandatory to have their unified understanding in order to solve high-$T_c$ superconductivity. 

Thus, in order to understand high-$T_c$ materials, it is clearly insufficient just to realize the superconducting phase in AdS/CFT. It does not seem an easy task to understand all these phases at once by AdS/CFT. Rather, first we had better try to understand each phases separately in the AdS/CFT framework. 

Although holographic \SCs\ correspond to the GL theory, one can discuss the Cooper pair instability if one considers bulk fermions instead of bulk scalars \cite{Hartman:2010fk}. Also, using the bulk fermions, one can realize Fermi liquid phase and non-Fermi liquid phase (see, \eg, Ref.~\cite{Faulkner:2010da} for a review). If one can construct a gravity dual which covers all these phases, it would be useful to understand high-$T_c$ superconductivity. 

\section{\titlesummary}

\begin{itemize}
\item 
Large-$\Nc$ gauge theories have rich phase structures like condensed-matter systems. Studying such phase structures may be useful even in condensed-matter physics.
\item 
As examples, we discuss first-order and second-order phase transitions and superconductivity in AdS/CFT.
\item 
A \bh with compact horizon can undergo a first-order phase transition. A hairy \bh can undergo a second-order phase transition. 
\item
An ultimate goal is to understand \hightc\ superconductivity in AdS/CFT.
\end{itemize}

\titlenewterms

\begin{multicols}{2}
\noindent
Hawking-Page transition\\
reference spacetime\\
Casimir energy\\
holographic superconductors\\
probe limit
\end{multicols}

\section{Appendix: More about \HSC\ \advanced}\label{sec:HSC_more}

The matter field equations are given by
\begin{align}
D^2\Psi - m^2 \Psi &=0~, 
\\
\nabla_N F^{MN} &= j^M~,
\\
j^M &:=  -i\left\{ \Psi^\dag (D^M\Psi) - \Psi (D^M\Psi)^\dag \right\}~.
\end{align}
They are similar to the GL equations of motion for \SCs, but the above equations are bulk equations. For example, $j^M$ is the bulk matter current, not the boundary current $J^\mu$. The boundary current $J^\mu$ is extracted from the slow falloff of $A_M$. 

We take the gauge $A_u=0$. We consider the solution of the form $\Psi=\Psi(u)$ and $A_0=A_0(u)$. The $u$-component of the Maxwell equation gives $ 0 = j_u \propto \Psi^\dag\Psi' - \Psi\Psi^\dag{}'$, so the phase of $\Psi$ must be constant. We take $\Psi$ to be real without loss of generality. The rest of bulk equations reduce to Eqs.~\eqref{eq:HSC_matter_explicit1}-\eqref{eq:HSC_matter_explicit2}.

\head{Critical temperature}

Near $\Tc$, the scalar field remains small, 
and one can expand matter fields as a series in $\epsilon$, where $\epsilon$ is the deviation parameter from the critical point:
\begin{align}
\Psi &= \epsilon^{1/2} (\psi_1 + \epsilon \psi_2 + \cdots)~,\\
A_0 &= \phi_0 + \epsilon \phi_1 + \cdots~,\\
A_x &= \epsilon (a_1 + \epsilon a_2 +\cdots)~.
%
\end{align}
At leading order in $\epsilon$, the $A_0$ solution is given by $\phi_0=\mu(1-u)$. At next order, the $\Psi$ equation becomes
\be
\frac{u^2}{h}\left(\frac{h}{u^2} \psi_1'\right)' +\left\{ \left(\frac{\mu}{\Ts}\frac{1-u}{h}\right)^2 +\frac{2}{u^2 h} \right\} \psi_1 = 0~,
\label{eq:scalar_1st}
\ee
where $\Ts:=4\pi T/3$ and we set $L^2m^2=-2$. We impose the following boundary conditions:
\begin{itemize}
\item The horizon $u\rightarrow1$: regularity.
\item The AdS boundary $u\rightarrow0$: $\psi_1$ asymptotically behaves as 
\be
\psi_1 \sim \psi_1^{(0)} u + \cpsi \bra\calO\ketj\, u^2~,
\quad (u\rightarrow0)~,
%
\ee
and we impose $\psi_1^{(0)}$=0 (spontaneous condensate)%
\footnote{For simplicity, we consider only the ``standard quantization" and do not consider the ``alternative quantization" (\sect{massive_more}).}. 
\end{itemize}
The problem has a nontrivial solution only for particular values of $T/\mu$, which determines $\Tc$. 
 
This problem is essentially the same as the quasinormal mode problem in \sect{QNM}. In both cases, we impose the vanishing slow falloff condition at the AdS boundary. The differences are
\begin{itemize}
\item Quasinormal mode: a nontrivial solution exists only for a particular $\omega/q$.
\item This problem: a nontrivial solution exists only for a particular $T/\mu$.
\end{itemize}
The vanishing slow falloff problem determines the location of poles, and here we interpret the pole as the diverging susceptibility. This is because 
\be
\chi = \frac{\del \bra \calO \ketj}{\del \psi_1^{(0)}} \propto \frac{\bra \calO \ketj}{\psi_1^{(0)}} \rightarrow \infty
%
\ee
when a spontaneous condensate develops. In both cases, we are interested in poles in the Green's function (or the response function), and they signal the presence of light degrees of freedom (hydrodynamic modes or a massless order parameter).

We solve \eq{scalar_1st} by a power series expansion around the horizon. Suppose that our differential equation takes the form
\be
\psi_1'' + \frac{p(u)}{u-1}\psi_1' + \frac{q(u)}{(u-1)^2}\psi_1 = 0~.
\label{eq:model_eq}
\ee
We expand the solution by a power series expansion around the horizon:
\be
\psi_1(u) = \sum_{n=0}^\infty \coeff_n (u-1)^{n+\lambda}~,
\label{eq:series_sol}
\ee
and expand the functions $p(u)$ and $q(u)$, \eg, 
\be
p(u) = \sum_{n=0}^\infty p_n (u-1)^n~.
\label{eq:series_coeff}
\ee
Substitute Eqs.~\eqref{eq:series_sol} and \eqref{eq:series_coeff} into \eq{model_eq}. At the lowest order, one gets
\be
\{ \lambda(\lambda-1)+\lambda p_0+q_0 \} \coeff_0 = 0~.
%
\ee
This is the indicial equation to determine $\lambda$. Since we have a linear differential equation, the coefficient $\coeff_0$ is undetermined, and we set $\coeff_0=1$. [The computation here determines only $\Tc$. In order to obtain the magnitude of the condensate, one has to solve coupled matter equations \eqref{eq:HSC_matter_explicit1}-\eqref{eq:HSC_matter_explicit2}.]
For our problem \eqref{eq:scalar_1st}, the indicial equation gives $\lambda(\lambda-1)+\lambda = \lambda^2=0$, and we henceforth set $\lambda=0$. 

Proceeding further, one obtains $\coeff_n$ recursively:
\be
\coeff_n = -\frac{1}{n(n-1)+np_0+q_0} \sum_{k=0}^{n-1} \{ k(k-1) + kp_{n-k} + q_{n-k} \} \coeff_k~.
\label{eq:series_recursion}
\ee
For our problem, the first few series from \eq{series_recursion} are
\be
\psi_1(u)=1 + \frac{2}{3}(u-1) 
- \left\{ \frac{2}{9}+\frac{1}{36} \left(\frac{\mu}{\Ts} \right)^2 \right\} (u-1)^2 + \cdots~, 
\quad (u\rightarrow1)~.
\label{eq:condenstate_horizon}
\ee
One truncates the series after a large number of terms $n=N$. One can check the accuracy as one goes to higher sums. One typically needs to compute the sum up to $N\approx100$. 

In order to impose the $u\rightarrow0$ boundary condition, expand the solution around $u\rightarrow0$:
\be
\psi_1(u) = \left\{ \sum_n^N (-)^n \coeff_n \right\} +  \left\{ \sum_n^N (-)^{n-1} n \coeff_n \right\} u +O(u^2)~.
%
\ee
We impose the vanishing slow falloff condition $\psi_1^{(0)} =0$, so the $O(u)$ term must vanish:
\be
\psi_1^{(0)} = \sum_{n=0}^N (-)^{n-1} n \coeff_n = 0~.
\label{eq:BC_bdy}
\ee
[The $O(1)$ term converges to zero after a large number of partial sums.] This is a polynomial equation for $T/\mu$ which is satisfied only for particular values of $T/\mu$. As one approaches from high temperature, the first solution one encounters gives $\Tc$. On the other hand, the condensate which comes from the $O(u^2)$ term is nonvanishing at $\Tc$.

One can solve \eq{BC_bdy} numerically and obtains
\be
\frac{\Ts_c}{\mu} \approx 0.24608 
\quad \text{or} \quad
\frac{\Tc}{\mu} \approx 0.058747~. 
%
\ee

\head{DC conductivity}

Here, we discuss the diverging DC conductivity more. The presence of an $\omega$-independent term in $\vevA{x}$ signals the $1/\omega$ pole, so it is enough to consider the $\omega\rightarrow0$ limit of \eq{Ax_eom_4}. At leading order in $\epsilon$, the $A_x$ equation gives
\be
\frac{1}{h}(h a_1')' = 0~.
%
\ee
This equation is easily solved, and a regular solution is trivial: $a_1=c_1$, where $c_1$ is a constant%
\footnote{This is most easily seen in the tortoise coordinate $u_*$, where $\del_{u_*} = h \del_u$. Then, the solution is given by $a_1 = c_1 + c_2 u_*$, where $c_1$ and $c_2$ are integration constants. In the tortoise coordinate, the horizon is located at $u_*\rightarrow-\infty$, so $c_2=0$ from regularity.}.
The solution represents only the slow falloff, which implies that the fast falloff does not have an $\omega$-independent term at this order. This also shows that the conductivity has no $1/\omega$ pole in the high-temperature phase. 

At next order, the $A_x$ equation gives 
\be
(h a_2')'  = 2a_1 \frac{\psi_1^2}{u^2}~.
%
\ee
Then, the fast falloff is given by
\be
\left. a_2' \right|_{u=0}
= 2a_1 \int^0_1 du\, \frac{\psi_1^2}{u^2}~.
%
\ee
The integrand is positive definite, so the integral is nonvanishing. The integral is also finite since $\psi_1^2/u^2 \sim u^2$ as $u\rightarrow0$ and $\psi_1^2/u^2 \sim 1$ as $u\rightarrow1$. Thus, in the low-temperature phase, the fast falloff has an $\omega$-independent term, and the conductivity has the $1/\omega$ pole .

\endofsection

\ifx\nameofpaper\undefined
  \usepackage{macro_natsuume}
  \def\beginsection{\section*}
  \def\endofsection{\end{document}}
  \input draft_header.tex
\else
  \def\beginsection{\chapter}
  \def\endofsection{ }
\fi

\newcommand{\prerequisite}[1]{\textit{Advanced section prerequisite: #1}}
\newcommand{\ketc}{\ket_c}

\chapter{Exercises}\label{chap:exercise}
\begin{prob}
Find errors in this textbook as many as possible, and report to the author.
\end{prob}

\section*{Chap.~3}


\begin{prob}
\textbf{Thermodynamic quantities of higher-dimensional Schwarzschild black holes}\\
Verify Eqs.~\eqref{eq:temp_Sch_d} and \eqref{eq:entropy_Sch_d}.
\end{prob}

\begin{prob}
\textbf{Thermodynamic quantities of the RN black holes}\\
Verify Eqs.~\eqref{eq:temp_RN} and \eqref{eq:entropy_RN}.
\end{prob}

\section*{Chap.~5}

\begin{prob}
\textbf{Maxwell theory in $d$-dimensions} \\
\prerequisite{\sect{weyl_inv} } \\
Compute the energy-momentum tensor and its trace for
\be
\action = - \frac{1}{4e^2} \int d^dx\, F^{\mu\nu}F_{\mu\nu}~.
%
\ee
The trace does not vanish except $d=4$. But the trace vanishes up to a total derivative:
\be
T^{\mu}_{~\mu} = \frac{4-d}{4e^2} F^2 = \frac{4-d}{2e^2} \del_\mu(A_\nu F^{\mu\nu})~,
%
\ee
where we use the equation of motion in the last equality. Then, the Maxwell theory is scale invariant according to \sect{weyl_inv}. Comparing with \eq{virial}, $K^\mu = (4-d) A_\nu F^{\mu\nu}/(2e^2)$.

However, $K^\mu$ is not a divergence, so the Maxwell theory is scale invariant but is not conformal invariant when $d\neq4$.

\end{prob}

\begin{prob}
\textbf{Local Weyl invariant scalar theory in $d$-dimensions} \\
\prerequisite{\sect{weyl_inv} } \\
Show that $T_{\mu\nu}$ \eqref{eq:EM_scalar_conf} is traceless when $\xi=(d-2)/4(d-1)$.
\end{prob}

\section*{Chap.~6}

\begin{prob}
\textbf{The $H^2$, AdS$_2$, and dS$_2$ metric}\\
Verify Eqs.~\eqref{eq:metric_h2}, \eqref{eq:metric_ads2}, and \eqref{eq:metric_ds2}.
\end{prob}

\begin{prob}
\textbf{The AdS$_2$ metric in various coordinates}\\
Verify Eqs.~\eqref{eq:ads2_static}, \eqref{eq:ads2_conf}, and \eqref{eq:ads2_poincare}.
\end{prob}

\begin{prob}
\textbf{\poincare\ patch}\\
Derive the region coved by \poincare\ coordinates, or the \poincare\ patch, in the AdS$_2$ spacetime  (\fig{AdS2_Penrose}). To do so, one has to know the relation between conformal coordinates $(\tg, \theta)$ and \poincare\ coordinates $(t,r)$. First, write $r$ in terms of embedding coordinates $(X,Y,Z)$, and rewrite the expression in terms of conformal coordinates.
\end{prob}

{\color{blue} 
\textit{Solution:}
\begin{align}
Z-Y &= r \\
&= \frac{1}{\cos\theta}(\cos\tau-\sin\theta)>0~,
%
\end{align}
An inspection of this relation reproduces \fig{AdS2_Penrose}.
}

\begin{figure}[tb]
\centering
\subfigure[]{
\scalebox{0.55}{ \includegraphics{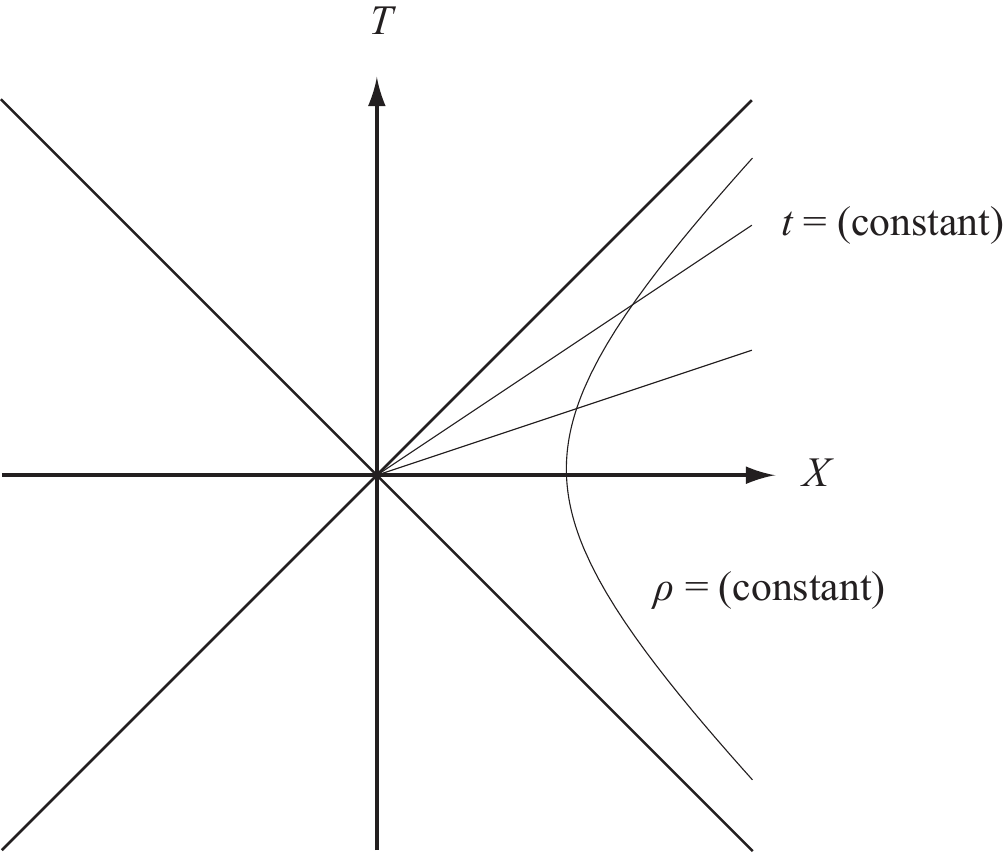} }
\label{fig:rindler}} \qquad
\subfigure[]{
\scalebox{0.55}{ \includegraphics{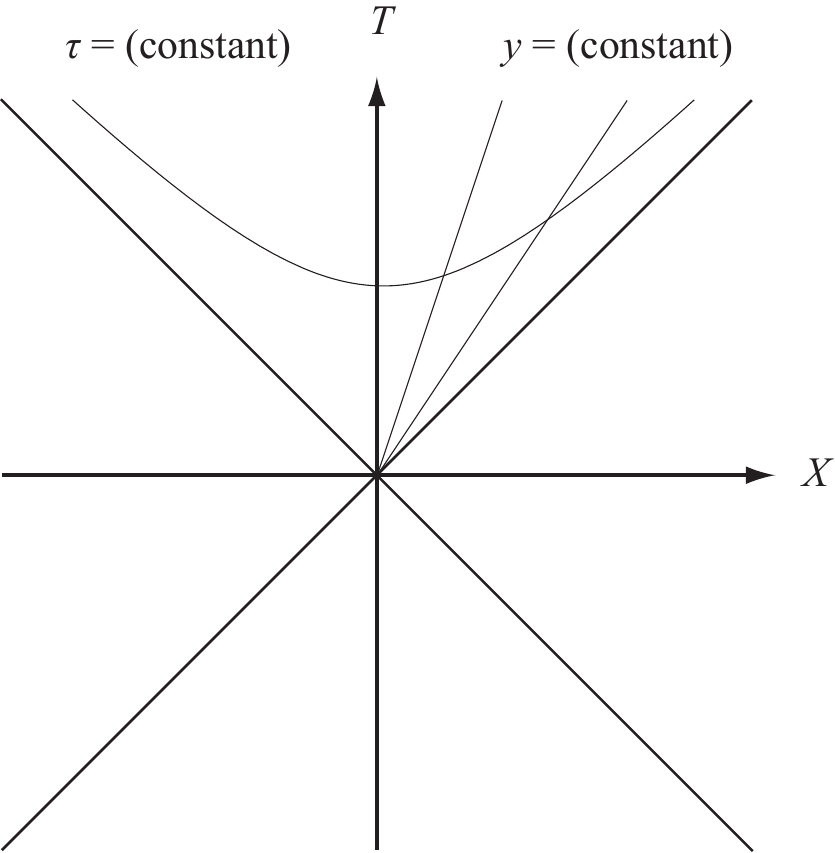} }
\label{fig:milne}}
\vskip2mm
\caption{Flat spacetime in disguise.}
\end{figure}

\begin{prob}
\label{prob:rindler}
\textbf{Flat spacetime in disguise 1: Rindler spacetime}\\
\noindent
(a) Consider the Minkowski spacetime:
\be
ds^2 = -dT^2+dX^2~.
%
\ee
Show that the coordinate transformation
\be
\begin{split}
T &= \rho \sinh t~, \\
X &=\rho\cosh t
\end{split}
\label{eq:rindler_transf}
\ee
leads to 
\be
ds^2 = - \rho^2 dt^2 + d\rho^2~.
\label{eq:rindler2}
\ee
This is the Rindler spacetime. The $\rho=\text{(constant)}$ lines are hyperbolas, and the $t=\text{(constant)}$ lines are straight lines. They cover the region $X>|T|$ of the Minkowski spacetime (\fig{rindler}).

A further coordinate transformation $\rho=2r^{1/2}$ leads to \eq{rindler}:
\be
ds^2 = -r dt^2 + \frac{dr^2}{r}
%
\ee
(up to a trivial scaling of $t \rightarrow t/2$).

\noindent
(b) Note that the Euclidean continuation of \eq{rindler2} takes the same form as the near-horizon metric \eqref{eq:polar}. Then, the near-horizon geometry of the Schwarzschild \bh should become the Rindler spacetime. Verify this explicitly. 

In the near-horizon limit, the coordinate transformation to the Kruskal coordinates \eqref{eq:kruskal} reduces to \eq{rindler_transf} with $T \propto v$ and $X \propto u$.

\noindent
(c) Suppose the constant $\rho=\rho_0$ line represents the world-line of an observer. One can check that his proper time $\tau$ is given by $\tau = \rho_0 t$. Compute the covariant acceleration $a^\mu = dx^\mu/d\tau$ and verify that $a^2 = \eta_{\mu\nu} a^\mu a^\nu$ is constant. Namely, the world-line represents a uniform accelerated observer with $a^2=1/\rho_0^2$.

\end{prob}

\begin{prob}
\textbf{Flat spacetime in disguise 2: Milne universe}\\
Another flat spacetime in disguise is obtained by the coordinate transformation
\begin{align}
T &= \tau \cosh y~, \\
X &= \tau \sinh y~.
%
\end{align}
Show that the metric in $(\tau, y)$ coordinates is given by
\be
ds^2 = - d\tau^2 + \tau^2 dy^2~.
\label{eq:milne}
\ee
This is known as the \keyword{Milne universe}. The metric \eqref{eq:milne} can be also obtained by the double Wick rotation from the Rindler spacetime \eqref{eq:rindler2}. 

The constant $y=y_0$ trajectory represents a boosted observer with Lorentz factor $ \gamma = \cosh y $ ($y := \tanh^{-1} v$ is known as \keyword{rapidity}.) 

The Milne universe may be regarded as a cosmological expansion. The constant $y=y_0$ trajectory represents the observer who experiences a universe expanding from the origin. 
%
%
The Milne universe is often used in the analysis of the QGP time evolution (\probset{bjorken}). In this sense, QGP can be regarded as a cosmological expansion.
\end{prob}

\section*{Chap.~7}

\begin{prob}
\textbf{Thermodynamic quantities of SAdS$_{p+2}$ black holes}\\
The SAdS$_{p+2}$ black hole is given by
\begin{align}
ds_{p+2}^2 &= - f dt^2  + \frac{dr^2}{f} + \left(\frac{r}{L}\right)^2 d\bmx_{p}^2~, \\
f &= \left(\frac{r}{L}\right)^2 \left\{ 1 - \left( \frac{r_0}{r} \right)^{p+1} \right\}~.
%
\end{align}
Following \sect{sads_thermo}, compute the Hawking temperature of the black hole. Compute the other thermodynamic quantities using two methods below.\\
(a) (quick and dirty) 
Following \sect{sads_thermo}, compute $s$ from the area law, compute $\varepsilon$ from the first law, and compute $P$ and $F$ from the Euler relation.\\
(b) (\prerequisite{\sect{SAdS_free}})
Following \sect{SAdS_free}, compute $F$ from the on-shell action (bulk, Gibbons-Hawking, and counterterm actions).
\end{prob}

{\color{blue} 
\textit{Solution:}
\be
F = -\frac{V_p}{16\pi G_{p+2} L} \left( \frac{r_0}{L} \right)^{p+1}.
%
\ee
}

\section*{Chap.~8}

\begin{prob}
\label{prob:wilson_ads_soliton}
\textbf{Wilson loop in a confining phase}\\
\prerequisite{\sect{ads_soliton} } \\
Consider the AdS soliton geometry \eqref{eq:AdS_soliton}. Following \sect{wilson_details}, compute the quark potential, or the Wilson loop on the $t'$-$x$ plane, and verify \eq{wilson_ads_soliton}.
\end{prob}

{\color{blue} 
\textit{Solution:}
It is convenient to express results in dimensionless quantities. The quark separation $\calR$ is given by
\be
\frac{\calR}{l}
= \frac{2}{\pi}(1-\epsilon)^{1/4}
\int_1^\infty \frac{dy}{ \sqrt{(y^4-1)(y^4-1+\epsilon)} }~,
%
\ee
%
%
where $\epsilon:=1-(r_0/\rmin)^4$.
The potential energy is given by
\be
l \Delta E = \left(\frac{L}{l_s}\right)^2 \frac{1}{ (1-\epsilon)^{1/4} }
\left\{  
\int_1^\infty\left(\frac{y^4}{ \sqrt{(y^4-1)(y^4-1+\epsilon)} } -1 \right) dy - 1+ (1-\epsilon)^{1/4}
\right\}~,
%
\ee
where $\Delta E :=E-E_0$. 

When $\epsilon \simeq 1$, $\calR \ll l$, \ie, the quark separation is shorter than the $S^1$ radius. In this case, one obtains the Coulomb potential. When $\epsilon \simeq 0$, $\calR \gg l$, and one obtains the confining potential. 
}

\begin{prob}
\label{prob:wilson_debye}
\textbf{Wilson loop at finite temperature} 
\cite{Rey:1998bq2,Brandhuber:1998bs2}\\
Consider the SAdS$_5$ geometry. Following \sect{wilson_details}, compute the quark potential, or the Wilson loop on the $t$-$x$ plane, and verify \eq{wilson_debye}.
\end{prob}

{\color{blue} 
\textit{Solution:}
The quark separation $\calR$ is given by
\be
T \calR
= \frac{2}{\pi}(1-\epsilon)^{1/4}\sqrt{\epsilon}
\int_1^\infty \frac{dy}{ \sqrt{(y^4-1)(y^4-1+\epsilon)} }~.
%
\ee
%
%
The potential energy is given by
\be
\frac{\Delta E}{T} = \left(\frac{L}{l_s}\right)^2 \frac{1}{ (1-\epsilon)^{1/4} }
\left\{  
\int_1^\infty \left(\sqrt{ \frac{y^4-1+\epsilon}{y^4-1} } -1 \right)dy - 1+ (1-\epsilon)^{1/4}
\right\}~.
\label{eq:potential_debye}
\ee
%
%

When $\epsilon \simeq 1$, $T\calR \ll 1$ (low temperature). In this case, one obtains the Coulomb potential. Now, decrease $\epsilon$ or increase temperature. A close inspection of \eq{potential_debye} shows that $\Delta E$ vanishes at some $\epsilon<1$. For higher temperature, the free quark configuration becomes energetically favorable. 
}

\begin{prob}
\label{prob:drag_force}
\textbf{Quark drag force}
\cite{Herzog:2006gh2,Herzog:2006se}\\
\noindent
Consider the SAdS$_5$ geometry, and consider a string moving at constant velocity $v$ in the $x$-direction. The string feels a drag force in the plasma. We compute the friction coefficient $\mu$ defined by
\be
\frac{dp}{dt} = -\mu p~,
\label{eq:def_friction}
\ee
where $p=Mv\gamma$ is the relativistic momentum with quark mass $M$.

\noindent
(a) Following \sect{wilson_details}, take the static gauge
($\sigma^0 = t, \sigma^1 = r$),
and consider the stationary solution of the form
\be
x(t,r)=vt+x(r)~, \quad x(r\to\infty)=0~.
%
\ee
Write down the string action.

\noindent
(b) The canonical momentum is given by

\be
\pi^a_{~x} := \frac{\del L}{\del(\del_a x^\mu)}~.
%
\ee
In our problem, the Lagrangian does not contain $x$, so $\pi^1_{~x} = \del L/\del x'~(' := \del_r)$ is a constant. Rewrite the $\pi^1_{~x}$ equation in terms of $x'$. In this case, the string has no turning point and must hang down to the horizon. This condition determines $\pi^1_{~x}$. 

\noindent
(c) The momentum change of the string is given by%
\footnote{
From the conservation law 
$\del_a \pi^a_{~x}=0$,
$\del_t p= \int d\sigma\, \del_0 \pi^0_{~x} =  -\int d\sigma\, \del_1 \pi^1_{~x} = - \pi^1_{~x}$.
}
\be
\frac{dp}{dt} = -\pi^1_{~x}~.
%
\ee
Using the result of (b), determine $\mu M$ and the configuration $x(r)$.

We consider a stationary solution, so the string does not really change the momentum. Actually, there is a force acting on the quark. The force is an electric field and keeps the quark moving with constant velocity $v$ against the drag force.

\end{prob}

{\color{blue} 
\textit{Solution:}
The induced metric is given by
\be
ds^2 = g_{00} dt^2 + g_{xx} (\dot{x} dt+x' dr)^2+ g_{rr} dr^2
\quad (\dot{} := \del_t~, ' := \del_r)~,
%
\ee
so 
\be
\det h_{ab} = g_{00}g_{rr}+g_{xx}g_{rr}v^2+g_{00}g_{xx} x'^2~.
%
\ee
Then, the action becomes
\be
\action = - \frac{1}{2\pi l_s^2} \int d^2\sigma \sqrt{ -\det h_{ab} }
= - \frac{\calT}{2\pi l_s^2} \int dr\, \sqrt{ \frac{h - v^2 + \left(\frac{r}{L}\right)^4 h^2 x'^2}{h} }~.
\label{eq:action_drag}
\ee
$\pi^1_{~x}$ is given by
\be
\frac{\del L}{\del x'} 
= - \frac{1}{2\pi l_s^2} \left(\frac{r}{L}\right)^4 h^{3/2} 
\frac{x'}{ \sqrt{h - v^2 + \left(\frac{r}{L}\right)^4 h^2 x'^2}}
=: \frac{C}{2\pi l_s^2}~,
%
\ee
which is rewritten as 
\be
x'^2 = \frac{ C^2 (h-v^2) }{ \left(\frac{r}{L}\right)^4h^2 \left\{ \left(\frac{r}{L}\right)^4h-C^2 \right\} }~.
\label{eq:config_drag}
\ee
In \eq{config_drag}, both the numerator and the denominator are positive for large $r$ and negative near $r=r_0$. Since $x'^2>0$, both the numerator and the denominator must have a zero at the same $r=r_c$ and must change sign. This condition determines $C$, and $\pi^1_{~x}$ becomes
\be
\pi^1_{~x} =\frac{1}{2\pi l_s^2}\left(\frac{r_0}{L}\right)^2 v \gamma~.
%
\ee
Comparing with \eq{def_friction}, one gets
\be
\mu M = \frac{1}{2\pi l_s^2}\left(\frac{r_0}{L}\right)^2 
= \frac{\pi}{2}\sqrt{\lambda} T^2~.
%
\ee
The string configuration is given by
\begin{align}
x(r) &= -vr_0^2 L^2 \int_{r_0}^r \frac{dr}{r^4-r_0^4} \\
&= \frac{L^2}{2r_0} v \left\{ \frac{\pi}{2} 
- \tan^{-1}\left(\frac{r}{r_0}\right) 
- \coth^{-1}\left(\frac{r}{r_0}\right) \right\}~.
%
\end{align}

}

\begin{figure}[tb]
\centering
\scalebox{1}{ \includegraphics{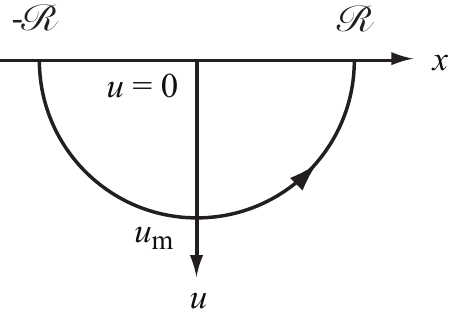} }
\vskip2mm
\caption{The AdS$_3$ geodesic}
\label{fig:geodesic}
\end{figure}

\begin{prob}
\textbf{Geodesic in pure AdS}\\
In \chap{wilson}, we evaluated the world-sheet action, but one often evaluates the other \keyword{world-volume} actions. As a simple example, consider the world-line action, or the particle action. 

Consider a pure AdS spacetime 
\be
ds^2 = L^2 \frac{+d\tE^2+d\bmx^2+du^2}{u^2}~,
%
\ee
where the Euclidean formalism is used here. Finding a geodesic is very similar to Wilson loop computations. Instead of the string action, consider the particle action
\be
\action = + m \int ds~.
%
\ee
Following \sect{wilson_details}, find a spacelike geodesic $x=x(u)$ with \fig{geodesic} configuration.
\end{prob}

{\color{blue} 
\textit{Solution:}
The action becomes 
\be
\action = m \int du\, \frac{L}{u} \sqrt{x'^2+1}~.
%
\ee
Then, the configuration is determined by
\be
x'^2 = \frac{y^2}{1-y^2}~, \qquad(y:=u/u_m)~,
%
\ee
which can be easily integrated. One gets $u_m=\calR$, and the geodesic is a semi-circle $x^2+u^2=\calR^2$. 

Then, the action becomes
\begin{align}
\Sos &= 2mL \int_0^1 \frac{dy}{y\sqrt{1-y^2}} \\
&\simeq 2mL \ln \frac{2\calR}{\epsilon}.
%
\end{align}
The result diverges, so we introduced the UV cutoff $u=\epsilon$.

This result has several applications. First, this gives an approximate two-point correlation function of an operator $\calO$ with $\Delta \gg 1$. From \eq{conf_weight_scalar}, $\Delta \gg 1$ implies $mL \gg 1$. Then, the scalar field is approximated by a massive point particle, and the two-point function becomes
\begin{align}
\bra \calO(\calR)\calO(-\calR) \ketc &\simeq e^{-\Sos} \\
&\simeq \frac{1}{R^{2mL}} \simeq \frac{1}{R^{2\Delta}}~,
%
\end{align}
where the subscript ``$c$" represents the connected function. The result has the correct scaling dimension $2\Delta$.

Another application is the computation of the \keyword{holographic entanglement entropy} in AdS$_3$/CFT$_2$ \cite{Nishioka:2009un}.
}

\section*{Chap.~9}

\begin{prob}
\textbf{Density matrix and entropy} \\
From the density matrix $\rho$, the \keyword{von Newman entropy} is defined by
\be
S = -\text{tr} \left[ \rho \ln \rho \right]~.
%
\ee
Using the canonical ensemble $\rho=e^{-\beta H}/Z$, show that $S$ indeed reduces to the thermodynamic entropy $S=\ln Z+ \beta H$.

For illustration, take the spin-1/2 systems in the text. For a pure ensemble, $S=0$. When $\rho$ is proportional to the identity operator $I$, the ensemble is called \keyword{maximally mixed}. The ensemble which contains  which contains $\kets{+}$ and $\kets{-}$ with equal weights is maximally mixed. In this case, $S=\ln2$. In general, $S \leq \ln d_{\calH}$, where $d_{\calH}$ is the dimensionality of the Hilbert space, and the equality is valid for a maximally mixed ensemble.
\end{prob}

\begin{prob}
\label{prob:bjorken}
\textbf{Bjorken flow (perfect fluid)} \\
We consider the time evolution of QGP using hydrodynamics. A heavy-ion collision is often approximated by a $(1\!+\!1)$-dimensional model $ds^2=-dT^2+dX^2$, where only the longitudinal beam direction $X$ is taken into account, and the transverse directions $(Y,Z)$ are ignored. The time evolution goes as follows (\fig{bjorken}):
\begin{itemize}
\item 
Because of ultra-relativistic nature, heavy-ions approximately move along light-cones $T=\pm X$.
\item
They collide at $(T,X)=(0,0)$ and pass through each other leaving behind excited quarks and gluons with various velocities. 
\item
The excited particles interact each other and reach a local equilibrium after some proper time $\tau_0$. (Time scales are determined by each particle's proper time $\tau$.) The particles then form QGP, and one is able to use hydrodynamics. 
\end{itemize}

\begin{figure}[tb]
\centering
\scalebox{0.55}{ \includegraphics{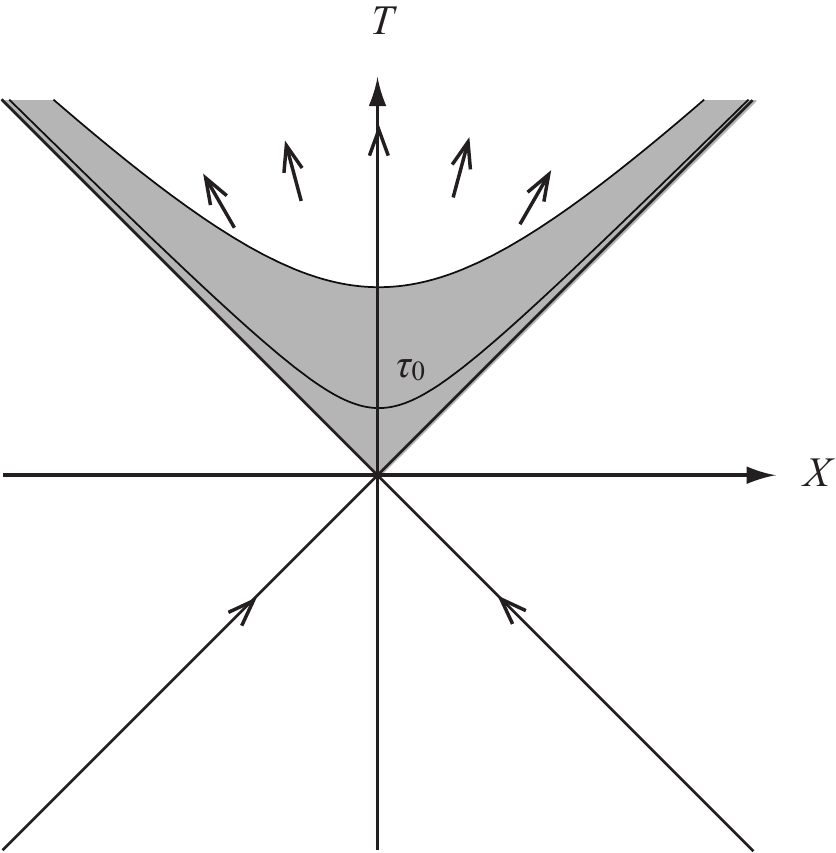} }
\vskip2mm
\caption{The Bjorken flow. The shaded region represents excited particles and form QGP after $\tau_0$.}
\label{fig:bjorken}
\end{figure}%

A simple hydrodynamic model of the QGP evolution is known as the \keyword{Bjorken flow} \cite{Bjorken:1982qr}. In this model, each particle velocity is assumed to be $v=X/T$ or $u^\mu = (T,X)/\tau$. This configuration is boost-invariant, namely a different position of particles is simply Lorentz-boosted. It is also assumed that the fluid velocity $u^\mu(x)$ is given by the particle velocity there. In this exercise, we also assume a scale-invariant fluid where $T^\mu_{~\mu}=0$. We first consider the perfect fluid case. 

It is convenient to use the Milne coordinates \eqref{eq:milne} since it is the fluid rest frame. Adding transverse directions, 
\be
ds^2 = - d\tau^2 + \tau^2 dy^2 + dY^2 + dZ^2~.
\label{eq:milne2}
\ee

\noindent
(a) Show that a Lorentz boost acts as $y \rightarrow y+\text{constant}$, so the boost invariance means that physical quantities are $y$-independent. We will see that this is indeed the case. 

\noindent
(b) Write down $T^{\mu\nu}$ components in the Milne coordinates.

\noindent
(c) Derive hydrodynamic equations (continuity equation and Euler equation) for the Bjorken flow from the conservation equation $\nabla_\mu T^{\mu\nu}=0$. Note that the covariant derivative appears because the Milne coordinates do not take the Minkowski form. Solve hydrodynamic equations, and determine the time evolution of the energy density $\varepsilon$.

\noindent
(d) Determine the $\tau$-dependence of $s$ and $T$ using the Euler relation $\varepsilon+P=Ts$ and the first law $d\varepsilon = T ds$.
\end{prob}

{\color{blue} 
\textit{Solution:}
A Lorentz boost is given by $\tilde{x}^\mu = \Lambda^\mu_{~~\nu} x^\nu$, where
\be
\Lambda^\mu_{~~\nu} =
	\begin{pmatrix}
	  \cosh y & \sinh y  \\
	  \sinh y & \cosh y
	\end{pmatrix}
%
\ee
(We consider only the $(T,X)$-directions.) Two successive boosts give 
\be
\Lambda^\mu_{~~\rho}(y) \Lambda^\rho_{~~\nu}(y') =
	\begin{pmatrix}
	  \cosh (y+y') & \sinh (y+y')  \\
	  \sinh (y+y') & \cosh (y+y')
	\end{pmatrix}~,
%
\ee
so a boost acts as $y \rightarrow y+\text{constant}$.

The $T^{\mu\nu}$ components are given by
\begin{align}
 T^{\mu\nu} &= (\varepsilon+P)u^\mu u^\nu + P \eta^{\mu\nu} 
 \nonumber \\
 &\stackrel{RF}{=} 
	\begin{pmatrix}
	  \varepsilon & 0 & 0 & 0 \\
	  0 & \frac{P}{\tau^2} & 0 & 0 \\
	  0 & 0 & P & 0 \\
	  0 & 0 & 0 & P \\
	\end{pmatrix}~.
%
\end{align}
The Euler equation reduces to
\be
\del_y P = 0~,
%
\ee
so the pressure is $y$-independent or boost invariant. It then follows that the other thermodynamic quantities are all boost invariant. The continuity equation becomes
\be
\del_\tau \varepsilon + \frac{4}{3} \frac{\varepsilon}{\tau} = 0~,
%
\ee
where $\epsilon=3P$ is used. Integrating the equation and using thermodynamic relations, one gets
\begin{align}
\varepsilon &\propto \tau^{-4/3}~,\\
s &\propto \tau^{-1}~,\\
T &\propto \tau^{-1/3}~.
%
\end{align}
}

\begin{prob}
\textbf{Bjorken flow (viscous fluid)} \\
Now, consider a viscous fluid case for the Bjorken flow. We again consider a scale-invariant fluid.

\noindent
(a) Write down $T^{\mu\nu}$ components in the Milne coordinates. One needs to evaluate $\nabla_\mu u_\nu$. 

\noindent
(b) Derive hydrodynamic equations from $\nabla_\mu T^{\mu\nu}=0$. 

\noindent
(c) Solving the continuity equation as a power series expansion in $1/\tau$, determine the large-$\tau$ behavior of $\varepsilon$.
\end{prob}

{\color{blue} 
\textit{Solution:}
The $T^{\mu\nu}$ components are given by
\be
T^{\mu\nu} 
\stackrel{RF}{=} 
	\begin{pmatrix}
	  \varepsilon & 0 & 0 & 0 \\
	  0 & \frac{P}{\tau^2} - \frac{4\eta}{3\tau^3} & 0 & 0 \\
	  0 & 0 & P + \frac{2\eta}{3\tau^3}  & 0 \\
	  0 & 0 & 0 & P + \frac{2\eta}{3\tau^3} \\
	\end{pmatrix}~.
%
\ee
The $\nabla_\mu T^{\mu y}=0$ equation reduces to
\be
\del_y T^{yy} \propto \del_y P - \frac{4}{3\tau} \del_y \eta =0~,
%
\ee
so a boost-invariant $P$ and $\eta$ satisfy the equation. The continuity equation becomes
\be
\del_\tau \varepsilon + \frac{4}{3} \frac{\varepsilon}{\tau} -  \frac{4}{3} \frac{\eta}{\tau^2} = 0~.
%
\ee
The large-$\tau$ behavior is given by
\be
\varepsilon \propto \tau^{-4/3} - 2 \eta_0 \tau^{-2} + \cdots~,
%
\ee
where $\eta_0$ is a constant which is related to $\eta$ as
\be
\eta = \frac{\eta_0}{\tau}~.
%
\ee
$\eta$ depends on $\tau$ so is not a constant because $\eta$ is temperature-dependent:
\be
\eta \propto T^3 \propto \tau^{-1}~.
%
\ee
}

\section*{Chap.~10}

\begin{prob}
\label{prob:fieldop_Maxwell}
\textbf{Field/operator correspondence for the Maxwell field}\\
Consider the Maxwell field in the asymptotically AdS geometry:
\be
\action =  -\frac{1}{4} \int d^{p+2}x\, \sqrt{-g} F_{MN}^2~.
\ee
Following Sects.~\ref{sec:scalar} and \ref{sec:current}, derive the field/operator correspondence for the Maxwell field. Namely, evaluate the on-shell Maxwell action and compute $\cA$ in Eqs.~\eqref{eq:response_rho_ads} or \eqref{eq:response_current_ads}. The holographic renormalization is unnecessary for the Maxwell field.
\end{prob}

{\color{blue} 
\textit{Solution:}
\be
\bra J^x \ketj = (p-1) \vevA{x} \sourceA{x}~.
%
\ee
or $\cA=p-1$. (We set $r_0=L=1$.) One can rewrite the result as
\be
A_x \sim \sourceA{x} + \frac{1}{p-1} \bra J^x \ketj\, u^{p-1}~,
\quad (u\rightarrow0)~.
%
\ee
The scaling dimension of the current is $[J]=p$. The scaling dimension of the electric field $\sourceE = -\del_t \sourceA{x}$ is $[E]=2$. From the Ohm's law $J^x = \sigma \sourceE$, the scaling dimension of the conductivity is $[\sigma]=p-2$.
}

\begin{prob}
\label{prob:field_op_Lifshitz}
\textbf{Field/operator correspondence in the Lifshitz geometry}
\cite{Kachru:2008yh2,Taylor:2015glc} \\
The Lifshitz geometry is given by 
\begin{align}
ds^2 &= -r^{2z}dt^2 + r^2 d\bmx_p^2 + \frac{dr^2}{r^2} \\
&= - \frac{dt^2}{u^{2z}} + \frac{d\bmx_p^2}{u^2} + \frac{du^2}{u^2}~.
%
\end{align}
The geometry is invariant under the anisotropic scaling
\be
t \rightarrow a^z t~, \bmx \rightarrow a \bmx~, u \rightarrow au~. \quad
(\text{or~}  r \rightarrow r/a)
%
\ee

\noindent
(a) Consider the massless scalar field. Following \sect{scalar}, derive the field/operator correspondence: solve the equation of motion asymptotically, evaluate the on-shell action, and obtain the one-point function.

\noindent
(b) Consider the massive scalar field. Following \sect{massive_scalar}, solve the equation of motion asymptotically, and obtain the BF bound.

\noindent
(c) Derive the field/operator correspondence for the Maxwell field $A_x$.\\
\end{prob}

{\color{blue} 
\textit{Solution:}
For the massless scalar field,
\be
\phi \sim \source + \frac{1}{z+p} \bra \calO \ketj\, u^{z+p}~,
\quad (u\rightarrow0)~.
%
\ee
The scaling dimension of $\bra \calO \ketj$ is $\Delta_+ = z+p$. This is because the perturbed action of the boundary theory $ \delta \action_\text{QFT} = \int d^{p+1}x\, \source \calO $ involves $d^{p+1}x$ which scales as $d^{p+1}x \rightarrow a^{z+p}d^{p+1}x$.

For the massive scalar field,
\begin{alignat}{2}
\phi &\sim\source u^{\Delta_-} + \cphi \bra\calO\ketj\, u^{\Delta_+}~,
&\quad& (u\rightarrow0)~,
 \\
\Delta_\pm &= \frac{z+p}{2} \pm \sqrt{\frac{(z+p)^2}{4}+m^2}~,&&
\end{alignat}
so the BF bound is given by
\be
m^2 \geq -\frac{(z+p)^2}{4}~.
%
\ee

For the Maxwell field,
\be
A_x \sim \sourceA{x} + \frac{1}{z+p-2} \bra J^x \ketj\, u^{z+p-2}~,
\quad (u\rightarrow0)~.
%
\ee
The scaling dimensions are $[J] = z+p-1$, $[E]=z+1$, and $[\sigma]=p-2$.
}

\begin{prob}
\textbf{Temperature dependence of physical quantities in the Lifshitz geometry}\\
 We use the radial coordinate $r$ in this exercise. Consider the asymptotically Lifshitz black hole. Suppose that the horizon radius of the \bh satisfies the scaling $r_0 \rightarrow r_0/a$ like the SAdS black hole. Also, assume that $r_0$ or temperature $T$ is the only scale in the dual field theory. \\
 
\noindent
(a) Following \sect{SAdS}, obtain the temperature dependence of thermodynamic quantities.  Assume the area law, the first law, and the Euler relation. 
 
\noindent
(b) Similarly, using the result of \probset{field_op_Lifshitz}(c), obtain the temperature dependence of the conductivity $\sigma$.

\end{prob}

{\color{blue} 
\textit{Solution:}
\begin{align}
T &\simeq r_0^z~, \\
s &\simeq r_0^p \simeq T^{p/z}~, \\
\varepsilon &\simeq P \simeq T^{(z+p)/z}~, \\
\sigma &\simeq T^{(p-2)/z}~. 
%
\end{align}
You do not need the explicit form of the \bh metric to know the dependence, but of course the biggest assumption is the scaling.

Including numerical coefficients, one can also show that $z\varepsilon=pP$ instead of $\varepsilon=pP$, which is a consequence of the anisotropic scaling.
}

\section*{Chap.~12}

\begin{prob}
\label{prob:SAdS4_diff}
\textbf{Diffusion constant and conductivity for SAdS$_4$ black hole} \\
\prerequisite{\sect{SAdS5_diff} } \\
Consider the Maxwell field in the SAdS$_4$ background. Following \sect{SAdS5_diff}, obtain the conductivity $\sigma$ and the diffusion constant $D$ from the Maxwell vector mode computation. This exercise computes $\sigma$ for the holographic \SC\ in \chap{phase} in the high-temperature phase (in the probe limit).

\noindent
(a) Derive the $A_x$ equation from the Maxwell equation. Solve the equation of motion up to $O(\omega)$. Expand the solution asymptotically keeping the terms of $O(u)$.  One may use \textit{Mathematica} for this problem.

\noindent 
(b) Obtain $\sigma$ from \eq{fast_vs_cond_ads5}. Also, obtain $\chi_T$ and $D$. \\

\end{prob}

{\color{blue} 
\textit{Solution:} 
The $A_x$ equation is written as
\be
\frac{1}{h}(h A_x')' + \frac{\omega^2}{\Ts^2 h^2} A_x = 0~,
%
\ee
where $h=1-u^3$ and $\Ts:=4\pi T/3=r_0/L^2$. This is \eq{Ax_eom_4} with $\Psi=0$. The solution asymptotically behaves as 
\be
A_x = \sourceA{x} \left\{ 1 + \frac{i\omega}{\Ts} u + \cdots \right\}~, 
\qquad (u\rightarrow 0)~.
%
\ee

$\vevA{x} = i \omega/\Ts$, so 
\be
\sigma = \frac{\cA}{\Ts} = 1~.
\label{eq:cond_ads4} 
\ee
The factor $\cA$ has been computed in \probset{fieldop_Maxwell}. Recovering dimensionful quantities, $\cA=\Ts$. [We set the normalization factor $\alpha=1$ in \eq{action_Maxwell}.] Finally,
\begin{align}
\chi_T &= \cA = \Ts~, \\
D &= \frac{\sigma}{\chi_T} = \frac{1}{\Ts} = \frac{3}{4\pi T}~.
%
\end{align}
}

\begin{prob}
\label{prob:2nd}
\textbf{Second-order transport coefficients for the \Nfour\ SYM} 
\cite{Baier:2007ix3} \\
\prerequisite{Sects.~\ref{sec:Lorentzian_prescription} and \ref{sec:action_tensor} } \\
One can derive second-order transport coefficients \eqref{eq:transport_causal} from the tensor mode response $\delta \bra \tau^{xy}  \ket $.

\noindent
(a) Following \sect{action_tensor}, compute the on-shell tensor mode action up to $O(\omega^2,q^2)$. You need to reevaluate the counterterm action. \\

{\color{blue} 
\textit{Solution:} 
We set $r_0=L=16\pi G_5=1$ in this problem. 
\be
\Sos = \left. V_4 + \left(-\frac{1}{2} + \frac{q^2-\omega^2}{4u^2} \right) \phim\cdot\phip+\frac{1}{2u^3}\phim\cdot\phip'  \right|_{u=0}~
%
\ee
instead of \eq{tensor_on_shell1}.\\
}

\noindent
(b) Following \sect{tensor_mode_sol}, solve the tensor mode equation of motion up to $O(\omega^2,q^2)$. Expand the solution asymptotically keeping the terms of $O(u^4)$. One may use \textit{Mathematica} for this problem.\\

{\color{blue} 
\textit{Solution:}
Write the solution as $\phi_k = \source_k f_k(u)$. Then, 
\be
f_k = 1 - \frac{q^2-\omega^2}{4} u^2 + \frac{1}{4} \left\{i\omega + \frac{1}{2}q^2 - \frac{1-\ln 2}{2} \omega^2 \right\} u^4 + \cdots~, \qquad (u\rightarrow 0)~.
%
\ee
}

\noindent
(c) Following \sect{Lorentzian_prescription}, substitute the solution into the on-shell action and obtain 
$\delta \bra \tau^{xy}  \ket $. Comparing with the Kubo formula \eqref{eq:kubo_causal}, obtain $\kappa$ and $\tau_\pi$. \\

{\color{blue} 
\textit{Solution:} 
In the convention of \sect{Lorentzian_prescription}, the fluctuation-dependent on-shell action becomes
\begin{align}
%
\Sos[\source] &= \int (dk)\, \phim^{(0)} \left\{
\left. \Fp \right|_{u=0} - \left.\Fp \right|_{u=1} 
\right\} \phip^{(0)}~,
\\
\left. 2\Fp \right|_{u=0} &= \left. \left( -1 + \frac{q^2-\omega^2}{2u^2} \right) f_{-k}f_{k} 
+ \frac{1}{u^3} f_{-k}f_{k}' \right|_{u=0}
\\
& = -1 + i\omega + \frac{1}{2}q^2 - \frac{1-\ln 2}{2} \omega^2 + \cdots~.
\label{eq:def_Fk2}
%
\end{align}
Recall that the response is written as
\be
\bra \calO_k  \ketj = \left. 2\Fp \right|_{u=0} \source_k~.
\tag{\ref{eq:Lorentzian_prescription}}
%
\ee
Comparing with the Kubo formula, one gets
$P=\eta=\kappa=1$, and
$\tau_\pi =(2-\ln2)/2$.
}
\end{prob}

\textit{Further suggestions:} Similarly, compute transport coefficients for SAdS$_4$ and SAdS$_7$ black holes \cite{Natsuume:2008iy}:
\begin{alignat}{3}
P&=\eta=1, \tau_\pi =  1-\frac{\ln 3 -\frac{\sqrt{3} \pi}{9}}{2} &
\qquad &(\text{SAdS$_4$}) \\
P&=\eta=1, \tau_\pi = 1-\frac{\ln 3 +\frac{\sqrt{3} \pi}{9}}{4}~,~ &\kappa =\frac{1}{2}
\qquad &(\text{SAdS$_7$}) 
%
\end{alignat}
The result of $\tau_\pi$ can be summarized using the harmonic number $H_n = \sum_{k=1}^n \frac{1}{k}$ as follows \cite{Natsuume:2008gy}:
\be
\tau_\pi = \frac{1}{2} + \frac{H_{\frac{2}{p+1}}}{p+1}~.
%
\ee

\section*{Chap.~13}

\begin{prob}
\label{prob:factorization}
\textbf{Large-$\Nc$ factorization}\\
Consider a two-point correlation function of an operator%
\footnote{More precisely, we consider the so-called single-trace operators such as $\calO \simeq \text{tr} F^2$. In this book, we consider only single-trace operators.} 
$\calO$
\be
\bra \calO\calO \ket = \bra\calO\ket \bra\calO\ket + \bra \calO\calO \ketc~,
%
\ee
where the subscript ``$c$" represents the connected correlation function. Argue that the first term is $O(\Nc^4)$ whereas the second term is $O(\Nc^2)$ using AdS/CFT. 

This property is known as \keyword{large-$\Nc$ factorization}. Although large-$\Nc$ gauge theories are strongly-coupled, gauge invariant operators are weakly-coupled, and their interactions are suppressed by $1/\Nc$. They are classical and behaves as $c$-numbers in the large-$\Nc$ limit.
\end{prob}

\newcommand{\Sosd}{\underline{I}}

{\color{blue} 
\textit{Solution:}
This behavior essentially comes from the fact that the Newton's constant appears only as the overall coefficient of the bulk action. Let us introduce the dimensionless metric $ds^2 =L^2 ds^2_\text{dimensionless}$. Then, the bulk action becomes
\begin{align}
\action &= \frac{1}{16\pi G_5} \int d^5x \sqrt{-g} R + \cdots 
\\
&= \frac{L^3}{16\pi G_5} S_\text{dimensionless}
\simeq \Nc^2 S_\text{dimensionless}~.
%
\end{align}
The action is proportional to $\Nc^2$
which implies
\begin{align}
\bra \calO \ketnoj &
= \left.\frac{\delta Z}{\delta \source}\right|_{\source=0} 
\simeq \Nc^2~.\\
\bra \calO\calO \ketc &
= -\left.\frac{\delta^2 W}{\delta \phi^{(0)2}}\right|_{\source=0} 
\simeq \Nc^2~,
%
\end{align}
where $Z = \bra e^{-\source\calO} \ket$, $W$ is the generating function of connected correlation functions given by $Z=e^{-W}$ (it is essentially the bulk on-shell action), and we use the Euclidean formalism.
Thus,
\begin{alignat}{2}
\bra\calO\calO\ket  
 = &\bra \calO \ket \bra \calO \ket 
 + &\bra \calO\calO \ketc~. \\
 &\quad\Nc^4 & \Nc^2 \qquad\nonumber
\end{alignat}

The argument here is formal in character. But, for example, we computed the one-point function of the energy-momentum tensor:
\be
\bra T_{\mu\nu}\ket 
 \simeq \Nc^2~.
%
\ee
We also computed the one-point function in the presence of the source which is related to the two-point function via the linear response theory:
\be
\delta \bra T_{xy}\ket
 \simeq \bra T_{xy}T_{xy}\ketc
 \simeq \Nc^2~.
%
\ee
%
%
(Here, we were sloppy about the differences among various correlators/Green's functions. We have mostly computed retarded Green's functions, but all others can be obtained once the retarded Green's function is known.)
}

\begin{prob}
\label{prob:ginzburg}
\textbf{Ginzburg criterion}\\
The large-$\Nc$ factorization does not hold for field theory in general, but a similar property holds when the number of spatial dimensions are large enough. The mean-field theory utilizes this property. Again, compare connected and disconnected correlation functions. We would like to know when the following condition is satisfied:
\be
\bra m(\bmx) \ket \bra m(\bm{0}) \ket \gg \bra m(\bmx) m(\bm{0}) \ketc~.
%
\ee
Show that the condition leads to the Ginzburg criterion:
\be
m^2 \xi^{\ds} \gg T\chi_T~,
%
\ee
In terms of critical exponents, one gets \eq{critical_dim}:
\be
\ds \geq d_\text{UC} = \frac{2\beta+\gamma}{\nu}~.
%
\ee
\end{prob}

{\color{blue} 
\textit{Solution:}
Integrating the equation over the region with size $\xi$, one gets
\begin{align}
\int_0^\xi d\bmx\, \bra m(\bmx) \ket \bra m(\bm{0}) \ket
 &\simeq m^2 \xi^{\ds}~,
\\
\int_0^\xi d\bmx\, \bra m(\bmx) m(\bm{0}) \ketc
 &\simeq \int_0^\infty d\bmx\, T \chi(\bmx) = T\chi_T~.
%
\end{align}
}

\section*{Chap.~14}

\begin{prob}
\textbf{Hawking-Page transition of SAdS$_4$ black hole}\\
The SAdS$_4$ black hole with $S^2$ horizon is given by
\be
ds_4^2 =  - \left( \frac{r^2}{L^2} + 1 - \frac{r_0^3}{L^2r} \right) dt^2 
+ \frac{dr^2}{ \frac{r^2}{L^2} + 1 - \frac{r_0^4}{L^2r} } 
+ r^2 d\Omega_2^2~.
%
\ee
Compute the free energy using two methods below and discuss the Hawking-Page transition.

\noindent
(a) Following \sect{HP_S3}, compute free energy using the reference spacetime method.

\noindent
(b) (\prerequisite{\sect{SAdS_free}}) Following \sect{SAdS_free}, compute free energy. Is there a Casimir energy in this case?
\end{prob}

{\color{blue} 
\textit{Solution:}
\be
F =  - \frac{\ro}{4G_4 L^2} (\ro^2-L^2)~.
%
\ee
}

\begin{prob}
\textbf{Free energy of Schwarzschild black hole} \\
\prerequisite{\sect{SAdS_free} } \\
Compute the free energy of the four-dimensional Schwarzschild black hole and verify \eq{free_Sch}. 

First, following \sect{SAdS_free}, compute the Gibbons-Hawking action. (The on-shell bulk action vanishes.) In order to obtain a finite free energy, use the reference spacetime method in \sect{HP_S3} instead of the counterterm method. Use the flat spacetime for the reference spacetime.
\end{prob}

\begin{prob}
\textbf{Critical temperature in alternative quantization} \\
\prerequisite{Sects.~\ref{sec:massive_more} and \ref{sec:HSC_more} } \\
In the text, we consider the asymptotic behavior
\be
\psi_1 \sim \psi_1^{(0)} u + \cpsi \bra\calO_2\ketj\, u^2~,
\quad (u\rightarrow0)~.
%
\ee
This is the so-called standard quantization. But when
\be
- \frac{9}{4} \leq m^2 \leq - \frac{5}{4} \quad(\text{for~} p=2)~,
%
\ee
(For the holographic \SC, $m^2=-2$, so this condition is satisfied), the slow falloff is also normalizable so that one can exchange the role of the external source and the operator. Namely, one can consider
\be
\psi_1 \sim \cpsi \bra\calO_1\ketj\, u + \psi_1^{(0)} u^2~,
\quad (u\rightarrow0)~.
%
\ee
This is the so-called alternative quantization. The scaling dimensions are $[\calO_2]=2$ and $[\calO_1]=1$. 

Following \sect{HSC_more}, compute the critical temperature in the alternative quantization by imposing $\psi_1^{(0)}=0$. One may use \textit{Mathematica}.
\end{prob}

{\color{blue} 
\textit{Solution:}
\begin{align}
\frac{\Ts_c}{\mu} &\approx 0.893~, \quad(\text{for } \calO_1)~, \\ 
\textit{cf. } \frac{\Ts_c}{\mu} &\approx 0.246~, \quad(\text{for } \calO_2)~. 
%
\end{align}
In general, the critical temperature rises as the scaling dimension of the order parameter becomes smaller.
}

\endofsection

\begin{prob}
\textbf{}\\
\end{prob}

{\color{blue} 
\textit{Solution:}
}





\begin{thebibliography}{99}

%
%

\bibitem{Maldacena:1997re2}
  J.~M.~Maldacena,
  ``The Large N limit of superconformal field theories and supergravity,''
  Adv.\ Theor.\ Math.\ Phys.\  {\bf 2 } (1998)  231
  [hep-th/9711200].

\bibitem{MTW}
C.~W.~Misner, K.~S.~Thorne and J.~A.~Wheeler, 
\textit{Gravitation} 
(W. H. Freeman, 
1973).

\bibitem{Aharony:2000ti}
O.~Aharony, S.~S.~Gubser, J.~Maldacena, H.~Ooguri and Y.~Oz,
``Large N field theories, string theory and gravity,''
Phys.\ Rept.\  {\bf 323} (2000) 183
[hep-th/9905111].

\bibitem{Green:1987sp}
  M.~B.~Green, J.~H.~Schwarz and E.~Witten,
  \textit{Superstring theory}
  (Cambridge Univ.\ Press, 1987).
    
\bibitem{Polchinski:1998rq}
  J.~Polchinski,
  \textit{String theory}
  (Cambridge Univ.\ Press, 1998).

\bibitem{Zwiebach:2004tj}
  B.~Zwiebach,
  \textit{A first course in string theory, second edition}
  (Cambridge Univ.\ Press, 2009).

\bibitem{Becker:2007zj}
  K.~Becker, M.~Becker and J.~H.~Schwarz,
  \textit{String theory and M-theory: A modern introduction}
  (Cambridge Univ.\ Press, 2007).

\bibitem{Kiritsis:2007zz}
  E.~Kiritsis,
  \textit{String theory in a nutshell}
  (Princeton Univ.\ Press, 2007).
  
\bibitem{Johnson:2003gi}
  C.~V.~Johnson,
  \textit{D-branes}
  (Cambridge Univ.\ Press, 2003).   

\bibitem{Schutz:1985jx}
B.~F.~Schutz,
\textit{A first course in general relativity, second edition}
(Cambridge Univ.\ Press, 2009).

\bibitem{Zee:2013dea}
A.~Zee,
\textit{Einstein Gravity in a Nutshell}
(Princeton Univ.\ Press, 2013).

\bibitem{Wald:1984rg}
R.~M.~Wald,
\textit{General relativity}
(The Univ.\ of Chicago Press, 1984).

\bibitem{Ryder:1985wq}
  L.~H.~Ryder,
  \textit{Quantum Field Theory, second edition} 
  (Cambridge Univ.\ Press, 1996).
  
\bibitem{Zee:2003mt}
  A.~Zee,
  \textit{Quantum field theory in a nutshell}
  (Princeton Univ.\ Press, 2010).

\bibitem{Srednicki:2007qs}
  M.~Srednicki,
  \textit{Quantum field theory}
  (Cambridge Univ.\ Press, 2007).
  
\bibitem{Peskin:1995ev}
  M.~E.~Peskin and D.~V.~Schroeder,
  \textit{An Introduction to quantum field theory}
  (Addison-Wesley, 1995).
  
\bibitem{Weinberg:1995mt}
  S.~Weinberg,
  \textit{The Quantum theory of fields}, Vol. 1-3
  (Cambridge Univ.\ Press, 1995-2000).



\bibitem{Papantonopoulos:2011zz}
\textit{From Gravity to Thermal Gauge Theories: The AdS/CFT Correspondence}, 
eds. E. Papantonopoulos, Lecture Notes in Physics 828 (Springer, 2011).

\bibitem{Horowitz:2012nnc}
\textit{Black Holes in Higher Dimensions}, 
eds. G. Horowitz (Cambridge University Press, 2012).

\bibitem{DeWolfe:2013cua}
  O.~DeWolfe, S.~S.~Gubser, C.~Rosen and D.~Teaney,
  ``Heavy ions and string theory,''
  arXiv:1304.7794 [hep-th].

\bibitem{Hartnoll:2009sz}
  S.~A.~Hartnoll,
  ``Lectures on holographic methods for condensed matter physics,''
  Class.\ Quant.\ Grav.\  {\bf 26} (2009) 224002
  [arXiv:0903.3246 [hep-th]].
  
\bibitem{McGreevy:2009xe}
  J.~McGreevy,
  ``Holographic duality with a view toward many-body physics,''
  Adv.\ High Energy Phys.\  {\bf 2010} (2010) 723105
  [arXiv:0909.0518 [hep-th]].

\bibitem{Iqbal:2011ae}
  N.~Iqbal, H.~Liu and M.~Mezei,
  ``Lectures on holographic non-Fermi liquids and quantum phase transitions,''
  arXiv:1110.3814 [hep-th].

\bibitem{CasalderreySolana:2011us}
  J.~Casalderrey-Solana, H.~Liu, D.~Mateos, K.~Rajagopal and U.~A.~Wiedemann,
  \textit{Gauge/String Duality, Hot QCD and Heavy Ion Collisions} (Cambridge Univ.\ Press, 2014)
  [arXiv:1101.0618 [hep-th]].

\bibitem{Ammon:2015wua}
  M.~Ammon and J.~Erdmenger,
  \textit{Gauge/gravity duality : Foundations and applications}
  (Cambridge Univ.\ Press, 2015).

\bibitem{Zaanen:2015oix}
  J.~Zaanen, Y.~W.~Sun, Y.~Liu and K.~Schalm,
  \textit{Holographic Duality in Condensed Matter Physics}
  (Cambridge Univ.\ Press, 2015).

\end{thebibliography}

\begin{thebibliography}{99}

%
%

\bibitem{Bekenstein:1996pn}
  J.~D.~Bekenstein,
  ``Black hole hair: 25 - years after''
  [gr-qc/9605059].

\bibitem{Heusler:1996ft}
  M.~Heusler,
  ``No hair theorems and black holes with hair,''
  Helv.\ Phys.\ Acta {\bf 69} (1996) 501
  [gr-qc/9610019].

\bibitem{'tHooft:1993gx}
  G.~'t Hooft,
  ``Dimensional reduction in quantum gravity,''
  gr-qc/9310026.

\bibitem{Susskind:1994vu}
  L.~Susskind,
  ``The World as a hologram,''
  J.\ Math.\ Phys.\  {\bf 36} (1995) 6377
  [hep-th/9409089].

\bibitem{Wald:1994yp}
R.~M.~Wald,
\textit{Quantum field theory in curved space-time and black hole thermodynamics}
(The Univ.\ of Chicago Press, 1994).

\bibitem{Gibbons:1976ue}
  G.~W.~Gibbons and S.~W.~Hawking,
  ``Action Integrals and Partition Functions in Quantum Gravity,''
  Phys.\ Rev.\  {\bf D15 } (1977)  2752.

\bibitem{Zurek:1985gd}
  W.~H.~Zurek and K.~S.~Thorne,
  ``Statistical Mechanical Origin Of The Entropy Of A Rotating, Charged Black
  Hole,''
  Phys.\ Rev.\ Lett.\  {\bf 54} (1985) 2171.

\bibitem{Horowitz:1996qd}
  G.~T.~Horowitz,
  ``The Origin of black hole entropy in string theory,''
  gr-qc/9604051.

\bibitem{Dabholkar:2012zz}
  A.~Dabholkar and S.~Nampuri,
  ``Quantum black holes,''
  Lect.\ Notes Phys.\  {\bf 851} (2012) 165
  [arXiv:1208.4814 [hep-th]].
  

\bibitem{HE}
S.~W.~Hawking and G.~F.~R.~Ellis, 
\textit{The large scale structure of spacetime} 
(Cambridge Univ.\ Press, 1973).

\bibitem{Gregory:1993vy}
  R.~Gregory and R.~Laflamme,
  ``Black strings and p-branes are unstable,''
  Phys.\ Rev.\ Lett.\  {\bf 70} (1993) 2837
  [hep-th/9301052].

\bibitem{Horowitz:2012nnc2}
\textit{Black Holes in Higher Dimensions}, 
eds. G. Horowitz (Cambridge University Press, 2012).
  
\bibitem{Harmark:2007md}
  T.~Harmark, V.~Niarchos and N.~A.~Obers,
  ``Instabilities of black strings and branes,''
  Class.\ Quant.\ Grav.\  {\bf 24} (2007) R1
  [hep-th/0701022].

\bibitem{Cardoso:2006ks}
  V.~Cardoso and O.~J.~C.~Dias,
  ``Rayleigh-Plateau and Gregory-Laflamme instabilities of black strings,''
  Phys.\ Rev.\ Lett.\  {\bf 96} (2006) 181601
  [hep-th/0602017].
 
\end{thebibliography}

\begin{thebibliography}{99}

%
%

\bibitem{Hatsuda:2007rt}
  T.~Hatsuda,
  ``Bulk and spectral observables in lattice QCD,''
  arXiv:hep-ph/0702293.

\bibitem{qgp}
  K.~Yagi, T.~Hatsuda and Y.~Miake,
  \textit{Quark-Gluon Plasma: From Big Bang To Little Bang}
  (Cambridge Univ.\ Press, 2005).

\bibitem{press}
\url{http://www.bnl.gov/newsroom/news.php?a=1303}

\bibitem{Csernai:2006zz}
  L.~P.~Csernai, J.~.I.~Kapusta and L.~D.~McLerran,
  ``On the Strongly-Interacting Low-Viscosity Matter Created in Relativistic Nuclear Collisions,''
  Phys.\ Rev.\ Lett.\  {\bf 97} (2006) 152303
  [nucl-th/0604032].
  
\bibitem{'tHooft:1973jz}
  G.~'t Hooft,
  ``A Planar Diagram Theory for Strong Interactions,''
  Nucl.\ Phys.\  B {\bf 72} (1974) 461.

\bibitem{coleman}
S.~Coleman, 
\textit{Aspects of symmetry} 
(Cambridge Univ.\ Press, 1985).

\end{thebibliography}

\begin{thebibliography}{99}

%
%

\bibitem{Donnachie:2002en}
  S.~Donnachie, H.~G.~Dosch, O.~Nachtmann and P.~Landshoff,
  \textit{Pomeron physics and QCD} (Cambridge Univ.\ Press, 
  2002).

\bibitem{Chew:1962eu}
  G.~F.~Chew and S.~C.~Frautschi,
  ``Regge Trajectories And The Principle Of Maximum Strength For Strong Interactions,''
  Phys.\ Rev.\ Lett.\  {\bf 8 } (1962)  41.

\bibitem{Dai:1989ua}
  J.~Dai, R.~G.~Leigh and J.~Polchinski,
  ``New Connections Between String Theories,''
  Mod.\ Phys.\ Lett.\ A {\bf 4} (1989) 2073.

\bibitem{Polchinski:1995mt}
  J.~Polchinski,
  ``Dirichlet Branes and Ramond-Ramond charges,''
  Phys.\ Rev.\ Lett.\  {\bf 75} (1995) 4724
  [hep-th/9510017].

\bibitem{Sakai:2004cn}
  T.~Sakai and S.~Sugimoto,
  ``Low energy hadron physics in holographic QCD,''
  Prog.\ Theor.\ Phys.\  {\bf 113} (2005) 843
  [hep-th/0412141].
  
\bibitem{Freedman:2012zz}
  D.~Z.~Freedman and A.~Van Proeyen, \textit{Supergravity} (Cambridge Univ.\ Press, 2012).

\bibitem{Polchinski:1992ed}
  J.~Polchinski,
  ``Effective field theory and the Fermi surface,'' TASI1992
  [hep-th/9210046].

\bibitem{Polyakov:1981rd}
  A.~M.~Polyakov,
  ``Quantum Geometry of Bosonic Strings,''
  Phys.\ Lett.\ B {\bf 103} (1981) 207.
 
\bibitem{Natsuume:1992ky}
  M.~Natsuume,
  ``Nonlinear sigma model for string solitons,''
  Phys.\ Rev.\ D {\bf 48} (1993) 835
  [hep-th/9206062].

\bibitem{Srednicki:2007qs2}
  M.~Srednicki,
  \textit{Quantum field theory}
  (Cambridge Univ.\ Press, 2007).

\bibitem{Polchinski:1987dy}
  J.~Polchinski,
  ``Scale and conformal invariance in quantum field theory,''
  Nucl.\ Phys.\  B {\bf 303} (1988) 226.

\bibitem{Gubser:1998bc}
  S.~S.~Gubser, I.~R.~Klebanov and A.~M.~Polyakov,
  ``Gauge theory correlators from noncritical string theory,''
  Phys.\ Lett.\  {\bf B428 } (1998)  105
  [hep-th/9802109].
  
\bibitem{Witten:1998qj}
  E.~Witten,
  ``Anti-de Sitter space and holography,''
  Adv.\ Theor.\ Math.\ Phys.\  {\bf 2 } (1998)  253
  [hep-th/9802150].

\bibitem{Kiritsis:2007zz2}
  E.~Kiritsis,
  \textit{String theory in a nutshell}
  (Princeton Univ.\ Press, 2007).

\bibitem{Wald:1984rg2}
R.~M.~Wald,
\textit{General relativity}
(The Univ.\ of Chicago Press, 1984).

\bibitem{Baier:2007ix}
  R.~Baier, P.~Romatschke, D.~T.~Son, A.~O.~Starinets and M.~A.~Stephanov,
  ``Relativistic viscous hydrodynamics, conformal invariance, and holography,''
  JHEP {\bf 0804} (2008) 100
  [arXiv:0712.2451 [hep-th]].

\bibitem{Horowitz:1991cd}
  G.~T.~Horowitz and A.~Strominger,
  ``Black strings and P-branes,''
  Nucl.\ Phys.\ B {\bf 360} (1991) 197.

\end{thebibliography}

\begin{thebibliography}{99}

%
%

\bibitem{York:1986it}
  J.~W.~York, Jr.,
  ``Black hole thermodynamics and the Euclidean Einstein action,''
  Phys.\ Rev.\  {\bf D33 } (1986)  2092.
  
\bibitem{Gubser:1996de}
  S.~S.~Gubser, I.~R.~Klebanov and A.~W.~Peet,
  ``Entropy and temperature of black 3-branes,''
  Phys.\ Rev.\  {\bf D54 } (1996)  3915
  [hep-th/9602135]. 
 
\bibitem{Balasubramanian:1999re}
  V.~Balasubramanian and P.~Kraus,
  ``A Stress tensor for Anti-de Sitter gravity,''
  Commun.\ Math.\ Phys.\  {\bf 208 } (1999)  413 
  [hep-th/9902121].
  
\bibitem{Skenderis:2002wp}
  K.~Skenderis,
  ``Lecture notes on holographic renormalization,''
  Class.\ Quant.\ Grav.\  {\bf 19} (2002) 5849
  [hep-th/0209067].

\end{thebibliography}

\begin{thebibliography}{99}

%
%

\bibitem{Susskind:1998dq}
  L.~Susskind and E.~Witten,
  ``The Holographic bound in anti-de Sitter space,''
  hep-th/9805114.
  
\bibitem{Peet:1998wn}
  A.~W.~Peet and J.~Polchinski,
  ``UV / IR relations in AdS dynamics,''
  Phys.\ Rev.\ D {\bf 59} (1999) 065011
  [hep-th/9809022].


\bibitem{Rey:1998ik}
  S.~-J.~Rey and J.~-T.~Yee,
  ``Macroscopic strings as heavy quarks in large N gauge theory and anti-de Sitter supergravity,''
  Eur.\ Phys.\ J.\ C {\bf 22} (2001) 379
  [hep-th/9803001].

\bibitem{Maldacena:1998im}
  J.~M.~Maldacena,
  ``Wilson loops in large N field theories,''
  Phys.\ Rev.\ Lett.\  {\bf 80} (1998) 4859
  [hep-th/9803002].


\bibitem{Erickson:2000af}
  J.~K.~Erickson, G.~W.~Semenoff and K.~Zarembo,
  ``Wilson loops in N=4 supersymmetric Yang-Mills theory,''
  Nucl.\ Phys.\ B {\bf 582} (2000) 155
  [hep-th/0003055].

\bibitem{Drukker:2000rr}
  N.~Drukker and D.~J.~Gross,
  ``An Exact prediction of N=4 SUSYM theory for string theory,''
  J.\ Math.\ Phys.\  {\bf 42} (2001) 2896
  [hep-th/0010274].
  
 
\bibitem{Polchinski:2001tt}
  J.~Polchinski and M.~J.~Strassler,
  ``Hard scattering and gauge/string duality,''
  Phys.\ Rev.\ Lett.\  {\bf 88} (2002) 031601
  [hep-th/0109174].


\bibitem{Rey:1998bq}
  S.~-J.~Rey, S.~Theisen and J.~-T.~Yee,
  ``Wilson-Polyakov loop at finite temperature in large N gauge theory and anti-de Sitter supergravity,''
  Nucl.\ Phys.\ B {\bf 527} (1998) 171
  [hep-th/9803135].
  
\bibitem{Brandhuber:1998bs}
  A.~Brandhuber, N.~Itzhaki, J.~Sonnenschein and S.~Yankielowicz,
  ``Wilson loops in the large N limit at finite temperature,''
  Phys.\ Lett.\ B {\bf 434} (1998) 36
  [hep-th/9803137].

\bibitem{Bak:2007fk}
  D.~Bak, A.~Karch and L.~G.~Yaffe,
  ``Debye screening in strongly coupled N=4 supersymmetric Yang-Mills plasma,''
  JHEP {\bf 0708} (2007) 049
  [arXiv:0705.0994 [hep-th]].


\bibitem{Matsui:1986dk}
  T.~Matsui and H.~Satz,
  ``$J/\Psi$ Suppression by Quark-Gluon Plasma Formation,''
  Phys.\ Lett.\  {\bf B178 } (1986)  416.


\bibitem{Liu:2006ug}
  H.~Liu, K.~Rajagopal and U.~A.~Wiedemann,
  ``Calculating the jet quenching parameter from AdS/CFT,''
  Phys.\ Rev.\ Lett.\  {\bf 97} (2006) 182301
  [hep-ph/0605178].

\bibitem{Herzog:2006gh}
  C.~P.~Herzog, A.~Karch, P.~Kovtun, C.~Kozcaz and L.~G.~Yaffe,
  ``Energy loss of a heavy quark moving through N=4 supersymmetric Yang-Mills plasma,''
  JHEP {\bf 0607} (2006) 013
  [hep-th/0605158].

\bibitem{CasalderreySolana:2006rq}
  J.~Casalderrey-Solana and D.~Teaney,
  ``Heavy quark diffusion in strongly coupled N=4 Yang-Mills,''
  Phys.\ Rev.\ D {\bf 74} (2006) 085012
  [hep-ph/0605199].

\bibitem{Gubser:2006bz}
  S.~S.~Gubser,
  ``Drag force in AdS/CFT,''
  Phys.\ Rev.\ D {\bf 74} (2006) 126005
  [hep-th/0605182].

\bibitem{Gubser:2008as}
  S.~S.~Gubser, D.~R.~Gulotta, S.~S.~Pufu and F.~D.~Rocha,
  ``Gluon energy loss in the gauge-string duality,''
  JHEP {\bf 0810} (2008) 052
  [arXiv:0803.1470 [hep-th]].

\bibitem{Chesler:2008wd}
  P.~M.~Chesler, K.~Jensen and A.~Karch,
  ``Jets in strongly-coupled N = 4 super Yang-Mills theory,''
  Phys.\ Rev.\ D {\bf 79} (2009) 025021
  [arXiv:0804.3110 [hep-th]].

\bibitem{Drukker:1999zq}
  N.~Drukker, D.~J.~Gross and H.~Ooguri,
  ``Wilson loops and minimal surfaces,''
  Phys.\ Rev.\ D {\bf 60} (1999) 125006
  [hep-th/9904191].

\bibitem{Horowitz:1998ha}
  G.~T.~Horowitz and R.~C.~Myers,
  ``The AdS / CFT correspondence and a new positive energy conjecture for general relativity,''
  Phys.\ Rev.\ D {\bf 59} (1998) 026005
  [hep-th/9808079].

\end{thebibliography}

\begin{thebibliography}{99}

%
%

\bibitem{callen}
H.~B.~Callen, \textit{Thermodynamics and an introduction to thermostatistics} (John Wiley \& Sons, 1985).


\bibitem{fluid_text}
L.~D.~Landau and E.~M.~Lifshytz, \textit{Fluid mechanics, second edition} (Elsevier, 1987).


\end{thebibliography}

\begin{thebibliography}{99}

%
%
 
\bibitem{Gubser:1998bc2}
  S.~S.~Gubser, I.~R.~Klebanov and A.~M.~Polyakov,
  ``Gauge theory correlators from noncritical string theory,''
  Phys.\ Lett.\  {\bf B428 } (1998)  105
  [hep-th/9802109].
  
\bibitem{Witten:1998qj2}
  E.~Witten,
  ``Anti-de Sitter space and holography,''
  Adv.\ Theor.\ Math.\ Phys.\  {\bf 2 } (1998)  253
  [hep-th/9802150].
  
\bibitem{Breitenlohner:1982bm}
  P.~Breitenlohner and D.~Z.~Freedman,
  ``Positive Energy in anti-De Sitter Backgrounds and Gauged Extended Supergravity,''
  Phys.\ Lett.\ B {\bf 115} (1982) 197.
  

  
\bibitem{Son:2002sd}
  D.~T.~Son and A.~O.~Starinets,
  ``Minkowski-space correlators in AdS/CFT correspondence: Recipe and
  applications,''
  JHEP {\bf 0209} (2002) 042
  [arXiv:hep-th/0205051].
  
\bibitem{Herzog:2002pc}
  C.~P.~Herzog and D.~T.~Son,
  ``Schwinger-Keldysh propagators from AdS/CFT correspondence,''
  JHEP {\bf 0303} (2003) 046
  [arXiv:hep-th/0212072].

\bibitem{Skenderis:2008dh}
  K.~Skenderis and B.~C.~van Rees,
  ``Real-time gauge/gravity duality,''
  Phys.\ Rev.\ Lett.\  {\bf 101} (2008) 081601
  [arXiv:0805.0150 [hep-th]].

\bibitem{Klebanov:1999tb}
  I.~R.~Klebanov and E.~Witten,
  ``AdS/CFT correspondence and symmetry breaking,''
  Nucl.\ Phys.\  B {\bf 556} (1999) 89
  [arXiv:hep-th/9905104].

\end{thebibliography}

\begin{thebibliography}{99}

%
%

  
\bibitem{Behrndt:1998jd}
  K.~Behrndt, M.~Cvetic and W.~A.~Sabra,
  ``Nonextreme black holes of five-dimensional N=2 AdS supergravity,''
  Nucl.\ Phys.\ B {\bf 553} (1999) 317
  [hep-th/9810227].

\bibitem{Kraus:1998hv}
  P.~Kraus, F.~Larsen and S.~P.~Trivedi,
  ``The Coulomb branch of gauge theory from rotating branes,''
  JHEP {\bf 9903} (1999) 003
  [hep-th/9811120].

\bibitem{Cvetic:1999xp}
  M.~Cvetic {\it et al.},
  ``Embedding AdS black holes in ten-dimensions and eleven-dimensions,''
  Nucl.\ Phys.\ B {\bf 558} (1999) 96
  [hep-th/9903214].

\bibitem{Bekenstein:1971hc}
  J.~D.~Bekenstein,
  ``Nonexistence of baryon number for static black holes,''
  Phys.\ Rev.\ D {\bf 5} (1972) 1239.

\bibitem{Cai:1998ji}
  R.~-G.~Cai and K.~-S.~Soh,
  ``Critical behavior in the rotating D-branes,''
  Mod.\ Phys.\ Lett.\ A {\bf 14} (1999) 1895
  [hep-th/9812121].

\bibitem{Cvetic:1999ne}
  M.~Cvetic and S.~S.~Gubser,
  ``Phases of R charged black holes, spinning branes and strongly coupled gauge theories,''
  JHEP {\bf 9904} (1999) 024
  [hep-th/9902195].


\bibitem{Marolf:2011zs}
  D.~Marolf,
  ``Black holes and branes in supergravity,''
  arXiv:1107.1022 [hep-th].
  
\bibitem{Aharony:2008ug}
  O.~Aharony, O.~Bergman, D.~L.~Jafferis and J.~Maldacena,
  ``N=6 superconformal Chern-Simons-matter theories, M2-branes and their gravity duals,''
  JHEP {\bf 0810} (2008) 091
  [arXiv:0806.1218 [hep-th]].
  
\bibitem{Drukker:2010nc}
  N.~Drukker, M.~Marino and P.~Putrov,
  ``From weak to strong coupling in ABJM theory,''
  Commun.\ Math.\ Phys.\  {\bf 306} (2011) 511
  [arXiv:1007.3837 [hep-th]].

\bibitem{Itzhaki:1998dd}
  N.~Itzhaki, J.~M.~Maldacena, J.~Sonnenschein and S.~Yankielowicz,
  ``Supergravity and the large N limit of theories with sixteen supercharges,''
  Phys.\ Rev.\ D {\bf 58} (1998) 046004
  [hep-th/9802042].

\bibitem{Huijse:2011ef}
  L.~Huijse, S.~Sachdev and B.~Swingle,
  ``Hidden Fermi surfaces in compressible states of gauge-gravity duality,''
  Phys.\ Rev.\ B {\bf 85} (2012) 035121
  [arXiv:1112.0573 [cond-mat.str-el]].


\bibitem{Pilch:2000ue}
  K.~Pilch and N.~P.~Warner,
  ``N=2 supersymmetric RG flows and the IIB dilaton,''
  Nucl.\ Phys.\ B {\bf 594} (2001) 209
  [hep-th/0004063].
  
\bibitem{Buchel:2003ah}
  A.~Buchel and J.~T.~Liu,
  ``Thermodynamics of the N = 2* flow,''
  JHEP {\bf 0311} (2003) 031
  [arXiv:hep-th/0305064].

  
\bibitem{Polchinski:2000uf}
  J.~Polchinski and M.~J.~Strassler,
  ``The string dual of a confining four-dimensional gauge theory,''
  hep-th/0003136.


  
\bibitem{Klebanov:2000hb}
  I.~R.~Klebanov and M.~J.~Strassler,
  ``Supergravity and a confining gauge theory: Duality cascades and $\chi$SB-resolution of naked singularities,''
  JHEP {\bf 0008} (2000) 052
  [hep-th/0007191].



\bibitem{Gubser:2001ri}
  S.~S.~Gubser, C.~P.~Herzog, I.~R.~Klebanov and A.~A.~Tseytlin,
  ``Restoration of chiral symmetry: A supergravity perspective,''
  JHEP {\bf 0105} (2001) 028
  [arXiv:hep-th/0102172].





  
\bibitem{Kachru:2008yh}
  S.~Kachru, X.~Liu and M.~Mulligan,
  ``Gravity Duals of Lifshitz-like Fixed Points,''
  Phys.\ Rev.\ D {\bf 78} (2008) 106005
  [arXiv:0808.1725 [hep-th]].


\bibitem{Spradlin:2001pw}
  M.~Spradlin, A.~Strominger and A.~Volovich,
  ``Les Houches lectures on de Sitter space,''
  hep-th/0110007.

\bibitem{Bredberg:2011hp}
  I.~Bredberg, C.~Keeler, V.~Lysov and A.~Strominger,
  ``Cargese Lectures on the Kerr/CFT Correspondence,''
  Nucl.\ Phys.\ Proc.\ Suppl.\  {\bf 216} (2011) 194
  [arXiv:1103.2355 [hep-th]].

\bibitem{Bredberg:2010ky}
  I.~Bredberg, C.~Keeler, V.~Lysov and A.~Strominger,
  ``Wilsonian Approach to Fluid/Gravity Duality,''
  JHEP {\bf 1103} (2011) 141
  [arXiv:1006.1902 [hep-th]].

\bibitem{Klebanov:2002ja}
  I.~R.~Klebanov and A.~M.~Polyakov,
  ``AdS dual of the critical O(N) vector model,''
  Phys.\ Lett.\ B {\bf 550} (2002) 213
  [hep-th/0210114].



\bibitem{Son:2009tf}
  D.~T.~Son and P.~Surowka,
  ``Hydrodynamics with Triangle Anomalies,''
  Phys.\ Rev.\ Lett.\  {\bf 103} (2009) 191601
  [arXiv:0906.5044 [hep-th]].
  
\bibitem{Nakamura:2009tf}
  S.~Nakamura, H.~Ooguri and C.~-S.~Park,
  ``Gravity Dual of Spatially Modulated Phase,''
  Phys.\ Rev.\ D {\bf 81} (2010) 044018
  [arXiv:0911.0679 [hep-th]].

\end{thebibliography}

\begin{thebibliography}{99}

%
%

\bibitem{nist}
\url{http://webbook.nist.gov}


\bibitem{Natsuume:2010ky}
  M.~Natsuume and M.~Ohta,
  ``The shear viscosity of holographic superfluids,''
  Prog.\ Theor.\ Phys.\  {\bf 124 } (2010)  931
  [arXiv:1008.4142 [hep-th]].

\bibitem{Erdmenger:2010xm}
  J.~Erdmenger, P.~Kerner and H.~Zeller,
  ``Non-universal shear viscosity from Einstein gravity,''
  Phys.\ Lett.\  {\bf B699 } (2011)  301
  [arXiv:1011.5912 [hep-th]].
  

\bibitem{Jeon:1994if}
  S.~Jeon,
  ``Hydrodynamic transport coefficients in relativistic scalar field theory,''
  Phys.\ Rev.\ D {\bf 52} (1995) 3591
  [hep-ph/9409250].

\bibitem{Jeon:1995zm}
  S.~Jeon and L.~G.~Yaffe,
  ``From quantum field theory to hydrodynamics: Transport coefficients and effective kinetic theory,''
  Phys.\ Rev.\ D {\bf 53} (1996) 5799
  [hep-ph/9512263].
  
\bibitem{Arnold:2000dr}
  P.~B.~Arnold, G.~D.~Moore and L.~G.~Yaffe,
  ``Transport coefficients in high temperature gauge theories. 1. Leading log results,''
  JHEP {\bf 0011} (2000) 001
  [hep-ph/0010177].

\bibitem{Huot:2006ys}
  S.~C.~Huot, S.~Jeon and G.~D.~Moore,
  ``Shear viscosity in weakly coupled N = 4 super Yang-Mills theory compared to QCD,''
  Phys.\ Rev.\ Lett.\  {\bf 98} (2007) 172303
  [hep-ph/0608062].
  


\bibitem{Policastro:2001yc}
  G.~Policastro, D.~T.~Son and A.~O.~Starinets,
  ``The shear viscosity of strongly coupled N = 4 supersymmetric Yang-Mills
  plasma,''
  Phys.\ Rev.\ Lett.\  {\bf 87} (2001) 081601
  [arXiv:hep-th/0104066].

\bibitem{Policastro:2002se}
  G.~Policastro, D.~T.~Son and A.~O.~Starinets,
  ``From AdS/CFT correspondence to hydrodynamics,''
  JHEP {\bf 0209} (2002) 043
  [arXiv:hep-th/0205052].
  

\bibitem{Kovtun:2003wp}
  P.~Kovtun, D.~T.~Son and A.~O.~Starinets,
  ``Holography and hydrodynamics: Diffusion on stretched horizons,''
  JHEP {\bf 0310} (2003) 064
  [arXiv:hep-th/0309213].

\bibitem{Buchel:2003tz}
  A.~Buchel and J.~T.~Liu,
  ``Universality of the shear viscosity in supergravity,''
  Phys.\ Rev.\ Lett.\  {\bf 93} (2004) 090602
  [arXiv:hep-th/0311175].
  

\bibitem{Herzog:2002fn}
  C.~P.~Herzog,
  ``The hydrodynamics of M-theory,''
  JHEP {\bf 0212} (2002) 026
  [arXiv:hep-th/0210126].

  
\bibitem{Mas:2006dy}
  J.~Mas,
  ``Shear viscosity from R-charged AdS black holes,''
  JHEP {\bf 0603} (2006) 016
  [arXiv:hep-th/0601144].

\bibitem{Son:2006em}
  D.~T.~Son and A.~O.~Starinets,
  ``Hydrodynamics of R-charged black holes,''
  JHEP {\bf 0603} (2006) 052
  [arXiv:hep-th/0601157].

\bibitem{Saremi:2006ep}
  O.~Saremi,
  ``The Viscosity bound conjecture and hydrodynamics of M2-brane theory at finite chemical potential,''
  JHEP {\bf 0610} (2006) 083
  [hep-th/0601159].

\bibitem{Maeda:2006by}
  K.~Maeda, M.~Natsuume and T.~Okamura,
  ``Viscosity of gauge theory plasma with a chemical potential from AdS/CFT correspondence,''
  Phys.\ Rev.\ D {\bf 73} (2006) 066013
  [arXiv:hep-th/0602010].

\bibitem{Benincasa:2006fu}
  P.~Benincasa, A.~Buchel and R.~Naryshkin,
  ``The Shear viscosity of gauge theory plasma with chemical potentials,''
  Phys.\ Lett.\ B {\bf 645} (2007) 309
  [hep-th/0610145].


\bibitem{Mateos:2006yd}
  D.~Mateos, R.~C.~Myers and R.~M.~Thomson,
  ``Holographic viscosity of fundamental matter,''
  Phys.\ Rev.\ Lett.\  {\bf 98} (2007) 101601
  [hep-th/0610184].

\bibitem{Karch:2002sh}
  A.~Karch and E.~Katz,
  ``Adding flavor to AdS/CFT,''
  JHEP {\bf 0206} (2002) 043
  [arXiv:hep-th/0205236].


\bibitem{Janik:2006ft}
  R.~A.~Janik,
  ``Viscous plasma evolution from gravity using AdS/CFT,''
  Phys.\ Rev.\ Lett.\  {\bf 98} (2007) 022302
  [hep-th/0610144].

  

    

  

\bibitem{Song:2010mg}
  H.~Song, S.~A.~Bass, U.~Heinz, T.~Hirano and C.~Shen,
  ``200 A GeV Au+Au collisions serve a nearly perfect quark-gluon liquid,''
  Phys.\ Rev.\ Lett.\  {\bf 106} (2011) 192301
   [Erratum-ibid.\  {\bf 109} (2012) 139904]
  [arXiv:1011.2783 [nucl-th]].

\bibitem{Song:2012ua}
  H.~Song,
  ``QGP viscosity at RHIC and the LHC - a 2012 status report,''
  Nucl.\ Phys.\ A {\bf 904-905} (2013) 114c
  [arXiv:1210.5778 [nucl-th]].
  

\bibitem{Meyer:2007ic}
  H.~B.~Meyer,
  ``A calculation of the shear viscosity in SU(3) gluodynamics,''
  Phys.\ Rev.\  D {\bf 76} (2007) 101701
  [arXiv:0704.1801 [hep-lat]].

\bibitem{Nakamura:2004sy}
  A.~Nakamura and S.~Sakai,
  ``Transport coefficients of gluon plasma,''
  Phys.\ Rev.\ Lett.\  {\bf 94} (2005) 072305
  [arXiv:hep-lat/0406009].
  
\bibitem{Karsch:2006sf}
  F.~Karsch,
  ``Properties of the quark gluon plasma: A lattice perspective,''
  Nucl.\ Phys.\  A {\bf 783} (2007) 13
  [arXiv:hep-ph/0610024].

\bibitem{Bazavov:2009zn}
  A.~Bazavov, 
  {\it et al.},
  ``Equation of state and QCD transition at finite temperature,''
  Phys.\ Rev.\ D {\bf 80} (2009) 014504
  [arXiv:0903.4379 [hep-lat]].

\bibitem{Nishioka:2007zz}
  T.~Nishioka and T.~Takayanagi,
  ``Free Yang-Mills versus toric Sasaki-Einstein,''
  Phys.\ Rev.\ D {\bf 76} (2007) 044004
  [hep-th/0702194].
  
\bibitem{Kovtun:2004de}
  P.~Kovtun, D.~T.~Son and A.~O.~Starinets,
  ``Viscosity in strongly interacting quantum field theories from black hole
  physics,''
  Phys.\ Rev.\ Lett.\  {\bf 94} (2005) 111601
  [arXiv:hep-th/0405231].

\bibitem{Cremonini:2011iq}
  S.~Cremonini,
  ``The Shear Viscosity to Entropy Ratio: A Status Report,''
  Mod.\ Phys.\ Lett.\ B {\bf 25} (2011) 1867
  [arXiv:1108.0677 [hep-th]].


\bibitem{Buchel:2004di}
  A.~Buchel, J.~T.~Liu and A.~O.~Starinets,
  ``Coupling constant dependence of the shear viscosity in N=4 supersymmetric Yang-Mills theory,''
  Nucl.\ Phys.\  {\bf B707 } (2005)  56
  [hep-th/0406264].

\bibitem{Buchel:2008sh}
  A.~Buchel,
  ``Resolving disagreement for $\eta/s$ in a CFT plasma at finite coupling,''
  Nucl.\ Phys.\  {\bf B803 } (2008)  166
  [arXiv:0805.2683 [hep-th]].

\bibitem{Brigante:2007nu}
  M.~Brigante, H.~Liu, R.~C.~Myers, S.~Shenker and S.~Yaida,
  ``Viscosity Bound Violation in Higher Derivative Gravity,''
  Phys.\ Rev.\  D {\bf 77} (2008) 126006
  [arXiv:0712.0805 [hep-th]].


\bibitem{Kovtun:2003vj}
  P.~Kovtun and L.~G.~Yaffe,
  ``Hydrodynamic fluctuations, long-time tails, and supersymmetry,''
  Phys.\ Rev.\  D {\bf 68} (2003) 025007
  [arXiv:hep-th/0303010].

\bibitem{CaronHuot:2009iq}
  S.~Caron-Huot and O.~Saremi,
  ``Hydrodynamic Long-Time tails From Anti de Sitter Space,''
  JHEP {\bf 1011 } (2010)  013
  [arXiv:0909.4525 [hep-th]].

\bibitem{Anninos:2010sq}
  D.~Anninos, S.~A.~Hartnoll and N.~Iqbal,
  ``Holography and the Coleman-Mermin-Wagner theorem,''
  Phys.\ Rev.\  {\bf D82 } (2010)  066008
  [arXiv:1005.1973 [hep-th]].
  
\bibitem{Natsuume:2010bs}
  M.~Natsuume and T.~Okamura,
  ``Dynamic universality class of large-$N$ gauge theories,''
  Phys.\ Rev.\  {\bf D83 } (2011)  046008
  [arXiv:1012.0575 [hep-th]].


\bibitem{muller}
I. M\"{u}ller, 
``Zum Paradoxon der W\"{a}rmeleitungstheorie," 
Z.\ Phys.\ {\bf 198} (1967) 329. 

\bibitem{Israel:1976tn}
  W.~Israel,
  ``Nonstationary Irreversible Thermodynamics: A Causal Relativistic Theory,''
  Annals Phys.\  {\bf 100} (1976) 310.

\bibitem{Israel:1979wp}
  W.~Israel and J.~M.~Stewart,
  ``Transient relativistic thermodynamics and kinetic theory,''
  Annals Phys.\  {\bf 118} (1979) 341.

\bibitem{Baier:2007ix2}
  R.~Baier, P.~Romatschke, D.~T.~Son, A.~O.~Starinets and M.~A.~Stephanov,
  ``Relativistic viscous hydrodynamics, conformal invariance, and holography,''
  JHEP {\bf 0804} (2008) 100
  [arXiv:0712.2451 [hep-th]].

\bibitem{Bhattacharyya:2008jc}
  S.~Bhattacharyya, V.~E.~Hubeny, S.~Minwalla and M.~Rangamani,
  ``Nonlinear Fluid Dynamics from Gravity,''
  JHEP {\bf 0802} (2008) 045
  [arXiv:0712.2456 [hep-th]].

\bibitem{Natsuume:2007ty}
  M.~Natsuume and T.~Okamura,
  ``Causal hydrodynamics of gauge theory plasmas from AdS/CFT duality,''
  Phys.\ Rev.\  D {\bf 77} (2008) 066014
  [arXiv:0712.2916 [hep-th]].


\bibitem{hiscock_lindblom2}
  W.~A.~Hiscock and L.~Lindblom,
  ``Generic instabilities in first-order dissipative relativistic fluid theories,"
  Phys.\ Rev.\ D {\bf 31} (1985) 725.

\bibitem{hiscock_lindblom3}
  W.~A.~Hiscock and L.~Lindblom,
  ``Linear plane waves in dissipative relativistic fluids,"
  Phys.\ Rev.\ D {\bf 35} (1987) 3723.


   

\bibitem{Horowitz:1999jd}
  G.~T.~Horowitz and V.~E.~Hubeny,
  ``Quasinormal modes of AdS black holes and the approach to thermal equilibrium,''
  Phys.\ Rev.\ D {\bf 62} (2000) 024027
  [hep-th/9909056].

\end{thebibliography}

\begin{thebibliography}{99}

%
%

\bibitem{critical_text1}
J.~Cardy, \textit{Scaling and renormalization in statistical physics} (Cambridge Univ.\ Press, 1996).

\bibitem{critical_text2}
H.~Nishimori and G.~Ortiz, 
\textit{Elements of Phase Transitions and Critical Phenomena} (Oxford Univ.\ Press, 2011).

\bibitem{Coleman:1973ci}
  S.~R.~Coleman,
  ``There are no Goldstone bosons in two-dimensions,''
  Commun.\ Math.\ Phys.\  {\bf 31} (1973) 259.

\bibitem{Mermin:1966fe}
  N.~D.~Mermin and H.~Wagner,
  ``Absence of ferromagnetism or antiferromagnetism in one-dimensional or
  two-dimensional isotropic Heisenberg models,''
  Phys.\ Rev.\ Lett.\  {\bf 17} (1966) 1133.

\bibitem{hohenberg_halperin}
P.~C.~Hohenberg and B.~I.~Halperin, 
``Theory of dynamic critical phenomena," 
Rev.\ Mod.\ Phys.\  {\bf 49} (1977) 435.

\bibitem{Maeda:2008hn}
  K.~Maeda, M.~Natsuume and T.~Okamura,
  ``Dynamic critical phenomena in the AdS/CFT duality,''
  Phys.\ Rev.\ D {\bf 78} (2008) 106007
  [arXiv:0809.4074 [hep-th]].

\bibitem{Polchinski:1992ed2}
  J.~Polchinski,
  ``Effective field theory and the Fermi surface,'' TASI1992
  [hep-th/9210046].

\bibitem{tinkham}
M.~Tinkham, {\it Introduction to superconductivity: second edition} (Dover Publications, 2004). 

\bibitem{hussey}
N.~E.~Hussey, 
``Phenomenology of the normal state in-plane transport properties of high-$T_c$ cuprates,"
J.\ Phys: Condens.\ Matter {\bf 20} (2008) 123201
[arXiv:0804.2984 [cond-mat.supr-con]].

\end{thebibliography}

\begin{thebibliography}{99}

%
%


\bibitem{Hawking:1982dh}
  S.~W.~Hawking and D.~N.~Page,
  ``Thermodynamics of Black Holes in anti-De Sitter Space,''
  Commun.\ Math.\ Phys.\  {\bf 87 } (1983)  577.

\bibitem{Witten:1998zw}
E.~Witten,
``Anti-de Sitter space, thermal phase transition, and confinement in  gauge theories,''
Adv.\ Theor.\ Math.\ Phys.\  {\bf 2} (1998) 505 
[arXiv:hep-th/9803131].

\bibitem{Balasubramanian:1999re2}
  V.~Balasubramanian and P.~Kraus,
  ``A Stress tensor for Anti-de Sitter gravity,''
  Commun.\ Math.\ Phys.\  {\bf 208 } (1999)  413 
  [hep-th/9902121].
  
\bibitem{BD}
  N.~D.~Birrell and P.~C.~W.~Davies,
  \textit{Quantum fields in curved space}
  (Cambridge Univ.\ Press, 1982).




  
\bibitem{Gubser:2008px}
  S.~S.~Gubser,
  ``Breaking an Abelian gauge symmetry near a black hole horizon,''
  Phys.\ Rev.\  D {\bf 78} (2008) 065034
  [arXiv:0801.2977 [hep-th]].
  
\bibitem{Hartnoll:2008vx}
  S.~A.~Hartnoll, C.~P.~Herzog and G.~T.~Horowitz,
  ``Building a Holographic Superconductor,''
  Phys.\ Rev.\ Lett.\  {\bf 101} (2008) 031601
  [arXiv:0803.3295 [hep-th]].

\bibitem{herzog}
\url{http://insti.physics.sunysb.edu/~cpherzog/superconductor/index.html}

\bibitem{Maeda:2009wv}
  K.~Maeda, M.~Natsuume and T.~Okamura,
  ``Universality class of holographic superconductors,''
  Phys.\ Rev.\  D {\bf 79} (2009) 126004
  [arXiv:0904.1914 [hep-th]].

\bibitem{Coleman:1973ci2}
  S.~R.~Coleman,
  ``There are no Goldstone bosons in two-dimensions,''
  Commun.\ Math.\ Phys.\  {\bf 31} (1973) 259.

\bibitem{Mermin:1966fe2}
  N.~D.~Mermin and H.~Wagner,
  ``Absence of ferromagnetism or antiferromagnetism in one-dimensional or
  two-dimensional isotropic Heisenberg models,''
  Phys.\ Rev.\ Lett.\  {\bf 17} (1966) 1133.
  
\bibitem{Weinberg:1986cq}
  S.~Weinberg,
  ``Superconductivity For Particular Theorists,''
  Prog.\ Theor.\ Phys.\ Suppl.\  {\bf 86} (1986) 43.


\bibitem{Nakano:2008xc}
  E.~Nakano and W.~-Y.~Wen,
  ``Critical magnetic field in a holographic superconductor,''
  Phys.\ Rev.\ D {\bf 78} (2008) 046004
  [arXiv:0804.3180 [hep-th]].

\bibitem{Albash:2008eh}
  T.~Albash and C.~V.~Johnson,
  ``A Holographic Superconductor in an External Magnetic Field,''
  JHEP {\bf 0809} (2008) 121
  [arXiv:0804.3466 [hep-th]].

\bibitem{Hartnoll:2008kx}
  S.~A.~Hartnoll, C.~P.~Herzog and G.~T.~Horowitz,
  ``Holographic Superconductors,''
  JHEP {\bf 0812} (2008) 015.
  [arXiv:0810.1563 [hep-th]].

\bibitem{Albash:2009ix}
  T.~Albash and C.~V.~Johnson,
  ``Phases of Holographic Superconductors in an External Magnetic Field,''
  arXiv:0906.0519 [hep-th].

\bibitem{Montull:2009fe}
  M.~Montull, A.~Pomarol and P.~J.~Silva,
  ``The Holographic Superconductor Vortex,''
  Phys.\ Rev.\ Lett.\  {\bf 103} (2009) 091601
  [arXiv:0906.2396 [hep-th]].

\bibitem{Maeda:2009vf}
  K.~Maeda, M.~Natsuume and T.~Okamura,
  ``Vortex lattice for a holographic superconductor,''
  Phys.\ Rev.\ D {\bf 81} (2010) 026002
  [arXiv:0910.4475 [hep-th]].


\bibitem{Herzog:2008he}
  C.~P.~Herzog, P.~K.~Kovtun and D.~T.~Son,
  ``Holographic model of superfluidity,''
  Phys.\ Rev.\ D {\bf 79} (2009) 066002
  [arXiv:0809.4870 [hep-th]].
  
\bibitem{Gubser:2008pf}
  S.~S.~Gubser and A.~Nellore,
  ``Low-temperature behavior of the Abelian Higgs model in anti-de Sitter space,''
  JHEP {\bf 0904} (2009) 008
  [arXiv:0810.4554 [hep-th]].

\bibitem{Denef:2009tp}
  F.~Denef and S.~A.~Hartnoll,
  ``Landscape of superconducting membranes,''
  Phys.\ Rev.\ D {\bf 79} (2009) 126008
  [arXiv:0901.1160 [hep-th]].


\bibitem{Gubser:2008zu}
  S.~S.~Gubser,
  ``Colorful horizons with charge in anti-de Sitter space,''
  Phys.\ Rev.\ Lett.\  {\bf 101} (2008) 191601
  [arXiv:0803.3483 [hep-th]].

\bibitem{Gubser:2008wv}
  S.~S.~Gubser and S.~S.~Pufu,
  ``The Gravity dual of a p-wave superconductor,''
  JHEP {\bf 0811} (2008) 033
  [arXiv:0805.2960 [hep-th]].

\bibitem{Benini:2010pr}
  F.~Benini, C.~P.~Herzog, R.~Rahman and A.~Yarom,
  ``Gauge gravity duality for d-wave superconductors: prospects and challenges,''
  JHEP {\bf 1011} (2010) 137
  [arXiv:1007.1981 [hep-th]].

\bibitem{Hartman:2010fk}
  T.~Hartman and S.~A.~Hartnoll,
  ``Cooper pairing near charged black holes,''
  JHEP {\bf 1006} (2010) 005
  [arXiv:1003.1918 [hep-th]].


 

    
  
\bibitem{Faulkner:2010da}
  T.~Faulkner, N.~Iqbal, H.~Liu, J.~McGreevy and D.~Vegh,
  ``From Black Holes to Strange Metals,''
  arXiv:1003.1728 [hep-th].

\end{thebibliography}

\begin{thebibliography}{99}

%
%

\bibitem{Rey:1998bq2}
  S.~-J.~Rey, S.~Theisen and J.~-T.~Yee,
  ``Wilson-Polyakov loop at finite temperature in large N gauge theory and anti-de Sitter supergravity,''
  Nucl.\ Phys.\ B {\bf 527} (1998) 171
  [hep-th/9803135].

\bibitem{Brandhuber:1998bs2}
  A.~Brandhuber, N.~Itzhaki, J.~Sonnenschein and S.~Yankielowicz,
  ``Wilson loops in the large N limit at finite temperature,''
  Phys.\ Lett.\ B {\bf 434} (1998) 36
  [hep-th/9803137].
  
\bibitem{Herzog:2006gh2}
  C.~P.~Herzog, A.~Karch, P.~Kovtun, C.~Kozcaz and L.~G.~Yaffe,
  ``Energy loss of a heavy quark moving through N=4 supersymmetric Yang-Mills plasma,''
  JHEP {\bf 0607} (2006) 013
  [hep-th/0605158].

\bibitem{Herzog:2006se}
  C.~P.~Herzog,
  ``Energy Loss of Heavy Quarks from Asymptotically AdS Geometries,''
  JHEP {\bf 0609} (2006) 032
  [hep-th/0605191].

\bibitem{Nishioka:2009un}
  T.~Nishioka, S.~Ryu and T.~Takayanagi,
  ``Holographic Entanglement Entropy: An Overview,''
  J.\ Phys.\ A {\bf 42} (2009) 504008
  [arXiv:0905.0932 [hep-th]].

\bibitem{Bjorken:1982qr}
  J.~D.~Bjorken,
  ``Highly Relativistic Nucleus-Nucleus Collisions: The Central Rapidity Region,''
  Phys.\ Rev.\ D {\bf 27} (1983) 140.
  
\bibitem{Kachru:2008yh2}
  S.~Kachru, X.~Liu and M.~Mulligan,
  ``Gravity Duals of Lifshitz-like Fixed Points,''
  Phys.\ Rev.\ D {\bf 78} (2008) 106005
  [arXiv:0808.1725 [hep-th]].

\bibitem{Taylor:2015glc}
  M.~Taylor,
  ``Lifshitz holography,''
  Class.\ Quant.\ Grav.\  {\bf 33} (2016) no.3,  033001
  [arXiv:1512.03554 [hep-th]].
  
\bibitem{Baier:2007ix3}
  R.~Baier, P.~Romatschke, D.~T.~Son, A.~O.~Starinets and M.~A.~Stephanov,
  ``Relativistic viscous hydrodynamics, conformal invariance, and holography,''
  JHEP {\bf 0804} (2008) 100
  [arXiv:0712.2451 [hep-th]].

\bibitem{Natsuume:2008iy}
  M.~Natsuume and T.~Okamura,
  ``A Note on causal hydrodynamics for M-theory branes,''
  Prog.\ Theor.\ Phys.\  {\bf 120} (2008) 1217
  [arXiv:0801.1797 [hep-th]].

\bibitem{Natsuume:2008gy}
  M.~Natsuume,
  ``Causal hydrodynamics and the membrane paradigm,''
  Phys.\ Rev.\ D {\bf 78} (2008) 066010
  [arXiv:0807.1392 [hep-th]].
  
\end{thebibliography}
\end{document}